\pdfoutput=1
\documentclass[11pt,american]{yalephd}
\usepackage[LGR,T1]{fontenc}
\usepackage[latin9]{inputenc}
\usepackage{geometry}
\usepackage{color}
\usepackage{array}
\usepackage{float}
\usepackage{booktabs}
\usepackage{units}
\usepackage{textcomp}
\usepackage{mathrsfs}
\usepackage{multirow}
\usepackage{amsmath}
\usepackage{amssymb}
\usepackage{graphicx}
\usepackage{setspace}
\usepackage{wasysym}
\usepackage{esint}
\usepackage{nomencl}
\providecommand{\printnomenclature}{\printglossary}
\providecommand{\makenomenclature}{\makeglossary}
\makenomenclature

\makeatletter

\DeclareRobustCommand{\greektext}{%
  \fontencoding{LGR}\selectfont\def\encodingdefault{LGR}}
\DeclareRobustCommand{\textgreek}[1]{\leavevmode{\greektext #1}}
\ProvideTextCommand{\~}{LGR}[1]{\char126#1}

\newcommand{\lyxmathsym}[1]{\ifmmode\begingroup\def\b@ld{bold}
  \text{\ifx\math@version\b@ld\bfseries\fi#1}\endgroup\else#1\fi}

\let\SF@@footnote\footnote
\def\footnote{\ifx\protect\@typeset@protect
    \expandafter\SF@@footnote
  \else
    \expandafter\SF@gobble@opt
  \fi
}
\expandafter\def\csname SF@gobble@opt \endcsname{\@ifnextchar[
  \SF@gobble@twobracket
  \@gobble
}
\edef\SF@gobble@opt{\noexpand\protect
  \expandafter\noexpand\csname SF@gobble@opt \endcsname}
\def\SF@gobble@twobracket[#1]#2{}
\providecommand{\tabularnewline}{\\}

\renewcommand\[{\begin{equation}} 
\renewcommand\]{\end{equation}}
\usepackage{setspace}
\usepackage[font={onehalfspacing, it}]{caption}
\usepackage[margin=10pt,font={footnotesize},labelfont=bf,labelsep=colon]{caption}
\usepackage{geometry} 
\usepackage{graphicx} 
\usepackage{dcolumn}
\usepackage{bm}
\usepackage{amsmath}
\usepackage{amsfonts}
\usepackage{amssymb}
\usepackage{appendix}
\usepackage{comment}
\usepackage{hyperref}
\hypersetup{colorlinks=true,linkcolor=black, citecolor=blue, urlcolor=cyan}
\interfootnotelinepenalty=\@M

\@ifundefined{showcaptionsetup}{}{%
 \PassOptionsToPackage{caption=false}{subfig}}
\usepackage{subfig}
\makeatother

\usepackage{babel}
\begin{document}
\title{Sub-GeV Dark Matter Searches and Electric Field Studies for the LUX and LZ Experiments} 
\author{Lucie Tvrzníková} 
\advisor{Professor Daniel N. McKinsey} 
\date{May 2019}
\frontmatter

\begin{abstract}

Abundant evidence from cosmological and astrophysical observations
suggests that the Standard Model does not describe 84\% of the matter
in our universe. The nature of this dark matter (DM) remains a mystery
since it has so far eluded detection in the laboratory. To that end,
the Large Underground Xenon (LUX) experiment was built to directly
observe the interaction of DM with xenon target nuclei. LUX acquired
data from April 2013 to May 2016 at the Sanford Underground Research
Facility (SURF) in Lead, South Dakota, which led to publications of
many world-leading exclusion limits that probe much of the unexplored
DM parameter space.

This manuscript describes two novel direct detection methods that
used the first LUX dataset to place limits on sub-GeV DM. The Bremsstrahlung
and Migdal effects consider electron recoils that accompany the standard
DM-nucleus scattering, thereby extending the reach of the LUX detector
to lower DM masses. The spin-independent DM-nucleon scattering was
constrained for four different classes of mediators for DM particles
with masses of 0.4-5~GeV/c$^{2}$.

The detector conditions changed significantly before its final 332
live-days of data acquisition. The electric fields varied in a non-trivial
non-symmetric manner, which triggered a need for a fully 3D model
of the electric fields inside the LUX detector. The successful modeling
of these electric fields, described herein, enabled a thorough understanding
of the detector throughout its scientific program and strengthened
its sensitivity to DM.

The LUX-ZEPLIN (LZ) experiment, the successor to LUX, is a next-generation
xenon detector soon to start searching for DM. However, increasingly
large noble liquid detectors like LZ are facing challenges with applications
of high voltage (HV). The Xenon Breakdown Apparatus (XeBrA) at the
Lawrence Berkeley National Laboratory was built to characterize the
HV behavior of liquid xenon and liquid argon. Results from XeBrA will
serve not only to improve our understanding of the physical processes
involved in the breakdown but also to inform the future of noble liquid
detector engineering. 

\end{abstract}

\maketitle

\thispagestyle{empty}
\makecopyright{2019}

\singlespacing

\tableofcontents{}

\chapter*{Acknowledgements}

It is impossible to acknowledge all the people to whom I owe a debt
of gratitude for what I have learned, so I would like to extend the
most heartfelt thank you to anyone whom I might have forgotten to
mention below. Thank you to everyone who has helped me better understand
how the world works and thank you to those who will do so in the future.

First and foremost, I would like to thank my advisor, Dan McKinsey.
His passion for physics, incredible breadth of knowledge, focus, and
optimism have been inspirational. I am extremely grateful for his
trust and support that enabled me to grow into a confident researcher
with a deep appreciation for the physics of dark matter and beyond.
As part of his research group, I was surrounded by fantastic people
at Yale; Ethan, Scott He., Kevin O., Markus, and Blair were all amazing
teachers, mentors, and friends. In particular, I would like to thank
Ethan for pushing me to understand physical phenomena at a fundamental
level and for being such a great mentor in the lab. My fellow grad
students provided so much support on both coasts: Evan, Elizabeth,
Nicole, and Brian, I could not have wished for better colleagues.
I would also like to acknowledge the support and camaraderie of my
cohort at Yale that made the first two years of grad schools not just
manageable, but actually enjoyable despite the incessant mountain
of homework and quals. 

I am indebted to my dissertation committee, Professors Priya Natarajan,
Meg Urry, and Steve Lamoreaux for their time and feedback.

Grad school took an unexpected turn when I traveled cross-country
to settle at UC~Berkeley. I am very thankful for our Berkeley crew
that created a new research home on the other coast - Jeremy, Kate,
Kelsey, Scott K., Vetri, Andy, Reed, Junsong, Quentin, Simon, Peter,
Kevin L., Gil, and Mike. I would also like to acknowledge the engineering
and support staff at LBNL for all their help. 

The LUX experiment was possible due to the tireless effort of the
smart and kind collaborators with whom I had the privilege to work.
Many talented people taught me so much about LUX and LZ; to mention
a few, thanks to Aaron, Curt, Scott Ha., Richard, Wing, Sally, (and
so many others!). Thank you to the many wonderful collaborators for
making my time in South Dakota enjoyable both above- and under- ground.
Thanks to Dave at SURF, and Robyn for keeping the 4850' always sunny.

I would  like to acknowledge Dan A. and Harry for their guidance,
support, and extended conversations about many delicate topics during
the early times of the LZ Equity \& Inclusion committee. And of course,
thanks to Rachel.

My quest to find dark matter started before graduate school, and I
would like to thank those who set me on the hunt for rare events in
the first place. Otokar Dragoun's excitement for neutrinos initially
sparked my interest in the world of rare event searches during the
summer of 2010 at the Czech Academy of Sciences in \v{R}ež. Ben Monreal
encouraged me to apply to Yale while I worked with him during my year
at UCSB; I hope we will cross paths again on Project 8. Finally, Alex
Murphy welcomed me into his research group at the University of Edinburgh
while I was working on my master's thesis at St Andrews. This kick-started
my journey with LUX and LZ that I was so lucky to begin.

Thank you to my wonderful New Haven roommates, Glorili and Emily for
providing sunshine and happiness during the uncertain days of grad
school. Also, thanks to everyone at CrossFit 03/HCC and Grassroots
CrossFit for keeping me sane during the peak of my experimental days.

I want to thank my family who supported me throughout all of my life,
including the 22 years spent in school. Thank you for your encouragements,
and thanks Dad for instilling my interest in learning and understanding.
Finally, thank you, Geoff, for being there with and for me throughout
this endeavor. Your company, wisdom, support, (math skills), willingness
to listen; I am endlessly thankful for all of it.

\listoffigures

\listoftables

\printnomenclature[2cm]{}

\pagebreak{}

\doublespacing 
\mainmatter

\chapter{\textsc{Quest for the missing mass}}

The missing mass in our universe is one of the critical mysteries
of contemporary physics. Despite the success of the Standard Model
(SM) of particle physics developed in the 1970's, the nature of 84\%
of the matter remains a mystery. The existence of dark matter (DM)
was first postulated in 1933 by Fritz Zwicky~\cite{Zwicky1933} to
account for the missing mass in orbital velocities of galaxy clusters.
A satisfactory explanation of this missing matter requires the existence
of a ``dark'' particle lacking electromagnetic interactions. Nowadays,
DM enjoys plenty of attention from both theoretical and experimental
physicists. There is abundant evidence for DM from cosmological and
astrophysical observations, but it keeps eluding detection, which
is the focus of this dissertation. 

This chapter first presents an overview of the evidence supporting
the existence of dark matter, along with DM candidates, and the possible
detection methods with a focus on direct detection. The remainder
of this work then discusses techniques developed in support of the
quest for dark matter.

\section{Evidence for dark matter}

The amount of observational evidence for the existence of DM has steadily
increased over the last several decades. This section reviews observational
studies of the composition of our universe pointing toward the evidence
of dark matter\footnote{Despite DM being the most common explanation for the observations
described in this chapter, there are efforts to use modified gravity
to explain these observations~\cite{modifiedGravity,Bellini:2015xja}.}. The evidence comes at various scales: cosmological (Section~\ref{subsec:Cosmological-scale}),
the scale of clusters (Section~\ref{subsec:Gravitational-lensing}),
or at the local galactic scale (Section~\ref{subsec:Rotational-velocity-of}). 

\subsection{Cosmological scale \label{subsec:Cosmological-scale}}

The standard Big Bang model is based on three fundamental astronomical
observations: the homogeneity and isotropy of the universe on large
scales, the blackbody spectrum of the cosmic microwave background
(CMB)\nomenclature{CMB}{Cosmic Microwave Background} radiation and
the Hubble law. Based on those observations, the Friedmann-Lemaitre-Robertson-Walker
(FLRW)\nomenclature{FLRW}{Friedmann-Lemaitre-Roberson-Walker} metric
can be used to describe our universe in comoving spherical coordinates
$\left(r,\theta,\phi\right)$:
\begin{equation}
ds^{2}=dt^{2}-a^{2}(t)\left[\frac{dr^{2}}{1-kr^{2}}+r^{2}(d\theta^{2}+\sin^{2}\theta d\phi^{2})\right]\label{eq:FLWR}
\end{equation}
where $a(t)$ is a scale factor that represents the time-dependent
expansion of the universe and relates the changing proper distance
$d(t)$ between a pair of objects

\[
d(t)=a(t)d_{0}
\]
where $d_{0}$ is the distance at a reference time $t_{0}$. By definition
$a(t_{0})=1$ at present time $t=t_{0}$. Different geometries of
space are described by the constant $k$, with $k=1$ for elliptical,
$k=0$ for Euclidean, and $k=-1$ for hyperbolic space. The measurement
and description of those two quantities, $k$ and $a$, is the main
focus of cosmology.

Combining the FLRW metric in Equation~\ref{eq:FLWR} and Einstein's
field equations yields an equation for the Hubble parameter, defined
as the rate of change of the scale factor with time:
\begin{align}
H\left(t\right)^{2} & \equiv\left(\frac{\dot{a}\left(t\right)}{a\left(t\right)}\right)^{2}\label{eq:-12}\\
 & =\frac{8\pi G_{N}}{3}\rho-\frac{k}{a^{2}}\label{eq:Hubble}
\end{align}
where $G_{N}$ is the Newton gravitational constant, and $H$ is the
Hubble parameter. This equation depends linearly on
\[
\rho=\rho_{M}+\rho_{R}+\rho_{\Lambda}
\]
which is the average density of the universe with contributions from
matter $\rho_{M}$, radiation $\rho_{R}$, and the cosmological constant
$\rho_{\Lambda}$, the value of the energy density of the vacuum of
space. The total energy density evolves as a function of $a$ and
can be found by solving the equation of state of the fluid filling
the universe $P=P(\rho)$:
\[
\frac{d(\rho a^{3})}{da}=-3Pa^{2}.
\]
This results in $\rho_{M}\propto a^{-3}$ for matter, $\rho_{R}\propto a^{-4}$
for radiation, and $\rho_{\Lambda}\propto\mathrm{const}.$ for cosmological
constant. 

The observation of redshifts of Type Ia supernovae (SNe)\nomenclature{SNe}{Supernovae}
at large distances in 1998~\cite{riess,perlmutter} found evidence
for an accelerating universe caused by a dark energy component. Furthermore,
since the path of light depends on the curvature of the universe,
analysis of the CMB discovered that $k$ is very close to 0, consistent
with a flat universe. Therefore, using Equation~\ref{eq:Hubble}
we can define a critical density $\rho_{cr}$. The results of this
theoretical framework combined with observations can be used to calculate
the critical density

\begin{align}
\rho_{cr} & =\frac{3H_{0}^{2}}{8\pi G_{N}}\label{eq:-13}\\
 & \equiv\frac{\rho}{\Omega}.\label{eq:-14}
\end{align}
which is estimated to be about five atoms of monatomic hydrogen per
cubic meter\footnote{For comparison, the average density of ordinary matter in the universe
is believed to be $\unit[\sim0.2]{atoms/m^{3}}$.}. Here $\Omega$ is the energy density in units of the critical density
with 
\[
\Omega=\sum_{i}\Omega_{i}
\]
where the sum is over all the different species of material in the
universe (baryonic matter, dark matter, radiation and cosmological
constant $\Lambda$), and $\Omega=1$ is equivalent to a flat universe
with $k=0$~\cite{tegmark}. By comparing observations and calculations
for the constituent parts of $\Omega$, we can conclude that the universe
today is most likely dominated by cold particles, i.e., those that
had non-relativistic velocities when they decoupled from the thermal
plasma after Big Bang~\cite{ryden2006introduction,iocco2009primordial,sarkar1996big}.
This theoretical framework, known as ``Lambda Cold Dark Matter''
($\Lambda$CDM)\nomenclature{$\Lambda$CDM}{Lambda Cold Dark Matter}
model is the cosmological equivalent of the Standard Model of particle
physics, which enables us to place constraints on the dark matter
density. 

There is a wealth of evidence from cosmology and astronomy that can
be used to constrain the amount of dark matter in the universe~\cite{DelPopolo:2013qba,DAmico2011};
the remainder of this section highlights some of the methods used
to place constraints on the amount of dark matter in the universe. 

\subsubsection{The cosmic microwave background}

\begin{figure}[p]
\begin{centering}
\includegraphics[scale=0.42]{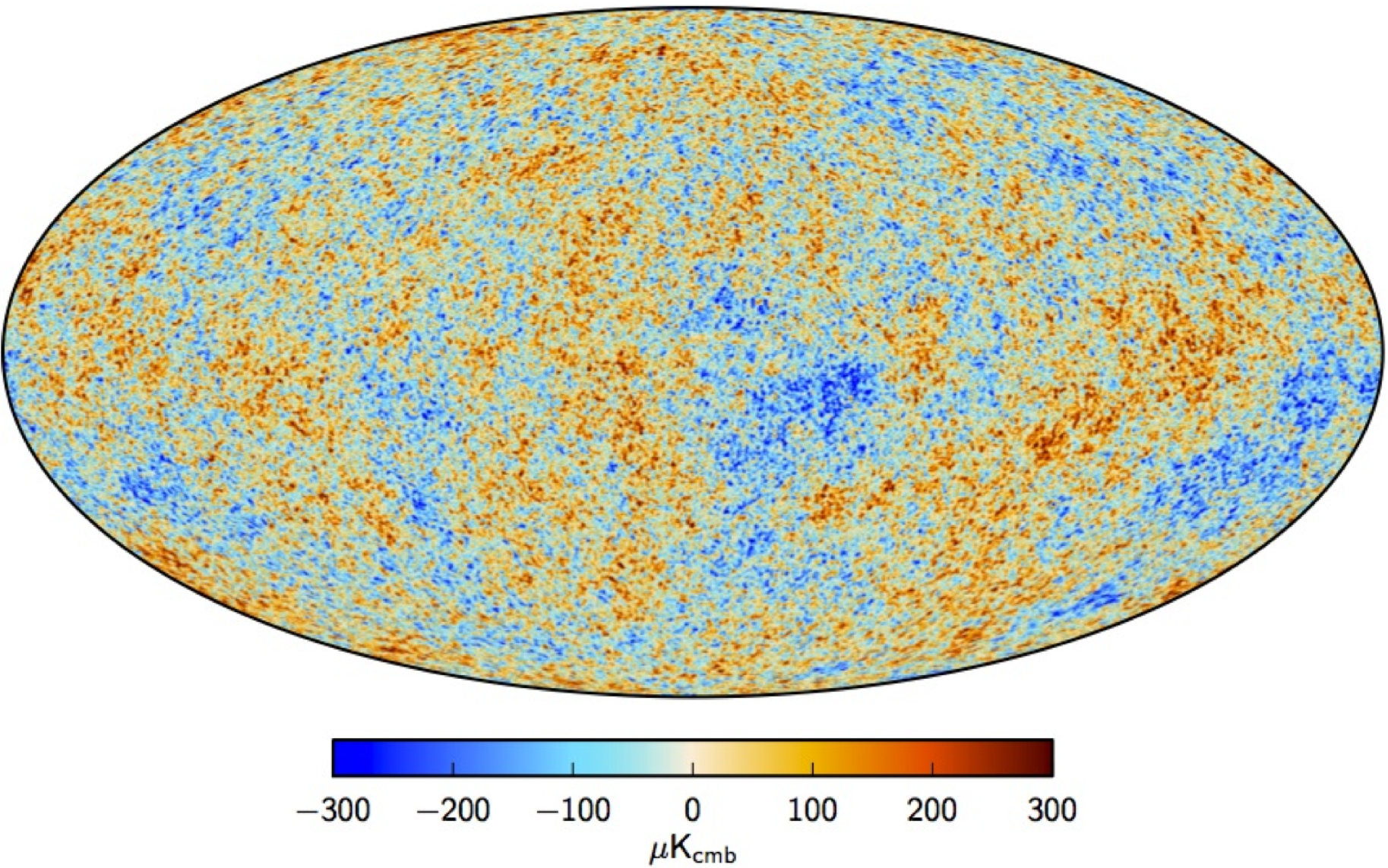}
\par\end{centering}
\caption[CMB temperature fluctuations measured by Planck]{A map of the temperature fluctuations of the CMB as seen by the Planck
Satellite. Figure from~\cite{Adam:2015rua}.\label{fig:Planck}}

\vspace{0.7cm}
\begin{centering}
\includegraphics[scale=0.48]{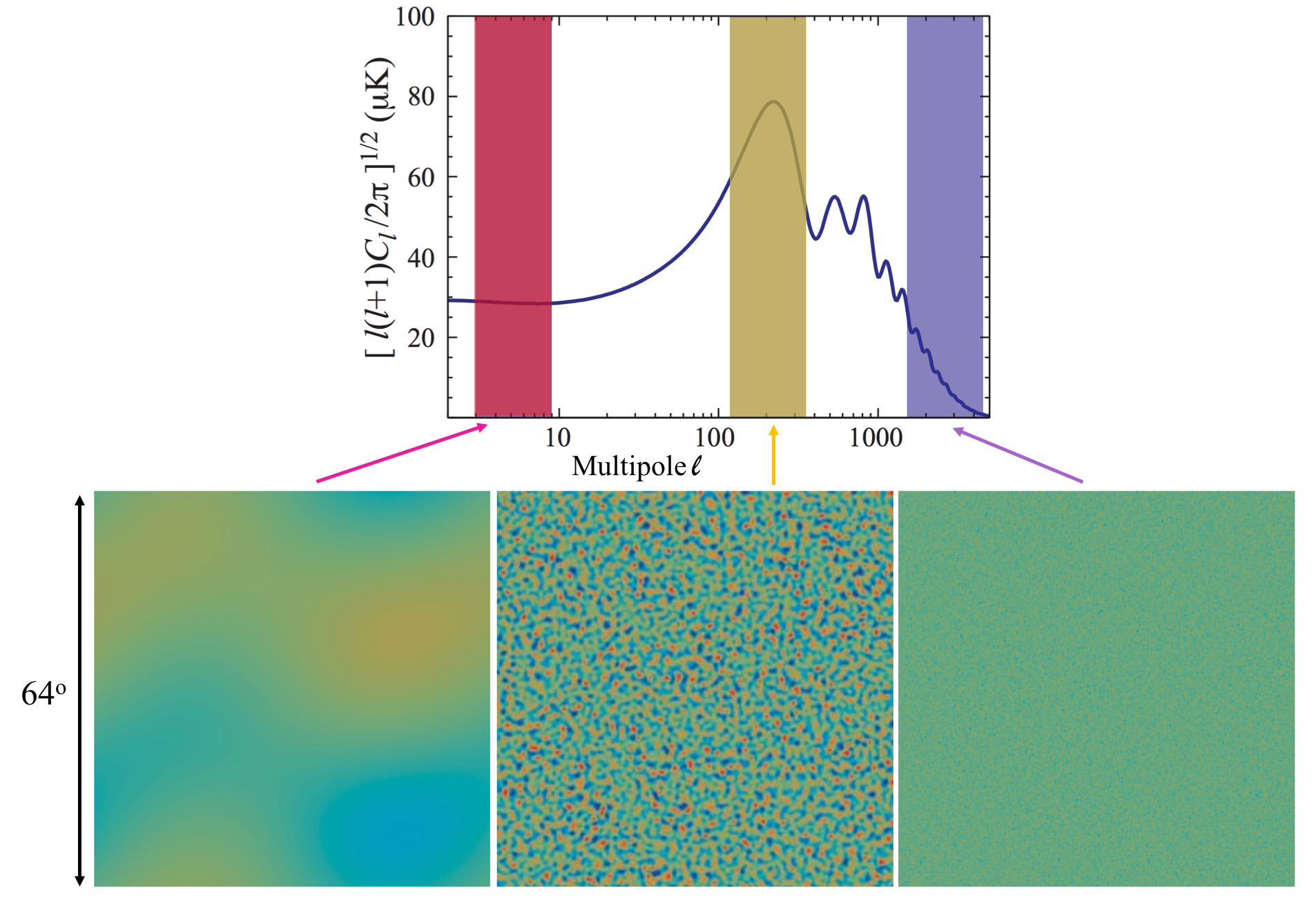}
\par\end{centering}
\caption[From temperature fluctuations to the power spectrum]{From temperature fluctuations to the power spectrum. A band filter
was applied to a temperature fluctuation map (such as the one shown
in Figure~\ref{fig:Planck}) to illustrate the power spectrum at
three different regimes. This figure was made using simulations. Figure
modified from~\cite{Hu:2008hd}.\label{fig:From-temperature-fluctuations}}
\end{figure}

The most powerful constraints on the free parameters in $\Lambda$CDM
come from the CMB, which originates from background radiation of decoupled
photons in the early universe. CMB has a blackbody spectrum~\cite{COBE}
with $T=2.726\,\mathrm{K}$ and is known to be isotropic at $10^{-5}$~level~\cite{Smoot:1992td}.
The small anisotropies $\Delta T/T$ shown in Figure~\ref{fig:Planck}
can be used to constrain cosmological parameters as shown in Figure~\ref{fig:From-temperature-fluctuations}.
Since the CMB blackbody spectrum peaks at $\lambda\sim\unit[2]{mm}$,
a wavelength readily absorbed by water molecules in the atmosphere,
it is studied using satellite missions such as COBE, WMAP, or Planck. 

\begin{figure}[p]
\begin{centering}
\includegraphics[scale=0.45]{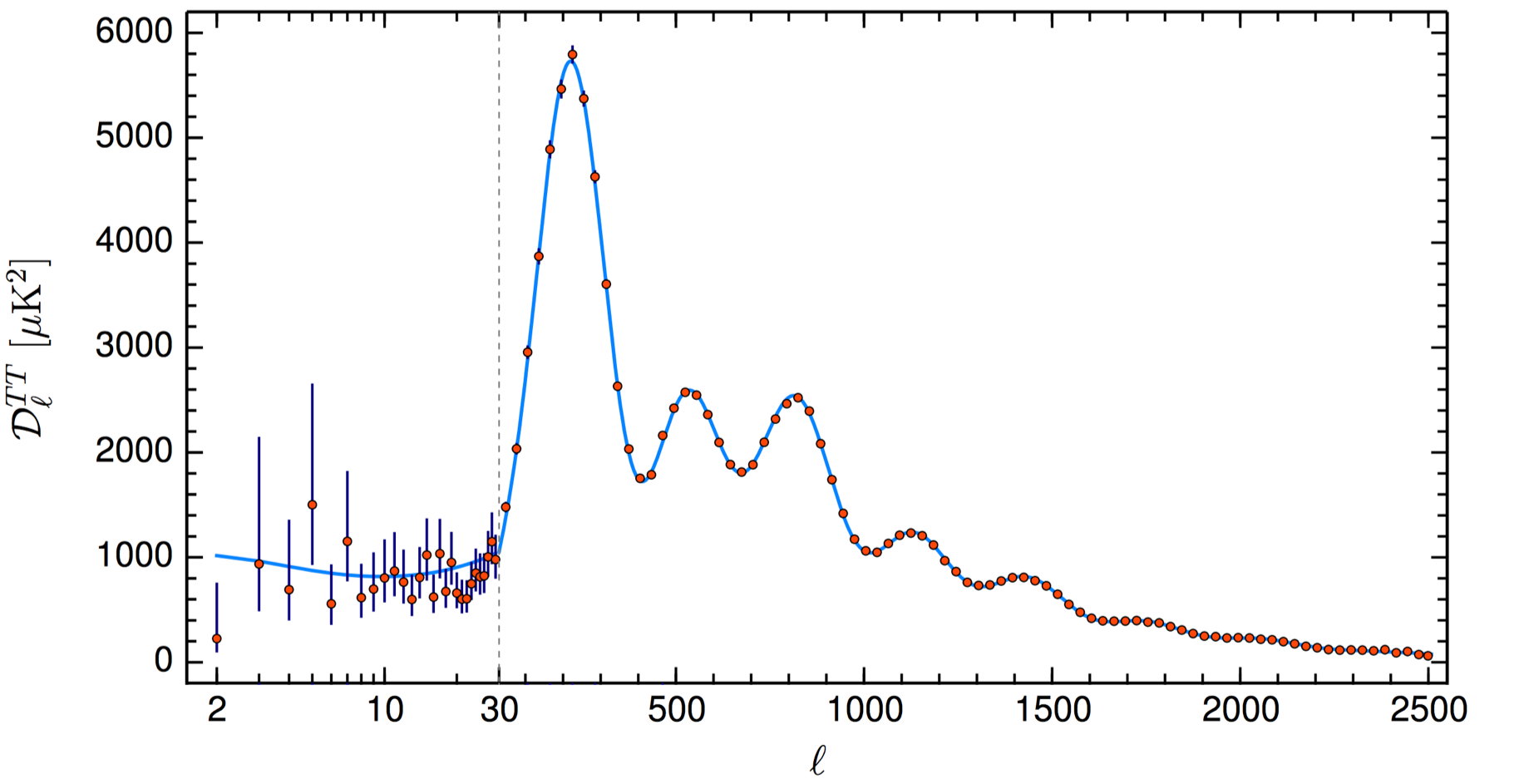}
\par\end{centering}
\caption[Temperature fluctuations of the CMB from the Planck Satellite]{The temperature power spectrum measured by Planck (red dots). The
solid blue line shows the prediction from the best-fit \textgreek{L}CDM
theoretical spectrum. The vertical gray dashed line indicates a scale
change on the horizontal axis from logarithmic to linear. The location
of the first peak gives information about the total amount of matter
in the universe, which is dependent on the speed of sound of the photon-baryon
fluid before decoupling, while the second peak provides information
about the total amount of baryonic matter in the universe. Figure
from~\cite{Aghanim:2018eyx}. \label{fig:planck-fit}}

\vspace{0.7cm}
\begin{centering}
\includegraphics[scale=0.37]{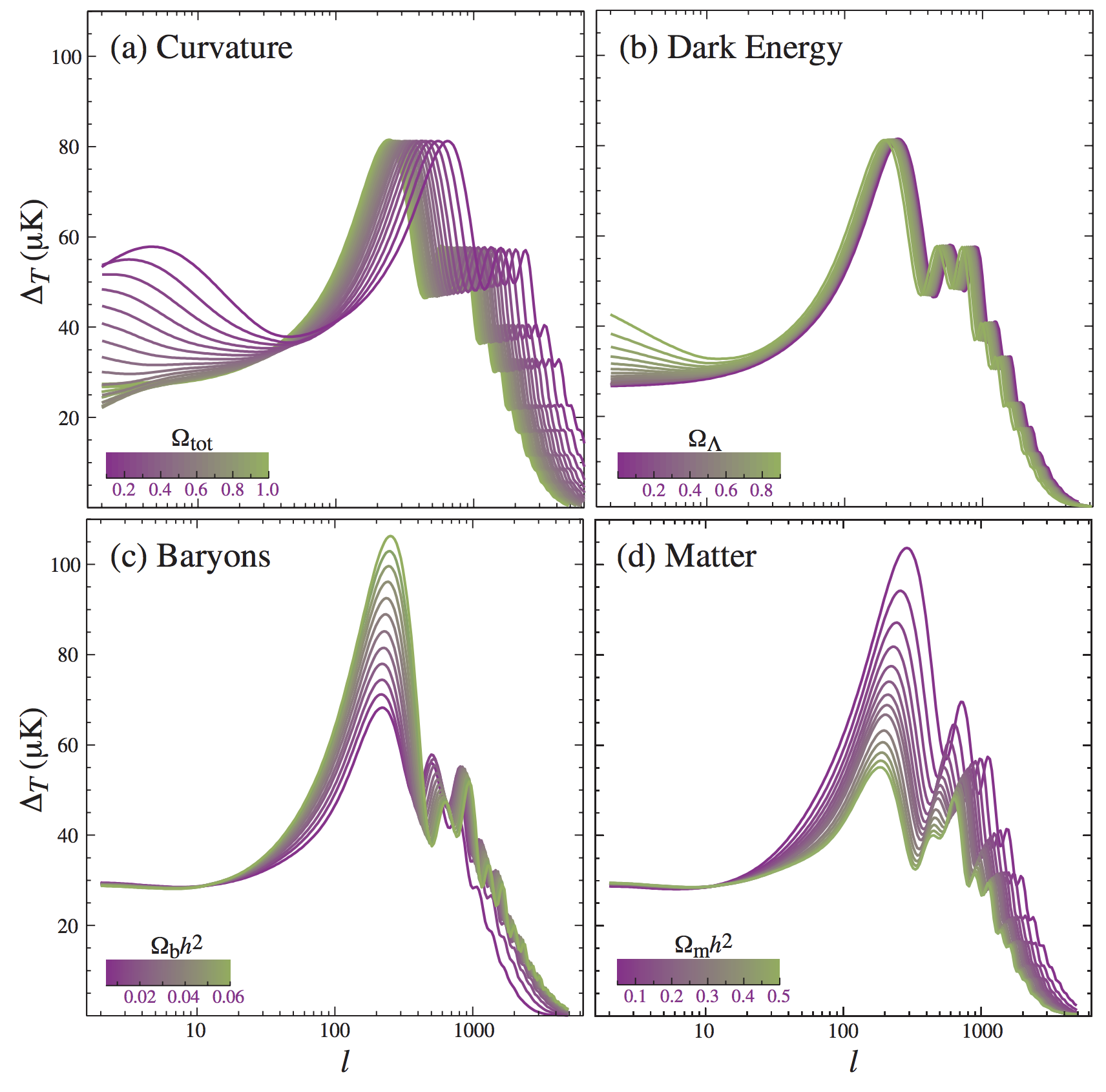}
\par\end{centering}
\caption[The sensitivity of the acoustic temperature spectrum to cosmological
parameters]{The sensitivity of the acoustic temperature spectrum to a)~the curvature~$k$,
b)~the dark energy~$\Omega_{\Lambda}$, c)~the baryon density~$\Omega_{b}$,
and d)~the matter density~$\Omega_{m}$. Models were varied around
$\Omega_{tot}=1$. Figure from~\cite{Hu:2001bc}.\label{fig:omega_variations}}
\end{figure}

The temperature map can be expanded using spherical harmonics $Y_{lm}(\theta,\phi)$.
On small scales where the curvature of space can be neglected, a Fourier
analysis can be performed and the spherical harmonics $l$ become
the Fourier wavenumbers. The power spectrum can be fitted as shown
in Figure~\ref{fig:planck-fit} and characterized as a function of
the multipole number~$l$. The angular wavelength $\theta=2\pi/l$,
so large multipole moments $l\sim10^{2}$ correspond to a few degrees
of separation on the sky. The precise location of each peak is sensitive
to different parameters of the cosmological model as illustrated in
Figure~\ref{fig:omega_variations}.

By combining the results of the best fit values from the CMB, along
with many of the observations described below, the Planck collaboration
constrained the main parameters of the $\Lambda$CDM model. The resulting
values of the different density components in our universe are presented
in Table~\ref{tab:Planck_data}.

\begin{table}
\begin{centering}
\begin{tabular}{llr}
\hline 
Parameter & Symbol & Value\tabularnewline
\hline 
\hline 
Baryonic matter density & $\Omega_{b}$ & $0.0490\pm0.0003$\tabularnewline
Cold dark matter density & $\Omega_{c}$ & $0.2606\pm0.0020$\tabularnewline
Total matter density & $\Omega_{m}=\Omega_{b}+\Omega_{c}$ & $0.3111\pm0.0056$\tabularnewline
Dark energy density & $\Omega_{\Lambda}$ & $0.6889\pm0.0056$\tabularnewline
\hline 
\end{tabular}
\par\end{centering}
\caption[Selected constraints on the density parameters in the $\Lambda$CDM
model]{Selected constraints on the density parameters in the $\Lambda$CDM
model as reported by the Planck collaboration~\cite{Aghanim:2018eyx}.
Note that some of those are derived quantities, rather than direct
parameters of the model. Uncertainties are given at the 68\% interval.\label{tab:Planck_data}}
\end{table}

\subsubsection{Baryon acoustic oscillations}

Acoustic oscillations of the baryon-photon fluid from the pre-recombination
fluid are imprinted in the large-scale structures of the universe.
These oscillations provide a standard length scale by looking at the
correlation between large-scale objects. After inflation, higher dark
matter density regions attracted matter, which started forming structures
by gravitationally attracting surrounding matter. As the baryon density
increased, the increase of radiation pressure resulted in acoustic
waves,  which propagated until decoupling. This created hotter regions
where matter contracted, and cooler regions, which resulted in photon
emission. Between the formation of those perturbations and the epoch
of recombination, the sound waves with different wavelengths will
complete a different number of oscillations. This translates time
into a characteristic length scale: the maximum distance that an acoustic
wave traveled before the epoch of recombination provides a ``standard
ruler'' for length scale in cosmology.

\begin{figure}[t]
\begin{centering}
\includegraphics[scale=0.4]{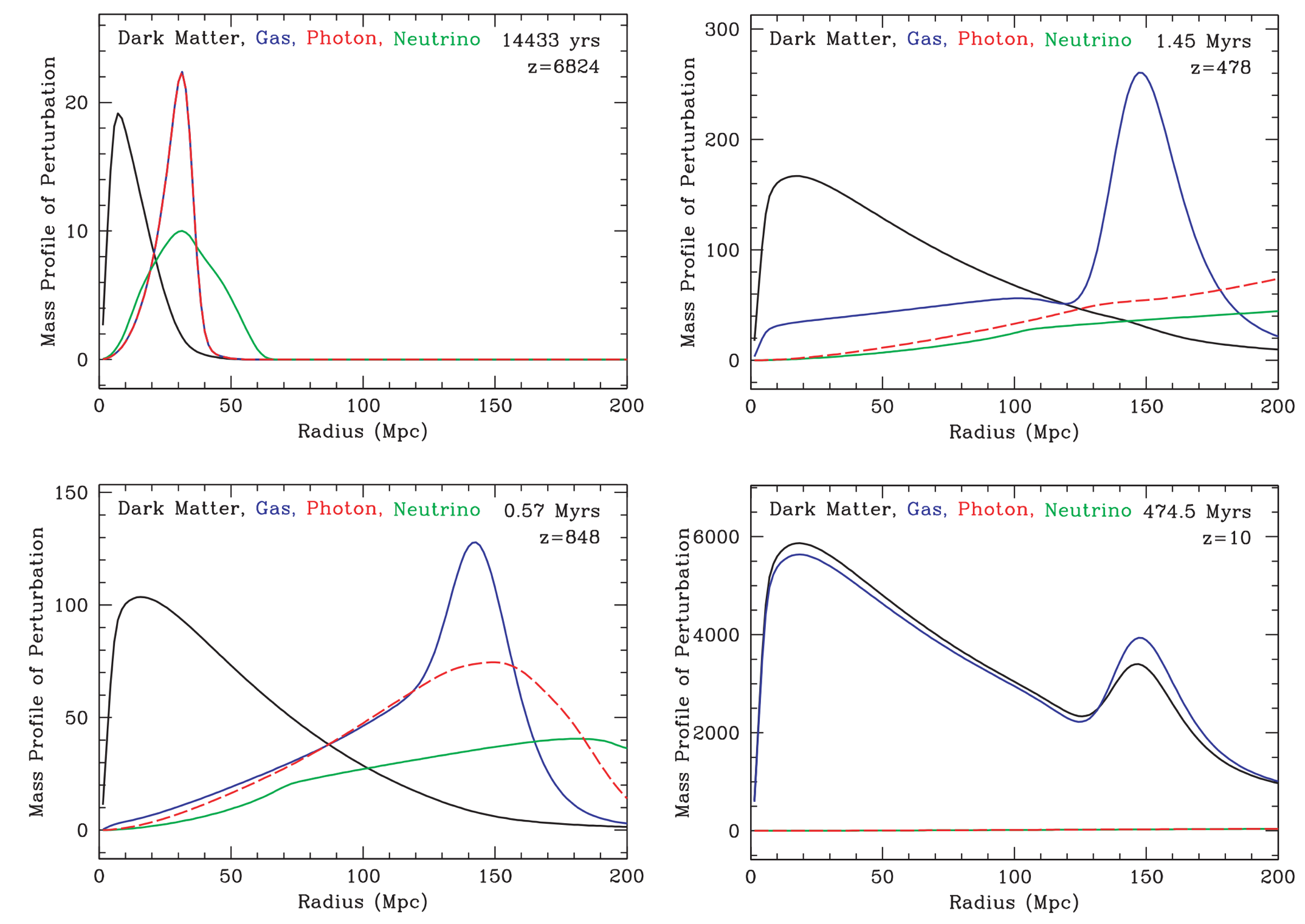}
\par\end{centering}
\caption[Simulated thermal history of the universe]{Simulated thermal history of the universe with dark matter (black),
gas (baryons, blue), and photons (red). An acoustic wave drives photons
and baryons together in a tightly bound plasma. After decoupling,
photons stream away while the baryon wave stalls at a length scale
of 150~Mpc, still visible today. Figure modified from~\cite{Eisenstein:2006nj}.\label{fig:BAO}}
\end{figure}

The baryon acoustic oscillations (BAO)\nomenclature{BAO}{Baryon Acoustic Oscillations}
measure the comoving separation of large-scale objects (such as galaxies)
at the sound horizon of the time of photon-baryon decoupling. An excess
clustering at a particular length scale can then be used to constrain
the DM density, as illustrated in Figure~\ref{fig:BAO}. This effect
has been measured using a spectroscopic survey of luminous red galaxies~\cite{Eisenstein:2005su}
from the Sloan Digital Sky Survey~\cite{York:2000gk}. The shape
of the correlation function obtained from the distribution of galaxies
in the universe provides a measurement of the matter density $\Omega_{m}$.
This constraint from BAO combined with others is shown in Figure~\ref{fig:BBN abundance}.

\subsubsection{Big Bang nucleosynthesis}

Big Bang nucleosynthesis (BBN)\nomenclature{BBN}{Big Bang nucleosynthesis}
is a sequence of nuclear reactions that led to the synthesis of light
elements such as D, $^{3}$He, $^{4}$He, and $^{7}$Li during the
first hundreds of seconds after the Big Bang. Before the epoch of
nucleosynthesis, the universe was hot and dense enough for all the
particles to be in equilibrium. As the expansion proceeded and temperature
decreased some particles were able to separate from the thermodynamic
equilibrium with the plasma. First to ``freeze out'' from the plasma
(drop out of thermal equilibrium) were neutrinos at $t\approx\unit[0.1]{s}$
with temperatures $\sim\unit[2-3]{MeV}$ since the neutral current
weak interactions became too slow to keep up with the expansion rate.
At $t\approx\unit[1]{s}$ with $T\sim\unit[1]{MeV}$ the charged-current
weak interactions became too slow to maintain neutron-proton equilibrium
and the n/p density ratio froze out at $\mathrm{n/p}\sim1/7$. 

\begin{figure}[t]
\begin{centering}
\includegraphics[scale=0.52]{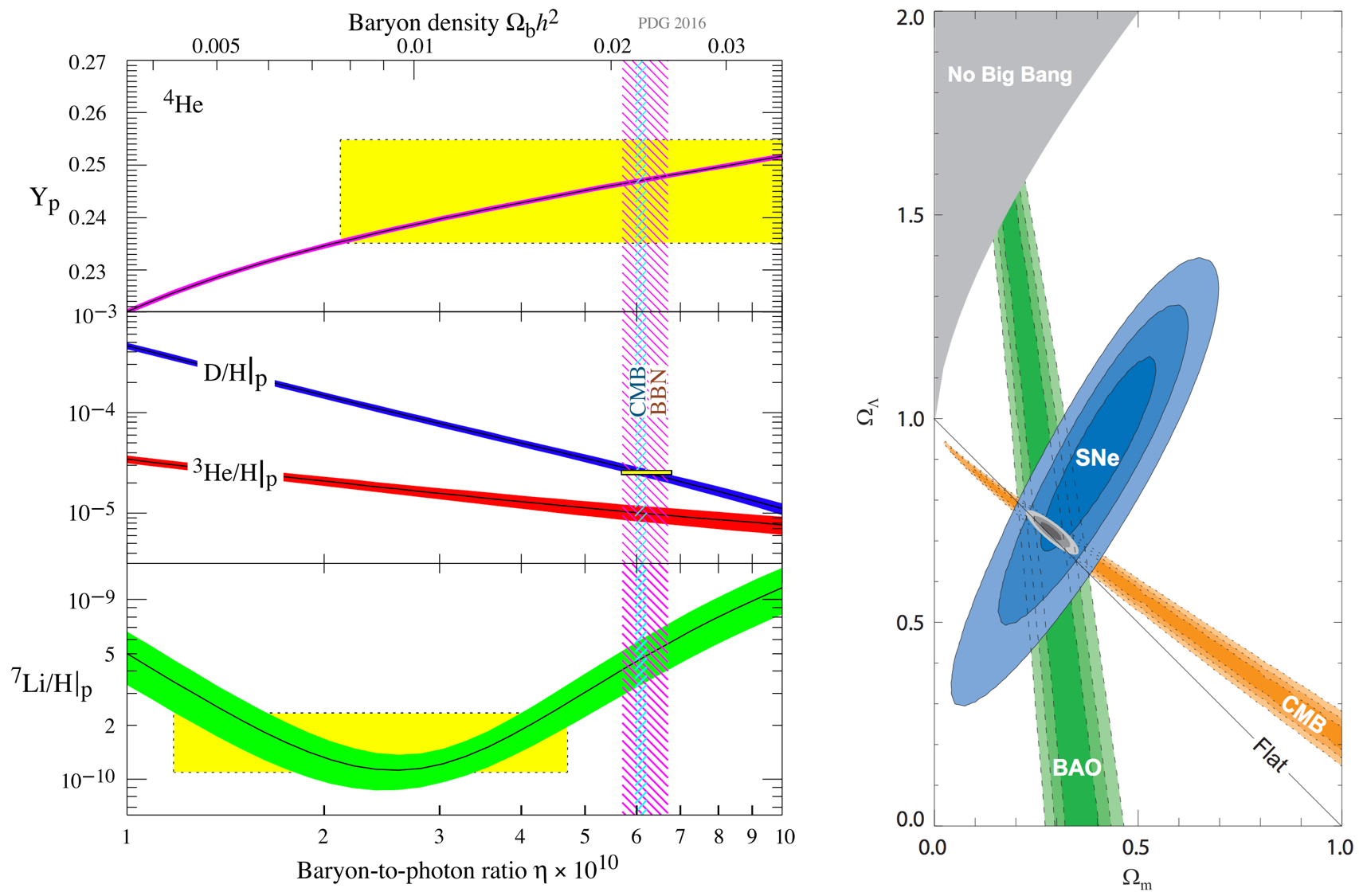}
\par\end{centering}
\caption[Constraints $\Omega_{b}$, $\Omega_{m}$, and $\Omega_{\Lambda}$
obtained from BBN, CMB, BAO, and Supernovae]{\textbf{Left:} The abundances of $^{4}$He, D, $^{3}$He, and $^{7}$Li
as predicted by BBN. The bands show the 95\% CL range, boxes indicate
the observed light element abundances, and the narrow vertical band
is a result from CMB measurements of the cosmic baryon density. The
wider band indicates the BBN concordance range. Figure from~\cite{PDG2016}.
\textbf{Right:} Constraints on $\Omega_{m}$ and $\Omega_{\Lambda}$
obtained from CMB, BAO~\cite{Eisenstein:2005su}, and SNe. The contours
show 68.3 \%, 95.4 \% and 99.7\% confidence level. Figure from~\cite{CosmoConstraints}.
\label{fig:BBN abundance}}
\end{figure}

At $t\gtrsim100\,\mathrm{s}$ the universe expanded and cooled enough
so that photons were no longer able to photo-dissociate deuterium
(D)\nomenclature{D}{Deuterium}, which marks the start of nucleosynthesis.
The formation of D strongly depends on $\eta$, the baryon-to-photon
ratio $\eta=\frac{n_{N}}{n_{\gamma}}$ that gives the total number
of nucleons, bound or free, as a fraction of photons. BBN eventually
stops at $t\gtrsim1000\,\mathrm{s}$ since the Coulomb barrier becomes
too large for nuclear fusion and nucleosynthesis to occur~\cite{boesgaard1985big}.

The baryon-to-photon ratio is also crucial for determining the density
of baryons $\Omega_{b}$; the importance of deuterium as a ``baryometer''
was recognized for the first time in the 1970's. The equilibrium abundance
of D is not well known observationally because it is not clear how
much of the dark matter in the universe is in the form of nucleons,
but the upper bound to the primordial D abundance can be obtained
by measuring the quasar Lyman-$\alpha$ lines to find the baryon density
$\Omega_{b}$~\cite{sarkar1996big,schramm1997big}. This constrains
the number of baryons and hence baryonic DM\footnote{The cosmic density of luminous matter is below the observed quantity
for baryons. Therefore, there might be baryons that are optically
dark, such as in the form of Massive Astrophysical Compact Halo Objects
(MACHOs)\nomenclature{MACHO}{Massive Astrophysical Compact Halo Object}
or brown dwarfs. However, microlensing results have shown that <25\%
of dark halos could be due to baryonic dark matter~\cite{galactic_DM,macho}. }, implying that the majority of DM must be non-baryonic since $\Omega_{m}\gg\Omega_{b}$. 

Figure~\ref{fig:BBN abundance} shows the observed and predicted
primordial abundances of light elements compared to results from the
CMB. The agreement\footnote{The disagreement in the case of $^{7}$Li is likely caused by the
lack of understanding of stellar processes, rather than the $\Lambda$CDM
model~\cite{steigman}.} of the two models confirms the success of the $\Lambda$CDM model. 

\subsection{Gravitational lensing\label{subsec:Gravitational-lensing}}

\begin{figure}
\begin{centering}
\includegraphics[scale=0.27]{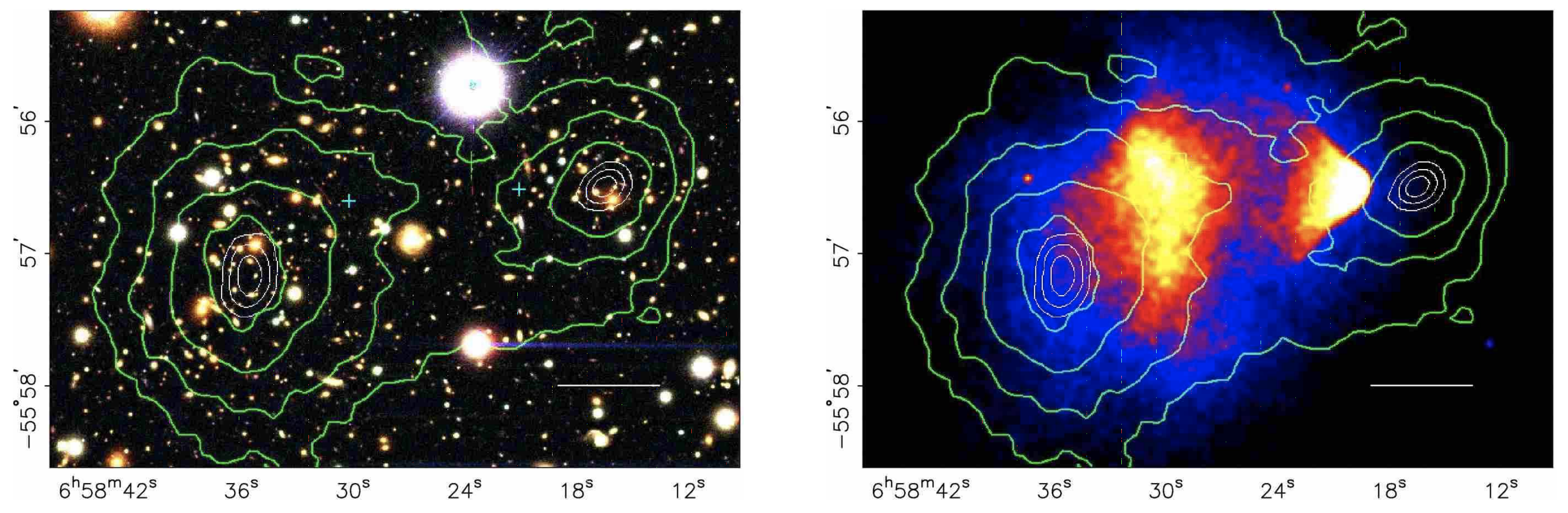}
\par\end{centering}
\caption[Two images of the Bullet cluster]{Two images of the merging cluster 1E0657\textminus 558, known as
Bullet cluster, from the Hubble Space Telescope (left), illustrating
the distribution of galaxies, and from the Chandra X-ray Observatory
(right), illustrating the intracluster plasma. The green contours
show the reconstructed mass from weak gravitational lensing. The white
bar indicates 200~kpc. Figures from~\cite{Clowe:2006eq}.\label{fig:bullet-cluster}}
\end{figure}

The existence of dark matter can also be inferred from collisions
of galaxy clusters. The most famous example of this is the Bullet
Cluster 1E0657\textminus 558. Gravitational lensing leverages the
fact that massive objects in the foreground bend light from a bright,
distant source. A particular type of gravitational lensing, known
as weak lensing, determines the presence of an intermediate mass through
correlations in the observed ellipticity of the distorted objects.
This method was used to measure the distribution of mass in the Bullet
Cluster, shown as green outlines in Figure~\ref{fig:bullet-cluster}.
Along with the images from the Hubble Space Telescope used for the
weak lensing measurements, researchers analyzed the x-ray spectra
of the cluster recorded by the Chandra X-ray Observatory. The inferred
mass from those two techniques produced disparate results. It implies
that the x-ray emitting ordinary matter clustered together in the
center during the collision, while the two cores of the galaxies passed
through one another with little interaction, suggesting that weakly-interacting
dark matter forms the majority of the cluster's mass.

There are now many examples of non-gravitational interactions of dark
matter in colliding galaxy clusters. In 2015, Reference \cite{Harvey:2015hha}
examined 72 collisions using the Chandra and Hubble Space Telescopes.
The observed offset of the center of mass for the gas and stars from
the DM confirmed the existence of DM at 7.6$\sigma$ significance.

\subsection{Rotational velocity of spiral galaxies\label{subsec:Rotational-velocity-of}}

Further evidence for DM comes from observations. Fritz Zwicky was
the first to infer the existence of dark matter from measurements
of the velocity dispersion of galaxies in the Coma cluster. By applying
the virial theorem to the gravitational potential $U$ and rotational
kinetic energy $\left\langle T\right\rangle =-\frac{1}{2}\left\langle U\right\rangle $,
he found that the mass-to-light ratio significantly outweighed the
mass deduced from the system's luminosity.

\begin{figure}
\begin{centering}
\includegraphics[scale=0.4]{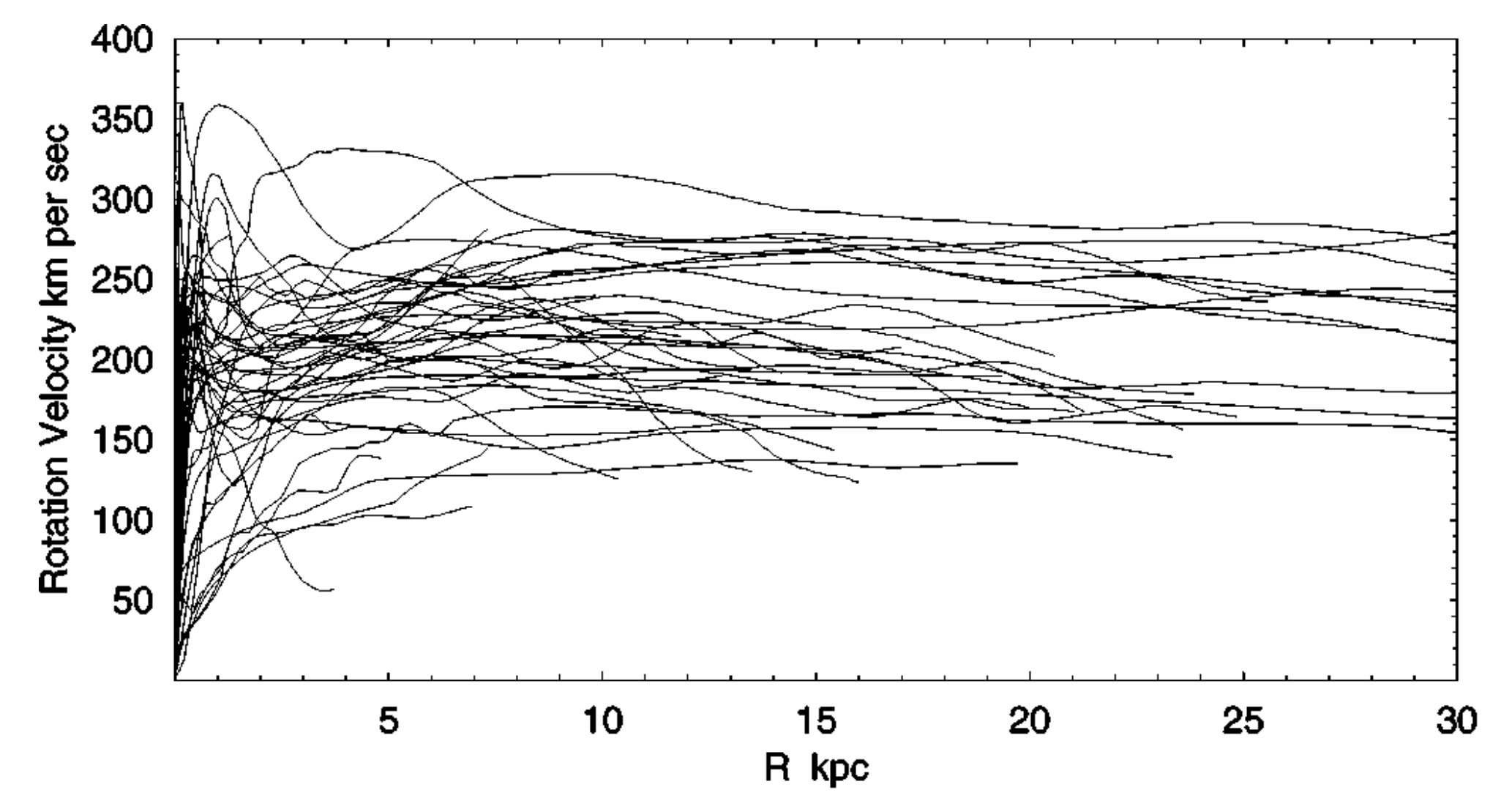}
\par\end{centering}
\caption[A compilation of rotation curves of spiral galaxies from various sources]{A compilation of rotation curves of spiral galaxies from various
sources. The flatness of the curves at large radii indicates the presence
of a DM halo. Figure from~\cite{Sofue:2000jx}.\label{fig:rotational-curves}}
\end{figure}

Another piece of evidence came much later, in the 1970's-1980's from
observations of galactic rotation curves by V. Rubin \textit{et al.}~\cite{Rubin:1980zd}.
The rotational velocity $v$ of a point-like object on a Keplerian
orbit with radius $r$ around a galaxy should follow the equation
\begin{align}
v(r) & =\sqrt{\frac{GM(r)}{r}}\label{eq:-11-1}
\end{align}
derived from Newton's second law, where $G$ is the gravitational
constant and $M$ is the reduced mass. However, the observed velocity
outside the visible part of the galaxies does not decline as predicted
by the equation but becomes approximately constant as shown in Figure~\ref{fig:rotational-curves},
which compiles rotation curves from many galaxies. This suggests the
existence of a DM halo with mass density $\rho(r)\propto1/r^{2}$
and $M(r)\propto r$ and gives a lower bound to the DM mass density
$\Omega_{DM}\gtrsim0.1$~\cite{Navarro:1995iw}. Note that $\rho$
needs to fall faster at some radius, but when this happens is as yet
unknown. 

\section{Dark matter candidates}

There is a plethora of DM candidate particles, and finding a suitable
candidate requires physics beyond the SM. Since $\Omega_{m}\gg\Omega_{b}$
this suggests that most matter in the universe must be non-baryonic,
candidates for DM particles must be stable on cosmological scales,
interact only weakly (or not at all) with electromagnetic radiation,
and be non-relativistic at the time of freeze-out, or ``cold.''
Such candidates include, for example, primordial black holes, Weakly
Interacting Massive Particles (WIMPs)\nomenclature{WIMP}{Weakly Interacting Massive Particle},
Strongly Interacting Massive Particles (SIMPs), axions, or hidden
DM. Constraints on DM are usually calculated assuming that a single
species constitutes all of DM, but it is possible that a combination
of these candidates is responsible for $\Omega_{c}$. Reviews of possible
DM candidates can be found, for example, in~\cite{fengReview,Cline:2018fuq,cushman2013snowmass}.
Below we focus on the WIMP and the axion, two of the more popular
candidate particles.

\subsection{Axions}

Axions are motivated by the fact that the violation of the charge
conjugation parity (CP)\nomenclature{CP}{Charge conjugation Parity}
symmetry has not been observed in strong interactions in quantum chromodynamics
(QCD)\nomenclature{QCD}{Quantum Chromodynamics}. The solution from
Peccei and Quinn (PQ)~\cite{Peccei:1977hh} introduces a pseudoscalar
field $a$ where the particle associated with the symmetry breaking
is the axion~\cite{PhysRevLett.40.223,PhysRevLett.40.279}. Axion
is an extremely light, weakly interacting particle produced non-thermally
that can also serve as a well-motivated dark matter candidate. Several
constraints bound the axion mass $m_{a}$; the most stringent limit
derived from the observed length of the neutrino pulse from Supernova
1987a requires $m_{a}\lesssim\unit[10]{meV}$~\cite{Raffelt:2006cw}.

\begin{figure}
\centering{}\includegraphics[scale=0.2]{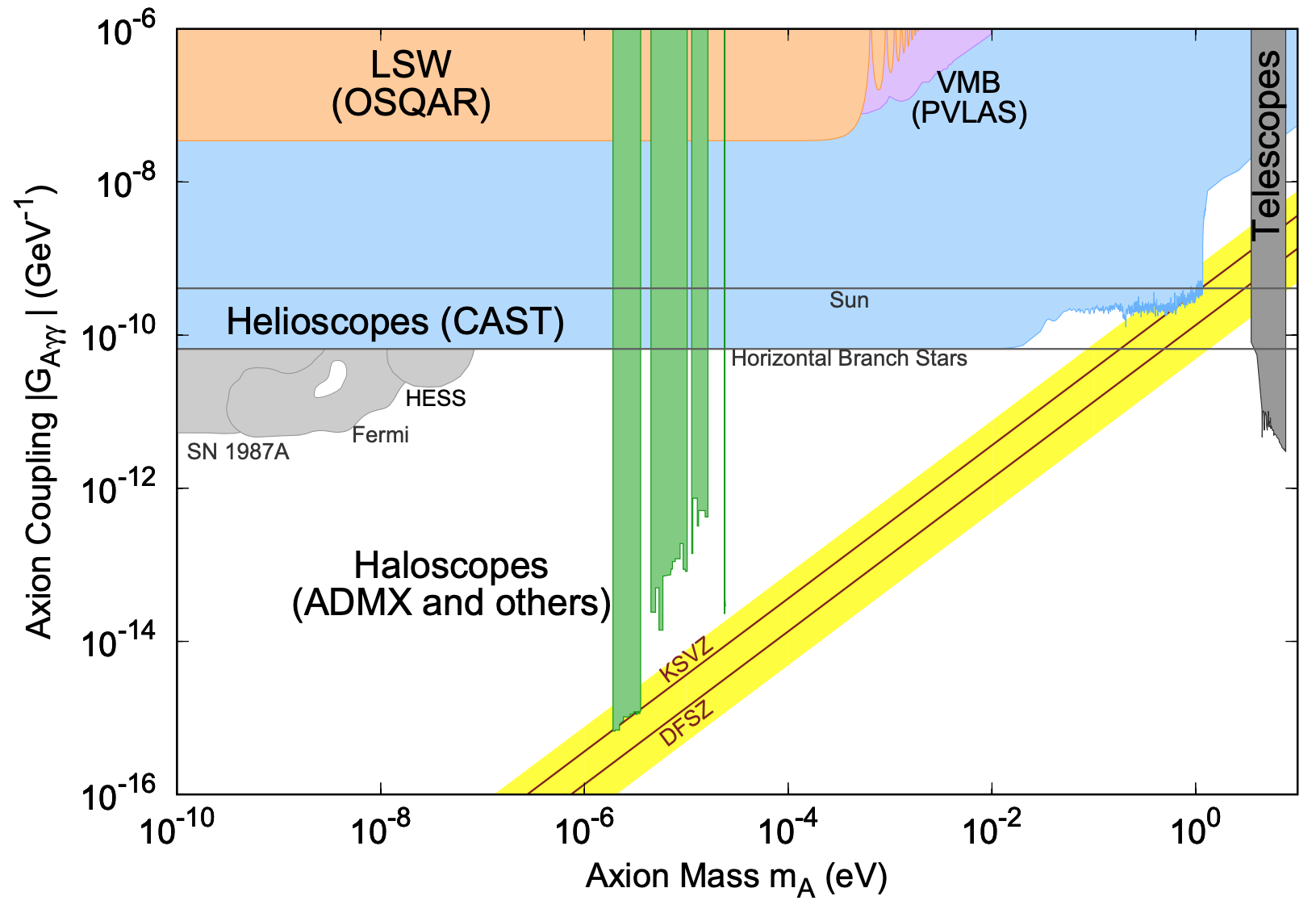}\caption[Exclusion plot from axion searches]{Exclusion plot from axion searches. KSVZ~\cite{Shifman:1979if}
and DFSZ~\cite{DINE1981199} refer to archetype theoretical predictions
for axion coupling models. LSW refers to the current leading limit
from a ``light shining through walls'' experiment~\cite{Ballou:2015cka}
while VMB stands for vacuum magnetic birefringence looking for a change
in the polarization of light caused by the Primakoff effect~\cite{Zavattini:2005tm}.
Figure from~\cite{PDG2016}.\label{fig:Exclusion-plot-axions}}
\end{figure}

Various experiments are searching for axions both directly and indirectly
as illustrated in Figure~\ref{fig:Exclusion-plot-axions}. Axions
are theorized to interact with gluons, fermions, and photons. However,
most experiments search for axions through the Primakoff effect where
a pseudoscalar is converted into a photon in the presence of an electromagnetic
field~\cite{Sikivie:1983ip} with a coupling strength given by $g_{a\gamma\gamma}$.
The resulting photon is collinear with the incoming axion, and the
photon's energy is equal to the axion's total energy in a time-independent
magnetic field. Direct experiments use for example haloscopes~\cite{Boutan:2018uoc,Zhong:2018rsr}
to search for a resonant radio frequency corresponding to the Compton
wavelength of the axion or helioscopes which use a strong magnetic
field to convert axions produced in the Sun into x-rays~\cite{CAST}.
Indirect astrophysical and cosmological searches consider stellar
evolutions to place upper bounds on the axion coupling. 

While axions remain to be observed directly, their theoretical motivation
for resolving the CP problem while serving as an attractive DM particle
makes them an attractive candidate for both contemporary and planned
searches. For a more in-depth review of axions consult for example~\cite{Asztalos:2006kz,brubaker2018first}. 

\subsection{WIMPs}

WIMPs ($\chi$) are non-relativistic particles with an expected mass
between $\mathrm{GeV/c^{2}-\unit[100]{TeV/c^{2}}}$. Their present
relic density can be calculated assuming that WIMPs were in thermal
and chemical equilibrium with the plasma after the Big Bang. As the
universe cooled down, the production of particles with masses above
the temperature of the universe diminished. Eventually, the production
of cold dark matter stopped, and the particles started to annihilate
with one another until the expansion rate of the universe exceeded
the rate of annihilation. This freeze-out\footnote{Freeze-out is not the only possible mechanism for DM production. Other
production mechanisms include ``freeze-in,'' which assumes the DM
interacts so weakly it was never in thermal equilibrium in the early
universe or asymmetric DM, which assumes the existence of a dark sector,
similar to the matter-antimatter asymmetry that gives origin to baryonic
abundance~\cite{Cline:2018fuq}.} fixed the DM density. 

\begin{figure}
\begin{centering}
\includegraphics[scale=0.5]{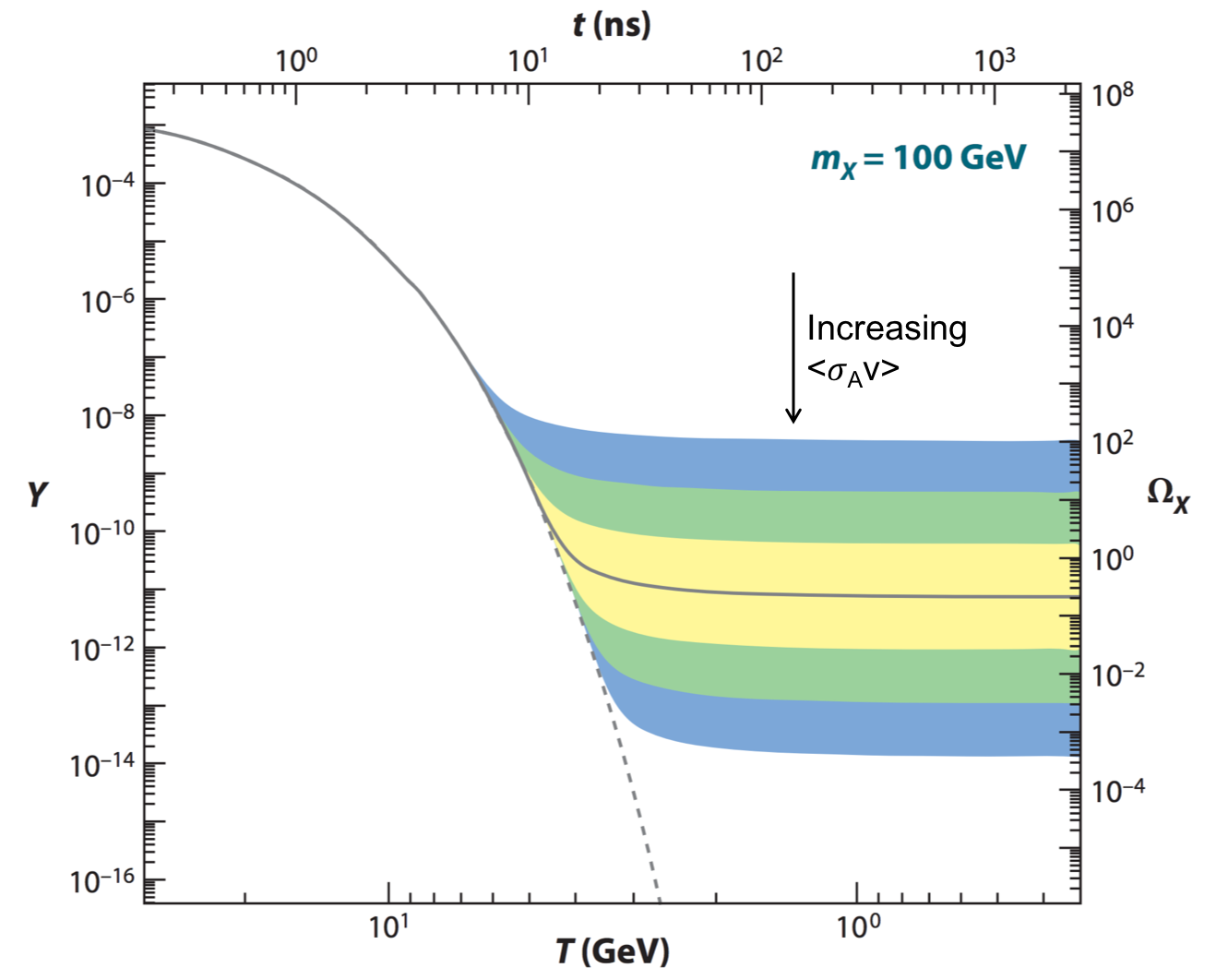}
\par\end{centering}
\caption[Comoving number density of WIMPs over time]{Comoving number density $Y$ of WIMPs over time $t$ with $\left\langle \sigma_{A}v\right\rangle $
that produces the correct relic density $\Omega_{\chi}$ (solid gray
line). The contours of the colored bands differ from this value by
1, 2, and 3 orders of magnitude. The dashed gray line shows the number
density of WIMPs that remains in thermal equilibrium. Figure modified
from~\cite{fengReview}.\label{fig:WIMP-miracle}}
\end{figure}

The relic density of WIMPs today can be obtained from the Boltzmann
equation:
\begin{equation}
\Omega_{\chi}h^{2}\simeq\frac{T^{3}}{M_{Pl}^{3}\left\langle \sigma_{A}v\right\rangle }\label{eq:WIMPs}
\end{equation}
where $h$ is the dimensionless Hubble parameter, $h=H_{0}/\left(\unit[100]{km\cdot s^{-1}\cdot Mpc^{-1}}\right)$,
$T$ is the current temperature of the CMB, $M_{Pl}$ is the Planck
mass, $c$ is the speed of light, $\sigma_{A}$ is the total annihilation
cross section of a pair of WIMPs into SM particles and $v$ is the
relative velocity between two WIMPs~\cite{PDG2016,gaitskell2004direct}.
A solution to Equation~\ref{eq:WIMPs} yields 
\begin{align}
\Omega_{\chi}h^{2} & \approx\frac{\unit[3\times10^{-27}]{cm^{3}\cdot s^{-1}}}{\left\langle \sigma_{A}v\right\rangle }\label{eq:WIMPs-1}\\
 & \approx\frac{\unit[0.1]{pb\cdot c}}{\left\langle \sigma_{A}v\right\rangle }.
\end{align}
This result is roughly independent of the WIMP mass and results in
fewer relic particles after freeze-out for a higher annihilation rate
as illustrated in Figure~\ref{fig:WIMP-miracle} for a $\unit[100]{GeV/c^{2}}$.
For a particle with electroweak scale interactions, the relic abundance
in Equation~\ref{eq:WIMPs-1} naturally generates WIMPs with a relic
density consistent with that required for DM, $\Omega_{c}$. This
is sometimes referred to as the \textquotedblleft WIMP miracle.\textquotedblright{} 

The currently best-motivated WIMP candidates are superparticles in
the minimal supersymmetric model (MSSM)\nomenclature{MSSM}{Minimal Supersymmetric Model},
which is the simplest supersymmetry (SUSY)\nomenclature{SUSY}{Supersymmetry}
extension to the SM. SUSY was proposed to unify the electroweak, strong
and gravitational forces, and it matches each particle of the SM to
its partner with a different spin. Some of the candidates for WIMP
from the MSSM are the gravitino, sneutrino, or neutralino (a mix of
gaugino and higgsino~\cite{Ellis:1983ew}). These are stable, electrically
neutral, and are expected to be massive enough to contribute to the
DM abundance substantially. For more information about WIMPs consult
for example~\cite{Jungman:1995df,bertone2005particle,Arcadi2018,roszkowski}.

\section{Detection strategies}

\begin{figure}
\begin{centering}
\includegraphics{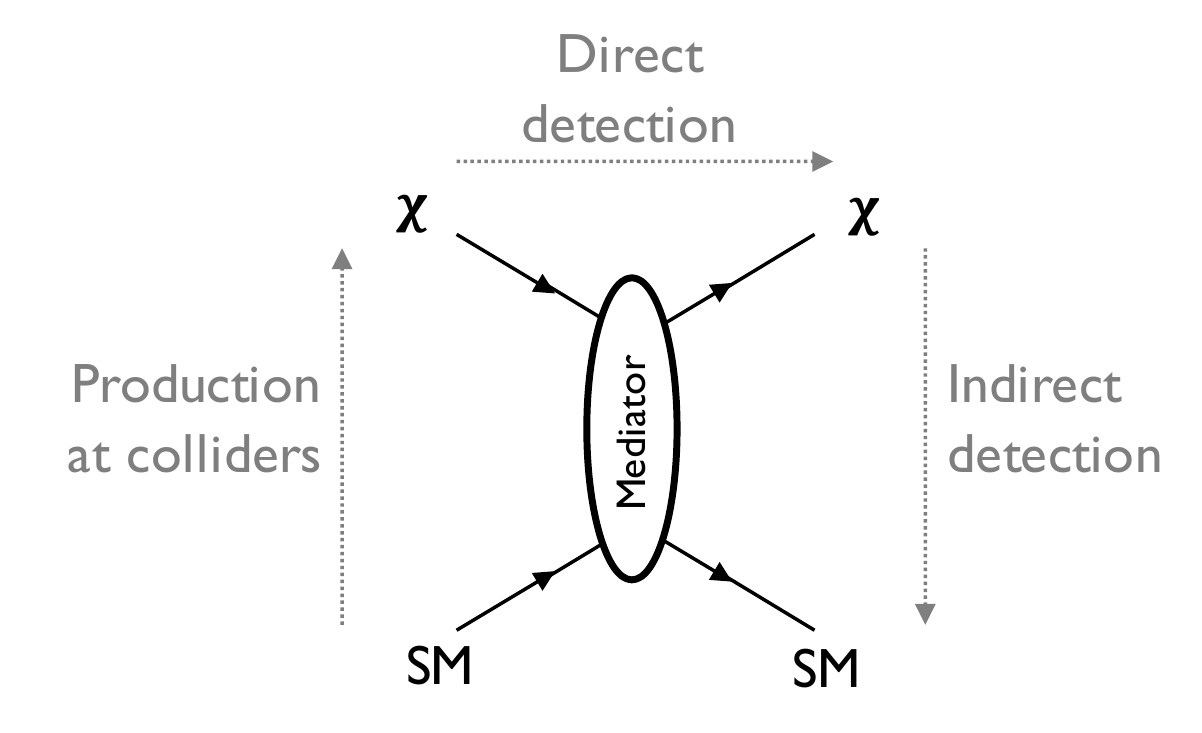}
\par\end{centering}
\caption[Schematics of the various detection methods]{Schematics of the various detection methods. The DM particle is denoted
by $\chi$.\label{fig:Detection-techniques}}

\end{figure}

Given the existence of DM, the experimentalist asks how it can be
detected. The detection techniques can be broadly classified into
three main categories as outlined in Figure~\ref{fig:Detection-techniques}:
production, indirect detection, and direct detection. The discussion
of those three techniques follows, with a focus on direct detection
strategies. The primary goal of these detectors is to probe the properties
of the DM particles, like their interaction mechanism or scattering
rate.

\subsection{Production}

Production of DM can be achieved using accelerators such as the Large
Hadron Collider (LHC)\nomenclature{LHC}{Large Hadron Collider} at
the European Organization for Nuclear Research, better known as CERN.
The pair production shown in Figure~\ref{fig:Detection-techniques}
does not leave any observable signal in the detectors. Instead, experiments
like ATLAS and CMS search for higher order interactions: $p\bar{p}\rightarrow\chi\bar{\chi}+\mathrm{jets}$,
$\gamma$s, or other particles that can serve as a trigger signaling
an event. Reconstruction of event energies enables inference of missing
energy and momentum after a collision. This could potentially signal
a DM particle leaving the detector. However, the interpretation of
these results requires an assumption of a DM model on the nature of
the mediator and the corresponding operator. Excess beyond the SM
has yet to be observed~\cite{Aaboud2018,Beltran2010}.

\subsection{Indirect}

Indirect detection methods look for DM annihilation products where
DM density is expected to be high, such as at the center of the Sun,
in the center of our galactic halo, or in the center of nearby galaxies.
DM annihilation can produce a number of particles; indirect searches
look for gamma rays, neutrinos, and positron annihilation remnants.
Most of these searches are subject to many backgrounds resulting from
physical processes that are poorly understood, which means that it
is challenging to obtain reliable background simulations~\cite{Strigari:2013iaa}.
An advantage of indirect searches is their ability to probe the WIMP
annihilation cross section $\left\langle \sigma_{A}v\right\rangle $
directly. If a single particle constitutes DM, the value has been
calculated to be $\left\langle \sigma_{A}v\right\rangle \simeq\unit[2.2\times10^{-26}]{cm^{3}/s}$
for DM masses above $\unit[10]{GeV/c^{2}}$~\cite{Steigman:2012nb},
often referred to as the thermal relic cross section.

There are multiple different types of indirect DM searches; consult,
for example~\cite{Slatyer:2017sev,Gaskins:2016cha} for an in-depth
review. Satellites, such as Fermi-LAT, can be used to search for a
monoenergetic line in the $\gamma$ spectrum usually from dwarf galaxies
as those have relatively little astrophysical backgrounds~\cite{Ackermann:2012qk}.
Interactions of $\gamma$-rays in the atmosphere create a cascade
of secondary particles that produce \v{C}erenkov radiation detectable
in ground-based imaging Air \v{C}erenkov telescopes. There are numerical
backgrounds present in these searches that require extensive simulations~\cite{hess}.

DM annihilation can also create high energy neutrinos. Those can be
detected by the IceCube~\cite{Icecube,Aartsen:2013dxa}, ANTARES~\cite{ADRIANMARTINEZ201669},
or Super-Kamiokande~\cite{Desai:2004pq}experiments, that look for
\v{C}erenkov light generated in ice or water. Unlike the terrestrial
$\gamma$-ray excess searches, detection of muon tracks would be a
strong indication of DM due to a lack of other production processes.
DM could scatter in the Sun causing them to lose enough energy to
be captured gravitationally. As they sink to the core, they can annihilate
and produce an excess of neutrinos~\cite{Wikstrom:2009kw}. These
neutrino detectors have placed strong limits on spin-dependent (SD)
scattering cross section, but unlike direct detection limits, their
results are strongly dependent on assumptions for the DM annihilation
process and are, therefore, strongly model-dependent.

Finally, detection of a positron excess can be a complementary probe
of DM. Despite several experiments observing a positron excess~\cite{FermiLAT:2011ab,Adriani:2013uda,Accardo:2014lma},
it is more likely that nearby pulsars provide the dominant contribution
of this signal~\cite{HAWC}.

\subsection{Direct}

Direct detection searches look for the interaction of the Milky Way
DM halo with nuclei in detectors. Assuming that WIMPs interact via
the weak force, their scattering will cause a recoil of the atomic
nucleus in the target medium. This recoil produces three signals in
the detectors: ionization, scintillation, and heat or phonons. Most
DM detectors then look for one of, or a combination of, those three
signals, which enables the measurement of the event rate as a function
of recoil energy. In comparison, most backgrounds caused energy depositions
of radioactive decays or other charged particles result in an electron
recoil in the target medium, a feature used for background discrimination.
Here we focus our attention on the interaction of WIMPs, but that
is indeed not the only type of DM that researchers are trying to detect
directly. Even the detectors built primarily to search for WIMPs conduct
many different analyses for other types of DM. Additionally, different
detectors are optimized to search for other particles, such as dedicated
axion searches~\cite{Brubaker:2016ktl,SLOAN201695}. 

\subsubsection{WIMP event rates in detectors\label{subsec:WIMP-event-rates}}

\begin{figure}
\begin{centering}
\includegraphics[scale=0.25]{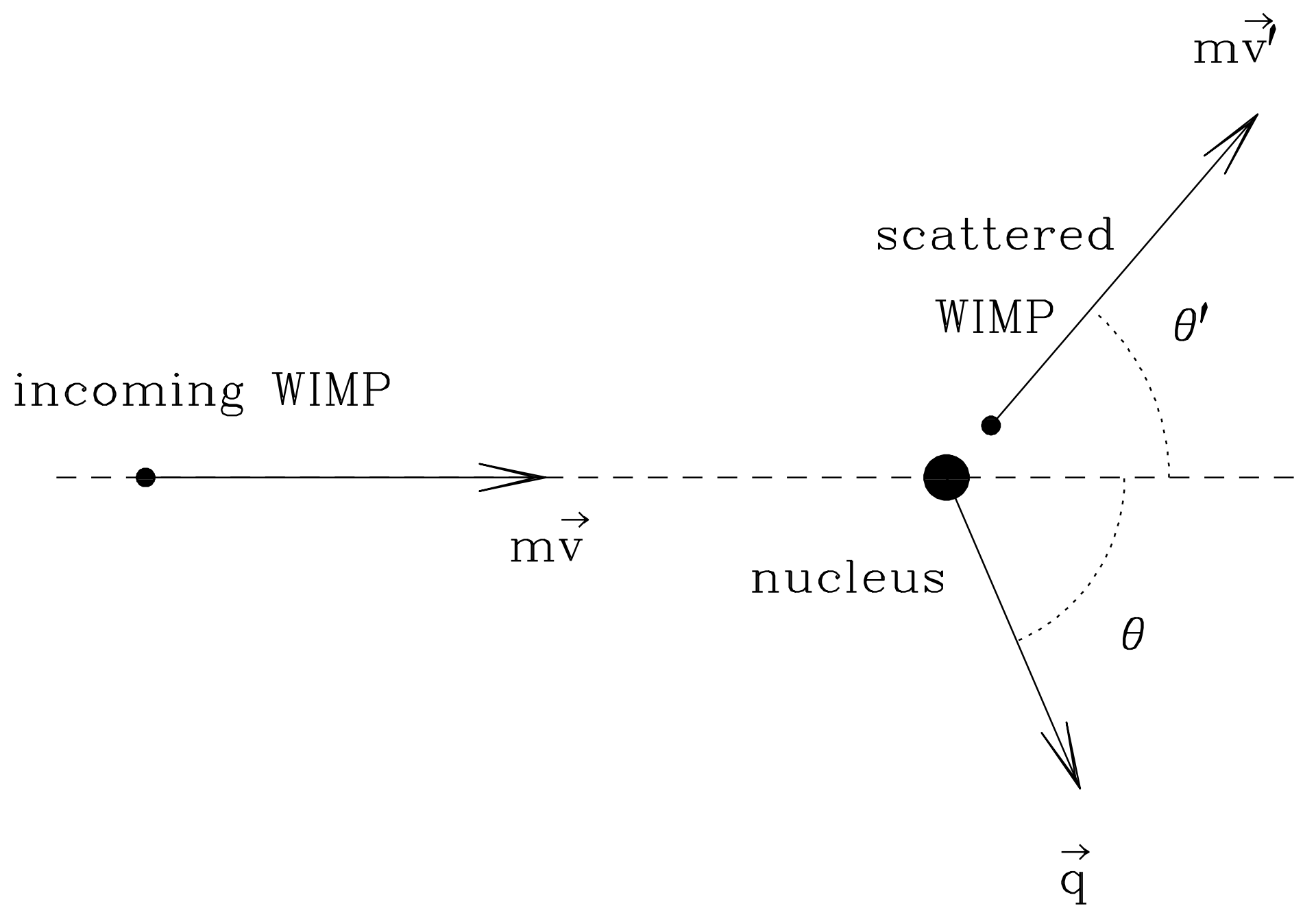}
\par\end{centering}
\caption[Kinematics of elastic WIMP-nucleus scattering]{Kinematics of elastic WIMP-nucleus scattering. Figure from~\cite{Gondolo:2002np}.\label{fig:Kinematics-of-elastic}}
\end{figure}

The energy transfer during a non-relativistic elastic WIMP-nucleus
interaction, illustrated in Figure~\ref{fig:Kinematics-of-elastic},
is given by 
\[
E_{R}=2\frac{\mu_{N}}{m_{\chi}}v^{2}\cos^{2}\left(\theta\right)
\]
where $m_{\chi}$ is the WIMP mass, $\mu_{N}=m_{\chi}m_{N}/\left(m_{\chi}+m_{N}\right)$
is the WIMP-nucleus reduced mass, $\theta$ is the initial direction
of the WIMP, and $v$ is the velocity of WIMP in the lab frame~\cite{riffard:tel-01258830}. 

The quantity of interest is the number of events caused by WIMPs at
a given recoil energy, so we wish to calculate the rate of WIMP-nucleus
scattering as a function of energy $dR/dE_{R}$ in the detector. This
differential rate (the number of events per unit time per unit energy)
is the product of the number of nuclei in the detector, the WIMP flux,
and the WIMP-nucleus cross section. The rate depends on the coupling
strength of a WIMP to SM, while the size of the energy deposition
depends on the kinematics of the DM-nucleus interaction. Below we
focus on the physics aspects of this calculation. A thorough derivation
of the WIMP-nucleus cross section can be found, for example, in~\cite{Lewin:1995rx,Jungman:1995df}.
Additionally, Chapter~\ref{chap:Searching-for-sub-GeV-dm} presents
an overview of scattering rates for sub-GeV DM interactions caused
by Bremsstrahlung and the Migdal effect expected in the LUX detector. 

\subsubsection*{WIMP velocity}

To find the WIMP flux, consider that the Earth is moving around the
galaxy through a DM halo with local density $\rho_{\chi}$, composed
of WIMPs with mass $m_{\chi}$. To calculate the WIMP flux through
the Earth, we assume a DM halo with an isotropic Maxwellian velocity
distribution\footnote{Models that describe the halo as an isotropic isothermal sphere are
referred to as the ``standard halo model.'' This assumption is also
supported by cosmological N-body simulations~\cite{Ling:2009eh}.} $f(\mathbf{v})$. The distribution is truncated at the Milky Way
galaxy's escape velocity $v_{esc}$ since WIMPs traveling faster than
this speed will not be gravitationally bound to the galaxy. The dominant
component of the Earth's velocity is the motion of the Sun around
the Galactic center. The resulting velocity distribution has been
calculated in~\cite{PhysRevD.74.043531,McCabe:2017rln}, and can
also be seen in Equation~\ref{eq:velocity-dist}. Additionally, the
Earth's velocity varies periodically due to its rotation about the
Sun. This effect is expected to produce an annual modulation in the
WIMP flux, peaked around June\footnote{Some detectors attempt to detect this annual modulation, see Sections~\ref{subsec:Scintillating-crystals}
and \ref{subsec:Directional-detection} for details.}.

A minimum WIMP velocity
\[
v_{min}=\sqrt{\frac{E_{R}m_{\chi}}{2\mu_{N}^{2}}}
\]
is needed for a scattering event to occur; lighter WIMPs require higher
velocities in order to be detected. From a kinematic standpoint, the
best-suited target material is the one whose nucleus mass is closest
to the WIMP mass. This is one reason why xenon detectors have an advantage
in heavy WIMP detection, while detectors with lighter target materials
like He, O, Si take the lead when it comes to probing lighter DM~\cite{DM_helium}.

\subsubsection*{WIMP-nucleus cross section}

Next, consider the WIMP-nucleus cross section $\sigma$. It is conventionally
expressed as the product of $\sigma_{0}$ at the coherent scattering
limit in which the WIMP interacts with the entire nucleus assuming
$q=0$ momentum transfer and a nuclear form factor $F^{2}\left(q\right)$.
The form factor accounts for the loss of coherence in the scattering
amplitudes per nucleon due to $q>0$, and generally has an exponential
form. It becomes relevant when the wavelength $h/q$ of the momentum
transfer is on the same scale as the effective nuclear radius $r_{n}$.
For example, for xenon $r_{n}\simeq\unit[6]{fm}$, which corresponds
to recoil energy of 5 keV. For spin-independent (SI) interactions,
the form factor is given by the Fourier transform of the density of
nucleon distribution in the nucleus. A widely-used nuclear form factor
for DM calculations is the Helm form factor~\cite{PhysRev.104.1466},
also known as the Woods-Saxon potential\footnote{The Helm factor is not the only analytical solution for this problem;
other form factors can be used in the calculations~\cite{formFactors}.}. For further discussion of SI form factors, refer to Sections~\ref{subsec:Heavy-scalar-mediator},
\ref{subsec:Light-scalar-mediator}, and~\ref{subsec:Vector-mediator}.
Calculations of form factors for SD scattering are more intriguing
since they require a knowledge of the quark contents of the nucleus
and are discussed in~\cite{Engel:1991wq,Belanger:2008sj}. 

WIMPs can either couple to the mass of the nucleus, referred to as
SI scattering, mediated by a scalar, or to the spin of the target
nucleus, referred to as SD scattering, mediated by an axial-vector\footnote{The SI and SD are the simplest DM coupling scenarios but are model
dependent. However, if DM is more complex one can expect a diverse
spectrum for its structure and interactions. Effective Field Theory
considers a full set of possible operators. This includes momentum-
and velocity-dependent operators in an attempt to relate DM-nucleus
response functions to more general underlying effective theory operators
that mediate the DM interaction. For more information consult for
example,~\cite{Fitzpatrick:2012ix,Fitzpatrick:2012ib,PhysRevC.89.065501,catena}.}. The SI interaction cross section is given by~\cite{Gondolo}:

\[
\sigma_{SI}=\frac{4}{\pi}\mu_{N}\left[Zf_{p}+\left(A-Z\right)f_{n}\right]^{2}
\]
where $Z$ is the atomic number and $A$ is the atomic mass of the
target nucleus (so $A-Z$ is the number of neutrons), and $f_{p}$
and $f_{n}$ are the effective scalar couplings of WIMPs to protons
and neutrons, respectively. Since $f_{n}\simeq f_{p}$ in most theoretical
models, the SI cross section can be simplified to
\[
\sigma_{SI}\simeq\frac{4}{\pi}\mu_{N}^{2}A^{2}\left|f_{p}\right|^{2}.
\]
This can be rewritten as 
\begin{equation}
\sigma_{SI}\simeq\sigma_{n}\frac{\mu_{N}^{2}}{\mu_{n}^{2}}A^{2}\label{eq:WIMP-nucleon}
\end{equation}
to factor out the effect of the target material. Here $\mu_{n}$ is
the WIMP-nucleon reduced mass, and the target-independent spin-independent
cross section of a WIMP on a single nucleon is given by
\[
\sigma_{0}^{SI}=\frac{4}{\pi}\mu_{n}^{2}f_{p}^{2}.
\]

The SD cross section is
\[
\sigma_{SD}=\frac{32}{\pi}G_{F}^{2}\mu_{N}^{2}\frac{J+1}{J}\left[\left\langle S_{p}\right\rangle a_{p}+\left\langle S_{n}\right\rangle a_{n}\right]^{2}
\]
where $G_{F}$ is the Fermi constant, $J$ is the total spin of the
target nucleus, and $\left\langle S_{p,n}\right\rangle $ and $a_{p,n}$
refer to the expectation values of the proton and neutron group spins
and SD couplings, respectively. Notice that the SI scattering benefits
from higher-mass nuclei, while the SD scattering is related to target
nuclei with an unpaired nucleon\@. 

\begin{figure}
\begin{centering}
\includegraphics[scale=0.25]{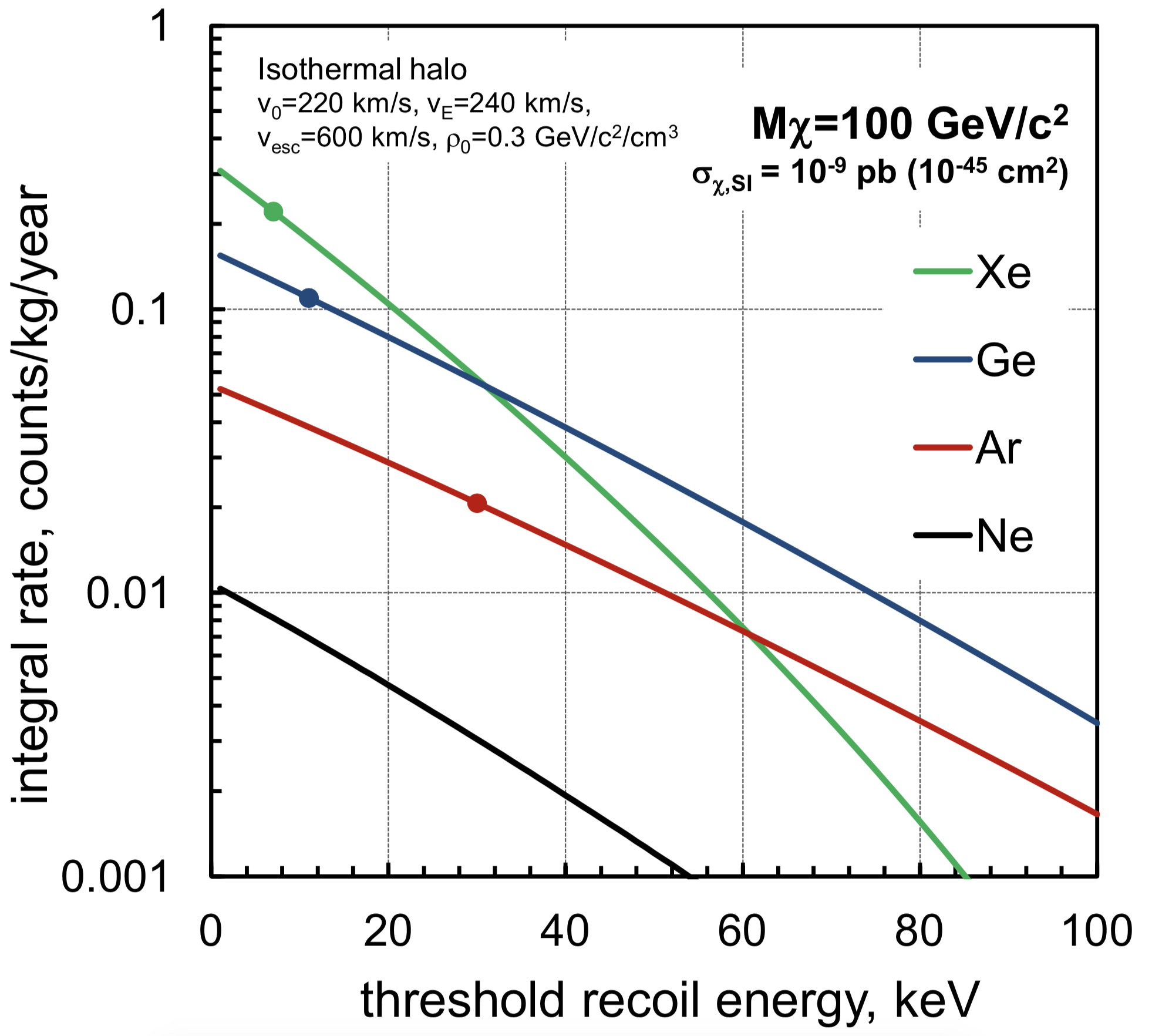}
\par\end{centering}
\caption[Predicted integral spectra for WIMP-nucleus elastic scattering]{Predicted integral spectra for WIMP-nucleus elastic scattering assuming
perfect detector energy resolution for various materials: Xe ($A=131$),
Ge ($A=73$), Ar ($A=40$), and Ne ($A=20$). The spectra are calculated
for a $\unit[100]{GeV/c^{2}}$ WIMP with $\sigma=\unit[10^{-45}]{cm^{2}}$
interaction cross section per nucleon and the halo parameters shown
in the figure. The circles on each curve indicate the typical WIMP
search threshold for each material. Figure from~\cite{Chepel}.\label{fig:Predicted-integral-spectra}}
\end{figure}

Combining the above yields the standard WIMP scattering event rate~\cite{Gondolo}

\begin{equation}
\frac{dR}{dE_{R}}=\frac{\rho_{\chi}}{2m_{\chi}\mu_{N}}\sigma_{0}F^{2}\left(E_{R}\right)\intop_{v_{min}}^{v_{esc}}\frac{f(\mathbf{v})}{v}d^{3}v.\label{eq:elastic-rate}
\end{equation}
As discussed above, the DM distribution is included in $\rho_{\chi}$
and the velocity integral, while the nuclear and particle physics
aspects are included within $\sigma_{0}F^{2}\left(E_{R}\right)$.
Integrating Equation~\ref{eq:elastic-rate} yields the number of
expected events in a detector for a given choice of target, as illustrated
in Figure~\ref{fig:Predicted-integral-spectra}. The low event rates
mean that in order to see DM, detectors need to focus on background
reduction. The exponential tail caused by the form factor results
in detector technologies pushing detection threshold to low energies.
Additionally, present-day experiments use various targets since each
will be most sensitive to a different WIMP mass. 

Even though DM has not yet been detected directly, the non-detection
results, presented as limits on WIMP-nucleon interactions (see, for
examples Figures~\ref{fig:final_exclusion}, \ref{fig:limits-scalar},
or~\ref{fig:limits-vector}), are helpful since they constrain the
theoretical DM models.

\section{Direct dark matter detection technologies}

\begin{figure}[t]
\begin{centering}
\includegraphics[scale=0.32]{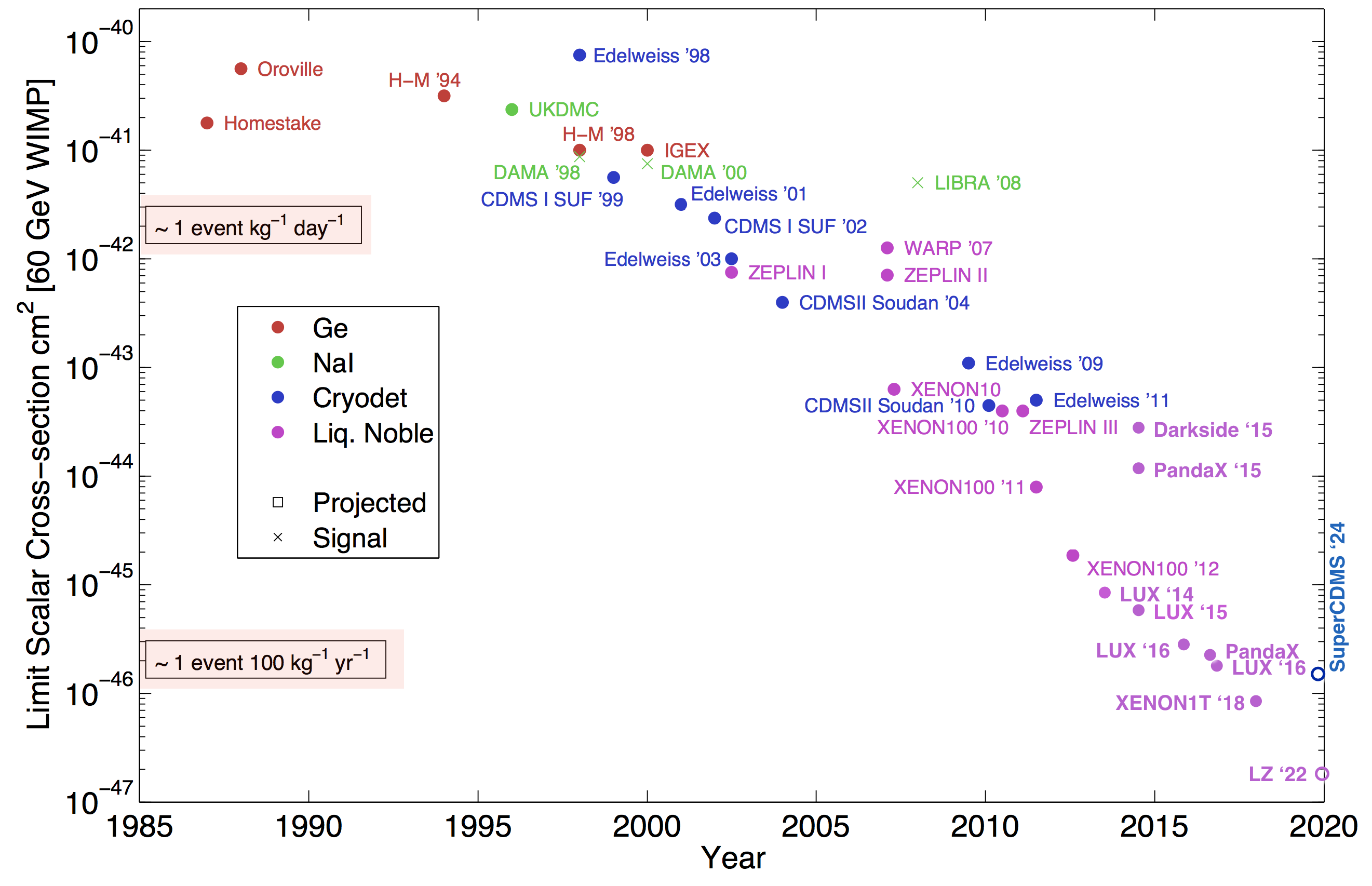}
\par\end{centering}
\caption[Evolution of the WIMP-nucleon spin-independent cross section limit]{Evolution of the WIMP-nucleon spin-independent cross section limit
for a WIMP mass of $\unit[60]{GeV/c^{2}}$. Different colors correspond
to various detector technologies. Empty markers indicate predicted
performance. Liquid noble detectors have been the leading technology
for direct dark matter detection. Figure from~\cite{Gatiskell2017}.\label{fig:moores-law}}
\end{figure}

Given the WIMP interaction rates outlined in the previous section,
various technologies can be leveraged to search directly for these
particles here on Earth. Detectors primarily focus on interactions
of WIMPs with nuclei by elastic scattering, which requires detecting
nuclear recoil energies in the $1-100\,\mathrm{keV}$ range~\cite{Lewin:1995rx}
while being sensitive to cross sections of order $10^{-35}-10^{-48}$~cm$^{2}$.
These very low energies and cross sections constitute an experimental
challenge, but the field has achieved steady progress since the 1990's
as shown in Figure~\ref{fig:moores-law}: sensitivity of detectors
looking for SI WIMP-nucleus interactions has been doubling every 18
months. 

\begin{figure}[t]
\begin{centering}
\includegraphics[scale=0.63]{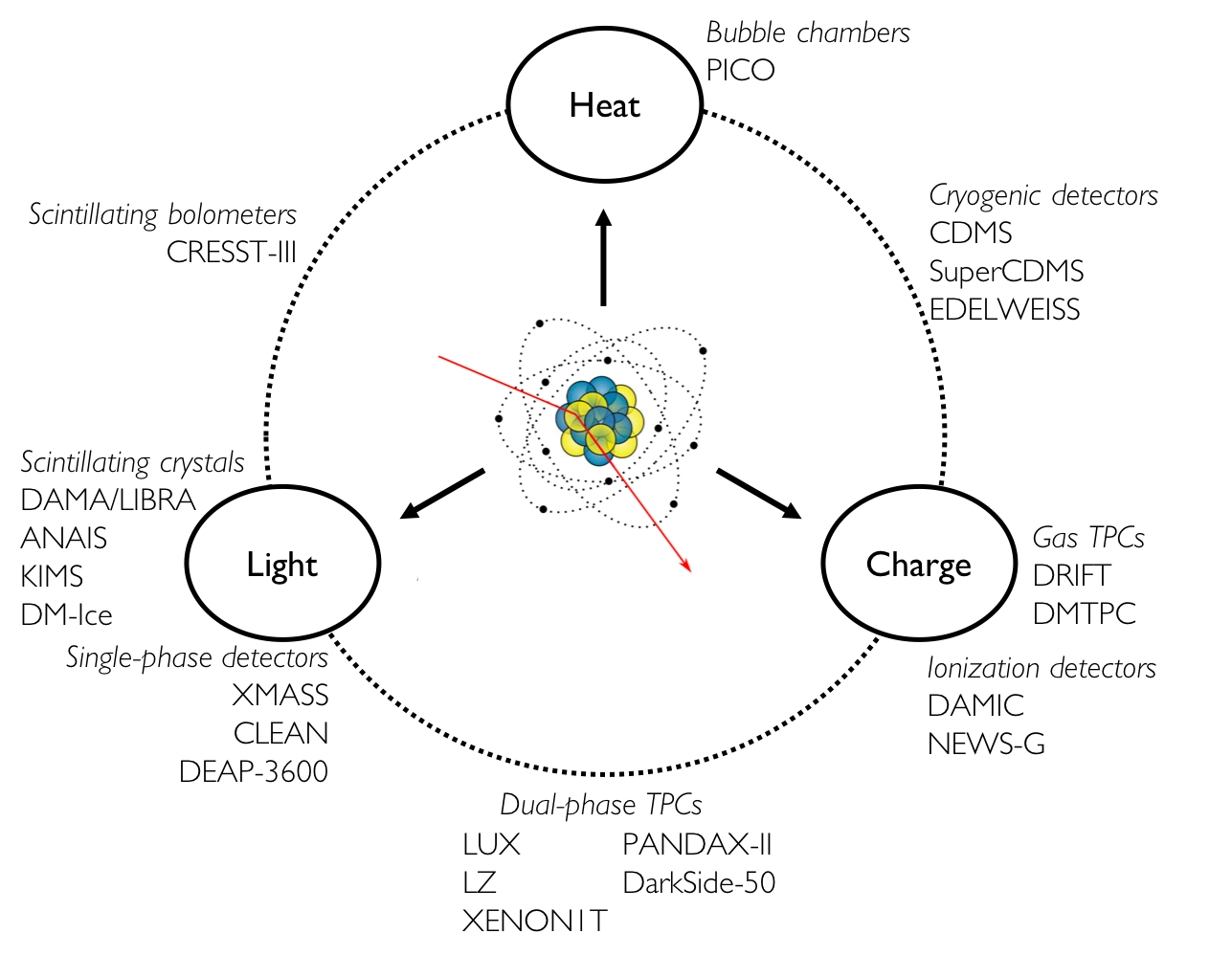}
\par\end{centering}
\caption[Possible signals in direct detection and a sample of experiments]{Possible signals in direct detection and a sample of experiments
leveraging various channels.\label{fig:Possible-signals}}

\end{figure}

A variety of materials are used in detector construction to leverage
their diverse benefits: lighter detectors using Si or Ge lead the
way in the search for lighter WIMPs, while for WIMP masses $\apprge\unit[5]{GeV/c^{2}}$
the most stringent limits are set by xenon detectors. Broad complementarity
of searches is important both to probe a spectrum of DM candidates,
but in case of detection, independent confirmations from various independent
sources will be necessary. This section briefly discusses some of
the detector methods currently used for direct DM detection, with
a focus on technologies that have reached the best limits to date;
Figure~\ref{fig:Possible-signals} shows a graphical overview of
the detector technologies discussed. For more information on direct
detection experiments, see for example~\cite{COOLEY201492,Liu:2017drf,Feng:2014uja,Undagoitia:2015gya}.

Despite various technologies used in the detectors discussed, for
the majority of detectors background reduction is of utmost importance;
detectors use passive shielding, low radioactivity materials, and
environment, which frequently results in their underground location.
Limits obtained by many of the collaborations described below are
also included in Figure~\ref{fig:limits-scalar} in Chapter~\ref{chap:Searching-for-sub-GeV-dm}
that focuses on sub-GeV DM searches with the LUX detector.

\subsection{Noble gas detectors}

Noble liquid detectors are the current leaders in sensitivity to WIMP
masses $\lesssim\unit[5]{GeV}$. Four noble gasses have been considered
for DM detection: helium~\cite{Hertel:2018aal}, neon~\cite{MCKINSEY2007152},
argon, and xenon. Only argon and xenon detectors have thus far placed
limits on DM interactions. The most successful xenon detectors are
two-phase time projection chambers (TPC), of which LUX is one example,
described in great detail in Chapter~\ref{chap:The-LUX-experiment}.
This approach was pioneered by the ZEPLIN~\cite{Akimov:2011tj} and
XENON10 experiments \cite{xenon10}. The LUX detector was the first
to achieve sub-zeptobarn ($\unit[<1\times10^{-45}]{cm^{2}}$) sensitivity~\cite{Akerib:2013tjd}.
Currently, there is a tight race between three collaborations building
detectors with increasing target mass: PandaX \cite{Cui:2017nnn}
located at the JinPing underground laboratory (CJPL) in China, XENON1T~\cite{Aprile:2018dbl}
and its planned upgrade XENONnT located in the Gran Sasso National
Laboratory (LNGS) in Italy, and LZ~\cite{Akerib:2018lyp}, discussed
in Chapter~\ref{chap:High-voltage-development-LZ}, planned to start
operations in 2019 at the Sanford Underground Research Facility in
the USA. Looking ahead, the DARWIN (DARk matter WImp search with liquid
xenoN) collaboration is considering a detector with up to 40 tons
of target mass~\cite{darwin}. 

An alternative to the two-phase TPC is the single phase detector.
Since single-phase detectors only use scintillation signal, position
reconstruction of interaction vertices is a significant challenge.
One such xenon detector is currently operational, the XMASS (Xenon
detector for Weakly Interacting MASSive Particles) experiment at the
Kamioka Observatory in Japan. Its PMTs cover more than 62\% of the
inner surface in $4\pi$ resulting in large photoelectron yield~\cite{MINAMINO2010448,Abe:2018mxq}. 

Argon is frequently used as an alternative to xenon due to its lower
cost, which allows construction of more massive detectors leading
to more exposure. However, the presence of the radioactive isotope
of $^{39}$Ar hinders the performance of these detectors in DM searches;
this limitation can be partially overcome by using underground argon~\cite{Agnes:2015ftt,Galbiati:2007xz,Back:2012pg}
or through distillation~\cite{Back:2012tx} to reduce the $^{39}$Ar
content. There are currently three detectors using LAr for DM detection:
the DarkSide collaboration uses a two-phase TPC technology~\cite{AGNES2015456},
while DEAP-3600 (Dark Matter Experiment using Argon Pulse-shape)~\cite{KUZNIAK2016340}
and MiniCLEAN (Cryogenic Low-Energy Astrophysics with Noble liquids)~\cite{JuiJen:2017juc}
utilize the single-phase technology. Unlike in xenon, the argon scintillation
long-lived triplet and the short-lived singlet states have very different
decay times. This enables pulse shape discrimination which can achieve
background reduction by $10^{-7}$~\cite{AGNES2015456}. This means
that good background rejection can be achieved using only the scintillation
signal. A thorough review of technological advances for both light
and charge can be found in~\cite{Chepel}.

\subsection{Cryogenic solid state detectors}

Cryogenic detectors operate below 100 mK and aim to detect small increases
in temperature. They provide the leading limits for WIMP masses $\lesssim\unit[5]{GeV/c^{2}}$
by detecting phonons produced within the crystal lattice by scattering
of individual DM particles. The CDMS (Cryogenic Dark Matter Search)
collaboration was the first to deploy such a detector using germanium
and silicon crystals initially at a shallow site at the Stanford Underground
Facility~\cite{Akerib:2010pv} and then at the Soudan Underground
Laboratory in Minnesota, USA~\cite{CDMSII}. The detectors are sensitive
to both the phonon (heat) and ionization signals and provide excellent
discrimination of electron recoils via ionization/heat ratio and pulse
shape discrimination (with a misidentification rate of less than 1
in $10^{6}$). The phonons are detected using transition edge sensors\footnote{Recently, this technology has demonstrated single-electron sensitivity~\cite{doi:10.1063/1.5010699}.}
while an electric field is set up to collect the ionization charge.
The SuperCDMS collaboration uses advanced technology with improved
identification of background events near the surfaces of the crystals~\cite{SuperCDMS,Agnese:2015nto}.
SuperCDMS is planning an increase in detector mass and exposure, as
well as relocation to a deeper lab (SNOLAB in Sudbury, Canada) in
order to reduce backgrounds and search for WIMP masses between 0.5
and $\unit[7]{GeV/c^{2}}$~\cite{PhysRevD.95.082002}. 

The leading limits for low mass WIMPs below $\unit[1.8]{GeV/c^{2}}$
were obtained by the CRESST (Cryogenic Rare Event Search with Superconducting
Thermometers) collaboration. The CRESST-II experiment used 300 g of
CaWO$_{4}$ crystals operating at $\sim\unit[10]{mK}$ located at
LNGS. By detecting phonons and scintillation signal with an energy
threshold of $\unit[\sim300]{eV}$, they were the first to place nuclear
recoil cross section limits on DM with masses below $\unit[1]{GeV/c^{2}}$~\cite{Angloher:2015ewa}.
The upgraded detector with smaller crystals and an improved threshold
of $<\unit[100]{eV}$, CRESST-III has released preliminary improved
limits compared to its predecessor~\cite{Petricca:2017zdp}. The
collaboration recently placed limits on DM masses down to $\unit[140]{MeV/c^{2}}$
using a prototype 0.5~g sapphire (Al$_{2}$O$_{3}$) detector operating
on the surface~\cite{Angloher:2017sxg,Strauss:2017cuu}. The EDELWEISS
collaboration is also using cryogenic germanium detectors for phonons
and ionization signals for DM searches~\cite{edelweissIII,Armengaud:2012pfa}. 

\subsection{Superheated liquid detectors}

Bubble chambers and superheated droplet detectors provide the leading
limits for SD WIMP-proton interactions\footnote{Xenon detectors hold the best limits on SD WIMP-neutron interactions~\cite{Akerib:2017kat}.}.
Only energy depositions from nuclear recoils or $\alpha$ events with
large enough energy will create a bubble in a superheater liquid detector,
which can then be photographed and acoustically recorded. The use
of acoustics allows discrimination of $\alpha$ background events
that also have enough energy for bubble nucleation. Since ER will
not deposit enough energy to trigger a formation of bubbles, this
technique features strong ER background discrimination. However, this
technique requires one to perform a scan over various operating temperatures
and pressures to produce an energy spectrum, rather than measuring
energy on a per-event basis. The PICO collaboration is currently operating
two chambers and has tested two different liquids C$_{3}$F$_{8}$
and CF$_{3}$I in SNOLAB~\cite{PICO2L,PICO60}. 

A recent R\&D effort at Northwestern University used a 30~g xenon
bubble chamber to detect nuclear recoils using both simultaneous scintillation
and bubble nucleation. This chamber is instrumented with a PMT, and
a near-IR camera and a piezoelectric acoustic transducer to detect
bubbles. The detector combines an improved electron recoil rejection
compared to CF$_{3}$I with an energy resolution of a liquid scintillator.
The improved electron recoil rejection is likely caused by the scintillation
production suppressing bubble nucleation by electron recoils, while
nuclear recoils remain unaffected~\cite{Baxter:2017ozv}. This approach
may allow both energy reconstruction (by measuring scintillation)
and ER background rejection (by detecting bubble formation).

\subsection{Scintillating crystals\label{subsec:Scintillating-crystals}}

The most notable example of the use of scintillating crystals is the
DAMA/LIBRA experiment that observed an annual modulation in the event
rate in 250~kg of scintillating inorganic NaI(Tl) crystals that they
attribute to a DM signal~\cite{Bernabei2013,Bernabei:2015xba}. This
experiment is a successor to the DAMA/NaI experiment which observed
an annual modulation throughout 1995-2002 with $\sim\unit[100]{kg}$
of NaI(Tl) crystals~\cite{Bernabei:2002pz}. Their claim is not compatible
with the WIMP hypothesis and has been refuted by many other experiments'
null results using other detector materials. The majority of the DM
community is skeptical of the DAMA/LIBRA results, see for example~\cite{Ralston:2010bd,nygren2011testable,Davis:2014cja,McKinsey:2018xdb}.

Nevertheless, efforts to test their results using NaI detectors are
underway. The SABRE (Sodium Iodide with Active Background REjection)
collaboration plans to operate one detector each on the northern (at
LNGS) and southern hemispheres (at the Stawell gold mine in Victoria,
Australia). This will reduce the chance for false positives from possible
seasonal systematic effects~\cite{TOMEI2017418}. The COSINE-100
experiment is a joint effort between the DM-Ice~\cite{deSouza:2016fxg}
and KIMS~\cite{kims} collaborations located at the Yangyang Underground
Laboratory in Korea.  Its first 106~kg of low-background NaI(Tl)
has been taking physics data since September 2016~\cite{Adhikari:2017esn}.
Lastly, ANAIS\textendash 112 (Annual modulation with NaI Scintillators)
is a dark matter search using 112.5~kg of NaI(Tl) scintillators located
at the Canfranc Underground Laboratory in Spain~\cite{Amare:2016fmp,Coarasa:2017aol}.

\subsection{CCD chips}

The DAMIC (DArk Matter In CCDs) experiment uses high resistivity charge-coupled
silicon detectors (known as CCD chips) to record ionization signals
produced in the silicon pixels~\cite{BARRETO2012264}. They have
recently released a result using 23.2~g of material with 0.6~kg-day
exposure at SNOLAB for WIMP masses below $\unit[10]{GeV/c^{2}}$~\cite{Aguilar-Arevalo:2016ndq}.
A larger detector, DAMIC100 with 100~g of material is being deployed
at the same place, and a 1-kg detector DAMIC-M has been approved for
installation at the Laboratoire Souterrain de Modane (LSM) in France. 

\subsection{Spherical proportional counters}

The NEWS-G (New Experiments With Spheres-Gas) is a large copper vessel
filled with a mixture of Ne and CH$_{4}$ with a small high voltage
sensor in its center. This creates a graded electric field in the
detector that causes an avalanche process if an ionization electron
is generated. Therefore, the detector features a very low energy detection
threshold. The detector is located at LSM and recently published new
limits~\cite{Arnaud:2017bjh}. 

\subsection{Directional detection\label{subsec:Directional-detection}}

Directional detectors are trying to detect the ``WIMP wind'' from
the motion of Earth through the galaxy using low-pressure gas TPCs.
The low-pressure gas allows 3D reconstruction of particle tracks;
particles that pass through the detector in directions other than
the expected WIMP wind are deemed background. Additional background
discrimination comes from track length. One of the main drawbacks
of these detectors is their low-pressure nature since a large ($\unit[1]{m^{3}}$)
detector contains only $\unit[\sim100]{g}$ of target mass making
it difficult to reach large exposures. This type of detectors was
pioneered by the DRIFT (Directional Recoil Information From Tracks)
collaboration. Rather than drifting ionization charge, the detector
drifts negative ions that captured liberated electrons in particle
interactions. The DRIFT detector was installed in the Boulby mine
in the UK and was filled with either CS$_{2}$ or CF$_{4}$ gas mixtures
and currently has achieved the leading limit among the directional
detectors for SD WIMP-nucleon scattering for WIMP masses above $\unit[10]{GeV}/c^{2}$~\cite{BATTAT20151}.
A detector upgrade with a larger active volume is underway~\cite{DRIFTII}.
The DMTPC m$^{3}$ experiment~\cite{dmtpc} uses light and charge
readout to measure the directional anisotropy of nuclear recoils.
Most collaborations working on directional detection are planning
to join their forces in the CYGNUS collaboration~\cite{cygnus}. 

\subsection{Overview of the status of the field}

\begin{figure}[t]
\centering{}\includegraphics[scale=0.475]{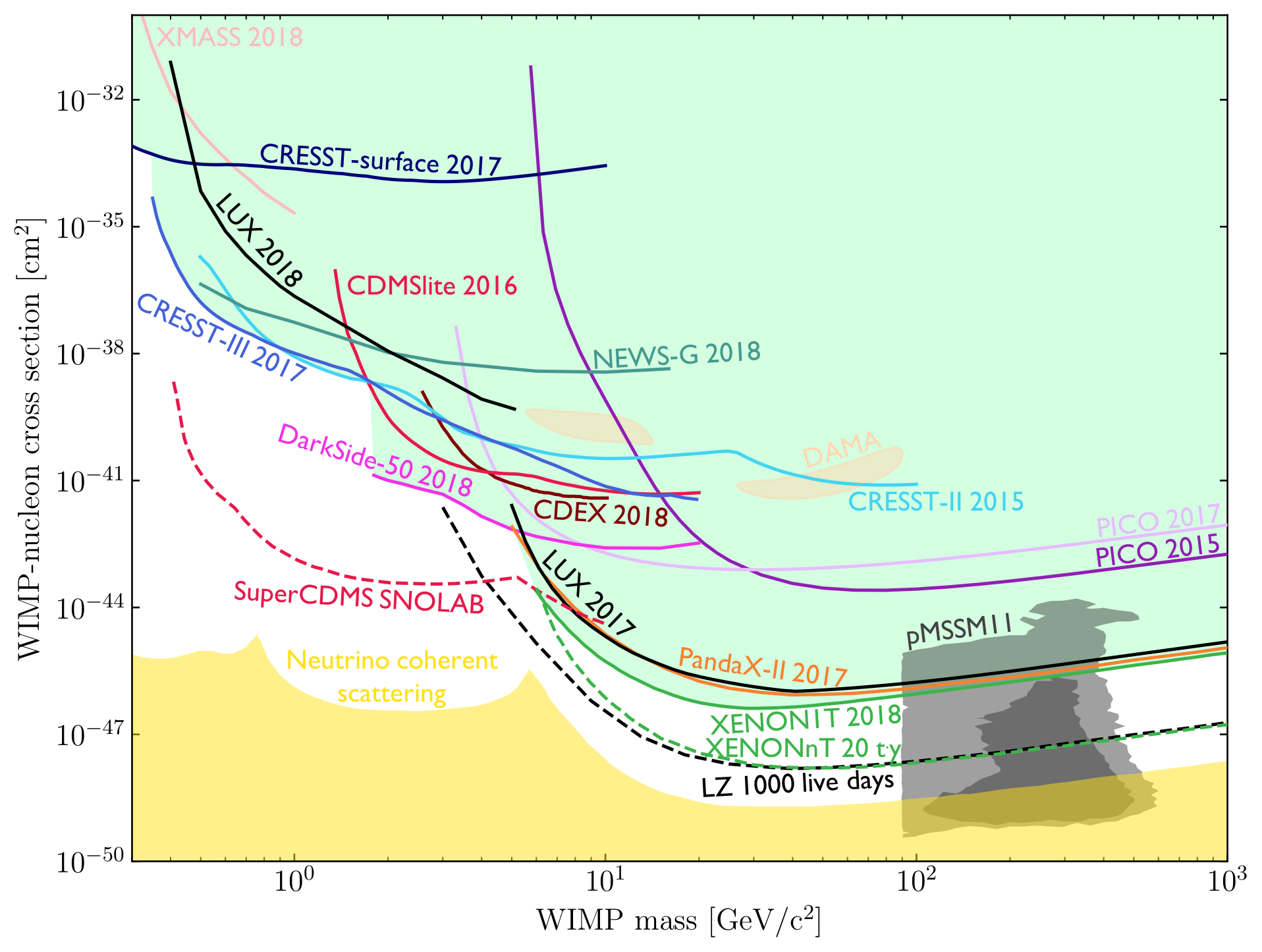}\caption[A compilation of WIMP-nucleon spin-independent cross section limits]{A compilation of WIMP-nucleon spin-independent cross section limits
(solid curves), DM signal hints (closed contours), and projections
(dashed curves). Limits for a heavy mediator along with limits from
the spin-independent analyses of LUX~\cite{Akerib:2017kat,Akerib:2018subGeV}
(black), PandaX-II~\cite{Cui:2017nnn} (orange), XENON1T~\cite{Aprile:2018dbl}
(green), CDEX-10~\cite{Jiang:2018pic} (maroon), CDMSlite~\cite{Agnese:2015nto}
(red), CRESST-II~\cite{Angloher:2015ewa} (cyan), CRESST-III~\cite{Petricca:2017zdp}
(blue), CRESST-surface~\cite{Angloher:2017sxg} (navy), DarkSide-50~\cite{Agnes:2018ves}
(magenta), NEWS-G~\cite{Arnaud:2017bjh} (teal), PICO C$_{3}$F$_{8}$
2017 \cite{Amole:2017dex} (purple), PICO CF$_{3}$I 2015~\cite{PICO60}
(lavender), and XMASS~\cite{Abe:2018mxq} (pink). The dashed curves
show projections for LZ~\cite{Akerib:2018lyp} (black), XENONnT \cite{xenon1tDesign}
(green), and SuperCDMS SNOLAB~\cite{Agnese:2016cpb} (red). Also
shown is a hint for a WIMP signal from DAMA~\cite{Savage:2008er}
(orange). The yellow region indicates a background from coherent neutrino-nucleus
scattering~\cite{Ruppin:2014bra}. The gray region shows predictions
from pMSSM11~\cite{Bagnaschi:2017tru,MasterCode}. \label{fig:limits-plot}}
\end{figure}

Figure~\ref{fig:limits-plot} shows a compilation of limits and signal
hints for WIMP-nucleon spin-independent interactions from some of
the experiments discussed above, as well as projections from future
experiments. As detectors become more sensitive, they approach an
irreducible background from coherent neutrino-nucleus scattering,
which limits their discovery potential. The so-called ``neutrino
floor'' is shown for a xenon target with no other backgrounds, 100\%
efficiency, an energy threshold of 3~eV, and a 1000~ton-year exposure
as calculated in~\cite{Ruppin:2014bra}. Also shown is the prediction
region from the phenomenological MSSM (pMSSM11) constrained using
the MasterCode~\cite{Bagnaschi:2017tru,MasterCode}. The MasterCode
compiles current experimental data and allows fits to different versions
of MSSM.

The non-detection of WIMPs with masses larger than a few GeV/c$^{2}$
has led to an increase of detection efforts in the sub-GeV region.
This includes both a focus on the development of new detector technologies
(e.g.,~\cite{Hertel:2018aal}) and a focus of current experiments
on increasing their sensitivity to lower mass DM as discussed in Chapter~\ref{chap:Searching-for-sub-GeV-dm}.

\section{Summary and outline }

The existence of DM is well motivated, with many potential candidate
particles and detection mechanisms. The rest of this manuscript dives
into the various aspects of direct detection of DM using a two-phase
xenon time projection chamber. The detector technologies are quickly
maturing, thanks to a large number of R\&D efforts. This enables scaling
of the detectors to achieve unprecedented low background, which might
lead to convincing evidence for positive DM signal in the future.

Chapter~\ref{chap:The-LUX-experiment} provides an overview of the
LUX detector, which led the field for several years. The remainder
of this dissertation then discusses original work related to direct
DM detection. Chapters~\ref{chap:efield-modeling} and~\ref{chap:Searching-for-sub-GeV-dm}
are devoted to original work pertaining to LUX: Chapter~\ref{chap:efield-modeling}
discusses 3D modeling of electric fields of the LUX detector and Chapter~\ref{chap:Searching-for-sub-GeV-dm}
presents limits on sub-GeV DM using the 2013 LUX dataset. Chapter~\ref{chap:High-voltage-development-LZ}
starts with an overview of the LZ detector, the successor to LUX,
and then discusses design and construction of a couple of noble liquid
purity monitors used in high voltage research and development (R\&D).
Finally, Chapter~\ref{chap:XeBrA} describes the design, construction,
and data acquisition of the Xenon Breakdown Apparatus (XeBrA), designed
to study the breakdown behavior of noble liquids. Additionally, since
a diverse body of researchers is essential for success, Appendix~\ref{chap:LZ-E=000026I}
presents a synopsis of the Equity \& Inclusion work done as part of
the LZ collaboration.

\chapter{\textsc{The LUX experiment}\label{chap:The-LUX-experiment}}

The Large Underground Xenon (LUX)\nomenclature{LUX}{Large Underground Xenon}
experiment was a two-phase liquid-gas xenon time projection chamber
(TPC)\nomenclature{TPC}{Time Projection Chamber} containing 370 kg
of xenon. LUX was formed by $\sim100$ collaborators from 24 institutions
in the USA, UK, and Portugal. This chapter covers the basics of two-phase
TPC operations, the design of the LUX detector, highlights of some
its subsystems and calibration techniques, and provides an overview
of the data analyses leading to several world-leading exclusion limits
for WIMP-nucleon elastic scattering.

\section{Two-phase time projection chambers}

Originally, single-phase gas filled projection chambers were developed
for high-energy physics in 1974 by David R. Nygren at Lawrence Berkeley
National Laboratory (LBNL)\nomenclature{LBNL}{Lawrence Berkeley National Laboratory}\footnote{By coincidence, the XeBrA experiment I built, discussed in Chapter~\ref{chap:XeBrA},
is located in a lab space at LBNL that was inherited from D. Nygren. }~\cite{Nygren:1976fe}. The design was modified and applied to a
single-phase liquid argon TPC in 1977 by Carlo Rubbia~\cite{Rubbia:1977zz}.
Over time, thanks to their particle detection capabilities, both single
and two-phase noble gas TPCs have been deployed throughout the experimental
direct dark matter field using xenon~\cite{Akerib:2016vxi,Aprile:2017iyp,Cui:2017nnn}
and argon~\cite{Agnes:2015ftt,JuiJen:2017juc,Marchionni:2010fi,Amaudruz:2017ekt}.
However, the development and refinement of two-phase TPCs unlocked
an opportunity for improved detection and exclusion mechanisms. Nowadays,
two-phase xenon TPCs provide the most stringent limits on WIMP candidates
with masses of $\sim5$-100,000 GeV/c$^{2}$ due to their low intrinsic
backgrounds, position reconstruction and discrimination between electron
and nuclear recoils, which enables excellent background rejection. 

\begin{figure}[t]
\centering{}\includegraphics[scale=0.5]{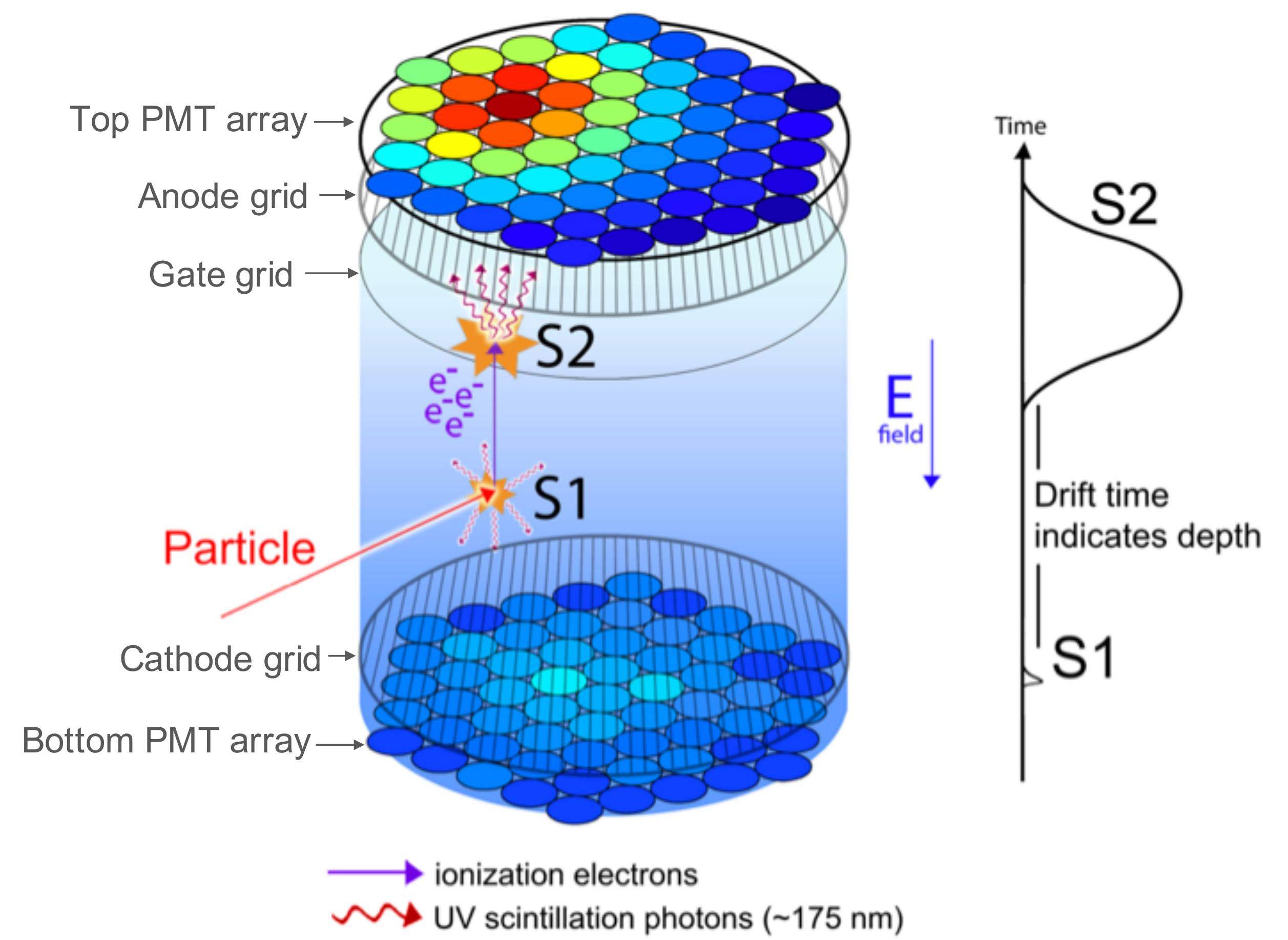}\caption[TPC detection principle]{Illustration of detection principles of a two-phase TPC. A particle
interacts with a xenon atom and produces prompt scintillation photons
(S1) that get detected by the PMT arrays and ionization electrons.
The electron cloud drifts against the applied electric field to the
top of the detector, where it creates a secondary proportional scintillation
signal (S2) in the gas. The S2 signal is detected by the top PMT array
and enables $\left(x,y\right)$ position reconstruction on the order
of millimeters. The depth of the event is given by the time separation
between the S1 and S2 signals. The amount of liquid xenon enclosed
between the gate and cathode grids and the detector walls is called
the active volume. Figure by Carlos Faham. \label{fig:TPC-detection-principle}}
\end{figure}

In a two-phase TPC, scintillation and ionization signals produced
by excitations of xenon atoms and electron-ion pairs can be detected.
This provides a crucial feature of TPCs as dark matter detectors since
this light-to-charge ratio is far lower for recoiling nuclei, caused
either (ideally) by WIMPs or by neutron scattering, than for recoiling
electrons, which are caused by the vast majority of background interactions.
Figure~\ref{fig:TPC-detection-principle} illustrates the operation
of the LUX detector. First, a particle traveling through a detector
collides with a xenon atom and deposits energy in the form of prompt
scintillation (light), ionization (charge), and heat. The scintillation
light is detected within 100~ns using sensitive light detectors known
as photomultiplier tubes (PMTs)\nomenclature{PMT}{Photomultiplier Tube}
at the top and bottom of the detector volume. This primary signal
is referred to as S1. The ionization electrons liberated during the
collision drift to the top of the detector in the presence of an electric
field. The electrons are extracted out of the liquid xenon using a
stronger electric field established by a pair of grids. This accelerating
cloud of electrons creates a secondary proportional scintillation
signal (S2), also known as electroluminescence, as it travels through
a thin layer of gaseous xenon. 

The TPC provides a wealth of information for each scattering event:
its energy, the event location within the detector, and whether it
is an electron or nuclear recoil. The total energy deposited in the
event can be estimated from a linear combination of the sizes of the
S1 and S2 signals. The location of the event can be reconstructed
from the S1 and S2 signals: the $\left(x,y\right)$ position of the
event is given by the distribution of S2 photons incident on the top
PMT array with up to millimeter resolution and the $z$ coordinate
is given by the time separation of the S1 and S2 signals, given by
the electron drift time\footnote{For the S1 signal, the PMT rise time as discussed in Section~\ref{subsec:Detector-design}
is $\mathcal{O}\left(\mathrm{ns}\right)$ while for the S2 signal
the electron drift time is $\mathcal{O}\left(\mathrm{\mu s}\right)$,
making them easy to distinguish. This can be seen in Figure~\ref{fig:raw_signal}.}. Furthermore, the ratio of the S2 and S1 signals differs for electron
recoils (ER)\nomenclature{ER}{Electron Recoil} and nuclear recoils
(NR)\nomenclature{NR}{Nuclear Recoil}. In ERs, an incident particle
scatters off a bound electron, or a $\beta$ decay occurs. The majority
of ERs are generated from background events, mostly stemming from
radioactivity such as $\beta$, x-rays, and $\gamma$ radiation. It
is expected that WIMPs would cause NRs, along with interactions from
background neutrons. Distinguishing NRs from ERs is, therefore, a
crucial tool for improving dark matter searches. 

As might be clear, many pieces need to come together to enable dark
matter searches. First, a dark matter search entails building the
detector (Section~\ref{subsec:Detector-design}) and thoroughly understanding
its operations and operating conditions (Sections~\ref{subsec:Xenon-circulation-and}
and~\ref{subsec:Performance-monitoring}). This starts by interpreting
interactions occurring inside the detector by understanding xenon
microphysics (Section~\ref{subsec:Liquid-xenon-physics}), describing
efficiencies of detection of light and charge generated by the interaction
(Section~\ref{subsec:Calibrations}), performing signal processing
(Section~\ref{subsec:Electronics-and-data}), interpreting the data
correctly and conducting appropriate analysis (Section~\ref{subsec:Data-processing})
before generating the final result. The construction and signal interpretation
is the bulk of the work in detector science, but appropriate software
modeling, accurate simulation of the detector itself, backgrounds
present and particle interactions are also important (Section~\ref{sec:Simulations-and-backgrounds}).
Information for the background simulation is obtained via a thorough
radioactive screening campaign of the detector materials and parts.
This information can be implemented in the detector's 3D simulation
software, and with a simulation of xenon physics, all the models needed
for the final result can be built. Together all those efforts contribute
to the great science done by the LUX experiment over the years (Section~\ref{sec:LUX-WIMP-search}).

\subsection{Liquid xenon physics\label{subsec:Liquid-xenon-physics}}

Liquid xenon (LXe)\nomenclature{LXe}{Liquid Xenon} provides several
advantages for dark matter detection. This section first gives a synopsis
of xenon qualities that make it a suitable element for dark matter
searches and then provides a characterization of xenon's response
to particle interaction and energy deposition reconstruction techniques.

\begin{figure}
\begin{centering}
\includegraphics{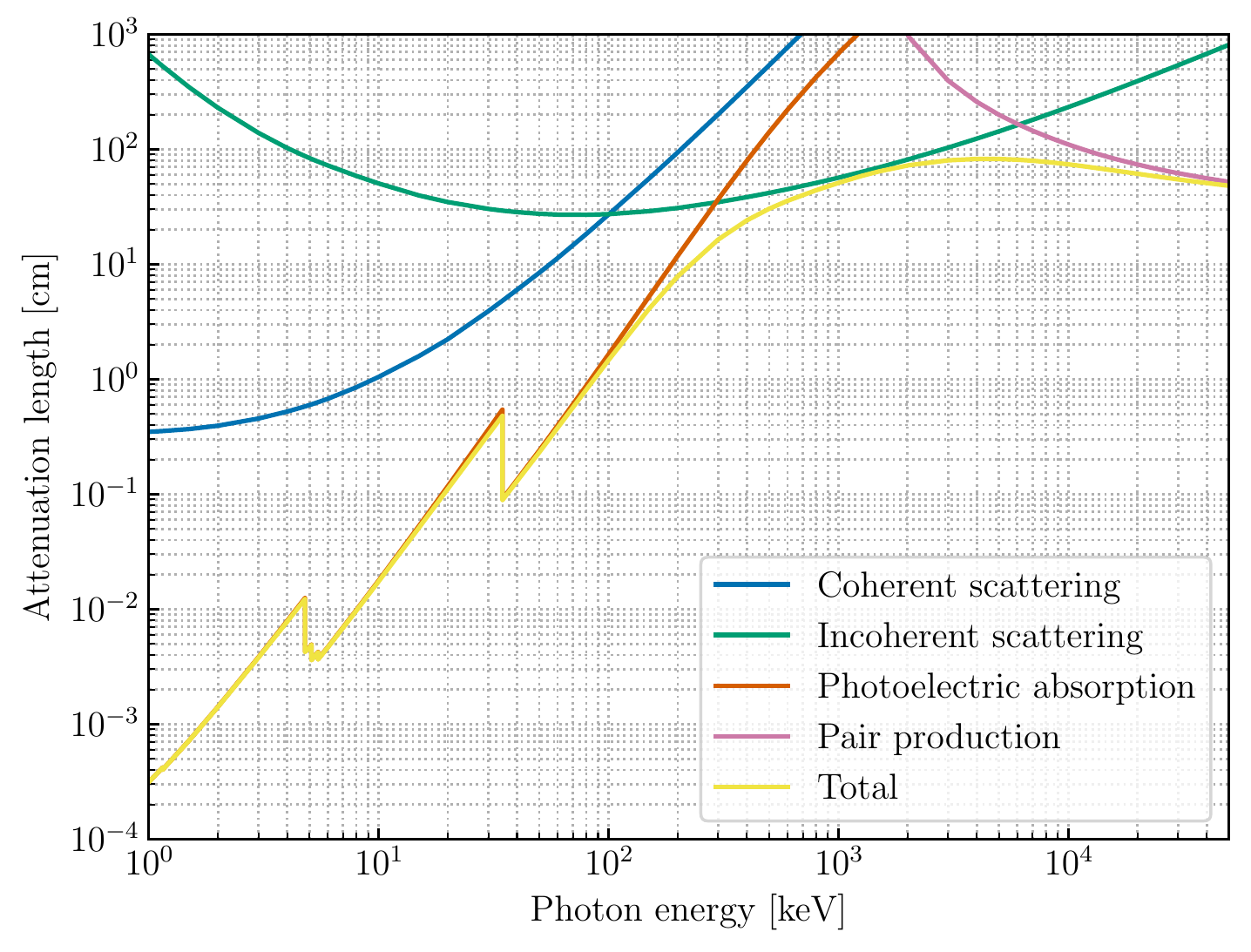}
\par\end{centering}
\caption[Photon attenuation length in xenon]{Photon attenuation length in xenon assuming $\rho_{Xe}=\unit[2.95]{g/cm^{3}}$.
Total is the maximum attenuation length from all effects combined;
the photoelectric effect dominates at low photon energies. Data from~\cite{photon_scattering}.\label{fig:Photon-cross-section}}
\end{figure}

Xenon\footnote{The name derived from the Greek \textit{xenos} for ``the stranger''
and was discovered in 1898 in a liquefied air sample by Scottish chemist
William Ramsay and English chemist Morris William Travers~\cite{de2003atomic}.} is the heaviest of the stable noble gases with atomic number $Z=54$
and atomic mass $A=\unit[131.293(6)]{g}$. Xenon's high density of
$\sim2.95$~g/cm$^{3}$ at $\unit[170]{K}$ enables compact detector
design and allows for efficient self-shielding, creating an ultra-low
background inner detector volume (also referred to as ``fiducial
volume'') suitable for rare event searches. The most significant
impact of self-shielding against typical external gamma-rays is realized
in the outermost few centimeters of xenon, as illustrated in Figure~\ref{fig:Photon-cross-section},
and $\beta$-rays penetrate even shorter distances. The high mass
number $A$ also results in good sensitivity to the spin-independent
(SI)\nomenclature{SI}{Spin-Independent} WIMP-nucleon interaction,
and a 48\% abundance of neutron-odd isotopes enables sensitivity to
the spin-dependent (SD)\nomenclature{SD}{Spin-Dependent} WIMP-neutron
interaction; the list of all known natural xenon isotopes is in Table~\ref{tab:Xenon-isotopic-abundance}.
Xenon has essentially no long-lived naturally abundant radioactive
isotopes, except for the $2\nu\beta\beta$ decay of $^{136}$Xe with
a measured half-life of $2.2\times10^{21}$ years~\cite{Albert:2013gpz,KamLAND}.
Xenon also has only moderate cryogenic requirements with a triple
point of $\unit[161]{K}$, and commercial gas-phase xenon purifiers
(getters) are available for the removal of electronegative and light-absorbing
impurities such as O$_{2}$ and N$_{2}$. Furthermore, xenon is transparent
to its scintillation light with wavelength $\lambda=\unit[178]{nm}$
(7 eV), since this energy is lower than the first ionization potential
of xenon, which is 12.1~eV.

\begin{table}
\begin{centering}
\begin{tabular}{lrr}
\hline 
Isotope & Atomic mass/u & Natural abundance\tabularnewline
\hline 
\hline 
$^{124}$Xe & 123.906 & 0.09\%\tabularnewline
$^{126}$Xe & 125.904 & 0.09\%\tabularnewline
$^{128}$Xe & 127.904 & 1.91\%\tabularnewline
$^{129}$Xe & 128.905 & 26.40\%\tabularnewline
$^{130}$Xe & 129.904 & 4.07\%\tabularnewline
$^{131}$Xe & 130.905 & 21.23\%\tabularnewline
$^{132}$Xe & 131.904 & 26.91\%\tabularnewline
$^{134}$Xe & 133.905 & 10.44\%\tabularnewline
$^{136}$Xe & 135.907 & 8.86\%\tabularnewline
\hline 
\end{tabular}\caption[Xenon isotopic atmospheric abundance]{Atmospheric abundance of known natural xenon isotopes. Data from~\cite{de2003atomic}.\label{tab:Xenon-isotopic-abundance}}
\par\end{centering}
\end{table}

As mentioned above, a particle interaction in LXe deposits energy
into three channels: heat, light, and charge. This energy deposition
leaves xenon atoms in both excited and ionized states. Within picoseconds,
the excited Xe atoms combine with neighboring Xe atoms in their ground
state to form excitons, short-lived diatomic molecules Xe$_{2}^{*}$:
\begin{equation}
\mathrm{Xe^{*}+Xe\rightarrow Xe_{2}^{*}.}\label{eq:scint1}
\end{equation}
Excitons can exist either in their singlet state with an electron
spin quantum number $s=0$ (corresponding to the state $\frac{1}{\sqrt{2}}\left(\left|\uparrow\downarrow\right\rangle -\left|\downarrow\uparrow\right\rangle \right)$)
or a triplet state with a spin quantum number $s=1$ (corresponding
to 3 different states: $\frac{1}{\sqrt{2}}\left(\left|\uparrow\downarrow\right\rangle +\left|\downarrow\uparrow\right\rangle \right)$;
$\left|\uparrow\uparrow\right\rangle $; $\left|\downarrow\downarrow\right\rangle $).
Both states de-excite by emitting a $\unit[178]{nm}$ vacuum-ultraviolet
(VUV)\nomenclature{VUV}{Vacuum Ultraviolet} photon, with a lifetime
of $\unit[24]{ns}$ for the singlet state and $\unit[3.1]{ns}$ for
the triplet state\footnote{The difference in exciton lifetime can also be used to discriminate
between ER and NR using only the S1 signal, a technique commonly used
in argon TPCs since the singlet (7~ns) and triplet (1,600~ns) lifetimes
are very different~\cite{Lippincott:2008ad}. However, LUX sampled
data at 10~ns, and even though pulse shape discrimination is not
currently used in the WIMP search data analysis, the abilities of
xenon pulse shape discrimination were explored in~\cite{Akerib:2018kjf}.}~\cite{Mock:2013ila} and a FWHM of 13 nm~\cite{xe_exciton}:
\begin{equation}
\mathrm{Xe_{2}^{*}\rightarrow2\,Xe\,+\,}h\nu.\label{eq:scint2}
\end{equation}

The ionized Xe atoms will also recombine with nearby Xe atoms on a
time scale of picoseconds. Then, in the absence of an applied electric
field, the excitons form excited Xe atoms by capturing nearby free
electrons:
\begin{align}
\mathrm{Xe^{+}+Xe} & \rightarrow\mathrm{Xe_{2}^{*}}\label{eq:-22}\\
\mathrm{Xe_{2}^{+}+}e^{-} & \rightarrow\mathrm{Xe^{**}+Xe}\label{eq:-23}\\
\mathrm{Xe^{**}} & \rightarrow\mathrm{Xe^{*}\,+\,heat.}
\end{align}
The excited Xe atoms release addition scintillation photons according
to Equations~\ref{eq:scint1} and~\ref{eq:scint2}. This recombination
process can be suppressed by the application of an electric field,
which leads to a quenching of the primary scintillation yield.

The energy $E$ deposited by a particle is given by 
\begin{equation}
E=fW\left(n_{i}+n_{ex}\right)\label{eq:EW}
\end{equation}
where the quenching factor $f$ accounts for a varying fraction of
energy lost to atomic motion (heat) and the work function $W=\unit[13.7\pm0.2]{eV}$~\cite{dahl2009physics}.
This is multiplied by the total number of quanta produced along the
track of the recoiling particles, which is the sum of $n_{i}$ (the
number of electron-ion pairs), and $n_{ex}$ (the number of excitons
created), as discussed below. For ER the fraction of energy dissipated
by heat is constant with energy, so $f_{ER}\equiv1$~\cite{Takahashi:1975zz}
and the energy heat loss is included in the value of $W$. For NR
the recoiling nucleus will generate secondary recoils below the excitation
threshold $W$. The microphysics of this process is described by the
Lindhard theory~\cite{lindhard1963integral}. The quenching factor
for nuclear recoils varies with deposited energy. This means that
in the LUX detector, an NR signal is smaller than an ER signal of
the same energy. 

The ratio of excitons to ions produced along the track of the recoiling
particle
\[
\alpha\equiv\frac{n_{ex}}{n_{i}}
\]
can be approximated as roughly constant with $\alpha_{NR}\sim1$~\cite{dahl2009physics}
and $\alpha_{ER}\sim0.2$~\cite{Doke:2002oab}. However, following
ionization a fraction $r$ of the initial electron-ion pairs recombines
and forms additional excitons, enhancing the S1 signal. This recombination
depends on LXe density, the applied electric field, and particle energy.
At 180 V/cm the recombination peaks at $\unit[\sim12]{keV}$ as measured
by the LUX detector~\cite{Akerib:2016qlr,dahl2009physics,PhysRevB.68.054201,PhysRevB.76.014115,Akerib:2015wdi}.
Therefore, the quantities directly measurable in the detector are
the de-excitation photons $n_{\gamma}$ from all the excitons formed
and the electrons that escape recombination $n_{e}$:
\begin{align}
n_{\gamma} & =n_{ex}+rn_{i}=\left(\alpha+r\right)n_{i}\label{eq:n_gamma}\\
n_{e} & =\left(1-r\right)n_{i}.\label{eq:n_e}
\end{align}
The detected S1 and S2 signals are given by 
\begin{align}
S1= & \,g_{1}n_{\gamma}\label{eq:}\\
S2= & \,g_{2}n_{e}\label{eq:-1}
\end{align}
where the italicized scintillation ($S1$) and ionization ($S2$)
signals are measured in units of detected photons (phd)\nomenclature{phd}{Photons Detected}
in the TPC. These pulse sizes are corrected for geometrical effects
in the detector and electron lifetime in LXe~\cite{Baudis:2013bba}
in the data processing chain described in Section~\ref{subsec:Data-processing}.
The units of detected photons differ from the more commonly used unit
of photoelectrons (phe)\nomenclature{phe}{Photoelectron} by a small
factor representing the probability of a single photon to produce
multiple phe in a PMT photocathode~\cite{Faham:2015kqa}. The detector
efficiencies $g_{1}$ and $g_{2}$ are the signal gains for $S1$
and $S2$ in units of phd/quantum. For S1 photons $g_{1}$ is the
product of geometrical light collection efficiency and the averaged
PMT quantum efficiency, as it represents the probability of a scintillation
photon to produce at least one phe on the PMT photocathode. For S2
photons, $g_{2}$ depends on the electroluminescence photon yield,
the efficiency $\epsilon$ with which drifted electrons are extracted
from the liquid into the gas phase, and the average pulse size of
a single extracted electron (SE)\nomenclature{SE}{Single Electron}:
\[
g_{2}=\epsilon\times SE.
\]
 In the LUX detector $g_{1}\sim0.1$ and $g_{2}\sim12$ as derived
from calibrations.

Equations~\ref{eq:n_gamma} and~\ref{eq:n_e} show that, ignoring
quenching effects, all excitons and electron-ion pairs result in either
an escaping photon or an electron with perfect efficiency, i.e., $n_{\gamma}+n_{e}=n_{ex}+n_{i}$.
Although the light and charge yields vary with energy due to dependence
on recombination, the proportion of energy that goes into the combination
of ionization and scintillation signal remains constant. Therefore,
the combined energy from $S1$ and $S2$ can be used to reconstruct
the true energy of an event. As such, Equation~\ref{eq:EW} can be
rewritten as 
\begin{align}
E & =fW\left(n_{\gamma}+n_{e}\right)\label{eq:-2}\\
 & =fW\left(\frac{S1}{g_{1}}+\frac{S2}{g_{2}}\right).\label{eq:EW_S1S2}
\end{align}
This equation describes a response of a specific TPC with given operating
conditions, unlike Equation~\ref{eq:EW}, which describes the general
response of LXe an energy deposit. However, while reconstructing events
in the detector, it may not be known whether a particular event is
the result of an electron or nuclear recoil. If the energy is reconstructed
under the assumption of an ER with $f=1$, units of $\unit{keV_{ee}}$
are used where ``ee'' stands for ``electron equivalent.'' Similarly,
if the energy calculation assumes an NR, units of $\unit{keV_{nr}}$
(often written without the subscript) are used. The following equation
can be used to convert between those two units
\[
\mathrm{keV_{ee}=}\left(\frac{12.6}{73}\right)\mathrm{keV_{nr}}^{1.05}.
\]
 It was obtained from a fit to data by NEST v2.0~\cite{nest2.0}.

\begin{figure}
\begin{centering}
\includegraphics[scale=0.3]{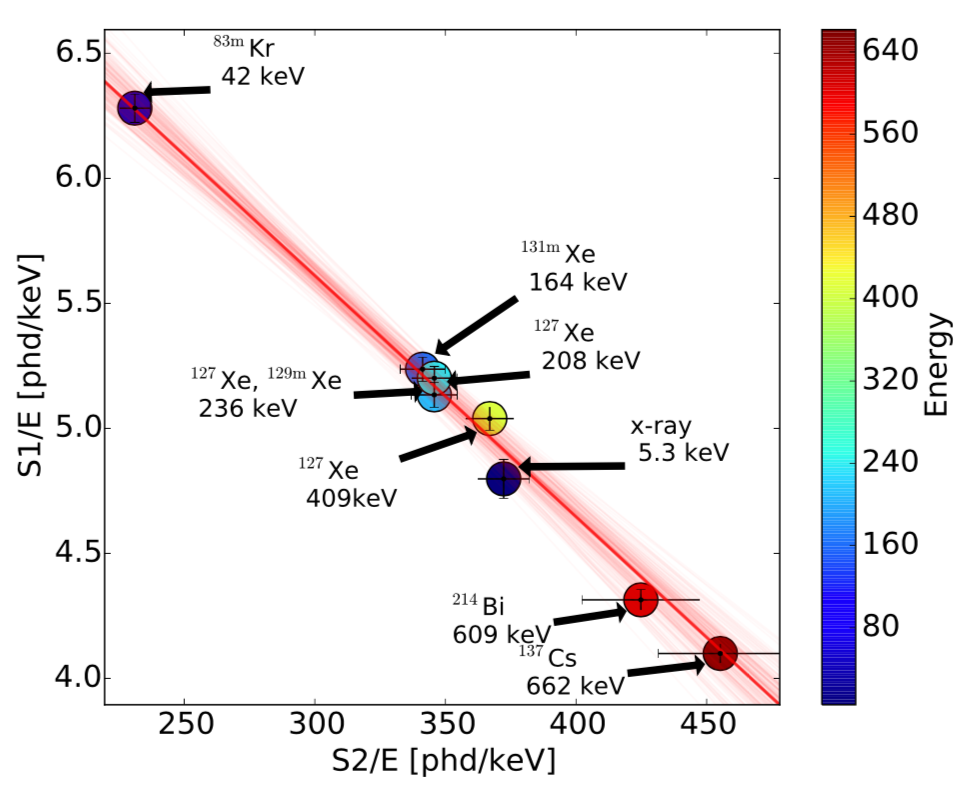}
\par\end{centering}
\caption[Doke plot from WS2013]{Doke plot used to calculate $g_{1}$ and $g_{2}$ in the LUX 2013
reanalysis. Each point represents a different line source collected
at a drift field of 180 V/cm. Values of $g_{1}=0.117\pm0.003$ and
$g_{2}=12.1\pm0.8$ were found. The further to the lower right a point
is on the plot, the less recombination (less S1 and more S2). Figure
from~\cite{Akerib:2017vbi}.\label{fig:Doke-plot}}

\end{figure}

The mean $S1$ and $S2$ response at two or more known energies can
be used to calculate $g_{1}$ and $g_{2}$ by requiring that the combined
energy from $S1$ and $S2$ reconstructs the true energy of a calibration
source of known energy. Those values are described by a line $y=mx+b$
where $x=\left\langle S2\right\rangle /E$ and $y=\left\langle S1\right\rangle /E$
as shown in Figure~\ref{fig:Doke-plot}. A fit to the slope $m=-g_{1}/g_{2}$
and $y$-intercept $b=g_{1}/W$ are then used to measure $g_{1}$
and $g_{2}$. Further details about this so-called ``Doke plot analysis''
can be found in~\cite{Akerib:2017vbi,Akerib:2015rjg}. A more detailed
review of the operation principles and the most essential instrumentation
aspects of two-phase TPCs can be found in~\cite{Chepel}. 

\section{The LUX detector}

\subsection{Detector design\label{subsec:Detector-design}}

\begin{figure}
\begin{centering}
\includegraphics[scale=0.77]{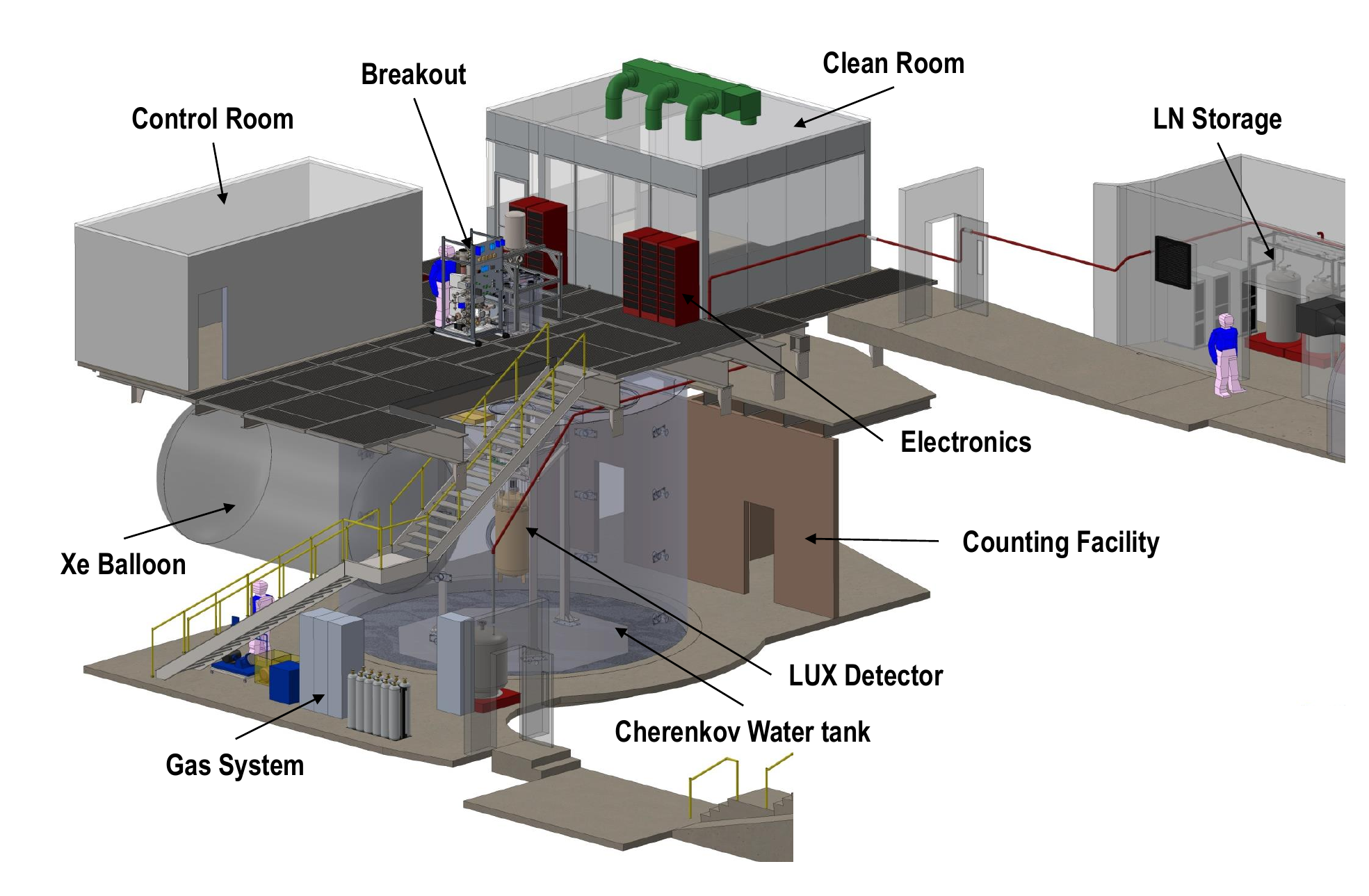}
\par\end{centering}
\caption[The layout of the underground LUX laboratory]{The layout of the underground LUX laboratory.\label{fig:Lab-layout-1}}
\end{figure}

The LUX detector was a two-phase TPC located a mile underground in
the Davis cavern\footnote{There is a long history of rare event searches in Lead. Most notably,
while SURF was still the Homestake gold mine, Ray Davis and his collaborators
were the first to measure the flux of solar neutrinos in the 1960s~\cite{Davis:1968cp,bahcall}.
This earned him a Nobel prize and the cavern where he did his pioneering
solar neutrino experiment was named in his honor.} at the Sanford Underground Research Facility (SURF)\nomenclature{SURF}{Sanford Underground Research Facility}
in Lead, South Dakota. The layout of the underground laboratory and
the support systems needed to operate the detector is shown in Figure~\ref{fig:Lab-layout-1}.
As for many other rare event searches, one of the main design goals
was the reduction of the rate of background events to ensure that
the detector was not swamped with an unwanted signal. The underground
location with $\unit[1478]{m}$ (4850~ft or 4300~m water equivalent)
of rock overburden provides shielding from cosmogenic radiation. To
that end, the detector was also located inside of a water tank, and
careful material assays were performed prior to detector assembly
to establish a low-background environment for the detector. Furthermore,
the realities of our world such as cost, ease of operations, and safety
were considered in the detector design. This section highlights selected
parts of the detector; a more detailed description of the design choices
for the LUX detector can be found in~\cite{Akerib:2012ys}, and in-depth
descriptions of the individual subsystems can be found in the various
LUX theses mentioned throughout the text.

\begin{figure}
\begin{centering}
\includegraphics[angle=90,scale=0.58]{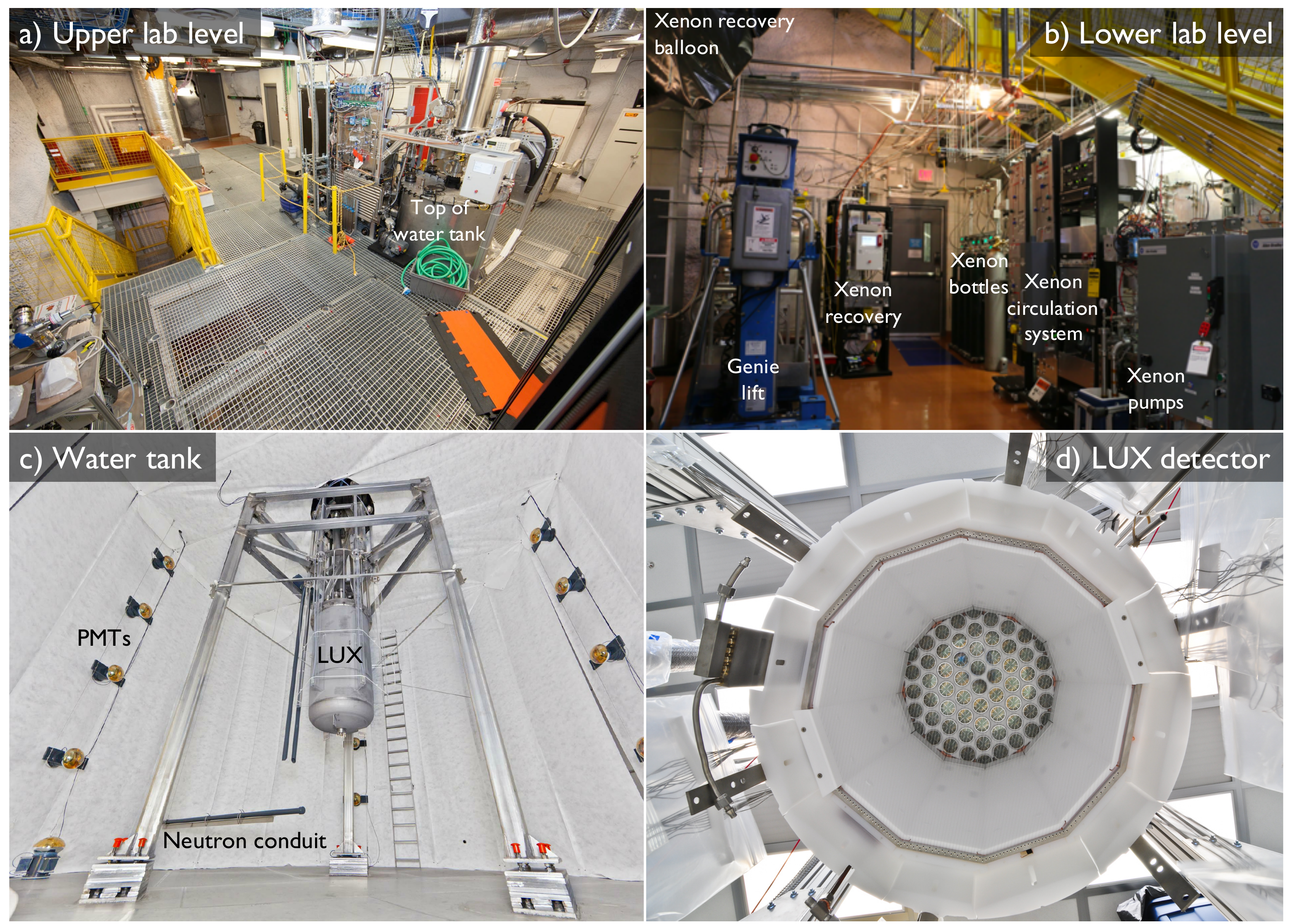}
\par\end{centering}
\caption[The LUX detector]{The LUX detector. \textbf{a)} and \textbf{b)} Layout of the underground
laboratory. \textbf{c)} The inside of the LUX water tank with the
PMTs, neutron conduit used for calibrations, acrylic source tubes,
and the LUX cryostat. The neutron conduit can move up and down along
the side of the cryostat. \textbf{d)} The active volume of the LUX
detector with the upper PMT array and the reflective PTFE panels visible.\label{fig:LUX-and-lab}}
\end{figure}

The detector was located in a $\unit[270,000]{liter}$, $\unit[5.9]{m}$
tall, cylindrical water tank with $\unit[7.6]{m}$ diameter filled
with de-ionized water to block $\gamma$-rays and neutrons emanating
from the cavern rock. The water tank was lined with reflective Tyvek
sheets and was instrumented with 20~Hamamatsu R7081 10'' PMTs that
detected \v{C}erenkov light from cosmic muons to veto events. A 20-tonne
steel inverted pyramid under the water tank provided additional $\gamma$
shielding to the detector. Figure~\ref{fig:LUX-and-lab} shows the
upper and lower floors of the Davis Cavern, the inside of the water
tank, and the detector itself.

\begin{figure}[t]
\begin{centering}
\includegraphics[scale=0.78]{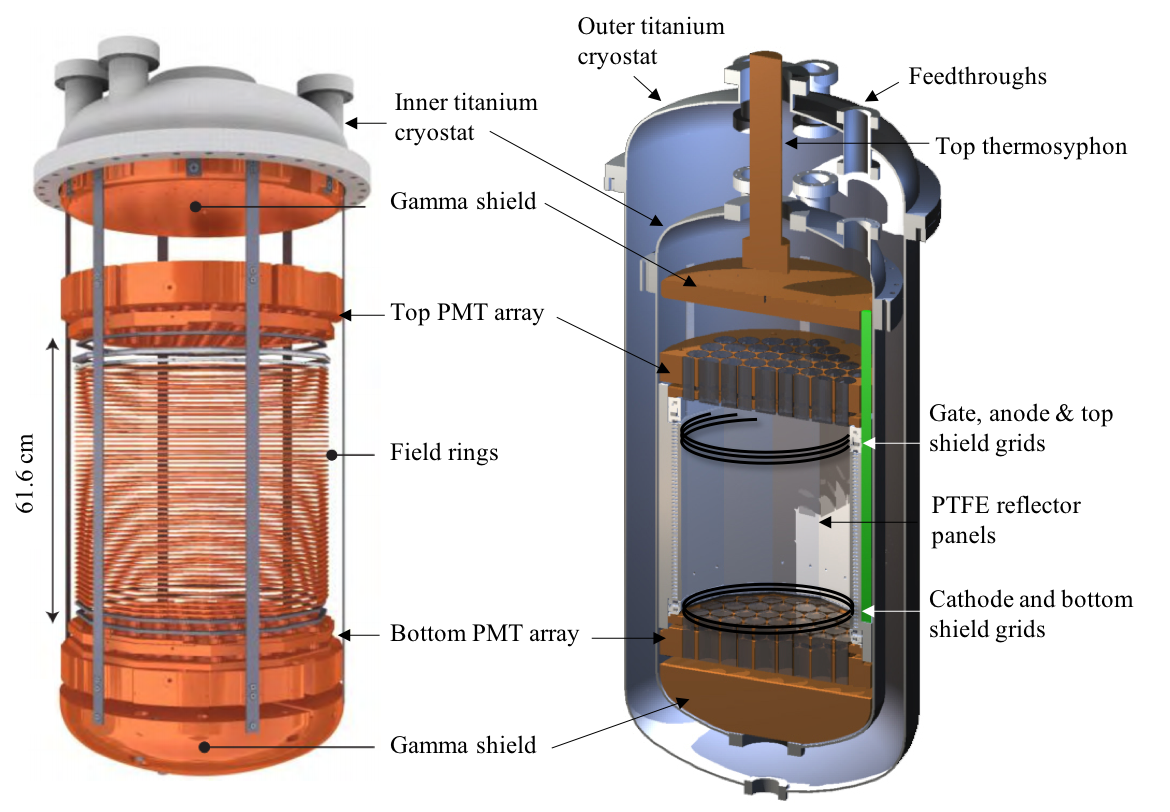}
\par\end{centering}
\caption[Annotated rendering of the LUX detector]{Annotated CAD renderings of the LUX detector. \textbf{Left:} LUX
detector metal internals supported from the top flange of the inner
cryostat. \textbf{Right:} A cross-sectional view of the LUX cryostat.
The green line illustrates the path of the cathode HV cable. Figures
modified from~\cite{Akerib:2012ys}.\label{fig:LUX-model}}
\end{figure}

The inner LUX cryostat was wrapped in insulating materials and nested
inside a larger outer cryostat, both made from low-background titanium~\cite{Akerib:2011rr}.
The inner cryostat was $\unit[101]{cm}$ tall, $\unit[62]{cm}$ in
diameter, and contained $\unit[370]{kg}$ of LXe with $\unit[251]{kg}$
of active mass, where active mass refers to LXe enclosed between the
gate and cathode grids and the detector polytetrafluoroethylene (PTFE)\nomenclature{PTFE}{Polytetrafluoroethylene}
walls. The active volume had a shape of a regular dodecagonal prism
enclosed by twelve PTFE faces with $\unit[47.3]{cm}$ diameter, as
measured at -100$^{\circ}$C between parallel opposite faces. In $z$,
the active volume was bounded by the cathode and gate wire grids.
This is illustrated in the computer-aided design (CAD)\nomenclature{CAD}{Computer-Aided Design}
rendering of the detector's cross section in Figure~\ref{fig:LUX-model}.

Given this high-level understanding of the design of the LUX detector,
the following sections unpack the inner workings and performance of
the individual components.

\subsubsection{High voltage }

In order to establish the electric field needed to drift the S2 electrons,
the detector contained five grids in total. The bottommost and topmost
grids shielded the bottom and top PMT arrays from the high fields
inside the detector. The detector drift field was set by cathode and
gate grids separated by $\unit[48.3]{cm}$. Between those two grids,
to ensure field uniformity inside the active volume, 48 copper field-shaping
rings were located behind the PTFE panels as illustrated in Figure~\ref{fig:LUX-model}.
The rings were separated vertically by $\unit[1]{cm}$, and the gate-cathode
voltage was graded evenly via a resistor chain. The anode grid was
located $\unit[1]{cm}$ above the gate grid to create the so-called
extraction region~\cite{Edwards:2017emx} held at a high electric
field in order to pull electrons out of the LXe and into the xenon
gas to create the S2 signal. A detailed description of the grid geometry
and detector's electric fields will be discussed in Chapter~\ref{chap:efield-modeling}.

\begin{figure}[!th]
\begin{centering}
\includegraphics[scale=0.74]{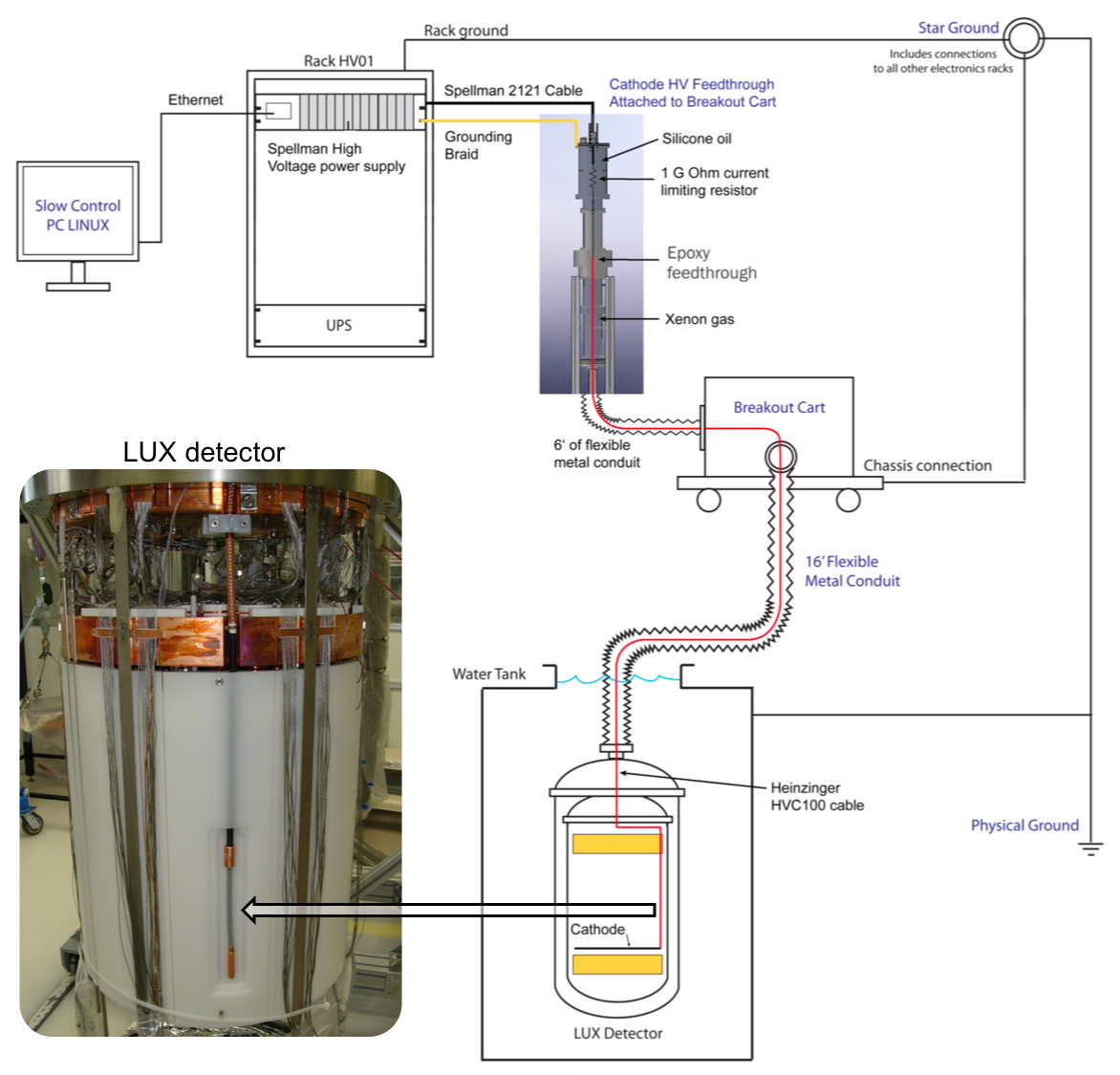}
\par\end{centering}
\caption[LUX cathode high voltage subsystem]{LUX cathode high voltage subsystem. A high voltage power supply monitored
via the slow control system is fed via an in-house designed feedthrough
into the xenon space. The red line indicates the high voltage cable
that is submerged in xenon gas until it reaches the main detector
volume where it comes down on the side of the detector through a cutout
in the ultra-high-molecular-weight polyethylene surrounding the PTFE
walls. Figure modified from~\cite{pease2017rare}.\label{fig:LUX-cathode-HV}}

\end{figure}

The cathode was biased to a high voltage (HV)\nomenclature{HV}{High Voltage}
to set up electric fields inside the detector. Since HV delivery to
liquid noble detectors has been a persistent challenge as will be
discussed in Chapter~\ref{chap:XeBrA}, careful design and testing
preceded the deployment of the feedthrough. Due to the assembly constraints
of the LUX experiment, the cathode HV feedthrough was located outside
the detector at the end of a long flexible conduit as shown in Figure~\ref{fig:LUX-cathode-HV}.
The feedthrough acts both as a ``plug'' preventing xenon gas escaping
the detector and a connection between the air- and the detector-side
cables. The HV feedthrough consists of a conductive cable embedded
in a specially designed epoxy\footnote{Stycast 2850FT Blue}-based
mass developed at Yale University. The advantage of the feedthrough
being located far from the TPC is the ability to use slightly radioactive
feedthrough materials, which would not be possible should the feedthrough
be located closer to the active LXe volume. This feedthrough was successfully
tested up to -100 kV by terminating the distant end of the cable in
silicone transformer oil as discussed in great detail in~\cite{pease2017rare}.
Despite the feedthrough's successful tests and its successful performance
throughout the LUX era, the detector's cathode voltage was limited
to -10 kV during the scientific dark matter search. This limiting
voltage was due to light production from the cathode and gate grids,
as discussed in Section~\ref{subsec:Grid-conditioning-campaign}.

\subsubsection{Light readout\label{subsec:Light-readout}}

The detector was instrumented with 122~PMTs detecting the primary
and secondary scintillation signals. These low-radioactivity PMTs~\cite{Akerib:2012da}
were mounted in copper blocks; 61 were located above the LXe in xenon
gas and 61 at the bottom of the detector and submerged in LXe. The
PMTs were manufactured by Hamamatsu, model R8778, with a typical quantum
efficiency at $\unit[178]{nm}$ of 33\%. This quantum efficiency includes
the probability of photon transmission through the synthetic silica
(SiO$_{2}$) window, its absorption in the Rb-Cs-Sb bialkali photocathode,
and the chance of the photon absorption exciting an electron that
gets detected. The PMT dynode chain then amplified the collected electrons
to mV-scale signals. 

The PMTs were operated with gains of $\sim4\times10^{6}$ for an average
single photoelectron (sphe)\nomenclature{sphe}{Single Photoelectron}
pulse area of $\sim\unit[16]{mV\cdot ns}$ at the PMT output. The
analog chain amplification and shaping increase the sphe pulse area
to $\unit[\sim151]{mV\cdot ns}$ and full-width (FWHM) of $\unit[54]{ns}$
at the digitizer. The PMT response limited the rise time of an S1
pulse to $\unit[\sim6]{ns}$. The effective time constant of the xenon
excimer relaxation defined the pulse decay constant. The pulse width
of an S2 event varies with depth due to the diffusion of the electron
cloud drifting to the top of the detector. More information about
the commissioning and performance of the LUX PMTs can be found in~\cite{faham2014prototype}.

\subsection{Xenon circulation and cryogenics\label{subsec:Xenon-circulation-and}}

Maintaining long electron drift lengths is crucial for proper detector
operation. Thus, the detector had stringent requirements regarding
xenon purity since electronegative impurities, such as O$_{2}$ and
N$_{2}$, can absorb scintillation light and capture drifting charge
thus decreasing the apparent size of the S2 signal. Small quantities
of those impurities are present in commercial xenon, but can also
be introduced via air leaks and from the continuous outgassing of
material inside the detector\footnote{In a two-phase TPC, electron lifetimes need to be sufficiently long
to enable charge drift to the liquid level. However, in liquid xenon
electronegative impurities at the levels of about 10$^{-9}$ g/g reduce
the electron attenuation length to several tens of centimeters depending
on the electric field and higher levels can severely limit or thoroughly
impair proper detector operation~\cite{LEONARD2010678}.}. Therefore, it is necessary to continuously clean the xenon during
detector operations.

Xenon circulation and purification were done in the gas phase since
there is not a practical solution for removal of N$_{2}$ impurities
from liquid xenon. Xenon was liquified in the detector, then evaporated
on its way out, purified in the gas phase and recondensed back in
the detector. Two double-diaphragm KNF pumps\footnote{Model PM-23480-150-3-1.2}
were installed in parallel to achieve a circulation rate of 25 standard
liters per minute (SLPM)\nomenclature{SLPM}{Standard Liter Per Minute},
corresponding to a turnover of xenon of $\unit[\sim229]{kg/day}$.
Two additional identical pumps were also obtained so that routine
pump maintenance would not interrupt detector operations. The main
xenon flow rate was controlled using two Brooks mass flow controllers
(MFCs)\nomenclature{MFC}{Mass Flow Controller} operating between
$\unit[1-50]{SLPM}$. Furthermore, four smaller-capacity MFCs were
located on purge lines that connect the ``dead ends'' of detector
conduits to the rest of the circulation system to prevent a buildup
of dirty xenon gas. A SAES MonoTorr\footnote{Model PS4-MT15-R1} heated
zirconium getter operated continuously was used to clean the gaseous
xenon from non-noble impurities. 

The concentration of impurities in LUX was monitored in two ways.
First, the electron drift lifetime was monitored directly by looking
at the characteristic S2 pulse size vs. the depth of an event in the
detector. Second, the LUX detector had a novel \textit{in situ} sampling
system developed for the LUX and EXO-200~\cite{Albert:2017owj} experiments
to regularly measure concentrations of impurities in xenon from various
parts of the detector (conduits, detector in/out, getter in/out).
The sampling system used a liquid nitrogen bath to freeze a small
sample of xenon in a U-bend cold trap. The composition of the cold
gas above the frozen xenon was then measured with a residual gas analyzer,
which provides measurements of minute traces of impurities. The LUX
system demonstrated sensitivity to less than 1 part per trillion of
krypton and other gasses~\cite{Dobi:2011vc,balajthy2018}.

LUX used a system of four thermosyphons to maintain xenon in the liquid
phase in the detector with exceptional temperature stability. A thermosyphon
is a passive gravity-assisted heat exchanger based on convection.
The top of the thermosyphon, the condenser, was coupled to a liquid
nitrogen reservoir, and the bottom, the evaporator, was in contact
with the heat load. Nitrogen condensed at the top and dripped down
to the cold head due to gravity. There, it evaporated and rose back
up to the condenser, transferring heat from the detector to the liquid
nitrogen condenser. In LUX each thermosyphon consisted of a closed
system of stainless steel tubing pressurized with gaseous nitrogen.
Two of the thermosyphons were able to deliver up to $\unit[400]{W}$
of cooling power while the other two in contact with the sides of
the detector could deliver up to $\unit[100]{W}$ of cooling power.
Each cold head was also equipped with a heater to regulate temperature
as necessary. 

\begin{figure}[t]
\begin{centering}
\includegraphics[scale=0.46]{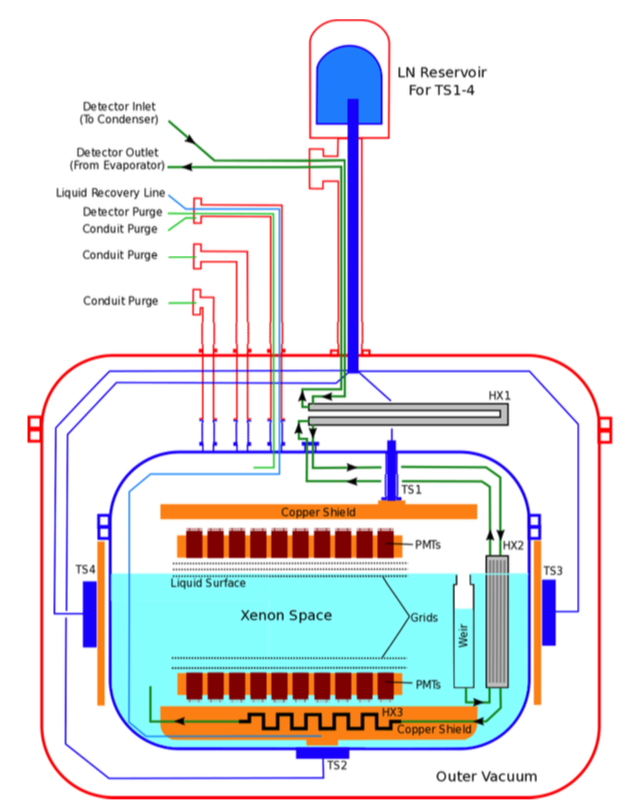}
\par\end{centering}
\caption[Schematics of the cryogenics and xenon circulation inside LUX]{Schematics of the cryogenics and xenon circulation inside the detector
including cabling, heat exchangers (HX), and conduits for thermosyphons
(TS). Bold green lines highlight the main xenon flow path. Figure
from~\cite{larsen2016effective}.\label{fig:LUX-cryogenics}}

\end{figure}

There were three heat exchangers in the xenon circulation path designed
to reduce the cooling power needed to condense the xenon. A spillover
weir set the LXe level. Liquid from the weir reservoir drained into
the evaporator portion of the main heat exchanger HX2 shown in Figure~\ref{fig:LUX-cryogenics}
depicting detector cryogenics and LXe circulation. HX2 operated in
dual phase and consisted of a rectangular evaporator with five vertical
tubes that served as a condenser that turned LXe to a cold gas. Next
in the path of xenon exiting the detector was heat exchanger HX1,
which consisted of two concentric pipes designed to pre-cool the incoming
xenon gas by transferring heat to the outgoing cold xenon gas. Finally,
HX3 was a long channel engraved in the copper $\gamma$ radiation
shield in the bottom of the detector providing further cooling and
equalizing the temperature of the incoming liquid to the temperature
of the main LXe bath.

In normal circumstances, xenon was stored in eight gas cylinders filled
using a compressor. Since the cost of 1 kg of xenon ranges from $\$1,000-\$3,500$,
several layers of recovery systems were implemented to avoid xenon
loss in case of emergencies. The storage and recovery vessel (SRV)\nomenclature{SRV}{Storage and Recovery Vessel}
was a high-pressure vessel able to capture the entirety of the LUX
xenon. The liquid nitrogen jacket of the SRV was always kept full
to enable recovery via cryopumping, and even if all the nitrogen evaporated,
the SRV was rated to hold all the xenon at high pressure. In case
of a catastrophic failure preventing recovery to the SRV, burst discs
and relief devices on the SRV and other parts of the circulation system
were connected to a $\unit[70]{m^{3}}$ polyester geomembrane recovery
balloon with enough volume to store all xenon at low pressure at room
temperature. Further details about LUX cryogenics and circulation
can be found in~\cite{larsen2016effective,Bradley:2012fsa,phelps2014lux}.

\subsection{Performance monitoring\label{subsec:Performance-monitoring}}

The LUX detector was fitted with numerous instruments to monitor and
stabilize conditions within the cryostat. Sixty-three $\unit[100]{\Omega}$
thin film platinum resistance temperature detectors (RTDs)\nomenclature{RTD}{Resistance Temperature Detector}
were used to monitor the temperature in both the inner and outer cryostats.
A variety of pressure sensors, manual pressure gauges, and a capacitance
manometer were also used throughout the detector. Furthermore, six
parallel wire sensors monitored the liquid level; the capacitance
of each wire pair depended on the length of wire submerged in the
liquid, allowing readout of the overall height of the liquid level.
Additionally, three parallel plate sensors were placed 120~degrees
apart between the gate and anode grids to ensure liquid surface uniformity
across the extraction region. Advantech Adam 6015 modules fed the
output voltage of the RTDs, pressure sensors, and level gauges to
a slow control database. The slow control system enabled multiple-user
access to detector conditions monitoring and automated alarms settings
to notify operators of unusual detector behavior. The Automated Controlled
Recovery System (ACRS)\nomenclature{ACRS}{Automated Controlled Recovery System}
triggered xenon recovery procedure in case of an emergency.

Throughout the WIMP search, the xenon vessel maintained thermal stability
of $\Delta T$~<~0.5~K and pressure stability of $\Delta P<\unit[10^{-2}]{bar}$~\cite{Akerib:2016vxi}.

\subsection{Electronics and data acquisition\label{subsec:Electronics-and-data}}

The LUX data acquisition (DAQ)\nomenclature{DAQ}{Data Acquisition}
system was designed to distinguish >95\% of sphe at 5~sigma above
baseline noise fluctuations and to prevent saturation of events with
energies smaller than $\unit[100]{keV_{ee}}$. The DAQ can operate
with a maximum trigger rate of $\unit[1.5]{kHz}$ before incurring
dead-time. With 122~PMT channels, each producing 14~bits of data
at 100~MHz, the three years of raw data would require 2~exabytes\footnote{That is $10^{18}$~bytes. If one were to store this amount of data
on consumer-grade hard drives assuming a price of \$50/TB that would
correspond to \$100~million in hard drive costs alone, significantly
more than the cost of the entire LUX detector.} of storage. Therefore, it is crucial to design a triggering system
that saves only buffers with information of potential interest. In
LUX, the risk of losing valuable data due to a lack of a trigger condition
was minimized by digitizing the trigger information alongside the
PMT waveforms. To achieve that the PMT readout was split into analog
and digital electronics. 

\begin{figure}
\centering{}\includegraphics[scale=0.26]{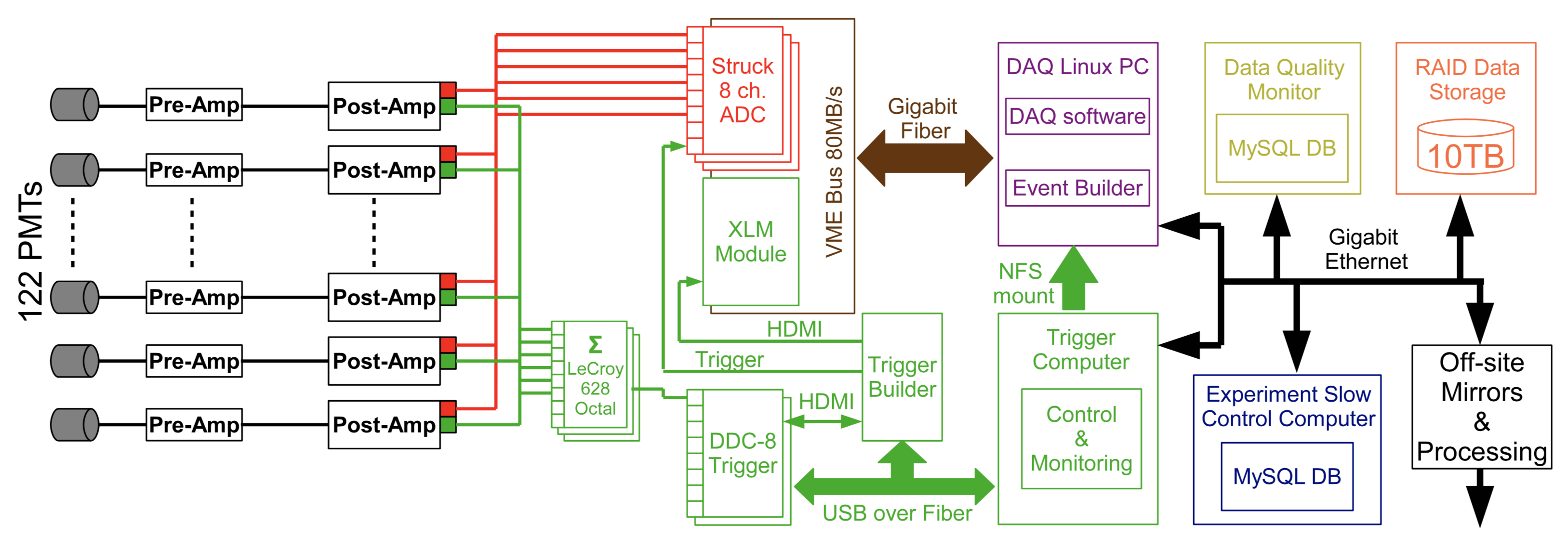}\caption[Overview of the signal and data processing in LUX.]{Overview of the signal and data processing in LUX. The PMT signals
are amplified, shaped and then digitized. Reduced quantities of the
trigger system and the waveform data are stored off-line for verification
and analysis. Figure from~\cite{Akerib201657}.\label{fig:LUX_FPGA}}
\end{figure}

Signals from the PMTs were pre-amplified immediately after leaving
the xenon space, while post-amplification provided the final signal
gain and shape. The signals were summed into 16~groups (8~groups
per array) with no adjacent PMTs belonging to the same sum before
being split into two copies for the trigger system. One copy of the
waveforms was passed through a field programmable gate array (FPGA)\nomenclature{FPGA}{Field Programmable Gate Array}-based
trigger system (shown as a green data stream in Figure~\ref{fig:LUX_FPGA}). 

\begin{figure}
\begin{centering}
\includegraphics[scale=0.28]{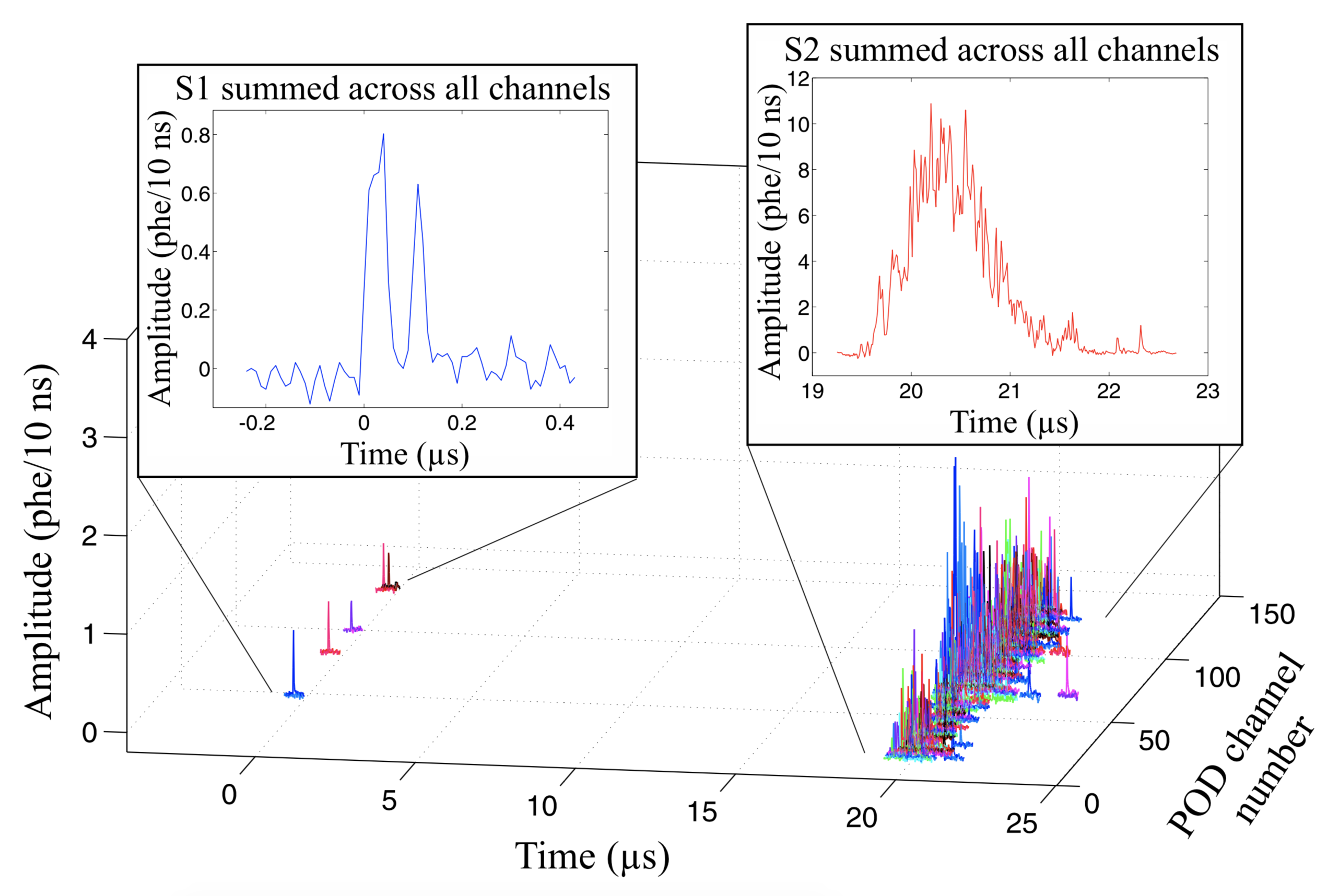}
\par\end{centering}
\caption[Example of a low-energy event as seen in raw data]{Example of a $\unit[1.5]{keV_{ee}}$ low-energy event as seen in
raw data. Each POD channel number corresponds to an individual PMT.
Figure from~\cite{Akerib201657}.\label{fig:raw_signal}}

\end{figure}

The second copy was digitized at $\unit[100]{MHz}$ by analog to digital
converters (ADCs)\nomenclature{ADC}{Analog to Digital Converter}
and written to disk (red data stream in Figure~\ref{fig:LUX_FPGA}).
In order to manage the required data storage space, the ADCs were
controlled by an FPGA that operated in a ``pulse only digitization''
(POD)\nomenclature{POD}{Pulse Only Digitization} mode. In the POD
mode, a threshold was set to $\unit[1.5]{mV}$ and only pulses rising
above this threshold were digitized, thereby rejecting baseline noise.
The trigger was applied offline when compressing the raw data into
blocks of event-level data. The samples were written to a memory bank
from the buffer until an end threshold was reached and a fixed number
of samples prior to and immediately following this range were appended.
Figure~\ref{fig:raw_signal} shows an example event. Further information
about signal and data processing in LUX can be found in~\cite{Akerib:2017vbi,Akerib201657,Akerib:2011ix,druszkiewicz2017digital}.

The data went through several stages of refinement. First, the least
filtered, un-triggered, raw data for each channel corresponding to
ADC memory bank downloads were saved. These data were then built into
events where an algorithm collected and saved all PODs occurring within
$\unit[0.5]{ms}$ of a potential event as determined by the trigger.
These triggered, raw data were saved into event-level files and passed
on to the data processing stage for analysis.

\subsection{Data processing\label{subsec:Data-processing}}

The LUX data processing framework (DPF)\nomenclature{DPF}{Data Processing Framework}
extracted essential information from the event-level files and saved
the output as reduced quantities (RQs)\nomenclature{RQ}{Reduced Quantity}.
These RQs include information about the number of identified pulses,
trigger timestamp, pulse information, etc. Thanks to the modular nature
of the DPF, the list of modules used for individual data processing
runs can be easily swapped in and out. The list of modules used with
their specific configurations and calibration constants were provided
to the framework and stored in a MySQL database (called LUG) and associated
with a unique identifier. This ensured that data for any particular
analysis were processed with the same settings. The stored detector
calibration constants included PMT gains, $\left(x,y,z\right)$ spatial
calibration maps for S1 and S2 pulses, electron lifetime, light response
function for position reconstruction, and more.

\begin{figure}
\begin{centering}
\includegraphics[scale=0.32]{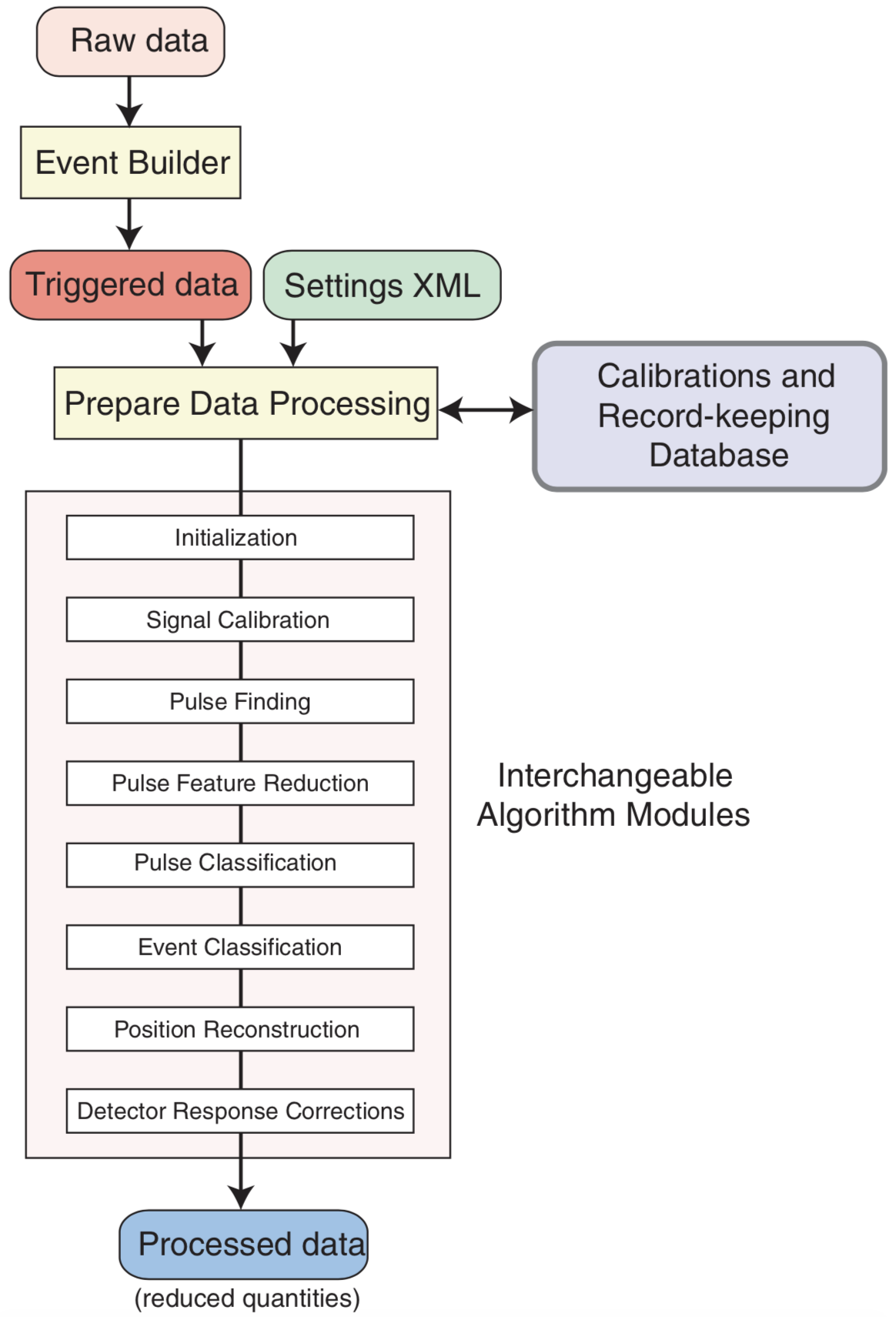}\includegraphics[scale=0.2]{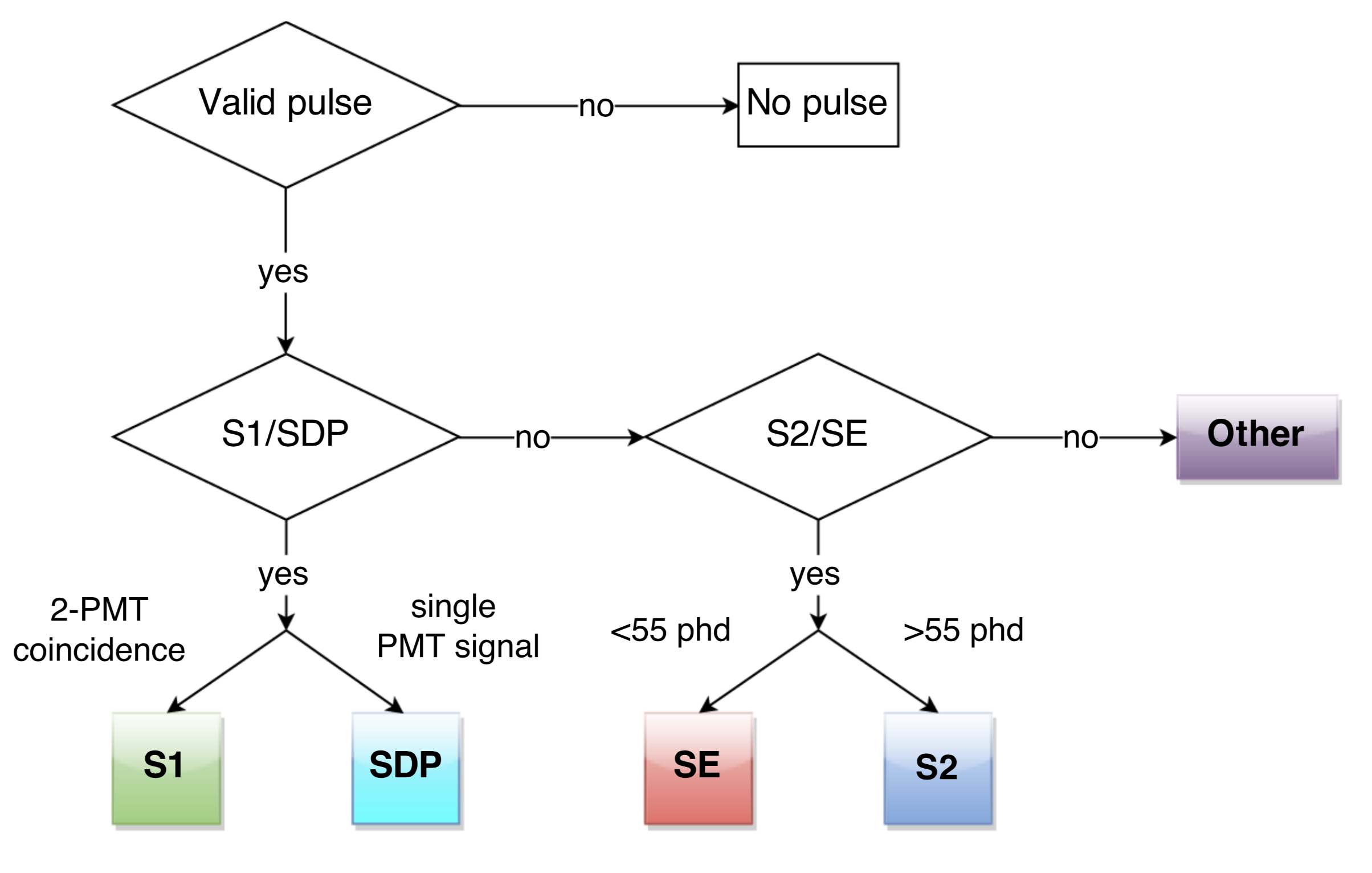}
\par\end{centering}
\caption[Schematic view of the LUX Data Processing Framework and pulse classifier]{\textbf{Left:} Schematic view of the LUX Data Processing Framework.
\textbf{Right:} Pulse classification decision tree. Figures from~\cite{Akerib:2017vbi}.\label{fig:DPF}}
\end{figure}

Figure~\ref{fig:DPF} shows a schematic representation of the journey
of raw data from the DAQ to the reduced quantities used in higher
level analyses. The primary goal of the RQs is to find and identify
different valid pulses in the acquired waveforms to be used in future
analyses. A sliding boxcar filter finds and separates S1-, S2-, and
other-type pulses returning their start and end times. Figure~\ref{fig:DPF}
illustrates the decision tree used for pulse classification. A two-PMT
coincidence is required for an S1 signal in order to reject photons
from spontaneous PMT photocathode emission. An S2-type signal needs
to be larger than 5~phd and can be further classified into single
electrons (SE) or S2 signals. The event type of interest for the WIMP
search analysis is a single scatter, or ``golden,'' event; this
type of event is required to have a single valid S1 signal followed
by a single S2 signal with at least 55~phd in size. About 95\% of
triggers are rejected by the golden event requirement. An extensive
hand-scanning campaign was conducted to cross-check the efficacy of
the gold event filter. Further information about the LUX DPF can be
found in~\cite{Akerib:2017vbi,nehrkorn2018testing}.

\subsubsection{Position reconstruction}

The $\left(x,y\right)$ position of an event in the LUX detector is
recovered from the light of the electroluminescent S2 signal collected
by the individual PMTs in the top array. For the position reconstruction,
an algorithm called Mercury was used, originally developed for the
ZEPLIN-III dark matter experiment~\cite{Solovov:2011aa}. It is a
statistical-based algorithm that uses a maximum likelihood test. It
uses an empirical light response function (LRF)\nomenclature{LRF}{Light Response Function}
to characterize the response of each PMT as a function of the position
of photon emission. A major advantage of this method is that it only
needs measured data obtained from \textit{in situ} calibrations (rather
than simulations) to recover the position of interactions. 

\begin{figure}
\begin{centering}
\includegraphics[scale=0.31]{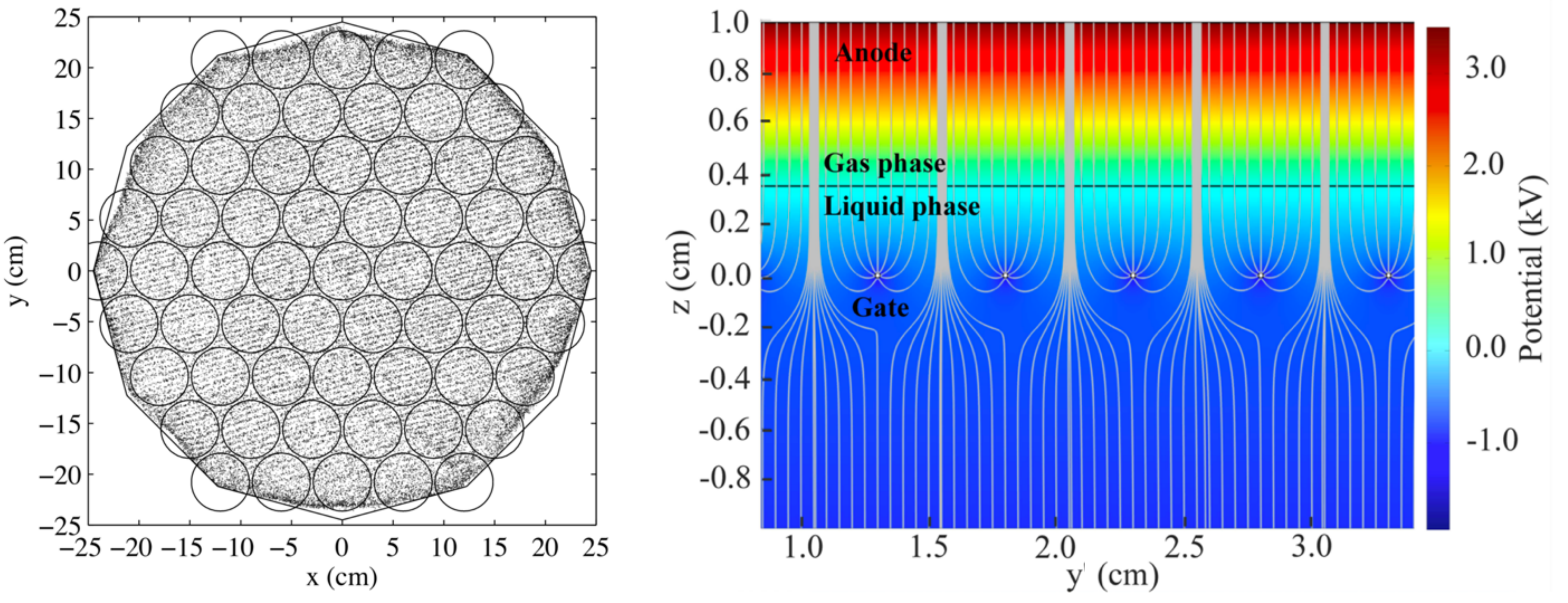}
\par\end{centering}
\caption[Position reconstruction using the Mercury algorithm]{\textbf{Left:} A scatter plot of reconstructed $\left(x,y\right)$
positions of uniformly distributed $\mathrm{^{83m}Kr}$ calibration
data for drift times between $\unit[4-10]{\mu s}$. Details about
$\mathrm{^{83m}Kr}$ are described in Section~\ref{subsec:Metastable-krypton-83}.
Black lines represent the PMTs and TPC inner walls. A striped pattern
with the pitch of gate grid wires is visible. Figure from~\cite{Akerib:2017riv}.
\textbf{Right:} Electric field simulation of gate regions showing
grids wires concentrating electron trajectories (gray lines) into
tights bands resulting in the effect shown on the left. The black
line illustrates the liquid level. Note that the field line density
does not illustrate field strength.\label{fig:position-reconstruction}}
\end{figure}

Due to the presence of PTFE reflectors around the detector volume,
the LRFs were implemented as a sum of two terms: an axial component
describing the light going to the PMT either directly or via a reflection
from the liquid surface and a polar component characterizing the light
reflected from the PTFE walls. This resulted in an observed position
resolution of $\unit[0.097\pm0.003]{cm}$ for pulses with an average
area of 22,000~phd and $\unit[2.24\pm0.04]{cm}$ for sphe. Figure~\ref{fig:position-reconstruction}
illustrates the performance of the position reconstruction software
using on $\mathrm{^{83m}Kr}$ calibration data. 

The time difference between the S1 and the S2 pulses provides information
about the event depth. In WS2013, where WS\nomenclature{WS}{WIMP Search}
stands for WIMP Search, the conversion between drift time and $z$
position was calculated assuming a uniform and constant electric field
in the active volume, using electron drift velocity $v_{d}=\unit[0.1518\pm0.0011]{cm/\mu s}$~\cite{Akerib:2017vbi}.
In WS2014-2016 I developed electric field maps as discussed in Chapter~\ref{chap:efield-modeling}
were used for this conversion. Further details about the position
reconstruction can be found in~\cite{Akerib:2017riv}.

\subsubsection{S2 coordinate system\label{subsec:S2-coordinate-system}}

The 3D position reconstruction of ionization vertices requires an
understanding of the electrons' path from the interaction site to
the point of detection in xenon gas. As described above, the $\left(x,y\right)$
position of an event in the detector is reconstructed using the Mercury
algorithm from the electroluminescence of the S2 signal in gaseous
xenon at high field. If the fields were perfectly perpendicular to
the liquid surface at all positions, the location of the event as
seen by the top PMT array and the real event location would be identical.
Although the field is mostly perpendicular to the liquid surface,
it is not perfectly the case everywhere. A non-zero radial component,
inducing a radially-inward electron drift, affects drifting electrons.
This radial component is due to the non-zero electrostatic transparency
of the field cage and field grids as will be discussed in detail in
Chapter~\ref{chap:efield-modeling}. 

\begin{figure}
\centering{}\includegraphics[scale=0.42]{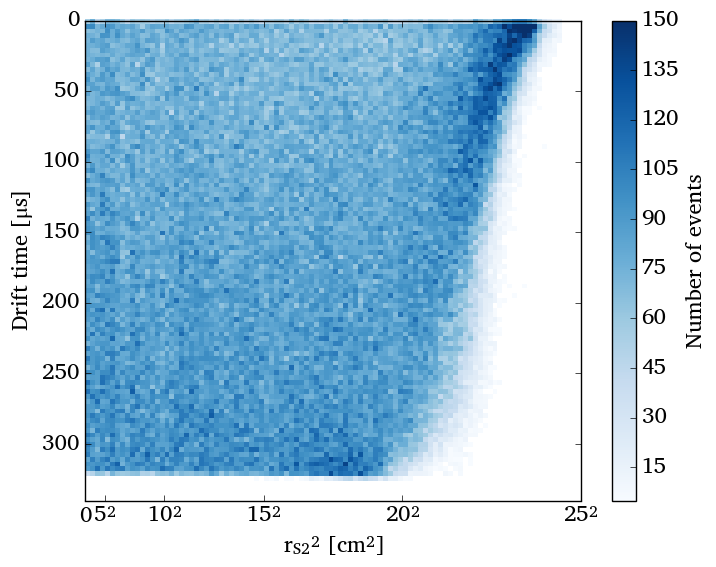}\caption[Distribution $\mathrm{^{83m}Kr}$ events in the detector]{2D histogram showing the reconstructed position of $\sim650,000$
uniformly distributed $\mathrm{^{83m}Kr}$ events in the active volume
of the LUX detector as a function of $r_{S2}^{2}$. This data comes
from a $\mathrm{^{83m}Kr}$ injection during WS2013 on 2013-05-10.
The higher event density in the upper right corner comes from a systematic
error in the Mercury algorithm in reconstructing events with low drift
time at high radius near the outermost PMT. This false artifact produced
by the Mercury algorithm is well known, far away from the fiducial
volume, and accounted for in analyses.\label{fig:krypton-1}}
\end{figure}

Therefore, there are two different coordinate systems used in the
analysis: the real coordinates and the S2, or reconstructed, coordinates.
The S2 coordinates ($x_{S2},\,y_{S2},\,\mathrm{drift\,time}$) represent
the event position as seen by the top PMT array with the event depth
given by the electron drift time. In these coordinates, all active
volume events have a slightly distorted edge as shown in Figure~\ref{fig:krypton-1}
due to field non-uniformities. A field correction needs to be applied
to translate the S2 coordinates into the real space $\left(x,\,y,\,z\right)$
coordinates. As described above, this was trivial in WS2013. However,
due to changing electric fields in WS2014-16, a special position reconstruction
algorithm was developed to map the Mercury algorithm coordinates into
real space coordinates using electric field models discussed in Chapter~\ref{chap:efield-modeling}. 

\subsection{Calibrations\label{subsec:Calibrations}}

A thorough characterization of particle interactions in the active
detector region is key to correctly identifying possible dark matter
particles. In LUX the primary goal is to identify unambiguous NR signals
composed of single-site scatters homogeneously distributed in the
active volume. It is essential to correctly identify NR interactions
in the detector and distinguish them from ERs. To achieve this, LUX
pioneered a number of novel calibration methods that were employed
extensively before, during, and after the scientific runs. Spectral
and monoenergetic sources, both internal and external, were used to
measure the detector's ER and NR response. 

The self-shielding ability of LXe that makes it such a great dark
matter target causes challenges during calibrations. Photons with
energies below the MeV scale are strongly attenuated in LXe as shown
in Figure~\ref{fig:Photon-cross-section}, and therefore external
$\gamma$ sources can only be used to calibrate the outer couple of
centimeters of LXe. Furthermore, high energy $\gamma$-rays such as
the ones from $^{137}$Cs are very likely to produce multiple-site
scatters; single-site scatters can be produced with $\unit[\sim1-10]{keV}$
energy deposition but only if the outgoing $\gamma$ can escape the
detector's active volume without any further scatters. During such
a calibration, the detector is swamped with edge events, causing saturation
effects due to the high event rates and making external $\gamma$-rays
not particularly useful for fiducial volume calibrations. 

However, $\gamma$ ray calibration sources are useful for position
reconstruction and background studies since they so strongly illuminate
detector walls. To that end, LUX had six air-filled acrylic source
tubes that extended down from the top of the water tank flush against
the sides of the outer cryostat as illustrated in Figure~\ref{fig:LUX-and-lab}.
Radioactive sources were deployed and retracted by a system of pulleys~\cite{phelps2014lux}.
Gamma-ray sources $^{57}$Co ($\unit[122]{keV}$ $\gamma$), $^{60}$Co
($\unit[1,173]{keV}$ and $\unit[1,332]{keV}$ $\gamma$s), $^{137}$Cs
($\unit[662]{keV}$ $\gamma$), and $^{228}$Th ($\unit[2,614]{keV}$
$\gamma$) were all available for ER response calibration via the
source tubes. 

For NR calibrations, AmBe (mix of an alpha-emitting $^{241}$Am +
$^{9}$Be with a mean energy of 4.2~MeV and maximum neutron energy
of 11~MeV) and $^{252}$Cf (emitting neutrons from spontaneous fission
emitting) were deployed in the source tubes. Later, calibrations were
performed using a deuterium-deuterium neutron beam generator located
outside the water tank was used to calibrate the LUX detector response
as discussed in Section \ref{subsec:DD}. The latter method is preferable
due to timing control over neutron emission and the mono-energetic
nature of the source.

However, the best way to calibrate ER events is by using radioactive
sources introduced to the active xenon volume itself. This runs counter
to the instinct of any reasonable dark matter detector operator due
to worries of LXe contamination. Therefore, the radioactive sources
introduced to the detector's active volume must be either short-lived
isotopes that will decay away rapidly or sources bound to a molecular
form that can be efficiently extracted by the getter and not halt
the dark matter search. LUX employed two such internal calibration
sources: a nearly monoenergetic metastable isotope of krypton, $\mathrm{^{83m}Kr}$,
and a low-energy beta decay isotope $^{3}$H, in the form of tritiated
methane CH$_{3}$T.

\begin{figure}
\begin{centering}
\includegraphics[scale=0.28]{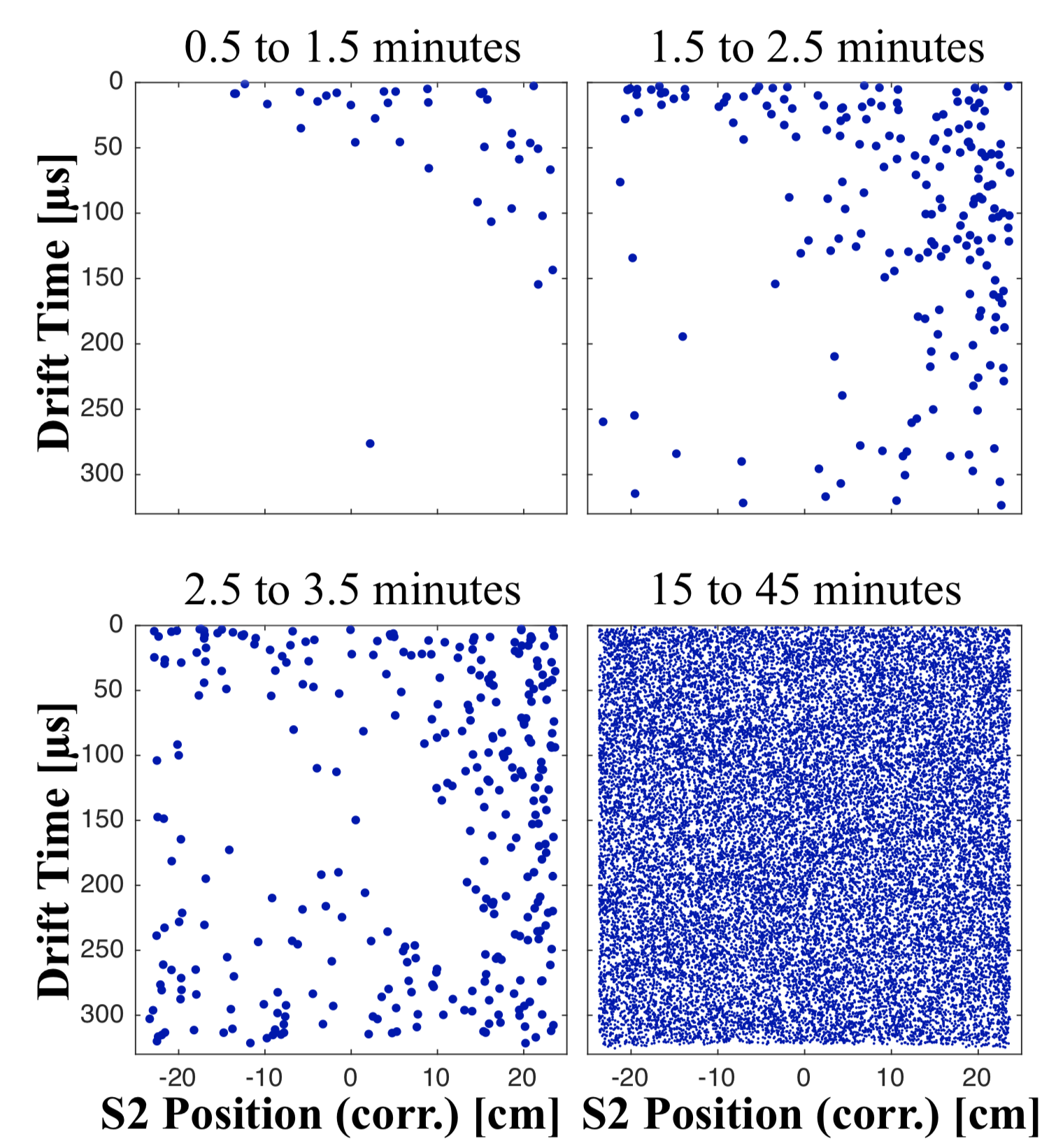}
\par\end{centering}
\caption[Internal source mixing in detector volume]{Positions of reconstructed $\mathrm{^{83m}Kr}$ event vertices in
four distinct time windows selected from a thin vertical slice in
the detector. Initially, a large scale clockwise flow is observed
along with turbulent mixing ($0.5-3.5$ minutes) until $\mathrm{^{83m}Kr}$
becomes uniformly distributed in LXe 15 minutes after injection. Figure
from~\cite{Akerib:2017eql}.\label{fig:kr-uniformity}}

\end{figure}

The internal source gases were introduced from a controlled, measured
volume via plumbing into the gas system circulation using a pressure
differential in the main xenon flow. Once condensed, the radioactive
isotopes mixed uniformly within the detector's active volume within
minutes as illustrated in the time lapse plots in Figure~\ref{fig:kr-uniformity}.

\subsubsection{Metastable krypton-83 ($\mathrm{^{83m}Kr}$)\label{subsec:Metastable-krypton-83}}

\begin{figure}
\centering{}\includegraphics[scale=0.36]{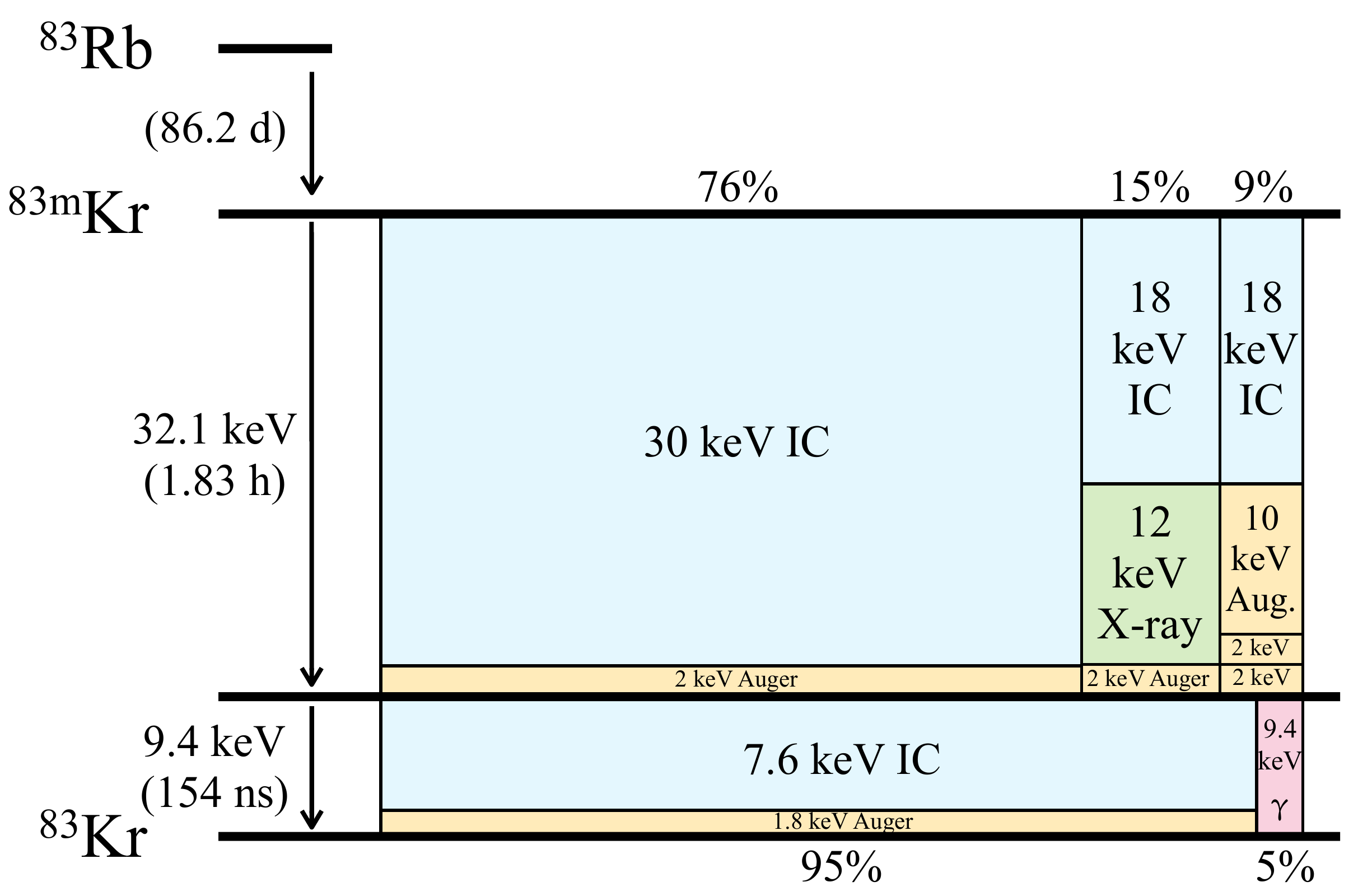}\caption[Decay scheme of $\mathrm{^{83m}Kr}$]{ Decay scheme of $\mathrm{^{83m}Kr}$. The width of each column is
proportional to the branching fraction of that decay mode. Energy
is partitioned between internal conversion electrons (blue), Auger
electrons (yellow), x-rays (green), and $\gamma$-rays (red). Figure
from~\cite{Akerib:2017eql}.\label{fig:kr_decay}}
\end{figure}

The metastable isotope of krypton-83, $\mathrm{^{83m}Kr}$, is a radioactive
element with a half-life of 1.86~hours that decays into its ground
state by emitting a 32.1~keV followed by a 9.4~keV conversion electron
154.4~ns later. Both of these transitions can occur in various ways
as illustrated in Figure~\ref{fig:kr_decay}, but most often they
result in an internal conversion (IC)\nomenclature{IC}{Internal Conversion}
followed by an emission of an Auger electron. Additionally, $\mathrm{^{83m}Kr}$
decay can also result in an emission of an x-ray or a $\gamma$-ray,
with a maximum photon energy of 12~keV. Since it is also a noble
gas, krypton is not removed by the getter. However, in its ground
state, $\mathrm{^{83}Kr}$ is not radioactive and only tiny quantities
are introduced into the detector during calibrations which do not
present any problem to the overall detector performance. LUX has made
extensive use of $\mathrm{^{83m}Kr}$ for calibration purposes. Due
to its inert nature and keV-scale decay energy, it is a natural choice
for calibrating liquid noble detectors. Furthermore, the short half-life
allows for a quick return to baseline trigger rate following injection
and since its total transition energy is well above the range for
WIMP scatters the risk of contaminating the search region is avoided.

$\mathrm{^{83m}Kr}$ was first used as a calibration source in high
energy and heavy ion detectors~\cite{DECAMP1990121,DeMin:1995tk,Eckardt:2001qq,Stiller:2013zba}
and in experiments measuring the endpoint of tritium spectrum~\cite{Asner:2014cwa,Belesev:2008zz,Picard1992,doi:10.1063/1.3671056}.
Initial demonstrations of $\mathrm{^{83m}Kr}$ injection and calibration
technique in liquid xenon, liquid argon, and liquid neon were performed
at Yale University~\cite{kastens2009calibration,Kastens:2009rt,Lippincott:2009ea}
with subsequent studies performed in other detectors~\cite{Manalaysay:2009yq,Aprile:2012an}.
The LUX detector was the first to use $\mathrm{^{83m}Kr}$ for calibrations
of a dark matter experiment \cite{Akerib:2017eql}. The source used
in the LUX detector was made using $^{83}$Rb produced in a proton
beam at Oak Ridge National Laboratory~\cite{MOGHISSI1971218}, deposited
on charcoal at Yale and inserted into the LUX injection system at
SURF. Since $^{83}$Rb decays to $\mathrm{^{83m}Kr}$ with an 86.2~day
half-life, several sources were manufactured during LUX operations.
Typical injections were $\unit[\sim10]{Bq}$, but depending on the
need the amount injected could be higher ($\unit[\sim100]{Bq}$) or
lower.

\begin{figure}
\centering{}\includegraphics[scale=0.58]{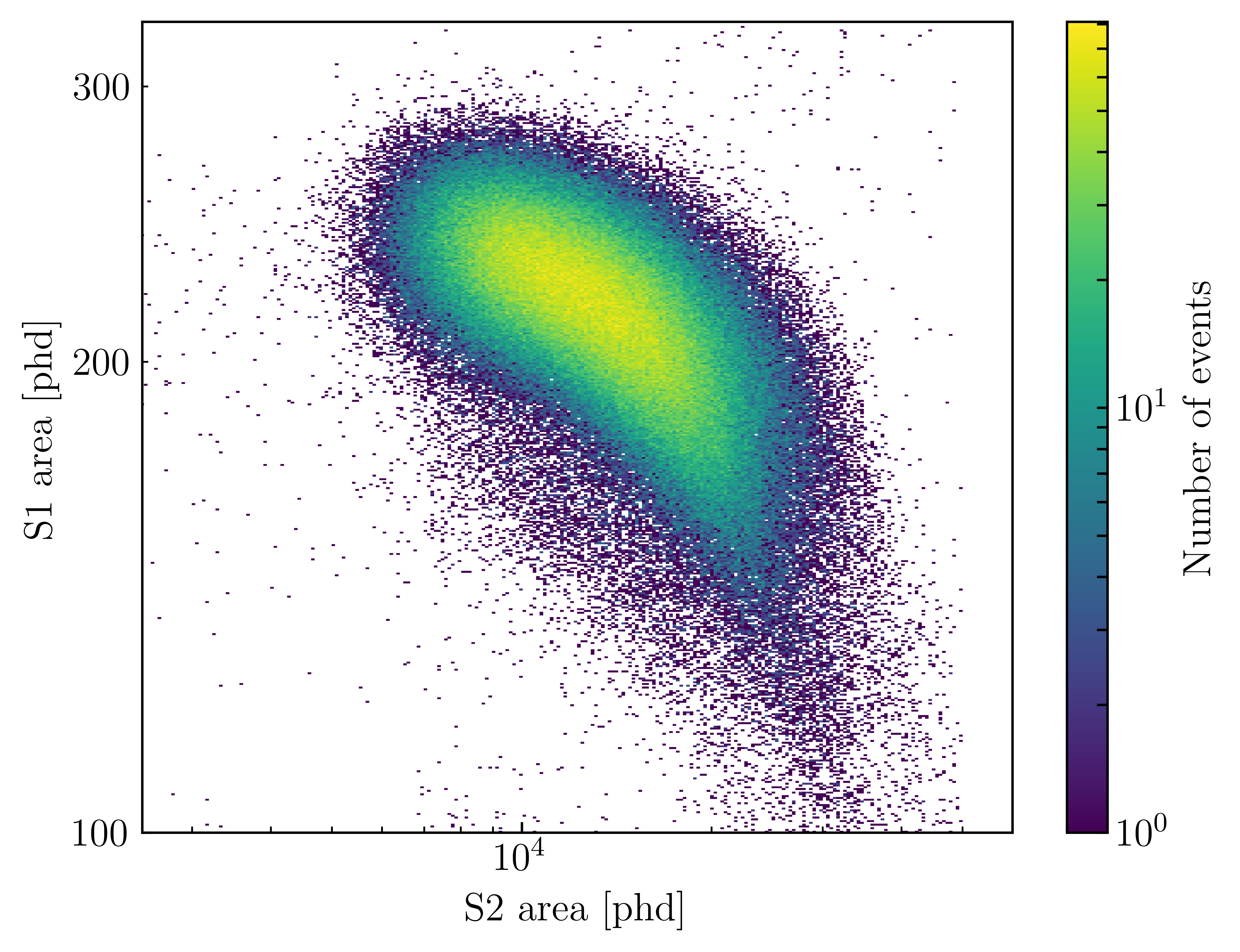}\caption[Selection of $\mathrm{^{83m}Kr}$ events in the detector]{Selection of $\mathrm{^{83m}Kr}$ calibration data in LUX from 2016-08-17
after cuts have been applied. \label{fig:krypton}}
\end{figure}

Figure \ref{fig:krypton} shows a selection of $\mathrm{^{83m}Kr}$
events in the LUX detector remaining after very loose data cuts were
applied. The selected events contain either 1 or 2 S1 signals at most
$\unit[10]{\mu s}$ apart happening before an S2 pulse. The corrected
S1 pulse area is constrained to $100<\mathrm{S1\,phd}<370$ and the
S2 uncorrected pulse area is constrained by $100<\mathrm{S2\,phd}<50,000$.

$\mathrm{^{83m}Kr}$ was used as a ``standard candle'' in LUX. There
were $\sim10^{6}$ $\mathrm{^{83m}Kr}$ events injected weekly. Krypton
atoms mix homogeneously with LXe making $\mathrm{^{83m}Kr}$ a perfect
candidate to use in developing 3D maps of the electric field in the
detector's active volume, position-dependent pulse-area corrections,
field corrections, and more. The application of $\mathrm{^{83m}Kr}$
for development of field maps of the LUX detector will be discussed
in great detail in Chapter~\ref{chap:efield-modeling}. 

\subsubsection*{Pulse-area corrections}

\begin{figure}
\begin{centering}
\includegraphics[scale=0.37]{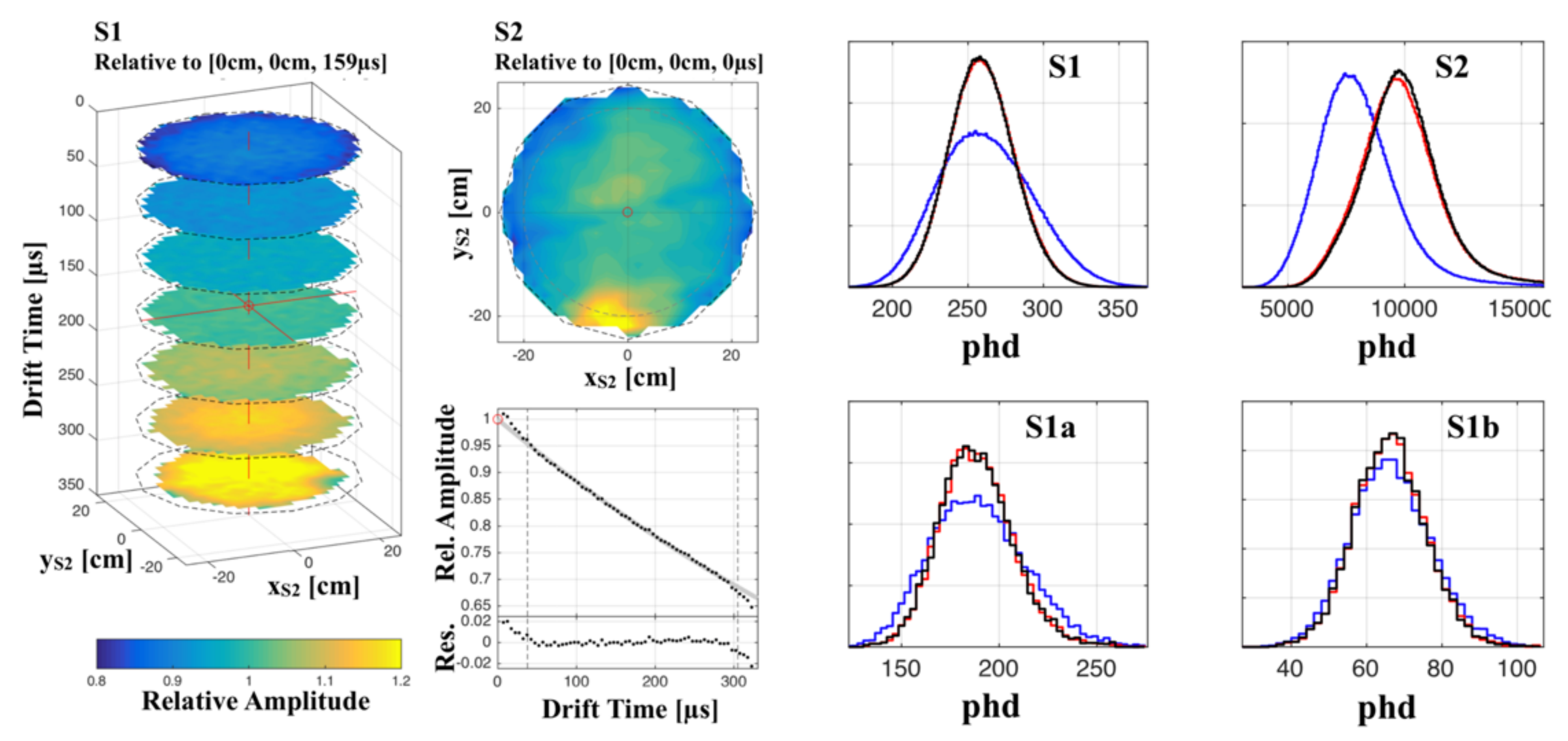}
\par\end{centering}
\caption[Pulse area correction of S1 and S2 signals using $\mathrm{^{83m}Kr}$]{\textbf{Left:} Maps of relative variation of $\mathrm{^{83m}Kr}$
S1 and S2 amplitudes as derived from injection on May 10, 2013. The
S1 and S2 amplitudes share the same color scale. Also shown is the
1D $z$ map correcting for electron lifetime. The $z$ correction
has an exponential fit (in gray) with $\tau_{e}=\unit[805.2]{\mu s}$.
The red circles in each plot represent the normalization points. \textbf{Right:}
The effect of applying the S1 and S2 amplitude corrections. The uncorrected
data is shown in blue; areas corrected for $z$-dependence variation
is shown in red; areas corrected in all three spatial coordinates
are black. The top row treats the two S1 pulses as a single pulse
with $\mathrm{t_{sep}<\unit[1.2]{\mu s}}$. The bottom row shows the
individual S1 peaks used $\left(\mathrm{t_{sep}>\unit[1.2]{\mu s}}\right)$
as a cross-check for S1 corrections. It can be seen that a full 3D
correction represents only a marginal improvement over a $z$-only
correction. Figures from~\cite{Akerib:2017eql}. \label{fig:krypcal}}
\end{figure}

S1 photon detection efficiency varies primarily with event depth,
which needs to be corrected for in the analyses. S1 photons emitted
near the bottom PMT array are $\sim50\%$ more likely to be detected
than those originating near the liquid surface due to the total internal
reflection of scintillation photons at the liquid-gas interface. Hence
the same number of initial photons will produce a different S1 signal
size in the PMT arrays. Figure \ref{fig:krypcal} shows this $z$
dependence on event location. To map this efficiency, the $\mathrm{^{83m}Kr}$
S1 amplitude is found for each bin in $\left(x_{\mathrm{S2}},\,y_{\mathrm{S2}},\,t_{\mathrm{S2}}\right)$
and a 3D S1 correction map is constructed as the inverse of these
amplitudes, normalized to the detector center at $\left(\unit[0]{cm},\,\unit[0]{cm},\,\unit[159]{\mu s}\right)$.
The efficiency-correction map is then applied as a linear interpolation
on the 3D grid. Since S1 correction maps did not vary much with date
a single large $\mathrm{^{83m}Kr}$ injection was used for the entire
WS2013. WS2014-16 required additional calibration data due to varying
electric fields.

For S2 pulse corrections, the $z$ and $\left(x,y\right)$ position
corrections are applied independently. The $z$ dependence results
from electron capture on electronegative impurities while drifting.
This correction is a simple exponential fit normalized to unity at
the liquid surface as illustrated in Figure~\ref{fig:krypcal}. The
correction is interpolated smoothly between $\mathrm{^{83m}Kr}$ injections.

Non-uniformities of the S2 signal in the $\left(x,y\right)$ plane
are due to variations of the liquid surface, a slight tilt of the
detector, and bowing or sagging of the wire grid plane. The corrections
are constructed by binning $\mathrm{^{83m}Kr}$ data on an $\left(x,y\right)$
grid and finding average S2 amplitudes for each 2D bin normalized
to the center of the liquid surface.

\subsubsection*{Field variations using S1a:S1b ratio}

As shown in Figure~\ref{fig:kr_decay}, $\mathrm{^{83m}Kr}$ proceeds
through two back-to-back transitions. Because the S2 signals have
FWHM\nomenclature{FWHM}{Full Width at Half Maximum} of $\unit[1.0-1.9]{\mu s}$
depending on $z$ position, the two decay steps are merged in the
relatively broad S2 signal. Due to the short S1 pulse width ($\unit[\sim100]{ns}$)
a significant fraction of $\mathrm{^{83m}Kr}$ S1 pulses are detected
individually in the PMT array. Those two S1 pulses are referred to
by their ordering as S1a (32.1 keV) and S1b (9.4 keV). The S1a and
S1b pulses can be used to study the dependence of electron-ion recombination
on the electric field. A weaker field allows more recombination, enhancing
the S1 signal and proportionally suppressing the S2 signal, while
a higher field has the opposite effect. The ratio of the two S1 amplitudes
varies with field alone since any detector gains and efficiencies
affect both amplitudes equally causing the S1b/S1a ratio to increase
with the electric field. This ratio technique for field amplitude
measurement was used in WS2014-16, where the field variations in $\mathrm{^{83m}Kr}$
recombination were more substantial and required careful treatment. 

Further details about the LUX $\mathrm{^{83m}Kr}$ calibration techniques
can be found in~\cite{Akerib:2017eql,knoche2016signal}.

\subsubsection{Tritium ($^{3}$H)\label{subsec:Tritium-(H)}}

Tritium has a 12.3 year half-life and $Q$ value\footnote{The $Q$ value is the total energy released in the nuclear decay,
given by the sum of kinetic energies of the emitted beta particle,
neutrino, and recoiling nucleus. Since the nucleus has a significantly
higher mass than the neutrino or the beta particle, the energy of
the emitted beta particle ranges from 0 to $\sim Q$.} $\unit[Q=18.6]{keV_{ee}}$. Its mean energy is $\unit[5.6]{keV_{ee}}$,
and the beta energy spectrum has a broad peak at $\unit[2.5]{keV_{ee}}$
both of which have a significant overlap with the WIMP NR range, so
it is perfectly tailored to the calibration of electron recoils of
a dark matter detector. 

$^{3}$H was introduced into the detector as tritiated methane gas
CH$_{3}$T and was mixed within the detector's active volume on a
similar timescale as $\mathrm{^{83m}Kr}$. The source was prepared
by an external vendor with an activity of $\unit[3.7]{MBq}$ per millimole
and inserted into the circulation in a 2.25-liter stainless steel
bottle mixed with 2 atmospheres of LUX-quality purified xenon. Prior
to the first injection, the ability of the LUX getter to efficiently
remove CH$_{3}$T from xenon gas was verified to prevent absorption
into plastics in the detector due to its long half-life.

\begin{figure}
\centering{}\includegraphics[scale=0.2]{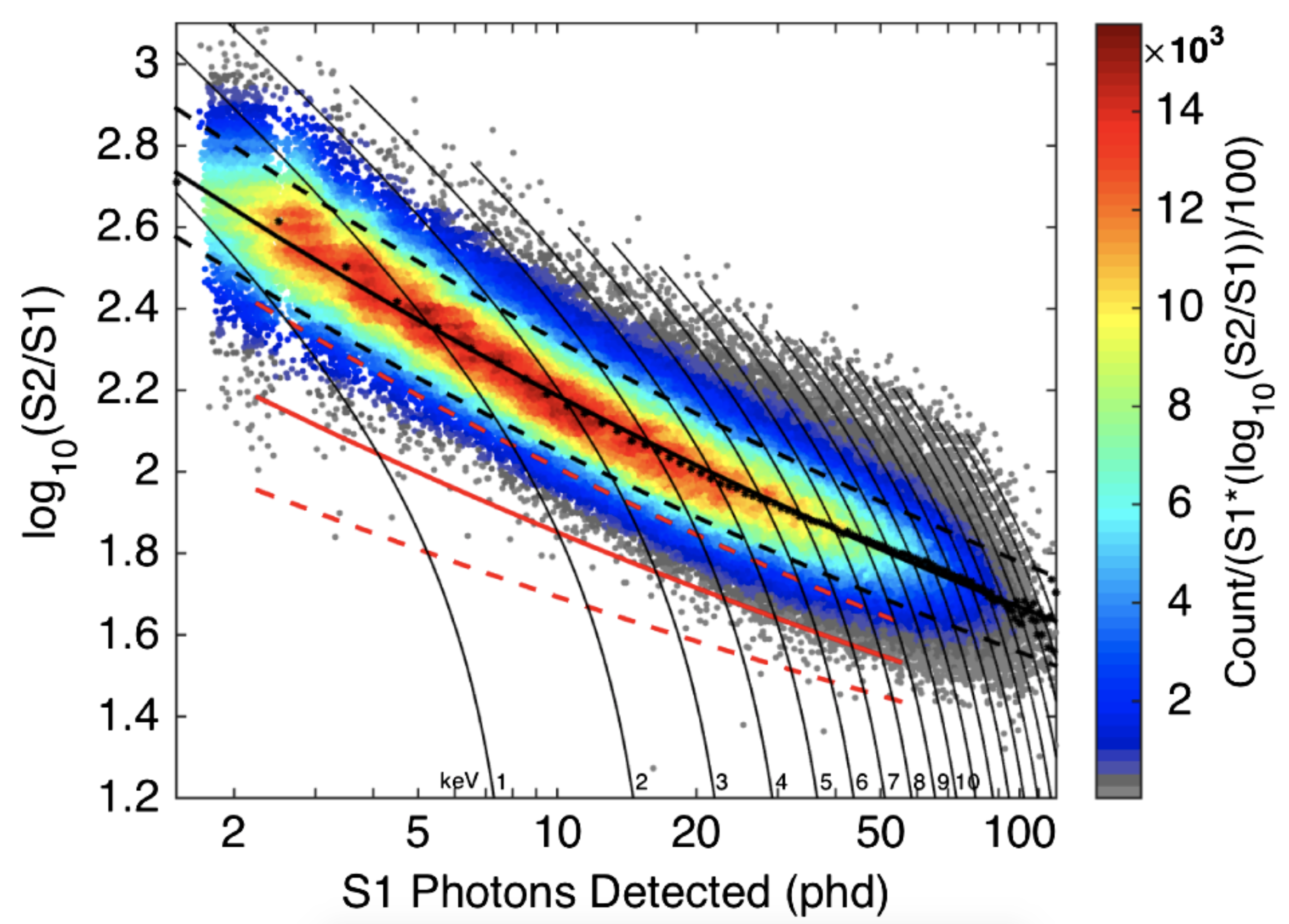}\caption[LUX electron recoil band from tritium data]{The LUX electron recoil band illuminated by $^{3}$H events at 180
V/cm. The tritium data is shown as a scatter plot in $S1$ vs. $\log(S2/S1)$
space with contours of constant ER energy (solid gray lines). Measured
Gaussian means in each S1 bin are shown with an empirical power line
fit (solid black line) along with the 10 and 90\% contours (dashed
black lines). Also shown is the equivalent NR band mean (solid red
line) along with the 10 and 90\% contours (dashed red lines) obtained
from deuterium-deuterium neutron generator data discussed in Section~\ref{subsec:DD}.
Figure from~\cite{Akerib:2015wdi}.\label{fig:ER-band-tritium}}
\end{figure}

\begin{figure}
\begin{centering}
\includegraphics[scale=0.28]{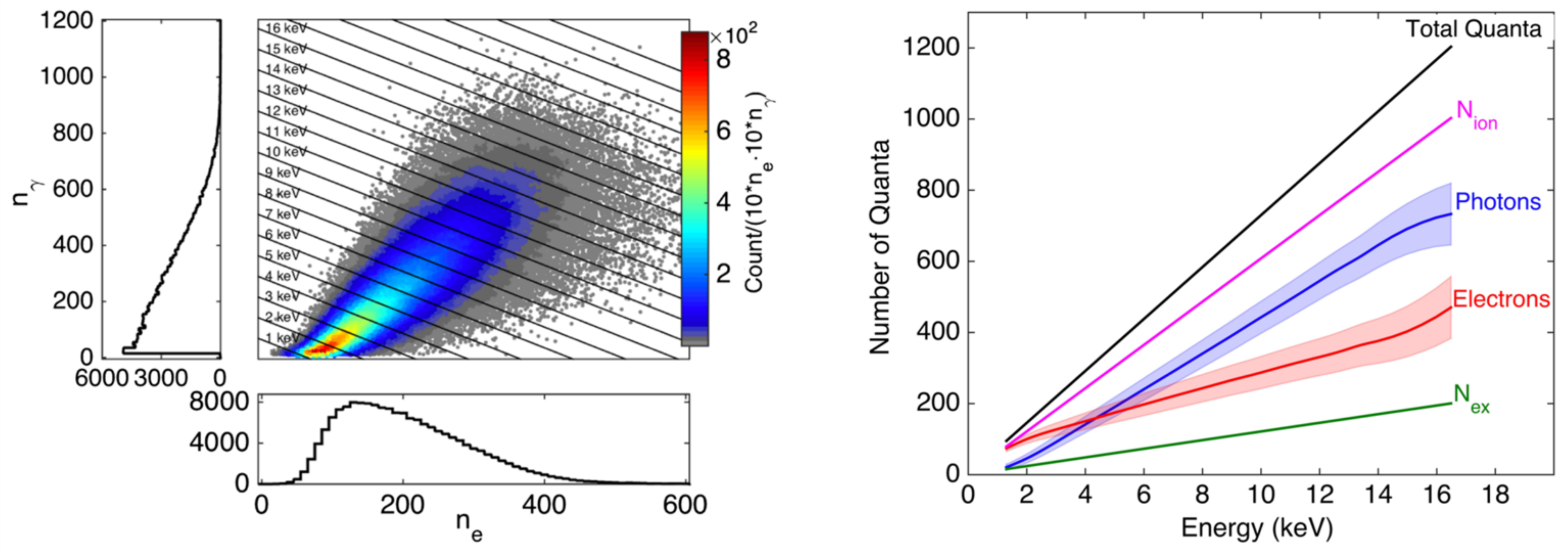}
\par\end{centering}
\caption[Characterization of tritium events in the LUX detector]{\textbf{Left:}~Scatter plot of $n_{e}$ vs $n_{\gamma}$ for 170,000
fiducial tritium events at 180 V/cm as measured by LUX. Lines of constant
energy according to Equation~\ref{eq:EW_S1S2} are shown assuming
$W=\unit[13.7]{eV}$. \textbf{Right:}~The underlying mean number
of quanta responsible for the shape of the ER band in LUX at 180 V/cm.
The bands indicate the correlated systematic errors on $g_{1}$ and
$g_{2}$. Figures from~\cite{Akerib:2015wdi}.\label{fig:tritium_energy}}
\end{figure}

\begin{figure}
\begin{centering}
\includegraphics[scale=0.29]{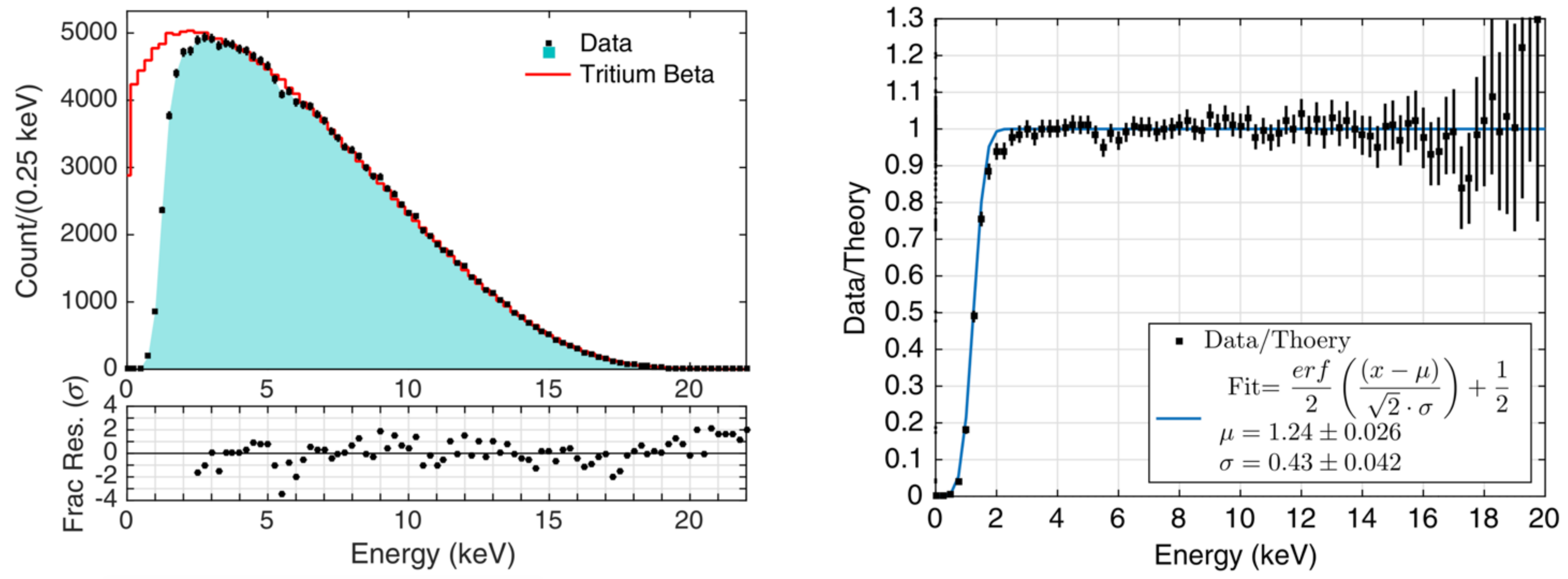}
\par\end{centering}
\caption[Tritium energy spectrum measured by LUX]{\textbf{Left:}~Tritium energy spectrum from LUX (black) compared
to a theoretical tritium spectrum convolved with detector resolution
(red) $\nicefrac{\sigma_{E}}{W}=\sqrt{\sigma^{2}\left(n_{\gamma}\right)+\sigma^{2}\left(n_{e}\right)}$.
The cut off of the data on the left is due to a finite detector resolution
at low energies. \textbf{Right:}~Ratio of measured tritium energy
spectrum (black data) and the theoretical one (blue line) for the
same dataset. A fit to an error function is also shown. Figures from~\cite{Akerib:2015wdi}.\label{fig:tritium-spectrum}}
\end{figure}

CH$_{3}$T was injected into the detector several times throughout
its operation to characterize the detector's ER response. This is
critical for correct mapping of the LUX background model to the signal
parameter space for analysis since the dominant background for the
WIMP search is leakage of ER events into the NR band in the $S1$
vs. $\log(S2/S1)$ space as illustrated in Figure~\ref{fig:ER-band-tritium}.
Figure~\ref{fig:tritium_energy} shows the observed spectrum in $n_{e}-n_{\gamma}$
space as measured by LUX using 170,000 $^{3}$H decays recorded in
the fiducial volume after the WS2013 dark matter search. Also shown
is the total number of electrons and scintillation photons after recombination
as a function of energy. Above 4 keV the number of electrons drops
below the number of photons, which is consistent with a substantial
recombination effect at these energies and an electric field of 180
V/cm. The tritium spectrum as measured by the detector and the ratio
of the data to the smeared theoretical spectrum is shown in Figure~\ref{fig:tritium-spectrum},
along with an empirical fit to an error function. The effective 50\%
energy threshold for ER events was found at $\unit[1.24\pm0.026]{keV}$.
Further details about the LUX $^{3}$H calibration technique can be
found in~\cite{Akerib:2015wdi,knoche2016signal,dobi2014measurement}.


\subsubsection{Deuterium-deuterium (DD) neutron beam \label{subsec:DD}}

\begin{figure}
\begin{centering}
\includegraphics[scale=0.3]{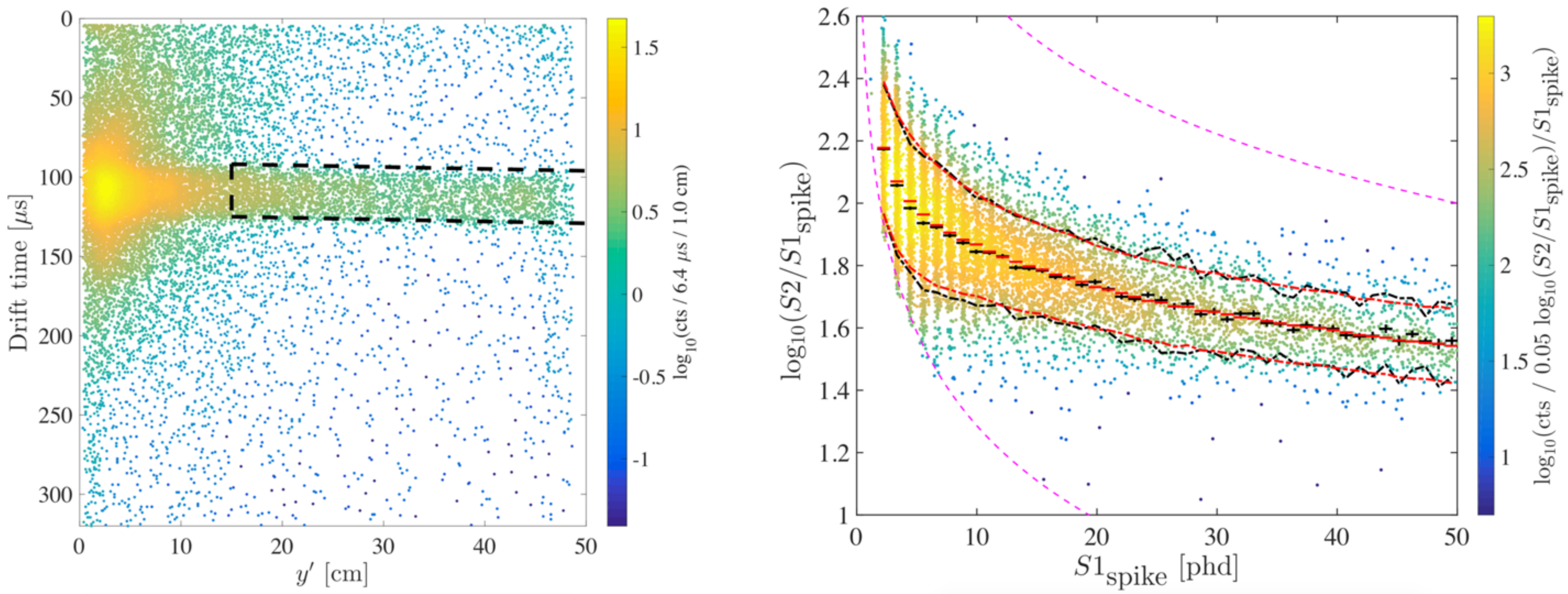}
\par\end{centering}
\caption[Calibration of LUX using deuterium-deuterium neutron beam]{\textbf{Left:}~Spatial distribution of single-scatter events from
DD calibration where $y'$ axis is parallel to the direction of the
neutron beam. The neutron beam pipe is aligned at $\unit[107]{\mu s}$.
Black dashed rectangle selects for monoenergetic scatters inside the
fiducial volume. A 12.6 cm total mean free path for neutrons makes
it more probable for a neutron to exit out of the top of the xenon
volume than the bottom. This plot contains the full 107.2 live-hours
of 2013 DD data. \textbf{Right:}~The LUX nuclear recoil band (black
line) and the corresponding Gaussian fit mean value for the simulated
NR band (red line). The 10 and 90\% contours are also shown for data
(dashed black) and simulation (dashed red). Figures from~\cite{Akerib:2016mzi}.\label{fig:dd}}
\end{figure}

Calibration of the NR response was accomplished with the scattering
of 2.45~MeV neutrons sourced by a deuterium-deuterium (DD)\nomenclature{DD}{Deuterium-Deuterium}
neutron fusion generator. An Adelphi Technology, Inc. DD108 generator
was located outside the water tank. A collimated beam of neutrons
was fired through an air-filled conduit suspended in the water to
penetrate the LXe space. The conduit is visible in Figure~\ref{fig:LUX-and-lab}
and could be raised or lowered alongside the detector to enable calibrations
at various $z$. The generator works by accelerating deuterium ions
into a deuterium-loaded target. To achieve a monoenergetic beam, only
neutrons emitted at a $90^{\circ}$ angle relative to the incoming
deuterium are selected from the $^{2}$H(d,n)$^{3}$H fusion reaction.
The air-filled conduit results in a tightly collimated beam visible
in the detector as shown in Figure~\ref{fig:dd}. Visible closer
to the wall are neutron events that have scattered outside of the
TPC and contaminate the monoenergetic data selected by the black dashed
box.

The collimated monoenergetic neutrons arriving to the TPC can scatter
off xenon nuclei with a known angular distribution and deposit varying
amounts of energy. Even though the two S1s from the interactions overlap,
the two S2 signals can be resolved if they are sufficiently separated
in $z$. This can be leveraged for measurements of neutron charge
and light yields in the detector. The charge yields are calculated
first by selecting for double scatter events where the reconstructed
vertex position enables calculation of the scattering angle at the
first interaction site. This kinematically constrains the deposited
energy of the first nuclear recoil, enabling the ionization yield
to be measured as a function of nuclear recoil energy from 0.7 to
24.2~keV$_{\mathrm{nr}}$. With this measurement, single-scatters
can be used to determine the light yield. In LUX the light yield was
measured as a function of recoil energy between $\unit[0.7-74]{keV_{nr}}$.
Figure~\ref{fig:dd} shows the NR band of the LUX detector found
both from data and simulations. Further details about the LUX DD calibration
technique can be found in~\cite{Akerib:2016mzi,verbus2016absolute}.

\section{Simulations and backgrounds\label{sec:Simulations-and-backgrounds}}

The LUX detector is sufficiently complex that a deterministic model
would require excessive computational resources. However, Monte Carlo
simulation methods with appropriate constraints generate the required
statistical confidence needed to achieve a thorough understanding
of the detector. This starts by modeling the detector itself; this
model is then used to simulate particle interactions that are used
to simulate signal and background models for the final analysis. In
LUX, the \textsc{Geant4}-based~\cite{AGOSTINELLI2003250} LUXS\textsc{im}~\textsc{\cite{AKERIB201263}}
software package was used featuring a very detailed reconstruction
of the LUX detector geometry inside the water tank based on CAD designs.
Energy deposits from particle interaction in LXe were then used to
model the production of VUV scintillation photons and ionization electrons.
These light and charge yields were modeled with the Noble Element
Simulation Technique (NEST)~\cite{NEST_Szydagis,NEST_lenardo}, a
semi-empirical collection of models based on past detector calibration
data. Finally, full ER and NR simulations datasets were created and
passed through the data-processing chain to provide estimates of the
expected WIMP NR signals and the predicted background spectra. Those
were then used in the probability density functions (PDFs)\nomenclature{PDF}{Probability Density Function}
used in the profile likelihood ratio analysis. Further details about
the use of LUXS\textsc{im} in LUX analysis can be found in~\cite{Akerib:2017vbi}.

Due to the rare-event nature of the search, minimizing backgrounds
was one of the primary concerns during detector design and construction.
There are several groups of backgrounds discussed in the remainder
of this section. Further details about both ER and NR backgrounds
are described in~\cite{Akerib:2017vbi,Akerib:2014rda}. 

\paragraph{ER backgrounds:}

ERs dominated backgrounds: the main sources were Compton scattering
$\gamma$-rays from detector components, and $\beta$ and electron-capture
decays within the LXe target. The bulk of the events originated in
the detector materials, such as products from the decays of $^{40}$K,
$^{60}$Co, and the many radioisotopes of the uranium and thorium
chains that emit high-energy $\gamma$-rays that can Compton scatter
once in the fiducial volume and leave. Other backgrounds arise from
the xenon target itself: $\beta$ decay backgrounds from $^{85}$Kr
(remaining in xenon after purification) and $^{214}$Pb (a radon daughter),
and electron capture decays of $^{127}$Xe and $^{37}$Ar. $^{127}$Xe
is created while xenon is exposed to cosmic rays and thermal neutrons
on the surface, but thanks to its half-life of 36 days $^{127}$Xe
was a significant source of backgrounds only during WS2013, where
it was also used for low-energy ER calibrations~\cite{Akerib:2017hph}. 

\paragraph{NR backgrounds:}

\begin{figure}
\begin{centering}
\includegraphics[scale=0.29]{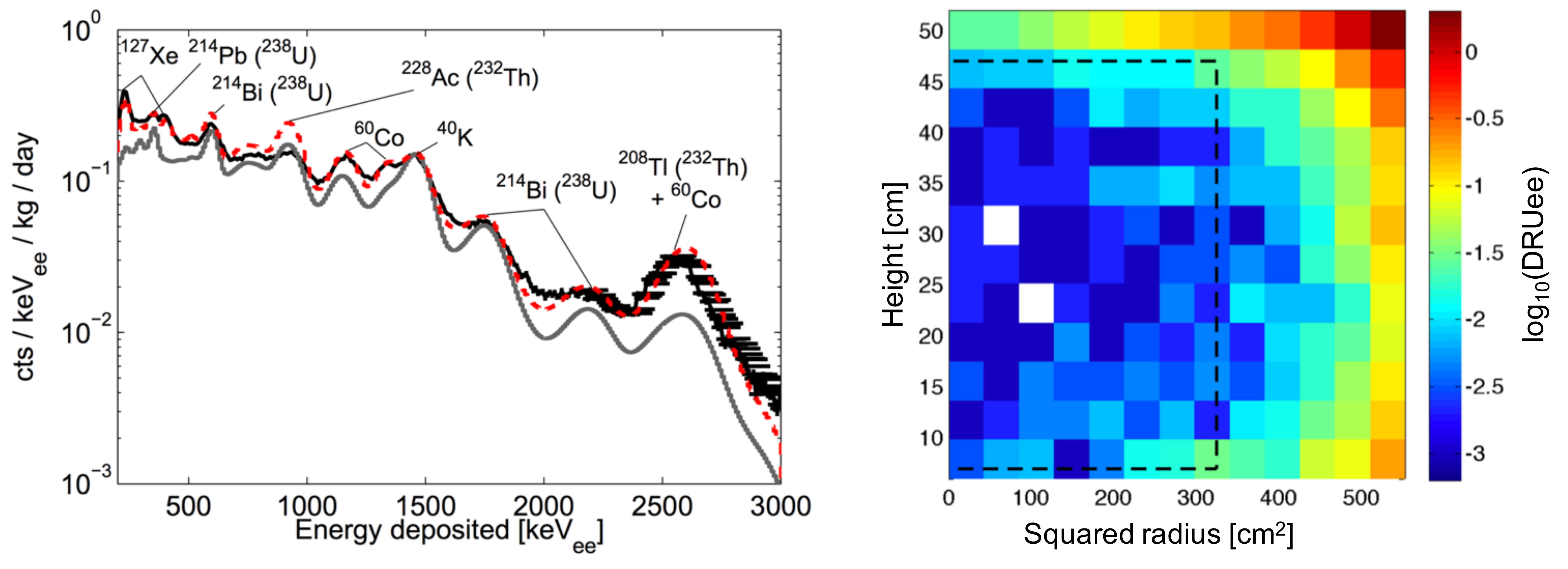}
\par\end{centering}
\caption[Backgrounds in the LUX detector]{\textbf{Left:} Measured $\gamma$ spectrum in the LUX active volume
without the top and bottom 2~cm (black) including both single- and
multiple-scatters. Two simulations spectra are also shown: a spectrum
based on counting measurements alone (gray) and a spectrum with best-fit
scaling for $^{238}$U, $^{232}$Th, $^{40}$K, and $^{60}$Co decays
(dashed red). \textbf{Right:} LUX $\gamma$ ray ER background density
in the range $\unit[0.9-5.3]{keV_{ee}}$ ($\sim\unit[2-30]{phe}$)
as a function of position from measured data. Rates are in units of
log$_{10}$(DRU$_{\mathrm{ee}}$) where DRU is the differential rate
unit of $\mathrm{events\cdot kg^{-1}\cdot day^{-1}\cdot keV_{ee}^{-1}}$.
The 118~kg fiducial volume used in the WS2013 is overlaid as the
black dashed contour. Figures from~\cite{Akerib:2014rda}.\label{fig:LUX-Backgrounds}}
\end{figure}

NR backgrounds arise from ($\alpha$, n) reactions or from cosmogenic
muon interactions in the water or cavern rock, but provide only a
minimal background for the search. Results from an extensive radioactivity
screening campaign with high-purity germanium detectors and neutron
activation form the LUX background model, providing input into the
Monte Carlo simulations. Figure~\ref{fig:LUX-Backgrounds} shows
the measured $\gamma$ ray spectrum in the LUX active volume with
identified peaks and the observed distribution of low-energy background
events in WS2013. 

\paragraph{Wall backgrounds:}

Further backgrounds arise from interactions occurring very close to
the detector's PTFE wall, referred to as ``wall events.'' These
events can exhibit a loss of charge to the PTFE, which results in
a low $\log_{10}(S2)$. In theory, these backgrounds should be excluded
by the fiducial cut, but due to the limited accuracy of position reconstruction,
some of these events statistically leak into the search region. Empirical
estimates are calculated using control samples of the data and included
in the background model. A detailed discussion of these backgrounds
can be found in~\cite{nehrkorn2018testing,lee2015mitigation,boulton2018}.

\paragraph{Random coincidence backgrounds:}

Random coincidence of unrelated S1-only and S2-only events can generate
apparent WIMP-like events. The isolated S1 rate from threshold to
50~phd is 1~Hz. This leads to a prediction of 1.1~background events
in the 95~days of WS2013~\cite{Akerib:2017vbi}. 

\paragraph{Neutrino backgrounds:}

Another background for WIMP detectors without directional sensitivity
are two kinds of neutrino interactions: $\nu-e^{-}$ neutral current
elastic scattering, where the neutrino interacts with atomic electrons,
and $\nu-A$ neutral current coherent elastic scattering, where the
neutrino interacts with the target nucleus~\cite{monroe2007,Billard:2013qya}
(observed recently for the very first time~\cite{Akimoveaao0990}).

Neutrino-electron scattering contributes to ER backgrounds. It is
dominated by low energy neutrinos, such as $pp$ solar neutrinos,
with contributions from the $^{7}$Be and CNO\footnote{Carbon\textendash nitrogen\textendash oxygen (CNO) cycle is a set
of fusion reactions that convert hydrogen to helium in stars.} chains. Coherent elastic neutrino-nucleus scattering results in NR
background. At low WIMP masses, $^{8}$B and $hep$ solar neutrinos\footnote{The $hep$ reaction $^{3}\mathrm{He}+p\rightarrow^{4}\mathrm{He}+e^{+}+\nu_{e}$
produces neutrinos with the highest endpoint (18.8 MeV) expected for
solar neutrinos.} are expected to contribute the most background events, while at masses
above $20$ GeV/c$^{2}$ contributions from diffuse supernova neutrinos
and atmospheric neutrinos will dominate~\cite{Akerib:2018lyp}.

Fortunately, in LUX these interactions are expected to have produced
only \ensuremath{\sim} 0.1 detectable NR events in the full exposure~\cite{Akerib:2015rjg}.
However, these will become an obstacle for the increasingly large
next-generation dark matter detectors, such as LZ.

\section{LUX WIMP search results\label{sec:LUX-WIMP-search}}

\begin{figure}[t]
\begin{centering}
\includegraphics[scale=0.58]{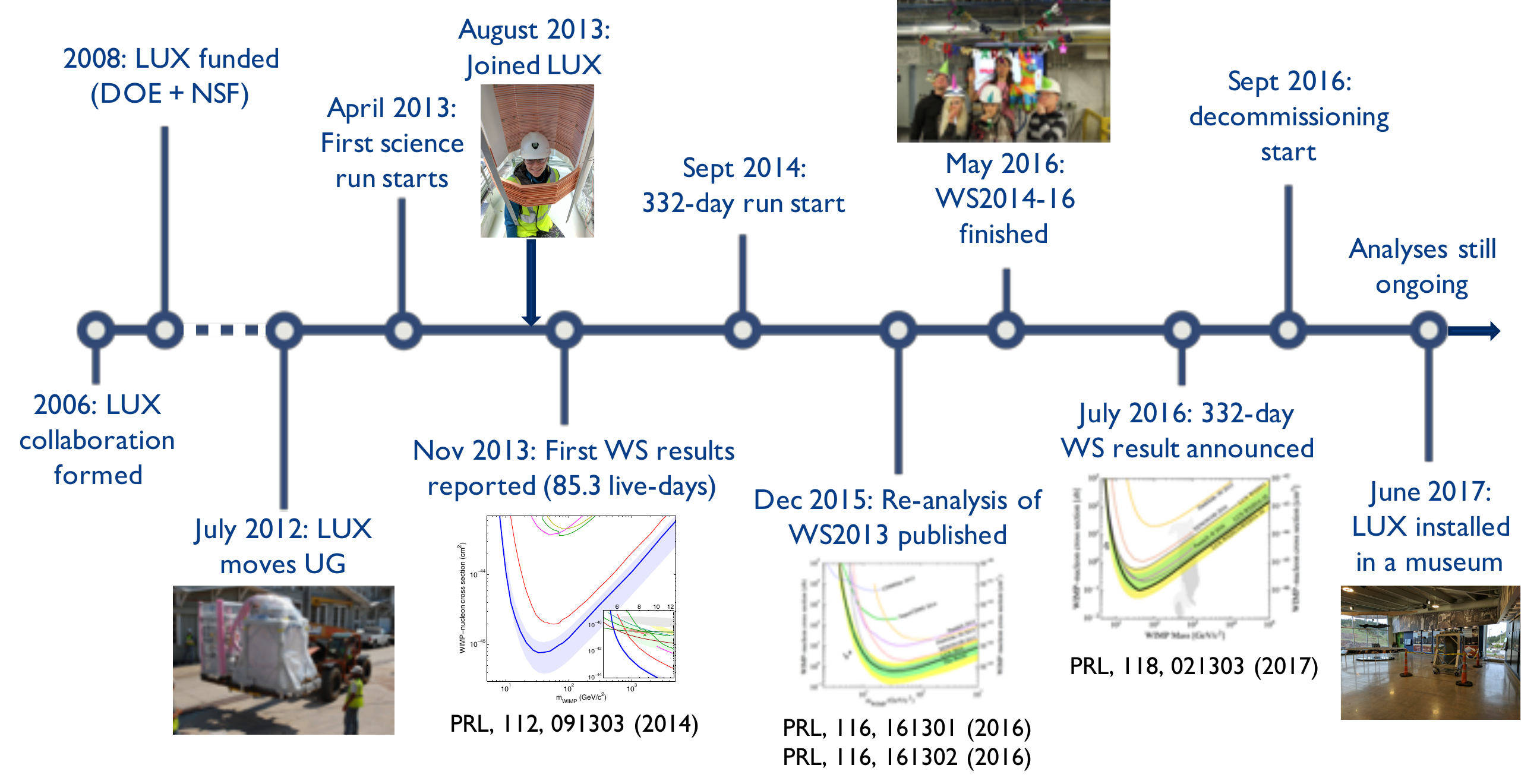}
\par\end{centering}
\caption[LUX timeline]{LUX timeline. References to results published in Physical Review
Letters (PRL)\nomenclature{PRL}{Physical Review Letters} are indicated
below the limits plots. \label{fig:LUX-timeline.}}
\end{figure}
\begin{table}
\bigskip{}
\begin{centering}
\begin{tabular}{l|>{\raggedright}p{4cm}>{\raggedright}m{7cm}}
\hline 
Name & Date & Goal\tabularnewline
\hline 
\hline 
Run1 & Early 2011\\
 & \multirow{1}{7cm}{First detector cool down with argon gas \\
in the surface lab}\tabularnewline
Run2 & Late 2011\\
 & \multirow{1}{7cm}{Full systems test in the water tank \\
in the surface lab}\tabularnewline
Run3 & April 24, 2013 -\\
September 1, 2013 & \multirow{1}{7cm}{WS2013, 95 live-days, \\
in Davis Cavern }\tabularnewline
Run4 & September 11, 2014 -\\
May 2, 2016 & \multirow{1}{7cm}{WS2014-16, 332 live-days, \\
in Davis Cavern }\tabularnewline
\hline 
\end{tabular}
\par\end{centering}
\caption[Overview of LUX data taking periods]{Overview of LUX data taking periods. Details regarding the detector
assembly, Run1, and Run2 can be found in~\cite{faham2014prototype}.
A grid conditioning campaign took place during an engineering run
between Runs 3 and 4 in spring 2014 in the Davis Cavern.\label{tab:Overview-of-LUX-runs}}
\end{table}

After the tremendous effort that was put into the LUX detector design,
assembly, and commissioning, the LUX detector began taking WIMP search
data in 2013 and continued its scientific quest until fall 2016 when
it was disassembled to make space for the LUX-ZEPLIN (LZ)\nomenclature{LZ}{LUX-ZEPLIN}
dark matter detector and moved to a museum exhibit at the Sanford
Lab visitor center in Lead, SD. Despite not finding WIMPs, the LUX
operations were a success. The absence of a dark matter signal combined
with its edge-cutting sensitivity enabled LUX to place world-leading
exclusion limits on WIMP properties. Exclusion limits are helpful
for theoretical modeling, evaluating past detection claims of dark
matter, and informing the next generation of dark matter experiments.
LUX also pioneered many novel calibration techniques now used commonly
in the field. Furthermore, the expertise developed by the LUX experiment
provides beneficial know-how for the design and construction of the
next-generation of dark matter detectors.

Figure~\ref{fig:LUX-timeline.} shows the timeline of the LUX experiment.
Before starting the main scientific dark matter search, the detector
was commissioned on the surface to ensure all systems were adequately
operating before the detector was sealed and moved underground. An
overview of the four main scientific runs is illustrated in Table~\ref{tab:Overview-of-LUX-runs}.
Data used for the main dark matter search analysis were collected
over two scientific runs - Runs 3 (WS2013) and Run4 (WS2014-16), which
led to the publication of three primary WIMP search results discussed
below. These are limits on isospin-invariant SI coupling as this is
the WIMP scenario that the LUX detector is the most sensitive to.
Additionally, scientific results for various other types of dark matter
have been published.

\subsection{Profile likelihood ratio\label{subsec:Profile-likelihood-ratio}}

To claim a discovery or place a limit, a frequentist significance
test using the profile likelihood ratio (PLR)\nomenclature{PLR}{Profile Likelihood Ratio}
statistic was used, a common choice among many recent particle physics
experiments. The following discussion provides a high-level overview
of this topic. For a more in-depth treatment consult References~\cite{nehrkorn2018testing,Patrignani:2016xqp,Cowan:2010js,Cranmer:2015nia,Feldman}
that were used to compile this section.

A hypothesis $H$ is a statement about the probability for the data,
often written $P\left(x|H\right)$. For purposes of discovering a
new signal, the WIMP signal plus background model constitutes the
null hypothesis $H_{0}$, which is tested against an alternative background-only
hypothesis $H_{1}$. In this context the null hypothesis is the test
hypothesis since one is trying to test the presence of a signal with
strength parametrized by $\sigma$, trying to reject it at a specified
confidence limit (CL)\nomenclature{CL}{Confidence Limit}. In the
analysis, $H_{0}$ and $H_{1}$ are a suite of Monte Carlo pseudo
experiments generated using modeled backgrounds ($H_{1}$) and backgrounds
with an expected signal ($H_{0}$). To set a limit, an acceptance
region in data space $\mathbf{x}$ is defined that contains 90\% of
experiments under $H_{0}$. The signal model $H_{0}$ is rejected
when the observed data falls outside that region. 

If the probability $P\left(\mathbf{x}|H\right)$ for data $\mathbf{x}$
is regarded as a function of the hypothesis $H$, it is called the
likelihood of $H$, written as $\mathscr{L}\left(H\right)$. Often
the hypothesis is characterized by one or more parameters $\theta$,
in which case $\mathscr{L}\left(\theta\right)\equiv P\left(\mathbf{x}|\theta\right)$
is called the likelihood function.

A region of data space $\mathbf{x}$, called the acceptance region
$w$, is specified such that there is no more than a given probability
$\alpha$ under $H_{0}$, called the confidence level of the test,
to find $\mathbf{x}\in w$. If the data are observed in the acceptance
region, $H_{0}$ is rejected. To maximize the power of a test of $H_{0}$
with respect to the alternative $H_{1}$, the Neyman-Pearson lemma~\cite{Neyman289}
states that the acceptance region $w$ should be chosen such that
for all data values $\mathbf{x}$ inside the acceptance region $w$
the ratio
\[
\lambda(\mathbf{x})=\frac{f\left(\mathbf{x}|H_{1}\right)}{f\left(\mathbf{x}|H_{0}\right)}
\]
is greater than a given constant, the value of which is determined
by the size of the test $\alpha$. In a dark matter search, one is
trying to minimize $\beta\equiv\int_{critical\,region}P\left(\mathbf{x}|H_{1}\right)\mathrm{d\mathbf{x}}$
at fixed $\left(1-\alpha\right)\equiv\int_{critical\,region}P\left(\mathbf{x}|H_{0}\right)\mathrm{d\mathbf{x}}=0.9$.

If a model contains systematic uncertainties, they can be handled
via nuisance parameters. Let $\sigma$ denote the model parameter
of interest (i.e., the WIMP-nucleon cross section for a given WIMP
mass in most LUX analysis) and $\theta$ denotes the nuisance parameters,
whose values are also unknown but not of primary concern. Some nuisance
parameters used in the LUX experiment were the Lindhard parameter
$k$ and the Gaussian constraints on various background event counts.
The likelihood of a dataset $\mathcal{D}$ as a function of these
model parameters is expressed as $\mathscr{L}\left(\mathcal{D}|\sigma,\theta\right)$
and is the probability given certain values of the model parameters.

A maximum likelihood estimator (MLE) $\hat{\theta}$ (written with
a single hat) is a function of the data used to estimate the value
of the parameter $\theta$. A conditional maximum likelihood estimate
(CMLE) $\hat{\hat{\theta}}(\sigma)$ (written with two hats) is the
value of $\theta$ that maximizes the likelihood function with $\sigma$
fixed in order to test a hypothesis value. Then the procedure for
choosing specific values of the nuisance parameters for a given value
of $\sigma$, the given dataset $\mathcal{D}$, and the global observable
is often referred to as ``profiling.'' Similarly, $\hat{\hat{\theta}}(\sigma)$
is often called the ``profiled value of $\theta$.'' Given these
definitions, the profile likelihood ratio defined as
\[
\lambda(\sigma_{test})=\frac{\mathscr{L}\left(H_{0}\right)}{\mathscr{L}\left(H_{1}\right)}=\frac{\mathscr{L}\left(\sigma_{test},\hat{\hat{\theta}}\right)}{\mathscr{L}\left(\hat{\sigma}_{test},\hat{\theta}\right)}
\]
depends explicitly on the parameter of interest $\sigma$ and is independent
of the nuisance parameters $\theta$ that have been eliminated via
profiling. The MLE $\hat{\sigma}$ for the parameter of interest $\sigma$
is obtained from a global fit (everything can float) of the likelihood
function to the data. The MLEs for the nuisance parameters $\hat{\theta}$
and the CMLEs $\hat{\hat{\theta}}$ for the nuisance parameters $\theta$
are obtained by maximizing the likelihood with the parameter of interest
fixed to $\sigma_{\mathrm{test}}$. The maximum likelihood $\mathscr{L}\left(\sigma_{test},\hat{\hat{\theta}}\right)$
is found by fixing $\sigma$ and optimizing all other parameters.
The maximum likelihood $\mathscr{L}\left(\hat{\sigma}_{test},\hat{\theta}\right)$
is found by optimizing $\sigma$. The PLR test statistic $q$ is then
defined as 
\[
q_{\sigma}\equiv-2\ln\left(\lambda\left(\sigma\right)\right).
\]
As the test statistic $\sigma\rightarrow\hat{\sigma}$ , $\lambda\rightarrow1$
and $q\rightarrow0$ (i.e., $q=0$ when $\sigma$ is the global best
fit to data). When another model does much better, $\lambda\ll1$
and $q$ is large. 

The hypothesis in question will determine the PDF $f\left(q_{\sigma}|\sigma\right)$
for the statistic. The significance of a discrepancy between the data
and what one expects under the assumption of the test hypothesis $\sigma$
is quantified by giving the $p$-value, defined as the probability
to find $q_{\sigma,\mathrm{obs}}$ in the region of equal or lesser
compatibility with $\sigma$ than the level of compatibility observed
with the actual data. This PDF (sometimes called a sampling distribution)
used to calculate the $p$-value of observed data $\mathcal{D}$ under
a test hypothesis $\sigma$ is 
\[
p\left(\sigma\right)=\int_{q_{\sigma,\mathrm{obs}}}^{\infty}f\left(q_{\sigma}|\sigma\right)dq.
\]

\begin{figure}
\begin{centering}
\includegraphics[scale=0.45]{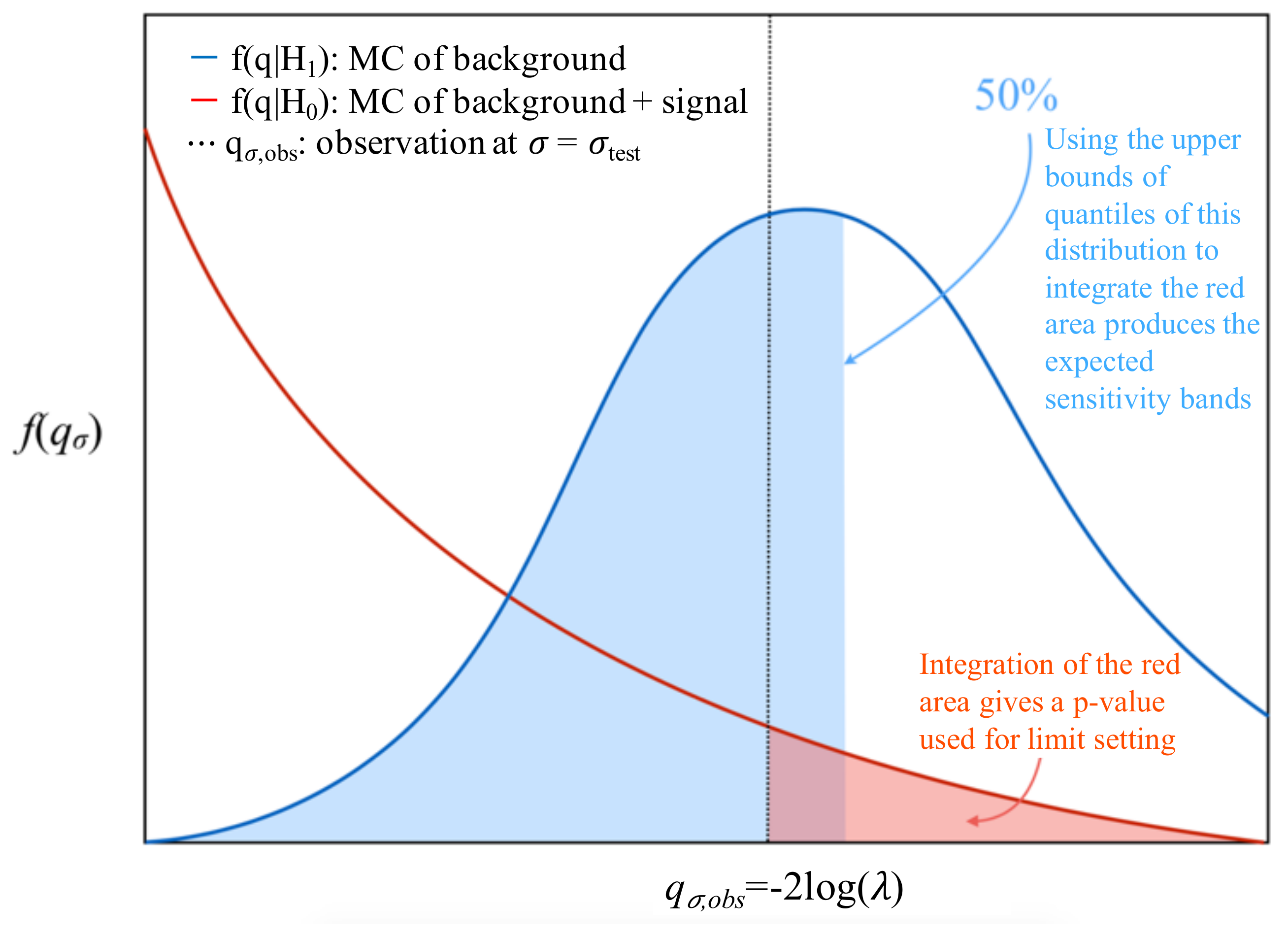}
\par\end{centering}
\caption[Illustration of a test statistic PDF]{Illustration of a test statistic PDF. The $q$ value calculated for
the observed data at a given value of $\sigma=\sigma_{\mathrm{test}}$
is $q_{\sigma,\mathrm{obs}}$. Figure modified from~\cite{nehrkorn2018testing}.
\label{fig:PDF-PLR-illustration}}
\end{figure}

Figure~\ref{fig:PDF-PLR-illustration} illustrates this calculation.
The red distribution is integrated to find the $p$-value for the
observed data under the test hypothesis $\sigma$ given $q_{\sigma,\mathrm{obs}}$,
the observed value of the test statistics. The blue distribution shows
$f\left(q_{\sigma}\right)$ for an ensemble of datasets where the
parameter of interested has some alternative value $\sigma_{\mathrm{alt}}$
(but $q_{\sigma}$ is still calculated by fixing the parameter of
interest to $\sigma$ in the numerator of $\alpha$). When setting
upper limits in LUX, this is used to calculate the expected limits,
i.e., the confidence interval generated by the typical signal-free
dataset (i.e., when the signal cross section is zero). The blue shaded
region contains 50\% of the distribution, and the rightmost edge is
the median value of $q_{\sigma}$ for datasets drawn from the $\sigma_{\mathrm{alt}}$.
The integral of $f\left(q_{\sigma}|\sigma\right)$ can be performed
from this median to calculate the typical $p\left(\sigma\right)$
for datasets where $\sigma_{\mathrm{alt}}$ is true. More generally,
other quantiles of $f\left(q_{\sigma}|\sigma\right)$ can be used
to determine the range of ``expected'' $p\left(\sigma\right)$ values
(i.e., where 0.023, 0.16, 0.5, 0.84, 0.977 quantiles of background
only trials $p$-values are below 0.1, which are the values traced
by the typical Brazil band).

\begin{figure}[!th]
\begin{centering}
\includegraphics[scale=0.35]{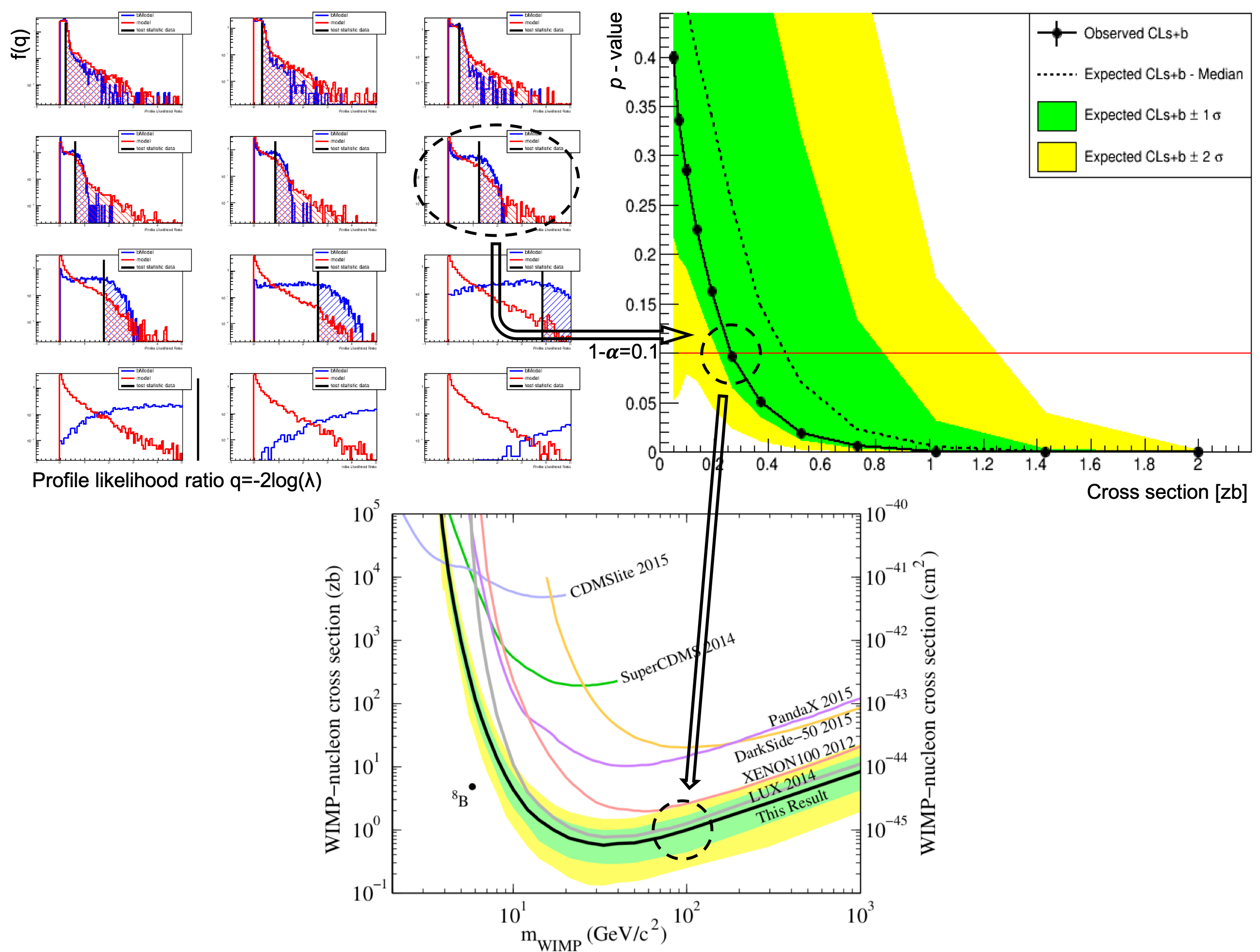}
\par\end{centering}
\caption[Limit setting using the PLR in LUX]{\textbf{Top left:} Nine examples of a test statistic PDF for a given
WIMP mass and varying cross-sections $\sigma_{\mathrm{test}}$ as
constructed by a suite of Monte Carlo pseudo-experiments. Black vertical
lines show $q_{\mathrm{obs}}$ determined from data. Each entry in
each of the nine plots corresponds to a value of $q$ as calculated
from a simulated dataset generated from either given the null hypothesis
where $\sigma=\sigma_{\mathrm{test}}$ (red) or an alternative hypothesis
$\sigma=0$ (blue). Note the increasing separation as $\sigma$ increases
since higher $q$ indicates a greater disagreement between the data
and the test hypothesis. i.e., pseudo experiments to the right of
the black line indicate the rejection of the signal + background hypothesis.
Each plot corresponds to a single black data point in the right-hand
plot. \textbf{Top right:} The frequentist hypothesis test inversion
to obtain the expected 90\% CL is shown for the test points on the
left. The $y$ axis corresponds to $p\equiv1-CL$. \textbf{Bottom:}
Upper limits on the spin-independent elastic WIMP-nucleon scattering
cross section at 90\% confidence limit from~\cite{Akerib:2015rjg}.
The 1- and 2-$\sigma$ ranges of background-only trials for this combined
result are shown in green and yellow, respectively creating the Brazil
band. The black arrows track a single point to illustrate the way
plots get populated.\label{fig:Limit setting}}
\end{figure}

The next step is to ``invert'' the test above to find a confidence
interval using the Neyman construction as illustrated in Figure~\ref{fig:Limit setting}.
For the observed data, $p\left(\sigma\right)$ was calculated over
a range of test hypothesis (top left), and the values of $\sigma$
were rejected when $p\left(\sigma\right)<1-\alpha$ (top right). Combining
results for many different masses results in the final limit curve
shown in the publications (bottom).

LUX used a constrained extended unbinned two-sided likelihood function.
A two-sided confidence interval allows a seamless transition between
limits and allowed region since the test hypothesis $\sigma$, and
the alternative hypothesis $\hat{\sigma}$ can have either more or
less signal strength, unlike a one-sided PLR test which restricts
the definition of incompatibility to mean $\hat{\sigma}<\sigma$~\cite{Cowan:2010js}.
As an example, the likelihood function used for the WS2013 result
was
\begin{align}
\mathscr{L_{\mathrm{WS2013}}\left(\mathcal{D}_{\mathrm{obs}}|\sigma,\theta\right)=} & \mathrm{Poiss}\left(n_{\mathrm{obs}}|n_{\mathrm{exp}}\right)\nonumber \\
 & \times\prod_{e=1}^{n_{\mathrm{obs}}}\frac{1}{n_{\mathrm{exp}}}\left[n_{x}\left(\sigma,k_{\mathrm{Lind.}}\right)f_{x}\left(\mathbf{x}_{e}|g_{2,DD}\right)+\sum_{b}n_{b}f_{b}(\mathbf{x}_{e})\right]\nonumber \\
 & \times\prod_{e\in\mathrm{constr.}}\mathrm{Gauss}\left(\theta_{i}|\theta_{i}^{\mathrm{exp}},\Delta\theta_{i}\right)\label{eq:-3}
\end{align}
where the parameter of interest is $\sigma$, the WIMP-nucleon elastic
scattering cross section, and $\theta$ are the nuisance parameters.
The number of events each with observables $\mathbf{x}$ in the dataset
$\mathcal{D_{\mathrm{obs}}}$ is $n_{\mathrm{obs}}$. In WS2013 the
observables were
\begin{align}
\mathbf{x} & =\left\{ r,z,S1,\log_{10}S2\right\} \label{eq:-4}
\end{align}
where $S1$ and $S2$ are the corrected pulse sizes and $\left(r,z\right)$
describe the reconstructed position of the event. The nuisance parameters
$n_{b}$ are for expected counts in each background population, $n_{s}$
is the expected number of signal events and $n_{\mathrm{exp}}=n_{s}+\sum n_{b}$
are the total expected counts and therefore the normalization of the
full model PDF. The first term is the Poisson probability of observing
$n_{\mathrm{obs}}$ events when expecting $n_{\mathrm{exp}}$, hence
``extended'' likelihood. The second term represents the weighted
sum of background PDFs $f_{b}(\mathbf{x})$ and signal PDF $f_{x}\left(\mathbf{x}_{e}|g_{2,DD}\right)$;
this is where the term unbinned comes in as it is a product of the
probabilities of each observed event, where the model is the weighted
sum of the signal and background component PDFs. The third term enforces
Gaussian constraints (hence ``constrained'') on a subset of the
nuisance parameters $n_{\mathrm{exp}},\,k_{\mathrm{Lind.}},\,g_{2,DD},\,n_{b},\,\theta_{i}$.

Furthermore, the LUX results were ``power constrained''~\cite{Cowan:2011an}
to some quantile of the expected distribution in case of a lucky downward
fluctuation in the background rate that might result in a confidence
interval which is too good, i.e., excluding cross section to which
the detector is insensitive for zero backgrounds. Further details
regarding both WS2013 and WS2014-16 likelihood functions and the application
of the PLR in the LUX analysis was discussed extensively in \cite{nehrkorn2018testing}. 

\subsection{WS2013\label{subsec:WS2013}}

The initial dark matter search consisted of 85.3~live-days using
118~kg of fiducial mass (out of the 250~kg of active LXe mass) of
WIMP search data acquired between April 21 and August 8, 2013. A non-blind
analysis was conducted on the WIMP search data, where only a minimal
set of data quality cuts, with high acceptance, was employed to reduce
the scope for bias. The first, world-leading limit for the LUX detector,
obtained using a rapid analysis from Run3 was reported in~\cite{Akerib:2013tjd}. 

\begin{table}
\begin{centering}
\begin{tabular}{lrr}
\hline 
Parameter & Value & Unit\tabularnewline
\hline 
\hline 
Live-time & 95 & days\tabularnewline
Fiducial radius & 20 & cm\tabularnewline
Fiducial drift time & {[}38, 305{]} & $\mu$s\tabularnewline
Fiducial mass  & $\unit[145.4\pm1.3]{}$ & kg\tabularnewline
Electric field  & $\unit[180\pm20]{}$ & V/cm\tabularnewline
$g_{1}$ & $\unit[0.117\pm0.003]{}$ & phd\tabularnewline
$g_{2}$ & $\unit[12.1\pm0.8]{}$ & phd\tabularnewline
$S1$ threshold & {[}1, 50{]} & phd\tabularnewline
$S2$ threshold & > 165 & phd\tabularnewline
\hline 
\end{tabular}
\par\end{centering}
\caption[Parameters used in WS2013]{Parameters used in WS2013. The fiducial cut in drift time corresponds
to a distance of $\unit[\sim48.6\lyxmathsym{\textendash}8.5]{cm}$
above the faces of the bottom PMTs in $z$. The fiducial mass was
calculated by counting tritium events and is consistent with 147~kg
expected from geometry. Values of $g_{1}$ and $g_{2}$ were extracted
from the Doke plot. \label{tab:WS2013-parameters}}
\end{table}

\begin{figure}
\begin{centering}
\includegraphics[scale=0.3]{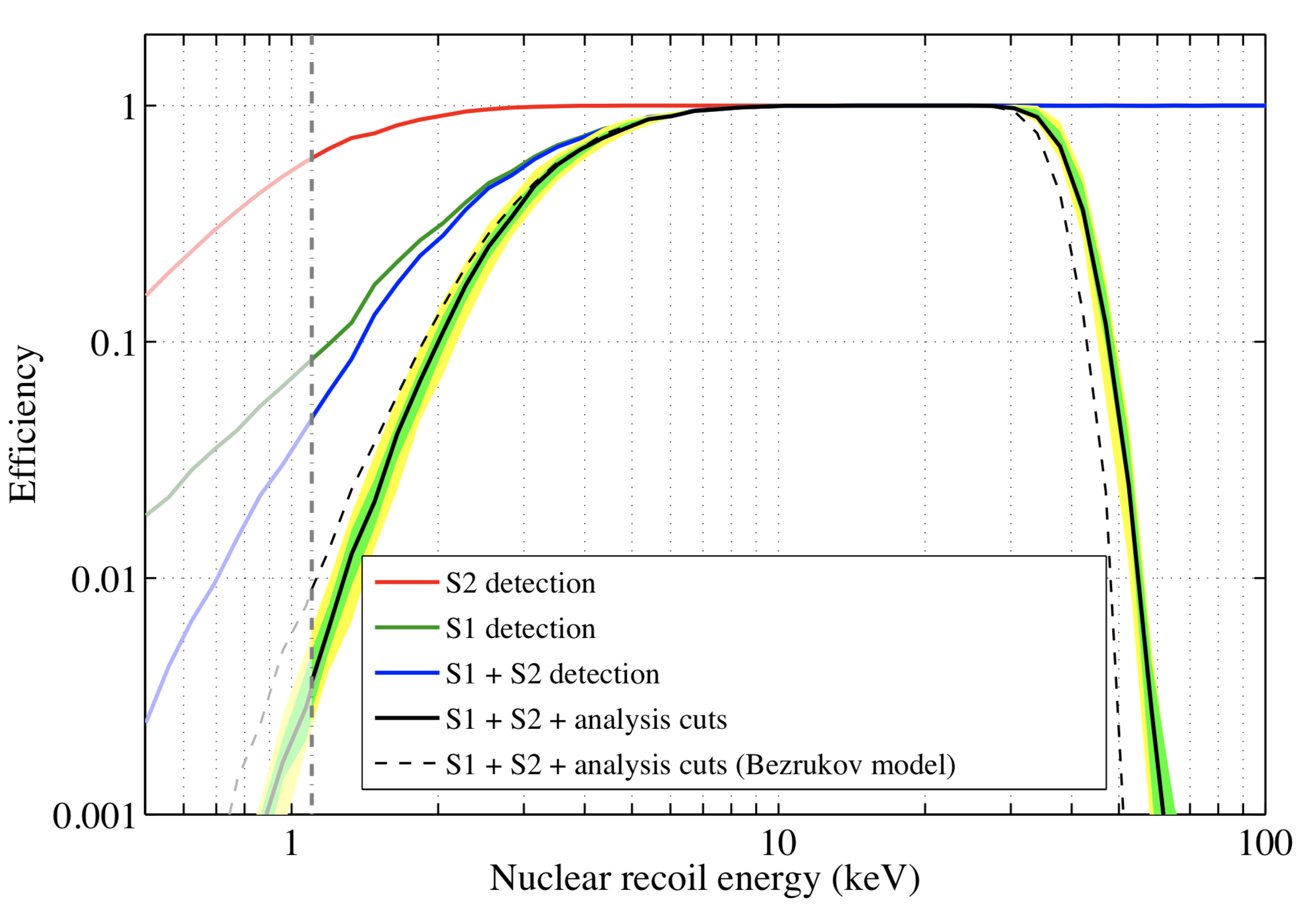}
\par\end{centering}
\caption[Detection efficiency of nuclear recoils in WS2013]{Detection efficiency of nuclear recoils in WS2013. The vertical gray
dashed line at 1.1 keV indicates the re-analysis threshold. The black
solid line shows the best fit of the Lindhard parametrization and
the green and yellow shaded regions span the 1- and 2-$\sigma$ uncertainty.
Black dashed line indicates an alternative, Bezrukov, NR parametrization.
Figure from~\cite{pease2017rare}.\label{fig:WS2013-efficiencies}}
\end{figure}

A more careful re-analysis of this dataset increased exposure to 95~live-days
of search data, lowered threshold below 3~keV due to the \textit{in
situ} DD NR calibrations, and improved event reconstruction and background
model. The improvement in re-analysis of Run3 data resulted in a publication
of another world-leading SI limit as described in~\cite{Akerib:2015rjg,Akerib:2017vbi}.
This result will be referred to as WS2013. Parameters of WS2013 are
summarized in Table~\ref{tab:WS2013-parameters}. The cathode and
gate grids established the electric field in the active xenon volume.
The electric field had a small radial component and thus was not entirely
uniform, due to field leakage through the cathode discussed in Chapter~\ref{chap:efield-modeling},
an effect that was corrected using the geometric spatial uniformity
of $\mathrm{^{83m}}$Kr. $S1$ pulses were required to have a two-PMT
coincidence in the range of 1-50 phd. A lower $S2$ threshold of 165
phd, corresponding to a factor of 6.7 the mean SE response, was applied
to mitigate random coincidence background from smaller, isolated $S2$s.
ER yields were obtained from $^{3}$H calibrations. The \textit{in
situ} DD NR calibrations provided measurements of scintillation yields
down to 1.1 keV and ionization yields of 0.7 keV. The resulting efficiency
plot is shown in Figure~\ref{fig:WS2013-efficiencies}. The expected
neutron background was $0.08\pm0.01$ NR events in the WIMP search
sample.

\begin{figure}
\begin{centering}
\includegraphics[scale=0.28]{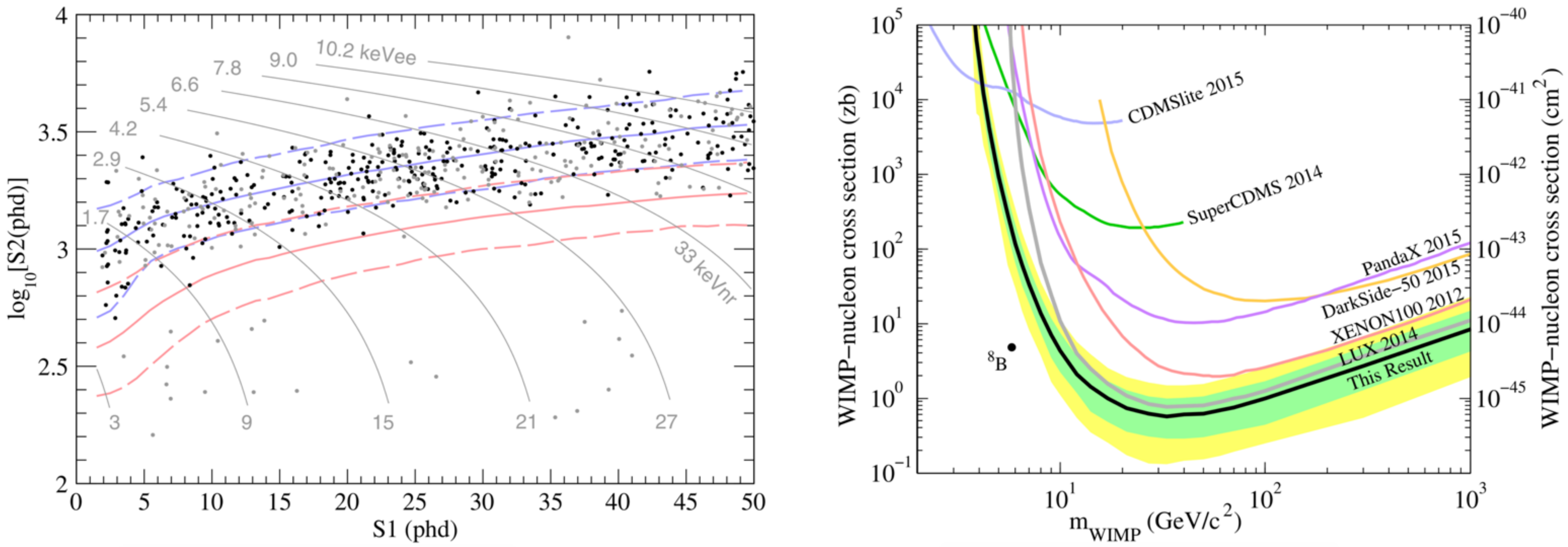}
\par\end{centering}
\caption[Results from WS2013]{\textbf{Left:} The final 591 data points used for WS2013 result.
Black data points are at $\unit[<18]{cm}$ radius, while points with
gray markers are at the edge of the fiducial volume with $r>\unit[18]{cm}$.
Also shown are the mean (solid) and the 10 and 90\% contours (dashed)
of the NR (red) and ER (blue) bands for a $\unit[50]{GeV/c^{2}}$
WIMP signal. Gray curves show the combined energy scale. \textbf{Right:}
Upper limits on the spin-independent elastic WIMP-nucleon scattering
cross section at 90\% confidence limit. The black curve shows the
LUX WS2013 result with expected 1- and 2-$\sigma$ sensitivity bands
in green and yellow, respectively. Figures from~\cite{Akerib:2015rjg}.\label{fig:WS2013-limit}}
\end{figure}

Search data analyzed in WS2013 were acquired between April 24-September
1, 2013. Figure~\ref{fig:WS2013-limit} shows the measured light
and charge of the 591 surviving events in the fiducial volume and
the upper limit on cross section for WIMP masses from 4-1,000~GeV/c$^{2}$
obtained from the double-sided PLR statistic used to test the signal
hypothesis. As is typical for these exclusion limits, at low WIMP
masses the signal spectrum falls off at smaller energies and due to
lower detector efficiency at these energies, causes the distinguishing
upturn of the curve. A decreased flux of WIMPs through the detector
causes the characteristic shape of the curve at high masses. Since
the total WIMP mass density is fixed at $\unit[0.3]{GeV/cm^{3}}$
in the signal model, increasing WIMP mass causes a decrease in number
density. Additionally, the highest studied DM mass of 1~TeV/c$^{2}$
is limited by a lack of thorough background simulations at high energies.

Many other results were also obtained using the WS2013 dataset, such
as a limit on spin-dependent scattering of WIMPs~\cite{Akerib:2016lao},
limits on axions and axion-like particles \cite{Akerib:2017uem},
studies on the effect of pulse shape discrimination in liquid xenon~\cite{Akerib:2018kjf},
a more general result using effective field theory, and limits on
sub-GeV dark matter discussed in Chapter~\ref{chap:Searching-for-sub-GeV-dm}.
Furthermore, many theses provide descriptions of various LUX components,
systems and subsystems, analysis approaches and more for Run3 in great
detail in~\cite{faham2014prototype,balajthy2018,larsen2016effective,phelps2014lux,dobi2014measurement,malling2014measurement,lee2015mitigation,verbus2016absolute,mock2014search,chapman2014first}.

\subsection{Grid conditioning campaign\label{subsec:Grid-conditioning-campaign}}

Between WS2013 and WS2014-16, from January to March 2014, the detector
grids were conditioned to improve the voltages at which the electrodes
could be biased. This was done to increase the magnitude of the applied
electric drift field in bulk LXe and to increase the efficiency for
extracting electrons from the liquid to the gas. As a result of conditioning,
a higher applied anode voltage enabled increased electron extraction
efficiency (i.e., the fraction of electrons which promptly cross the
liquid-gas interface) from $49\pm3\%$ in WS2013 to $73\pm4\%$ in
WS2014-16~\cite{Akerib:2016vxi}.

\begin{figure}
\begin{centering}
\includegraphics[scale=0.3]{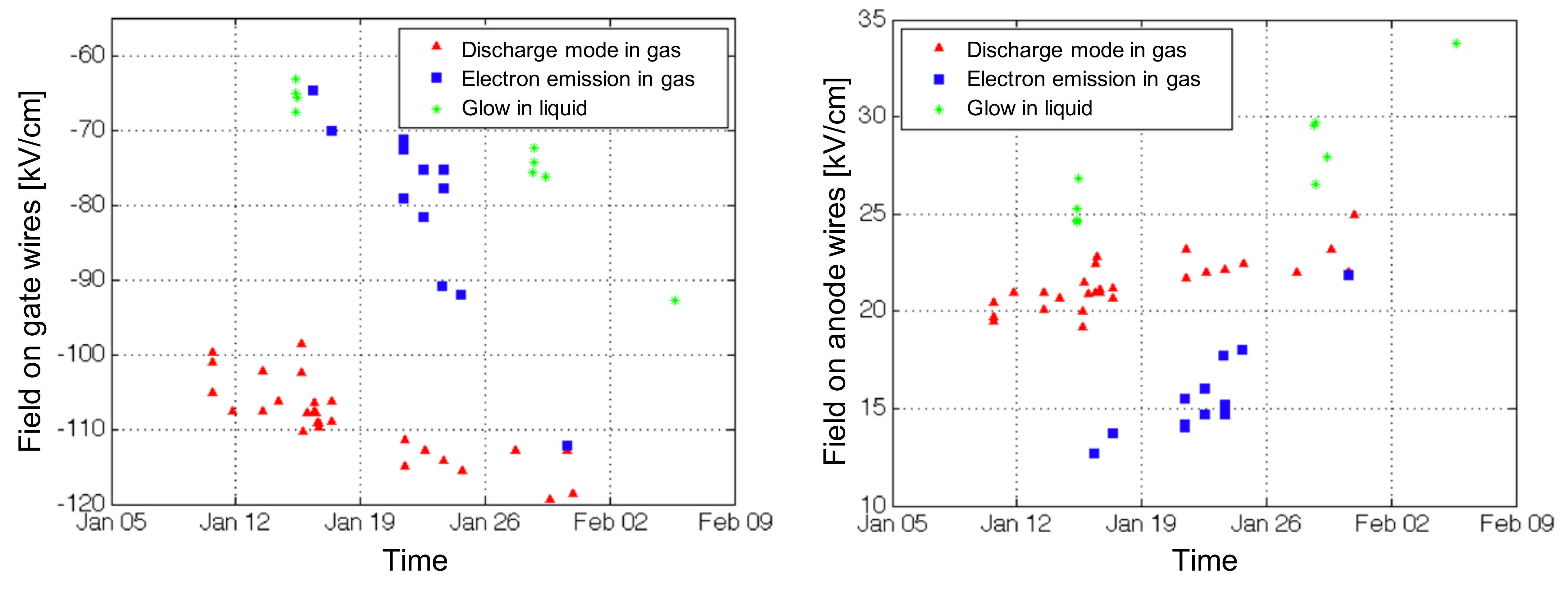}
\par\end{centering}
\caption[Effects of grid conditioning]{Electric fields measured on the gate (left) and anode (right) grid
wires throughout the 2014 conditioning campaign. The electric field
is plotted at the onset of electron emission in the gaseous xenon
(blue square), the onset of discharges in the gas (red triangle),
and the onset of glow, or visible photoelectrons on an oscilloscope,
while the gate grid is immersed in liquid xenon (green star). Figures
modified from~\cite{mannino2017measuring}.\label{fig:conditioning effects}}
\end{figure}

During the conditioning, potentials were held just above the onset
of sustained discharge and maintained for many days, akin to the burn-in
period often employed in room-temperature proportional counter commissioning~\cite{LHCb,ARNISON1990431,DEWULF1988109,AREFIEV198971}.
First, the potentials on gate and anode grids were raised to -5 kV
and 7 kV, respectively, while in cold xenon gas (180 K). Higher glow
and electron emission onset points were achieved following this conditioning
as shown in Figure \ref{fig:conditioning effects}. However, a persistent
high POD rate (i.e., the observed PMT pulse) and elevated amounts
of PMT noise were observed. This prompted a detector warm up in order
to speed up the PMT de-excitation. Subsequently, the potential on
the cathode was raised to -20 kV while in xenon gas at room temperature
(296 K). The detector was kept at room temperature for one month in
April 2014. The detector was then cooled down and filled with LXe
again starting on May 6, 2014, enabling the WIMP search data taking
to resume in September 2014. Further details about the conditioning
campaign can be found in~\cite{mannino2017measuring}.

\begin{figure}
\begin{centering}
\includegraphics[scale=0.55]{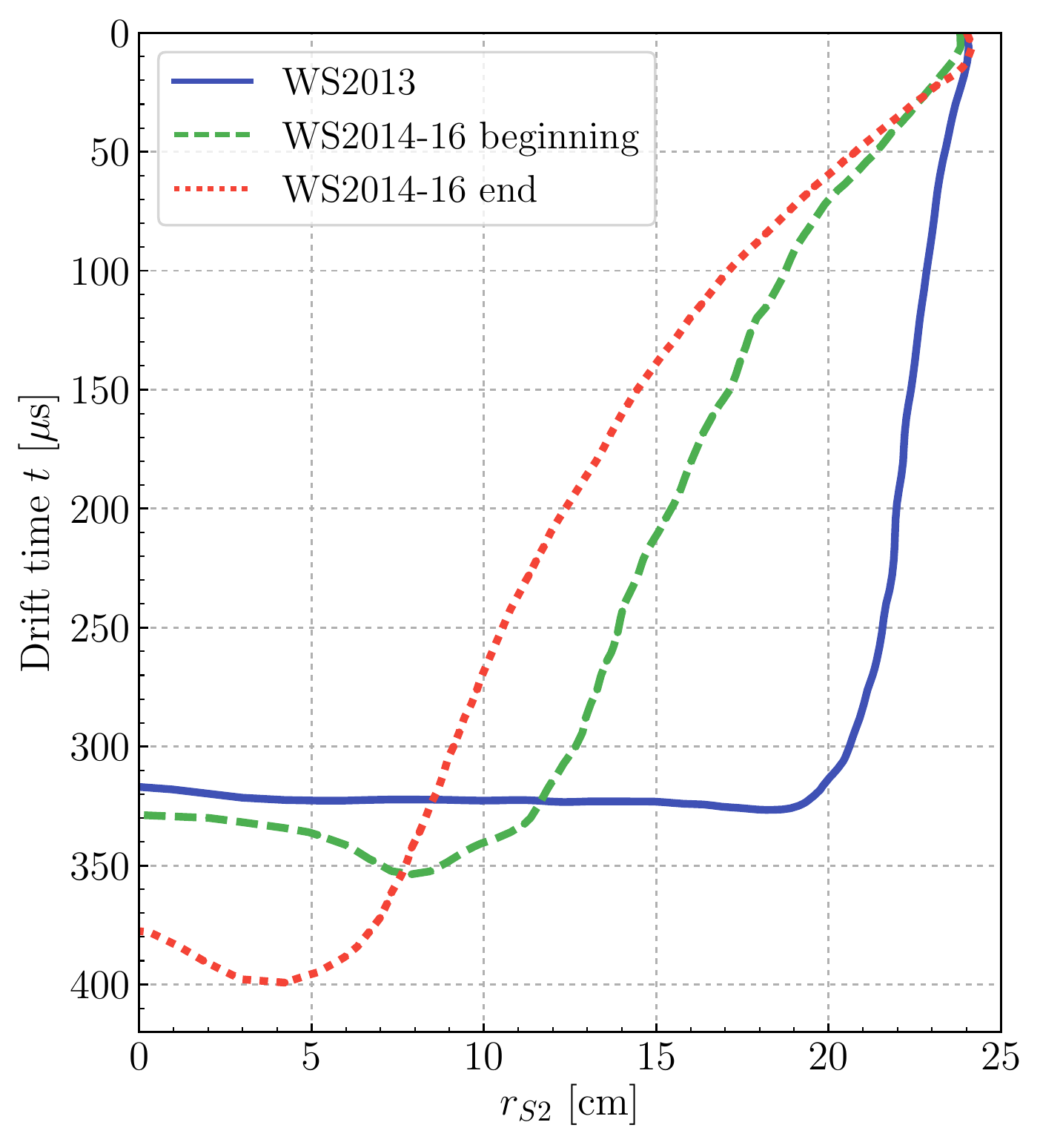}
\par\end{centering}
\caption[Contour plot illustrating detectors edges using $\mathrm{^{83m}Kr}$
calibration data]{Contour plot illustrating detector edges from the reconstructed positions
of the $\mathrm{^{83m}Kr}$ calibration data. Differences between
WS2013 (data from 2013-05-10, solid blue) and the time-dependent WS2014-16
(data from 2014-09-03 with a best fit charge of $\unit[-3.6]{\mu C/m^{2}}$,
dashed green and 2016-05-03 with a best fit charge of $\unit[-5.4]{\mu C/m^{2}}$,
dotted red) are seen. Due to the changing electric fields, $\mathrm{^{83m}Kr}$
calibration data from the later science run had events with greater
drift times.\label{fig:kr-contours}}

\end{figure}

Aside from the increased electron extraction efficiency, the conditioning
campaign unfortunately also resulted in some unwanted side effects.
In particular, as revealed by $\mathrm{^{83m}Kr}$ calibration data,
electron drift trajectories were significantly altered compared to
the regular near-vertical paths observed in WS2013 due to the electric
field variations of 50-600~V/cm. In WS2013, electrons starting from
just above the cathode grid at a radius of $r_{S2}\sim24$ cm traveled
slightly radially inwards during their upward drift to the gate due
to the field cage geometry alone and exited the liquid surface at
$r_{S2}\sim20$ cm. A similar observation was made in the XENON100
detector~\cite{mei2012direct}. However, in WS2014-16 electrons originating
from the same initial radius exhibited a significantly stronger radially
inward push and exited the liquid surface at $r_{S2}\sim10$ cm as
illustrated in Figure~\ref{fig:kr-contours}. The magnitude of this
effect varied in azimuth, depth and furthermore evolved in time, which
triggered a need for a full 3D model of the electric fields inside
the LUX detector throughout the year-long WS2014-16 data taking. This
effort was a crucial foundation for the analysis of the data since
knowing the electric field inside the detector is necessary for $(x,y)$
position reconstruction as the drift path of the electrons is field-dependent
and for event reconstruction due to the effect of the field on recombination.
I developed these field maps; details about this work has been published
in~\cite{Akerib:2017btb} and will be discussed extensively in Chapter~\ref{chap:efield-modeling}. 

\subsection{WS2014-16\label{subsec:WS2014-16}}

Following the conditioning campaign, the first data acquisition of
WS2014-16 began on September 11, 2014, and the final data acquisition
ended on May 2, 2016. The complications arising from the distorted
electric field as a result of the conditioning campaign necessitated
many analysis fixes. The WIMP search was interrupted by five $^{3}$H
and DD calibrations measuring the detector response to ERs and NRs
and monitoring the variation of detector conditions. The fiducial
volume was defined in the S2 coordinate space with a drift time cut
as in WS2013 but a more conservative radial cut including only 101
kg of LXe. An analysis tagging $^{210}$Pb chain events identified
the position of the detector's PTFE wall in S2 space as described
in~\cite{boulton2018}. 

\begin{figure}[p]
\begin{centering}
\includegraphics[scale=0.7]{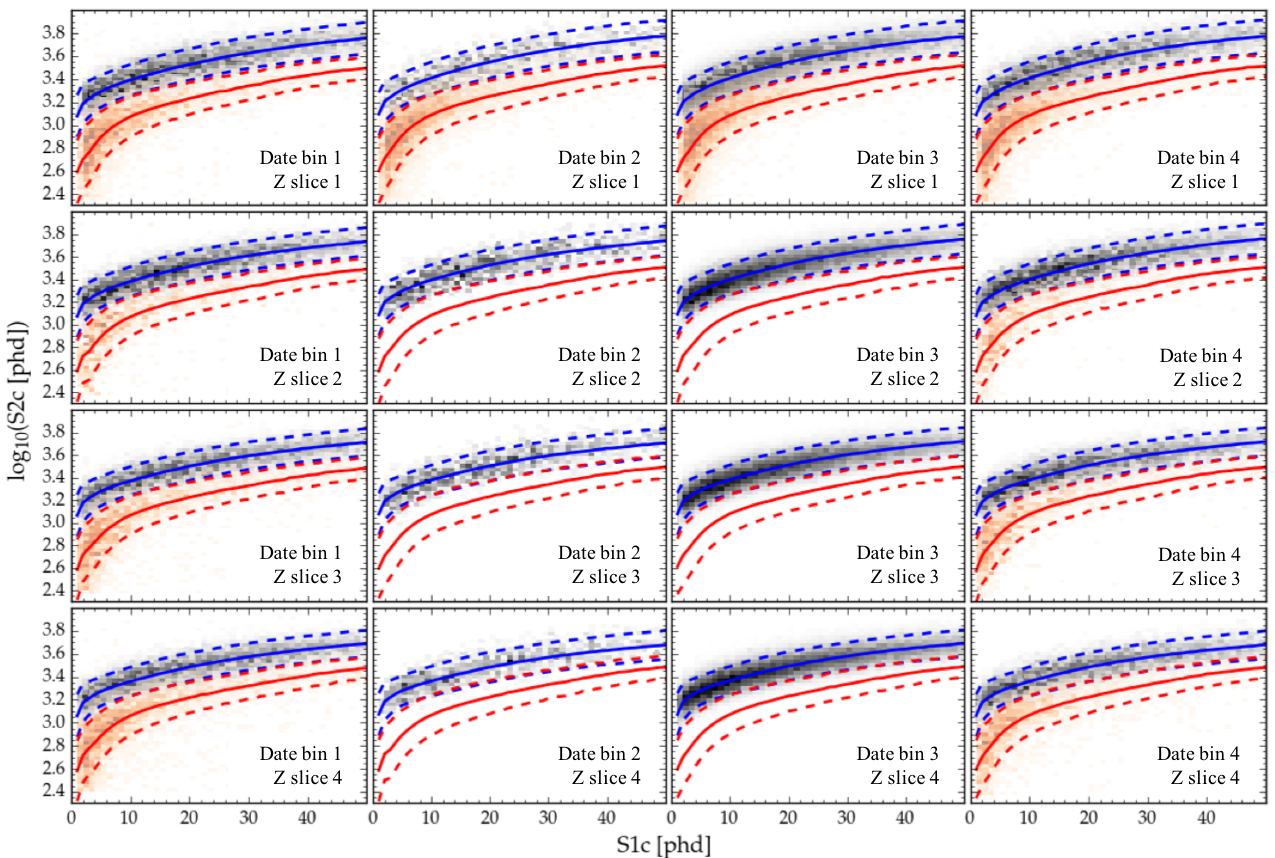}
\par\end{centering}
\caption[Calibrations used in WS204-16]{$^{3}$H (gray) and DD (orange) calibration data. The band medians
(solid lines) and 10- and 90\% contours (dashed lines) are shown for
ER (blue) and NR (red). They were calculated from NEST models of each
exposure segment. Note that NEST fitting was performed on ER calibration
data only since the resulting parameters from the NR model were found
to be in good agreement with the DD calibration data. Only the first
and last DD calibrations were performed at various $z$ positions.
Figure from~\cite{pease2017rare}.\label{fig:Run4_calibrations} }
\begin{centering}
\vspace{0.7cm}
\includegraphics[scale=0.29]{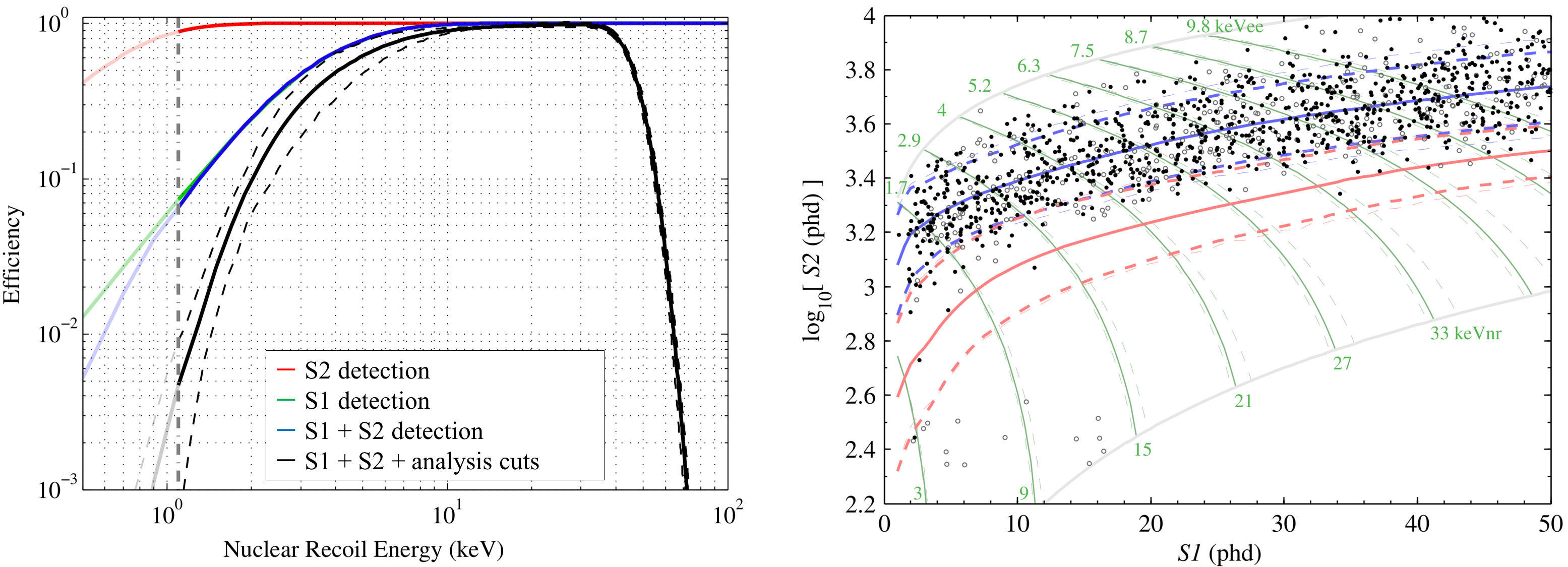}
\par\end{centering}
\caption[NR efficiencies and events for WS2014-16]{\textbf{Left:} The detection efficiencies for nuclear recoils in
WS2014-16. The vertical gray dashed line at 1.1 keV indicates the
re-analysis threshold. \textbf{Right:} Final 1,221 data points for
WS2014-16 passing all selection criteria used for the PLR analysis.
Fiducial events within 1 cm of the radial fiducial volume boundary
are shown as unfilled gray circles to convey their low WIMP-signal
probability relative to the background model. Also shown are the mean
(solid) and the 10 and 90\% contours (dashed) of the NR (red) and
ER (blue) exposure-weighted bands for a $\unit[50]{GeV/c^{2}}$ WIMP
signal. Green curves indicate the exposure-weighted energy scale.
The extrema boundaries from the 16 segments are shown in fainter shade.
Figures from~\cite{Akerib:2016vxi}.\label{fig:run4-efficiencies}}
\end{figure}

Since the light and charge yields in LXe depend on the electric field
strength, the large variations of electric field in WS2014-16 caused
the ER and NR bands to be substantially smeared throughout the entire
exposure, which would cause a reduction of ER-NR discrimination power.
Therefore in the analysis, the TPC was not treated as a single homogeneous
volume as in WS2013 but instead was segmented into parts within which
the field variations were smaller. In particular, the detector was
partitioned into 16 segments: four bins of drift time (related to
event depth) and four bins of date. Within each segment, the field
magnitude was considered to be constant and uniform - the field values
are chronicled in~\cite{nehrkorn2018testing}. Boundaries in date
were September 11, 2014; January 1, 2015; April 1, 2015; October 1,
2015; May 2, 2016 and the boundaries in drift time were 40, 105, 170,
235, 300 \textgreek{m}s. The choice of these divisions was driven
by the results of the electric field maps and the changing wall position
in S2 space and timing of calibration runs. The periodic calibrations
of $^{3}$H enabled tuning of the empirical NEST ER model for each
exposure segment. In practice, this meant varying detector gains as
well as Xe response parameters in order to reproduce the observed
bands in S1\textendash S2 space. The first and last DD calibrations
were performed at multiple heights to define the NR response as a
function of different field magnitude at different drift times. The
DD campaigns showed minimal variations of the mean and width of the
NR band, consistent with the literature~\cite{NEST_lenardo}. Consequently,
the best-fit parameters from the ER calibration data for each exposure
unit were applied to the NEST NR model. The fitted models are shown
in Figure~\ref{fig:Run4_calibrations}. 

An interpolation of the field map data capable of mapping arbitrary
points in real space to S2 space was implemented. This helped to construct
the probability models used to test background and signal hypotheses
against the WIMP search data. Since the field maps provided a 3D grid
of field values, a fast C3PO (Curt's Post-Processing Position Offset
Corrections)\nomenclature{C3PO}{Curt's Post-Processing Position Offset Corrections}
interpolation algorithm was used. The background model used in the
PLR had three classes of events: events of typical LXe light and charge
yield (constructed in a very similar way to WS2013), events affected
by proximity to the PTFE wall, and accidental coincidences of isolated
S1 and S2 pulses.

In order to minimize any biases, the WS2014-16 dataset was ``salted.''
Salting is an alternative to the more common approach of blinding
where the signal region of interest (i.e., the NR band in our case)
is hidden until the entire analysis is finalized. This provides many
pitfalls where many unanticipated detector pathologies can remain
unnoticed. Salting, also referred to as ``spiking,'' is common in
neutrino experiments, searches for fractional charge, and in gravitational
wave detection. LUX used a sequestered tritium calibration dataset
to insert $\mathcal{O}\left(100\right)$ events into the data stream.
The $^{3}$H events were selected to be uniformly distributed in the
LXe active volume and their S1 and S2 pulses were patched together
to mimic NR signals. Each event was then assigned a random ``luxstamp,''
a unique event identifier used by LUX, and eventually inserted into
the analysis data stream in the event-level file creation stage. A
library of salt events was stored in a database with restricted access.
Thanks to this approach, LUX analyzers were allowed to use all data
generated by the detector. Once the analysis was complete, a person
guarding the salt database revealed which of the events that passed
all analyses cuts were fake signal events.

\begin{figure}[t]
\begin{centering}
\includegraphics[scale=0.3]{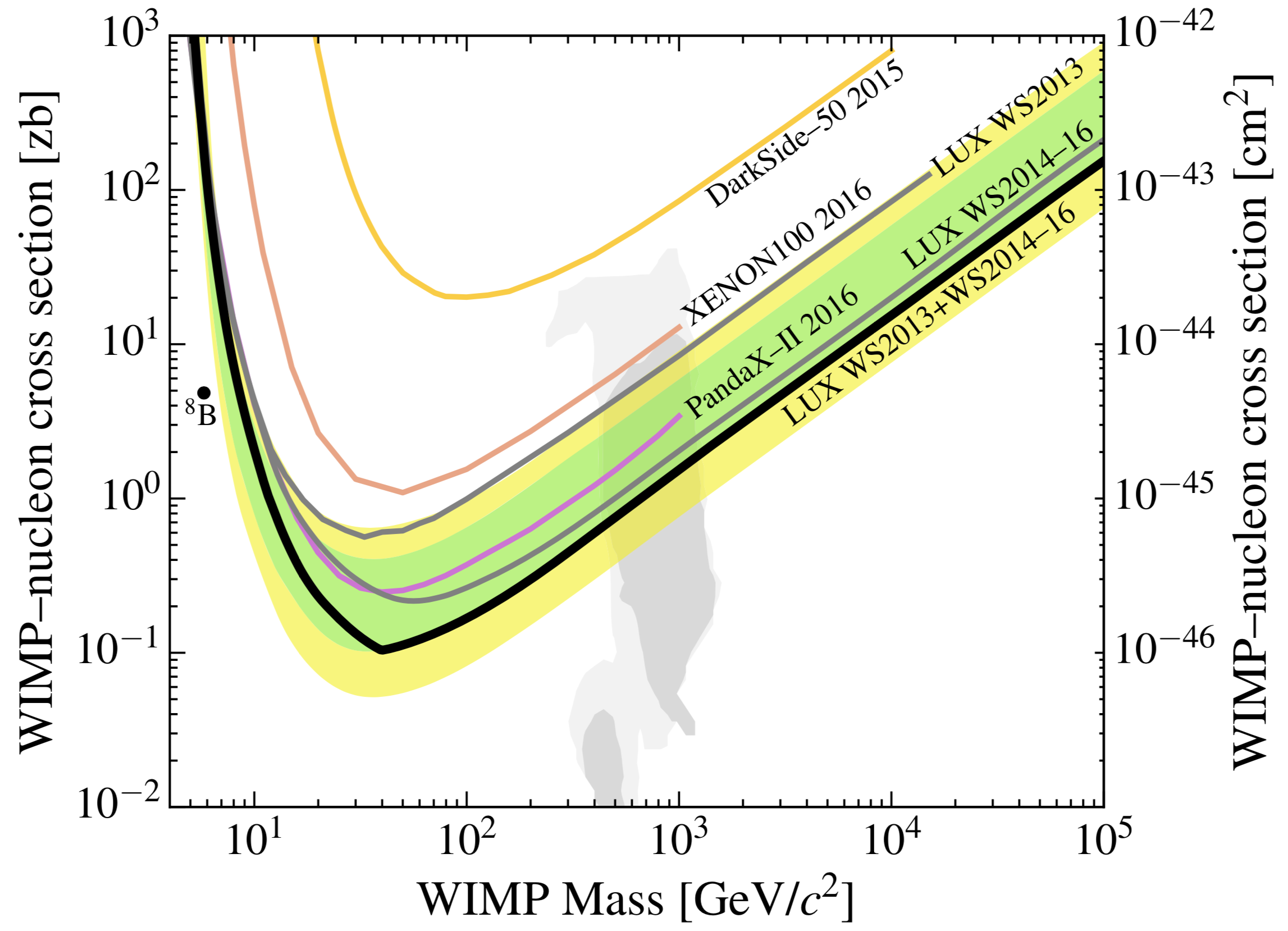}
\par\end{centering}
\caption[Upper limit on SI WIMP-nucleon scattering from the complete LUX exposure]{Upper limits on the spin-independent elastic WIMP-nucleon cross section
at 90\% confidence limit. The solid gray curves show the exclusion
curves from LUX WS2013 (95 live-days)~\cite{Akerib:2015rjg} and
LUX 2014-16 (332 live-days). The solid black line shows the limit
from these two combined to give the full LUX exclusion curve. The
1- and 2-$\sigma$ ranges of background-only trials for this combined
result are shown in green and yellow, respectively. The 1- (light)
and 2-$\sigma$ (dark) gray filled regions indicate the parameter
space favored by the fully constrained MSSM (CMSSM)~\cite{Bagnaschi2015,cmssm}
before the release of this result. The black $^{8}$B dot indicates
the maximum likelihood of coherent neutrino-nucleus scattering under
a WIMP only hypothesis assuming no energy threshold~\cite{Billard:2013qya}.
Figure from~\cite{Akerib:2016vxi}.\label{fig:final_exclusion}}
\end{figure}

Detection efficiencies for nuclear recoils along with the final dataset
with salt events removed are shown in Figure~\ref{fig:run4-efficiencies}.
In addition to the exclusion limit from WS2014-16 data alone, the
exposure from WS2013 was added as a separate 17th exposure segment,
which led to another world-leading limit at the time of publication
in~\cite{Akerib:2016vxi}; the final exclusion limit is shown in
Figure~\ref{fig:final_exclusion}. Limits on the spin-dependent elastic
scattering of WIMPs from the complete exposure were also published
in~\cite{Akerib:2017kat}. Outside of these publications, many details
of the analysis leading to the publication of the aforementioned results
have been discussed thoroughly in~\cite{pease2017rare,nehrkorn2018testing,knoche2016signal,boulton2018}.

\section{Summary}

The immense effort put into its detector design, construction, and
operations enabled LUX to collect plentiful data during its operations,
which led to many world-leading publications that excluded much of
the unexplored WIMP parameter space. LUX also pioneered many calibration
techniques now used throughout the field. Furthermore, there are still
many ongoing diligent analyses exploring parameter spaces relevant
to other dark matter and particle physics models. None of this would
have been possible without the tireless effort of all of the collaboration
members with whom I had the privilege to work. 

Despite its successful operations, LUX still left a significant amount
of the WIMP parameter space to be explored by the next generation
of dark matter detectors. Looking ahead, both the successful and the
less-so successful experiences of LUX are now being leveraged in the
design and construction of the LUX-ZEPLIN experiment discussed in
Chapter~\ref{chap:High-voltage-development-LZ}.

\subsection{Contributions to LUX}

I joined the LUX collaboration in September 2013, right before the
publication of the colla\-boration's first world-leading WIMP-nucleon
elastic scattering cross section limit. I traveled to SURF in Lead,
South Dakota several times as an on-site shifter. My first task underground
was to change a diaphragm of one of the two KNF circulation pumps,
a very on-topic introduction to the LUX detector, which was rewarded
by a box of Chubby Chipmunk chocolates, obviously setting up my time
on site to be a success thereafter. I was the Shift Manager and Detector
Operations Manager several times. I was trained as an operator of
the xenon gas system, liquid nitrogen, PMT, grid high voltage, cathode,
DAQ, xenon sampling, external radioactive source deployment, and internal
$\mathrm{^{83m}Kr}$ calibration systems. I was also trained as an
underground guide.

I participated in multiple LUX collaboration meetings and workshops
and partook in regular offsite analysis shifts ensuring consistent
data quality coming out of the detector. Even though the detector
was assembled before I joined the collaboration, I helped on the other
end - with the underground detector decommissioning and the detector
disassembly and museum prep for the Lead Visitor Center, which allowed
me to see the internal detector components that were so carefully
modeled in my electric field work.

My most significant contributions to the LUX collaboration were my
electric field modeling efforts. This work is discussed in detail
in Chapter~\ref{chap:efield-modeling} and in JINST publication~\cite{Akerib:2017btb},
where I am the corresponding author. It served as a foundation for
many critical analyses that led to the final world-leading SI and
SD limits. I also led an analysis searching for sub-GeV dark matter
using the novel Bremsstrahlung and Migdal effects. This work is described
in Chapter~\ref{chap:Searching-for-sub-GeV-dm} and in~\cite{Akerib:2018subGeV}
that has been accepted to Physical Review Letters for publication,
and for which I am the corresponding author.

\chapter{\textsc{3D modeling of electric fields in the LUX detector}\label{chap:efield-modeling}}

Electric fields in the LUX detector changed between WS2013 and WS2014-16.
The electric field during WS2013 was nearly uniform in the fiducial
volume, but in WS2014-16, the field became quite non-uniform. As discussed
in Section~\ref{subsec:Grid-conditioning-campaign}, after WS2013
the detector grids underwent conditioning done at very high voltages
with intentional continuous glow discharge over many-day periods.
This exposed the PTFE walls (that define the radial boundary of the
active volume) to VUV photons, which likely caused an in-situ creation
of electron-hole pairs in the PTFE surface, thus introducing a new
component to the electric fields during the second WIMP search period.
The work described in this chapter is published in JINST~\cite{Akerib:2017btb}
(WS2014-16) and Phys. Rev. D~\cite{Akerib:2017vbi} (WS2013). 

This conditioning campaign not only caused a distortion of the detector's
electric field; additionally, the field was observed to be changing
throughout WS2014-16 both with time and with azimuthal angle. Since
the conditioning altered the electric fields inside the detector in
a non-trivial non-symmetric manner, this triggered a need for a fully
3D model of the electric fields inside the LUX detector, varying with
date throughout the 23 month-long WS2014-16. Knowledge of the electric
fields inside the detector is necessary for $(x,y,z)$ position and
event reconstruction since the drift paths of the S2 electrons and
recombination are field-dependent. 

Mapping of the detector's electric field was possible due to weekly
$^{\mathrm{83m}}\mathrm{Kr}$ calibrations with decays uniformly distributed
throughout the detector's active volume. First, a 3D model of the
LUX detector was built for WS2013. The fields in the detector were
well understood at the time, so this allowed us to cross-check the
foundation of the modeling. With the basic model in place, a more
thorough and complex model was developed for WS2014-16. To monitor
the time-varying behavior in WS2014-16, electric field maps were created
on a monthly basis. This was done by fitting a model built in COMSOL
Multiphysics to the $^{\mathrm{83m}}\mathrm{Kr}$ data, varying the
hypothetical charge distributions among 42 spatial subdivisions of
the inner PTFE surface and performing fits using an altered Metropolis-Hastings
algorithm. The average PTFE charge density was observed to increase
in magnitude throughout the exposure, changing from $\unit[-3.6]{\mu C/m^{2}}$
to $\unit[-5.5]{\mu C/m^{2}}$. The electric field magnitude varied
from $\unit[\sim20-50]{V/cm}$ near the bottom of the detector to
$\unit[\sim500-650]{V/cm}$, while the mean value of the field of
$\unit[\sim200]{V/cm}$ remained mostly constant throughout the exposure.

The electric field in the gas region was modeled to be $\unit[\sim5345]{V/cm}$
in WS2013 and $\unit[\sim6555]{V/cm}$ throughout WS2014-16. This
value was assumed to be constant throughout WS2014-16, as charges
in the PTFE were shown to not significantly affect this value. 

A method describing the development of a 3D electric field model for
the LUX detector is outlined in this chapter. However, the method
can be applied to model electric fields inside any TPC. We start with
an a priori model of the electric fields in LUX based on the engineering
CAD designs. This model is then modified to include capabilities of
modeling charges in the PTFE walls. The results of the simulations
are compared to the $^{\mathrm{83m}}\mathrm{Kr}$ calibration data
until an agreement is achieved, which enables to map the evolution
of electric fields during WS2014-16. This method is then also applied
to study field discrepancies seen in WS2013. 

\section{Creating a 3D model of the LUX detector}

Electric field models were developed both with and without charge
densities in the PTFE panels to understand the detector behavior.
Prior to developing a model for the changing electric fields during
WS2014-16, a 3D model was developed without static charges in the
PTFE panels to confirm our understanding of the calibration data from
WS2013. The 3D model was built using the AC/DC Module of COMSOL Multiphysics
v5.0\textregistered ~\cite{comsolRef}, a commercially available
finite element simulation software. The lack of symmetries, such as
the regular dodecagonal nature of the detector or the parallel grids
that were rotated with respect to each other, ruled out the option
of using an axially symmetric electric field model as is usually done
in this field.

\subsection{LUX 3D model geometry }

\begin{figure}[t]
\begin{centering}
\includegraphics[scale=0.69]{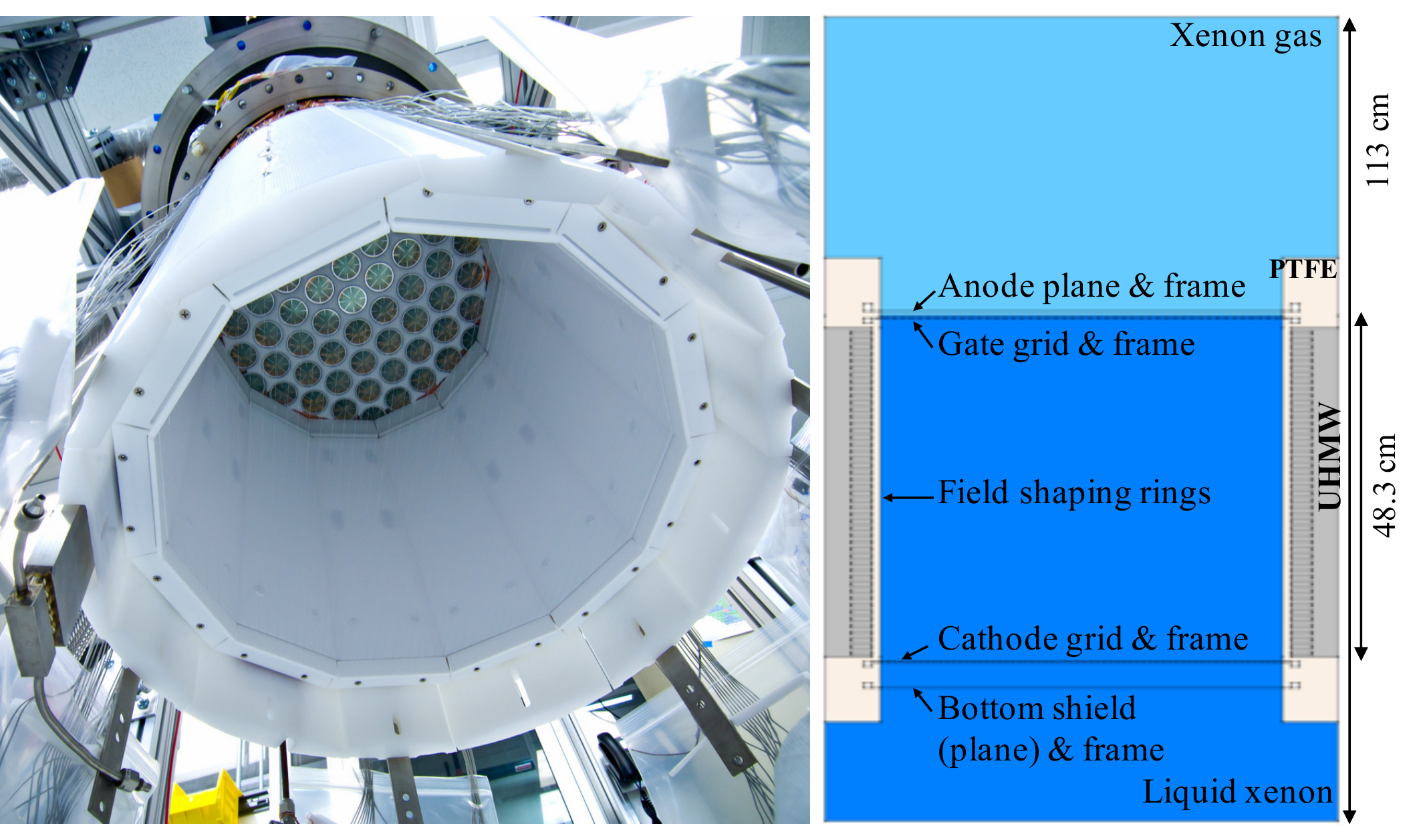}
\par\end{centering}
\caption[Illustration of LUX model geometry used in field modeling]{\textbf{Left:} Bottom view of the LUX detector showing the regular
dodecagonal prism shape, the PTFE panels, and the top PMT array. \textbf{Right:}
Cross section of the LUX model including material details. The radius
of the inner vacuum can is 30.16 cm while the width of each of 12
innermost PTFE panels is 12.67 cm so that the radius of the inscribed
circle (or the separation of parallel panels) is 23.65 cm. The liquid
level was assumed to be 3.6 mm above the gate grid in WS2013 and 4.4
mm in WS2014-16. \label{LUX geometry}}
\end{figure}

The detector 3D model used in the electric field simulations is based
on the LUX detector geometry as constructed in LUXS\textsc{im}, the
\textsc{Geant4}-based simulation framework used by the collaboration,
with CAD drawings used for reference. A full 3D model was needed due
to the detector\textquoteright s geometry. The active volume takes
the shape of a regular dodecagonal prism as shown in the left panel
of Figure~\ref{LUX geometry}. Four detector grids (bottom shield,
cathode, gate, top shield) are rotated by $60^{\circ}$ with respect
to each other (with bottom shield grid and gate being parallel to
each other), leaving a geometry that is not azimuthally or otherwise
symmetric. The 2D cross-section of the model used in the 3D simulations
is shown in the right panel of Figure~\ref{LUX geometry}; the figure
also illustrates some of the adopted simplifications. 

Due to the high degree of complexity of the detector, many simplifications
were adopted in the model compared to the real-life detector. Since
the focus of the modeling was on the active region of the detector,
the bottom shield and anode grids were modeled as planes imposing
electrostatic boundaries for the top and bottom of the detector. This
enabled the model to ignore many details included beyond those grids.
The field shaping rings inserted inside a volume of ultra-high-molecular-weight
polyethylene (UHMW)\nomenclature{UHMW}{Ultra-high-molecular-weight polyethylene}
were included, but the resistor chain connecting them, and many details
of the UHMW structure, as well as the details of the PTFE walls, were
omitted. This decreased the computational requirements for solving
the 3D model since these only result in very localized effects. The
relative DC dielectric constants of the materials used in the model
were: $\varepsilon_{\mathrm{liquid\,xenon}}=1.95$, $\varepsilon_{\mathrm{gaseous\,xenon}}=1.0$,
$\varepsilon_{\mathrm{PTFE}}=2.1$, $\varepsilon_{\mathrm{UHMW}}=2.3$.

There are three large metal objects inside the UHMW volume of potential
concern that were omitted in the LUX COMSOL model: the heat exchanger
(ground), the weir (also at ground) and the cathode cable (at -10/-8.5
kV). The detector was designed so that no parts in the UHMW region
would affect the electric field inside the detector's active volume.
This was cross-checked by building a second LUX model identical to
the one shown in Figure~\ref{LUX geometry} but with the heat exchanger,
weir and cathode cable included. The simulation confirmed that those
parts indeed do not affect the electric field inside the active region;
therefore these components were omitted in future models.

Since the model spans five orders of magnitude in length scale (from
28 \textgreek{m}m for the grid wire diameter to 1.1 m for the inner
vacuum cryostat height), the cathode and gate grid wires were modeled
as parallel lines with zero diameter to accelerate the convergence
of the model. As with all simplifications, simulations were done that
show that this wire diameter change had a negligible effect on the
resulting solution without a loss of field resolution (see also calculations
done in~\cite{mcdonald2003notes}), especially when using the ``Physics
mesh'' of the COMSOL meshing sequence. Further details about the
detector grids along with applied voltages for WS2013 and WS2014-16
are summarized in Table~\ref{Table: Grid property}.

\begin{table}[H]
\centering{}%
\begin{tabular}{l>{\raggedleft}p{0.9cm}>{\raggedleft}p{1.1cm}>{\raggedleft}p{1.4cm}>{\raggedleft}p{1cm}>{\raggedleft}p{1.5cm}>{\raggedleft}p{1.4cm}>{\raggedleft}p{1.9cm}}
\toprule 
Grid & $z^{\dagger}$ {[}cm{]} & Angle {[}deg{]} & Wire $\diameter$ {[}$\mu$m{]} & Pitch {[}mm{]} & Modeled as & WS2013 {[}kV{]} & WS2014-16 {[}kV{]}\tabularnewline
\midrule
\midrule 
Top shield & 58.6 & 135 & 50.8 & 5.00 & Absent & -1.0 & -1.0\tabularnewline
Anode & 54.9 & N/A & 28.4 & 0.25 & Plane & 3.5 & 7.0\tabularnewline
Gate & 53.9 & 15 & 101.6 & 5.00 & $\diameter0$ wires & -1.5 & 1.0\tabularnewline
Cathode & 5.6 & 75 & 206.0 & 5.00 & $\diameter0$ wires & -10.0 & -8.5\tabularnewline
Bottom shield & 2.0 & 15 & 206.0 & 10.00 & Plane & -2.0 & -2.0\tabularnewline
\bottomrule
\end{tabular}\caption[Grid properties and voltages used in field models]{Grid properties and voltages as relevant to the construction of the
electric field model used in WS2013 and WS2014-16 simulations, including
a description of geometric simplifications. $^{\dagger}$The vertical
distance from the face of the bottom PMT array, which is defined as
$z=0$, accounts for thermal contraction as appropriate. \label{Table: Grid property}}
\end{table}

\subsection{COMSOL Multiphysics}

COMSOL Multiphysics\textregistered ~\cite{comsolRef} is a finite
element analysis, solver, and simulation software\footnote{Many thanks to Scott Hertel for introducing me to COMSOL, an excellent
simulation software that accompanied me throughout my grad school
experience and helped me gain a deeper understanding of electric fields.
Also, thanks to Misha Guy from the Yale Science Research Software
Core, and the online COMSOL forum hero, Ivar Kjelberg for advice while
I was building the 3D LUX model.}. It is a cross-platform, commercially available, general-purpose
software. COMSOL Multiphysics can be used not only for conventional
physics-based problems but also allows incorporation of diverse coupled
physical phenomena using partial differential equations. 

The AC/DC module of COMSOL Multiphysics v5.0, released in 2014, was
used in this work. In its electrostatics interface, COMSOL solves
the differential equations introduced below. In electrostatics, Faraday's
law is simplified to 
\[
\nabla\times\mathbf{E}=-\frac{\partial\mathbf{B}}{\partial t}=0
\]
 where $\mathbf{B}$ is the magnetic field and $\mathbf{E}$ is the
electric field 
\[
\mathbf{E}=-\nabla\phi
\]
where $\phi$ is the scalar electric potential. The electrostatics
physics interface solves Gauss' law using the scalar electric potential
$\phi$ as the dependent variable: 
\[
\nabla\cdot\mathbf{D}=\rho
\]
where $\rho$ is the total charge and $\mathbf{D}$ is the electric
displacement field defined as 
\[
\mathbf{D}=\varepsilon_{0}\mathbf{E}+\mathbf{P}
\]
where $\varepsilon_{0}$ is the permittivity of free space and $\mathbf{P}$
is the macroscopic polarization density. In homogeneous, isotropic,
nondispersive, linear materials 
\begin{eqnarray*}
\mathbf{P} & = & \varepsilon_{0}\chi_{e}\mathbf{E}
\end{eqnarray*}
 where $\chi_{e}$ is the electric susceptibility. Hence in linear
materials, the constitutive relation for electric displacement is
\[
\mathbf{D}=\varepsilon\left(1+\chi_{e}\right)\mathbf{E}=\varepsilon_{0}\varepsilon_{r}\mathbf{E}=\varepsilon\mathbf{E}
\]
where $\varepsilon_{r}$ is the relative permittivity and $\varepsilon$
is a scalar permittivity of the material $\varepsilon=\varepsilon_{0}\varepsilon_{r}$.
In a dielectric material, the displacement field satisfies Gauss's
law
\[
\nabla\cdot\mathbf{D}=\rho-\rho_{b}=\rho_{f}
\]
where $\rho_{b}$ is the bound charge and $\rho_{f}$ is the free
charge.

\begin{figure}
\centering{}\includegraphics[scale=0.8]{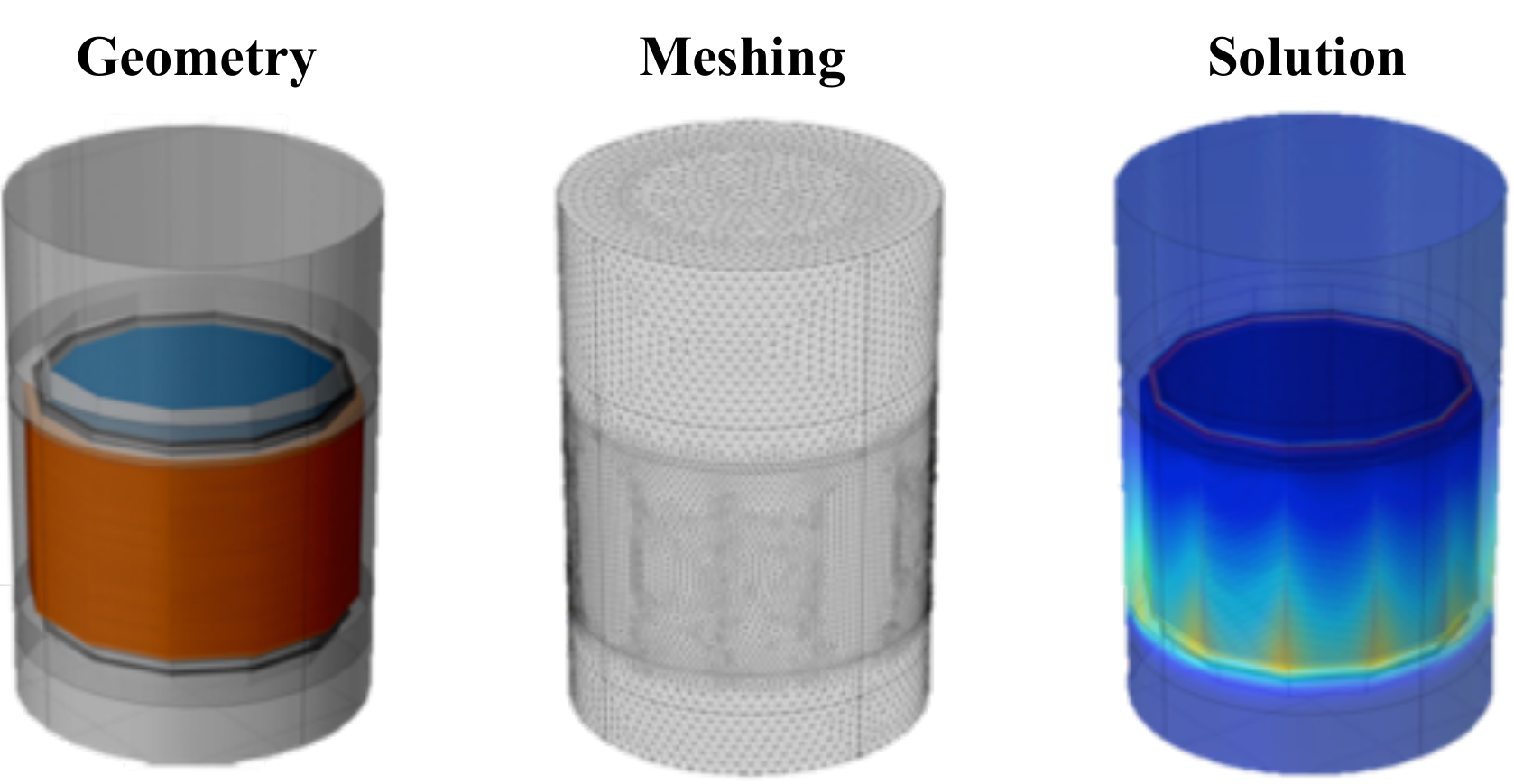}\caption[3D model of the LUX detector built using COMSOL Multiphysics]{Steps in building the 3D model of the electric fields in the LUX
detector using COMSOL Multiphysics v5.0. From left to right: First,
a geometry was built given materials and their dielectric constants.
Next, electrostatic conditions were defined by assigning voltages
to conductor boundaries and domains. COMSOL generated a mesh by discretizing
the space into tetrahedra and then solved the model. Using COMSOL's
adaptive mesh refinement allowed to improve meshing further while
minimizing solution error. The mesh of the LUX model had $\sim29,000,000$
mesh elements and took about 95 GB of RAM to find a solution. After
the model was created, an array of electric field values from the
active volume was exported for further analysis.\label{fig:COMSOL}}
\end{figure}

Once the detector geometry, materials, electrical dielectric properties,
and voltages were assigned to detector volumes and boundaries, this
defined the electrostatic conditions since COMSOL has built-in features
to compile these relationships. COMSOL then generated a mesh by discretizing
the space into tetrahedra, and a proposed field map was generated.
Using an adaptive mesh refinement~\cite{comsolBlog} allowed further
improvement of mesh quality while minimizing solution error. This
process is illustrated in Figure~\ref{fig:COMSOL}. 

\section{Modeling the LUX detector without charges\label{subsec:Modeling-LUX-run3}}

The electric field in the fiducial volume of the detector during WS2013
was assumed to be uniform in the analysis despite the slight radial
field component that pushed S2 electrons toward a smaller radius as
discussed below. There was also a small azimuthal contribution near
the walls outside of the fiducial volume, discussed in more details
in Section~\ref{subsec:scalloping}, which did not affect the WS2013
analysis. Overall, the field behavior was well understood and created
an opportunity to build, test, and understand the results of the initial
3D COMSOL Multiphysics LUX model.

\begin{figure}
\begin{centering}
\includegraphics[scale=0.22]{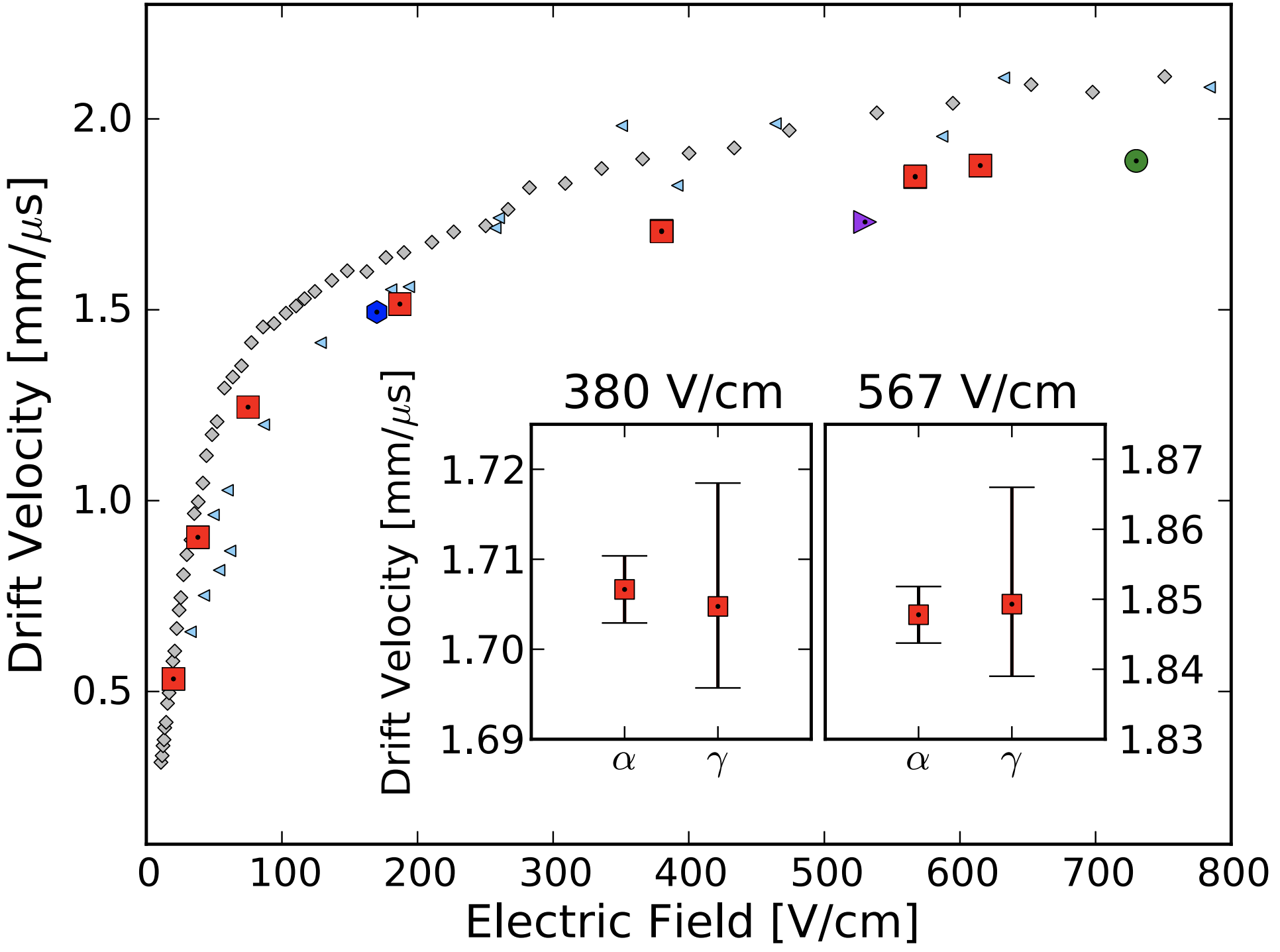}
\par\end{centering}
\caption[Electron drift velocity in LXe]{Electron drift velocity in LXe measured by the EXO-200 collaboration
(167 K, red square). Also shown are results from Miller et al. (163
K, cyan triangle)~\cite{PhysRev.166.871}, Gushchin et al. (165 K,
gray diamond)~\cite{gushchin1982electron}, XENON10 (177 K, green
circles)~\cite{Sorensen:2008kvp}, XENON100 (182 K, purple triangle)
\cite{APRILE201411}, and LUX (175 K, blue hexagon)~\cite{Akerib:2013tjd}.
Values used in this work are listed in Table~\ref{tab:Electron-velocity}.
Figure from~\cite{Albert:2016bhh}. \label{fig:Electron-drift-velocity-1}}
\end{figure}

The electric field values found in COMSOL Multiphysics were exported
and used to produce a simulation dataset of electron-like particles
in Python v2.7. This was accomplished by starting electron-like test
particles uniformly throughout the detector\textquoteright s active
volume and recording their intersection with the liquid surface after
drifting. These particles were propagated step-wise in 1 to $\unit[3]{\mu s}$
intervals using an interpolated velocity documented in Table~\ref{tab:Electron-velocity}.
This was necessary because the electron drift velocity in LXe varies
with the electric field, as shown in Figure~\ref{fig:Electron-drift-velocity-1}.
Once the electron-like test particles reached the liquid surface,
the simulated drift time $t\mathrm{_{sim}}$ and location of the simulated
S2 light production ($x_{S2_{sim}}$ , $y_{S2_{sim}}$) were compared
with the ($x_{S2}$, $y_{S2}$, $t$) distribution of $^{\mathrm{83m}}\mathrm{Kr}$
calibration data. 

\begin{figure}[p]
\begin{centering}
\includegraphics[scale=0.75]{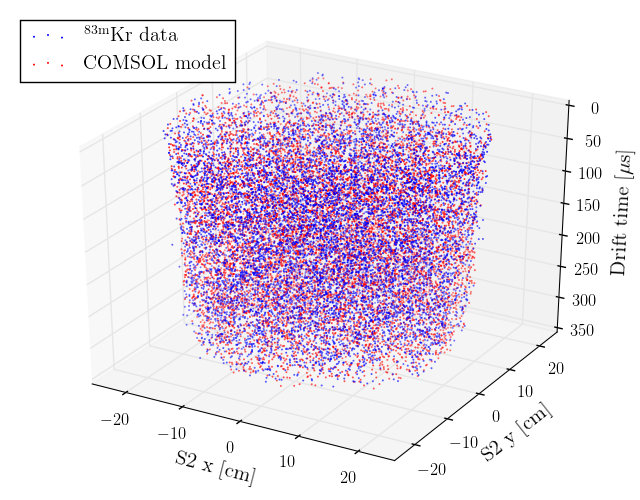}
\par\end{centering}
\caption[3D scatter plot comparing simulation and $^{\mathrm{83m}}\mathrm{Kr}$
data]{Scatter plot illustrating the 3D nature of the simulation. \label{Run3 model vs data}}
\begin{centering}
\vspace{1.5cm}
\includegraphics[scale=0.43]{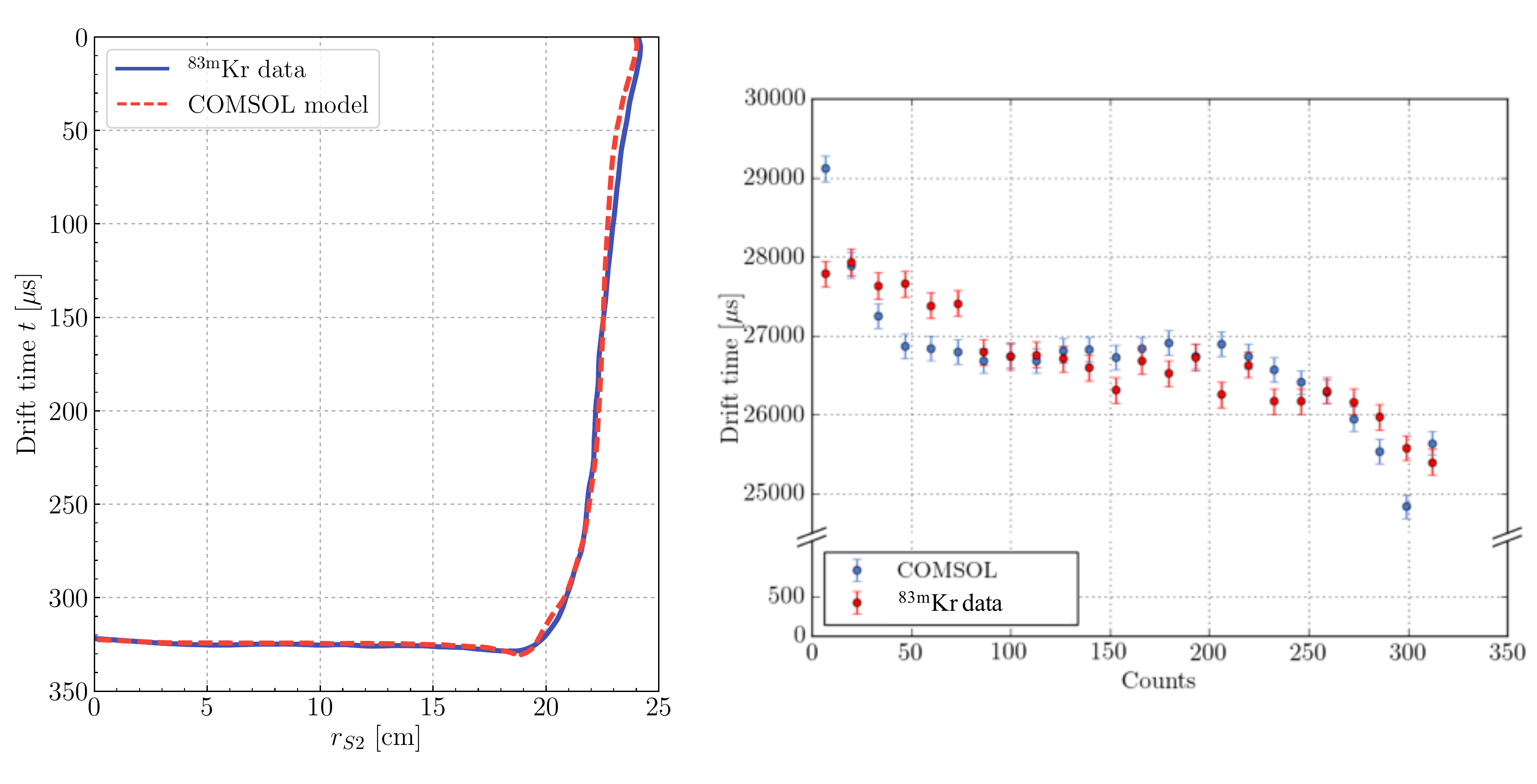}
\par\end{centering}
\caption[Plots illustrating agreement of edge location between data and simulation]{\textbf{Left:}~Contour plot showing histogram edges of the 3D field
model and $^{\mathrm{83m}}\mathrm{Kr}$ events averaged over all azimuthal
angles during WS2013. Data (solid blue) and simulation (dashed red)
contours agree without any fitting or tuning necessary. \textbf{Right:}~The
distribution of events at varying $z$ position agrees between the
$^{\mathrm{83m}}\mathrm{Kr}$ data and the simulation. \label{fig:run3_pocket}}
\end{figure}

Figure~\ref{Run3 model vs data} illustrates the 3D nature of both
datasets. Excellent agreement was seen as shown in Figure~\ref{fig:run3_pocket},
without a need to tune any aspect of the model to improve agreement
with the data. The 3D data match very well, which is more easily visualized
via a 2D summary. Similar to Figure~\ref{fig:kr-contours}, the contours
were created by finding the average number of events in all the non-empty
bins of a 2D histogram in $\left(r_{S2}^{2},t\right)$ and then drawing
a contour through those bins that contained half of the average number
of events. Therefore the contours approximate the edges of the detector\textquoteright s
active volume. The mean value of the electric field inside the fiducial
volume during WS2013 was $\unit[177\pm14]{V/cm}$ as calculated from
the COMSOL Multiphysics simulation. 

\begin{table}
\centering{}%
\begin{tabular}{l||rrrrrrrrr}
\hline 
Electric field {[}V/cm{]} & 20 & 38 & 51.8 & 68 & 105 & 182 & 375 & 727 & 1000\tabularnewline
Velocity {[}mm/$\mu$s{]} & 0.55 & 0.89 & 1.05 & 1.19 & 1.39 & 1.53 & 1.72 & 1.86 & 2.00\tabularnewline
\hline 
\end{tabular}\caption[Electron velocity in LXe used in simulations]{Electron velocity in LXe used in simulations~\cite{Albert:2016bhh,edwards2009zeplin_thesis,Sorensen:2008kvp}.\label{tab:Electron-velocity}}
\end{table}

\begin{flushright}
\begin{figure}
\begin{centering}
\includegraphics[scale=0.48]{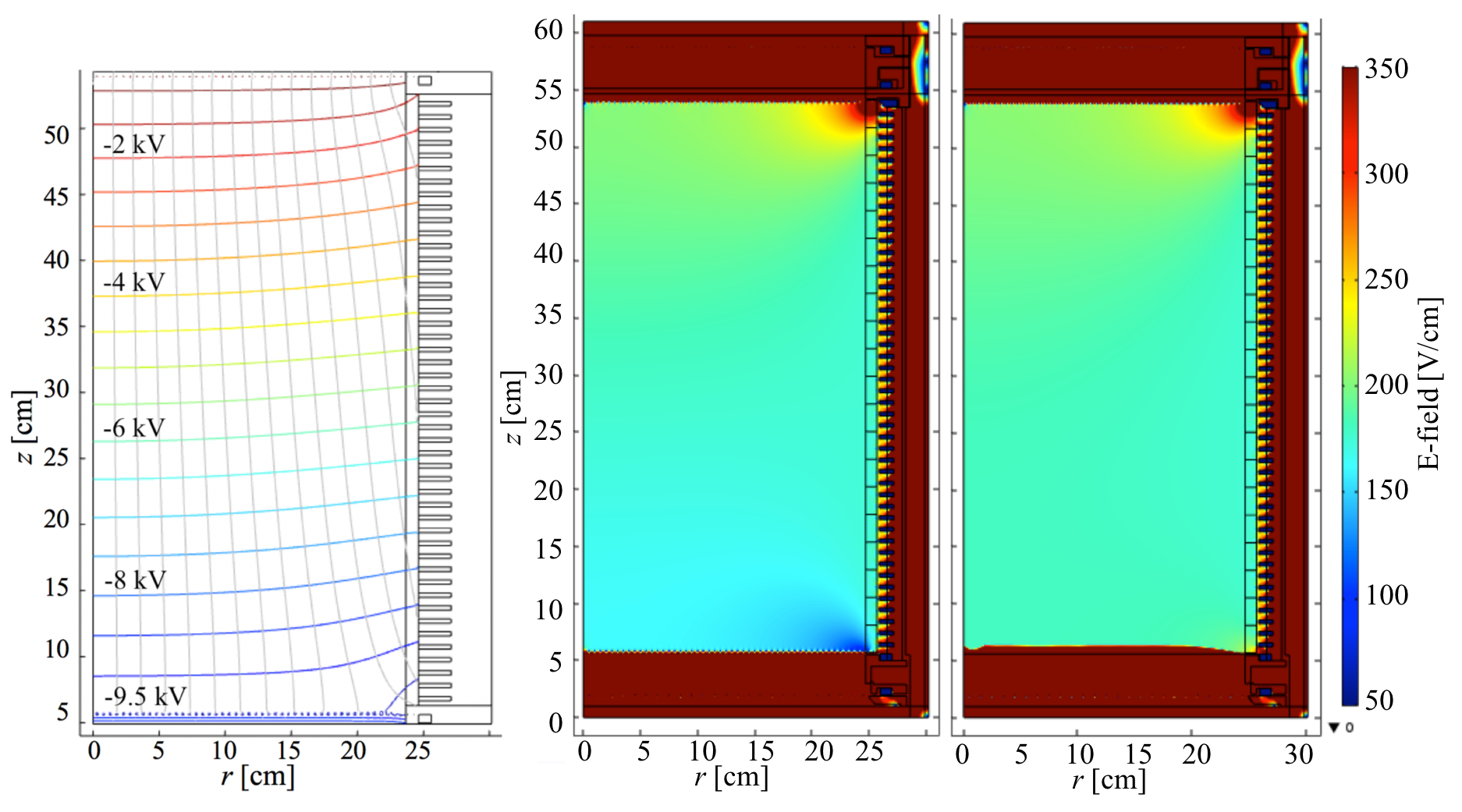}
\par\end{centering}
\caption[Results from an axially symmetric COMSOL Multiphysics model]{Results from an axially symmetric COMSOL Multiphysics model (for
illustration purposes.) \textbf{Left:}~Electric field lines and equipotentials
in the LUX detector during WS2013. \textbf{Center:}~Regions in the
corners of the detector show field leakage. A low-field pocket at
the high radius and low $z$ causes electrons originating in this
area to drift inward. This low-field pocket is predominantly due to
leakage of the much higher field in the reverse field region (region
between bottom shield grid and cathode), as well as the ground voltage
of the inner vacuum can. \textbf{Right:}~This leakage effect can
be mitigated by replacing the cathode grid with a plane with an equivalent
potential.\label{run3_field}}

\vspace{2cm}
\begin{centering}
\includegraphics[scale=0.5]{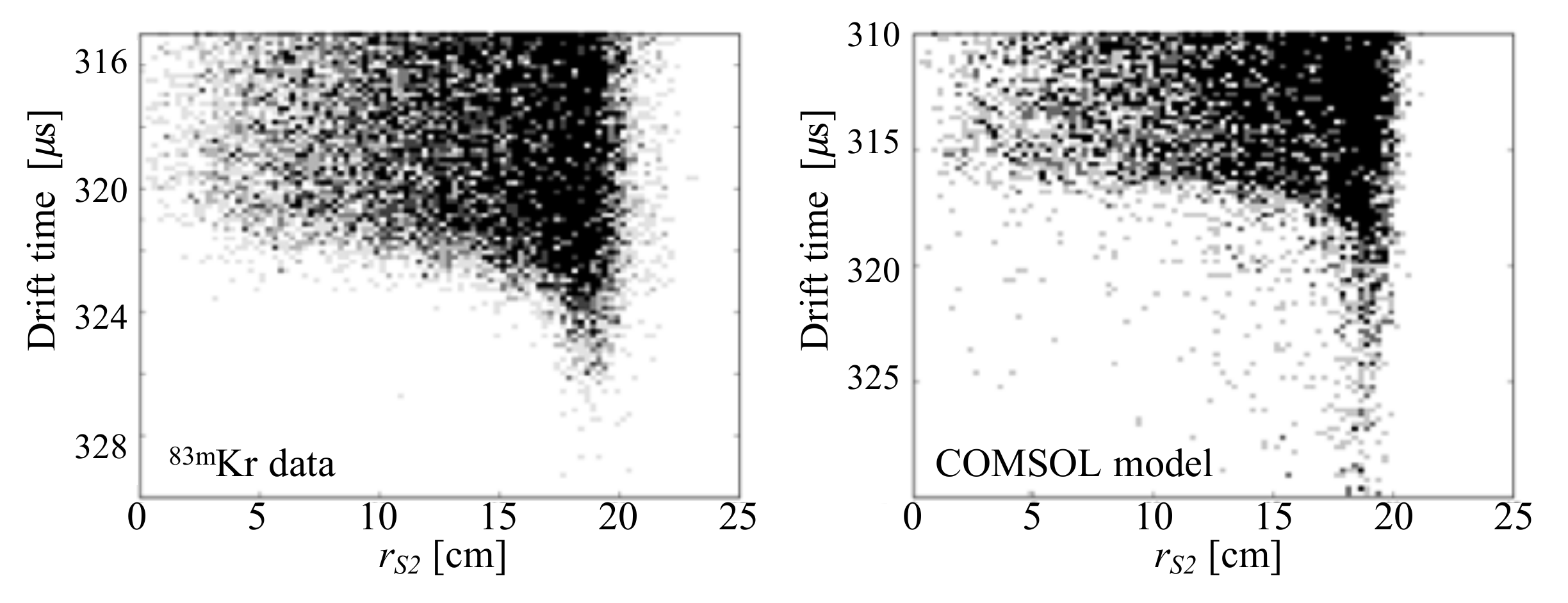}
\par\end{centering}
\caption[The low-field pocket in WS2013 showing the extended drift time in
the region]{Scatter plot of drift time vs. S2 radius for WS2013 $^{\mathrm{83m}}\mathrm{Kr}$
data (left) and COMSOL model (right) zoomed in to the low-field pocket
showing the extended drift time in the region. The slight quantitative
discrepancy is likely caused by a combination of an error on the simulated
electron velocity and slightly different dimensions of the real and
modeled detector. \label{fig:low_field_pocket}}
\end{figure}
\par\end{flushright}

The slight curve seen at a high radius and drift time in the reconstructed
S2 coordinates can be understood by looking at Figure~\ref{run3_field}.
This figure shows equipotentials and field lines from an axially symmetric
COMSOL Multiphysics LUX model built for visualization purposes. A
$^{\mathrm{83m}}\mathrm{Kr}$ event decaying in the bottom right-hand
corner of the active volume will follow the field lines shown and
end up at a slightly smaller radius than where it started. This slight
curvature has been seen before, for example in simulations of the
XENON100 experiment~\cite{mei2012direct}. The inward push is due
to a low-field pocket seen at the high drift time, high radius corner,
which is created by the reverse field region electric field (region
between the bottom shield grid and the cathode grid) and the ground
voltage of the inner vacuum can leaking through the cathode grid.
This low-field pocket goes away when the cathode grid wires are replaced
by a solid plane held at the same cathode voltage also shown in Figure~\ref{run3_field}.
The low-field pocket creates a mild concern for analysis: the electron-ion
recombination varies with field strength much more strongly at low
field. As a result, if the field is low enough, then high recombination
could potentially create a nearly S2-free detector response, and thus
a dead volume unaccounted for in the fiducial volume calculation.
However, events in this low-field pocket can be flagged in the data
using their long drift times, as shown in Figure~\ref{fig:low_field_pocket}. 

A similar field effect is seen near the top of the detector. The high-field
pocket causes a region of slightly enhanced $S2/S1$ ratio since high
electric field boosts S2 and suppresses S1 signals. Additionally,
events occurring at high radius experience a slight inward push. This
helps identify PTFE surface events since the radial position resolution
is improved by spreading the light over multiple PMTs. Without this
inward push, S2 signals of surface events would be primarily concentrated
on a single PMT.

\section{Modeling charges in the LUX detector}

\begin{figure}[t]
\begin{centering}
\includegraphics[scale=0.65]{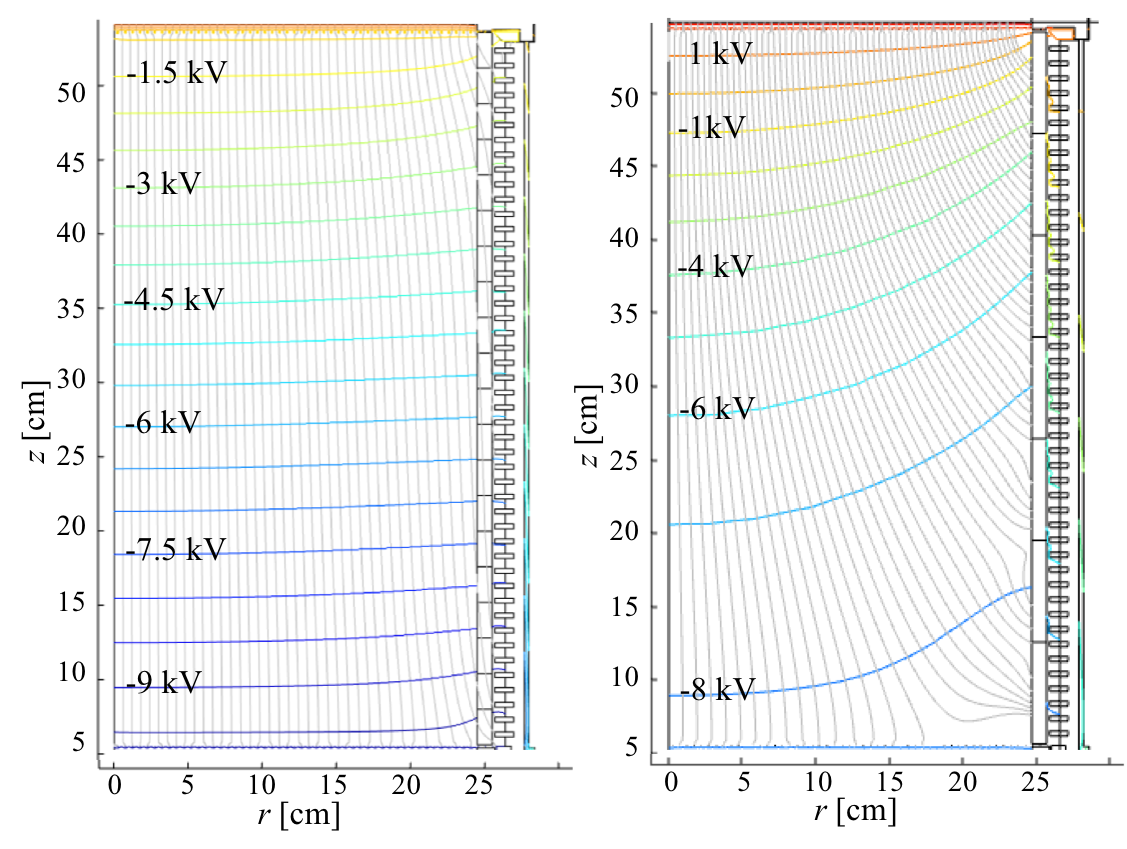}
\par\end{centering}
\caption[Comparison of field in WS2013 and WS2014-16]{Two axially symmetric COMSOL Multiphysics models built for visualization
purposes. \textbf{Left:}~Electric field lines and equipotentials
in the LUX detector during WS2013. A small radially-inward component
is seen, resulting from the geometry of the field cage and the grids.
\textbf{Right:}~Electric field lines implied by $^{\mathrm{83m}}\mathrm{Kr}$
from 2015-04-26 showing the much stronger radially-inward component
modeled by negative charge densities in the PTFE panels.\label{fig:field-comparison}}
\end{figure}

After the conditioning campaign that took place between WS2013 and
WS2014-16 as described in Section~\ref{subsec:Grid-conditioning-campaign},
a substantial change in the detector\textquoteright s electric field
occurred, which affected electron paths as shown in Figure~\ref{fig:kr-contours}.
This change in the field can be modeled with negative charge densities
present in the PTFE panels; the consequence of this accumulated charge
is illustrated in Figure~\ref{fig:field-comparison}. 

The conditioning campaign was performed in xenon gas and exposed the
PTFE wall to VUV photons, which created electron-hole pairs within
the PTFE surface that defines the radial boundary of the active volume
and serves as a reflector to VUV scintillation photons. The subsequent
observed time evolution of the PTFE charge is due to the higher mobility
of positive charges (holes) relative to the negative charges (electrons),
allowing the gradual removal of holes under the applied field and
leaving an on-balance negatively-charged surface. This behavior is
consistent with a non-uniform time-varying negative charge density
in the PTFE panels as described below. Due to PTFE being an excellent
insulator, it is very hard to get rid of those charges during the
operational lifetime of the detector and therefore it is critical
to understand their effects on the electric field. So in an (unfortunate)
way, LUX can also serve as a very sensitive static charge measuring
device.

\begin{figure}
\begin{centering}
\includegraphics[scale=0.46]{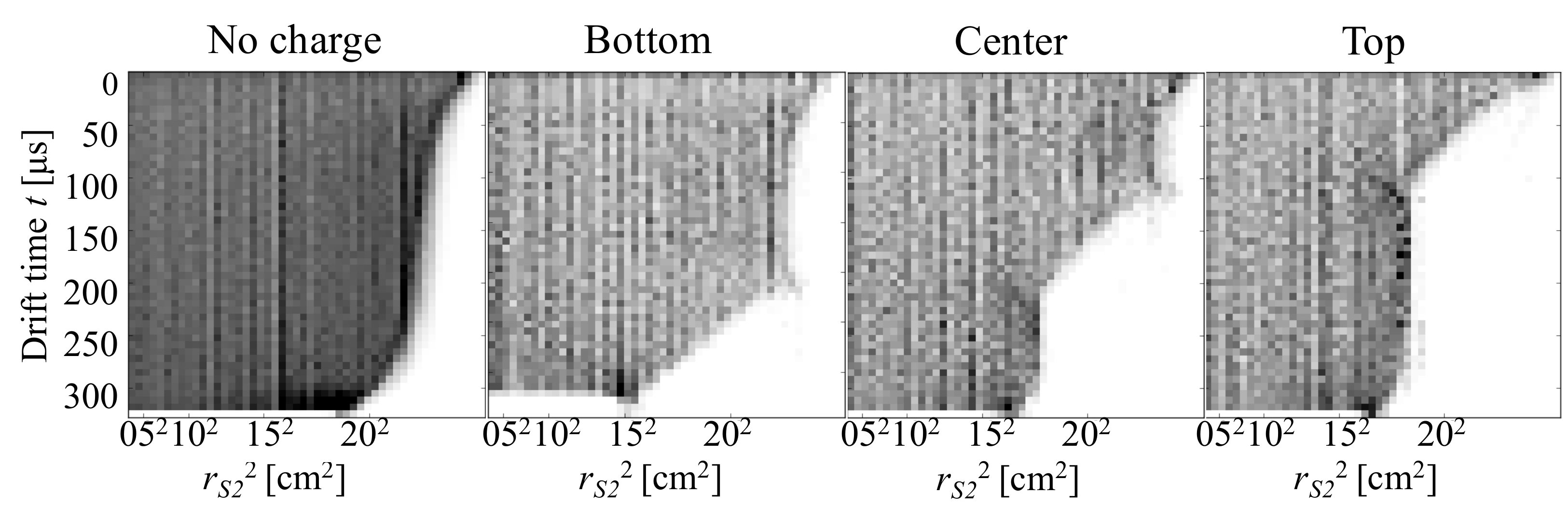}
\par\end{centering}
\caption[Effect of negative charge densities in the PTFE panels on events in
the detector]{Effect of negative charge densities in the PTFE panels on events
in the detector as simulated in the 3D COMSOL model. The four figures
illustrate COMSOL models with no charge in the PTFE panels and with
$\unitfrac[-1]{\mu C}{m^{2}}$ deposited at the lower, middle, and
upper segments of a PTFE panel causing a strong inward push for events
that originate near or below the charge site.\label{fig:charge_demo}}

\end{figure}

To gain intuition about the effects of charge in the PTFE panels on
electron paths, a simulation was performed by depositing $\unitfrac[-1]{\mu C}{m^{2}}$
of charge density in either the lower, middle, or upper third of a
PTFE panel; the resulting force causes inward displacement of electron
clouds that originate near or below the charged site, as demonstrated
in Figure \ref{fig:charge_demo}. 

\subsection{Alternative explanations for the field distortion }

\begin{figure}[t]
\begin{centering}
\includegraphics[scale=0.48]{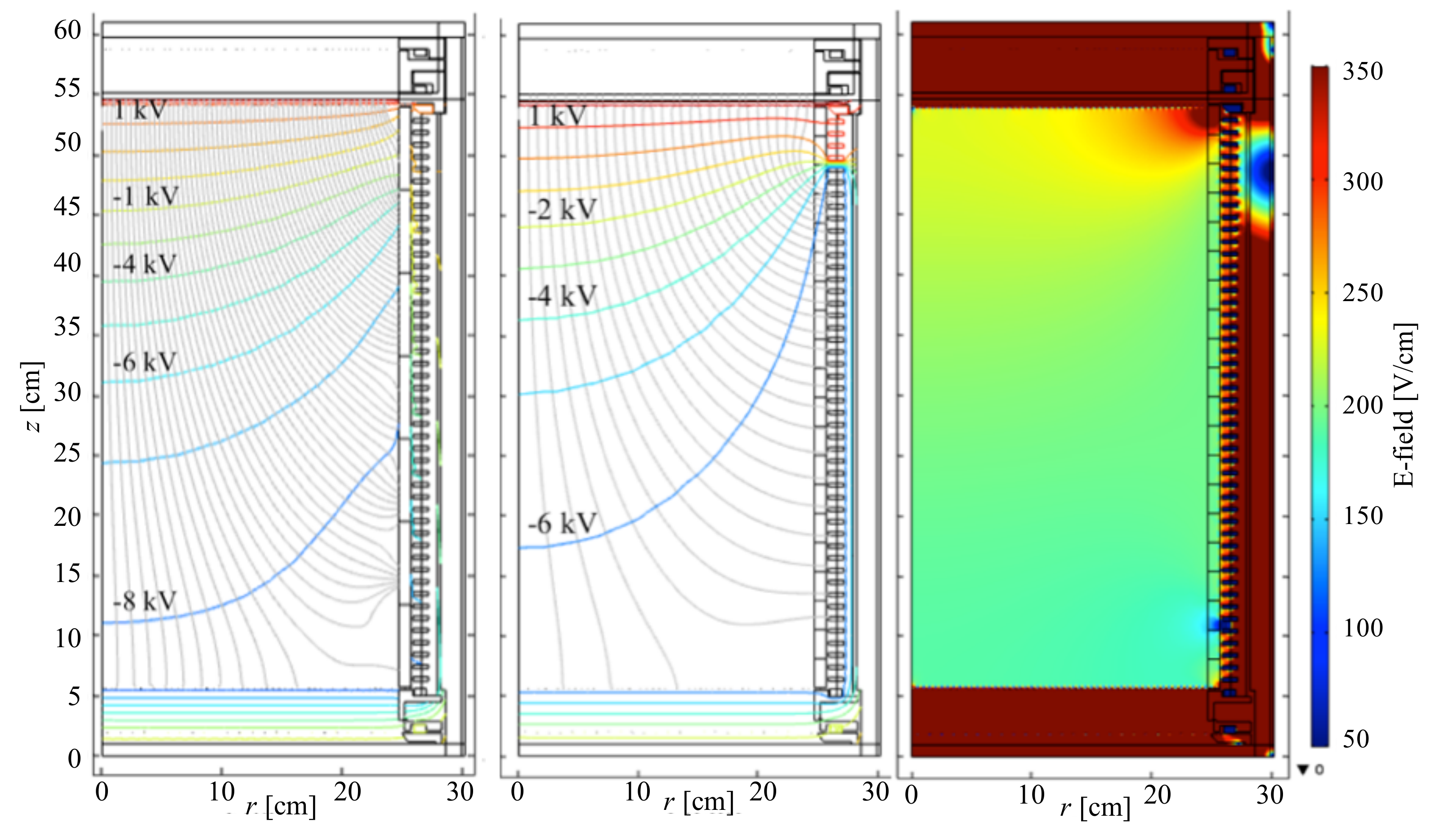}
\par\end{centering}
\caption[Models illustrating alternative explanations for the field distortion]{\textbf{Left:}~Electric field lines as modeled using $^{\mathrm{83m}}\mathrm{Kr}$
from 2015-07-22. \textbf{Center:}~Electric field lines caused by
an open circuit in the field shaping rings. In this case, the connection
between the 4$^{\mathrm{th}}$ and the 5$^{\mathrm{th}}$ top ring
is broken. A field distortion is obvious and not dissimilar to the
field observed in WS2014-16, however, this effect is not only ruled
out by our measurement of the resistance of the field-shaping ring
chain but would also not be time-dependent. \textbf{Right:}~Electric
field due to a shorted field ring. In this case, the 5$^{\mathrm{th}}$
field ring from the bottom is shorted. The effect on the field is
very localized and does not distort the field nearly as much as what
is observed in WS2014-16. All three images come from an axially symmetric
COMSOL Multiphysics WS2014-16 model built for illustration purposes
only.\label{run4_open_short}}
\end{figure}

Other than the charge evolution in the PTFE surface, there are other
plausible explanations for distorted fields inside a TPC. One possibility
would be that drifting electrons land on the PTFE surface and do not
get pulled toward the surface due to the high resistivity of PTFE.
This is unlikely since the detector is designed to pull electrons
up and away from the PTFE surface and because any negative charging
will prevent further charge from drifting toward the surface. Other
explanations might be a short or open circuit in the field-shaping
ring resistor chain. The field distortion caused by a shorted resistor
was simulated in the axially symmetric COMSOL model as shown in Figure~\ref{run4_open_short}.
A shorted resistor produces a small, very localized field deformity,
unlike the detector-wide field distortion observed in WS2014-16. This
effect would also not evolve throughout the detector operation. 

An open circuit was simulated in the axially symmetric COMSOL model
also shown in Figure~\ref{run4_open_short}. A field distortion is
distinct and not dissimilar to the field observed in WS2014-16. However,
depending on the location of the open circuit, it can create large,
dead, field-less regions in the detector. The resistance of the field-shaping
ring resistor chain was measured before WS2013 and then again before
WS2014-16. No significant change was observed, ruling out this explanation.
The possibility of an open circuit effect is also ruled out because
it would not create a time-dependent component. Additionally, $^{137}$Cs
calibrations~\cite{phelps2014lux} were performed in WS2014-16 using
the acrylic radioactive-source deployment tubes suspended along the
side of the detector, which showed events along the entire detector
height, inconsistent with the presence of field-less regions.

Therefore, the LUX collaboration concluded that charges in the PTFE
surfaces were the culprit for the varying electric fields. We next
describe the electrostatic properties of PTFE generally, before moving
to the central work of modeling these charges.

\subsection{Charges in PTFE\label{subsec:Charges-in-PTFE}}

Since the goal of this work was to model charges in the PTFE surface,
this section offers a brief overview of PTFE properties and their
relevance to this analysis.

PTFE is a type of synthetic fluoropolymer of tetrafluoroethylene also
known as Teflon\texttrademark~\cite{chemours} discovered by DuPont
Co. in 1938\footnote{The Teflon trade name is also used for other polymers with similar
compositions and properties such as perfluoroalkoxy alkane (PFA) and
fluorinated ethylene propylene (FEP).}$^{,}$\footnote{Fluorinated ethylene polypropylene (FEP) is another type of Teflon
worth mentioning. It is a transparent material, softer than PTFE,
frequently used is satellite construction. FEP has been well researched
at different temperatures due to a worry of charge buildup on the
satellites' surfaces causing sparks, which makes it a good material
to help build intuition about PTFE properties~\cite{paulmier2014radiation}.
Charge buildup is also a worry in other industries such as high voltage
DC power transmission lines and microelectronics.}. PTFE is a non-reactive, hydrophobic material with a very low coefficient
of friction, high dielectric constant, and good reflectivity for xenon
scintillation light. This makes it a very popular material in two-phase
dark matter detectors. However, fluorine is the most electronegative
element in the periodic table\footnote{$^{9}$F has 5 electrons in its 2p shell.}
making it a very attractive site for electrons.

\begin{figure}[t]
\begin{centering}
\includegraphics[scale=0.65]{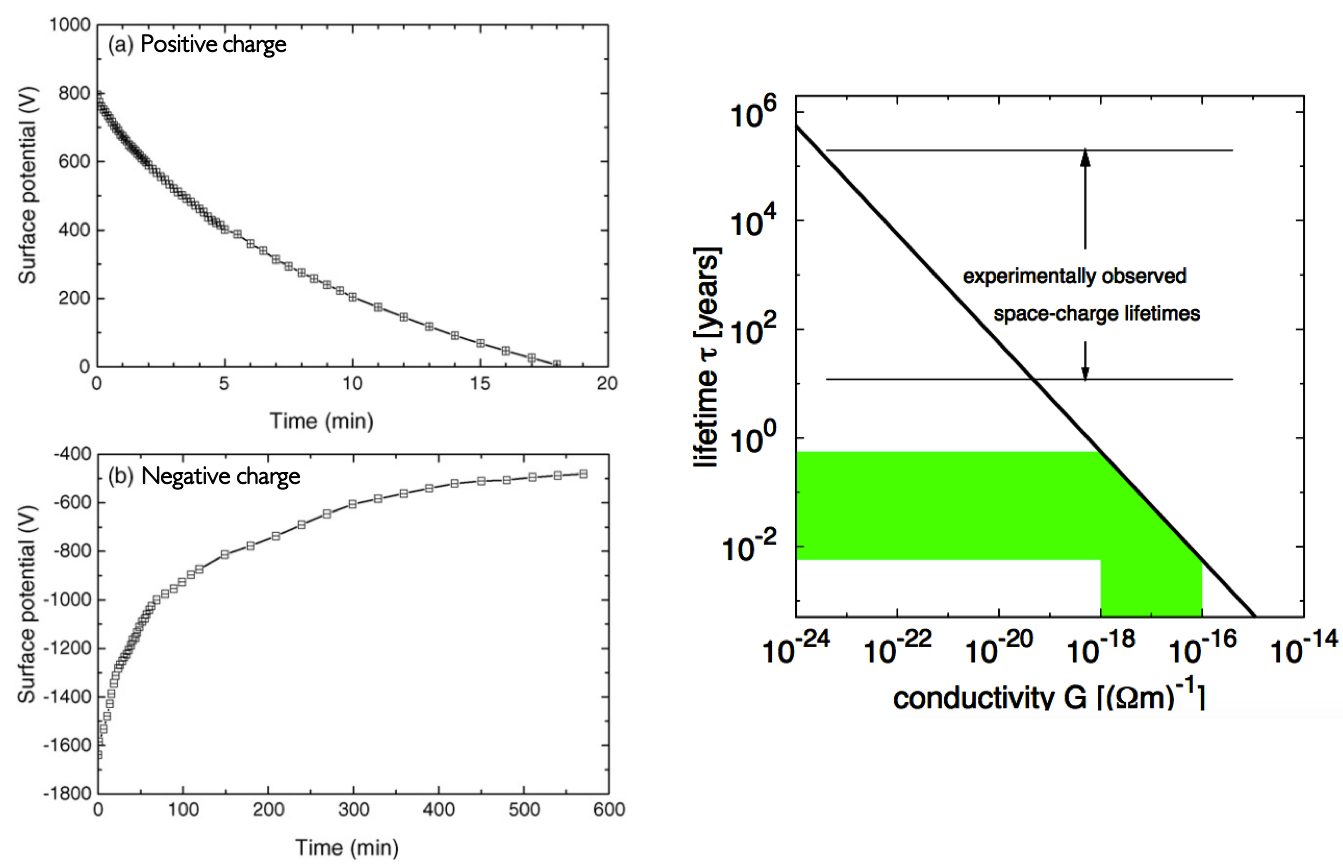}
\par\end{centering}
\caption[Electron-hole behavior in PTFE]{\textbf{Left:} The decay curve of the surface potential of PTFE samples
while charging positively (top) and negatively (bottom). Figure from~\cite{zhang2006surface}.
\textbf{Right:} Space-charge lifetime $\tau$ versus electrical conductivity
$G$, as given by Equation~\ref{eq:tau}. With experimental conductivities
between $\unit[10^{-16}-10^{-18}]{\left(\Omega\cdot m\right)^{-1}}$
(green shaded area), space charge should decay on a timescale of days
or months, rather than millennia. Figure from~\cite{mellinger2004charge}.
\label{fig:charge_decay}}
\end{figure}

PTFE is known as an excellent insulator. Due to its excellent charge
storage capacity, it is very popular as an electret\footnote{An electret is the electrostatic equivalent of a magnet. It is a polarized
material that generates an electrostatic field around itself thanks
to its semi-permanent electric charge.} in many applications, such as in microphones. Hence many studies
have been done investigating PTFE properties at room and elevated
temperature, but unfortunately not at the LXe temperatures of 175
K. Resistivity of PTFE has been measured to be $\rho\sim3\times\unit[10^{20}]{\Omega\cdot cm}$
at $\unit[5\times10^{-6}]{torr}$ at 25$^{\circ}$C~\cite{green2006ptferho}.
Charge lifetimes at room temperature have been measured to be $10^{3}-2.6\times10^{5}$~years~\cite{mellinger2004charge},
which is much longer than what can be inferred from the published
PTFE conductivities illustrated in Figure~\ref{fig:charge_decay},
a discrepancy that remains an open question, but does not affect results
presented in this work. 

The PTFE surface has been demonstrated to hold charges of at least
$\unit[0.1]{mC/m^{2}}$~\cite{kressmann1996electrets}, an order
of magnitude more than what was modeled to be observed in LUX. Peak
space charge densities of $\unit[2500]{C/m^{3}}$ have been reported
in corona-charged PTFE with bulk charge densities being 1-2 orders
of magnitude smaller~\cite{mellinger2004charge}. This puts the results
presented in this chapter within reasonable bounds.

The grid conditioning done in March 2014 was performed in cold xenon
gas. The far ultraviolet emission spectrum of xenon contains five
emission lines in the range $117-147$~nm~\cite{kanik1996far} corresponding
to energies of $10.6-8.4$ eV. The photoemission threshold energy
of PTFE is 10.6~eV while the VUV absorption spectrum shows an intense
peak at 7.7~eV likely corresponding to an excitation from the top
of the valence band to the bottom of the conduction band~\cite{seki1990electronic}.
Therefore, discharges during the conditioning campaign had sufficient
energy to liberate electron-hole pairs in PTFE\footnote{There are three different regimes in Teflon\texttrademark~corona
charging. First, there is a linear regime where the increase of the
surface potential is proportional to the charging time. In this case,
all of the incoming charges are deposited in the surface traps, and
the sample can be regarded as the equivalent of a charging capacitor.
After the filling of surface traps, charges are injected into the
traps in bulk in a sub-linear regime. Finally, in the saturation regime,
the displacement current is almost zero and the conduction current,
being only leakage current, approximately equals the total current~\cite{zhang1991constant}. }. Studies using multiple-needle-plane corona discharges on a PTFE
block show that charge created this way mainly exists in the surface
layer at $\unit[1-2]{\mu m}$. 

It has been shown that under an applied field, hole states exhibit
significantly higher mobility than electron states. The applied field
will gradually transport holes, and the surface charge will become
increasingly negative, asymptoting with the hole transport timescale.
One study found that after charging the PTFE block for 30~mins by
applying a voltage of $\unit[\pm5]{kV}$ at room temperature, the
half-life period of surface potential decay was 5~min for hole traps
and 150~min for electron traps, reflecting this distinct difference
in the surface layer of PTFE as shown Figure~\ref{fig:charge_decay}.
The energy levels of electron traps and hole traps were about $0.85-1.0$~eV
and $0.80-0.90$~eV, respectively~\cite{zhang2006surface}. A timescale
of hole transport in the LUX PTFE surface as seen in the high purity,
low-temperature LXe environment is shown in Figure~\ref{fig:Average-charge}.

In other research areas, PTFE is usually charged up using an electrode
contact, which injects electrons across the PTFE surface. A change
in conductivity is observed when electrons reach a grounded electrode.
This is called the transit time and has been computed for LUX in this
work. Another method of charging PTFE is by using a high energy electron
beam, which deposits electrons from tens of nm to tens of $\mu$m
deep into the surface depending on the beam energy. To mitigate the
amount of positive/negative charges in PTFE, the material can be heated
up while either blasting an electron gun or shining a VUV light until
an equilibrium is reached. Heating the material helps the charge mitigation
since both the charge lifetime $\tau$ and conductivity $G$ are temperature
dependent~\cite{zhang2006surface,mellinger2004charge}:
\[
\tau=\tau_{0}\exp\left(\frac{E_{trap}}{kT}\right)
\]
where $E_{trap}$ is the trap depth, $T$ is the temperature, $k$
is the Boltzmann constant and $\tau_{0}^{-1}$ set by molecular vibration
frequencies ($\unit[10^{13}]{Hz}$). 

To calculate the transit time, derivation from~\cite{mellinger2004charge}
can be followed. For the case of ohmic conductivity, current density
$j$ is given by 
\begin{equation}
j=-\frac{d\sigma}{dt}=GE\label{eq:surface_current}
\end{equation}
where $\sigma$ is the surface charge density, $G$ is the conductivity
and $E$ is the internal electric field. Assuming a thin space-charge
layer, the electric field in the sample is nearly homogeneous and
given by
\begin{equation}
E=\frac{\sigma}{\varepsilon_{0}\varepsilon_{r}}\label{eq:E_field}
\end{equation}
where $\varepsilon_{r}=2.1$ for PTFE. Solving the differential equation
obtained by inserting Equation~\ref{eq:E_field} into Equation~\ref{eq:surface_current}
yields an equation for the time decay of surface charge
\begin{equation}
\sigma=\sigma_{0}\exp\left(\nicefrac{-t}{\tau}\right)\label{eq:sigma}
\end{equation}
where $\sigma_{0}$ is the initial surface charge, and charge lifetime
$\tau$ is given by 
\begin{equation}
\tau=\frac{\varepsilon_{0}\varepsilon_{r}}{G}.\label{eq:tau}
\end{equation}
This result will be used in Section \ref{subsec:Charge-evolution}
to estimate the average charge density in LUX throughout WS2014-16.

\subsection{Modeling changing electric fields in WS2014-16\label{subsec:Modeling-Changing-Fields}}

\begin{figure}[!h]
\centering{}\includegraphics[scale=0.45]{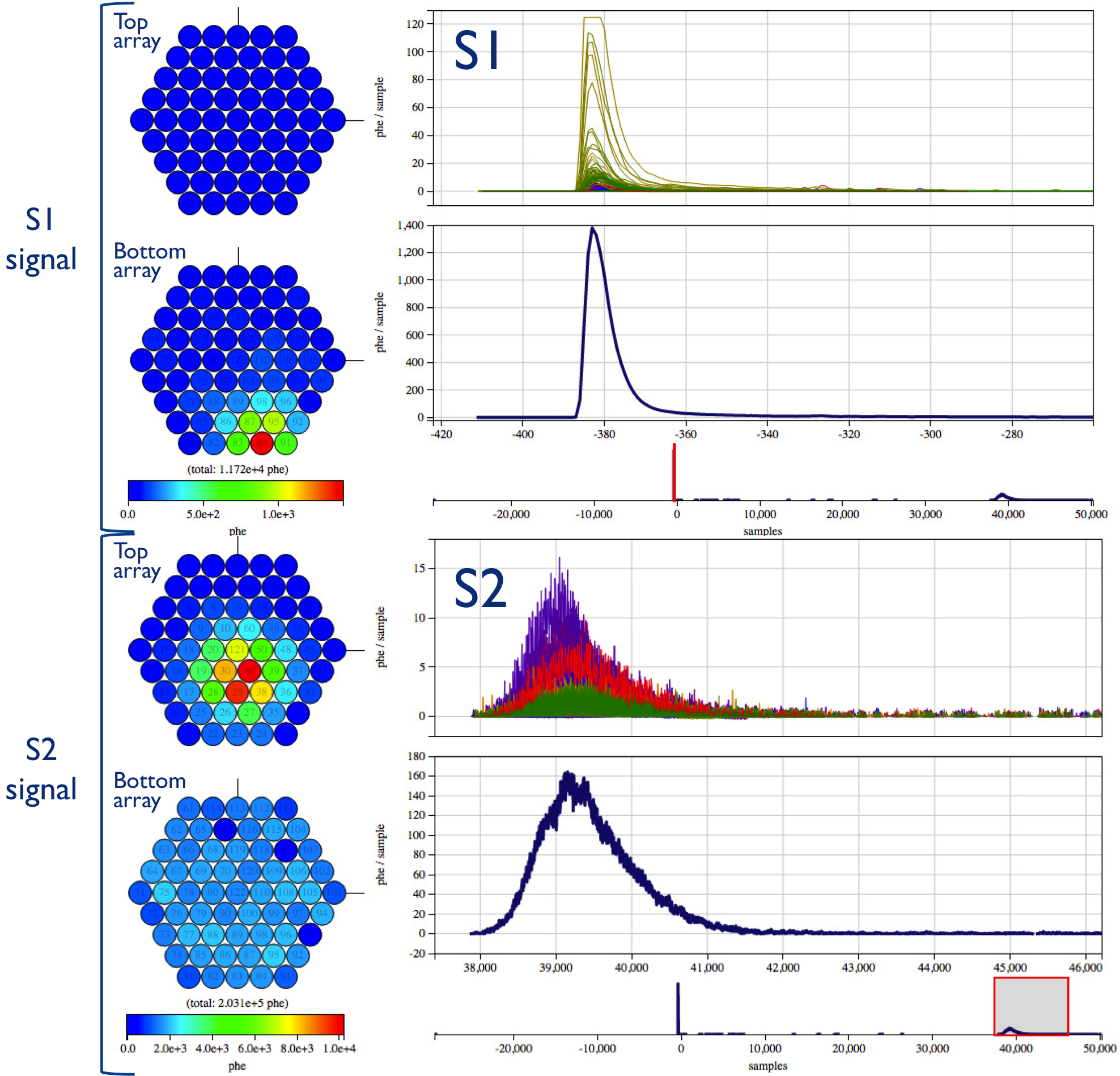}\caption[A detector event showing the strong field effect due to wall charges
in WS2014-16]{A detector event showing the strong field effect due to wall charges
from July 23, 2015, during WS2014-16. The top image shows an S1 pulse
near the detector wall above the cathode at $\unit[390]{\mu s}$,
while the bottom image shows the successive S2 pulse as it reaches
the liquid surface. The four plots on the right show the waveforms
as seen by each PMT (multiple rainbow colored lines) and the summed
average across the top and bottom PMT arrays (wide blue line) for
both the S1 and the S2 signal. The timeline is given in samples (1
sample = 10 ns). The schematics on the left shows the amount of light
seen by each PMT at the top and bottom arrays. From the location of
light seen by the PMT arrays, the event migration toward the center
of the detector on its way up can be observed. Images are taken from
Visualux, the LUX dataset preview software~\ref{visualux}.\label{visualux}}
\end{figure}

The effect of the inward push from the electric field is apparent
from an event from WS2014-16 shown in Figure~\ref{visualux}. Here
an event that originated near the bottom of the detector was pushed
inward while drifting. This phenomenon causes problems in analysis,
not only for the field-dependent recombination but also for position
reconstruction, since an electron cloud that starts near the center
and top of the detector has the same $\left(x,y\right)$ position
as a cloud that starts near the edge but at the bottom of the detector. 

\begin{figure}
\begin{centering}
\includegraphics[scale=0.6]{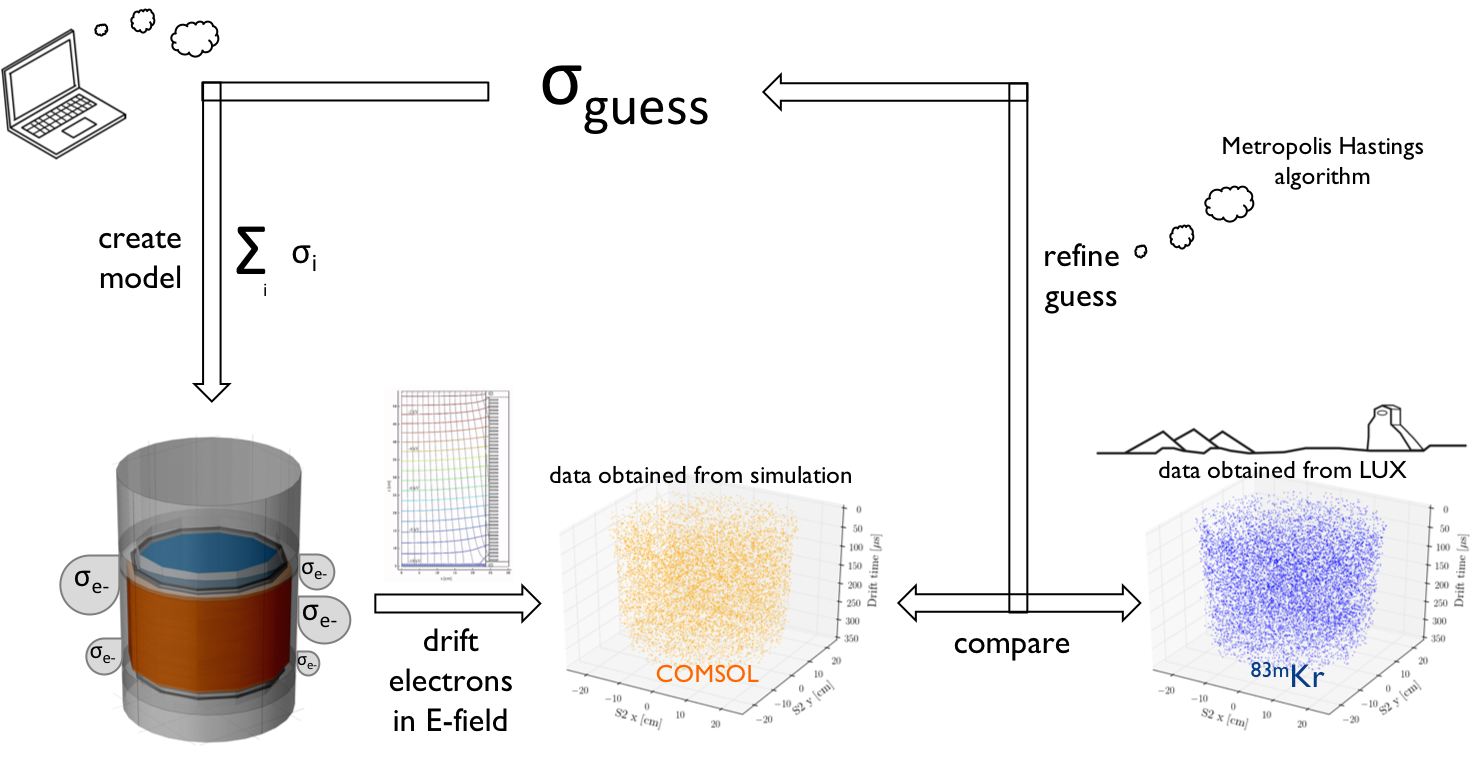}
\par\end{centering}
\caption[Illustration of steps involved in modeling electric fields during
WS2014-16]{Illustration of steps involved in modeling electric fields during
WS2014-16. First, a model of the LUX detector with 42 tiles was created
in COMSOL. The use of superposition principle then enabled placement
of any combination of charges $\Sigma_{i}\sigma_{i}$ on detector
walls (bottom left). Simulated particles were then drifted through
that charge distribution. The simulated particles (orange scatter
plot) were compared with real data from the detector (blue scatter
plot) and using the Metropolis-Hastings algorithm a more accurate
guess $\sigma_{\mathrm{guess}}$ was generated. This process was iterated
until convergence between simulation and data was observed. \label{fig:Illustration-of-steps-run4fields}}

\end{figure}

This effect was corrected in the analysis using field maps, the output
of this work. This section provides a detailed description of the
approach adopted to model the electric fields throughout WS2014-16
and resulting in those field maps. A summary of the contents of this
section is illustrated in Figure~\ref{fig:Illustration-of-steps-run4fields}.

\subsubsection{Principle of superposition\label{subsec:Principle-of-superposition}}

To model the detector behavior in WS2014-16, first, the charge-free
model described in Section~\ref{subsec:Modeling-LUX-run3} was modified
to include WS2014-16 grid voltages. Then 42 additional COMSOL Multiphysics
models were built, with voltages on all boundaries set to 0 V. The
12 PTFE panels surrounding the active volume were split into 42 tiles
by defining 6 angular and 7 vertical divisions of equal width and
height. Therefore there were 2 PTFE panels in each angular division.
Each one of the 42 models included $\unit[-1]{\mu C/m^{2}}$ of charge
on one of the 42 tiles. These 43 datasets (arising from 42 models
with charge and one without charge) served as basis vectors since
the charge on any tile could be scaled up or down as needed. Note
that this linear superposition of basis charges is valid since the
electric field arising from two isolated charges is equal to the sum
of the electric fields arising from each charge individually. Leveraging
the superposition principle, these models could be linearly combined
to produce electric field maps based on arbitrary charge distribution.
Those models were solved in COMSOL Multiphysics. The resulting 43
electric field vector maps were exported on a $\unit[1]{cm^{3}}$
grid to a Python environment where all further calculations were performed.

\begin{figure}
\begin{centering}
\includegraphics[scale=0.72]{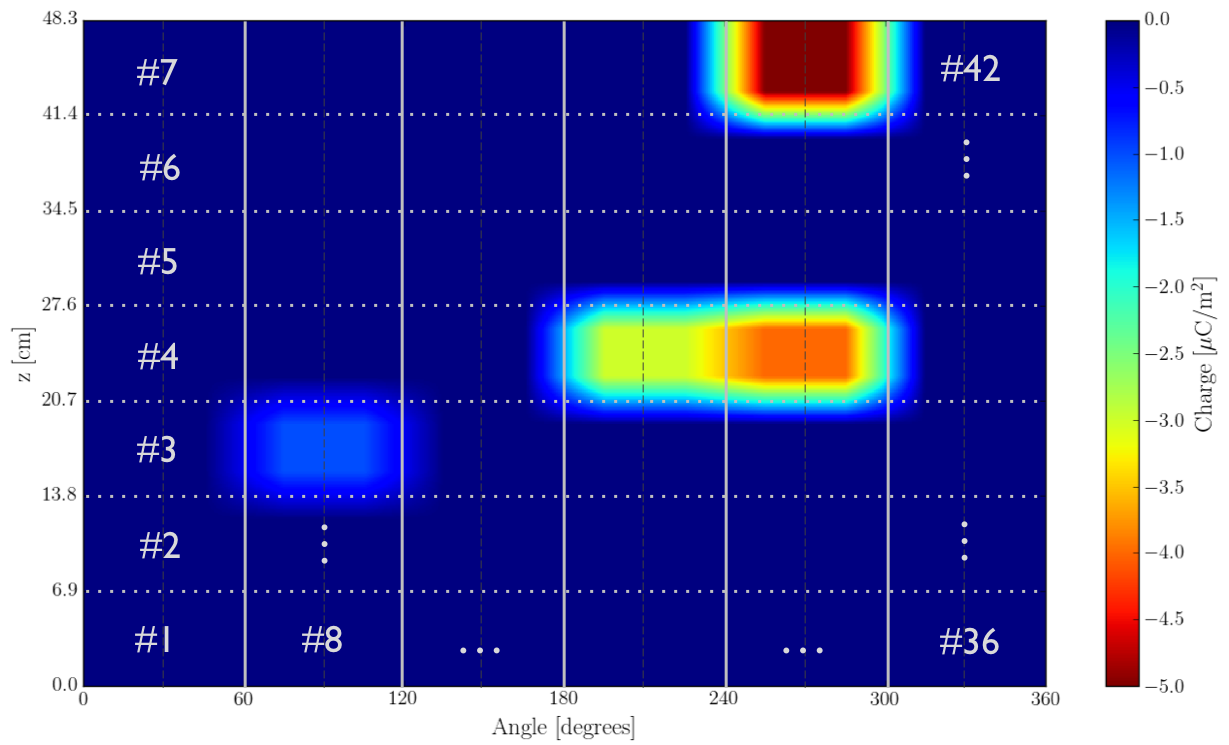}
\par\end{centering}
\caption[Illustration of the superposition principle]{Illustration of the PTFE wall with split into 42 tiles, where each
tile can contain a different amount of charge density. The smoothing
function described in Section~\ref{subsec:Smoothing-function} avoids
unphysical sharp drops between neighboring tiles with different charge
densities. \label{charge_demo}}
\end{figure}

Each one of the 42 tile models could be multiplied by a different
charge density to create any charge configuration desired as seen
in Figure~\ref{charge_demo}. All further calculations were done
in a Python environment to allow for easier and faster superposing.

The choice of the number of tiles used was optimized to balance two
opposing effects: more tiles result in a better real-life approximation,
but it also increases computational requirements due to the high number
of free parameters for minimization. To test this method, a simpler
model was initially created in which the PTFE panels were split only
into 15 tiles: 3 angular and 5 vertical divisions. However, the granularity
of the model was too large, which led to the development of a model
with a higher number of tiles. A model with a finer granularity was
not tested since even the 42-basis vector model used in the analysis
was quite complex given the available resources and time constraints,
and required manual adjustments to converge within the necessary time
window.

\subsubsection{Smoothing function\label{subsec:Smoothing-function}}

Unphysical behavior emerged when two neighboring tiles that constitute
the 42 basis vectors contained very different charges\footnote{The PTFE panels installed in the detector consist of 12 long vertical
pieces. Although unlikely, in theory, two adjacent panels could contain
different charge densities.}. To mitigate these sharp discontinuities, the charge on each tile
was set to fall off linearly both in the azimuthal and in $z$ directions.
This smoothing was defined such that in the azimuthal direction each
tile had a uniform $\unit[-1]{\mu C/m^{2}}$ charge density from $\nicefrac{1}{4}$
to $\nicefrac{3}{4}$ of tile width. This charge density then decreased
linearly to $\unit[0]{\mu C/m^{2}}$ through $\nicefrac{1}{4}$ of
the width of the neighboring tile. Therefore the charge density in
the azimuthal direction was continuous. The same technique was applied
in the $z$-direction except that the bottommost and topmost edge
tiles were smoothed in only the direction pointing toward the center
of the panel.

\subsubsection{Metropolis-Hastings algorithm}

With the charge density basis vectors defined, an iterative process
was used to find the true charge in the PTFE panels. First, a hypothesized
charge distribution was defined, and each one of the 42 charge basis
vectors was multiplied accordingly. The electric fields from each
basis vector simulation were combined to produce a hypothetical electric
field inside LUX. Simulated particles were drifted through this electric
field, and a likelihood function was used to compare the outcome to
$^{\mathrm{83m}}\mathrm{Kr}$ data. Then the initial charge hypothesis
was refined, and this approach was iterated using the Metropolis-Hastings
algorithm until convergence of results was observed. 

The Metropolis-Hastings algorithm is a Markov Chain Monte Carlo method
for sampling from a multi-dimensional distribution with a high number
of degrees of freedom. First, an initial guess was made for the amount
of charge on each tile, and the corresponding model was created. Each
subsequent guess was randomly sampled from a Gaussian distribution
centered at each charge density with a width of $\unit[\sim0.1]{\mu C/m^{2}}$.
The ideal step size was tuned to achieve acceptance of new charge
distributions of $\sim23\%$~\cite{chib1995understanding}, where
the acceptance ratio was defined as 
\[
\alpha=\frac{\mathcal{L_{\mathrm{new}}}}{\mathcal{L}_{\mathrm{old}}}.
\]
Here, $\mathcal{L_{\mathrm{new}}}$ and $\mathcal{L}_{\mathrm{old}}$
are the log likelihood functions for the current and previous steps,
respectively. 

To speed up the convergence at the beginning of the modeling efforts,
the amount of charge in the PTFE panels was adjusted manually to smooth
out charge deviations between neighboring vertical panels before using
the Metropolis-Hastings algorithm. As the dataset neared convergence,
a modified Metropolis-Hastings algorithm was used due to the high
number of degrees of freedom (42). Instead of varying all charges
simultaneously, as is usual for the algorithm, only one random charge
was varied at a time, which sped up convergence.

\subsubsection{Modeling detector observables\label{subsec:Modeling-detector-observables}}

To model the electric field, the simulation results were compared
to the field implied by $^{\mathrm{83m}}\mathrm{Kr}$ calibration
data to ensure its accuracy, since the real value of the electric
fields inside of the detector was unknown. An initial (but unsuccessful)
attempt to solve this problem involved drifting electron-like particles,
similar to the approach outlined in Section~\ref{subsec:Modeling-LUX-run3}.
However, $\sim100,000$ particles were needed (corresponding to particles
spaced $\unit[1]{cm^{3}}$ apart) to capture enough details of the
active volume, which was too computationally slow. This comparison
had to be done many times during the Metropolis-Hastings algorithm,
rendering the traditional method of drifting electrons through the
detector computationally too intensive, so an alternative method comparing
event densities was developed. The comparison was made first throughout
the entire detector active volume and at the final stages of refinement
only by comparing the edges of the distributions.

Instead, $^{\mathrm{83m}}\mathrm{Kr}$ events as seen in S2 space\footnote{The S2 coordinates ($x_{S2},\,y_{S2},\,\mathrm{drift\,time}$) represent
the event position as seen by the top PMT array with the event depth
given by the electron drift time. Refer to Section~\ref{subsec:S2-coordinate-system}
for more details.} were reverse-propagated through the modeled drift field to find the
initial homogeneous distribution. In practice, this was accomplished
by segmenting the $^{\mathrm{83m}}\mathrm{Kr}$ data into tetrahedra
within the active volume using Delaunay triangulation~\cite{delaunay1934sphere}
and recording the number of events enclosed by each segment. The vertices
of those tetrahedra were drifted \textquotedblleft back in time\textquotedblright{}
through the electric field; each vertex started at the liquid surface
and was drifted through the active volume, transforming S2 space to
real space. The new volume of each tetrahedron after drifting, $V_{tet}$,
was calculated and the number of $^{\mathrm{83m}}\mathrm{Kr}$ events,
$n_{S2}$, contained in each tetrahedron recorded. Since in real space
$^{\mathrm{83m}}\mathrm{Kr}$ events were uniformly distributed throughout
the detector, the number of events in each tetrahedron was proportional
to its volume 
\begin{eqnarray}
n_{real} & = & N_{Kr}*\frac{V_{tet}}{V_{TPC}}\label{eq:need-number-1}
\end{eqnarray}
where $N_{Kr}$ is the total number of $^{\mathrm{83m}}\mathrm{Kr}$
events in the detector and $V_{TPC}$ is the active volume of the
detector. If the simulated electric field captured the field in the
detector well, then $n_{real}$ and $n_{S2}$ had an equal number
of events. A $\chi^{2}$ was then calculated to compare data to simulation,
and the charge guess refined. This approach was iterated until convergence
of results was seen. The density comparison between $^{\mathrm{83m}}\mathrm{Kr}$
data and simulation is illustrated in the right panel of Figure~\ref{fig:edge-density}.

\subsubsection{Comparing edges}

\begin{figure}[H]
\begin{centering}
\includegraphics[scale=0.8]{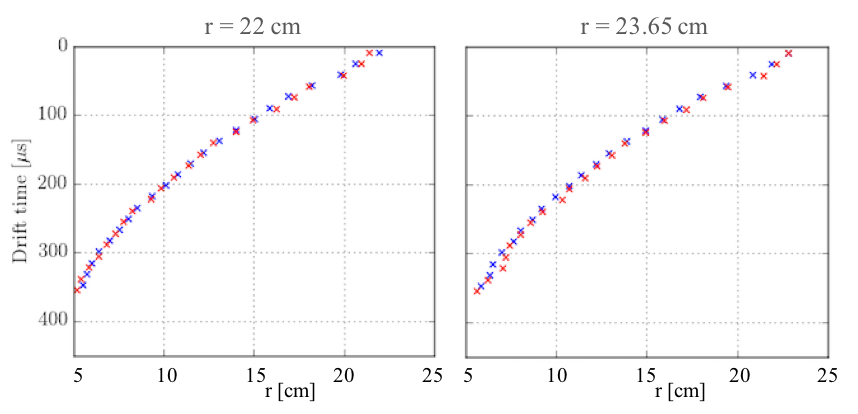}
\par\end{centering}
\caption[Comparison of wall position at different radii]{Comparison of wall position as found using $^{\mathrm{83m}}\mathrm{Kr}$
data (blue) and COMSOL model (red) at two different radii. Radius
$r=\unit[23.65]{cm}$ corresponds to the detector PTFE wall. Data
shown is from 2015-07-22 in $-15^{\circ}<\theta\leq15^{\circ}$ azimuthal
section of the detector.\label{fig:edges}}
\end{figure}

\begin{figure}[t]
\begin{centering}
\includegraphics[scale=0.55]{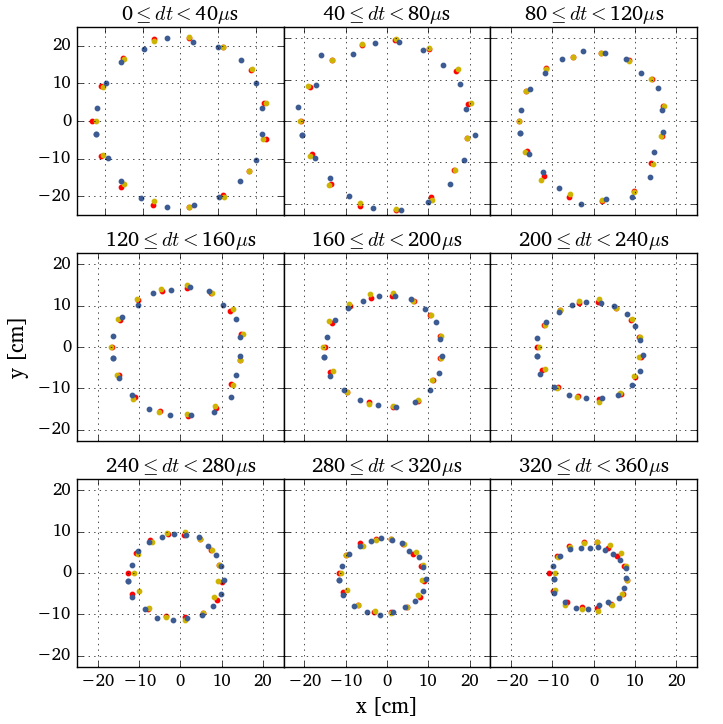}
\par\end{centering}
\caption[Comparison of wall position as found from various sources]{Wall position as found using three different methods for the period
of 2014-12-07 to 2015-03-19. Wall positions as found from the wall
model using $^{210}$Po (blue)~\cite{boulton2018}, 95\% of $^{\mathrm{83m}}\mathrm{Kr}$
from 2015-01-22 (red) and 95\% edge of the field map for 2015-01-22
(yellow); all three models agree.\label{fig:Wall-position}}
\end{figure}

\begin{figure}
\centering{}\includegraphics[scale=0.6]{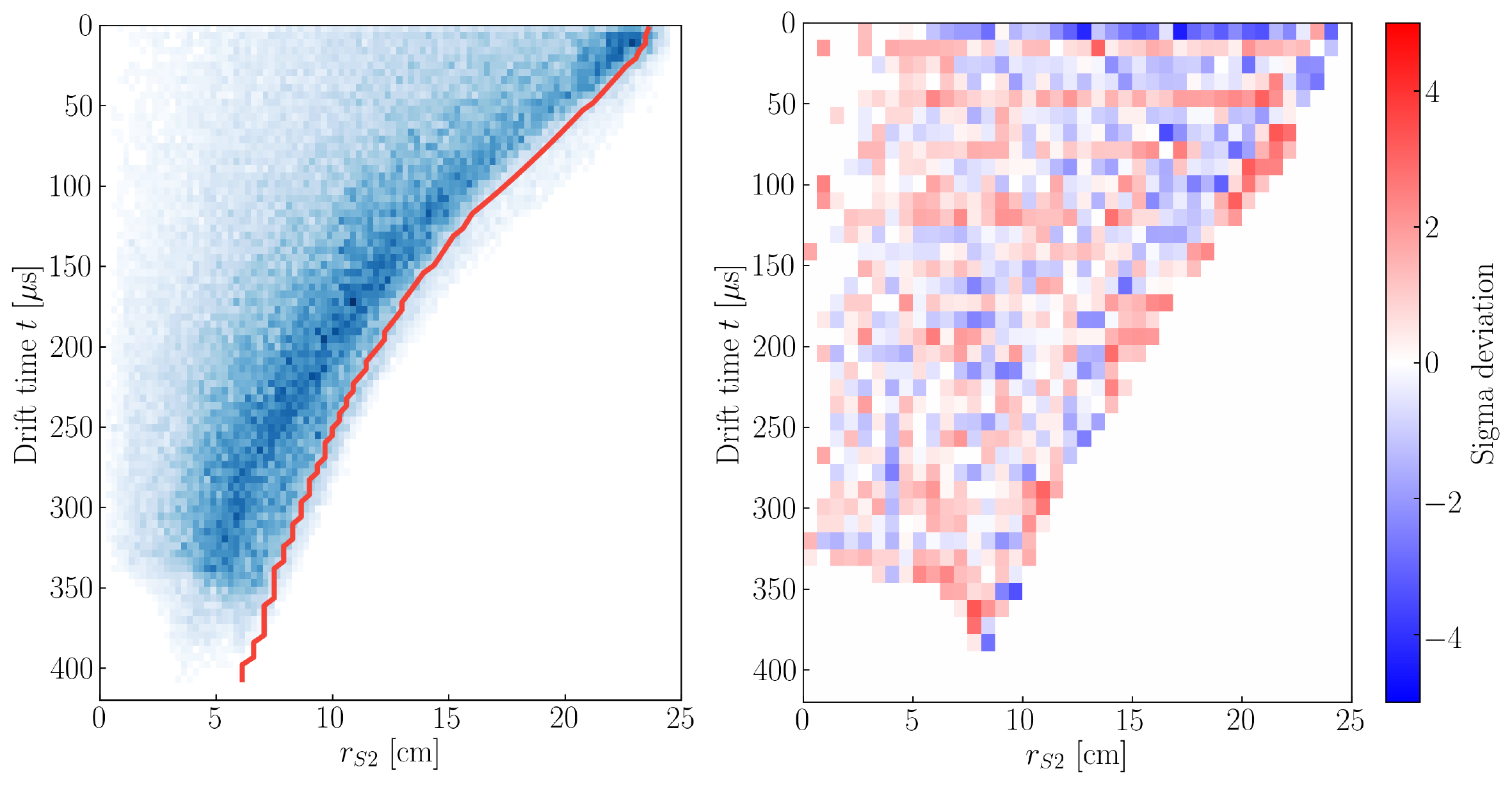}\caption[Illustration of wall agreement resulting in correct density distribution]{\textbf{Left:} Solid red contour at $r$ = 22 cm illustrates the
modeled edge of the $^{\mathrm{83m}}\mathrm{Kr}$ distribution (blue)
from 2016-02-22. \textbf{Right:} variation between normalized histograms
of $^{\mathrm{83m}}\mathrm{Kr}$ data and simulation from 2016-10-06
weighted by the error in each bin (using an azimuthal slice between
$120^{\circ}\leq\theta<180^{\circ}$).\label{fig:edge-density}}
\end{figure}

To complete the fine-tuning of the model, the edges of the $^{\mathrm{83m}}\mathrm{Kr}$
distribution were compared with data, in a method independent of the
$^{\mathrm{83m}}\mathrm{Kr}$ homogeneity in the detector. The detector
was split into 12 angular sections corresponding to the 12 PTFE panels.
Then negative charges were drifted to the liquid surface starting
along the center of each panel face. An offset radius (set at $r$
= 22 cm) was used due to a higher fidelity of field maps slightly
inward from the PTFE panels. The resulting simulated edge at the detector
wall ($r$ = 23.7 cm) agrees with the edge found both from the $^{\mathrm{83m}}\mathrm{Kr}$
data as shown in Figure~\ref{fig:edges} and from the location of
the wall as determined from $^{210}$Po background~\cite{Akerib:2018bgRun4}
(shown in Figure~\ref{fig:Wall-position}). It is worth noting that
even though only the detector wall was fitted, as shown in the left
panel of Figure~\ref{fig:edge-density}, the result correctly described
the density distribution inside the entire detector.

\subsubsection{Validation}

Before developing the simulation with 42 charge segments, a test model
was created using 3 angular and 5 $z$ slices resulting in 15 charge
segments. As an initial test, a dataset was simulated with known charge
distribution at the 15 sites. Then, starting with random charge distributions
on the 15 tiles, the simulation successfully converged to the expected
values. 

The model with 42 charge segments was first tested by making sure
the charge distribution successfully converged to a test dataset that
included no charges on the PTFE walls. However, in order to converge
to a model with a known charge distribution, the model with 42 free
parameters required much more computational power than the 15 parameter
model. As a result, the 42-tile model never reached a high enough
confidence in the algorithm convergence alone (as it would have taken
perhaps years to converge). In order to deliver results to the LUX
collaboration in a timely manner, the state of the algorithm was visually
inspected throughout the fitting procedure to make sure that the best-fit
charge distribution matched $^{\mathrm{83m}}\mathrm{Kr}$ data. In
the early stages of fitting, this visual inspection was used to slightly
modify charges in the neighboring tiles to avoid divergence of values
where those might become unphysical. 

\begin{figure}
\centering{}\includegraphics[scale=0.53]{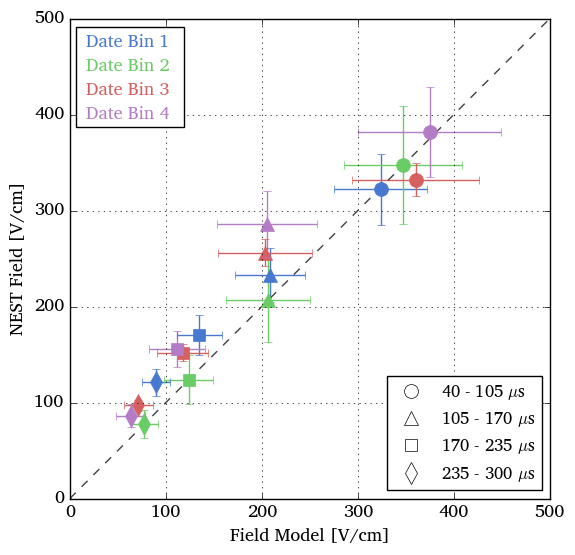}\caption[Field values from this work vs. NEST predictions]{Comparison of the electric field as modeled using the COMSOL modeling
method (\textquotedblleft Field Model\textquotedblright ) to the best
fit predicted by NEST (\textquotedblleft NEST Field\textquotedblright )
for the 16 detector models (four drift time spatial regions and four
time periods in date) used in the WS2014-16 analysis. Note that error
bars on field map data points indicate the spread of the field within
a bin. \label{fig:maps_vs_NEST}}
\end{figure}

As the exact amount of charge on the PTFE panels is unknown, it is
hard to characterize the accuracy of the field maps quantitatively.
The volume-averaged fields obtained in COMSOL models can be compared
to the NEST model fits that assume a uniform field within a given
drift time bin. Figure~\ref{fig:maps_vs_NEST} shows the comparison
of the field magnitudes obtained by NEST to the corresponding mean
field values obtained from the COMSOL studies for all 16 detector
models (four spatial regions and four time periods). The field estimates
obtained from the two methods show good agreement. This point deserves
emphasis because these two techniques for estimating electric field
magnitude are completely independent: the electrostatic field model
is based on the observed electron drift paths alone, while the NEST
fits are based on the S1 and S2 amplitudes and the known microphysics
of charge and light signal dependence on drift field. This agreement
provides confidence in the S1 and S2 signal reconstruction techniques
used in the LUX WIMP search analysis.

\subsubsection{Hand-off }

\begin{figure}[H]
\centering{}\includegraphics[scale=0.65]{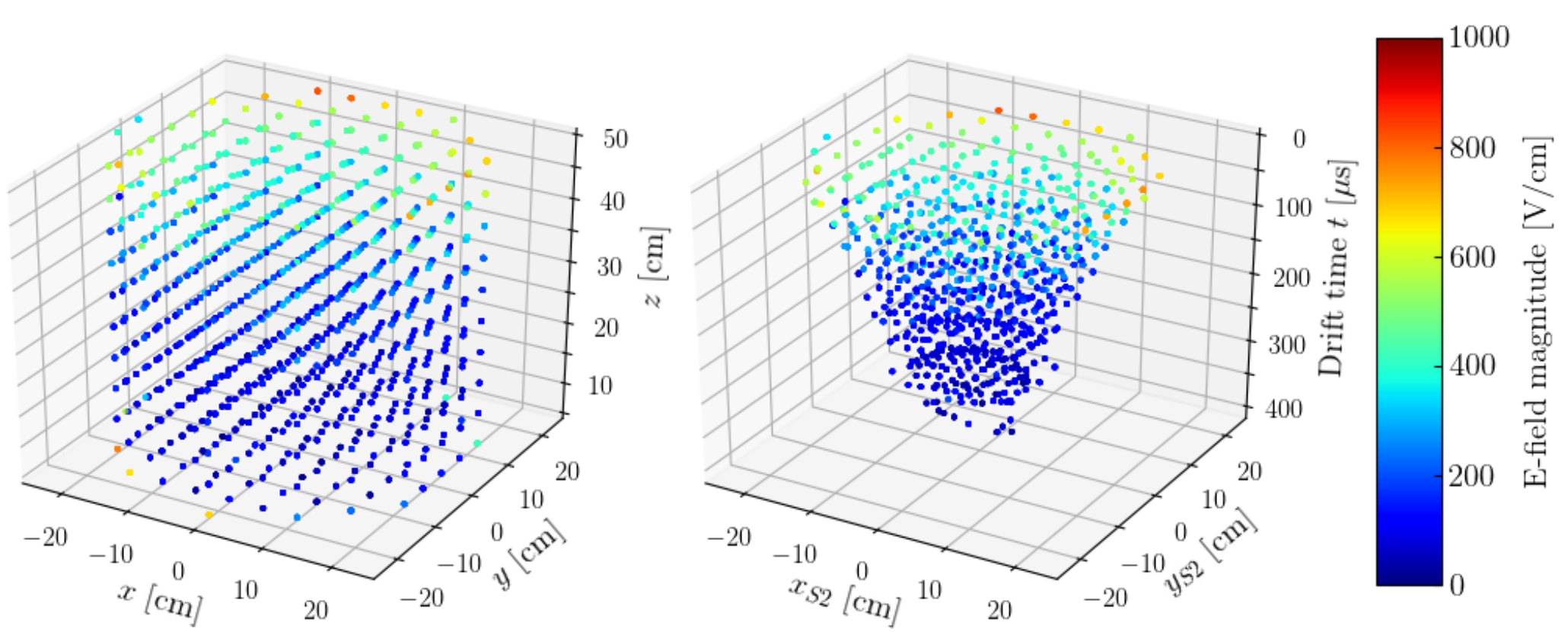}\caption[The transformation from the reconstructed S2 space to real space]{The transformation from the reconstructed S2 space (right) to real
space (left). The color scale shows the electric field strength at
each point using $^{\mathrm{83m}}\mathrm{Kr}$ data from 2015-08-26.\label{kr_over_time}}
\end{figure}

Once the final charge distribution was found, the information included
in the resulting field maps was provided on a $\unit[1]{cm^{3}}$
grid that included the detector real space coordinates $\left(x,y,z\right)$,
the electric field in each location, and the coordinates as seen by
the detector $\left(x_{S2},y_{S2},t\right)$. The transformation between
real space and the reconstructed S2 space can be seen in the two plots
in Figure~\ref{kr_over_time}. The field maps were then used in LUXS\textsc{im}
simulations, position reconstruction, and limit-setting code for LUX
searches for spin-independent and spin-dependent dark matter interactions~\cite{Akerib:2016vxi,Akerib:2017kat}.

\subsection{Time-dependence of electric fields during WS2014-16}

\subsubsection{Charge evolution\label{subsec:Charge-evolution}}

\begin{figure}
\begin{centering}
\includegraphics[scale=0.35]{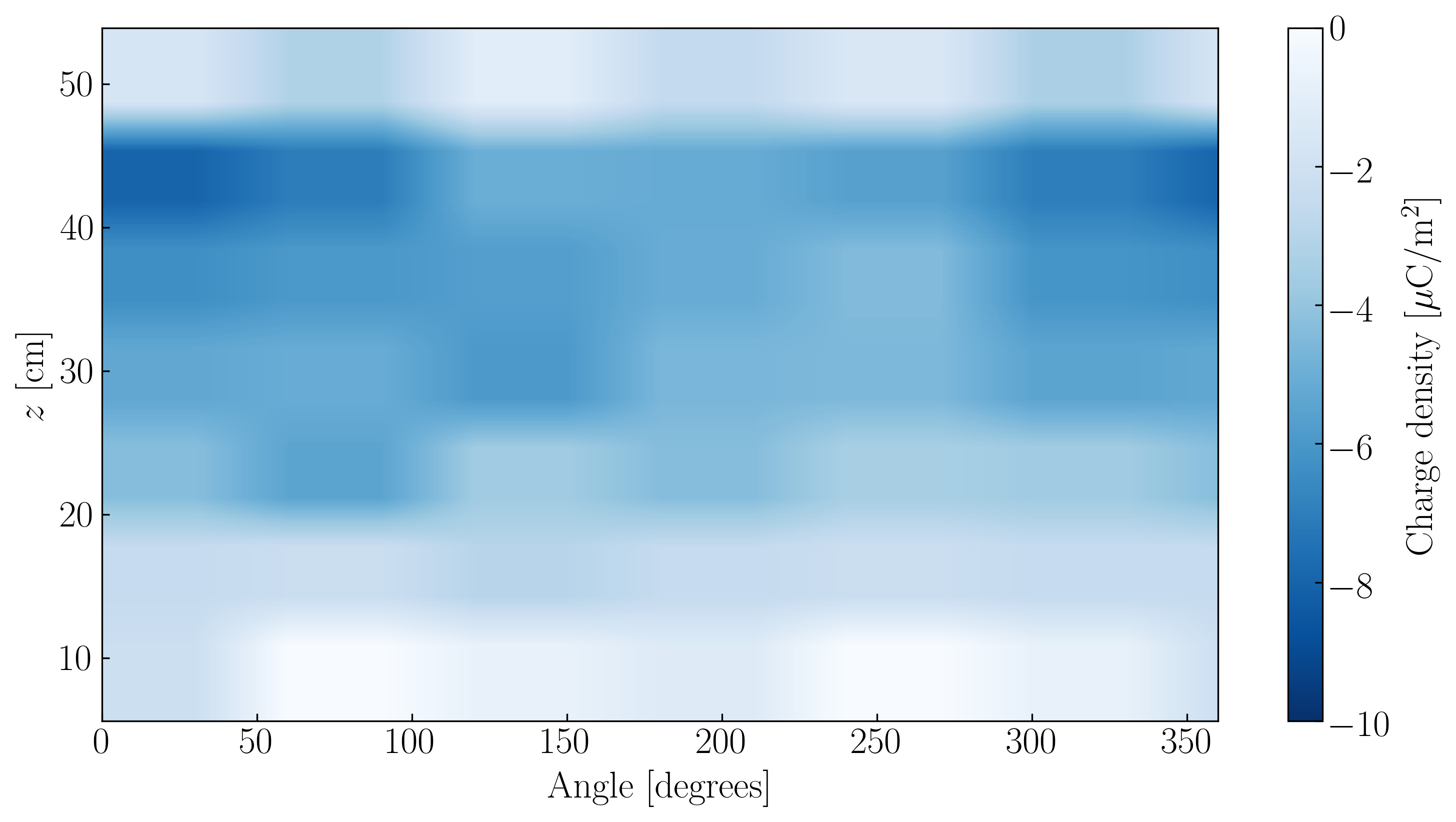}\includegraphics[scale=0.35]{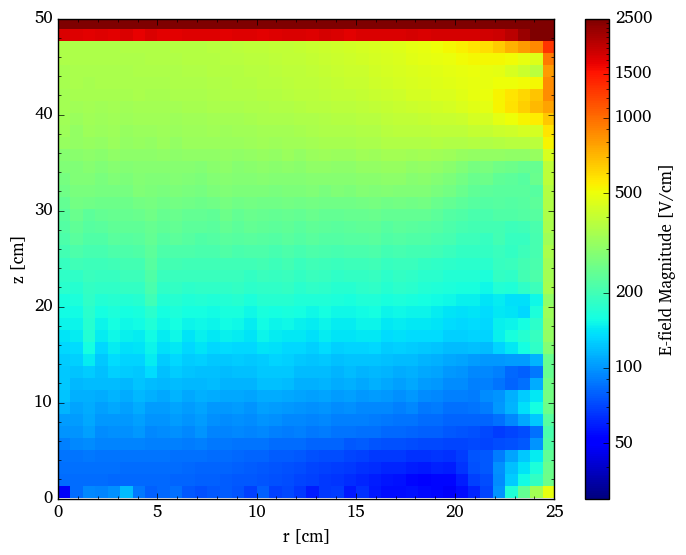}
\par\end{centering}
\caption[Charge density distribution in the PTFE walls and the corresponding
field]{\textbf{Left:} Charge density distribution in the detector\textquoteright s
PTFE panels obtained from fitting $^{\mathrm{83m}}\mathrm{Kr}$ data
from 2014-10-06. The 42 segments correspond to the smoothed tile charges
described in Section~\ref{subsec:Principle-of-superposition}. \textbf{Right:}
Modeled electric field for the same dataset. The ``bumps'' at high
$r$ are due to each panel having a slightly different charge and
the smoothing algorithm not being quite perfect. \label{fig:Charge-density-distribution}}
\end{figure}

\begin{figure}
\begin{centering}
\includegraphics[scale=0.6]{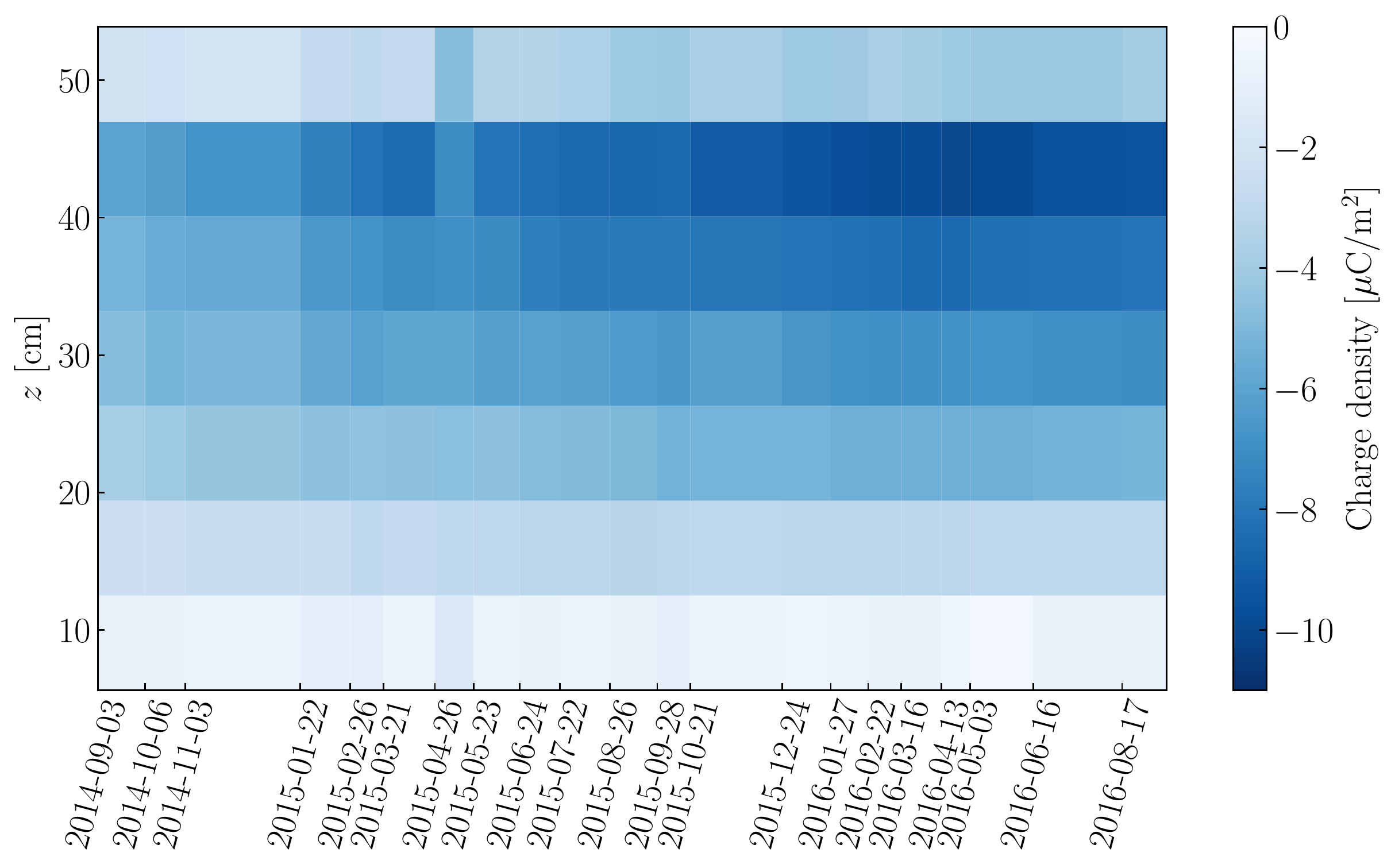}
\par\end{centering}
\caption[Charge density in each $z$ segment averaged over all azimuth in WS2014-16]{Charge density in each $z$ segment averaged over all azimuth throughout
WS2014-16. Cathode is at $z=0$ and gate at $z=\unit[48.3]{cm}$.
\label{fig:Charge-density-avg}}

\end{figure}

\begin{figure}
\begin{centering}
\includegraphics[scale=0.48]{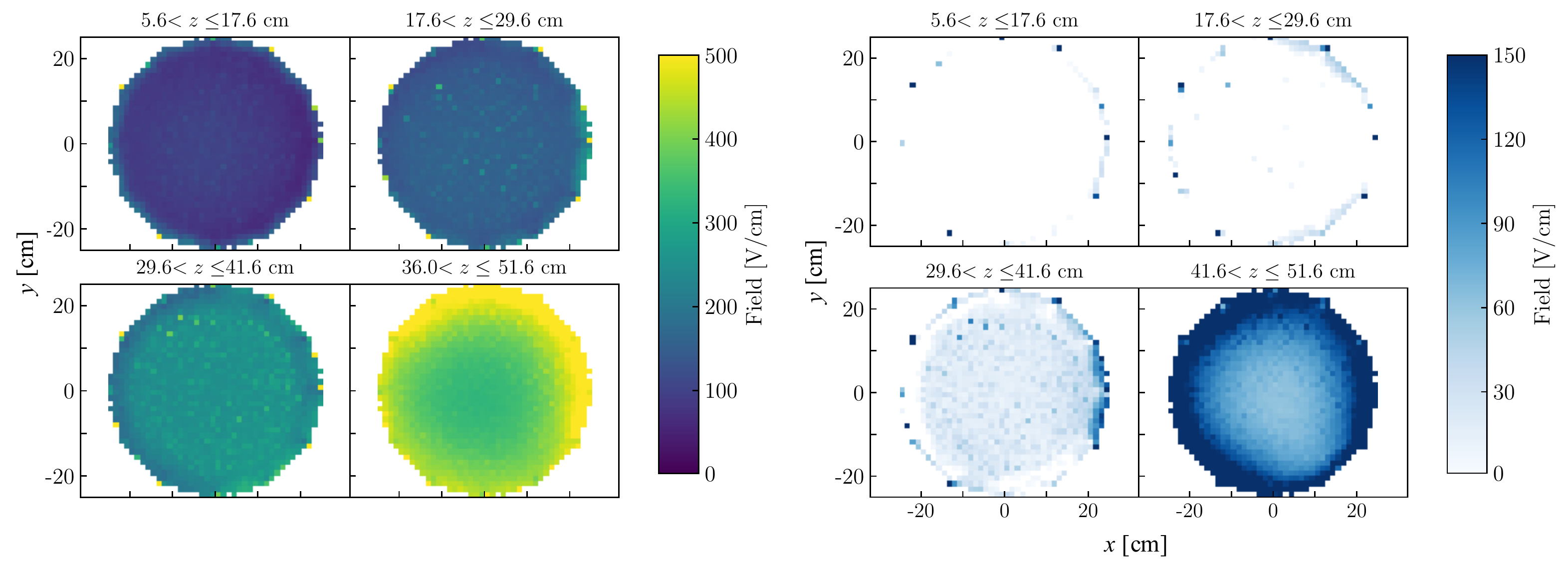}
\par\end{centering}
\caption[The electric field in different $z$ segments of the detector and
its time variation]{\textbf{Left:} The magnitude of electric field in different $z$
segments of the detector as modeled for 2014-09-03 dataset illustrating
the azimuthal non-uniformity. \textbf{Right:} Difference between the
electric field at the end of WS2014-16 (field map from 2016-05-03)
and the beginning (field map from 2014-09-03). The change in field
is not uniform throughout WS2014-16.\label{fig:e-field-var}}
\end{figure}

The method described here resulted in one field map for each month
of WS2014-16, corresponding to 21 field maps in total. As discussed
earlier, each field map was produced by adjusting the negative charge
distribution in the PTFE panels and fitting the modeled electric field
to the detector observables. One of the resulting charge distributions
in the PTFE panels and the corresponding electric field is shown in
Figure~\ref{fig:Charge-density-distribution}. The evolution of charge
throughout WS2014-16 is demonstrated in Figure~\ref{fig:Charge-density-avg}.
The varying charge results in a time-dependent electric field, also
changing in azimuth and $z$ as illustrated in Figure~\ref{fig:e-field-var}.
The variability in $z$ can be explained by the difference of electron
and hole mobility in the PTFE surfaces, while the non-uniformity in
azimuth arises from the individual 12 PTFE panels exposure to varying
intensity of VUV during the grid conditioning campaign.

\begin{figure}
\centering{}\includegraphics[scale=0.8]{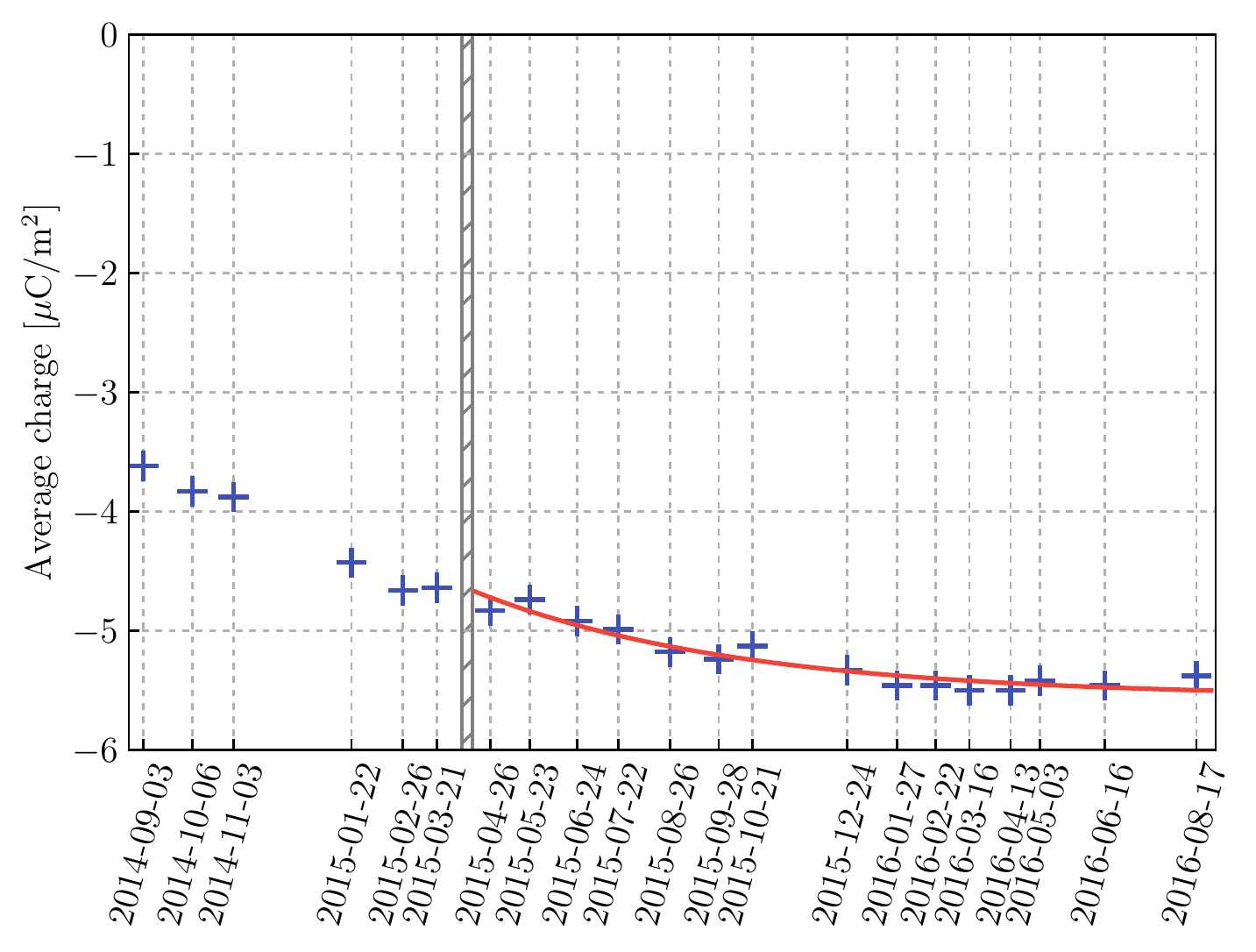}\caption[Charge density averaged over the entire PTFE panel surfaces in WS2014-16]{Increasingly negative charge density obtained from the field models
averaged over the entire PTFE panel surfaces for each month during
WS2014-16. The data points correspond to the modeled average charge
densities in the PTFE panels obtained from a small subset of the regular
$^{\mathrm{83m}}\mathrm{Kr}$ datasets. This data can be fitted with
an exponential function $\sigma=3.1\exp\left(-t/181\right)-5.6$ where
$t$ is in units of \textquotedblleft days since 2014-09-03.\textquotedblright{}
The gray dashed line indicates a week in April 2015 when the cathode
voltage was biased down to 0 V. \label{fig:Average-charge}}
\end{figure}

The most negative charge was found to be in the top $\nicefrac{1}{3}$
of the detector throughout all of WS2014-16, possibly due to hole
states in the PTFE being pulled toward the cathode region, depleting
the upper region first. Electron states may be moving upwards toward
the gate, but at a slower rate. The increasingly negative charge throughout
WS2014-16 averaged over the entire PTFE area is plotted in Figure~\ref{fig:Average-charge}.
The cathode voltage was biased down to 0~V from April 7 to April
15, 2015, to investigate its effects on the charge in the PTFE panels;
therefore, data from this period was not used in the analysis. Beginning
after the cathode re-bias to \textminus 8.5~kV the charge density
was fit with an exponential function corresponding to Equation~\ref{eq:sigma}
:
\[
\sigma=\sigma_{0}\exp\left(-t/\tau\right)+\sigma_{\infty}
\]
where $\sigma$ is the average surface charge density, $\sigma_{0}$
is the initial charge density, and $\sigma_{\infty}$ is the value
of the charge $\lim_{t\rightarrow\infty}\sigma$. $\tau$ is the charge
transit time. The best-fit values found were $\sigma_{0}=\unit[3.1]{\mu C/m^{2}}$,
$\tau=\unit[181]{days}=\unit[1.6\times10^{7}]{s}$, and $\sigma_{\infty}=\unit[\unit[-5.6]{\mu C/m^{2}}]{}$. 

There is a lack of research of PTFE properties at LXe temperatures,
so it is difficult to compare this number to the literature on PTFE
charging. This is made even more difficult because of the experimental
conditions used, such as surface cleanliness\footnote{A tiny amount of impurities likely affect the behavior of charges.
It is expected that a higher level of cleanliness results in slower
transit times.} and the electric field applied.

\begin{figure}[p]
\begin{centering}
\includegraphics[scale=0.39]{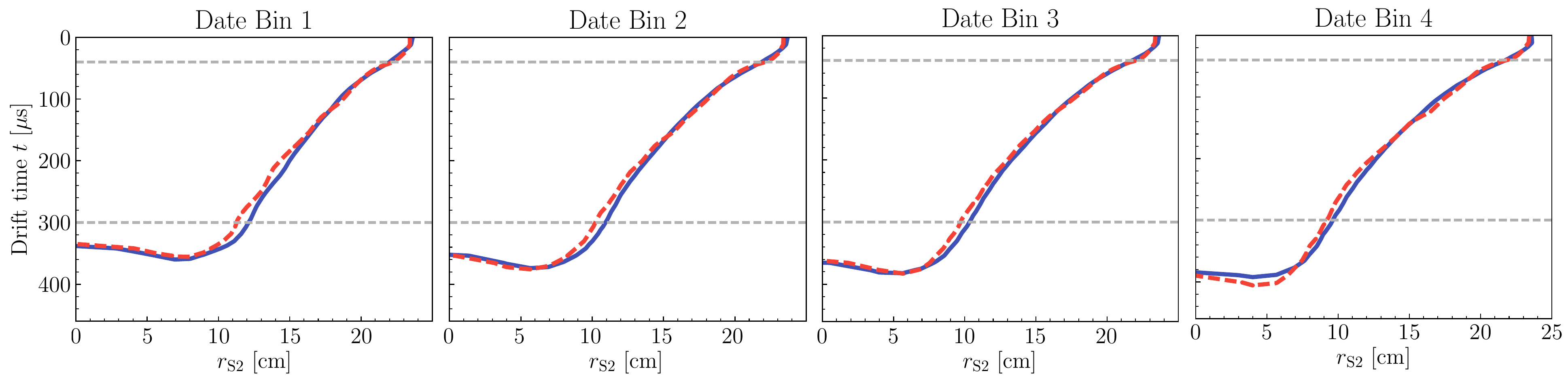}
\par\end{centering}
\caption[Comparison of the detector edge as obtained from data and simulation]{A comparison of the measured position of the detector wall and the
cathode from $^{\mathrm{83m}}\mathrm{Kr}$ (solid blue) to that predicted
by the best-fit electric field model (dashed red). Horizontal gray
dashed lines, at 40 and 300 $\mu$s, indicate the drift time extent
of the fiducial volume used in WS2014-16.\label{fig:ws2014-fields}}

\vspace{2cm}
\centering{}\includegraphics[scale=0.48]{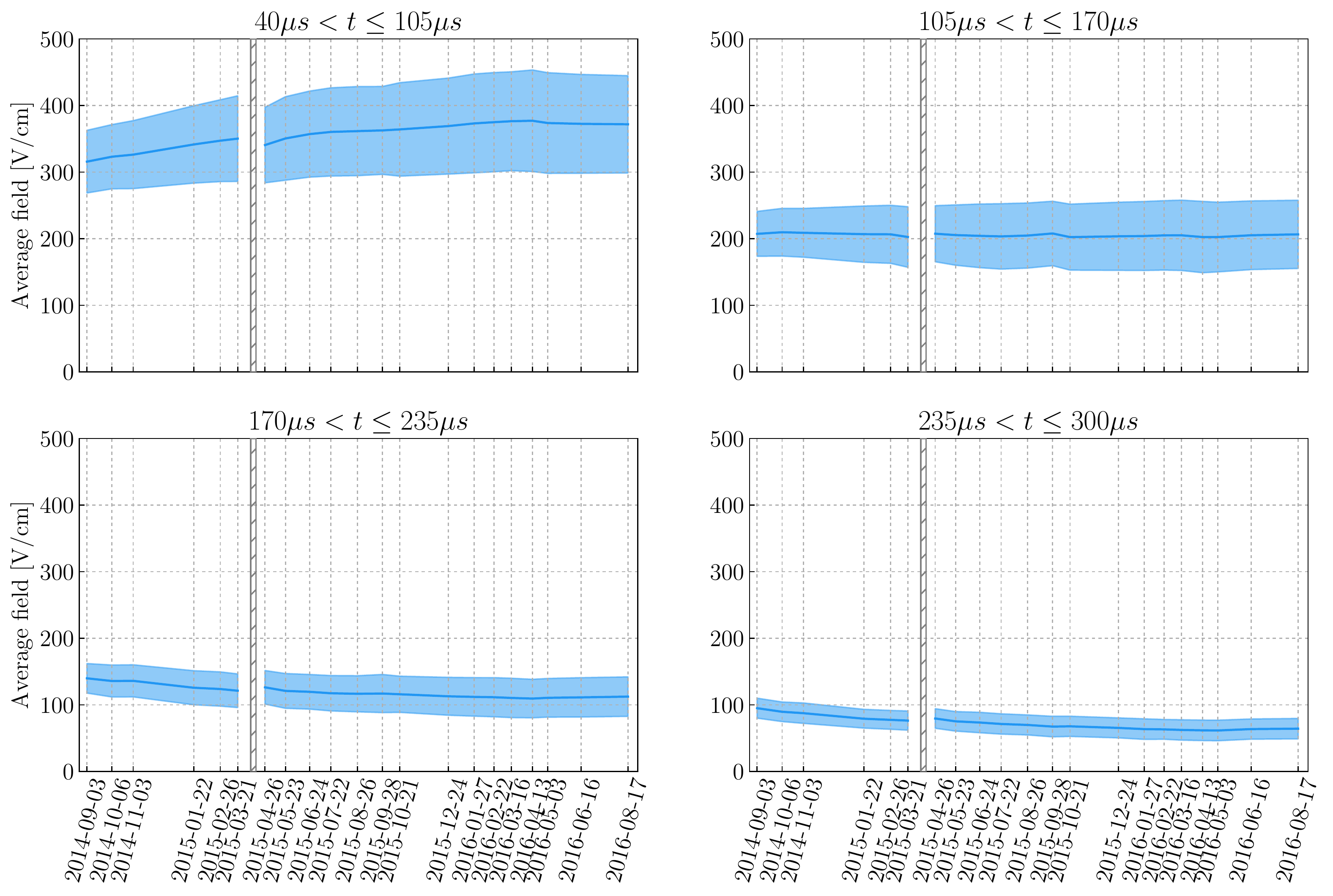}\caption[Average fields during WS2014-16 in the four drift time bins used in
the analysis]{Average fields during WS2014-16 in the four drift time bins used
in the analysis. Bands indicate the standard deviation of the field
in the given drift time bin. A radial cut at $r=\unit[20]{cm}$ and
a drift time cut at $\unit[40]{\mu s}<dt\leq\unit[300]{\mu s}$ were
applied. Gray dashed lines indicate a week in April 2015 when the
cathode voltage was biased down to 0 V.\label{field over time}}
\end{figure}

\begin{table}[p]
\begin{centering}
\begin{tabular}{l||>{\raggedleft}p{2cm}>{\raggedleft}p{3.2cm}>{\raggedleft}p{2cm}>{\raggedleft}p{2cm}}
\hline 
Date & Mean field {[}V/cm{]} & Standard deviation {[}V/cm{]} & Min field {[}V/cm{]} & Max field {[}V/cm{]}\tabularnewline
\hline 
\hline 
2013-05-10 & 178 & 14 & 125 & 207\tabularnewline
\hline 
\hline 
2014-09-03 & 194 & 89 & 45 & 487\tabularnewline
2014-10-06 & 195 & 95 & 42 & 494\tabularnewline
2014-11-03 & 196 & 96 & 38 & 503\tabularnewline
2015-01-22 & 196 & 107 & 32 & 553\tabularnewline
2015-02-26 & 197 & 111 & 43 & 560\tabularnewline
2015-03-21 & 196 & 113 & 28 & 568\tabularnewline
2015-04-26 & 197 & 106 & 44 & 576\tabularnewline
2015-05-23 & 197 & 113 & 30 & 576\tabularnewline
2015-06-24 & 199 & 117 & 21 & 584\tabularnewline
2015-07-22 & 199 & 120 & 26 & 596\tabularnewline
2015-08-26 & 199 & 121 & 29 & 606\tabularnewline
2015-09-28 & 201 & 121 & 27 & 608\tabularnewline
2015-10-21 & 199 & 123 & 24 & 607\tabularnewline
2015-12-24 & 200 & 126 & 24 & 621\tabularnewline
2016-01-27 & 201 & 129 & 19 & 635\tabularnewline
2016-02-22 & 202 & 130 & 24 & 636\tabularnewline
2016-03-16 & 202 & 131 & 20 & 641\tabularnewline
2016-04-13 & 201 & 132 & 18 & 662\tabularnewline
2016-05-03 & 200 & 130 & 18 & 637\tabularnewline
2016-06-16 & 201 & 129 & 25 & 632\tabularnewline
2016-08-17 & 201 & 128 & 23 & 631\tabularnewline
\hline 
\end{tabular}
\par\end{centering}
\caption[Field values throughout WS2014-16 in the fiducial volume]{Field values throughout WS2014-16 in the fiducial volume (radial
cut: $r\le\unit[20]{cm}$, drift time cut: $40\,\mu s<z<300\,\mu s$)
calculated on a $\unit[1]{cm^{3}}$ grid found from the model. Data
from 2013-05-10 is from WS2013 with radial cut $r\le\unit[20]{cm}$
and drift time cut $38\,\mu s<z<305\,\mu s$. \label{tab:Field-values}}
\end{table}

\subsubsection{Electric fields in WS2014-16 analysis}

As discussed in Section~\ref{subsec:WS2014-16}, for simplicity of
the analysis in WS2014-16 the data were divided into 16 bins, within
each of which the change in the wall position was not significant.
There were four bins in time, bounded by the following dates: September
11, 2014; January 1, 2015; April 1, 2015; October 1, 2015; May 2,
2016. Each of those bins was further split into four drift time bins
with boundaries of 40, 105, 170, 235, and 300 $\mu$s. These time
bins used in the PLR analysis are coarser than the monthly fields
maps. However, all field maps were used to inform a decision on the
boundaries of these four time bins. Only one field map was chosen
for each date bin; hence only four field maps were used in the WS2014-16
analysis. Each map was chosen to be close to the average of the electric
field values in the segment. The agreement between each field map
and a $^{\mathrm{83m}}\mathrm{Kr}$ distribution for each date bin
is shown in Figure~\ref{fig:ws2014-fields}.

Figure~\ref{field over time} shows the value of the electric field
throughout the WS2014-16 analysis; those values are also recorded
in Table~\ref{tab:Field-values}. The table shows a divergence of
the field extrema as the electric field magnitude varied from $\unit[\sim20-50]{V/cm}$
near the bottom of the detector to $\unit[\sim500-650]{V/cm}$ near
the top. The mean value of the field of $\unit[\sim200]{V/cm}$ remained
mostly constant throughout the exposure. The error on the values of
electric fields obtained in this work is likely $\sim\pm10\%$, the
field variation needed to agree with simulation-data comparisons done
at a later date based on charge and light signals. 

\section{Modeling small scale features\label{subsec:scalloping}}

\begin{figure}
\begin{centering}
\includegraphics[scale=0.46]{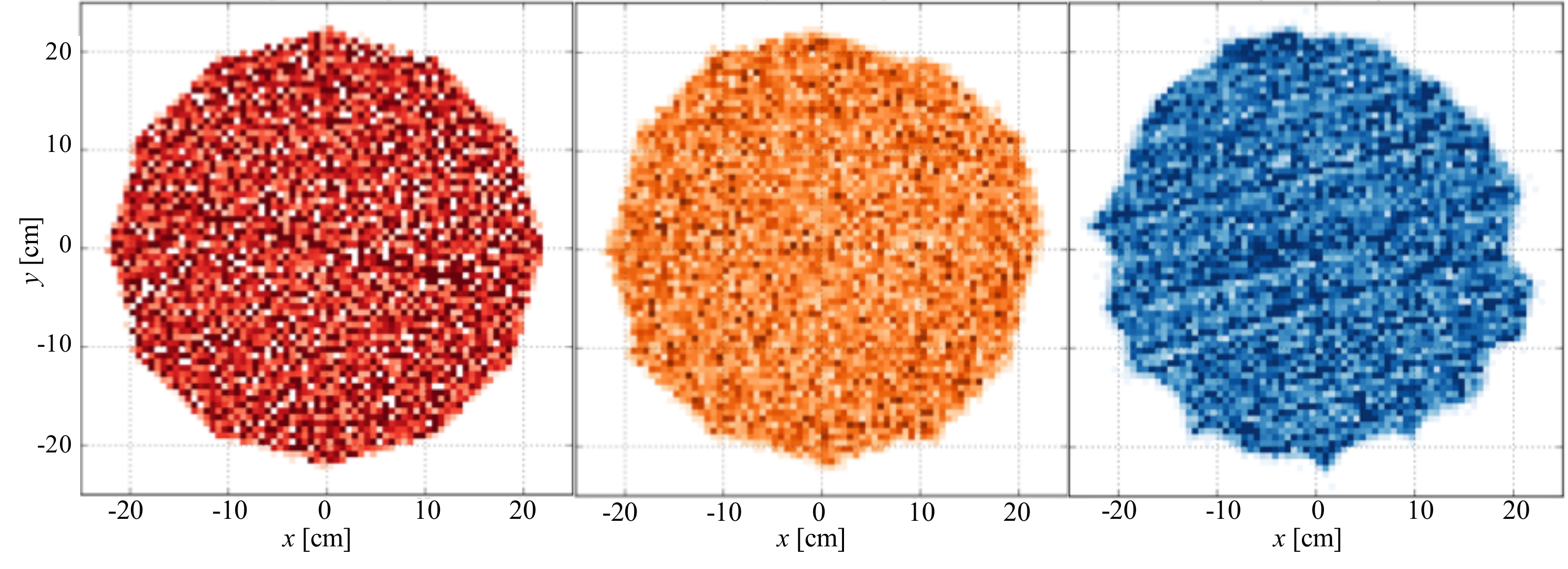}
\par\end{centering}
\caption[Scalloping in WS2013]{\textbf{Left:} Original COMSOL model with no charge showing a clear
dodecagonal cross-section. \textbf{Center:} Data from the detector
with visible scallops (most notably in the upper right quadrant and
near the bottom) using $^{\mathrm{83m}}\mathrm{Kr}$ from 2013-05-10
taken during WS2013. Also visible is the orientation of gate wires.
\textbf{Right:} COMSOL fit with charge included in the PTFE walls.
The poor fit is likely due to only having six angular slices since
the scalloping appears to be correlated with the individual detector
PTFE panels. All cross-sections are a slice at a drift time of $255\,\mathrm{\mu s}<t\leq290\,\mathrm{\mu s}$.\label{fig:scalloping}}

\vspace{2cm}
\begin{centering}
\includegraphics[scale=0.44]{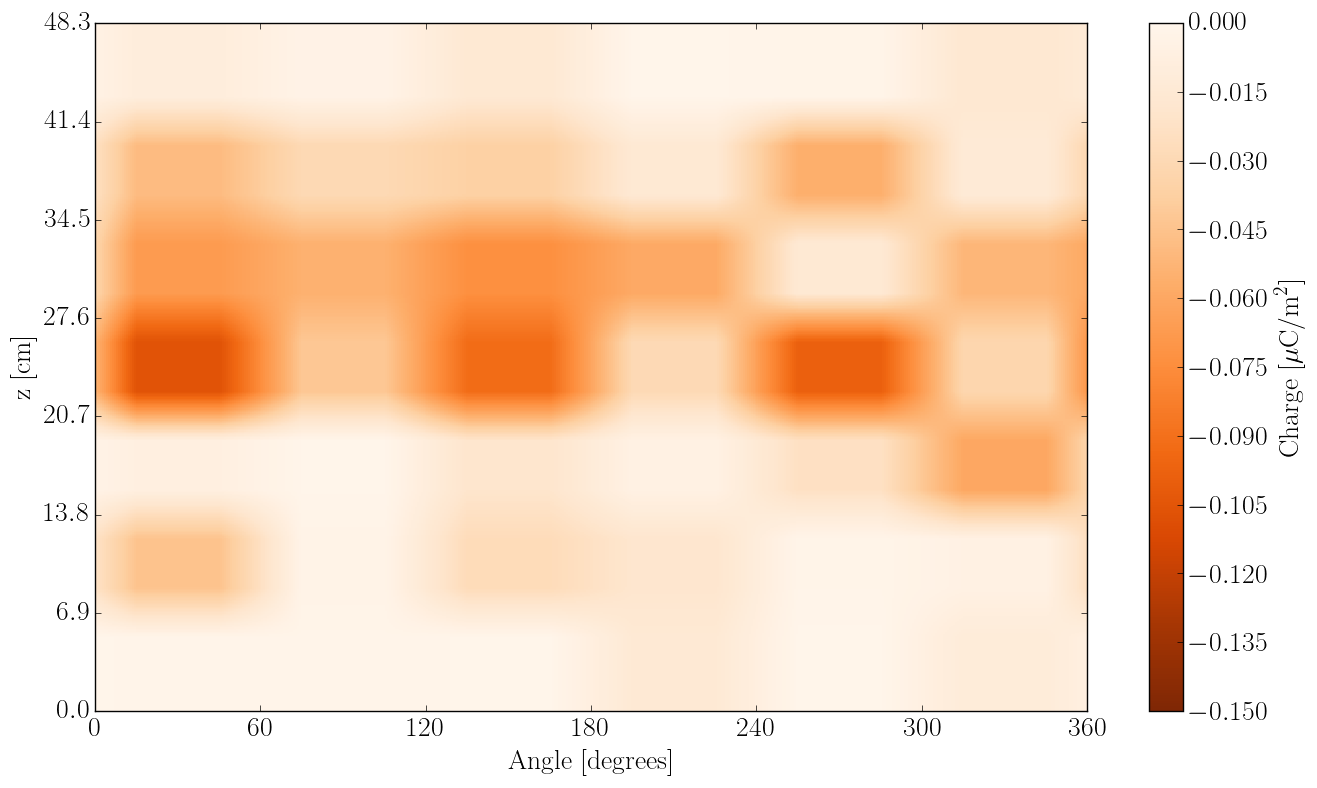}
\par\end{centering}
\caption[Charge distribution on the PTFE wall as modeled for 2013-05-10]{Charge distribution on the PTFE wall on 2013-05-10. The average charge
on the wall is $\unit[-0.03]{\mu C/m^{2}}$, consistent with triboelectric
charge densities. \label{fig:run3_charge}}
\end{figure}

Despite the mostly uniform electric field in WS2013, the $\left(x,y\right)$
locations of S2 pulses did not have a dodecagonal outline as one might
expect. Instead, small irregularities, referred to as ``scalloping,''
were observed along the edges of the detector that were not observed
in the COMSOL Multiphysics model without charge in the PTFE surfaces,
as shown in Figure~\ref{fig:scalloping}. 

Therefore, the same approach described in Section~\ref{subsec:Modeling-Changing-Fields}
to model electric fields in WS2014-16 was used to model this scalloping
effect observed in WS2013. Since the tiling was optimized for WS2014-16,
the model was unable to fully reproduce all the fine features, likely
due to the bundling of the two adjacent PTFE panels together. The
average fitted charge found in the PTFE panels was $\unit[-0.03]{\mu C/m^{2}}$
corresponding to $\unit[\sim2\times10^{5}]{e^{-}/mm^{2}}$, comparable
to triboelectric charge densities as calculated in Section~\ref{subsec:Static-charge-calculation}.
In comparison, the charge densities simulated in WS2014-16 reached
$\unit[-5.5]{\mu C/m^{2}}$. This illustrates the sensitivity of the
field modeling method to capture even very small charge densities.

However, the scalloping feature in WS2013 appears to be more correlated
with individual panels. Therefore the result could be improved by
creating 12 angular section in the model. It is worth noting that
increasing the number of angular sections would require a smaller
amount of vertical divisions in the model since using, for example,
12 angular sections and 7 vertical divisions creates 84 tiles, which
would more than double the already considerable computational complexity
required by the Metropolis-Hastings algorithm to achieve convergence. 

\subsection{Wire grid spacing}

\begin{figure}
\begin{centering}
\includegraphics[scale=0.47]{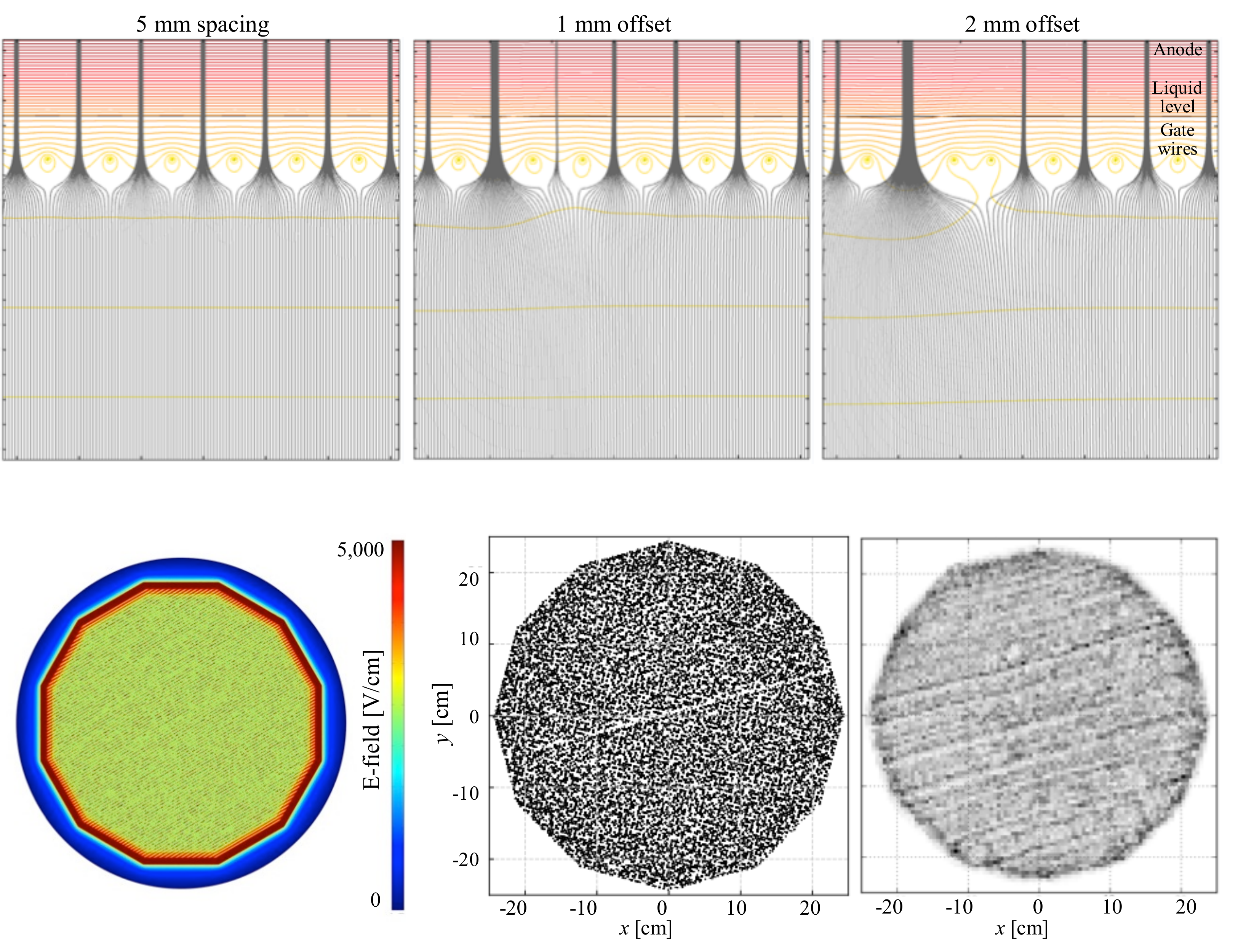}
\par\end{centering}
\caption[Gate wire simulations]{\textbf{Top:} Simulation of wire offset in the gate grid in WS2013
showing equipotentials at 50 V intervals (colored) and field lines
(gray). Only field lines originating in the fiducial volume are shown
indicating the path of electrons created in the detector. The left
diagram shows the ideally parallel wires while the other two diagrams
illustrate what happens when one wire is offset by 1 or 2 mm. A lateral
offset can significantly vary the number of events ``funneled''
between a wire pair, up to completely stealing events from one gap
to another. Figures modified from~\cite{Scott_wires}. \textbf{Bottom
left:} Electric field simulation from the COMSOL model as seen 1 mm
above gate grid. The field ``funnels'' the field lines between the
wires thus making the gate wires visible using the Mercury position
reconstruction algorithm~\cite{Akerib:2017riv}. Dark red is the
boundary of the grid frame, and the lighter red is where the grid
wires pass through the PTFE wall. \textbf{Bottom center:} Simulation
of particle distribution if one wire in the COMSOL model is shifted
by 2 mm, resulting in a redirection of events from one gap to another.
Events with drift time $<\unit[30]{\mu s}$ have been selected. \textbf{Bottom
right:} $^{\mathrm{83m}}\mathrm{Kr}$ event density with the gate
wires illuminated by the electrons. Events with drift time from $10\,\mathrm{\mu s}<t\leq100\,\mathrm{\mu s}$
were selected. The overdensity of events at high radius is an artifact
of the Mercury algorithm. \label{fig:wire_effect}}
\end{figure}

One of the possible explanations for the scalloping features is variations
in grid wire spacing. Figure~\ref{fig:wire_effect} illustrates the
effects of changing the distance between wires. Modifying this spacing
alters the density of field lines passing between each wire pair,
thus causing an effect throughout the length of the wire rather than
being confined to the edge.

\subsection{Static charge calculation\label{subsec:Static-charge-calculation}}

A simple back-of-the-envelope calculation can be performed to estimate
the theoretical amount of excess charge needed to be transferred from,
for example, a 1-g tissue paper, i.e., a Kimwipe, to the PTFE surface
in order for the tissue paper to be attached to the PTFE surface electrostatically\footnote{Paper and PTFE are on opposite sides of the triboelectric series -
PTFE likes to accept large negative charges while paper prefers positive
charges. }. Even though the electric charge found using charge modeling is not
due to electrostatic charge (if it were, these electrons would follow
the field lines and drift up to the gate rather than staying put),
this provides an insight to the quantity of charge needed to produce
such effects. The charge needed for the paper to hold on PTFE purely
electrostatically needs to be at least equal to the gravitational
force: 
\begin{eqnarray}
F & = & mg\label{eq:}\\
 & = & \unit[1\times10^{-3}]{kg}\times\unit[9.8]{m/s^{2}}=\unit[9.8\times10^{-3}]{N}\label{eq:-1}\\
 & = & \frac{1}{4\pi\varepsilon_{0}}\frac{q_{1}q_{2}}{r^{2}}.\label{eq:-2}
\end{eqnarray}
Electrons transferred to PTFE must be equal to the positive charges
left on the wipe, so $q_{1}q_{2}=q^{2}$ and $r$ is the distance
between point charges, which is assumed here to be only $\unit[\sim0.1]{mm}$.
Therefore 
\begin{eqnarray}
q & = & \sqrt{4\pi\varepsilon_{0}Fr^{2}}\label{eq:-3}\\
 & = & \sqrt{4\pi*8.85\times10^{-12}\times9.8\times10^{-3}\times\left(0.1\times10^{-3}\right)^{2}}\label{eq:-4}\\
 & = & \unit[1.0\times10^{-4}]{\mu C}.\label{eq:-5}
\end{eqnarray}
Assuming the mass of the tissue paper to be $\unit[\sim40]{g/m^{2}}$,
$\unit[1]{g}$ of tissue paper has an area of $\unit[\sim250]{cm^{2}}$.
Hence this charge corresponds to a surface charge density of $\unit[4\times10^{-7}]{\mu C/cm^{2}}=\unit[0.004]{\mu C/m^{2}}$.
Thus the charge observed in WS2013 in the PTFE panels surrounding
the active volume is only slightly larger than typical static charge.

\section{Future avenues of research}

Since the field maps laid the foundation for the rest of the WS2014-16
analysis, the timeline of this project was constrained in order to
provide the complete field maps on time to the rest of the LUX collaboration.
Therefore compromises were made throughout the project to abide by
the tight WS2014-16 analysis schedule. Nevertheless, the models developed
here show good agreement with charge, light, and event positions observed
in LUX data.

A significant factor in the quality of the models and the speed of
convergence was given by the number of divisions on the PTFE panels
used. The initial 15 tiles used in this work (3~angular and 5 vertical
sections, without the smoothing function) provided a model that was
too coarse. However, this worked as a proof of concept since the model
converged during tests to a known charge distribution on the PTFE
panels. The next model built consisted of 42 charge tiles (6 angular
and 7 vertical sections with the smoothing function). This allowed
for a better fidelity fit, however, the increased complexity of the
model due to the large number of free parameters caused a very slow
convergence of the Metropolis-Hastings algorithm, and a perfect fit
was never reached. Slightly different sectioning of the detector might
have been preferable, such as using only 4 angular and 6 or 7 vertical
slices, resulting in 24 or 28 tiles and thereby significantly decreasing
the number of free parameters while retaining a good resolution.

Better coding of the simulated drifting electrons would have also
improved fitting results. Improvements in time efficiency and therefore
accuracy could likely come from efforts focused on reducing the algorithmic
complexity class, effective parallelization of the computation, and
using a compiled programming language such as C/C++ rather than Python.

\section{Conclusion}

The above method for detailed modeling of the varying electric fields
inside the active volume of the LUX detector enabled understanding
of the electric fields. This was vital for event reconstruction and
hence served as the foundation of the LUX WIMP search analysis. The
fields observed in the bulk of the active volume of the detector during
WS2013 were reproduced without any tuning necessary. The COMSOL field
map modeling method described in this work is sensitive down to $\unit[-0.03]{\mu C/m^{2}}$,
enough charge density to model minor scalloping to the detector edge
as seen in WS2013. 

The time dependence of the electric field during WS2014-16 was likely
due to the creation of charge in the PTFE panels during detector grid
conditioning prior to the extended scientific run. The model of electric
fields was created by simulating the LUX detector in 3D using COMSOL
Multiphysics with the PTFE panels divided into 42 tiles. The use of
superposition principle then enabled placement of any combination
of hypothetical charges on the PTFE panels surrounding the active
region. The amount of charge on each tile was varied until convergence
on the distribution of $^{\mathrm{83m}}\mathrm{Kr}$ calibration data
within the detector was obtained. The resulting method allowed a faithful
reconstruction of electric fields in the LUX detector. The fields
during WS2014-16 varied in time, azimuth and depth; the method described
in this work was able to capture this evolution in all dimensions.
Twenty-one field maps were developed to accurately model the changing
electric field throughout the extended WS2014-16 with an average modeled
charge distribution in the PTFE panels varying from $\unit[-3.6]{\mu C/m^{2}}$
to $\unit[-5.5]{\mu C/m^{2}}$. 

The mean electric field values resulting from this COMSOL modeling
method were found to be in good agreement with electric field values
deduced from the NEST package, which fit the field-dependent scintillation
and ionization yields of the CH$_{3}$T calibration source. This work
also improved understanding of the cathode grid geometry on the field
leakage and its effect on axially symmetric models. The successful
modeling of electric fields, along with the robust calibration and
simulation programs developed by the LUX collaboration enabled a thorough
understanding of the LUX detector throughout all of its scientific
programs and~strengthened its sensitivity to WIMPs.

\chapter{\textsc{Searching for sub-GeV dark matter with the LUX detector}\label{chap:Searching-for-sub-GeV-dm}}

The two-phase xenon time projection chamber (TPC) is the leading technology
used to search for the weakly interacting massive particle (WIMP),
a favored dark matter \nomenclature{DM}{Dark Matter} (DM) candidate,
in the $\unit[5]{GeV/c^{2}}$ to $\unit[10]{TeV/c^{2}}$ mass range.
Despite substantial improvements in sensitivity over the recent years,
detecting DM remains an elusive goal~\cite{Akerib:2016vxi,Cui:2017nnn,Aprile:2017iyp}.
Consistent progress in ruling out WIMP parameter space has resulted
in a significant broadening of efforts, including focusing on lighter
particles as possible DM candidates. Currently, the intrinsic scintillation
properties of nuclear recoils prevent liquid xenon (LXe) TPCs to reach
sub-GeV DM masses, for analyses based on both S1 and S2 signals. Lower
thresholds can be achieved in S2-only studies~\cite{Aprile:2016wwo}
that drop the requirement for the scintillation (S1) signal since
charge yields are higher at low energies. However, this type of analysis
is currently limited by anomalous few-electron background, which prevents
construction of a complete background model, impairing dark matter
detection claims.

Recently, References~\cite{Kouvaris:2016afs,Ibe:2017yqa} proposed
novel direct dark matter detection methods that extend the reach of
LXe detectors to sub-GeV DM masses. They suggest that DM-nucleus scattering
can be accompanied by a signal that results in an electron recoil
at higher energies than the corresponding nuclear recoil in LXe detectors.
Since electron recoils (ER) produce a stronger signal than nuclear
recoils at low energies, this newly recognized channel enables liquid
xenon detectors to reach sub-GeV DM masses. In LUX the 50\% detection
efficiency for nuclear recoils is 3.3~keV, compared with 1.24~keV
for electron recoils.

This chapter discusses searches of sub-GeV DM in the LUX detector
using two different mechanisms: Bremsstrahlung, first proposed in~\cite{Kouvaris:2016afs},
and the Migdal effect as reformulated in~\cite{Ibe:2017yqa}. These
inelastic signals produce much stronger charge and light yields compared
to the traditional elastic recoil signal for DM candidates with masses
below $\unit[\sim5]{GeV/c^{2}}$. 

Bremsstrahlung considers emission of a photon from the recoiling atomic
nucleus. In the atomic picture, the process can be viewed as a dipole-emission
of a photon from a xenon atom polarized in the DM-nucleus scattering.
The Migdal effect describes the ionization of an atom as a result
of a sudden acceleration of a recoiling nucleus after a collision.
This creates an ER response alongside its nuclear recoil. Considering
photons from Bremsstrahlung or electrons from the Migdal effect allows
LUX to access sub-GeV DM parameter space. Both the S1 and the S2 signals
can be used in the analysis where the nuclear recoil signal would
not be visible otherwise, extending the reach of liquid xenon detectors
to lower masses of DM particles than has been achieved previously
in S1-S2 analyses. This analysis also provides sensitivity to lower
masses than S2-only analysis~\cite{Aprile:2016wwo} of nuclear recoils\footnote{S2-only can also be used to search for DM-electron interactions, constraining
DM with masses of 1-100~MeV~\cite{Akerib:2017uem}.}. 

This chapter first provides an overview of the scattering rate calculations
for the \linebreak{}
Bremsstrahlung and Migdal effects. Several different mediator scenarios
are considered for the two effects. The predicted scattering rates
are then applied in the analysis to place limits on DM-nucleon scattering
cross section. The results presented in this chapter have also been
accepted for publication to PRL~\cite{Akerib:2018subGeV}. 

\section{Bremsstrahlung}

As was first postulated in~\cite{Kouvaris:2016afs}, two-phase LXe
TPC can search for sub-GeV dark matter using an irreducible photon
emission in the form of bremsstrahlung from a polarized xenon atom,
caused by the displacement of the nucleus and electron charges after
the DM - nucleus interaction. The most common, elastic interaction
between DM $\chi$ with mass $m_{\chi}$ and nucleus~$N$ with mass~$m_{N}$
considers 
\begin{equation}
\chi+N\rightarrow\chi+N\left(E_{R}\right).\label{eq:-11}
\end{equation}
Here $E_{R}$ is the kinetic energy of the recoiling nucleus with
momentum $\mathbf{q}=\mathbf{p'_{\chi}}-\mathbf{p_{\chi}}$ given
by 
\begin{equation}
E_{R}=\frac{\left|\mathbf{q}\right|^{2}}{2m_{N}}\leq\frac{2\mu_{N}^{2}v^{2}}{m_{N}}\label{eq:-12}
\end{equation}
where $\mu_{N}$ is the DM-nucleus reduced mass
\[
\mu_{N}=\frac{m_{N}m\chi}{m_{N}+m_{\chi}}
\]
 and $v$ is the relative velocity. For an inelastic\footnote{Note that this is different from the inelastic dark matter interaction
which requires considerable momentum transfer and concerns electroweak
scale DM masses.} interaction 
\begin{equation}
\chi+N\rightarrow\chi+N\left(E_{R}^{'}\right)+\gamma\left(\omega\right)\label{eq:-13}
\end{equation}
where the available photon energy $\omega\leq\mu_{N}v^{2}/2$ as bounded
by the energy of the relative motion of DM and the target restricts
the kinetic energy of the recoiling nucleus $E_{R}\ll\omega_{max}$
for $m_{\chi}\ll m_{N}$. This larger energy deposition in photon
emission improves the sensitivity of LUX to nuclear recoils in the
sub-GeV DM mass regime. The emitted photon from the recoiling nucleus
has a continuous energy spectrum, exciting and ionizing surrounding
xenon atoms and resulting in an ER signal. However, this signal is
weaker compared to the standard nuclear elastic scattering since a
penalty is paid for going to the inelastic channel. Additionally,
in LXe TPCs most background events fall within the ER band. This means
that it will be subject to a higher amount of backgrounds while having
a significantly lower detection threshold. 

\begin{figure}
\begin{centering}
\includegraphics[scale=0.7]{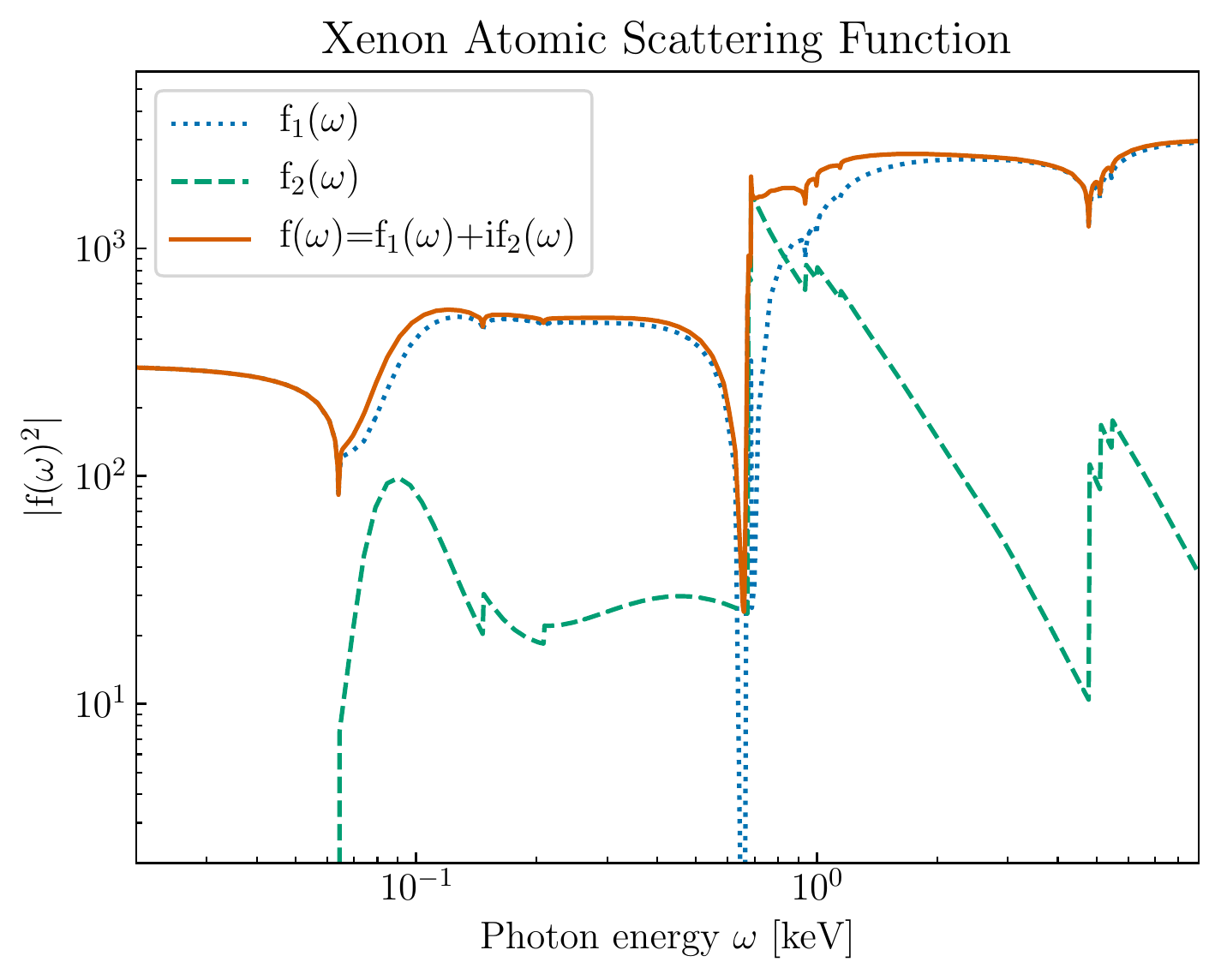}
\par\end{centering}
\caption[Xenon atomic scattering factor]{Xenon atomic scattering factor $f\left(\omega\right)=f_{1}\left(\omega\right)+if_{2}\left(\omega\right)$.
Data from~\cite{atomicFormFactor}. \label{fig:Xenon-atomic-scattering}}

\end{figure}

The scattering rates for Bremsstrahlung $\frac{dR}{d\omega}$ have
been derived~\cite{Kouvaris:2016afs}, and a short outline follows.
The differential cross section for a bremsstrahlung emission from
the nucleus in the presence of an electron cloud is given by
\begin{align*}
\frac{d^{2}\sigma}{d\omega dE_{R}} & =\frac{4\alpha}{3\pi\omega}\frac{E_{R}}{m_{N}}\left|f\left(\omega\right)\right|^{2}\times\frac{d\omega}{dE_{R}}\Theta\left(\omega_{\mathrm{max}}-\omega\right)\\
 & \propto\omega^{3}.
\end{align*}
Here $\alpha=1/137$ is the fine structure constant, characterizing
the strength of the electromagnetic interaction between elementary
charged particles. The atomic form (or scattering) factor $f\left(\omega\right)=f_{1}\left(\omega\right)+if_{2}\left(\omega\right)$
is the Fourier transform of the electron density of an atom which
measures the scattering amplitude of a wave by an isolated atom. Generally,
the electron density is assumed to be spherically symmetric so that
the value of the Fourier transform only depends on the distance from
the origin in reciprocal space. The case of xenon is illustrated in
Figure~\ref{fig:Xenon-atomic-scattering}. At large energies, $f_{1}\rightarrow Z\gg f_{2}$,
and the atomic states become irrelevant, where $Z$ is the atomic
number. At low photon energies, the process weakens as $\omega^{3}$,
which is typical for dipole emission created between the nucleus and
electrons. To simplify the atomic calculations, Reference~\cite{Kouvaris:2016afs}
assumes that the atom stays in the ground state. Therefore, the limits
presented below should be considered conservative. More detailed calculations
of the atomic processes are left for future research, potentially
leading to an increase in the photon signal.

To find an expression for the differential event rate seen in the
detector, the energy deposition $E_{R}$ is neglected since 
\begin{align}
E_{R,\mathrm{max}}=4\frac{m_{\chi}}{m_{N}}\omega_{\mathrm{max}}\ll\omega_{\mathrm{max}} & \quad\mathrm{for}\:m_{\chi}\ll m_{N}.\label{eq:-14}
\end{align}
Therefore, the photon energy is treated as the only detectable quantity
with
\begin{equation}
\frac{d\sigma}{d\omega}=\int_{E_{R,\mathrm{min}}}^{E_{R,\mathrm{max}}}dE_{R}\left(\frac{d\sigma}{dE_{R}d\omega}\right)\label{eq:-16}
\end{equation}
where the DM-nucleus recoil cross section is
\begin{equation}
\frac{d\sigma}{dE_{R}}=\frac{\sigma_{0}^{\mathrm{SI}}m_{N}}{2\mu_{N}^{2}v^{2}}F_{N}^{\mathrm{SI}}\left(E_{R}\right)F_{\mathrm{med}}\left(E_{R}\right).\label{eq:DM-nucleus-xsection}
\end{equation}
Here $\sigma_{0}^{\mathrm{SI}}$ is the usual DM-nucleon spin-independent
cross section 
\begin{equation}
\sigma_{0}^{SI}\simeq\sigma_{n}\left(\frac{\mu_{N}}{\mu_{n}}\right)^{2}A^{2}\label{eq:sigma0_SI}
\end{equation}
where $\sigma_{n}$ is the DM-nucleon elastic cross section, and $\mu_{n}$
is the DM-nucleon reduced mass (cf. Equation~\ref{eq:WIMP-nucleon}).
Note that $\sigma_{n}$ and $m_{\chi}$ form precisely the parameter
space that DM searches, including LUX, are attempting to probe. $F_{N}^{\mathrm{SI}}\left(E_{R}\right)$
is the nuclear form factor that is relevant when the wavelength $h/q$
is no longer large compared to the nuclear radius as discussed in
Section~\ref{subsec:WIMP-event-rates}. The form factor is unity
for small $q$, and it decreases with increasing $q$. The form factor
is a function of the dimensionless quantity $qr_{n}/\hbar$, where
$r_{n}$ is an effective nuclear radius and $\hbar=h/\left(2\pi\right)$
is the reduced Planck constant. Different approximations can be used
to calculate the nuclear form factor; this analysis adopts the Helm
form factor discussed in~\cite{Lewin:1995rx}. $F_{\mathrm{med}}$
depends on the mass of the particle mediating the interaction discussed
in Sections~\ref{subsec:Heavy-scalar-mediator} and~\ref{subsec:Light-scalar-mediator}. 

The differential cross section can be integrated to yield 
\[
\frac{d\sigma}{d\omega}=\frac{4\alpha\left|f\left(\omega\right)\right|^{2}}{2\pi\omega}\frac{\mu_{N}^{2}v^{2}\sigma_{0}^{\mathrm{SI}}}{m_{N}^{2}}\sqrt{1-\frac{2\omega}{\mu_{N}v^{2}}}\left(1-\frac{\omega}{\mu_{N}v^{2}}\right).
\]
The resulting event rate is then given by 
\begin{equation}
\frac{dR}{d\omega}=N_{T}\frac{\rho_{\chi}}{m_{\chi}}\int_{\left|\mathbf{v}\right|\geq v_{min}}d^{3}\mathbf{v}vf_{v}(\mathbf{v}+\mathbf{v_{e}})\frac{d\sigma}{d\omega}.\label{eq:-15}
\end{equation}
Here $N_{T}$ is the number of target nuclei per unit detector mass
and $\rho_{\chi}=\unit[0.3]{GeV/cm^{3}}$ is the local DM mass density.
$f_{v}\left(\mathbf{v}\right)$ is the DM velocity distribution represented
by a truncated Maxwellian 
\begin{equation}
f_{v}\left(\mathbf{v}\right)=\begin{cases}
\frac{1}{N_{esc}}\left(\frac{3}{2\pi\sigma_{v}^{2}}\right)^{3/2}e^{-3\mathbf{v^{2}}/2\sigma_{v}^{2}} & \mathrm{for}\,\left|\mathbf{v}\right|<v_{esc}\\
0 & \mathrm{otherwise}
\end{cases}\label{eq:velocity-dist}
\end{equation}
where 
\[
N_{esc}=\frac{\mathrm{erf}\left(z\right)-2z\exp\left(-z^{2}\right)}{\sqrt{\pi}}
\]
with the normalization factor\footnote{Distributions without an escape velocity where $v_{\mathrm{esc}}\rightarrow\infty$
result in $N_{\mathrm{esc}}=1$.} $z\equiv v_{esc}/v_{0}$. In this analysis, the escape speed $v_{esc}=\unit[544]{km/s}$,
$\mathbf{v}_{e}=\unit[232.4]{km/s}$ is the velocity of the Earth
relative to the Galactic rest frame, $v_{min}=\sqrt{2\omega/\mu_{N}}$,
and the most probable velocity $v_{0}=\sqrt{2/3}\sigma_{v}=\unit[220]{km/s}$~\cite{Savage:2008er}. 

The goal of the analysis presented in this chapter is to apply the
theoretical scattering rates from Bremsstrahlung presented in Equation~\ref{eq:-15}
and from the Migdal effect introduced in the next section to the data
acquired during LUX WS2013. Several different mediator scenarios are
considered. This analysis will determine whether LUX has observed
DM in the sub-GeV region.

\subsection{Heavy scalar mediator\label{subsec:Heavy-scalar-mediator}}

A mediator defines the coupling between the DM and Standard Model\nomenclature{SM}{Standard Model}
(SM) particles. Different mediators can enter Equation~\ref{eq:DM-nucleus-xsection}.
A scalar mediator couples to SM particles by mixing with the SM Higgs
boson making its coupling proportional to the particle's mass $A$
(equal interaction strength with protons and neutrons). 

The other factor in Equation~\ref{eq:DM-nucleus-xsection}, $F_{med}\left(E_{R}\right)$,
depends on the mass of the particle mediating the interaction. In
the heavy mediator limit $m_{med}\gg q$ and the mediator factor can
be well approximated as $F^{2}\simeq1$ at low energies. 

\subsection{Light scalar mediator\label{subsec:Light-scalar-mediator}}

\begin{figure}[t]
\centering{}\includegraphics{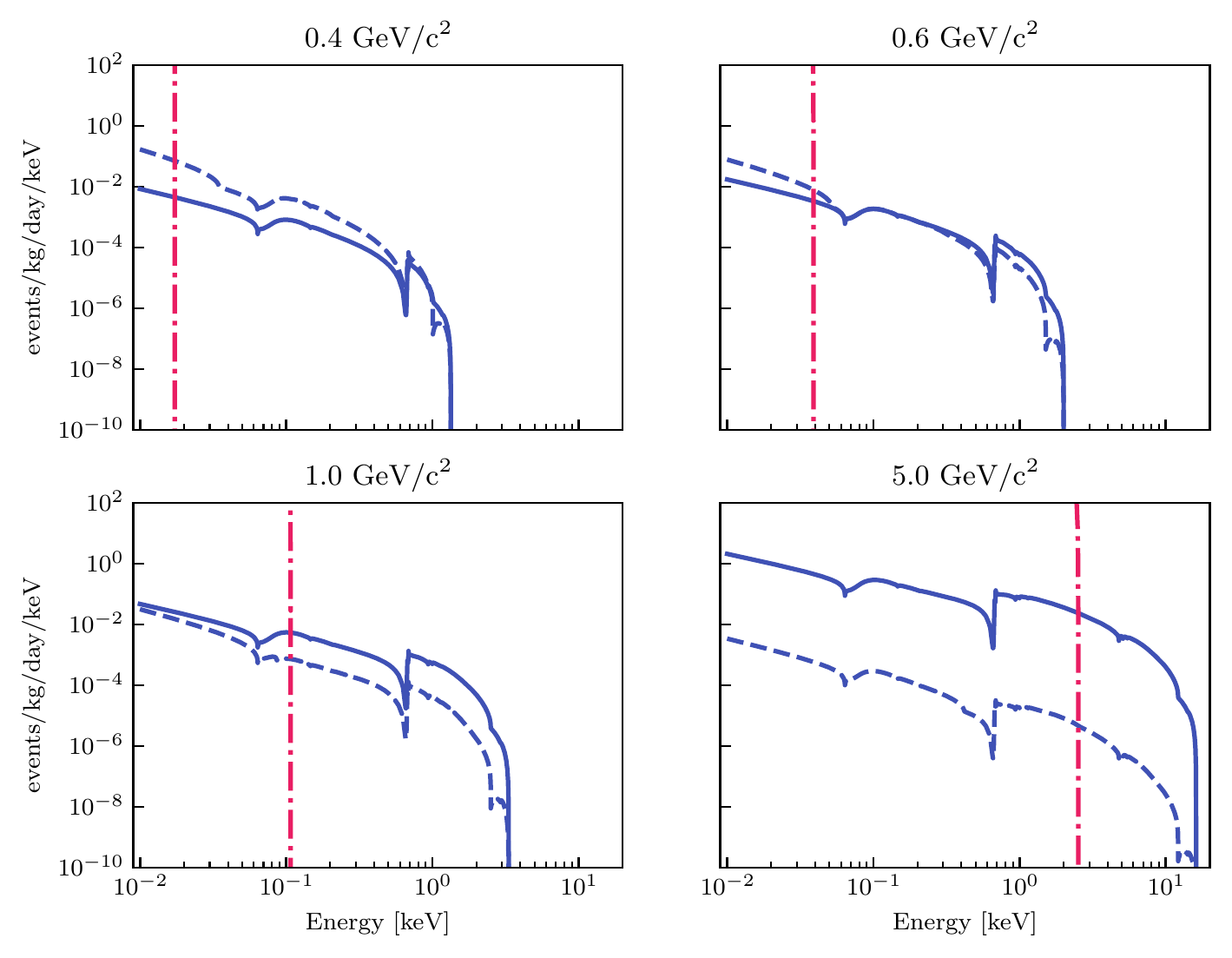}\caption[Scattering rates for Bremsstrahlung considering light and heavy scalar
mediators]{SI scattering rates in xenon for Bremsstrahlung that would be detected
as electron recoils in LUX assuming $\sigma=1\times\unit[10^{-35}]{cm^{2}}$.
Two different mediators are considered: a heavy scalar mediator ($m_{med}\gg\mathrm{MeV}$,
solid blue) and light mediator ($m_{med}\ll\mathrm{MeV}$ where $q_{\mathrm{ref}}=\unit[1]{MeV}$,
dashed blue). To illustrated the improvement in reach for LXe detectors
that Bremsstrahlung presents, the scattering rates resulting in nuclear
recoil calculated in Equation~\ref{eq:elastic-rate} are also shown
(dash-dot pink). The NR threshold in LUX is 1.1~keV. \label{fig:Brem-scattering-rates}}
\end{figure}

Aside from the common heavy scalar mediator case, a light scalar mediator
is considered where $m_{\mathrm{med}}\ll q$ results in $F_{\mathrm{med}}=q_{\mathrm{ref}}^{4}/q^{4}$.
We define $q_{\mathrm{ref}}=\unit[1]{MeV}$, a typical size of $q$
for $m_{\chi}\lesssim\unit[1]{GeV/c^{2}}$~\cite{McCabe:2017rln}.
The resulting event rates for both light and heavy scalar mediator
are shown in Figure~\ref{fig:Brem-scattering-rates} along with the
conventional elastic scattering rate. This plot illustrates the comparative
advantage that the Bremsstrahlung signal offers at low dark matter
masses: the nuclear recoils resulting from the SI scattering rate
are otherwise inaccessible due to the finite detector threshold (set
at 1.1~keV as discussed in Section~\ref{subsec:WS2013}).

\section{Migdal effect}

\begin{figure}
\begin{centering}
\includegraphics[scale=0.3]{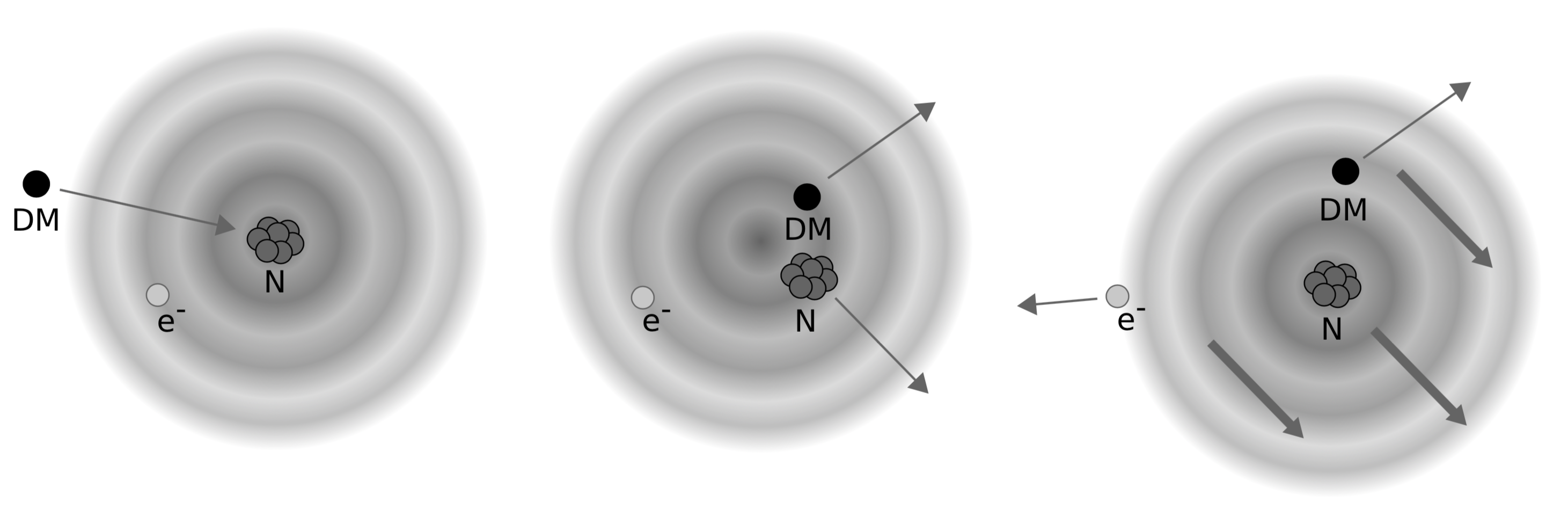}
\par\end{centering}
\caption[Illustration of a nuclear recoil causing a ``Migdal'' electron emission]{Illustration of a nuclear recoil causing an electron emission due
to the Migdal effect. After a DM particle scatters off a nucleus (left),
the Migdal effect assumes that the nucleus moves relative to its surrounding
electron cloud (center). Most electrons catch up with the nucleus,
but some electrons may be left behind and are emitted, leading to
ionization and excitation of the recoiling atom (right). Figure from~\cite{Dolan:2017xbu}.\label{fig:Migdal-illustration}}

\end{figure}

For nuclear recoils in liquid xenon, it is usually assumed that electrons
around the recoiling nucleus immediately follow the motion of the
nucleus so that the atom remains neutral. In reality, it takes some
time for the electrons to catch up, which may result in ionization
and excitation of the atoms~\cite{Ibe:2017yqa}, as illustrated in
Figure~\ref{fig:Migdal-illustration}. When A.B. Migdal originally
formulated the Migdal effect in 1941~\cite{migdal1941migdal}, he
assumed an impulsive force to justify his calculations~\cite{landau1981quantum}.
In his original approach, the final state ionization and excitation
are treated separately from the nuclear recoil, which makes energy-momentum
conservation and probability conservation somewhat obscure.

However, reference \cite{Ibe:2017yqa} has reformulated the approach
using atomic energy eigenstates for their calculation (rather than
the energy eigenstates of the isolated nucleus), avoiding the need
to resolve the complex time evolution of the nucleus-electron system.
In this way, the atomic recoil cross section is obtained coherently,
and energy-momentum conservation and probability conservation are
manifest in the final state. A DM-nucleus interaction is expected
to cause a non-vanishing overlap between the initial and final atomic
energy states, which result in the excitation and the ionization processes.
It should be noted that the Migdal effect is distinct from the ionization
and scintillation process since the Migdal effect takes place even
for a scattering of an isolated atom, while the latter occurs due
to the interatomic interactions in the xenon detector. Reference~\cite{Ibe:2017yqa}
contains the theoretical derivation and presents expected event rates
for the Migdal effect; only a brief summary of that manuscript is
presented below. 

\begin{figure}
\begin{centering}
\includegraphics[scale=0.4]{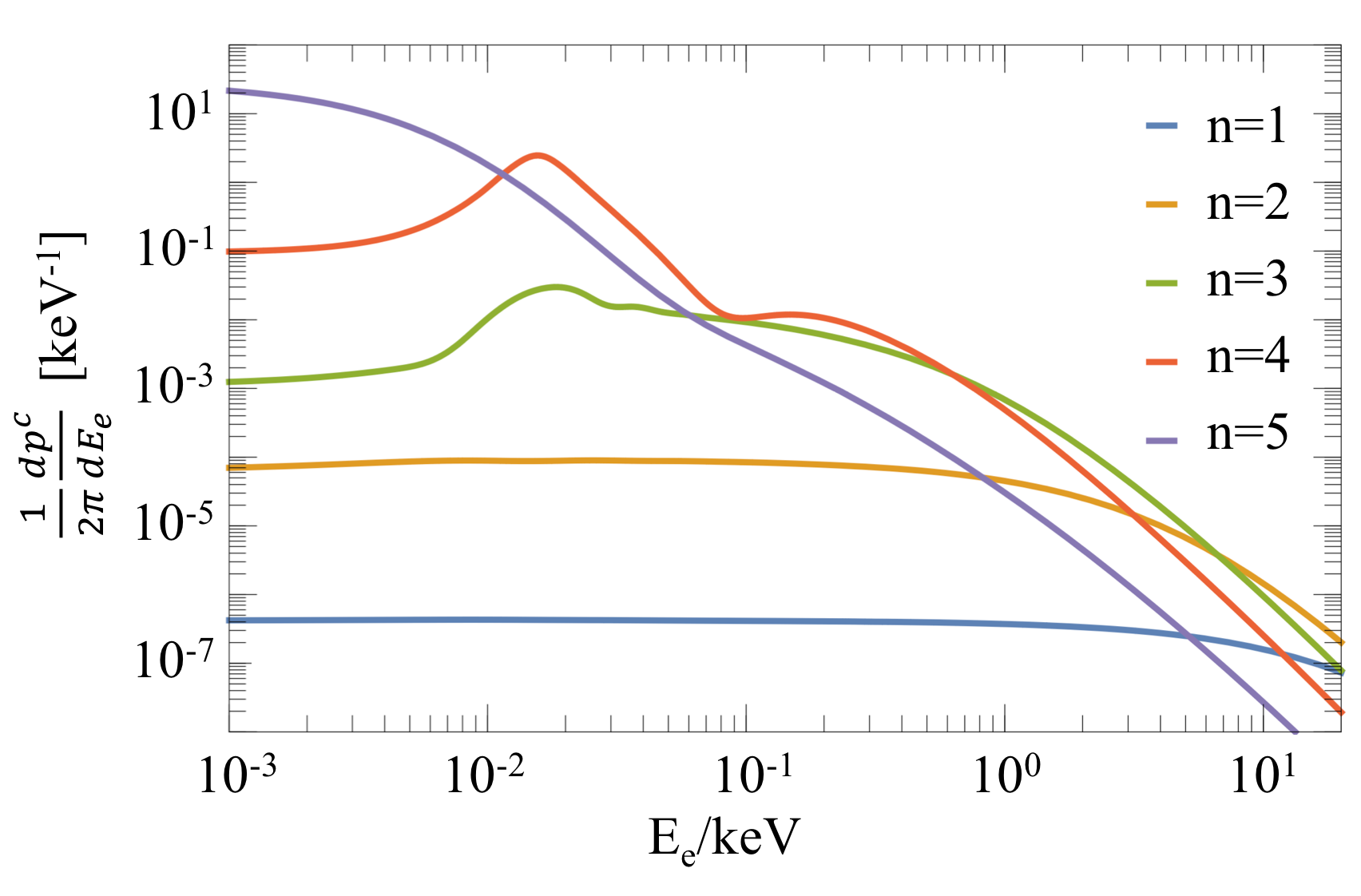}
\par\end{centering}
\caption[Differential ionization probabilities for isolated xenon atom]{Differential ionization probabilities as a function of the emitted
electronic energy $E_{e}$ for isolated xenon atom for $q_{e}=m_{e}v_{F}$
with the velocity after the recoil $v_{F}=10^{-3}$. The contributions
from different $\ell$ and the possible final states for a given $n$
are summed. Figure modified from~\cite{Ibe:2017yqa}.\label{fig:Differential-ionization-probabil}}

\end{figure}

The ionized electronic energy spectrum from an initial orbital $o_{k}$
is given as a product of the dark matter event rate for unit detector
mass $dR_{0}/dE_{R}dv_{\chi}$ and the differential ionization probability
$dp_{q_{e}}^{c}/dE_{e}$ of the bound states $\left(n,\ell\right)$
(shown in Figure~\ref{fig:Differential-ionization-probabil}):
\begin{align}
\frac{dR}{dE_{R}dE_{e}dv_{\chi}}\simeq & \frac{dR_{0}}{dE_{R}dv_{\chi}}\times\frac{1}{2\pi}\sum_{n,l}\frac{d}{dE_{e}}p_{q_{e}}^{c}\left(n\ell\rightarrow E_{e}\right)\label{eq:migdal-scattering}
\end{align}
where the dark matter event rate for unit detector mass
\begin{align}
\frac{dR_{0}}{dE_{R}dv_{\chi}}\simeq & \frac{1}{2}\frac{\rho_{\chi}}{m_{\chi}}\frac{1}{\mu_{N}^{2}}\tilde{\sigma}_{N}\left(q_{A}\right)\times\frac{\tilde{f}\left(v_{\chi}\right)}{v_{\chi}},\label{eq:-18-1}
\end{align}
$\tilde{f}\left(v_{\chi}\right)$ is the DM velocity distribution
integrated over the directional component normalized by $\int\tilde{f}_{\chi}\left(v_{\chi}\right)dv_{\chi}=1$,
and $\tilde{\sigma}_{N}=\sigma_{0}^{SI}F_{N}^{\mathrm{SI}}F_{\mathrm{med}}$.
The atomic recoil energy $E_{R}$ is given by 
\begin{align}
E_{R}\simeq & \frac{q_{A}^{2}}{2m_{A}}\label{eq:-19}\\
q_{e}\simeq & \frac{m_{e}}{m_{A}}q_{A}\label{eq:-20}
\end{align}
where $q_{A}$ is the momentum transfer of the recoil, $m_{e}$ is
the mass of the electron, and $m_{A}$ is the mass of the atom. The
minimum dark matter velocity for given $E_{R}$ and $\Delta E$
\[
v_{\chi,min}\simeq\frac{m_{N}E_{R}+\mu_{N}\Delta E}{\mu_{N}\sqrt{2m_{N}E_{R}}}
\]
which determines the kinematically allowed region of $E_{R}$.

\begin{figure}[t]
\centering{}\includegraphics[scale=0.57]{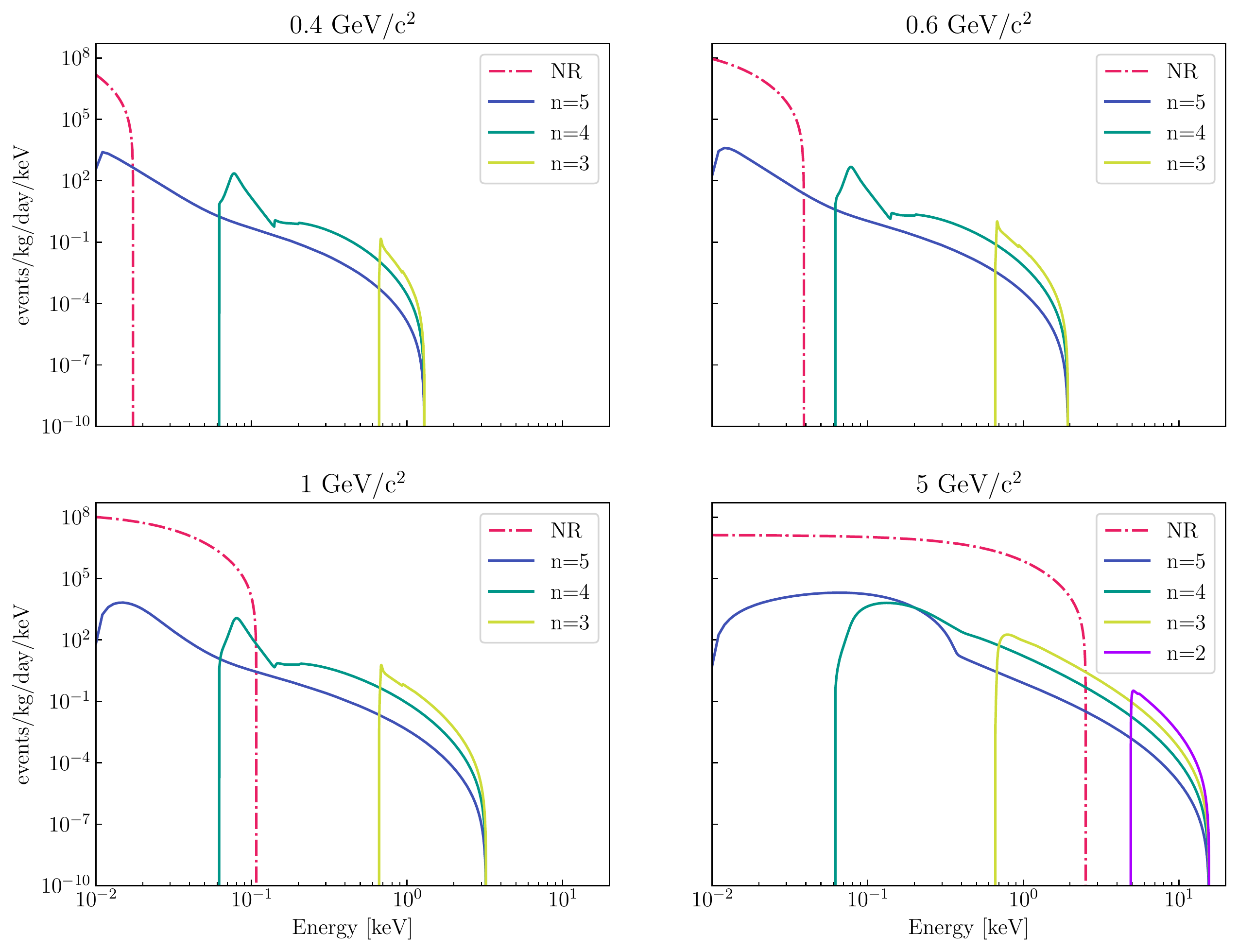}\caption[Scattering rates from Migdal effect with contributions from each electron
shell]{Scattering rates in xenon from the Migdal effect showing contributions
from electrons shells. Contributions from $n=1,\,2$ are not relevant
for the DM masses plotted here except at $\unit[5]{GeV/c^{2}}$. Since
the derivation assumes isolated atoms, the ionization spectrum from
the valence electrons, i.e., $n=5$, are not reliable. The analysis
thus uses only electrons from $n=3,\,4$. The DM-nucleus scattering
rates resulting in elastic nuclear recoil in LUX are also shown (dash-dot
pink). \label{fig:migdal-rates-allN}}
\end{figure}

Figure~\ref{fig:migdal-rates-allN} contains the scattering rates
derived in Equation~\ref{eq:migdal-scattering} for a given SI scattering
cross section of DM on nucleons, showing contributions from electrons
from the various atomic shells. Only electron injections caused by
ionization are included since the excitation probabilities into the
unoccupied binding energy levels are much smaller than the ionization
probabilities. Additionally, the derivation in~\cite{Ibe:2017yqa}
assumes isolated atoms, so additional considerations need to be made
for non-isolated atoms in liquid or crystals. In non-isolated atoms,
such as in liquid xenon, the energy levels of valence electrons in
$n=5$ states are affected by the ambient atoms by $\unit[\mathcal{O}\left(0.1\right)]{eV}$.
In comparison, the ionization rates from the inner orbitals are expected
to be less affected by ambient atoms since the binding energies of
inner orbitals are much larger than 0.1 eV. Thus, this analysis conservatively
ignores the contribution from electrons in $n=5$ states. The ionization
spectrum in Equation~\ref{eq:migdal-scattering} can be applied rather
reliably for electrons from the inner orbitals with $n=3,4$~\cite{Ibe:2017yqa}.
Electrons from $n=1$ require ionization energy of about 35 keV and
begin to contribute at DM masses of $\unit[\sim10]{GeV/c^{2}}$. At
lower masses, DM requires speeds higher than the escape velocity.
Contributions to the signal from $n=2$ electrons need ionization
energy of $\sim\unit[5]{keV}$ and become visible at DM masses of
$\unit[\sim2]{GeV/c^{2}}$, but the size of the signal is minimal.
Therefore, contributions to the signal model from $n=1,2$ orbitals
are neglected since their contribution is negligible\footnote{At $\unit[3]{GeV/c^{2}}$ electrons from $n=2$ increase the signal
yield by only $\sim0.5\%$ and at $\unit[5]{GeV/c^{2}}$ the signal
increases by $\sim2\%$.} at DM masses considered here. 

Figure~\ref{fig:migdal-rates-allN} also illustrates that the energy
from the ionization can be larger than the nuclear recoil energy from
the elastic scattering alone for light DM. For heavier dark matter,
$m_{\chi}>\unit[\mathcal{O}\left(10\right)]{GeV/c^{2}}$, on the other
hand, the Migdal effect is submerged below the conventional nuclear
recoil spectrum, and hence, does not affect the detector sensitivities.
In principle, the additional electronic energy injections affect the
S2/S1 ratio in the detector. 

\begin{figure}[th]
\begin{centering}
\includegraphics{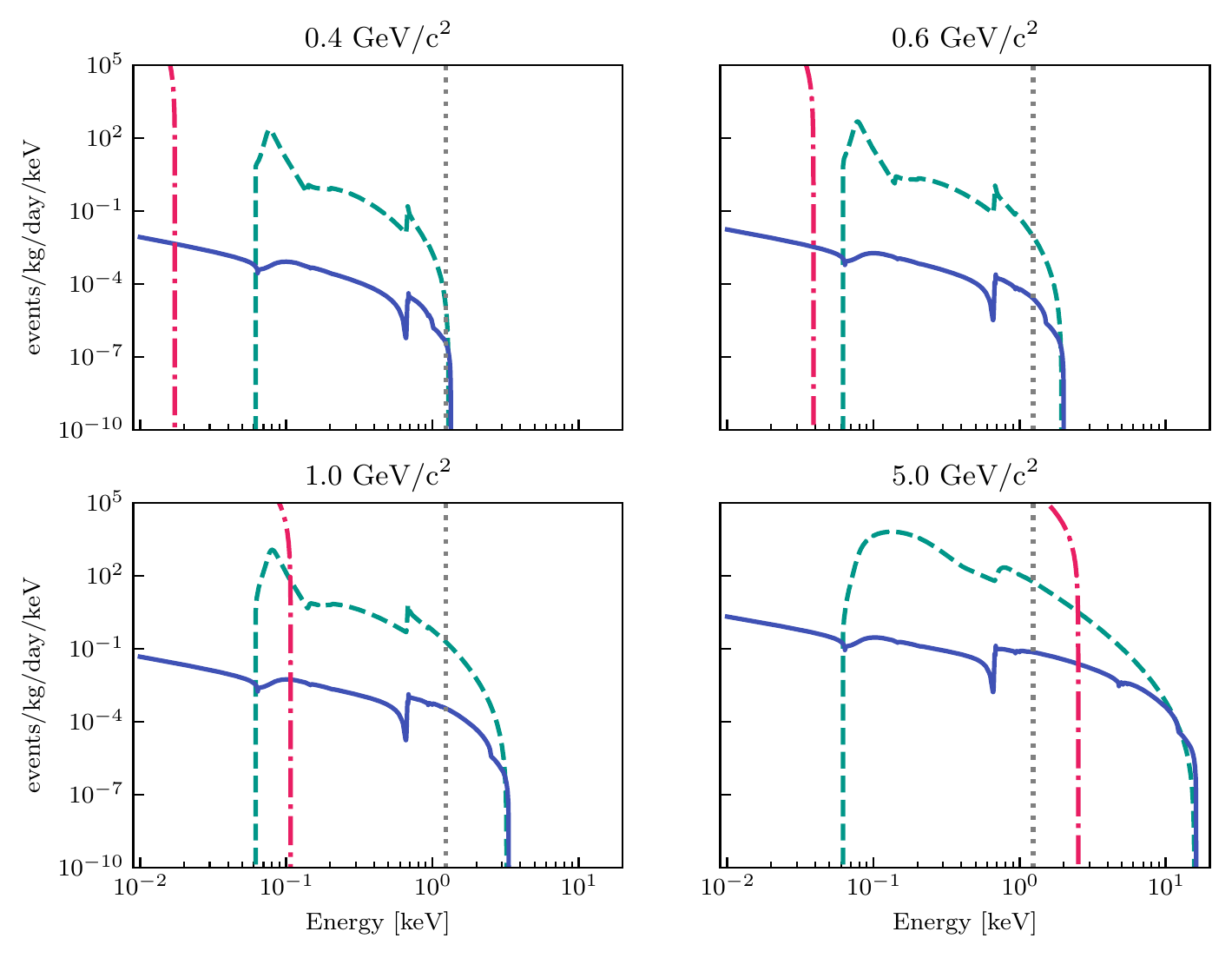}
\par\end{centering}
\caption[Scattering rates for Bremsstrahlung and Migdal effect]{Scattering rates in xenon for Bremsstrahlung (solid blue) and Migdal
effect (dashed teal) illustrating the suppression of Bremsstrahlung
by the mass of the nucleus resulting in lower scattering rates. The
DM-nucleus scattering rates resulting in elastic nuclear recoil in
LUX are also shown (dash-dot pink). Also shown is a signal cut off
at 1.24 keV (dotted gray) applied in the analysis, corresponding to
50\% efficiency of ER detection (see Section~\ref{subsec:Detector-response-to-ER}).
Note that 50\% efficiency for NR event detection occurs at 3.3~keV~\cite{Akerib:2015rjg}.
\label{fig:Brem-vs-Migdal}}

\end{figure}

It should be emphasized that both the nuclear and electron recoil
signals are present when considering the Bremsstrahlung and Migdal
effects. However, the nuclear recoil signal from sub-GeV DM is undetectable
by LUX, so only the electron recoil signal is used in this analysis.
Higher interaction rates in the region of interest are expected from
the Migdal effect than Bremsstrahlung as shown in Figure~\ref{fig:Brem-vs-Migdal}.

\subsection{Vector mediator\label{subsec:Vector-mediator}}

Unlike the scalar mediator whose coupling is proportional to the particle's
mass, the vector mediator couples to SM particles via mixing with
the SM photon, making its coupling proportional to the particle's
electromagnetic charge. These so-called dark photon models result
in a $Z^{2}$ enhancement for scattering on nuclei with charge number
$Z$ and have comparable couplings to protons and electrons~\cite{Dolan:2017xbu}. 

\begin{figure}
\centering{}\includegraphics[scale=0.7]{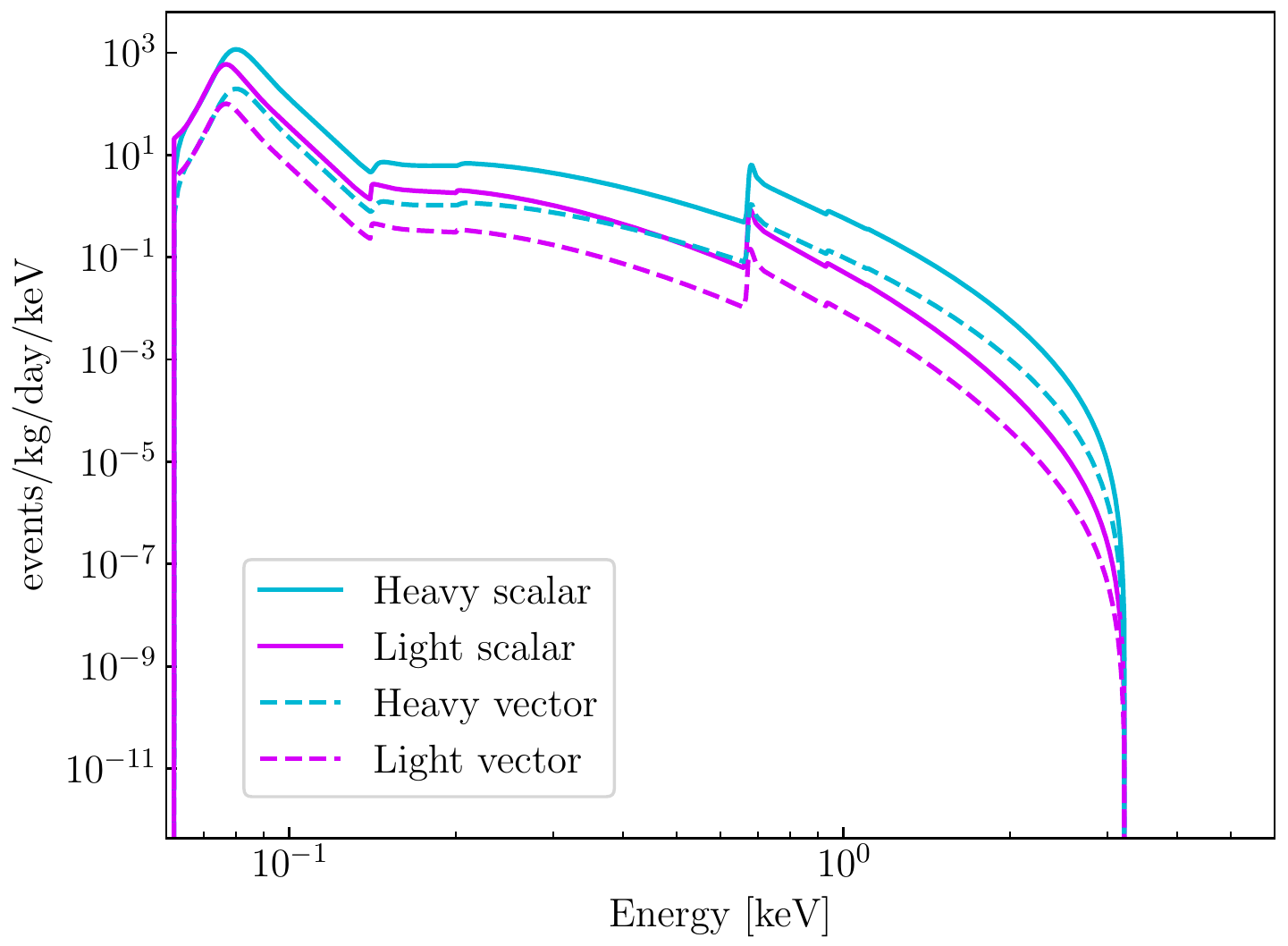}\caption[Scattering rates for various mediator cases for 1 GeV dark matter]{Scattering rates in xenon from Migdal effect for various mediator
cases at $\sigma=\unit[1\times10^{-35}]{cm^{2}}$ for $\unit[1]{GeV/c^{2}}$
dark matter. The scalar mediator's coupling is proportional to the
recoiling particle's mass $\left(\sigma_{N}\propto A^{2}\right)$,
while the vector mediator's coupling is proportional to the particle's
electromagnetic charge $\left(\sigma_{N}\propto Z^{2}\right)$. For
the heavy mediator $m_{med}\gg q$, while for the light mediator $m_{\mathrm{med}}\ll q$
where we define $q_{\mathrm{ref}}=\unit[1]{MeV}$ where $q$ is the
momentum transfer. \label{fig:Migdal_varying_mediators}}
\end{figure}

Additionally as mentioned in Section~\ref{subsec:Light-scalar-mediator},
both light and heavy mediator cases were considered. Overall this
results in up to 4 different limits for each of the signal models
as illustrated in Figure~\ref{fig:Migdal_varying_mediators} for
the case of the Migdal effect.

\section{Analysis}

LUX collected data during two scientific searches in 2013~\cite{Akerib:2013tjd,Akerib:2015rjg}
and in 2014-16~\cite{Akerib:2016vxi}. The work presented here uses
WIMP search data with a total exposure of 95 live-days using 118~kg
of LXe in the fiducial volume collected from April 24 to September
1, 2013, referred to as WS2013 presented in~\cite{Akerib:2015rjg}
and discussed in Section~\ref{subsec:WS2013}. Only single scatter
events (one S1 followed by one S2) are considered with analysis cuts
summarized in Table~\ref{tab:WS2013-parameters} with detector-specific
gain factors $g_{1}=0.117$ detected photons (phd) per photon emitted
and $g_{2}=\unit[12.2]{phd}$ per electron.

\subsection{Detector response to electron recoils\label{subsec:Detector-response-to-ER}}

\begin{figure}
\begin{centering}
\includegraphics[scale=0.52]{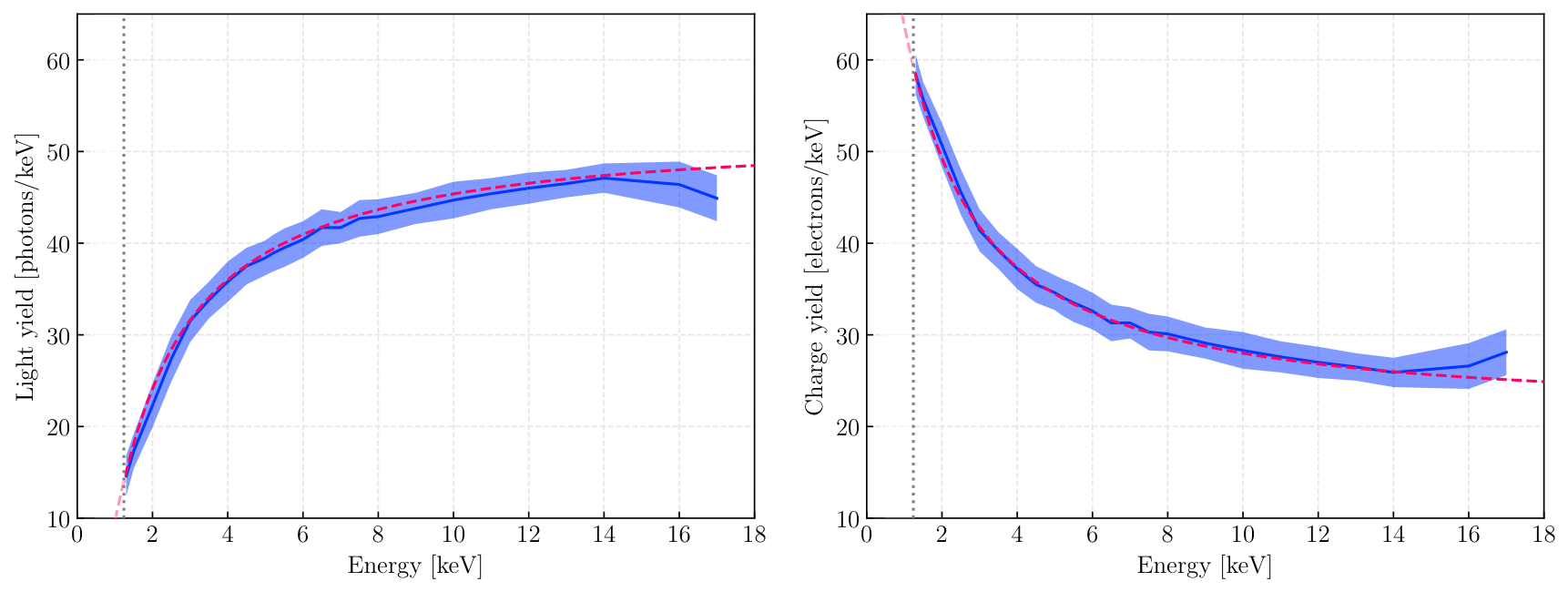}
\par\end{centering}
\caption[Light and charge yields of tritium ER events measured by LUX vs NEST
2.0]{The light (left) and charge yields (right) of tritium ER events as
a function of recoil energy measured \textit{in situ} by the LUX detector
at 180 V/cm (solid blue line) compared to NEST v2.0 (dashed pink line).
The bands indicate the 1 $\sigma$ uncertainties on the measurement.
The gray dotted line shows the 1.24 keV threshold implemented in the
analysis.\label{fig:ER_yields}}

\end{figure}

The response of the LUX detector to ERs was characterized using internal
tritium calibrations performed in December 2013, directly following
WS2013. Tritiated methane CH$_{3}$T was injected into the gas circulation
to achieve a spatially uniform distribution of events dissolved in
the detector\textquoteright s active region as described in~\cite{Akerib:2015wdi}
and discussed in Section~\ref{subsec:Tritium-(H)}. This direct calibration
is then used to build the signal model used in this analysis. Figure~\ref{fig:ER_yields}
shows excellent agreement between the ER yields from the \textit{in
situ} tritium calibrations and yields obtained from the Noble Element
Simulation Technique (NEST) package v2.0~\cite{nest2.0,NEST_Szydagis,Szydagis:2013sih,NEST_lenardo},
used to model the ER response in the signal model. The complementary
behavior between light and charge yields is due to recombination effects
described in~\cite{Akerib:2015wdi,Akerib:2016qlr}. Since this work
considers recoils at the lowest energies, it can be seen in Figure~\ref{fig:ER_yields}
that the analysis is limited by light production rather than charge
yields. Also, the efficiency for extracting electrons from liquid
to gas $\varepsilon=49\%\pm3\%$ while the aforementioned efficiency
to detect S1 photons $g_{1}=\unit[\left(0.117\pm0.003\right)]{phd}$
per photon. 

A 1.24 keV low-energy cutoff illustrated in Figure~\ref{fig:Brem-vs-Migdal}
was conservatively chosen to avoid uncertainties in efficiency at
low energy, where direct calibration data from tritium are not available.
1.24 keV corresponds to a 50\% efficiency of ER detection (cf. Figure~\ref{fig:tritium-spectrum}),
which imposes a lower limit on DM mass of $\unit[0.4]{GeV/c^{2}}$.
The upper DM mass boundary of $\unit[5]{GeV/c^{2}}$ was chosen as
that is the end of the reach of the current LUX limit from~\cite{Akerib:2017kat}
and at those higher masses, the traditional elastic recoils result
in a stronger signal than the Bremsstrahlung or Migdal effects. 

\begin{figure}[p]
\begin{centering}
\includegraphics[scale=0.7]{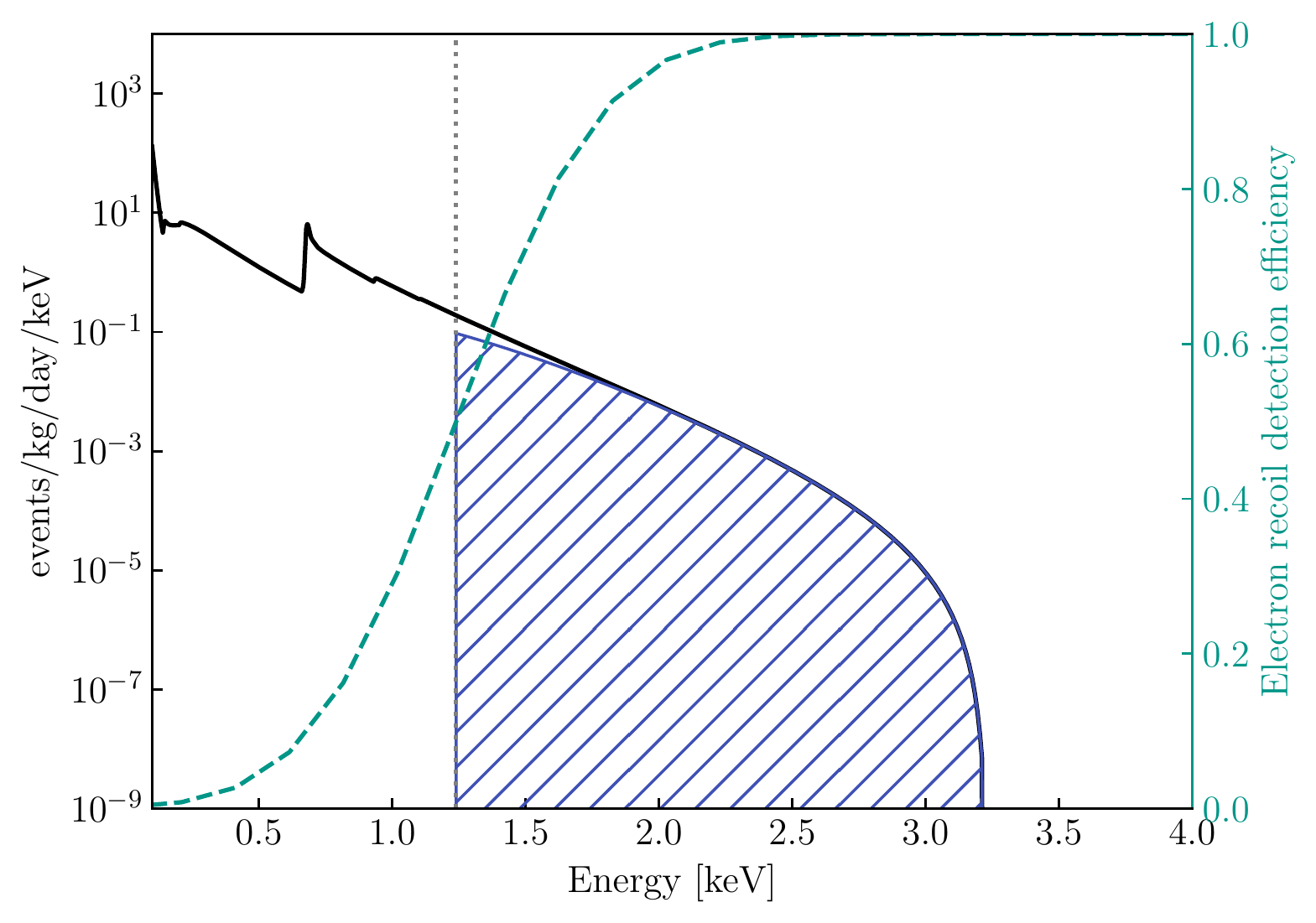}
\par\end{centering}
\caption[Scattering event rate from the Migdal effect with cuts]{Illustration of the DM-nucleus scattering event rate from the Migdal
effect with a heavy scalar mediator (solid black line) for $m_{\chi}=\unit[0.4]{GeV/c^{2}}$
at $\sigma=\unit[1\times10^{-35}]{cm^{2}}$. Also shown is the efficiency
from the \textit{in situ} tritium measurements performed by the LUX
detector (dashed teal line) and the event rate considered for this
analysis (hatched blue area) with tritium efficiency and a 1.24 keV
energy threshold (dotted gray line) applied. \label{fig:Migdal-event-rate-LUX}}

\vspace{0.9cm}
\centering{}\includegraphics[scale=0.27]{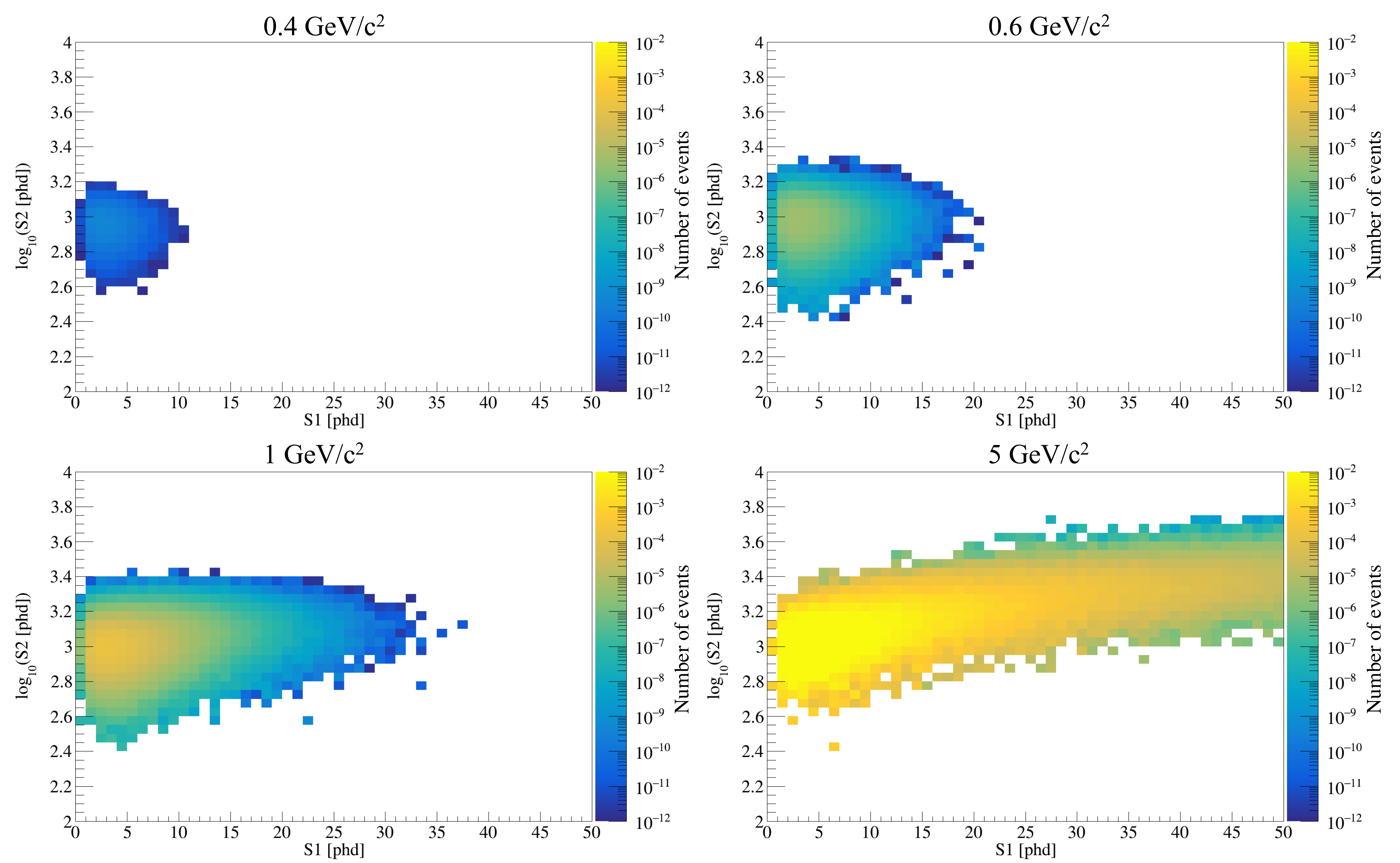}\caption[S1-log$_{10}$S2 signal from the Migdal effect as seen in LUX]{The expected signal from DM-nucleus interactions through Migdal effect
at $\sigma=\unit[1\times10^{-35}]{cm^{2}}$ projected onto a two-dimensional
space of log$_{10}$S2 as a function of S1 for four different DM masses.
Additional data quality cuts and detector efficiencies were applied.
Events in S1 bin smaller than 1 are omitted in the analysis. \label{fig:Migdal0-S1log10S2}}
\end{figure}

The expected event rate for a $\unit[1]{GeV/c^{2}}$ DM particle,
the detector ER efficiency, and the low-energy cutoff are illustrated
in Figure~\ref{fig:Migdal-event-rate-LUX}. The resulting signal
model is projected on the two-dimensional space of S1-log$_{10}$S2
with all analysis cuts applied as shown in Figure~\ref{fig:Migdal0-S1log10S2}
for various DM masses. S1 pulses are required to have at least a two-PMT
coincidence and be in the range of 1-50 phd. Since both S1 and S2
pulses are corrected for geometrical effects, S1 can have values below
2.0 phd even when the two-fold photon coincidence is satisfied. 

\begin{figure}
\begin{centering}
\includegraphics[scale=0.48]{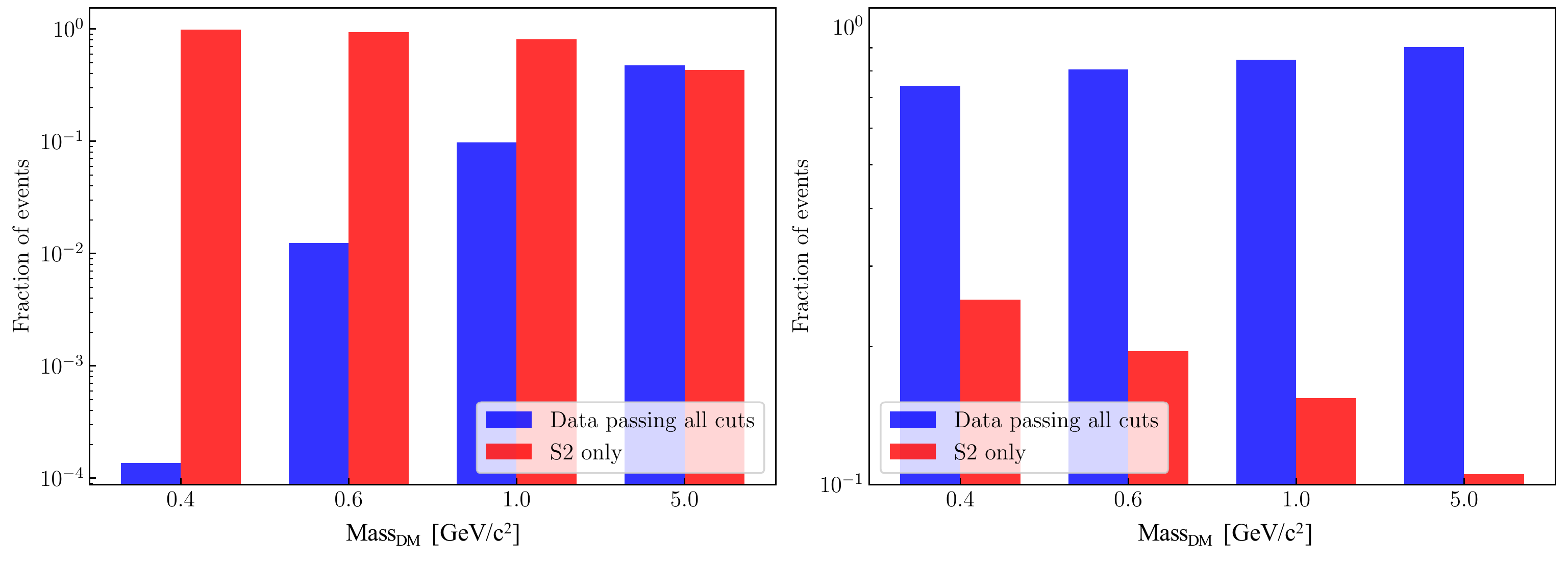}
\par\end{centering}
\caption[Total number of events expected from Bremsstrahlung and Migdal effects]{The number of events passing all cuts used in the analysis (blue)
expected from Bremsstrahlung (left) and Migdal (right). Also shown
is the total number of events resulting in an S2-only signal (red).
Each plot was normalized to the number of events without any analysis
cuts applied.\label{fig:Signal-ratios}}
\end{figure}

Figure~\ref{fig:Signal-ratios} shows the total number of events
expected from Bremsstrahlung and Migdal effects without any analysis
cuts applied, along with the number of events with only an S2 signal
and the total number of events passing all cuts used in the analysis.

\subsection{Background model}

The background model used in this work is identical to that used in~\cite{Akerib:2015rjg}.
An important distinction between~\cite{Akerib:2015rjg} and this
work is that the sub-GeV signal from both the Bremsstrahlung and the
Migdal effects would result in additional events within the ER classification,
as identified by the ratio of S2 to S1 size. Whereas the standard
WIMP search looks for events in the NR band and is essentially background-free,
only dominated by a small leakage of ER events, the sub-GeV signal
from these effects sees all ER events as background. The ER band is
populated significantly, with contributions from $\gamma$ rays and
beta particles from radioactive contaminations within the xenon, detector
instrumentation, and external environmental sources as described in~\cite{Akerib:2014rda}
and discussed in Section~\ref{sec:Simulations-and-backgrounds}.
Systematic uncertainties in background rates are treated via nuisance
parameters in the likelihood; these constraints are listed in Table~\ref{tab:WS2013-nuisance}. 

\begin{table}
\begin{centering}
\begin{tabular}{lrr}
\hline 
Parameter & Constraint & Fit value\tabularnewline
\hline 
\hline 
Low-$z$- origin $\gamma$ counts & $172\pm74$ & $165\pm16$\tabularnewline
Other $\gamma$ counts & $247\pm106$ & $228\pm19$\tabularnewline
$\beta$ counts & $55\pm22$ & $84\pm15$\tabularnewline
$^{127}$Xe counts & $91\pm27$ & $78\pm12$\tabularnewline
$^{37}$Ar counts & N/A & $12\pm8$\tabularnewline
Wall counts & $24\pm7$ & $22\pm4$\tabularnewline
\hline 
\end{tabular}
\par\end{centering}
\caption[Nuisance parameters in the WS2013 re-analysis likelihood model]{Nuisance parameters in the WS2013 re-analysis likelihood model. Constraints
are Gaussian with means and standard deviations indicated. Event counts
are after cuts and analysis thresholds. The best-fit model has zero
contribution from the signal PDF. In this case, the signal-model parameters
float to the central values of their constraints, and so are not listed.
Table from~\cite{Akerib:2015rjg}. \label{tab:WS2013-nuisance}}
\end{table}

\subsection{Light yield variation}

\begin{figure}[p]
\begin{centering}
\includegraphics[scale=0.3]{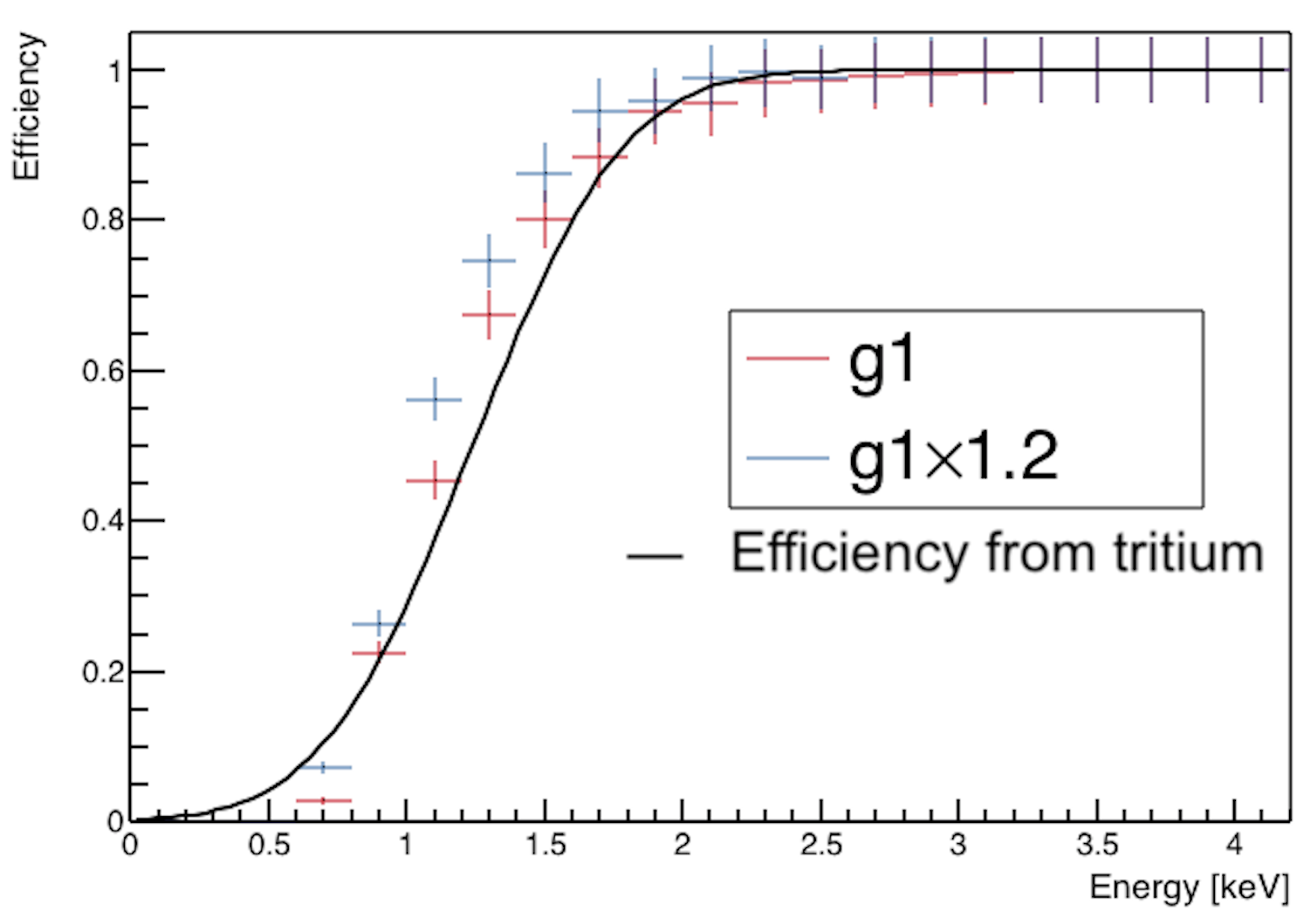}
\par\end{centering}
\caption[Effects of varying $g_{1}$]{Effects of varying $g_{1}$ using NEST v2.0 compared to the efficiency
fit from LUX tritium data (black line). Nominal $g_{1}=\unit[0.117]{phd}$
per photon is shown in red while the efficiency resulting from a $g_{1}$
increase by 20\% is shown in blue. \label{fig:g1_nest}}

\vspace{2cm}
\centering{}\includegraphics[scale=0.7]{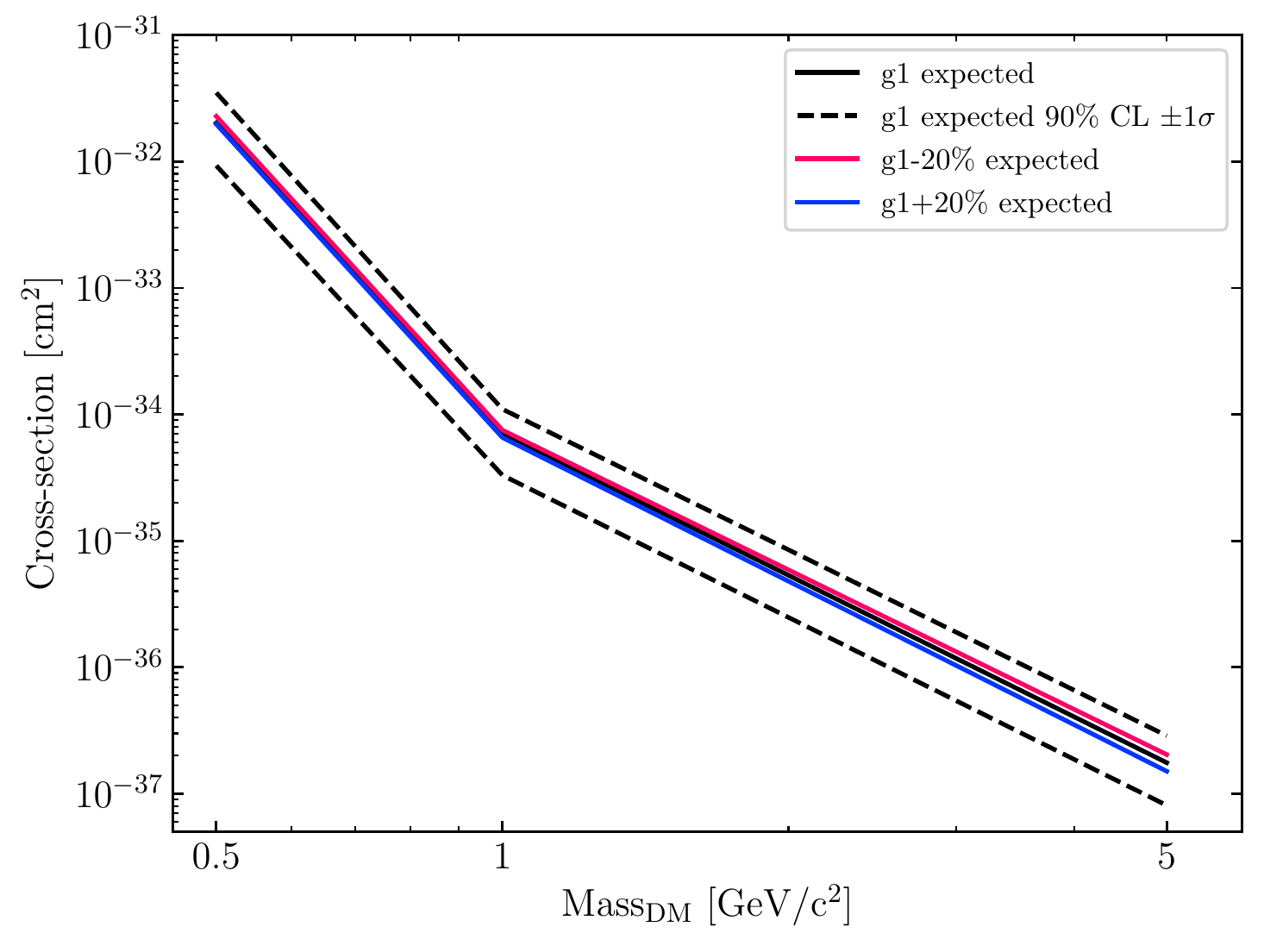}\caption[Effects of $g_{1}$ variation on limit setting]{Effects of varying $g_{1}$ for the case of a heavy scalar mediator
considering Bremsstrahlung signal model. The expected upper limit
on the spin-independent DM-nucleon cross section at 90\% C.L with
nominal $g_{1}=\unit[0.117]{phd}$ per photon is shown in black with
the $1\sigma$ range of background-only trials shown in dashed black.
The expected limit while varying $g_{1}$ by +20\% is shown in pink
and by -20\% is shown in blue.\label{fig:Effects-of-varying-g1}}
\end{figure}

Systematic uncertainties on the limit from light yield have been studied
but were not included in the final profile likelihood ratio (PLR)
statistic since their effects were negligible. The impact of $g_{1}$
variation on efficiency is illustrated in Figure~\ref{fig:g1_nest}.
For the tests, $g_{1}$ was varied by $\pm20\%$ at three different
masses resulting in 9\% difference at $0.\unit[5]{GeV/c^{2}}$, 6\%
difference at $\unit[1]{GeV/c^{2}}$, and 1\% difference at $\unit[5]{GeV/c^{2}}$.
This is because changing the light yield only affects a small fraction
of events passing all analysis cuts due to the applied energy cutoff.
The effect on the final limit is illustrated in Figure~\ref{fig:Effects-of-varying-g1}.
It should be noted that 20\% variation was chosen as a case study,
but $g_{1}$ is expected to vary by less than 20\% in the analysis
as the error on light yield obtained from the tritium measurements
ranges from 15\% at the 1.24 keV threshold to sub-1\% at higher energies.

\section{Results}

\begin{figure}[t]
\begin{centering}
\includegraphics[scale=0.4]{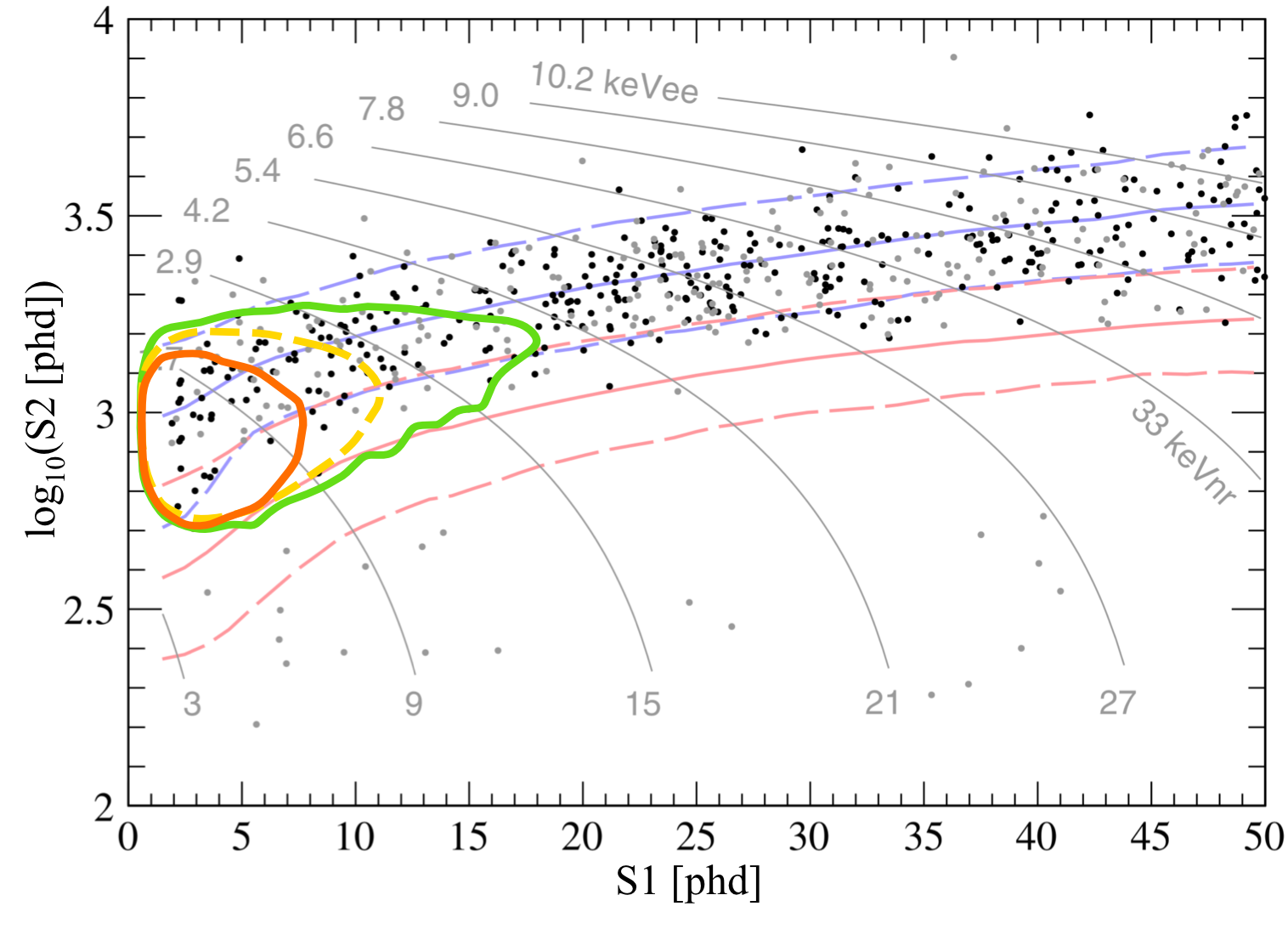}
\par\end{centering}
\caption[Contours containing 95\% of expected DM signal from the Migdal effect]{Contours containing 95\% of expected DM signal from the Migdal effect.
The solid contours are for a light vector mediator for $m_{\chi}=\unit[0.4]{GeV/c^{2}}$
(light blue) and $m_{\chi}=\unit[5]{GeV/c^{2}}$ (teal). Dashed yellow
contour assumes a heavy vector mediator and $m_{\chi}=\unit[1]{GeV/c^{2}}$.
The contours are overlaid on 591 events observed in the region of
interest from the 2013 LUX exposure of 95 live days and 145 kg fiducial
mass (cf. Reference~\cite{Akerib:2015rjg}). Points at <18 cm radius
are black; those at 18-20 cm are gray since they are more likely to
be caused by background events near the detector walls. Distributions
of uniform-in-energy electron recoils (blue) and an example $\unit[50]{GeV/c^{2}}$
WIMP signal (red) are indicated by 50$^{\mathrm{th}}$ (solid), 10$^{\mathrm{th}}$,
and 90$^{\mathrm{th}}$ (dashed) percentiles of S2 at given S1. Gray
lines, with ER scale of keVee at the top and Lindhard model NR scale
of keVnr at the bottom, are contours of the linear-combined S1-and-S2
energy estimator~\cite{Shutt:2007zz}.\label{fig:Migdal-contours}}
\end{figure}

The sub-GeV dark matter signal hypotheses were tested with a double-sided
PLR statistic described in Section~\ref{subsec:Profile-likelihood-ratio},
scanning over spin-independent DM-nucleon cross section at each DM
mass. For each DM mass, a scan was performed over cross sections to
construct a 90\% confidence interval, with the test statistic distribution
evaluated by Monte Carlo sampling using the\textsc{ RooStats} package~\cite{RooStats}.
For an illustration of the expected location of the signal in the
S1-log$_{10}$S2 detector space, contours for various DM masses with
different mediators in the case of the Migdal effect are overlaid
on the observed events from WS2013 shown in Figure~\ref{fig:Migdal-contours}. 

\begin{figure}
\begin{centering}
\includegraphics[scale=0.9]{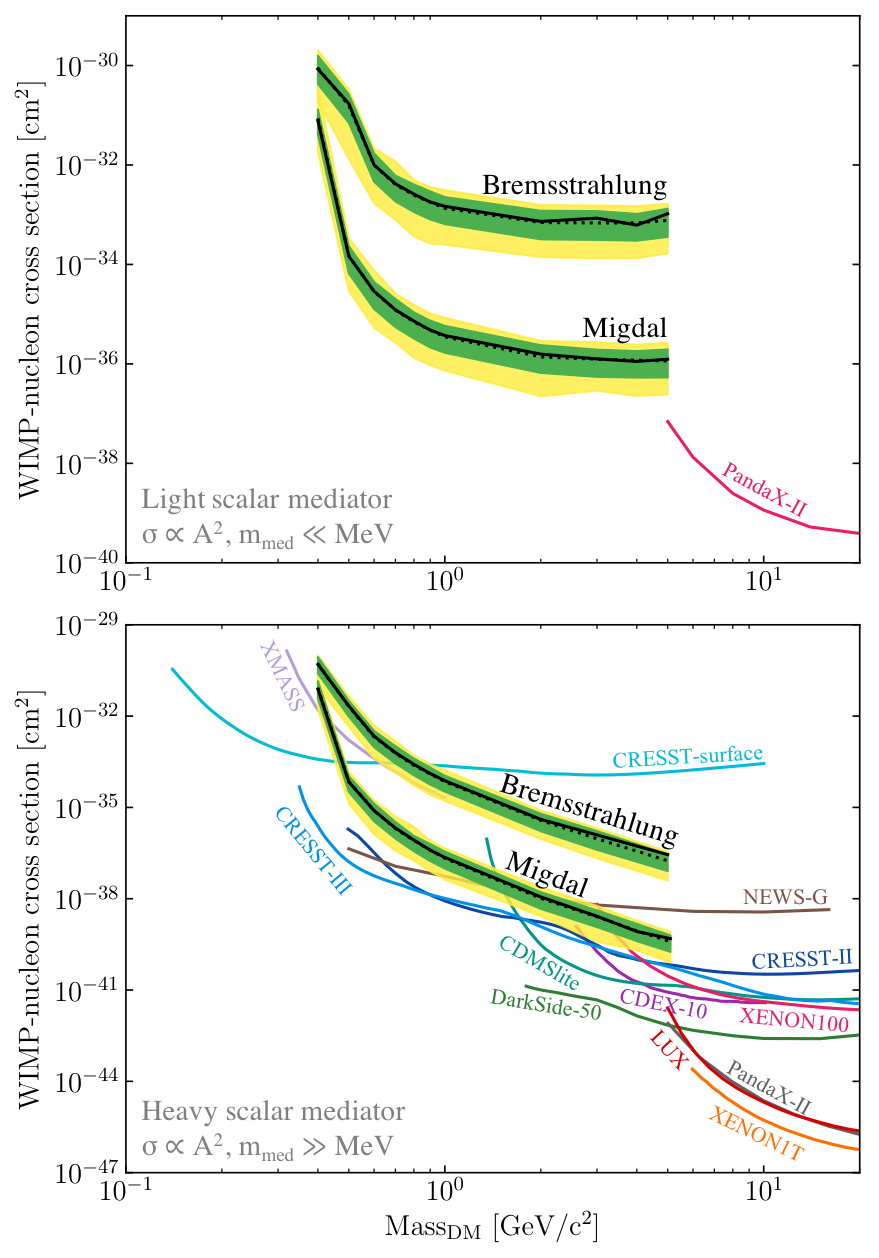}
\par\end{centering}
\caption[Upper limits on Bremsstrahlung and Migdal effects assuming a scalar
mediator]{Upper limits on the spin-independent DM-nucleon cross section at
90\% C.L. as calculated using Bremsstrahlung and Migdal effect signal
models assuming a scalar mediator (coupling proportional to mass).
The 1- and 2-$\sigma$ ranges of background-only trials for this result
are shown in green and yellow, respectively, with the median limit
shown as a black dashed line. The top figure shows limit for a light
mediator with $\unit[q_{\mathrm{ref}}=1]{MeV}$. Also shown is a limit
from PandaX-II~\cite{Ren:2018gyx} (pink). The bottom figure shows
limits for a heavy mediator along with limits from the spin-independent
analyses of LUX~\cite{Akerib:2017kat} (red), PandaX-II~\cite{Cui:2017nnn}
(gray), XENON1T~\cite{Aprile:2018dbl} (orange), XENON100 S2-only~\cite{Aprile:2016wwo}
(pink), CDEX-10~\cite{Jiang:2018pic} (purple), CDMSlite~\cite{Agnese:2015nto}
(teal), CRESST-II~\cite{Angloher:2015ewa} (dark blue), CRESST-III~\cite{Petricca:2017zdp}
(light blue), CRESST-surface~\cite{Angloher:2017sxg} (cyan), DarkSide-50~\cite{Agnes:2018ves}
(green), NEWS-G~\cite{Arnaud:2017bjh} (brown), and XMASS~\cite{Abe:2018mxq}
(lavender).\label{fig:limits-scalar}}
\end{figure}

\begin{figure}
\begin{centering}
\includegraphics[scale=0.9]{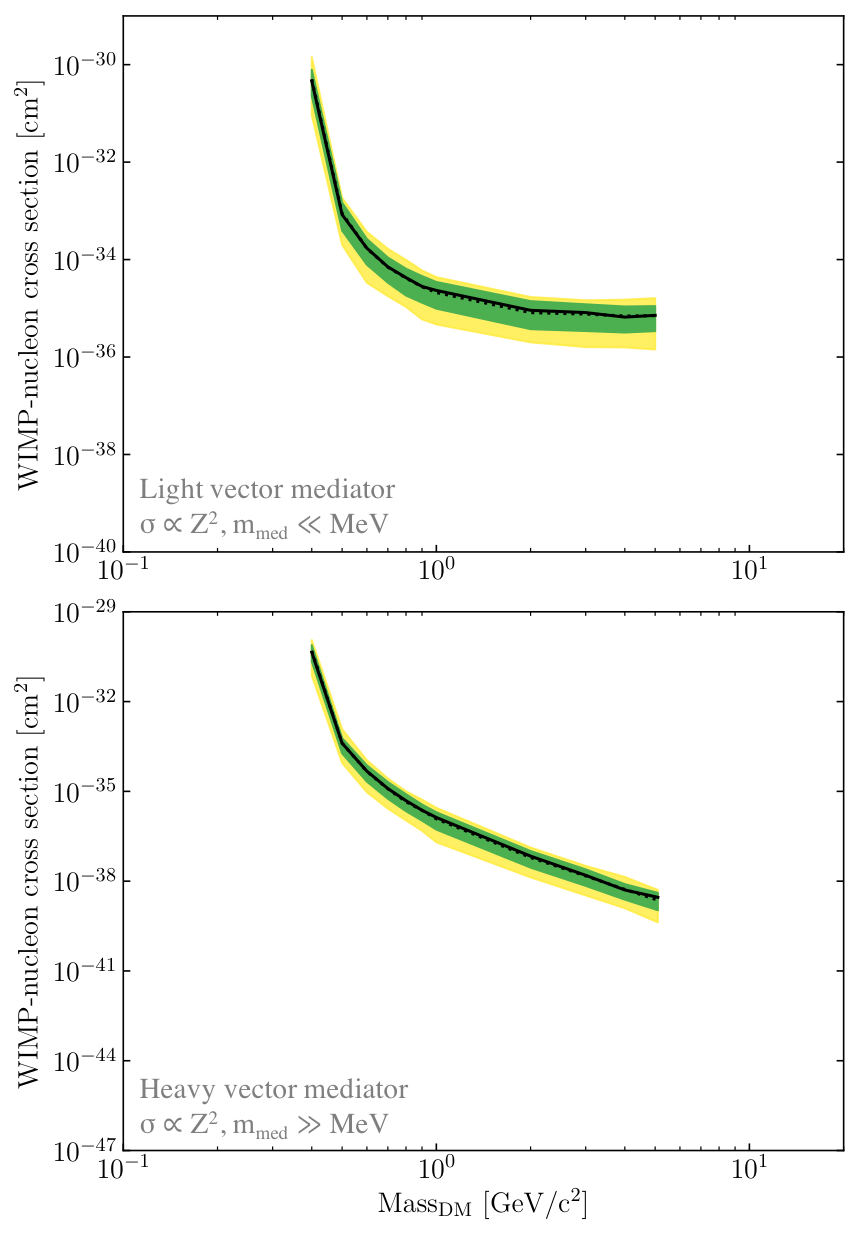}
\par\end{centering}
\caption[Upper limits on Migdal effect assuming a vector mediator]{Upper limits on the spin-independent DM-nucleon cross section at
90\% C.L. as calculated using the Migdal effect signal model assuming
a vector mediator (coupling proportional to charge). The 1- and 2-$\sigma$
ranges of background-only trials for this result are shown in green
and yellow, respectively, with the median limit shown as a black dashed
line. The top figure shows limit for a light vector mediator with
$\unit[q_{\mathrm{ref}}=1]{MeV}$. The bottom figure shows limit for
a heavy vector mediator.\label{fig:limits-vector}}
\end{figure}

Upper limits on cross section for DM masses from $\unit[0.4-5]{GeV/c^{2}}$
for both Bremsstrahlung and Migdal effect assuming both a light and
a heavy scalar mediator are shown in Figure~\ref{fig:limits-scalar}.
Upper limits for a light and a heavy vector mediator for the Migdal
effect only are shown in Figure~\ref{fig:limits-vector}. The observed
events are consistent with the expectation from background only. 


\subsection{Comments}

It should be noted that Bremsstrahlung and Migdal effects are anticipated
to be present in the DD calibration data (see Section~\ref{subsec:DD}),
but have not yet been observed. Even if observed, Bremsstrahlung and
Migdal effects would not be recognizable from each other in the analysis,
and in fact, the Migdal effect is expected to dominate due to its
higher scattering rates. In the current analysis, the electronic energy
injections from Bremsstrahlung and Migdal effects likely enhance the
S1-log$_{10}$S2 ratio of the NR signal that dominates. However, these
effects are taken into account by the detector NR calibration for
a given momentum transfer. 

For a detector with a lower NR detection threshold, an analysis could
place a limit considering contributions from both NR from elastic
scattering and the ER signals presented here. In that case, care should
be taken to avoid double-counting of signals. 

Note that the Migdal effect can also be used in the future to detect
coherent neutrino-nucleus scattering (C$\nu$NS) both from \textit{pp}
and $^{8}$B solar neutrinos, where the signal is otherwise below
threshold (cf. Figure~6 in~\cite{Ibe:2017yqa}).

\section{Conclusion}

The contributions from the Bremsstrahlung and Migdal effects extend
the sensitivity of the LUX detector to light DM masses previously
inaccessible via the elastic nuclear recoil detection method alone.
At low DM masses, the recoil signal from the Bremsstrahlung photon
and the electrons from Migdal effect is stronger and at higher energies
than the elastic NR scattering signal. The presence of these signals
in the ER channel is significant because LUX has a higher ER detection
efficiency at low recoil energies. Additionally, this search extends
the reach of LUX by an order of magnitude in mass without any changes
to the detector hardware and sets the first limits on the vector mediator
model among direct detection experiments.

This analysis places limits on spin-independent DM-nucleon scattering
cross sections assuming both scalar and vector, and light and heavy
mediators to DM masses of $\unit[0.4]{GeV/c^{2}}$ from $\unit[5]{GeV/c^{2}}$.
The final limits achieved using the Migdal effect are competitive
with limits from detectors dedicated to searches of light dark matter.
The observations from LUX are consistent with background only. Nevertheless,
this analysis is readily applicable to the next-generation DM detectors,
such as LZ~\cite{Akerib:2018lyp}, by extending the DM search to
sub-GeV masses.

\chapter{\textsc{High voltage development for the LZ experiment}\label{chap:High-voltage-development-LZ}}

The LUX-ZEPLIN (LZ) experiment, a successor to the LUX detector, is
a next-generation xenon TPC that will be used to search for dark matter.
As noble liquid detectors grow larger in size, new design challenges
arise, one of the most prominent being the higher cathode voltage
needed to reach the desired drift field. The next two chapters focus
on topics associated with high voltage and electric fields in these
detectors. In particular, this chapter discusses high voltage delivery
to the cathode of the LZ detector\footnote{\textbf{Why is the detector cathode biased to HV rather than the anode?}
Despite anodic surfaces providing better high voltage (HV) performance~\cite{Gerhold1989},
all two-phase xenon TPCs are designed to have their cathode, rather
than the anode, biased to HV. Negative HV delivered to the cathode
can be stepped down to the gate, and only a smaller positive HV can
be delivered to the anode. Biasing the cathode to HV rather than the
anode is mostly driven by the desire to minimize detector surfaces
that are set to HV while optimizing the use of LXe. Furthermore, the
space around the anode-gate region is one of the most technically
challenging since the extraction region needs to be at a high field
to enable the electroluminescence signal, while being in the proximity
of PMTs and needing to regulate liquid height, among other things.
Additionally, the anode is in gaseous xenon, and gas will break down
much more readily than liquid.} whose design, construction, and tests in liquid argon (LAr)\nomenclature{LAr}{Liquid Argon}
are underway at LBNL. Tests are performed in LAr instead of liquid
xenon (LXe) because LAr is inexpensive and can be thrown away after
each test, obviating the need for a complex and expensive xenon gas
handling system. Chapter~\ref{chap:XeBrA} describes the development,
construction, and results from the Xenon Breakdown Apparatus (XeBrA)
designed to characterize the breakdown behavior in LAr and LXe.

Since the purity of LAr affects dielectric breakdown behavior, it
is essential to provide a purity measurement in these tests. To that
end, I designed and built a compact purity monitor. The purity monitor
is a small double-gridded drift chamber fully submerged in LAr. Electrons
are extracted from a photocathode via photoelectric effect, and LAr
purity is measured as a function of the number of electrons that attach
to impurities. The purity monitor can be operated at a wide range
of electric fields which allows this purity monitor with a total length
of 7 cm to measure electron lifetimes of up to 1 ms. 

This chapter first provides a brief overview of the LZ experiment
(Section~\ref{sec:The-LUX-ZEPLIN-experiment}), the challenges of
HV design in two-phase TPCs (Section~\ref{subsec:LZ-High-voltage-delivery}),
and the development of the HV connection for the LZ detector (Section~\ref{subsec:LZ-high-voltage-connection}).
The second part of this chapter focuses on the design and operating
principles of liquid noble purity monitors (Section~\ref{sec:Liquid-argon-purity}).
After the successful construction and testing of the purity monitor
for LZ (Section~\ref{subsec:Purity-monitor-testing}), a second purity
monitor was developed for the XeBrA experiment (Section~\ref{sec:Purity-monitor-for-XeBrA}).
Additionally, a third purity monitor is currently under construction,
closely following the design of XeBrA's purity monitor, to be deployed
in an experiment studying optical reflectivity of PTFE in liquid xenon. 

\section{The LUX-ZEPLIN experiment\label{sec:The-LUX-ZEPLIN-experiment}}

The LUX-ZEPLIN (LZ) experiment was formed by a merger of two collaborations
- LUX (discussed in Chapters~\ref{chap:The-LUX-experiment},~\ref{chap:efield-modeling},
and~\ref{chap:Searching-for-sub-GeV-dm}) and ZEPLIN (ZonEd Proportional
scintillation in LIquid Noble gases)\footnote{ZEPLIN was a pioneering series of single- and two-phase xenon experiments
conducted at Boulby, UK.}~\cite{Lebedenko:2008gb,Akimov:2011tj,Akimov:2006qw}. The LZ collaboration
consists of $\sim250$ physicists and engineers from 31 institutions
in the USA, UK, Portugal, and South Korea. Its goal is to explore
the bulk of the region remaining above the neutrino floor with sensitivities
to spin-independent (SI) and spin-dependent (SD) cross sections of
$10^{-48}$ $\unit[\left(10^{-46}\right)]{cm^{2}}$ for a 40~GeV/c$^{2}$
WIMP mass after an exposure of 1,000~live-days. To achieve that,
LZ will feature many upgrades compared to its predecessors: a significantly
larger active xenon mass, a powerful nested veto system, new calibration
systems, and a rigorous cleanliness program including comprehensive
radio-assay and dust reduction programs. This section draws from the
recent LZ sensitivity prediction publication~\cite{Akerib:2018lyp};
further technical details about LZ design can be found in the Conceptual
Design Report from September 2014~\cite{Akerib:2015cja} and in the
Technical Design Report from March 2017~\cite{Mount:2017qzi}.

Similarly to LUX, the LZ detector will operate 1478~m (4850~ft)
below ground at the Sanford Underground Research Facility (SURF) in
the Davis Cavern, which has undergone many modifications to accommodate
the upgrades planned for the LZ detector. The LZ detector will be
located inside the water tank inherited from the LUX experiment. The
water tank will be filled with 228~tonnes of ultrapure water and
instrumented by PMTs surrounded by Tyvek diffuse reflectors to improve
light efficiency. As in LUX, the water tank provides shielding from
muons, neutrons, and $\gamma$-rays from the naturally occurring radiation
in the cavern rock and the PMTs.

\begin{figure}
\begin{centering}
\includegraphics[scale=0.65]{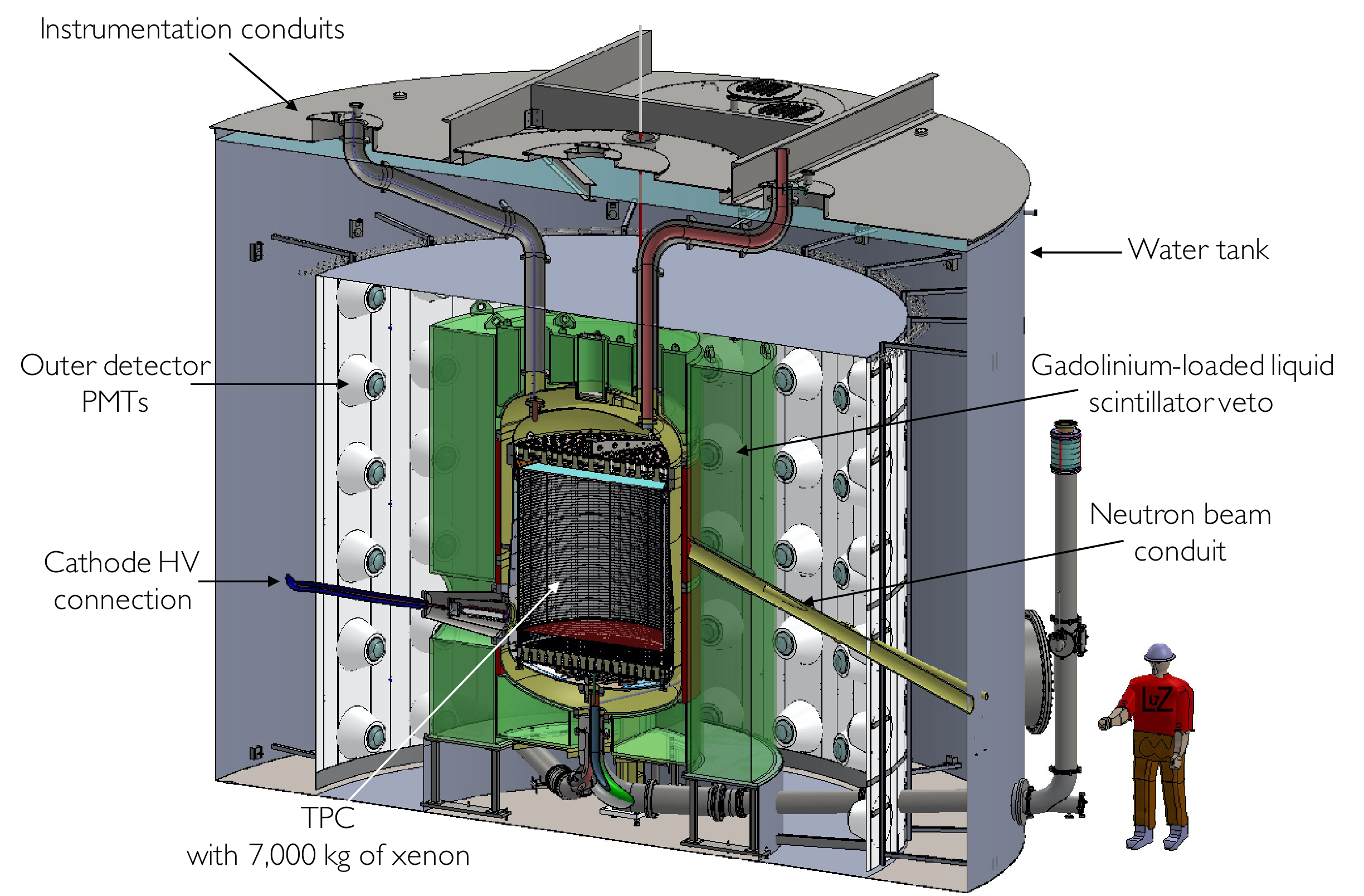}
\par\end{centering}
\caption[CAD rendering of the LZ detector]{A cutaway of a CAD rendering of the LZ detector. PMT and other cables
from the bottom of the TPC exit through a port at the bottom of the
cryostat. Figure modified from~\cite{Akerib:2018lyp}.\label{fig:LZ-detector}}
\end{figure}

The LZ detector will be housed inside outer and inner cryostats made
from titanium with some of the lowest ever reported intrinsic activities~\cite{Akerib:2017iwt}.
The TPC is cylindrical with both height and diameter equal to 1.46~m
as shown in Figure~\ref{fig:LZ-detector}. The 7~tonnes of active
xenon are viewed by 494~PMTs: 241 at the bottom and 253 at the top.
The active volume is surrounded by segmented, cylindrical, and highly
reflective PTFE panels that embed 57~field shaping rings. The xenon
circulation and purification strategies are based on the LUX system. 

\begin{figure}
\begin{centering}
\includegraphics[scale=0.12]{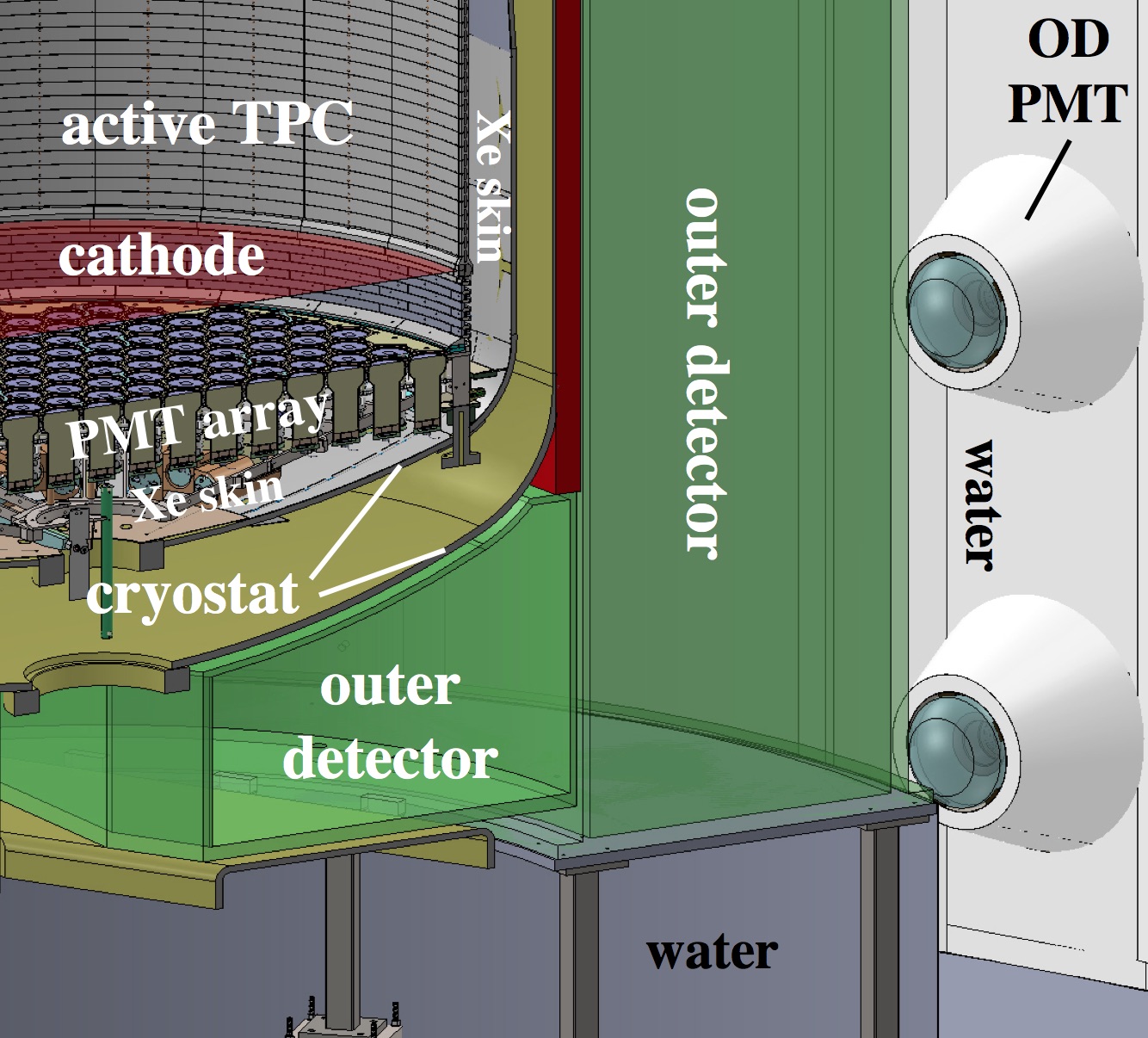}
\par\end{centering}
\caption[An expanded view of the lower right corner of the LZ cathode region]{An expanded view of the lower right corner of the LZ cathode region.
\textquoteleft OD PMT\textquoteright{} indicates the outer detector
photomultiplier tubes. The xenon skin region is observed by an independent
set of PMTs not depicted. Figure from~\cite{Akerib:2018lyp}.\label{fig:LZ-cathode-zoom}}
\end{figure}

One of the most notable LZ detector upgrades compared to the LUX detector
is the improved veto system. First, the active volume of the TPC is
surrounded by a xenon skin region illustrated in Figure~\ref{fig:LZ-cathode-zoom}
of a zoomed image of the cathode region. The xenon skin region is
composed of the outermost 2~tonnes of LXe located between the outside
of the PTFE panels and the inner titanium cryostat. The titanium walls
are lined with a thin layer of PTFE to improve light collection. The
xenon skin region is optically isolated from the main active volume
and is instrumented by PMTs. The xenon skin maximizes the use of LXe
inside the detector and helps reject scattered $\gamma$ rays that
could contribute to the backgrounds. 

\begin{figure}
\begin{centering}
\includegraphics[scale=0.43]{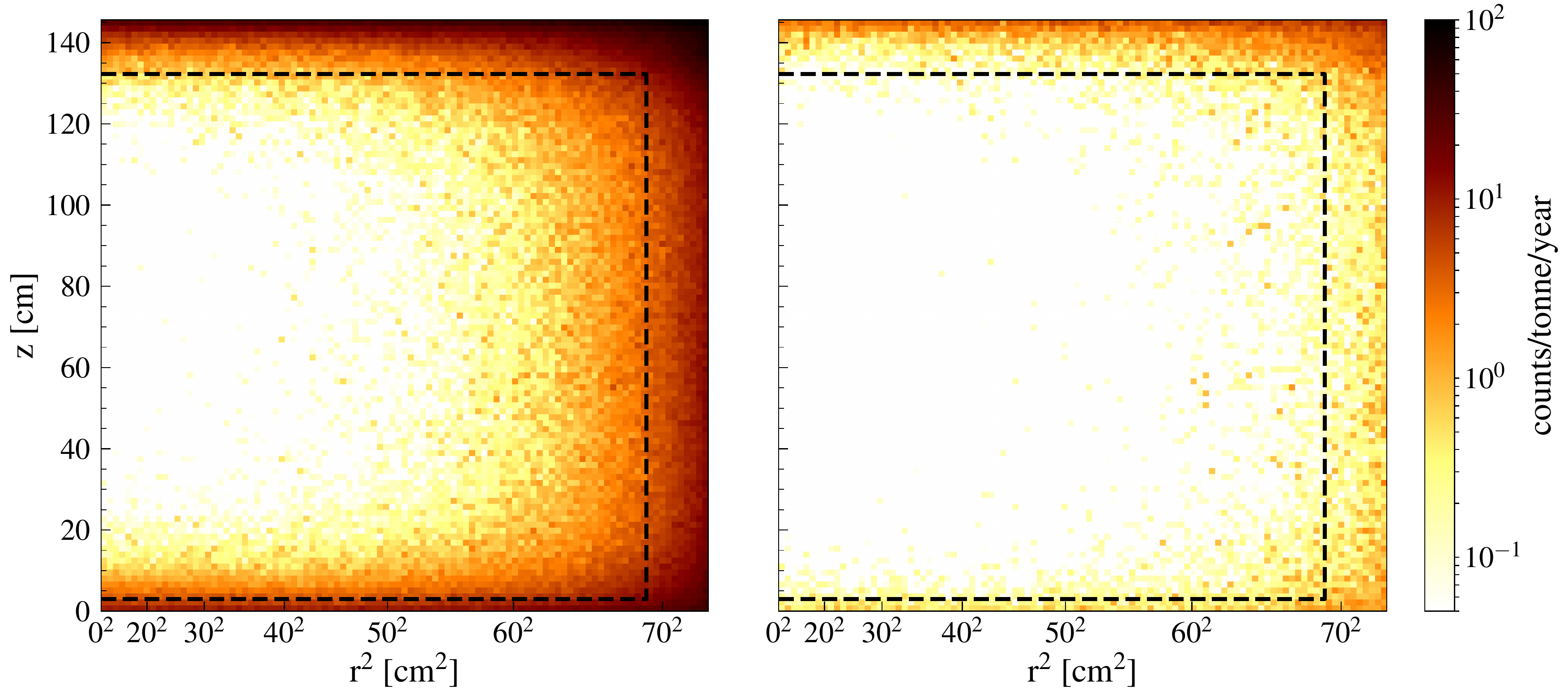}
\par\end{centering}
\caption[Single scatter events from all NR backgrounds in the LZ active volume]{Single scatter event distribution from all NR backgrounds in the
region of interest for a $\unit[40]{GeV/c^{2}}$ WIMP. The black dashed
line illustrates a 5.6 tonnes fiducial volume. \textbf{Left:} Event
distribution with no veto results in 12.31 counts/1,000 days. \textbf{Right:}
Only 1.03 counts/1,000 days remain after applying both xenon skin
and GdLS vetoes. Figure modified from~\cite{Akerib:2018lyp}.\label{fig:LZ-fiducial-bg}}

\end{figure}

The second veto component consists of 17 tonnes of gadolinium-loaded
liquid scintillator (GdLS)\nomenclature{GdLS}{Gadolinium-loaded Liquid Scintillator}
contained in 10 acrylic tanks viewed by the PMTs inside the water
tank. The GdLS veto is located just outside the outer cryostat as
depicted in green in Figures~\ref{fig:LZ-detector} and~\ref{fig:LZ-cathode-zoom}.
Its principal goal is to detect neutrons from $\left(\alpha,n\right)$
processes since single scatter neutrons represent the most dangerous
background for WIMP searches. If such a neutron single scatters inside
the TPC, the neutron will be thermalized and captured within $\sim\unit[30]{\mu s}$
in the GdLS. The Gd neutron capture results in $\unit[\sim8]{MeV}$
of $\gamma$ ray cascade, which can be detected by the skin or the
water tank PMTs, thus vetoing the event. The efficiency of this process
to identify a neutron that produces a single scatter event in LXe
above the analysis threshold is 96.5\%\footnote{This efficiency also includes neutrons that capture on the LAB hydrogen.}~\cite{Akerib:2018lyp,shaw2016dark}.
The entire veto system also makes possible an extended fiducial volume
cut as illustrated in Figure~\ref{fig:LZ-fiducial-bg}. Selected
parameters of the LZ detector are summarized in Table~\ref{tab:LZ-summary}.

\begin{table}[t]
\begin{centering}
\begin{tabular}{>{\raggedright}m{2cm}|l>{\raggedleft}p{4cm}>{\raggedleft}p{2.9cm}}
\hline 
Category & Parameter & Value & Note\tabularnewline
\hline 
\hline 
\multirow{4}{2cm}{Liquid xenon} & Total mass & 9,600 kg & \tabularnewline
 & TPC active mass & 7,000 kg & \tabularnewline
 & Skin mass & 2,000 kg & \tabularnewline
 & Fiducial mass & 5,600 kg & \tabularnewline
\hline 
\multirow{5}{2cm}{PMTs \\
(all from Hamamatsu)} & TPC & 253 (top) + \\
241 (bottom) & 3'' R11410-22\tabularnewline
 & Side skin (top) & 93 & 1'' R8520\tabularnewline
 & Side skin (bottom) & 20 & 2'' R8778\tabularnewline
 & Dome skin & 18 & 2'' R8778\tabularnewline
 & Water tank & 120 & 8'' R5912\tabularnewline
\hline 
\multirow{8}{2cm}{TPC dimensions (cold)} & Inner cryostat diameter & 1.58-1.66 m & \tabularnewline
 & Inner cryostat height & 1.83 m & \tabularnewline
 & Skin thickness & 40 mm (surface), \\
80 mm (cathode) & \tabularnewline
 & TPC diameter & 1,456 mm & \tabularnewline
 & Gate-anode & 13 mm & \tabularnewline
 & Cathode-gate & 1,456 mm & \tabularnewline
 & Bottom PMT-cathode & 137.5 mm & \tabularnewline
 & Field cage thickness & 15 mm & \tabularnewline
\hline 
\multirow{3}{2cm}{Electric fields} & Extraction region & 10.2 kV/cm & \tabularnewline
 & Drift field & 343 V/cm (baseline) & 686 V/cm (goal)\tabularnewline
 & Reverse field & 2.9 kV/cm (baseline) & 5.9 kV/cm (goal)\tabularnewline
\hline 
\multirow{3}{2cm}{Operations} & Electron lifetime & 850 $\mu$s & Lower limit\tabularnewline
 & Pressure & 1.8 bar & Pressure range\\
1.6 - 2.2 bar\tabularnewline
 & Equilibrium temperature & 175 K & \tabularnewline
\hline 
\end{tabular}
\par\end{centering}
\caption[Summary of key characteristics of the LZ detector]{Summary of key characteristics of the LZ detector. The inner cryostat
has a tapered radial profile in order to lower the magnitude of electric
fields around the cathode region near the bottom of the detector.
The ``Note'' column includes the diameter and model information
of the various PMT types used in the detector.\label{tab:LZ-summary}}
\end{table}

LZ will build on the extensive calibration experience from the LUX
detector, but its much larger size requires the development of additional
calibration methods. The goal of LZ is to detect WIMP-nucleus interactions
in LXe. To separate those events from background, it is imperative
to characterize the detector response to nuclear (NR) and electron
recoils (ER). This can be achieved through calibrations. Several internal
and external sources will be used for ER calibrations of the LXe response:
\begin{itemize}
\item Along with $\mathrm{^{83m}Kr}$ $\left(\unit[41.5]{keV}\right)$ that
was also used in LUX as described in Section~\ref{subsec:Metastable-krypton-83},
LZ will also use $\mathrm{^{131m}Xe}$ produced via decay of $\mathrm{^{131}I}$.
The $\mathrm{^{131m}Xe}$ nucleus decays into $\mathrm{^{131}Xe}$
by emitting a $\gamma$ with 163.9 keV and a half-life of 12 days. 
\item $^{3}$H ($\beta$ decay with Q $\unit[=18.6]{keV}$) and $^{14}$C
internal calibrations are also planned at several milestones throughout
data taking to provide a thorough mapping of ER light and charge yields.
Similarly to the tritium source, $^{14}$C will be injected as $^{14}$C-labeled
methane ($^{14}$CH$_{4}$). It transitions to $^{14}$N via a beta
decay with 5,730 year half-life and $\unit[Q=156]{keV}$. This calibration
was tested after the end of LUX dark matter search and is described
in~\cite{balajthy2018}. 
\item $^{228}\mathrm{Th}$ will be used to calibrate signal fidelity near
the edge of the TPC. $^{228}\mathrm{Th}$ is part of the thorium series,
which ends when $^{208}$Tl undergoes a beta decay with a half-life
of 3.1 minutes into $^{208}$Pb followed by a 2,615 keV $\gamma$
with 99.8\% branching fraction. External sources will be deployed
in 3 vertical stainless steel calibration tubes located in the vacuum
space between the inner and outer titanium cryostats. 
\end{itemize}
Similarly, various external calibrations are planned to measure detector
NR yields in LXe. 
\begin{itemize}
\item A DD (2,450 keV) generator, used in LUX and described in Section~\ref{subsec:DD},
will be used again. 
\item AmBe (maximum neutron energy of 11 MeV) and AmLi sources will be deployed
in the external calibration tubes. The AmLi source is preferable over
AmBe (that was used in LUX) due to its lower maximum neutron energy
of 1.5 MeV, resulting in an enhanced fraction of events at low recoil
energy of <10 keV. 
\item Additionally, photoneutron sources with well-defined kinematic endpoints
are being developed to calibrate low-energy NR and coherent scatters
from $^{8}$B neutrinos. One such photoneutron source is $^{88}$YBe,
which produces 1.836 MeV $\gamma$ along with 152~keV neutrons~\cite{Collar:2013xva},
resulting in an NR endpoint of $\unit[4.6]{keV}$. Several other photoneutron
sources are being considered as well. They will be deployed on-axis
on top of the outer cryostat vessel reaching just above the inner
cryostat. 
\end{itemize}
Furthermore, $^{220}\mathrm{Rn}$ will be used for xenon skin calibrations,
and separate calibrations are planned for the GdLS outer detector.
However, the calibration plan will likely change and evolve as the
needs of the detector become better known.

\begin{figure}
\begin{centering}
\includegraphics[scale=0.41]{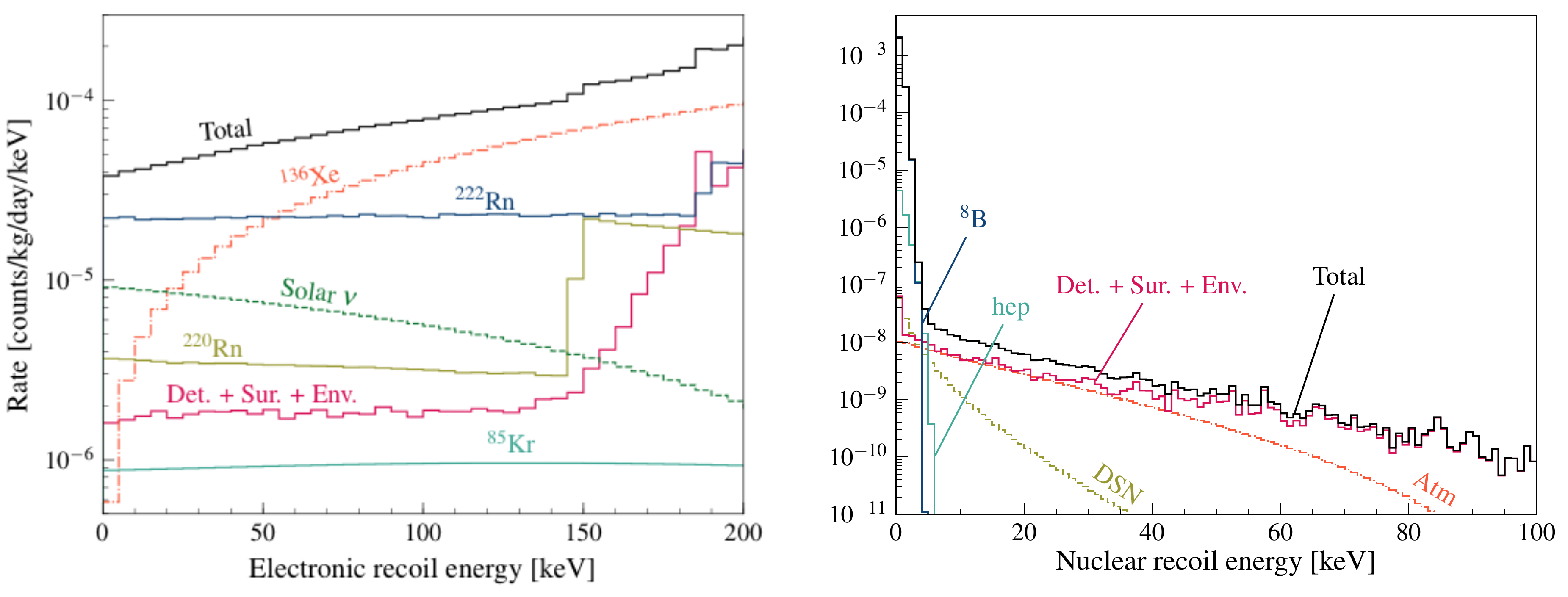}
\par\end{centering}
\caption[ER and NR background spectra in the 5.6 tonne fiducial volume]{ Background spectra in the 5.6 tonne fiducial volume for single scatter
events with neither a xenon skin nor a GdLS veto signal. No detector
efficiency or WIMP-search region cuts have been applied. \textbf{Left:}
ER.\textbf{ Right:} NR. Figures from~\cite{Akerib:2018lyp}.\label{fig:LZ-ER-NR-background}}

\end{figure}

LZ will use a simulation framework called BACCARAT based on the \textsc{Geant4}-based
LUXS\textsc{im} package. The anticipated amount of background events
for LZ is being carefully tracked and a table with an overview of
estimated backgrounds can be found in~\cite{Akerib:2018lyp}. Figure~\ref{fig:LZ-ER-NR-background}
shows the spectral contribution for ER and NR backgrounds in the fiducial
volume. As can be seen, radon presents the largest contribution to
the total number of background events, while atmospheric neutrinos
present the most significant contribution to NR counts. The expected
number of total background events after all cuts is 5.97~counts of
ER and 0.52~counts of NR in the region of interest. 

\begin{figure}
\begin{centering}
\includegraphics[scale=0.35]{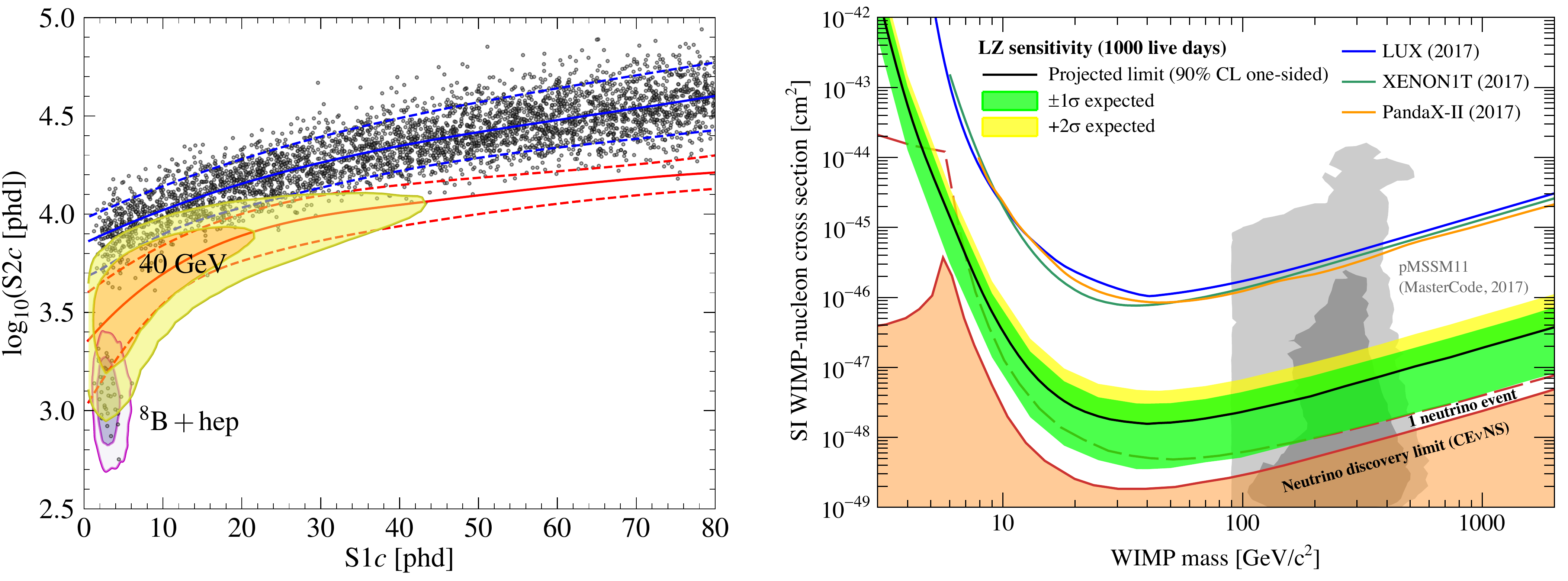}
\par\end{centering}
\caption[Expected results from 1,000~live-days with 5.6 tonne fiducial mass]{Expected results from 1,000~live-days with 5.6 tonne fiducial mass.
\textbf{Left:} LZ background-only simulated dataset. The mean (solid)
and the 10 and 90\% contours (dashed) indicate the NR (red) and ER
(blue) bands. Also shown (yellow shaded) are 1- and 2-$\sigma$ regions
for an expected $\unit[40]{GeV/c^{2}}$ WIMP signal and $^{8}$B and
$hep$ NR backgrounds (purple contours). \textbf{Right:} LZ projected
sensitivity to SI WIMP-nucleon elastic scattering for 1,000~live-days
and a 5.6 tonne fiducial mass. The lower shaded region bounded by
the red dashed line indicate the emergence of backgrounds from neutrino
coherent scattering~\cite{PhysRevD.90.083510,Billard:2013qya}. Gray
contoured areas show the favored regions from pMSSM11~\cite{Bagnaschi:2017tru}.
Figures from~\cite{Akerib:2018lyp}.\label{fig:LZ-limit}}
\end{figure}

Similarly to LUX, LZ will employ salting to prevent analysis biases,
and once the final dataset is established, PLR will be used to set
a limit. There are eleven types of background considered as nuisance
parameters in the PLR code. Figure~\ref{fig:LZ-limit} shows the
separation of the ER and NR bands with simulated 1,000~live-days
of background-only data and the expected limit if a WIMP signal is
not observed. 

\subsection{Modeling electric fields in the LZ detector\label{subsec:Modeling-electric-fields-LZ}}

The electric fields inside the TPC are set by the field shaping rings
built at LBNL and four mesh grids weaved on a custom-built loom at
SLAC National Accelerator Laboratory (SLAC)\nomenclature{SLAC}{SLAC National Accelerator Laboratory}.
The grid parameters are summarized in Table~\ref{tab:Grid-properties-LZ}. 

\begin{table}
\centering{}%
\begin{tabular}{l>{\raggedleft}p{1.5cm}>{\raggedleft}p{1.4cm}>{\raggedleft}p{1cm}>{\raggedleft}p{3.5cm}}
\toprule 
Grid & $z$ 

{[}cm{]} & Wire$\,\diameter$ {[}$\mu$m{]} & Pitch {[}mm{]} & Operating voltage {[}kV{]}\tabularnewline
\midrule
\midrule 
Anode & 161.65 & 100 & 2.5 & 5.75\tabularnewline
Gate & 160.35 & 75 & 5.0 & -5.75\tabularnewline
Cathode & 14.75 & 100 & 5.0 & -50.00\tabularnewline
Bottom shield & 1.00 & 75 & 5.0 & -1.50\tabularnewline
\bottomrule
\end{tabular}\caption[Grid properties and voltages as designed in LZ]{Grid properties and voltages as designed for the LZ detector. All
grids are wire meshes woven from 304 stainless steel from California
Fine Wire company. A linear voltage grading is established by a series
of resistors between cathode and gate. The face of the bottom PMT
array is defined at $z=0$. The liquid level is assumed to be 5 mm
above the gate grid.\label{tab:Grid-properties-LZ}}
\end{table}

Building a full 3D model of the detector, as was done for the LUX
detector (Chapter~\ref{chap:efield-modeling}), is not simple and
is not always necessary. Most often, a much simpler, axially symmetric
model will suffice to improve understanding the behavior of electric
fields inside a detector's active volume. However, most TPCs contain
grids with either parallel or mesh wires, which break axial symmetry.
It is essential to understand the transformation of the detector geometry
and its grids between a full 3D model and an axially symmetric model
to account for these changes since as was shown in Figure~\ref{run3_field},
grid transparency has a substantial effect on field leakage and therefore
on the electric fields in the detector's active volume. The study
below was conducted to inform the field modeling efforts of the LZ
experiment.

\subsubsection{Effects of modeling field grids in axially symmetric geometry}

\begin{figure}
\begin{centering}
\includegraphics[scale=0.7]{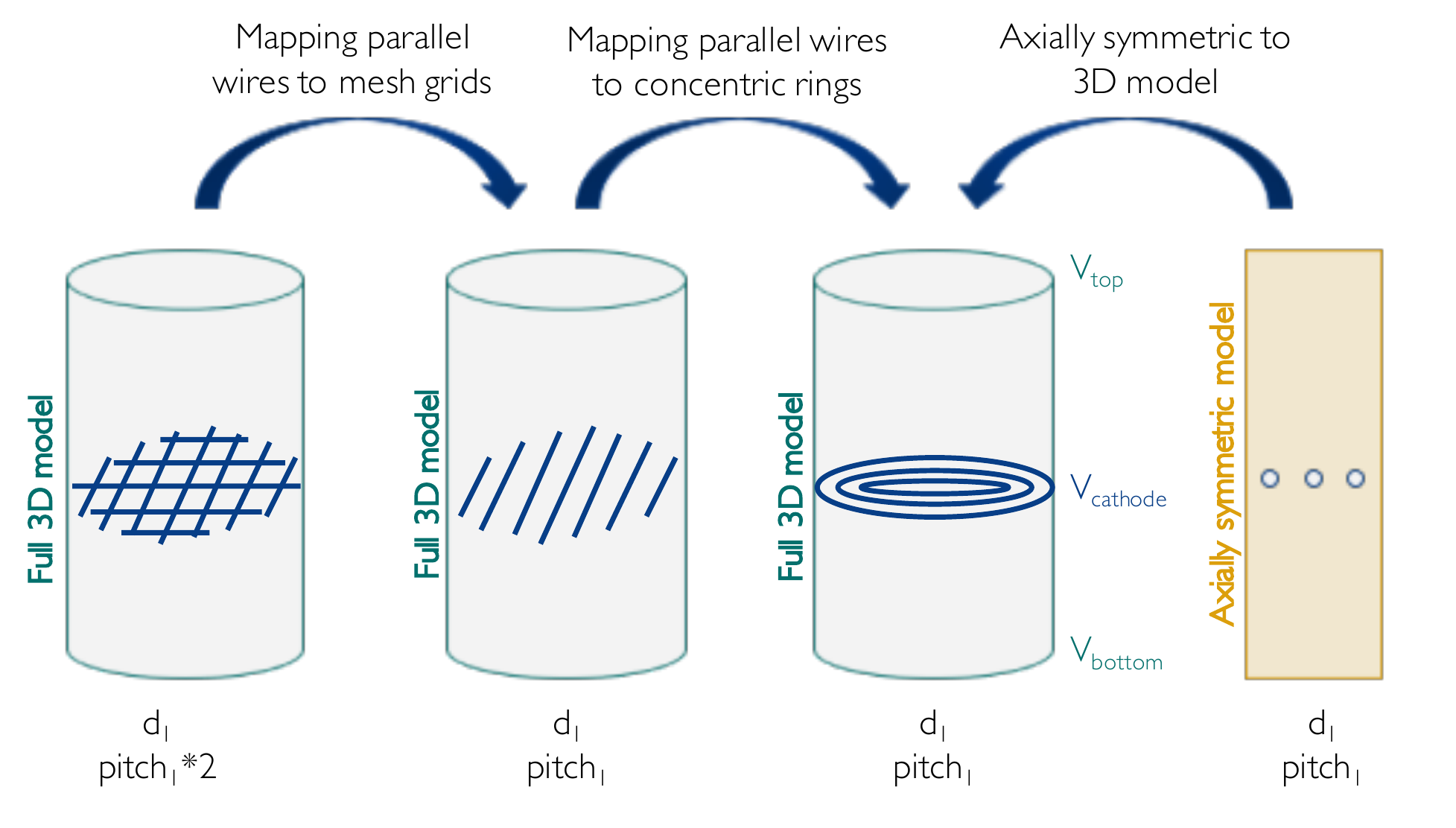}
\par\end{centering}
\caption[Illustration of different geometries modeled to study grid transparency
effects]{Illustration of different geometries modeled to study grid transparency
effects.\label{fig:grid-geometry-transformation}}
\end{figure}

Often optical transparency, defined as 1-diameter/pitch, is taken
as a proxy for electrostatic transparency, but as will be shown this
is not always a sound assumption. Therefore a study was conducted
to investigate the effects of various grid geometries on electrostatic
transparency. A simple 3D model was created in COMSOL Multiphysics
of a cylinder filled with LXe with dielectric constant $\varepsilon=1.85$
with a plane at the bottom with $V_{\mathrm{bottom}}=\unit[-1.5]{kV}$
at $z=\unit[0]{cm}$, a grid at $z=\unit[13.7]{cm}$ with $V_{\mathrm{cathode}}=\unit[-50]{kV}$,
and a plane on top at $z=\unit[26.9]{cm}$ with $V_{\mathrm{top}}=\unit[-46]{kV}$
corresponding to a field of $\unit[303]{V/cm}$ in the active volume.
Zero-charge boundary $n\cdot D=0$ was imposed around the edges of
the model. This mimics the settings in the center of the LZ detector
described in Section~\ref{sec:The-LUX-ZEPLIN-experiment}. Three
different grid geometries were modeled as illustrated in Figure~\ref{fig:grid-geometry-transformation}
each with the same grid diameter $d_{1}=\unit[100]{\mu m}$ but with
a varying pitch: a parallel wire grid with pitch $p_{1}=\unit[2.5]{mm}$,
a mesh wire grid with pitch $2\cdot p_{1}$, similar to the one used
in LZ, and a grid consisting of concentric rings of wires with pitch
$p_{1}$. The last geometry corresponds to the transformation of a
parallel wire grid into the solution of an axially symmetric model. 

These three geometries were chosen because the goal of these studies
was to determine whether an axially symmetric simulation would suffice
for investigations of electric fields inside the active volume of
the LZ detector. As LZ uses a mesh grid, representing that in an axially
symmetric model requires first to approximate the mesh grid as a parallel
grid (which is done by doubling the pitch) and then it is necessary
to investigate how the parallel grid compares to the concentric wire
grid modeled in the axially symmetric simulation.

\begin{figure}
\begin{centering}
\includegraphics[scale=0.63]{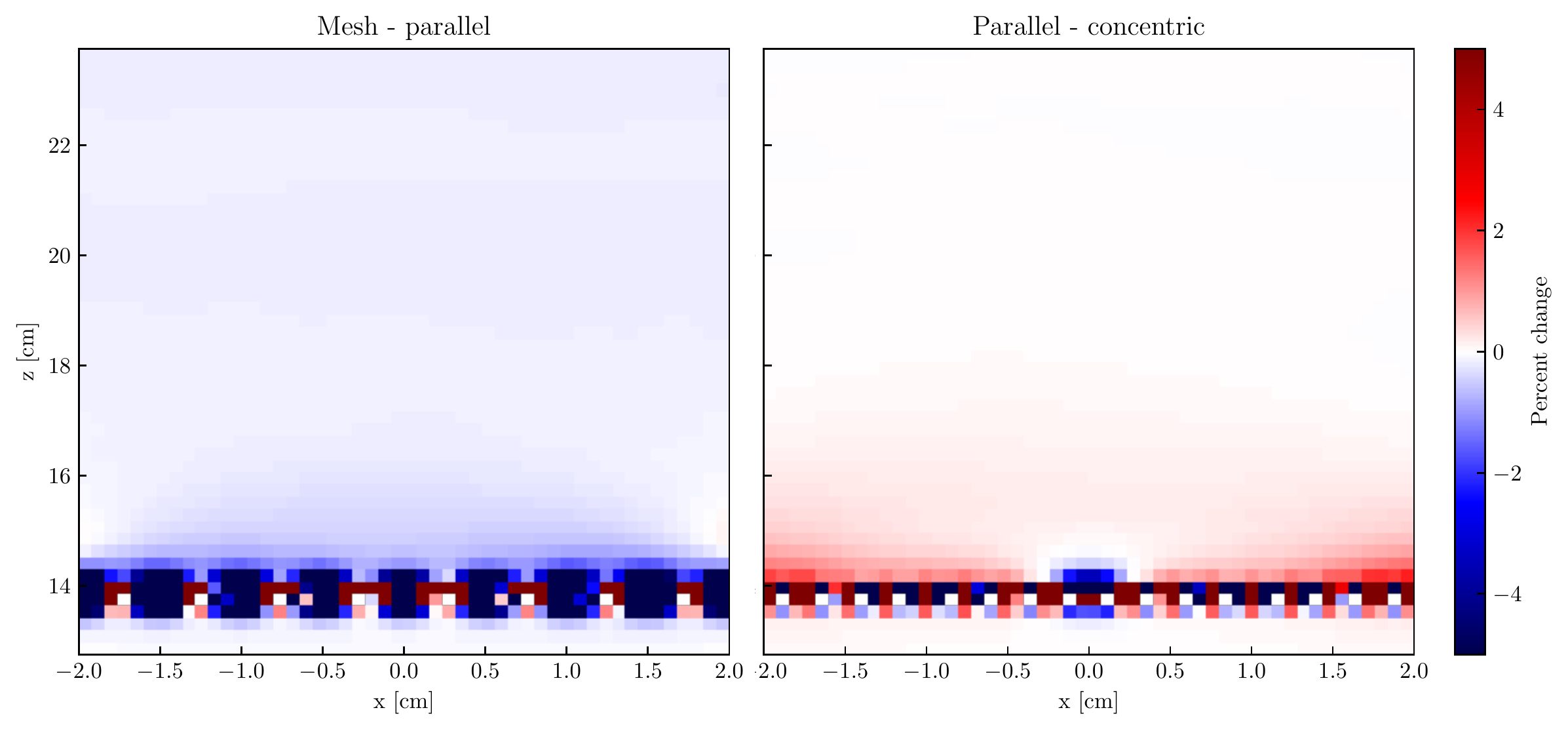}
\par\end{centering}
\caption[2D figure illustrating the field leakage between different grid geometries]{A 2D figure illustrating the field leakage difference between a mesh
and a parallel wire geometry (left) and a parallel and a concentric
wire geometry (right). Only a small $<2\%$ difference further than
$\sim\unit[1]{cm}$ away from the grids is visible.\label{fig:grid-field-diff}}
\end{figure}

Each of the three geometries was solved with the same boundary conditions
and the resulting fields were compared. A 2D figure  illustrating
the field difference between the mesh and parallel wire geometry and
between the parallel and concentric wire geometry is shown in Figure~\ref{fig:grid-field-diff}.
The field leakage in the active volume differs between the different
grid geometries, but appears to be $<2\%$ at $\sim\unit[1]{cm}$
away from the grids. Despite this small difference, due to the increase
in computational complexity for a full 3D model, the axially symmetric
models appear to be reasonable approximations. It should be noted
that the grid transparency increases with increasing radius so the
field leakage will be most pronounced at high radius close to detector
edges. 

\subsubsection{Understanding field leakage as a function of grid geometry\label{sec:Understanding-field-leakage}}

\begin{figure}
\centering{}\includegraphics[scale=0.95]{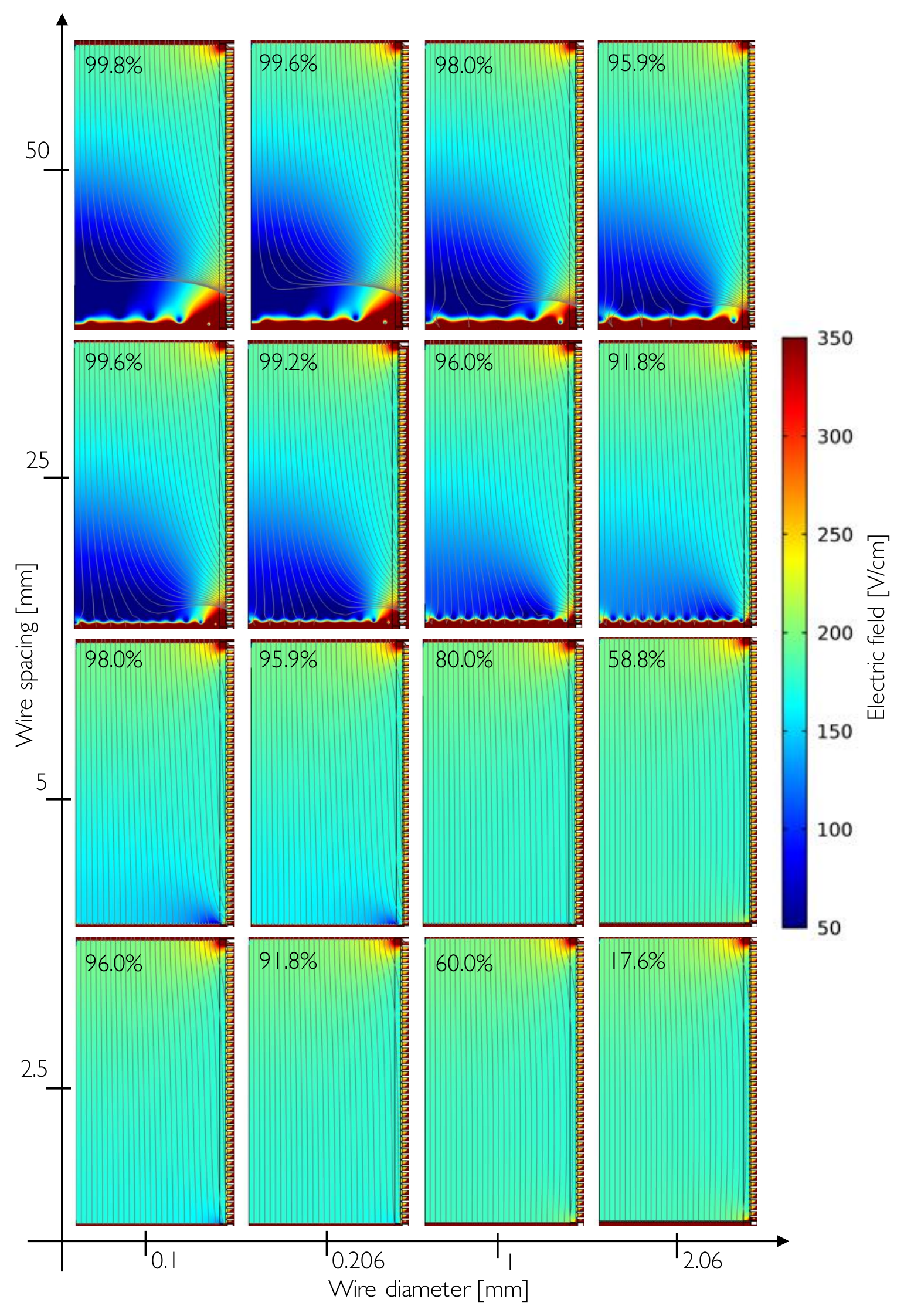}\caption[Effects of grid geometry on field leakage]{The wire diameters and pitch of the baseline axially symmetric model
used for visualization in WS2013 shown at $\left(0.206,\,5\right)$
was modified to illustrate effects of grid geometries on field leakage.
Optical transparency of each model is indicated in its upper left
corner and is defined as 1-diameter/pitch. The model at $\left(0.1,\,2.5\right)$
mimics the grid transparency of the LZ detector model described in
Section~\ref{subsec:Modeling-electric-fields-LZ}. \label{fig:many-LUXes}}
\end{figure}

A brief study using the well understood LUX axially symmetric model
was undertaken to visualize the effects of varying grid geometries
on grid transparency and the resulting field leakage. As shown in
Figure~\ref{fig:many-LUXes} the same optical transparency stemming
from various wire diameters and pitches (visible along the diagonal)
results in wildly varying electric fields inside the detector's active
volume. This illustrates that optical transparency can be used as
a proxy for electrostatic transparency only if the wire diameter and
pitch are not excessively modified.

These studies confirmed that an axially symmetric simulation with
wire diameter and pitches that respect the observations described
here are appropriate for the initial electric field model of the LZ
detector.

\begin{figure}[p]
\begin{centering}
\includegraphics[scale=0.58]{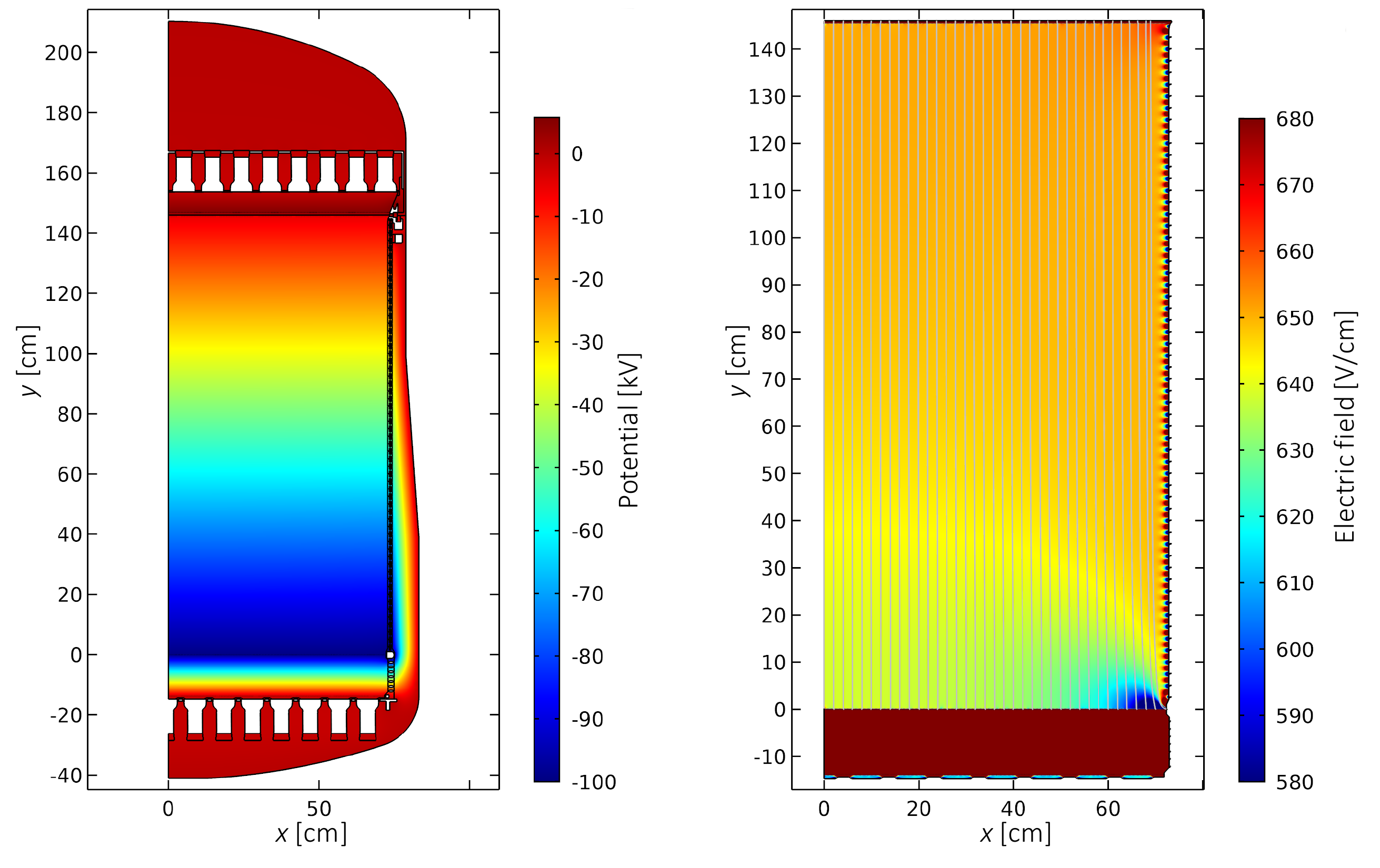}
\par\end{centering}
\caption[Electric field model of the LZ detector]{Model of the LZ detector with a cathode voltage of -100 kV. \textbf{Left:}~Potential
in the LZ detector. The region with one of the highest fields in the
detector can be seen between the cathode and the bottom shield grids
(here between $y\unit[\sim-15-0]{cm}$). It is referred to as reverse
field region since the electric field direction is opposite to the
active volume. \textbf{Right:}~Active volume of the LZ detector with
field lines shown in gray. The low and high field pockets also seen
in the LUX model (see Figure~\ref{run3_field}) are present as expected.
\label{fig:LZ-field-model}}
\begin{centering}
\vspace{2cm}
\includegraphics[scale=0.4]{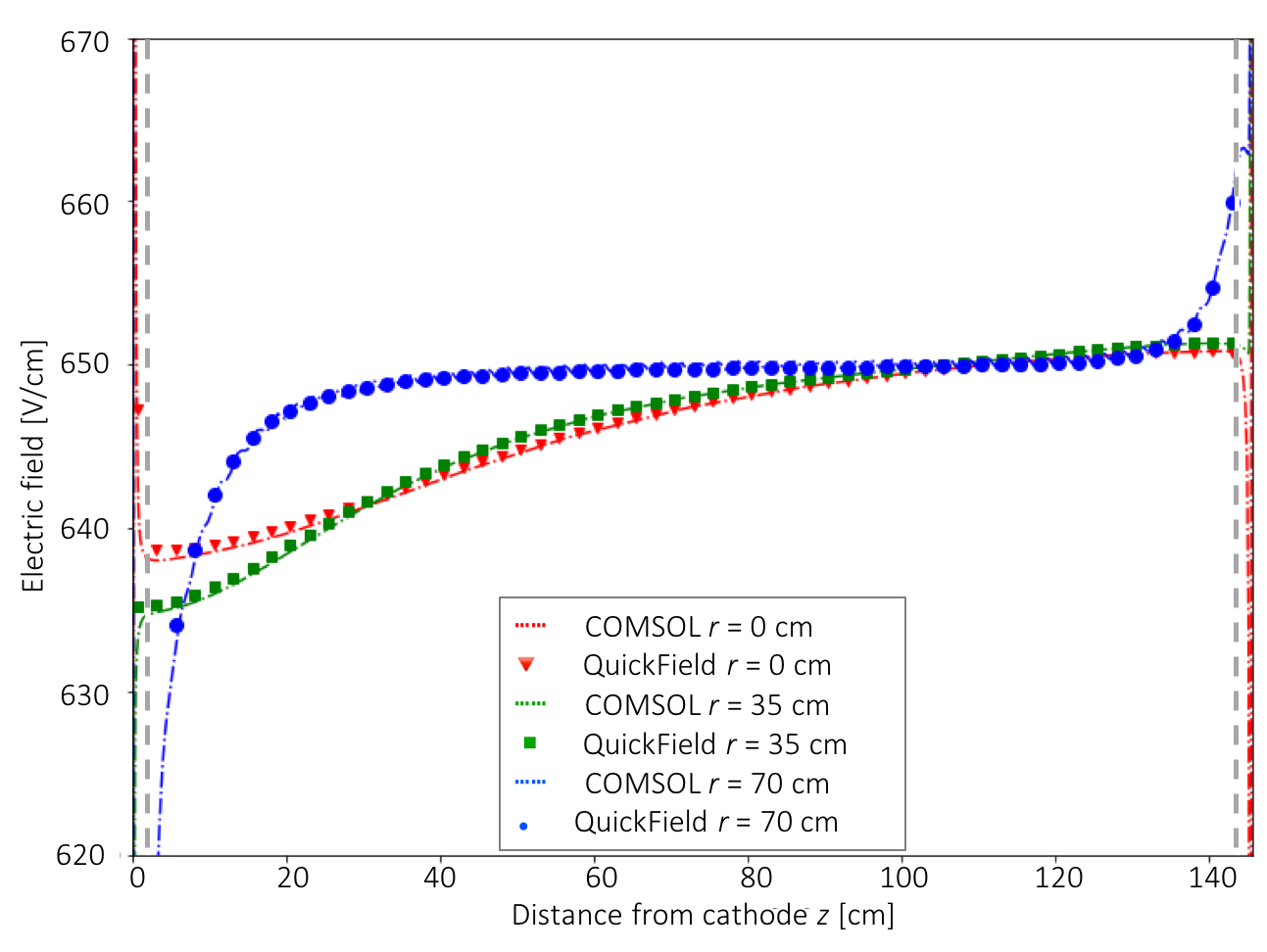}
\par\end{centering}
\caption[Comparison of electric fields as obtained from COMSOL and QuickField]{Comparison of the LZ electric fields as obtained from COMSOL and
QuickField models at three different vertical distances from the detector
center. The gray dashed lines indicate the drift time fiducial volume
boundary. The slight discrepancy at the low and high $z$ is due to
meshing anomalies in the QuickField simulation.\label{fig:comsol-quickfield}}
\end{figure}

\subsubsection{The LZ field model}

I built an axially-symmetric model in COMSOL Multiphysics that served
as an independent check for the electric field optimization studies
done by Reed Watson using the QuickField simulation package~\cite{quickfield}.
The resulting model is shown in Figure~\ref{fig:LZ-field-model}
and presents good agreement with simulations done in QuickField as
illustrated in Figure~\ref{fig:comsol-quickfield}.

\subsection{Hardship of high voltage in two-phase TPCs\label{subsec:LZ-High-voltage-delivery}}

A two-phase TPC works by detecting light and charge generated during
interactions in the active detector material. In order to detect the
charge generated by an interaction, the electric field needs to be
present to transport the charge to its detection site. In LUX and
LZ, the charge is detected by creating a secondary scintillation signal
via electroluminescence in xenon gas, but other methods can be used
as well. Generally, an electric field of $\mathcal{O}\left(100-1,000\right)$
V/cm in the bulk of LXe is required to achieve acceptable S2/S1 discrimination
and prevent event pile-up in the detector. Therefore, with the increasing
length of the drift region, it is necessary to deliver increasingly
higher voltages to achieve adequate electric field inside the active
volume. By design, in two-phase TPCs the cathode is biased at a negative
high voltage to enable electrons to drift toward the liquid-gas interface
where the field is further increased to facilitate electroluminescence.
Based on the length of the drift region, that means the cathode needs
to be biased to voltages $\mathcal{O}\left(10-100\right)$ kV with
only a small current needed to supply the TPC grading resistors. However,
this has proven to be a challenge in the past.

There are two sets of issues. First, there is the risk of a full dielectric
breakdown, which can cause irreparable damage to the detector hardware.
Second, there are concerns regarding elevated light production and
electron emission from regions with high fields since these typically
arise at lower voltages than a complete breakdown. Elevated light
production can cause an increased rate of low-energy S1-like signals,
while electron emission creates additional S2 signals. These can increase
the detector dead time by increasing the event rate and contributing
to the loss of WIMP sensitivity. Most commonly, the spurious light
and charge emission is caused by grid wires due to their small diameter,
but resistors, screws, and edges of miscellaneous parts at HV can
cause these too. Furthermore, increased electron emission can cause
material degradation, which can result in unexpected failure of a
detector part. 

There are no commercially available solutions to deliver high voltage
from a power supply located in the lab to the cathode of a liquid
noble TPC. Therefore, each collaboration develops a customized solution.
Ethan Bernard developed both the LUX and the LZ feedthroughs. The
development of these high voltage delivery systems is discussed in~\cite{pease2017rare}.
The design goal for the LUX HV feedthrough was -100~kV. The feedthrough
was successfully tested at that voltage in oil at Yale University.
However, the LUX cathode was never biased to that voltage during operations
due to the onset of light emission in the detector at lower voltages.
Instead, the detector operated at -10~kV and -8.5~kV during WS2013
and WS2014-16, respectively. Unfortunately, this is not unusual; other
detectors faced similar pains and were forced to operate at lower
fields than designed, usually due to the appearance of small light
pulses. The design goals and broken high voltage dreams of several
collaborations are summarized in Table~\ref{tab:HV-design-goals}.

\begin{table}
\begin{centering}
\begin{tabular}{>{\raggedright}p{2.5cm}|>{\raggedleft}p{2.5cm}>{\raggedleft}p{2cm}>{\raggedleft}p{2.5cm}>{\raggedleft}p{2cm}}
\hline 
Detector & Drift length {[}cm{]} & Drift field {[}V/cm{]} & Operating HV {[}kV{]} & Design HV {[}kV{]}\tabularnewline
\hline 
\hline 
ZEPLIN-III \cite{Akimov:2006qw,Akimov:2011tj} & 3.5 & 3,400 & -12.0 & -35\tabularnewline
XENON10 \cite{Aprile:2010bt} & 15.0 & 730 & -13.0 & N/A\tabularnewline
XENON100 \cite{APRILE2012573} & 30.5 & 530 & -16.0 & -30\tabularnewline
LUX \cite{Akerib:2015rjg,Akerib:2016vxi} & 48.0 & 50-600 & -8.5 & -100\tabularnewline
Panda-X II \cite{Cao2014,Cui:2017nnn} & 60.0 & 318 & -24.0 & -36\tabularnewline
XENON1T \cite{Aprile:2018dbl,xenon1tDesign} & 97.0 & 81 & -8.0 & -97\tabularnewline
LZ (projected)

\cite{Akerib:2018lyp} & 146.0 & 310 & -50.0 & -100\tabularnewline
\hline 
\end{tabular}\caption[Design goals and realities of two-phase TPCs HV]{Design goals and realities of two-phase TPC HV. The operating voltages
are reported from each detector's latest scientific runs. The design
HV refers to either the projected performance or the highest reported
value from TPC tests in the commissioning phases as available. Table
adapted from~\cite{Ethan_hv_table}. \label{tab:HV-design-goals}}
\par\end{centering}
\end{table}

In order to avoid the high voltage hardship experienced by the previous
generations of LXe dark matter experiments, LZ has invested much effort
into the development and testing of its high voltage systems. The
wires were tested at Imperial College London~\cite{Tomas:2018pny},
several tests of grids are being conducted at SLAC, and the cathode
HV feedthrough development started at Yale University and continued
at LBNL as discussed in the next section. 

\subsection{LZ high voltage cathode connection\label{subsec:LZ-high-voltage-connection}}

The central goal of high voltage delivery is maintaining cathode at
the appropriate voltage. This requires careful management of electric
fields around any conductors at high potential, which is particularly
intricate when it comes to the termination of the grounding shield
of the coaxial cable used for HV delivery. There are several options
for high voltage delivery from the power supply to the detector's
cathode. Most simply, conventional HV feedthroughs can be purchased,
as was done in the XeBrA detector described in Chapter~\ref{chap:XeBrA}.
However, conventional feedthroughs are often made from radioactive
materials or are unsuitable for use in cryogenic temperatures. Another
option is to avoid the delivery of high voltage altogether by creating
it within the detector~\cite{Clayton:2018sck}, an option favored
by experiments performed in liquid helium where thermal insulation
is important~\cite{Ito:2015hwa,ZHANG201631}, but also used in LAr
detectors~\cite{Argontube}. This solution is not optimal for LZ
where it is desirable to maximize the use of the LXe volume for WIMP
detection. 

The conservative design of the LZ feedthrough features a very careful
termination of the coaxial ground shield, which occurs in LXe near
the cathode. Because of the geometry, a high field region is unavoidable.
Therefore, the goal of the design is to minimize LXe exposure to high
fields to prevent spurious breakdowns. A smaller quantity of LXe is
preferable because small length-scales, in general, perform better
at high fields as will be discussed in great detail in Chapter~\ref{chap:XeBrA}.
From there, a linear voltage grading enforces smoother fields in this
critical region. Additionally, a leak-tight seal is required between
the cable and the cryostat to prevent the introduction of impurities
and xenon loss. Various cryogenic detectors face these issues, so
the LZ cathode connection design builds on experiences of previous
experiments. To learn more about the design and development of both
LUX and LZ feedthroughs, consult Reference~\cite{pease2017rare};
this section discusses the testing of the final design of the LZ cathode
HV connection. 

\begin{figure}[t]
\centering{}\includegraphics[scale=0.63]{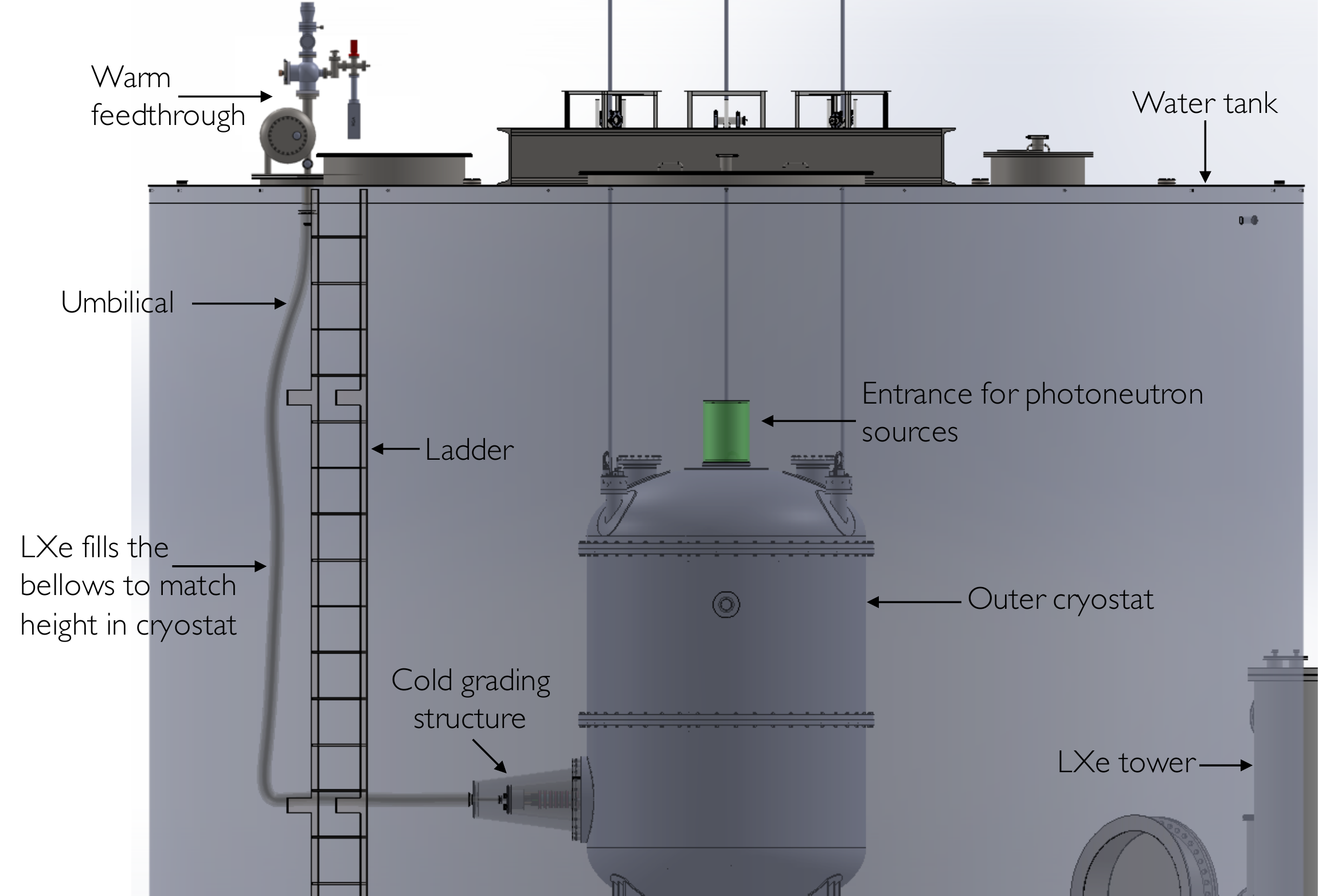}\caption[Overview of the HV delivery system from the lab to the detector]{Overview of the HV delivery system from the lab to the detector.
To thermally insulate xenon from the water tank, a 1'' diameter bellows
is nested inside a 3'' diameter bellows, which is used as a vacuum
jacket and will be filled with super-insulation. The inner bellows
connects to the xenon space of the inner vessel, and the evacuated
outer bellows connects to the vacuum space of the outer cryostat.
The vacuum space of the warm feedthrough will be monitored for gas
leaks by an RGA. \label{fig:LZ-HV-delivery}}
\end{figure}

Because of logistical, spatial, and design considerations, the LZ
cathode HV connection is made from the side of the detector as shown
in Figure~\ref{fig:LZ-HV-delivery}. In LUX, HV was delivered from
the top of the cryostat through the side of the TPC as shown in Figure~\ref{fig:LUX-cathode-HV}.
Having the cable come from the top makes the detector construction
more straightforward, but creates undesirable high fields within the
inner cryostat. If instead, the HV cable arrives to the detector from
the side near the cathode it avoids perturbing electric fields inside
the TPC. This approach also allows a larger volume of LXe surrounding
the cathode connection, increasing distances from HV to ground. In
fact, unlike LUX, LZ has no metal parts between the TPC and the inner
cryostat walls in an attempt to minimize electric fields in the fiducial
volume. For this reason, the LZ cryostat is larger near the bottom,
which then makes space for the skin veto. 

\begin{figure}
\begin{centering}
\includegraphics[scale=0.5]{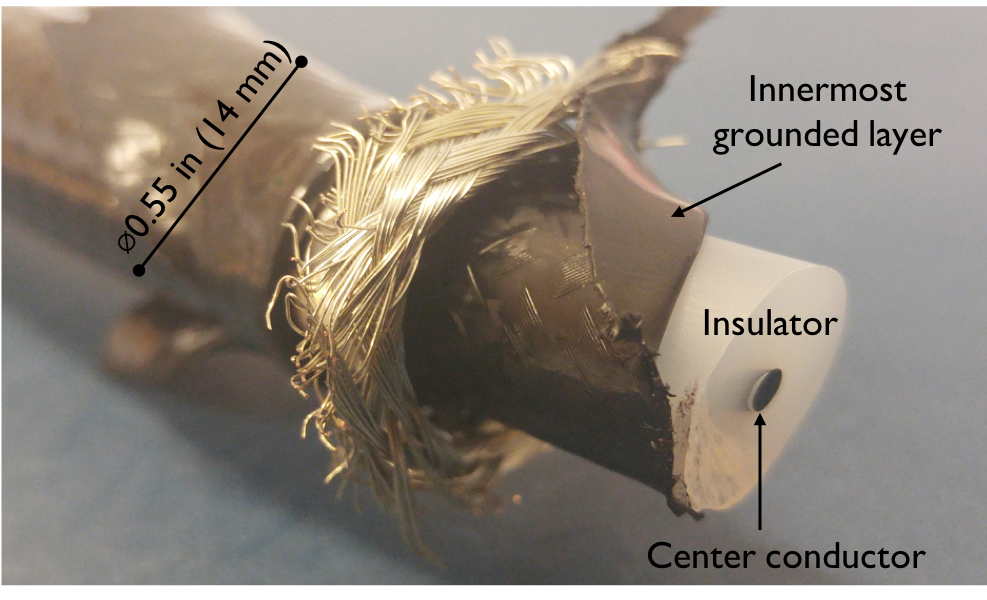}
\par\end{centering}
\caption[Cable from Dielectric Sciences model SK160318]{Coaxial cable from Dielectric Sciences model SK160318 rated to 150~kV.
The insulating layer is made from polyethylene and the innermost grounded
layer, and the conductive core are made from semiconductive polyethylene
with $\sim\unit[2800]{\nicefrac{\Omega}{ft}}$. The diameter of the
conductor is 0.08~in (2.0~mm), the diameter of the insulating dielectric
is 0.44~in (11.2~mm), and the thickness of the grounded layer is
0.01~in (0.3~mm). A single uninterrupted piece of cable ($\sim\unit[30]{m}$)
will connect the HV power supply directly to the cathode connection
structure. \label{fig:DS-cable}}
\end{figure}

An uninterrupted length of cable will connect from the LZ cathode
power supply\footnote{Spellman SL120N10} installed in the underground
lab to the cathode connection region inside the LZ cryostat. LZ uses
an all-plastic Dielectric Sciences cable model SK160318 rated up to
150~kV shown in Figure~\ref{fig:DS-cable}. There are two points
of interest on this path. First, there is a room temperature feedthrough
that makes a leak-tight seal between air and xenon. The warm feedthrough
features a double O-ring seal with a spool of extra cable needed to
allow for thermal contraction of the cable\footnote{The original feedthrough initially developed at Yale featured an epoxy-based
design, very similar to the one used in LUX. However, the initial
design goal for the LZ detector of -200~kV was lowered after it was
realized that higher drift fields do not necessarily provide better
ER rejection, and instead, data and NEST modeling suggest that sufficient
ER rejection in LZ can be achieved at drift fields comparable to those
achieved in LUX. This realization allowed LZ to shift to the smaller-diameter
Dielectric Sciences SK160318 cable, which enabled the O-ring feedthrough.}. 

\begin{figure}
\centering{}\includegraphics[scale=0.55]{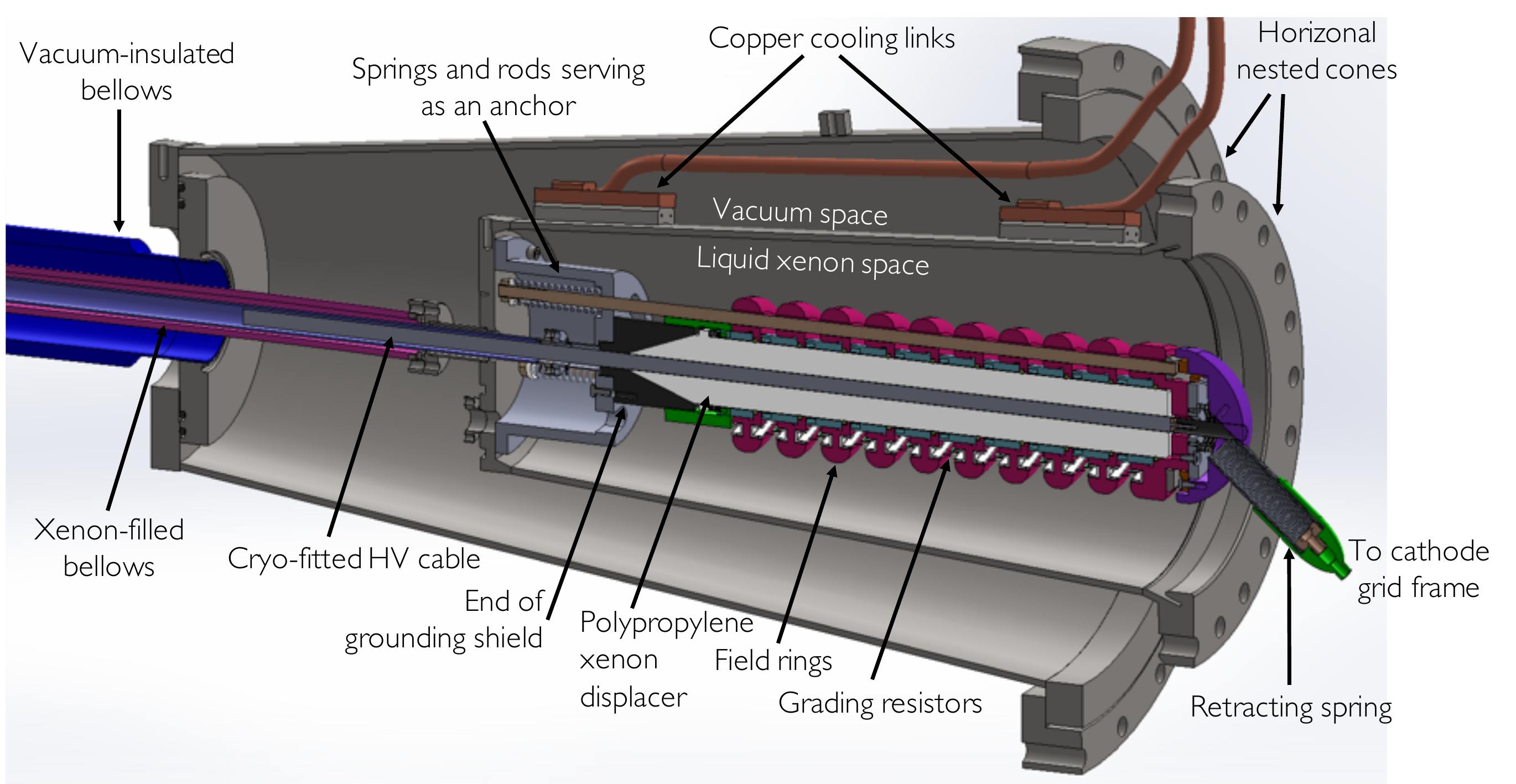}\caption[CAD rendering of LZ cathode HV connection region]{CAD rendering of the cold grading structure. The ground of the HV
cable is terminated within the stress cone (black) made from conductive
plastic. The field shaping rings are made from conductive plastic
and are linked together by linearly-graded resistors to taper off
the ground connection smoothly. The connection to the cathode frame
is made by a retracting spring whose flexibility ensures the connection
does not break during thermal cycling. \label{fig:LZ-cold-feedthrough}}
\end{figure}

\begin{figure}
\begin{centering}
\includegraphics[scale=0.7]{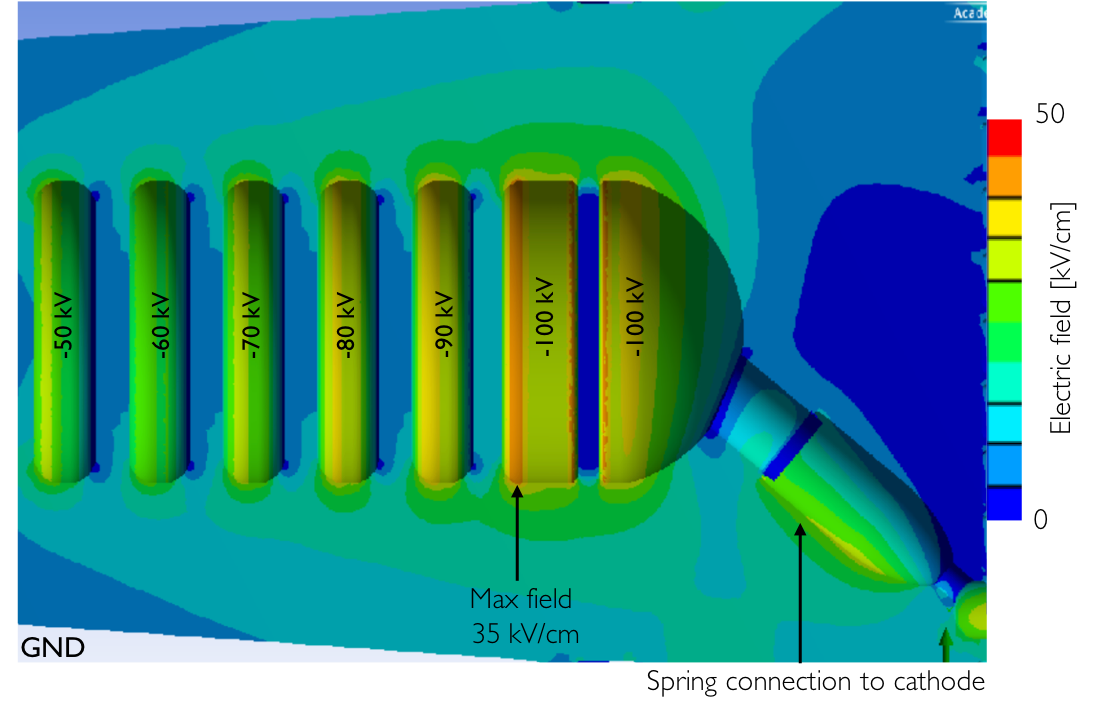}
\par\end{centering}
\caption[A simplified 3D electrostatic simulation of the HV cold grading structure]{A simplified 3D electrostatic simulation of the HV cold grading structure
done by Evan Pease in ANSYS Maxwell simulation software~\cite{ANSYS}.
Only the first six field shaping rings are shown. The spring connecting
the grading structure to the cathode has been simulated as a cylinder.
All the simulated parts have electric fields below 50 kV/cm, with
the highest field of 35 kV/cm occurring on the first field shaping
ring when -100 kV is applied to the cathode. \label{fig:LZ-cathode-field-sim}}
\end{figure}

Second, there is a cold connection near the cathode that provides
termination of the ground shield of the cable. Creating a connection
between the cathode and the HV delivery system that does not cause
excessively high electric fields, is secure, and can sustain thermal
cycling is often one of the critical parts of the HV design, due to
its complexity. A diagram of the cold connection used in LZ is shown
in Figure~\ref{fig:LZ-cold-feedthrough}. The cable is routed in
two bellows (flexible hoses) connecting the warm and cold grading
structures to a couple of cones attached to the side of the LZ detector.
The outer cone is attached to the outer cryostat that provides thermal
insulation to the inner cryostat. The inner cone is connected to the
inner cryostat and is filled with LXe. The grounding shield of the
HV cable is terminated inside the LXe cone. From there, HV is carefully
stepped up from ground to its maximum voltage (purple hemisphere on
the right) through a series of field shaping rings with linear voltage
grading. The electric field simulation of this region is shown in
Figure~\ref{fig:LZ-cathode-field-sim}. The grading structure accomplishes
to keep the electric field below 50 kV/cm in LXe, which is the LZ
design specification requirement.

The cable itself is cryofitted with polypropylene to exclude xenon
from the high field region. The cryofit ensures that the cable and
its expander undergo the same thermal contractions preventing xenon
leaking into the space\footnote{Even though the cable and the expander are made from different material,
polyethylene and polypropylene have very similar thermal properties.
Polypropylene is slightly stiffer and easier to machine, making it
an acceptable choice for this application.}. Three springs pull on rods holding the feedthrough to ensure structural
rigidity. The field shaping rings are made from conductive plastic
and are linked together by resistors that smoothly taper off the ground
connection to control electric field along the insulating surface
of the cable to prevent breakdown. The connection to the cathode is
made with a conductive spring, providing adequate mechanical compliance
and straightforward connection during assembly. Additionally, two
copper links attached to the HV cone cool the upper surface of the
cone below the LXe boiling temperature thus preventing accumulation
of bubbles. Xenon bubbles are undesirable because xenon gas has a
lower dielectric strength compared to LXe and therefore is more likely
to cause a breakdown. Furthermore, the difference in dielectric constants
enhances electric fields within the gas. The other ends of the thermal
links are connected to a thermosyphon that acts as a heat sink. 

\subsubsection{Feedthrough tests at LBNL\label{subsec:Feedthrough-tests-LBNL}}

\begin{figure}
\begin{centering}
\includegraphics[scale=0.39]{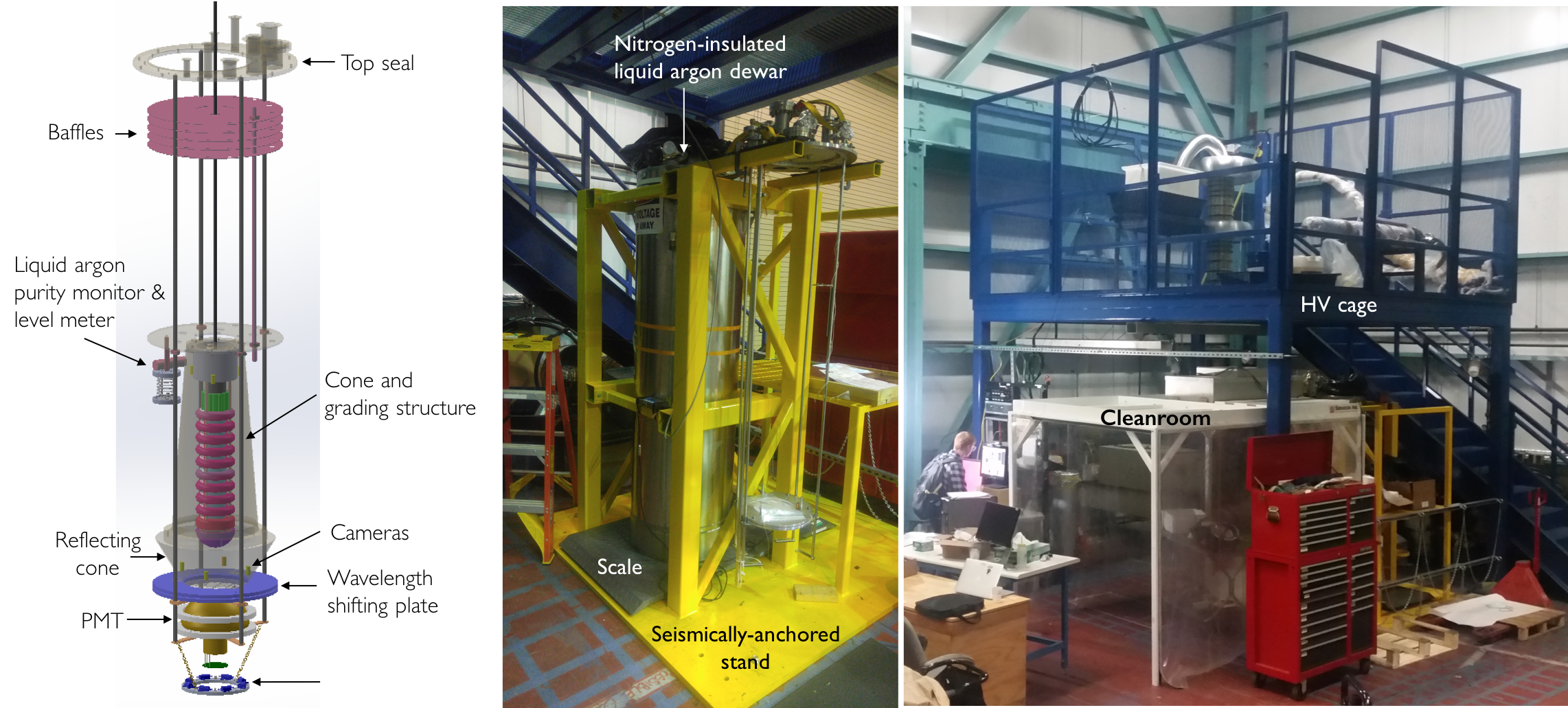}
\par\end{centering}
\caption[Setup used for testing the cone and grading structure of the LZ feedthrough]{Setup used for testing of the cone and grading structure of the LZ
feedthrough. \textbf{Left:} CAD rendering of the test structure. Baffles
on top improve thermal insulation of the inside volume. The boil-off
heater located near the bottom is used to accelerate LAr evaporation
once the tests are completed. \textbf{Center:} The test structure
is located inside a vacuum-insulated, nitrogen-jacketed dewar filled
with LAr during tests. The dewar is placed inside the seismically-anchored
yellow stand. The scale provides a measurement of the amount of LAr
in the dewar. \textbf{Right:} The test stand is located under the
HV cage in the platform (blue structure) built for these tests in
building 77A at LBNL. \label{fig:77A-cage}}

\end{figure}

To ensure that the cathode HV feedthrough performance meets design
expectations during 10 years of operation, extensive tests of the
cold grading structure are being conducted at LBNL. Several cryocycling
tests were performed of the HV cable and the xenon displacer cryofit
structure. Tests of the design of the grading structure are also underway
in the custom-built seismically-anchored HV cage shown in Figure~\ref{fig:77A-cage}.
Testing is performed in LAr since its HV performance is close to xenon\footnote{XeBrA, described in Chapter~\ref{chap:XeBrA}, is the first detector
to provide a direct comparison of the HV behavior in LAr and LXe.} at a fraction of the cost. The test system, therefore, does not require
expensive recirculation systems and other infrastructure needed to
prevent xenon loss. During the tests, the cone and grading structures
are submerged in LAr, and a high voltage is delivered to the structure
from the HV power supply located on the top of the blue grounded cage.
Discharges are monitored by charge-sensitive amplifiers, a set of
cameras, and a PMT located at the bottom of the test structure. Since
the 128~nm LAr scintillation light is not visible to the PMT, a tetraphenyl
butadiene \nomenclature{TPB}{Tetraphenyl Butadiene} (TPB)-coated
acrylic pane is located in front of it serving as a wavelength shifter.
TPB fluoresces promptly and efficiently, re-emitting the light with
a peak at 430~nm~\cite{Howard:2018tun}. A dedicated purity monitor
described in Section~\ref{sec:Liquid-argon-purity} is also installed
inside the cryostat since dielectric breakdown depends on the purity
of the noble liquid as discussed in Section~\ref{sec:HV-purity-dependence}.
LAr is discarded at the end of each testing cycle. Description of
the electrical simulation of the feedthrough is available in~\cite{pease2017rare}. 

Initial analysis shows that the feedthrough performs within the ranges
required for LZ. An upcoming publication about the feedthrough design,
testing, and performance is planned. 

\section{High voltage breakdown dependence on liquid noble purity\label{sec:HV-purity-dependence}}

\begin{figure}[h]
\begin{centering}
\includegraphics[scale=0.25]{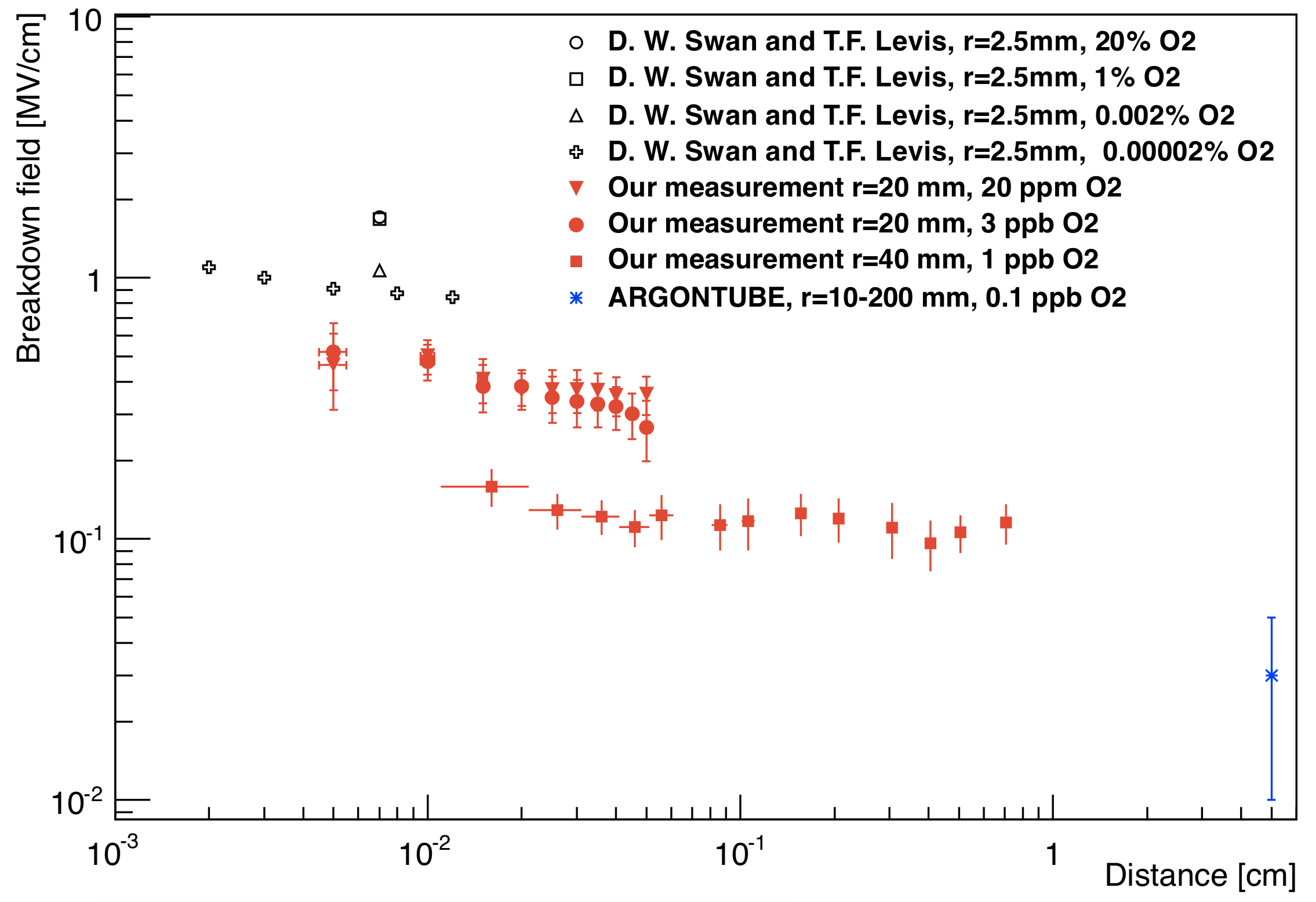}
\par\end{centering}
\caption[HV breakdown dependence at LAr purity]{Compilation of experimental data showing measurements of HV breakdown
at various LAr purities. The variable $r$ indicates the radius of
the cathode used in the tests. Figure from~\cite{blatter2014experimental}.\label{fig:purity_breakdown}}
\end{figure}

There is evidence that dielectric breakdown in LAr depends on the
noble liquid purity~\cite{blatter2014experimental,Auger:2015xlo,swan_LAr};
a smaller amount of impurities can trigger breakdowns at lower fields
compared to more polluted LAr as seen in Figure~\ref{fig:purity_breakdown}.
Therefore, in order to validate results from the HV feedthrough tests
described in Section~\ref{subsec:Feedthrough-tests-LBNL}, it is
important to know the purity of the LAr present in the dewar. Since
the setup itself cannot infer LAr purity directly, a standalone purity
monitor was needed for the purity measurements. The following section
details the design, construction, and performance of this purity monitor.

The purity monitor is an integral part of the system that allows a
fast and reliable measurement of electronegative contaminants both
in LAr or LXe. Three purity monitors have been built at Yale University
and LBNL. I developed the first purity monitor for the LZ cold grading
structure test. A close copy of the first purity monitor was built
for the XeBrA experiment, described in Chapter~\ref{chap:XeBrA},
by Glenn Richardson, undergraduate advisee from UC~Berkeley. Some
of the improvements and technical details of the XeBrA purity monitor
are outlined in Section~\ref{sec:Purity-monitor-for-XeBrA}. Furthermore,
a third purity monitor that very closely follows the design of the
XeBrA purity monitor is currently under construction for the Immersed
BRIDF\footnote{BRIDF stands for Bi-directional Reflectance Intensity Distribution.}
Experiment in Xenon (IBEX) optical experiment at LBNL. The subsequent
sections describe the design, construction, and results from these
purity monitors.

\section{Design and operating principles of liquid noble purity monitors\label{sec:Liquid-argon-purity}}

The purity monitor is a small double-gridded drift chamber fully submerged
in liquid. Electrons are extracted from a photocathode via the photoelectric
effect and drifted along electric field lines toward the anode. The
electrons cross two mesh grids on the way. The first grid shields
the rest of the detector from the electrons generated on the cathode.
The effect of the second grid is to shield the anode from the drifting
electrons so that a signal is induced on the anode only after the
electrons have crossed the grid, hence creating a well-defined signal.
The number of electrons decreases during this drift period as they
attach to impurities. This enables the determination of electron lifetime
$\tau$, which is a function of the ratio of electrons arriving at
the anode to the number of electrons emitted at the photocathode.
In practical use, the electron lifetime $\tau$ can also be converted
to the total oxygen-equivalent concentration of impurities.

This section provides a detailed overview of the construction of two
purity monitors along with their technical details and a theoretical
derivation of their operating principles. The purity monitor described
in this section was inspired by many purity monitors built by other
collaborations for both LAr and LXe as described in~\cite{bettini1991pm_initial,carugno1990pm_formula,arneodo2002electron,ferella2006study,li201620purifier,walkowiak2000drift},
among others. The purity monitor is a relatively inexpensive way to
measure liquid noble purity if used with a xenon flash lamp.

\subsection{Purity monitor system design}

\begin{figure}
\begin{centering}
\includegraphics[scale=0.65]{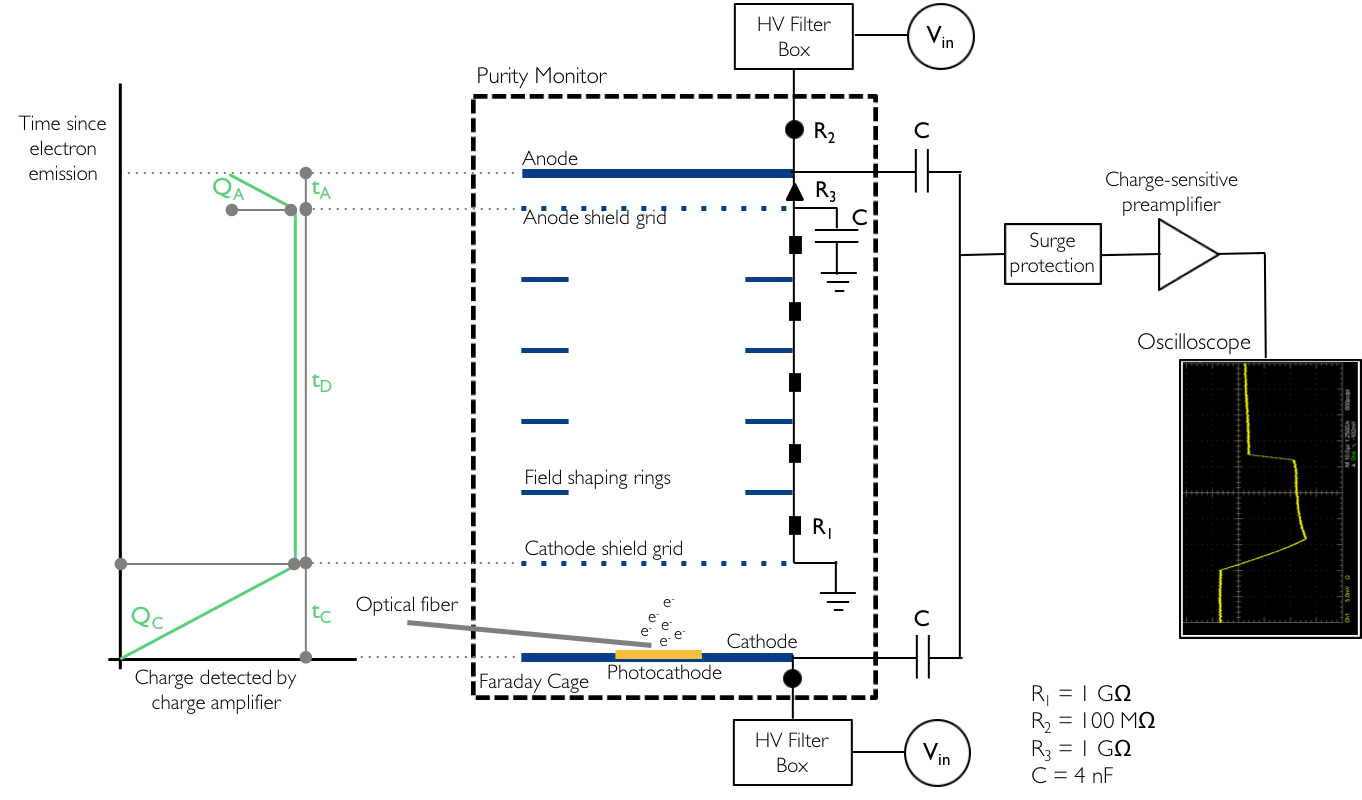}
\par\end{centering}
\caption[Schematics of the LZ purity monitor and its readout system]{Schematics of the purity monitor and its readout system. The plot
on the left illustrates the ideal signal detected by the purity monitor
while a real image from an oscilloscope is shown on the right. The
real signal includes the characteristic exponential RC decay of the
charge amplifier used to read out the signal visible during the drift
period of electrons, $t_{D}$. \label{fig:PrM-Schematics}}
\end{figure}

\begin{figure}
\begin{centering}
\includegraphics[scale=0.6]{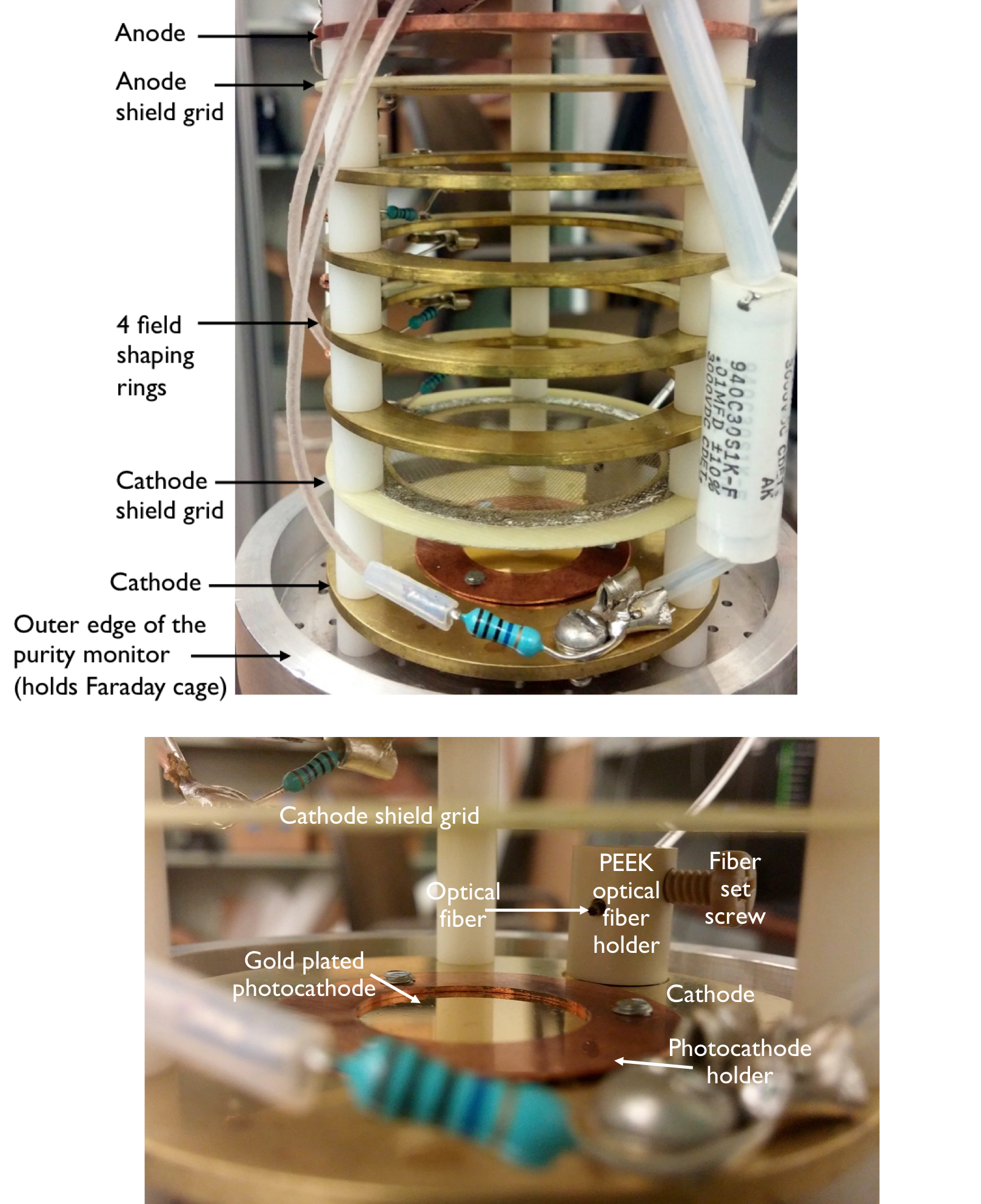}
\par\end{centering}
\caption[Image of the assembled purity monitor]{\textbf{Top:} Annotated image of the assembled purity monitor without
its Faraday cage. Shown is the early version with a plastic capacitor
(white cylinder). \textbf{Bottom:} Detail of the purity monitor cathode.
The PEEK fiber holder is set such that the optical fiber points at
the photocathode with a $10^{\circ}$ angle. The photocathode disk
can be changed and is attached to the cathode using a copper clamp.
\label{fig:Purity-monitor-real-life}}
\end{figure}

A schematic diagram of the purity monitor developed for the LZ HV
tests at LBNL is shown in Figure~\ref{fig:PrM-Schematics}. Electrons
are extracted from a $\unit[1]{\mu m}$ thick gold layer deposited
on a copper disk cathode using either a 266 nm laser or a xenon flash
lamp coupled to a silica optical fiber. The electrons then drift toward
the anode. The main drift region is 5-cm in length, enclosed by four
brass field shaping rings. The anode is made of copper. Both the cathode
and the anode are capacitively coupled to the same charge amplifier.
The purity monitor is assembled on three Delrin\footnote{Delrin is a commercial name for polyoxymethylene, also known as acetal
or polyformaldehyde.} rods, and the individual elements are held apart using Teflon spacers.
A photograph of the assembled purity monitor can be seen in Figure~\ref{fig:Purity-monitor-real-life}.

\begin{figure}
\begin{centering}
\subfloat[HV filter box circuit.]{\begin{centering}
\includegraphics[scale=0.3]{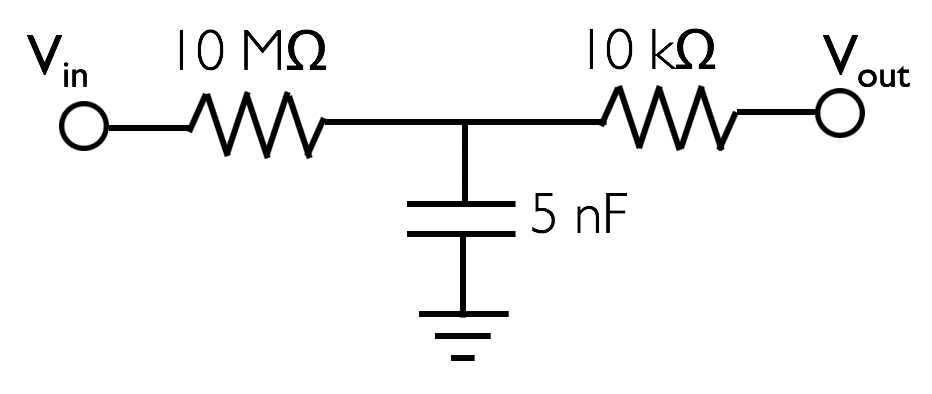}
\par\end{centering}
}\subfloat[Surge protection circuit.]{\begin{centering}
\includegraphics[scale=0.3]{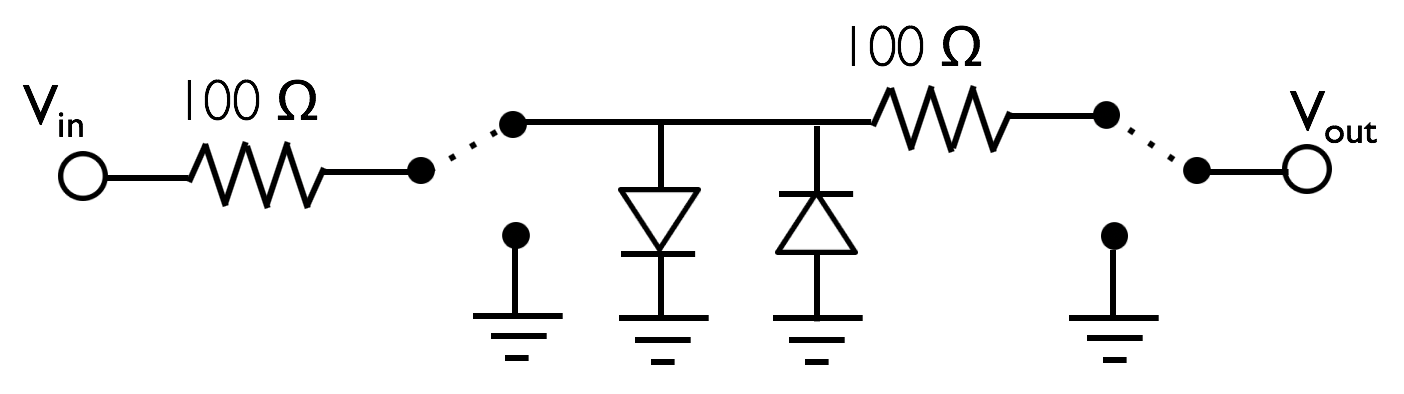}
\par\end{centering}
}
\par\end{centering}
\caption[Detailed circuit diagrams for selected electronics component in purity
monitor]{Detailed circuit diagrams for electronic components introduced in
Figure~\ref{fig:PrM-Schematics}.\label{fig:Electronics-detail}}
\end{figure}

As the electrons leave the cathode and move toward the cathode shield
grid, an induced cathode current is detected by the charge amplifier.
Once the electron cloud passes the anode shield grid, the charge amplifier
detects another current pulse as the cloud approaches the anode. HV
is supplied independently to the cathode and the anode through HV
filter boxes (circuit details are outlined in Figure~\ref{fig:Electronics-detail}).
The anode voltage is stepped down by a series of resistors to provide
voltage to the anode shield grid and the field shaping rings. Initially,
all three capacitors used in the purity monitor were made from metalized
polypropylene but were later changed for X7R ceramic type capacitors~\cite{teyssandier2010commercial_capacitors}
to reduce outgassing. The technical details of the purity monitor
are summarized in Table~\ref{tab:PrM-Detector-parameters}, and a
more detailed discussion of the individual components follows.

\begin{table}
\begin{onehalfspace}
\centering{}%
\begin{tabular}{lr}
\hline 
\textbf{\small{}Purity monitor geometry} & \tabularnewline
{\small{}Cathode \& anode diameter} & {\small{}57.0 mm}\tabularnewline
{\small{}Photocathode diameter} & {\small{}20.0 mm}\tabularnewline
{\small{}Cathode - cathode shield grid} & {\small{}15.0 mm}\tabularnewline
{\small{}Cathode shield grid - anode shield grid} & {\small{}51.7 mm}\tabularnewline
{\small{}Anode shied grid - anode} & {\small{}4.5 mm}\tabularnewline
{\small{}Cathode - anode distance} & {\small{}71.2 mm}\tabularnewline
{\small{}Number of field shaping rings} & {\small{}4}\tabularnewline
\hline 
\textbf{\small{}Shield grids} & \tabularnewline
{\small{}Grid wire diameter} & {\small{}50 $\mu$m}\tabularnewline
{\small{}Grid wire spacing} & {\small{}500 $\mu$m}\tabularnewline
\hline 
\textbf{\small{}Photocathode} & Au deposited on Cu\tabularnewline
{\small{}Deposit thickness} & {\small{}1.3 $\mu$m}\tabularnewline
W$_{\mathrm{Au}}$ & 5.1 eV = 243 nm\tabularnewline
\hline 
\textbf{\small{}Laser} & {\small{}CryLaS FGSS 266-Q4\_1k }\tabularnewline
{\small{}Wavelength} & {\small{}266 nm}\tabularnewline
{\small{}Pulse duration} & {\small{}$\leq1$ ns}\tabularnewline
{\small{}Pulse energy} & {\small{}$\geq12$ $\mu$J at 1 kHz}\tabularnewline
\hline 
\textbf{\small{}Flash lamp} & {\small{}Hamamatsu L9455-11 }\tabularnewline
{\small{}Spectral distribution} & {\small{}185 - 2000 nm}\tabularnewline
{\small{}Energy per flash} & {\small{}17.6 mJ}\tabularnewline
\hline 
\textbf{\small{}Optical fiber} & {\small{}JT Ingram Technologies Inc. }\tabularnewline
Material & Silica\tabularnewline
{\small{}Fiber core diameter} & {\small{}1 mm}\tabularnewline
{\small{}Static bend radius} & {\small{}27.5 cm}\tabularnewline
{\small{}Total length used} & {\small{}1 m outside + 60 cm inside}\tabularnewline
{\small{}Angle between fiber and photocathode} & {\small{}10$^{\circ}$}\tabularnewline
\hline 
\textbf{\small{}Charge amplifier} & {\small{}Cremat CR-110}\tabularnewline
{\small{}Decay time constant} & {\small{}140 $\mu$s}\tabularnewline
{\small{}Gain} & {\small{}1.4 V/pC}\tabularnewline
\hline 
\textbf{\small{}Oscilloscope} & {\small{}Tektronix TDS5034B}\tabularnewline
\hline 
\textbf{\small{}Design Electric Fields} & \tabularnewline
{\small{}Cathode - Cathode shield grid} & {\small{}250 V/cm}\tabularnewline
{\small{}Cathode shield grid - Anode shield grid} & {\small{}500 V/cm}\tabularnewline
{\small{}Anode shield grid - Anode} & {\small{}1,000 V/cm}\tabularnewline
\hline 
\end{tabular}\caption[Parameters of the LZ purity monitor]{Parameters of the purity monitor used for LZ HV tests in LAr.\label{tab:PrM-Detector-parameters}}
\end{onehalfspace}
\end{table}

\subsubsection{Electron extraction}

\begin{figure}
\begin{centering}
\includegraphics[scale=0.5]{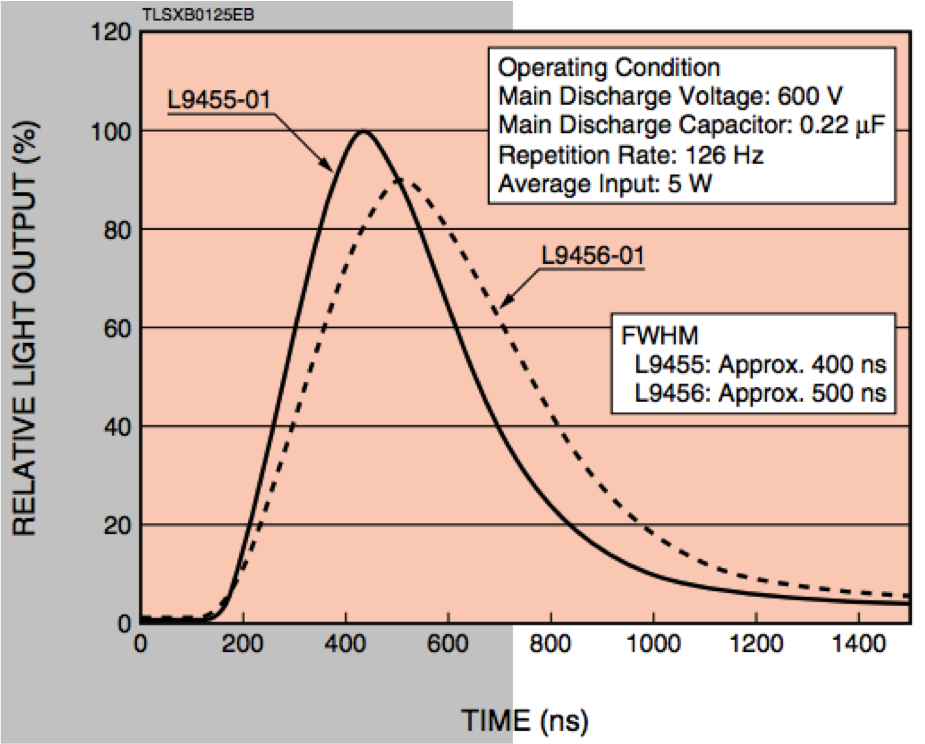}\includegraphics[scale=0.7]{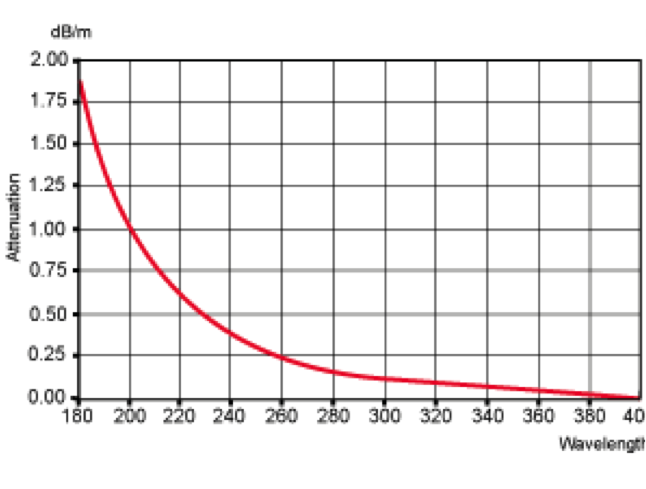}
\par\end{centering}
\caption[Characteristics of the light source and its transmission efficiency]{\textbf{Left:} Flash pulse waveform of the xenon flash lamp from
Hamamatsu L9455-11~\cite{xenonFlashLamp}. \textbf{Right: }Silica
fiber transmission for UV light~\cite{JTIngram}.\label{fig:fiber_property}}
\end{figure}

The photocathode consists of a gold-plated copper disk located in
the center of a brass cathode as seen in Figure~\ref{fig:Purity-monitor-real-life}.
Either a 266~nm (4.7~eV) laser or a xenon flash lamp can be coupled
to a silica fiber to extract electrons from the photocathode. The
spectral distribution of the xenon flash lamp ranges from $\lambda\sim\unit[185-2000]{nm}$
($6.7-0.6$~eV) with the emission peak located around $\lambda\sim\unit[240]{nm}$.
Figure~\ref{fig:fiber_property} shows the timing profile of the
xenon flash lamp pulse and the attenuation of the silica fiber used.
Properties of the laser, the xenon flash lamp, and the optical fiber
are summarized in Table~\ref{tab:PrM-Detector-parameters}. The laser
provides a monoenergetic peak and has a widely tunable pulse intensity,
while the flash lamp produces a much broader spectrum but at a fraction
of the cost of the laser. The broader spectrum also includes shorter
wavelengths, which is desired for this application, so the flash lamp
is the light source used for operations of all the purity monitors.

\begin{figure}
\begin{centering}
\includegraphics[scale=0.42]{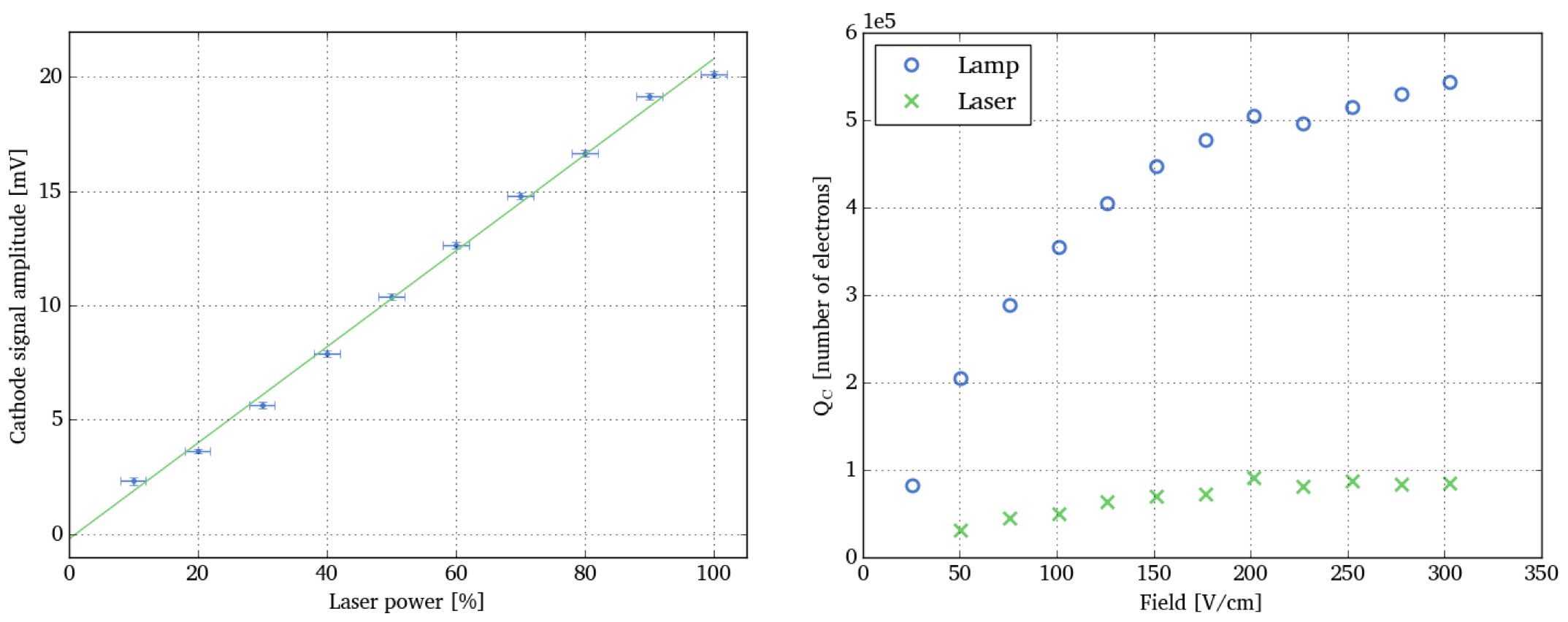}
\par\end{centering}
\caption[Characterization of laser behavior]{\textbf{Left:} Signal amplitude detected on the cathode as a function
of laser power. \textbf{Right: }Extracted charge $Q_{C}$ as a function
of the electric field between the cathode and the cathode shield grid
for laser (green crosses) and xenon flash lamp (blue circles).\label{fig:photoelectric-effect}}
\end{figure}

The work function of pure gold is $W_{Au}=\unit[5.3-5.5]{eV}$~\cite{rumble2018crc},
but values down to 4.7~eV have been reported when impurities are
present in the substrate~\cite{gold_workfunction}. The photoelectric
effect used to extract electrons from the cathode can be seen by varying
the laser power as shown in Figure~\ref{fig:photoelectric-effect}.
Also shown in the figure is the number of extracted electrons from
the photocathode as a function of the electric field in the cathode
region assuming a charge amplifier gain of 1.4~V/pC~\cite{creamat}.

A solid silver disc was tried as a photocathode since the work function
$W_{Au}\sim\unit[4.5]{eV}$ is lower than that of gold. However, no
signal was seen, likely due to a rapid formation of an oxidation layer
prior to submersion in LAr.


A drawback of the solid gilded metal cathode is that the optical fiber
needs to enter from the side, adding design complexity since the silica
fiber is not very flexible. An alternative is to use a semi-transparent
photocathode where gold is evaporated onto a sapphire disk, such as
was done in~\cite{li201620purifier}. This design was not tried but
would allow the optical fiber to enter the purity monitor on axis.

\subsubsection{Grids\label{subsec:Grid-Transparency-Correction}}

The shield grids have a wire diameter $d=\unit[50]{\mu m}$ and spacing
between wires $a=\mbox{\ensuremath{\unit[500]{\mu m}}}$ resulting
in an optical transparency of 83\%. To assure that no field lines
end on the shield grids, the field difference between each neighboring
region must be adequately large. For grids with parallel wires, a
100\% transparency to drifting electrons is achieved if the ratio
of neighboring electric fields is 
\[
\frac{E_{A}}{E_{B}}\geq\frac{1+\rho}{1-\rho}\left[\frac{1+\frac{a}{4\pi D_{A}}\left(\rho^{2}-4\ln\rho\right)}{1+\frac{a}{4\pi D_{B}}\left(\rho^{2}-4\ln\rho\right)}\right]
\]
where $E_{A}$ is the field after the grid, $E_{B}$ is the field
before the grid, $a$ is the grid pitch, $D_{A}$ and $D_{B}$ are
the lengths of regions $A$ and $B$, and 
\[
\rho=2\pi\frac{r}{a}
\]
where $r$ is the grid wire radius~\cite{bunemann1949design,bevilacqua2015frischGrid}. 

However, the grids in the purity monitor are crossed-wire meshes,
so a correction needs to be introduced. This correction is presented
in~\cite{bevilacqua2015frischGrid} as an empirical rule, where $\rho$
needs to be multiplied by two so that $\rho_{mesh}=2\rho$. Unfortunately,
this correction was only discovered after the testing of the purity
monitor was complete. As a result, all plots presented in this section
have a loss of charge to the shield grids unaccounted for, which slightly
reduces the estimated electron lifetime\footnote{A rough correction can be sought for this mistake in~\cite{bevilacqua2015frischGrid},
whose authors use a similar grid design. It can be deduced that about
95\% of electrons make it through our grids. Since the purity monitor
has two grids that electrons pass through, we estimate that 9\% of
electrons are lost due to grid transparency which can be used to correct
the lifetime calculation.}. 

This error was corrected before the deployment of the purity monitor
in the 77A test stand by replacing the $R_{3}=\unit[1]{G\Omega}$
resistor from Figure~\ref{fig:PrM-Schematics} between the anode
and the anode shield grid by a $\unit[2]{G\Omega}$ resistor and an
appropriate choice of bias voltages $V_{in}$. The new resistor results
in a field ratio $\mathrm{E_{A}/E_{B}=4.4}$.

\subsubsection{Anode feedthrough connection}

In order to achieve a wide range of purity measurements, the purity
monitor was designed to operate with up to 3 kV on the anode. It has
been empirically observed that argon and xenon gas exhibit low dielectric
strength. Therefore, particular care needs to be taken when using
HV feedthroughs in argon gas. A smooth-edged cylinder pictured on
Figure~\ref{fig:nub=000026level-sensor} was designed with $r=R/e$
where $r$ is the radius of the field-optimizing cylinder, and $R$
is the radius of the KF\footnote{KF is the ISO standard quick release vacuum flange. KF flanges can
be mated using a circular clamp and an elastomer O-ring mounted in
a metal centering ring. Further details about KF fittings can be found
in Appendix~\ref{chap:Intro-to-fittings}.} nipple serving as the HV feedthrough connection, and $e\simeq2.72$
is Euler's number. When assembled, the field-optimizing cylinder is
located in the center of the KF\nomenclature{KF/QF}{Klein Flange/Quick Flange}
nipple, and the HV cable then connects to the purity monitor attached
to the flange as shown on the picture. This field-optimizing cylinder
improved the voltage profile of the feedthrough, and the system was
able to deliver up to 3,700 V in gaseous argon. 

\subsubsection{Level sensor\label{subsec:Level-sensor}}

\begin{figure}
\begin{centering}
\includegraphics[scale=0.8]{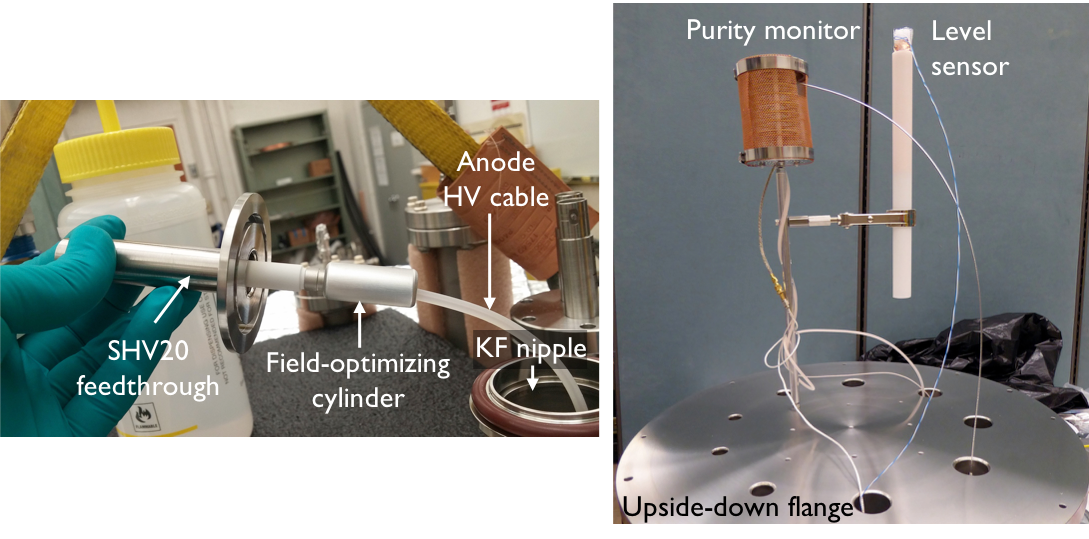}
\par\end{centering}
\caption[Feedthrough optimization and level sensor used in purity monitor tests]{\textbf{Left:} Field-optimizing cylinder designed to optimize electric
fields inside a KF40 nipple. \textbf{Right: }Purity monitor setup
alongside the liquid level sensor attached to a flange. Also visible
is the cutout in the Faraday cage for the optical fiber.\label{fig:nub=000026level-sensor}}
\end{figure}

The purity monitor needs to be fully submerged in LAr during operation.
To monitor the level of LAr in the experimental cryostat, a simple
level sensor was designed consisting of two rolled up layers of 2
each of a $\unit[50]{\mu m}$ copper sheet and a polypropylene plastic
mesh with 32\% transparency. This creates a rolled up parallel plate
capacitor with $C=\unit[2.1]{nF}$ while in air, vacuum or gaseous
argon and $C=\unit[2.8]{nF}$ while fully submerged in LAr. The level
sensor was only used to determine whether the purity monitor was fully
submerged in liquid argon. Figure~\ref{fig:nub=000026level-sensor}
shows the level sensor inserted in a Delrin tube to prevent unrolling
next to the purity monitor.

\subsubsection{Electronics}

\begin{figure}
\begin{centering}
\includegraphics[scale=0.7]{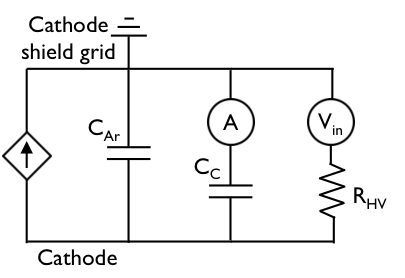}
\par\end{centering}
\caption[Simplified electronic circuit of the purity monitor cathode region]{Schematics of the electronics surrounding the purity monitor cathode.
The current source are electrons created via the photoelectric effect
and the charge amplifier can be approximated as an ammeter.\label{fig:cathode-circuit}}
\end{figure}

To ensure that electrons generated on the cathode travel through LAr
to the anode while the signal can also be sensed by the charge sensitive
preamplifier, the design requires that $Z_{C}\ll Z_{LAr}\ll R_{HV}$.
Here $R_{HV}=\unit[100]{M\Omega}$ ($R_{2}$ in Figure~\ref{fig:PrM-Schematics}).
$Z_{LAr}$ is the impedance of the LAr layer between the cathode and
the cathode shield grid:
\[
\left|Z_{LAr}\right|=\left|\frac{1}{j\omega C_{LAr}}\right|=\unit[3.5]{M\Omega}.
\]
The capacitance of LAr, $C_{LAr}$, is given by 
\begin{equation}
C_{LAr}=\varepsilon_{0}\varepsilon_{r\,LAr}\frac{A}{d}=\unit[2.3]{pF}\label{eq:capacitance_formula}
\end{equation}
where $\varepsilon_{0}=\unit[8.85\times10^{-12}]{F/m}$ is the permittivity
of vacuum, the relative permittivity of LAr is $\varepsilon_{r\,LAr}=1.5$,
and $\omega=2\pi f$ with $f\sim\unit[20]{kHz}$ corresponds to an
average drift time of electrons between the cathode and the cathode
shield grid with $t_{C}=\unit[50]{\mu s}$. $Z_{C}=\unit[2]{k\Omega}$
is the impedance of the coupling capacitor (C in Figure~\ref{fig:PrM-Schematics})
assuming $f=\unit[20]{kHz}$. A simplified electronic circuit of the
cathode region is shown in Figure~\ref{fig:cathode-circuit}.

There are three capacitors installed in the purity monitor. Two capacitors
prevent current from the cathode and anode power supplies from entering
the charge amplifier. The third capacitor holds the anode shield grid
at a stable voltage. The grids can only work as a shield if they can
be held at a stable voltage on fast timescales of $\mathcal{O}\left(\mu s\right)$
corresponding to the electron drift time. Without the third capacitor
in place, the voltage on the shield grid decreases as electrons approach
and the grid loses its function shielding anode from those drifting
electrons.

\paragraph{Signal read out: }

Both the cathode and the anode are coupled through a capacitor and
a surge protection box to a Cremat CR-110 charge sensitive pre-amplifier.
The light source is configured to emit pulses at $\sim\unit[10]{Hz}$,
and the resulting signal is connected directly into a Tektronix oscilloscope.
The oscilloscope averages all the signals used in the analysis and
the subsequent data analysis is performed using Python package v2.7.

\subsubsection{Electric field simulation}

\begin{figure}
\centering{}\includegraphics[scale=0.6]{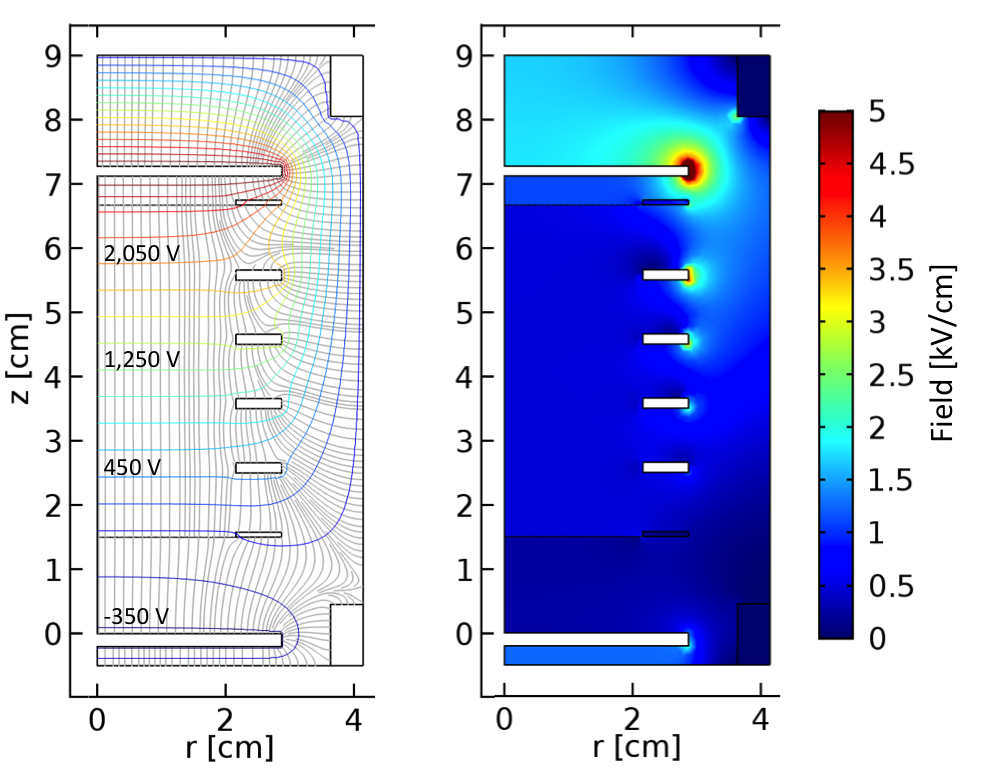}\caption[Simulations of the electric fields inside the purity monitor]{Simulations from an axially symmetric COMSOL model of the electric
fields inside the purity monitor\textbf{ Left:} Electric field lines
and equipotential contours for $E_{D}=\unit[500]{V/cm}$. \textbf{Right:}~Electric
field of the purity monitor with -375 V on the cathode, 2,500 V on
the anode shield grid and 3,000 V on the anode. The highest field
value is 10.8 kV/cm at the edge of the anode.\label{fig:Field-sim}}
\end{figure}

An axially symmetric electric field simulation of the purity monitor
was created in COMSOL Multiphysics, similar to those built in Chapter~\ref{chap:efield-modeling},
to confirm that electric fields within the region of interest were
uniform. The results of the simulation are shown in Figure~\ref{fig:Field-sim}.

\subsection{Purity monitor testing\label{subsec:Purity-monitor-testing}}

\begin{figure}[p]
\begin{centering}
\includegraphics[scale=0.6]{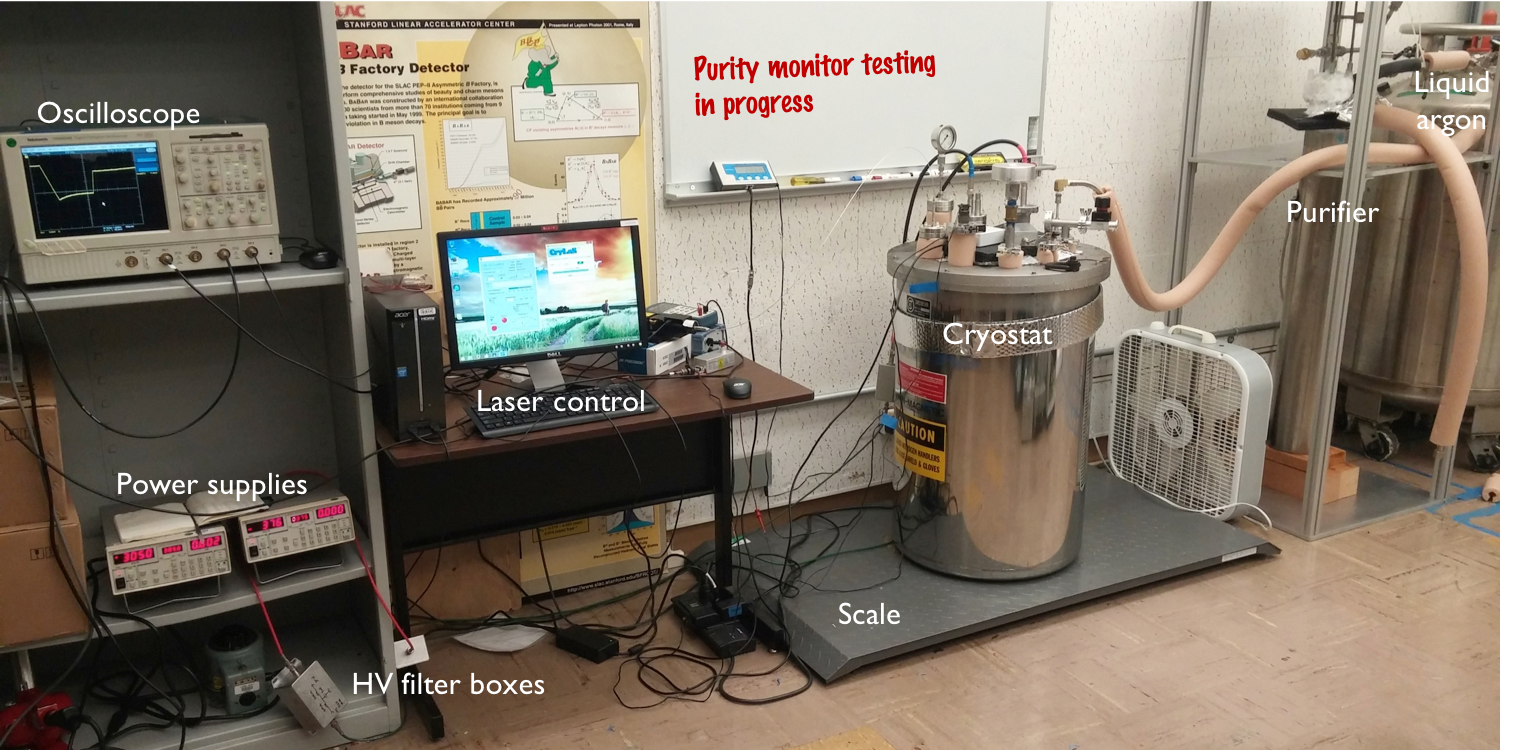}
\par\end{centering}
\caption[Lab layout for purity monitor testing in LAr]{Lab layout for purity monitor testing. Liquid argon from a commercial
dewar flows through the purifier into the cryostat shown in the center.
Also shown on the left are the power supplies and the data readout.\label{fig:Lab-layout}}

\vspace{2cm}
\begin{centering}
\includegraphics[scale=0.7]{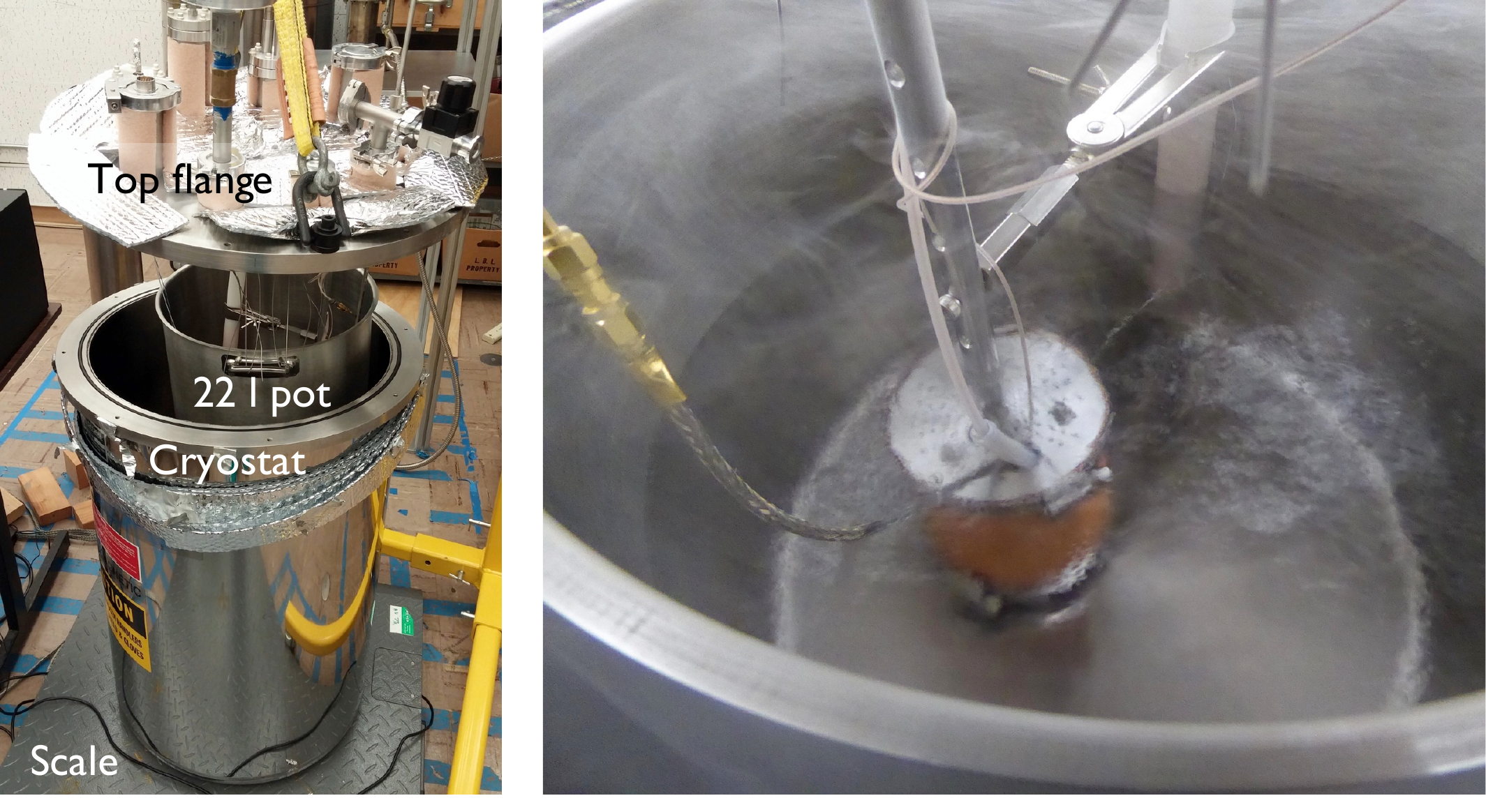}
\par\end{centering}
\caption[Purity monitor inside a 22 L stainless steel pot inside a cryostat]{\textbf{Left:} Purity monitor inside a 22 L stainless steel pot inside
a cryostat. The scale measured the amount of liquid argon in the cryostat.
\textbf{Right:} Purity monitor submerged in LAr. \label{fig:impurity-attachment}}
\end{figure}

The space needed at LBNL for purity monitor tests in LAr was set up
so that commercially available LAr flowed through a purifier into
a test cryostat as shown in Figure~\ref{fig:Lab-layout}. To reduce
the amount of LAr needed to fill the cryostat, the tests were conducted
inside a 22-liter stainless steel pot hanging from the flange of the
cryostat as shown in Figure~\ref{fig:impurity-attachment}. The cryostat
was evacuated before every fill to reduce impurities in LAr.

To simplify operations and to reduce costs, the testing setup did
not use a closed circulation system. Instead, commercially available
99.999\% pure liquid argon was used for the test, corresponding to
$\mathcal{O}\left(\mathrm{ppm}\right)$\footnote{ppm refers to parts per million\nomenclature{ppm}{parts per million}
$\left(10^{-6}\right)$, a pseudo-unit describing small values of
miscellaneous dimensionless quantities, such as a mole or a mass fraction.} of oxygen equivalent impurity concentration. Since the purity monitor
was designed to detect impurities $\mathcal{O}\left(\mathrm{ppb}\right)$\footnote{Similar as above, but parts per billion\nomenclature{ppb}{parts per billion}
$\left(10^{-9}\right)$.}, a purifier described in Section~\ref{subsec:Liquid-argon-purifier}
was used to reach the desired LAr purity. The purification is done
in-line with LAr flowing from a commercial LAr dewar through the purifier
directly into the well-insulated cryostat where it slowly evaporates
once the testing is over.

\subsection{Liquid argon purifier\label{subsec:Liquid-argon-purifier}}

The inline purifier is designed to remove O$_{2}$ and H$_{2}$O from
either gaseous or liquid argon. The filter, built at Yale University,
is filled with molecular sieve material and activated copper following
the design of~\cite{curioni2009regenerablePurifier}. Once the filter
particles are saturated with impurities, the filter can easily be
regenerated \textit{in situ} and used again at full capacity. Details
of this purifier are described below.

\paragraph{Molecular sieve\protect\footnote{The zeolite used was a 4~$\mathring{\mathrm{A}}$ SYLOSIV (Na$_{12}${[}(AlO$_{2}$)$_{12}$(SiO$_{2}$)$_{12}${]}xH$_{2}$0).
Zeolite translates to ``boiling stone'' from Greek. }:}

\begin{figure}
\begin{centering}
\includegraphics[scale=0.31]{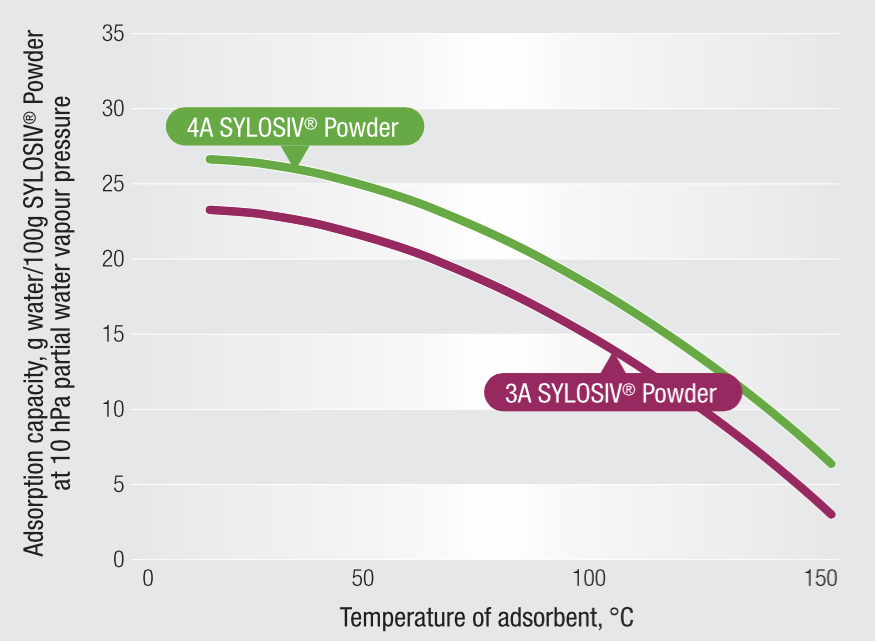}
\par\end{centering}
\caption[The adsorption capacity of molecular sieve at different temperatures]{The adsorption capacity of molecular sieve at different temperatures.
4A~SYLOSIV$^{\circledR}$ Powder was used in the purifier. Figure
from~\cite{sylosiv}.\label{fig:sylosiv}}
\end{figure}

The molecular sieve is a zeolite, a highly porous crystalline aluminosilicate
with three-dimensional, identical pores of precisely defined diameter.
The sieve has a high adsorption capacity for liquids and vapor. Due
to the molecule geometry the sieve can capture in its center molecules
small enough to permit passage through the pore, i.e. molecules with
diameter smaller than 4~$\mathring{\mathrm{A}}$ such as O$_{2}$
(d~=~3.5~$\mathring{\mathrm{A}}$), H$_{2}$O (d~=~2.7~$\mathring{\mathrm{A}}$),
N$_{2}$ (d~=~3.6~$\mathrm{\mathring{\mathrm{A}}}$) or CO$_{2}$
(d~=~3.3~$\mathrm{\mathring{\mathrm{A}}}$). Additionally, hydrophilic
substances experience stronger adsorption. The number of impurities
the molecular sieve is able to hold decreases with increasing temperature
as illustrated in Figure \ref{fig:sylosiv}.

\paragraph{Activated copper:}

Activated copper removes oxygen by exothermic copper oxidation 
\begin{align}
4\,\mathrm{Cu+O_{2}} & \rightarrow2\,\mathrm{Cu_{2}O}+\unit[2*156]{kJ/mol}\label{eq:-18}\\
2\,\mathrm{Cu}+\mathrm{O_{2}} & \rightarrow2\,\mathrm{CuO}+\unit[2*170]{kJ/mol}.\label{eq:-21}
\end{align}
The amount of bound oxygen is limited, but this process can be reversed
by flushing the purifier with hydrogen gas while providing heat according
to~\cite{schenk2015studiesLAr}: 
\begin{equation}
\mathrm{CuO}+\mathrm{H_{2}}\overset{\mathrm{heat}}{\rightarrow}\mathrm{Cu}+\mathrm{H_{2}O}+\mathrm{heat}\label{eq:water_formation}
\end{equation}

Activated copper was purchased from Engelhard Corporation as a composite
of alumina and copper oxide with 14 x 28 mesh beads\footnote{This means that 90\% of the particles will pass through a 14~mesh
sieve (particles are smaller than 1.4~mm), but will be retained by
a 28~mesh sieve (particles are larger than 0.7~mm).}. The alumina forms a very porous structure providing an inert support
to a thin layer of copper deposited on top.

\subsubsection{Purifier conditioning}

This type of purifier is fully regenerable. The filter can be regenerated
through evacuation while heated at $250^{\circ}$C for several hours
and then flowing hydrogen\footnote{Due to environmental health \& safety (EH\&S)\nomenclature{EH\&S}{Environmental Health \& Safety}
concerns we used a 98-2\% Ar-H mixture, which is non-flammable. Furthermore,
a small amount of hydrogen should be introduced at one time to prevent
overheating of the system (the water-forming reaction in Equation~\ref{eq:water_formation}
generating heat).} through the purifier. A detailed description of the regeneration
process that was followed is available in~\cite{li201620purifier,curioni2009regenerablePurifier}.
It was found that the most significant improvement in the regeneration
process was achieved by heating the purifier while evacuated, prior
to flowing the Ar-H mixture.

It is essential to monitor the temperature of the purifier throughout
its entire height since a large temperature gradient can develop if
the heating tape is not installed uniformly. This lesson was learned
when the bottom of the purifier was heated up beyond the recommended
$250^{\circ}$C, as was determined by the coloration of the stainless
steel container. Overheating\footnote{Since the melting temperature of zeolites is over $1000^{\circ}$
C, overheating the zeolite is less of a concern. } can cause the activated copper to anneal, thus greatly reducing the
exposed surface area that binds oxygen.

\subsection{Theoretical motivation for purity monitor operations}

The core of the purity monitor operations relies on accurate measurements
of electrons generated at the cathode and the number of electrons
not captured by impurities on the way to the anode. This section first
characterizes the attachment of electrons to impurities and then derives
the formula for calculation of electron lifetime from the current
detected on the purity monitor's cathode and anode.

\subsubsection{Electron attachment to impurities}

\begin{figure}
\begin{centering}
\includegraphics[scale=0.8]{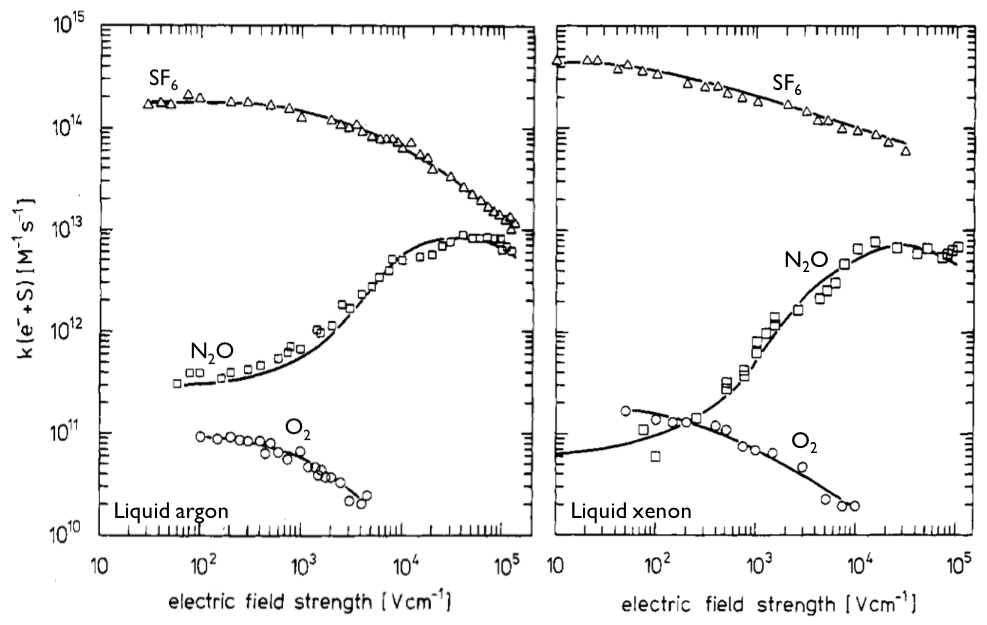}
\par\end{centering}
\caption[Field dependence of impurity attachment to several solutes in LAr
and LXe]{Rate constant for the attachment of electrons to several solutes.
It is expected that the main impurities present in LAr and LXe are
O$_{2}$, N$_{2}$, and $\mathrm{H_{2}O}$ (the latter two are not
included in the plot). \textbf{Left:} In LAr at 87 K. \textbf{Right:}
In LXe at 165 K. Figures from~\cite{bakale1976impurity_attachment}.\label{fig:impurity-field-dep}}
\end{figure}

Following the derivation from~\cite{buckley1989pm_grid_transparency}
the electron attachment to an impurity $S$ can be described by 
\[
e^{-}+\mathrm{S}\underset{k_{S}}{\rightarrow}\mathrm{S}^{-}
\]
with the rate constant $k_{S}$ 
\[
k_{S}=\int v\sigma(v)f(v)dv
\]
where $v$ is the instantaneous electron velocity (not the average
drift velocity), $f(v)$ is the velocity distribution of the electrons
and $\sigma(v)$ is the electron-impurity interaction cross-section.
This causes $k_{S}$ to vary as a function of an external electric
field as illustrated in Figure~\ref{fig:impurity-field-dep}. Assuming
that the number of electrons $N$ produced by the photoelectric effect
is much smaller than the impurity concentration $N_{S}$, the number
of electrons lost over time is 
\begin{equation}
\frac{dN}{dt}=-k_{S}N_{S}N.\label{eq:electrons-lost}
\end{equation}

For regular purity monitor operations, the number of impurities $N_{S}$
is assumed to be constant during data taking. Therefore integrating
Equation~\ref{eq:electrons-lost} leads to
\[
N=N_{0}e^{-t/\tau}
\]
where $\tau$ is the electron lifetime and is related to the impurity
concentration and rate constant by
\begin{equation}
\tau=\frac{1}{k_{S}N_{S}}.\label{eq:impurity_attachment}
\end{equation}
As a result of this process, electrons trapped by impurities create
negative ions that have a much smaller mobility\footnote{The charge-sensitive preamplifier is only sensitive to signals with
high-frequency components. Negative ions cannot be detected by the
charge amplifier because of their very long drift times.}. Depending on the purity of the liquid, a signal from the electrons
can then be measured on the cathode and the anode of the purity monitor.
The conspicuous decrease in the amplitude of the electron signal on
the anode due to the conversion of electrons into ions is used to
measure the presence of impurities in the intervening fluid.

\subsubsection{Electron lifetime to oxygen-equivalent impurity concentration conversion\label{subsec:Electron-Lifetime-to-ppb}}

In some publications, the purity of the liquid is stated as a function
of electron lifetime, while in others a unit of oxygen-equivalent
impurities is used. Equation~\ref{eq:impurity_attachment} can be
used to convert the electron lifetime $\tau$ to an equivalent concentration
of impurities in units of~mol/l. The following calculation follows
the community standard and assumes that oxygen accounts for 100\%
of impurities present, even though we expect nitrogen and other impurities
to be present in the liquid as well. The concentration of impurities
$N_{S}$ is related to the concentration $\rho$ in units of ppb by
\begin{eqnarray}
\rho & = & \frac{N_{S}}{N_{Ar}}\label{eq:-5}\\
\rho & = & \frac{1}{k_{S}N_{Ar}}\frac{1}{\tau}\label{eq:-6}
\end{eqnarray}
as defined in~\cite{buckley1989pm_grid_transparency}. $N_{Ar}$
is the number of moles in $V=\unit[1]{liter}$ of liquid argon 
\begin{align}
N_{Ar} & =\frac{\rho_{LAr}V}{M_{Ar}}=35\,\mathrm{mol}\label{eq:-17}
\end{align}
since the molar mass of argon is $M_{Ar}=\unit[39.95]{g/mol}$ and
LAr density $\rho_{LAr}=\unit[1.4]{g/cm^{3}}$ at its boiling point
of 87.3~K at 1~atm. The rate of attachment of impurities can be
obtained from Figure~\ref{fig:impurity-field-dep} with $k_{S}\sim\text{\ensuremath{\unit[7\times10^{10}]{l/\left(mol\cdot s\right)}}}$
for fields between $\unit[250-1000]{V/cm}$. Then from Equation~\ref{eq:-6}
\begin{align}
\rho_{LAr} & \sim\unit[\frac{408}{\tau}]{\frac{ppb}{\mu s}}\label{eq:-7}
\end{align}
 where ppb is defined as one oxygen molecule in every $10^{9}$ argon
atoms. 

Similarly, for LXe the rate of attachment of impurities $k_{S}\sim\text{\ensuremath{\unit[1\times10^{11}]{l/\left(mol\cdot s\right)}}}$
for fields between $\unit[250-1000]{V/cm}$. Therefore, the number
of moles in 1~l of LXe
\begin{align*}
N_{Xe} & =\frac{\rho_{LXe}V}{M_{Xe}}=\unit[22]{mol}
\end{align*}
 at the LXe boiling point of 165~K at 1~atm. Therefore, 
\begin{equation}
\rho_{LXe}\sim\unit[\frac{455}{\tau}]{\frac{ppb}{\mu s}}.\label{eq:lxe-oxygen-equivalen}
\end{equation}

\subsubsection{Electron lifetime derivation}

To find the formula for electron lifetime measured by the purity monitor,
the charge on the cathode and anode can be derived from the current
detected on those two electrodes as derived in this section. The current
as a function of time $I(t)$ measured at the cathode (anode) is determined
by the speed with which the electrons pass either the cathode or the
anode shield grids where $v=\frac{d}{t_{j}}$ and $j\epsilon\left\{ C,\,A\right\} $
stands for cathode (anode). The current detected at time $t$ is given
by
\[
I(t_{j})=N(t)e\frac{v}{d_{j}}=N(t)\frac{e}{t_{j}}
\]
where $N(t)$ is the number of electrons contributing to the pulse
as a function of time. There are 3 distinct time periods within the
purity monitor: $t_{C}$ is the time it takes the electrons to travel
from the cathode to the cathode shield grid, $t_{D}$ is the main
drift travel time between the cathode shield grid and the anode shield
grid and $t_{A}$ is the period between the anode shield grid and
the anode.

The charge sensed at the cathode $Q_{C}$ is given by integrating
the current between $t=0$ and $t=t_{C}$: 
\begin{equation}
Q_{C}=\intop_{0}^{t_{C}}I(t)dt=N_{0}e\frac{v}{d}\intop_{0}^{t_{C}}e^{-t/\tau}dt\label{eq:Qc}
\end{equation}
and the charge magnitude emitted from the photocathode is 
\[
Q_{0}=N_{0}e.
\]
Since the electron drift velocity is constant between the cathode
and the cathode shield grid, $\frac{v}{d}=\frac{1}{t_{C}}$ so Equation~\ref{eq:Qc}
becomes 
\[
Q_{C}=Q_{0}\frac{1}{t_{C}}\tau\left(1-e^{-t_{C}/\tau}\right).
\]

The number of surviving electrons $N_{D}$ crossing the anode shield
grid is 
\[
N_{D}=N_{0}e^{-t_{C}/\tau}e^{-t_{D}/\tau}
\]
hence the charge $Q_{A}$ detected at the anode is given by 
\[
Q_{A}=\intop_{0}^{t_{A}}i(t)dt=N_{D}e\frac{v}{d}\intop_{0}^{t_{A}}e^{-t/\tau}dt.
\]
Evaluating the integral gives 
\[
Q_{A}=Q_{0}\frac{1}{t_{A}}\tau\left(1-e^{-t_{A}/\tau}\right)e^{-t_{C}/\tau}e^{-t_{D}/\tau}.
\]
The electron lifetime can now be found by calculating the ratio $R$
of the charge detected at the cathode and the anode:
\begin{eqnarray}
R & = & \frac{Q_{A}}{Q_{C}}\label{eq:-8}\\
R & = & \frac{t_{C}}{t_{A}}\frac{\left(1-e^{-t_{A}/\tau}\right)}{\left(e^{t_{C}/\tau}-1\right)}e^{-t_{D}/\tau}\label{eq:-9}\\
R & = & \frac{t_{C}}{t_{A}}\frac{\sinh\left(\frac{t_{A}}{2\tau}\right)}{\sinh\left(\frac{t_{C}}{2\tau}\right)}e^{-\left(t_{D}+\frac{t_{A}+t_{C}}{2}\right)\frac{1}{\tau}}.\label{eq:-10}
\end{eqnarray}
Generally, it is safe to assume that $t_{A},t_{C}\ll\tau$ due to
the short distances between the cathode and cathode shield grid and
the anode and anode shield grid. In this limit the first part of Equation~\ref{eq:-10}
$\frac{t_{C}}{t_{A}}\frac{\sinh\left(\frac{t_{A}}{2\tau}\right)}{\sinh\left(\frac{t_{C}}{2\tau}\right)}\rightarrow1$.
Therefore, $R$ can be simplified to 
\[
R\simeq e^{-\left(t_{D}+\frac{t_{A}+t_{C}}{2}\right)\frac{1}{\tau}}
\]
hence 
\begin{equation}
\tau=-\frac{t_{W}}{\ln R}\label{eq:lifetime-eq}
\end{equation}
where $t_{W}=t_{D}+\frac{t_{A}+t_{C}}{2}$. This result is in agreement
with formulas presented in~\cite{bettini1991pm_initial,carugno1990pm_formula}.

\subsection{Data analysis}

\begin{figure}
\begin{centering}
\includegraphics[scale=0.39]{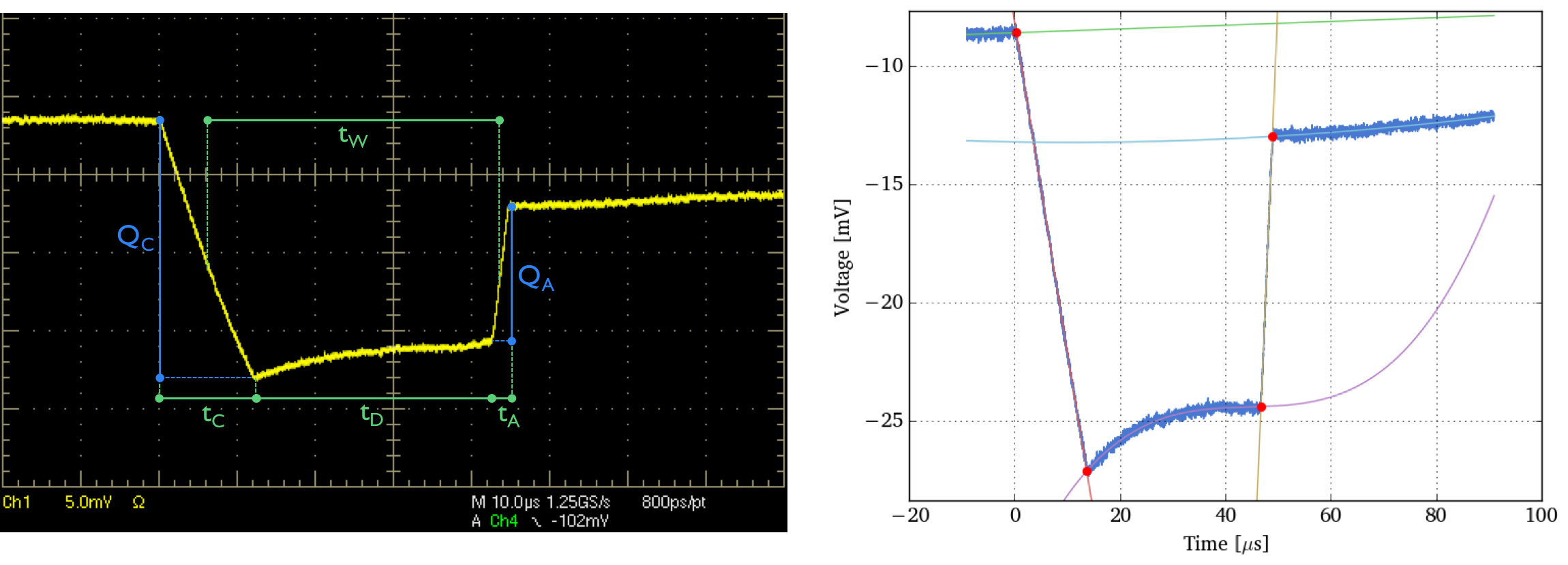}
\par\end{centering}
\caption[Annotated example of a purity monitor signal trace from an oscilloscope]{\textbf{Left:} Annotated example of a signal trace (yellow) from
the oscilloscope averaged over 600 light pulses. $Q_{C}$ and $Q_{A}$
(blue) are the integrated charges at the cathode and the anode; $t_{C}$,
$t_{D}$, and $t_{A}$ (green) are the transit times between cathode
and cathode grid, cathode grid and anode grid, and anode grid and
anode, respectively. $t_{W}$ is the FWHM of the overall signal. The
smaller amplitude of $Q_{A}$ relative to $Q_{C}$ indicates the presence
of impurities in LAr. The exponential decay apparent during $t_{D}$
arises from the RC decay constant of the charge amplifier. \textbf{Right:}
Example fit used to find boundary points for data analysis. The found
points are highlighted.\label{fig:signal-example}}
\end{figure}

An example of a signal trace from the LZ purity monitor fully submerged
in LAr is shown in~\ref{fig:signal-example}. The figure shows an
annotated screenshot of the trace and its consequent analysis performed
in Python. Each part of the trace was fitted with a $1^{\mathrm{st}}$,
$2^{\mathrm{nd}}$ or $3^{\mathrm{rd}}$ degree polynomial and intersections
of those curves help locate the boundary points for calculation of
$t_{C}$, $t_{D}$, $t_{A}$, $Q_{C}$, and $Q_{A}$ . This particular
trace was taken with the drift field at 500 V/cm and laser power at
90\% resulting in $t_{C}=\unit[13.4]{\mu s}$, $t_{D}=\unit[33.1]{\mu s}$,
$t_{A}=\unit[2.2]{\mu s}$, $Q_{C}=\unit[18.5]{mV}$, $Q_{A}=\unit[11.4]{mV}$
and $R=\frac{Q_{A}}{Q_{C}}=0.62$. Substituting these values in Equation~\ref{eq:lifetime-eq}
gives electron lifetime $\tau=\unit[85\pm3]{\mu s}$, which based
on Equation~\ref{eq:-7} corresponds to a concentration of oxygen
equivalent impurities in LAr of $\sim\unit[4.8]{ppb}$.

It should be noted that the calculations above assume a charge amplifier
with an infinite decay time. In reality, the characteristic RC decay
time constant is finite; the charge amplifier used in this work has
a decay time constant of $\unit[140]{\mu s}$. This modifies the purity
monitor signal as shown in Figure~\ref{fig:signal-example}. This
effect is visible in the segment of the signal where the drift time
is not perfectly horizontal but instead slowly decays with this time
constant. The cathode and anode signals are also affected by this
finite decay time constant, but because of their short duration, this
effect is not visible. For purity monitors with extremely long drift
times (where $t_{C}$ and $t_{A}$ are on the order of the decay time
constant), it might be necessary to calculate a correction. This can
be done by convoluting the current on the cathode (anode) with the
exponential decay of the charge amplifier to find the unaltered cathode
(anode) signal.

\subsubsection{Purity monitor resolution}

The error on lifetime calculation is given by~\cite{bettini1991pm_initial}:
\[
\frac{\sigma_{\tau}}{\tau}=\sqrt{\left(\frac{1}{\ln R}\frac{\sigma_{R}}{R}\right)^{2}+\left(\frac{\sigma_{t_{W}}}{t_{W}}\right)^{2}}
\]
assuming independence of $R$ and $t_{W}$. To find $\sigma_{R}$
and $\sigma_{t_{W}}$ dozens of measurements were recorded over a
short period with the same settings. The deviation of $R$ and $t_{W}$
was determined to be $\sigma_{R}=0.012$ and $\sigma_{t_{W}}=\unit[0.357]{\mu s}$.
To determine the error on lifetime accurately, multiple sets of measurements
should be taken at each of the different voltage settings. For this
analysis, the error was only determined at one voltage setting ($E_{D}=\unit[500]{V/cm}$
in the drift region). 

A wide set of electric fields can be used to operate the purity monitor
ranging from 25~V/cm in the cathode region up to 1200~V/cm in the
anode region corresponding to total electron drift times of $40-\unit[200]{\mu s}$.
For illustration Table~\ref{tab:voltage-setup} contains voltages,
electric fields, and transit times for two different operating modes.

\begin{table}
\begin{centering}
\subfloat{%
\begin{tabular}{lll}
\hline 
$V_{C}=-\unit[37.5]{V}$ & $E_{C}=\unit[25]{V/cm}$ & $t_{C}=\unit[100]{\mu s}$\tabularnewline
$V_{AG}=\unit[250]{V}$ & $E_{D}=\unit[50]{V/cm}$ & $t_{D}=\unit[102]{\mu s}$\tabularnewline
$V_{A}=\unit[300]{V}$ & $E_{A}=\unit[100]{V/cm}$ & $t_{A}=\unit[8]{\mu s}$\tabularnewline
\hline 
\end{tabular}}\qquad{}\subfloat{%
\begin{tabular}{lll}
\hline 
$V_{C}=\unit[-375]{V}$ & $E_{C}=\unit[250]{V/cm}$ & $t_{C}=\unit[13]{\mu s}$\tabularnewline
$V_{AG}=\unit[2500]{V}$ & $E_{D}=\unit[500]{V/cm}$ & $t_{D}=\unit[33]{\mu s}$\tabularnewline
$V_{A}=\unit[3000]{V}$ & $E_{A}=\unit[1000]{V/cm}$ & $t_{A}=\unit[2]{\mu s}$\tabularnewline
\hline 
\end{tabular}}
\par\end{centering}
\caption[Purity monitor settings for different operation modes]{Voltages, electric field magnitudes, and transit times for two different
operating modes of the purity monitor. Voltages on cathode, anode
grid, and anode are shown along with electric field magnitudes and
drift times in the three different purity monitor segments. The cathode
shield grid is always grounded.\label{tab:voltage-setup}}
\end{table}

\begin{figure}
\begin{centering}
\includegraphics[scale=0.41]{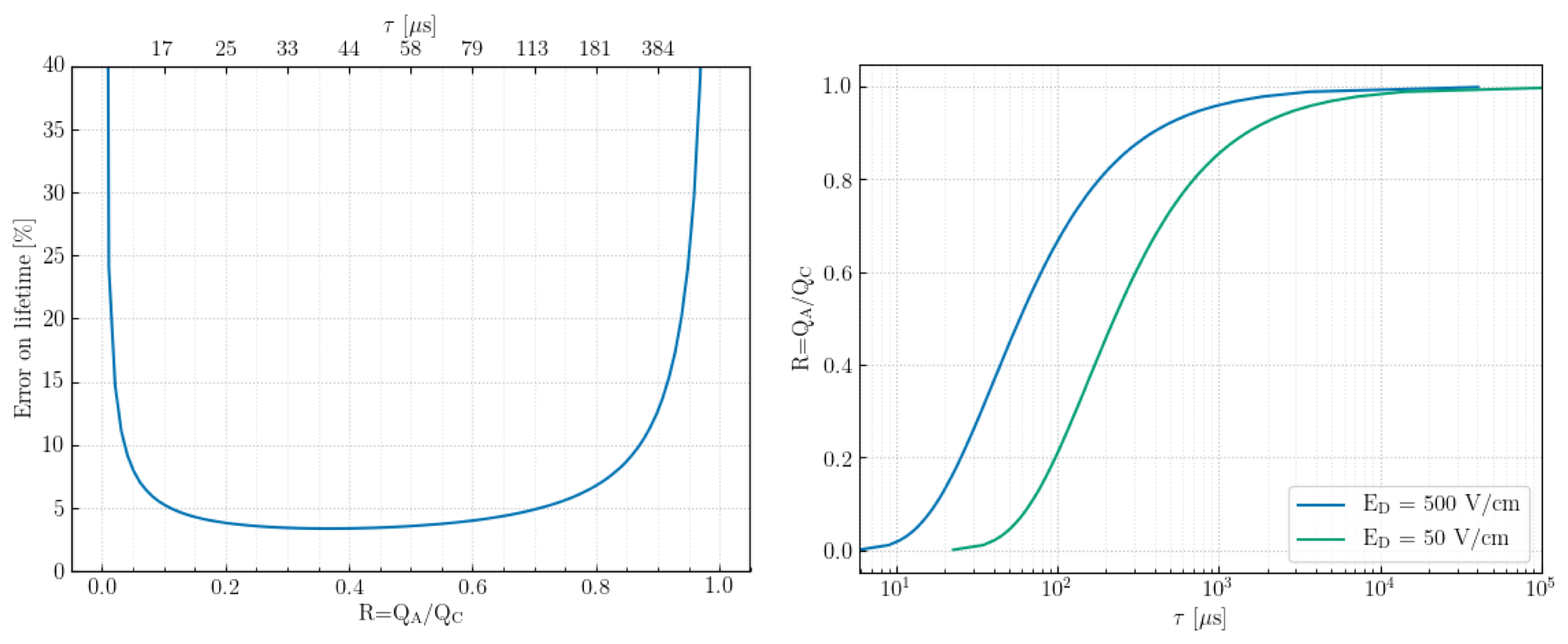}
\par\end{centering}
\caption[Characterization of purity monitor performance]{\textbf{Left:} The error on the lifetime versus the ratio $R=\nicefrac{\mathrm{Q_{A}}}{\mathrm{Q_{C}}}$
(bottom $x$ axis) for $E_{D}=\unit[500]{V/cm}$ and the estimated
electron lifetime (top $x$ axis) for $t_{W}=\unit[40.5]{\mu s}$.
\textbf{Right:} Ratio $R$ as a function of the electron lifetime
for two different field settings assuming $t_{W}=\unit[40.5]{\mu s}$
at $E_{D}=\unit[500]{V/cm}$ and $t_{W}=\unit[156]{\mu s}$ at $E_{D}=\unit[50]{V/cm}$.
The same error is assumed at $E_{D}=\unit[50]{V/cm}$ as was found
for $E_{D}=\unit[500]{V/cm}$. \label{fig:Error_plot}}
\end{figure}

Ratio $R=\nicefrac{Q_{A}}{Q_{C}}$ is strongly related to LAr purity.
When $R\rightarrow0$ or $R\rightarrow1$ the error on $\tau$ can
be substantial. This can be seen in Figure~\ref{fig:Error_plot}
which illustrates the error on the lifetime as a function of $R$
when the purity monitor has 500 V/cm in the drift region and assuming
the errors $\sigma_{R}$ and $\sigma_{t_{W}}$ quoted above. Figure~\ref{fig:Error_plot}
also shows the potential reach of the purity monitor for two different
voltage settings. These plots show that the current purity monitor
should be able to measure electron lifetime $\tau$ up to a millisecond
with less than 10\% error with $E_{D}=\unit[50]{V/cm}$ which corresponds
to $\sim\unit[0.4]{ppb}$ oxygen-equivalent impurities in LAr, well
within the range needed for LZ \& XeBrA.

\subsubsection{Electron lifetime dependence on electric field}

\begin{figure}[h]
\begin{centering}
\includegraphics[scale=0.36]{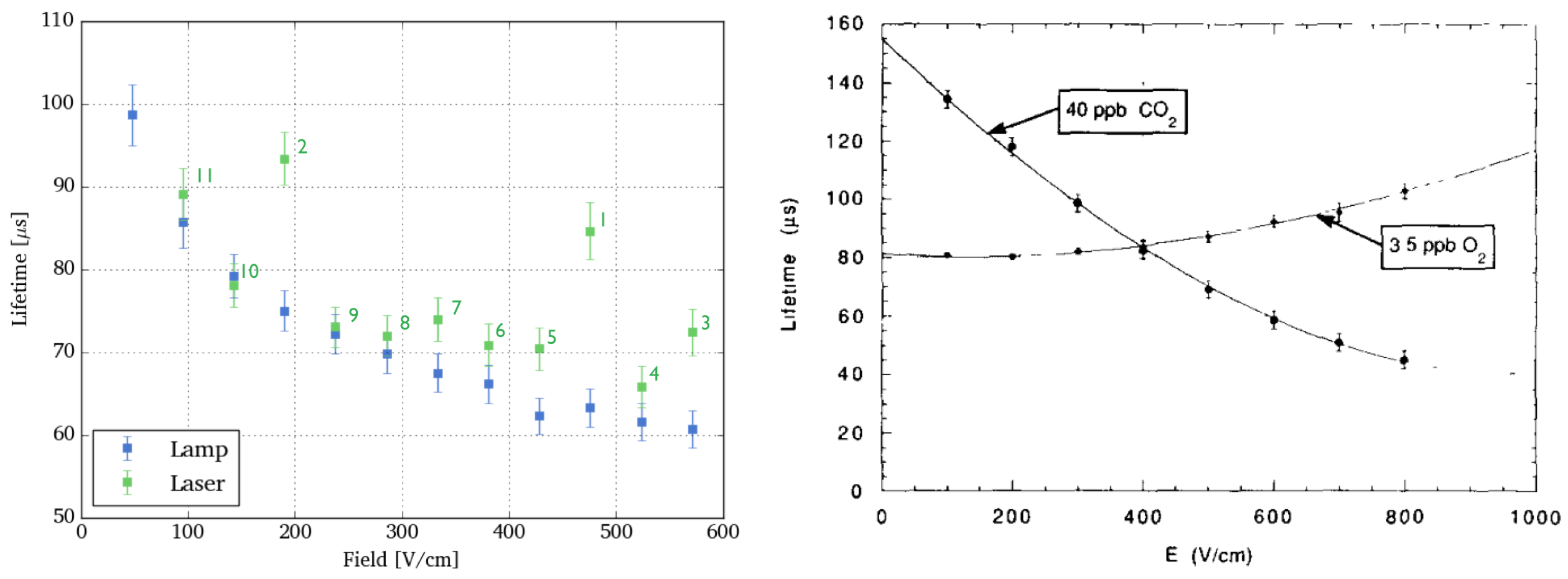}
\par\end{centering}
\caption[Electron lifetime as a function of electric field]{\textbf{Left:} Electron lifetime as a function of mean electric field.
The first laser measurement was taken at t = 0 with following measurements
taken at t = 12, 24, 36, 37, 40, 41, 44, 46, 48, 50 min. Measurements
taken with the xenon flash lamp were taken in order with increasing
field ranging from 56 - 77 min. \textbf{Right:} Electron lifetime
dependence on the electric field in purified LAr as a function of
different impurities~\cite{bettini1991pm_initial}.\label{fig:lifetime_field}}
\end{figure}

Figure~\ref{fig:lifetime_field} shows several measurements of LAr
purity performed over time at varying electric fields. Similar measurements
were also performed by~\cite{bettini1991pm_initial,bakale1976impurity_attachment}.
There is a clear dependence of electron lifetime on the electric field,
but this behavior is not yet thoroughly characterized in liquid argon
and xenon. 

It is unclear what are the dominant impurities present in LAr used
for the tests described here. However, this measurement suggests that
the purity monitor could be used to measure the effect of impurities
on electron lifetime as a function of electric field if known amounts
of impurities were injected into the system and the measurement was
repeated for each impurity at a variety of fields.

\begin{figure}[H]
\begin{centering}
\includegraphics[scale=0.41]{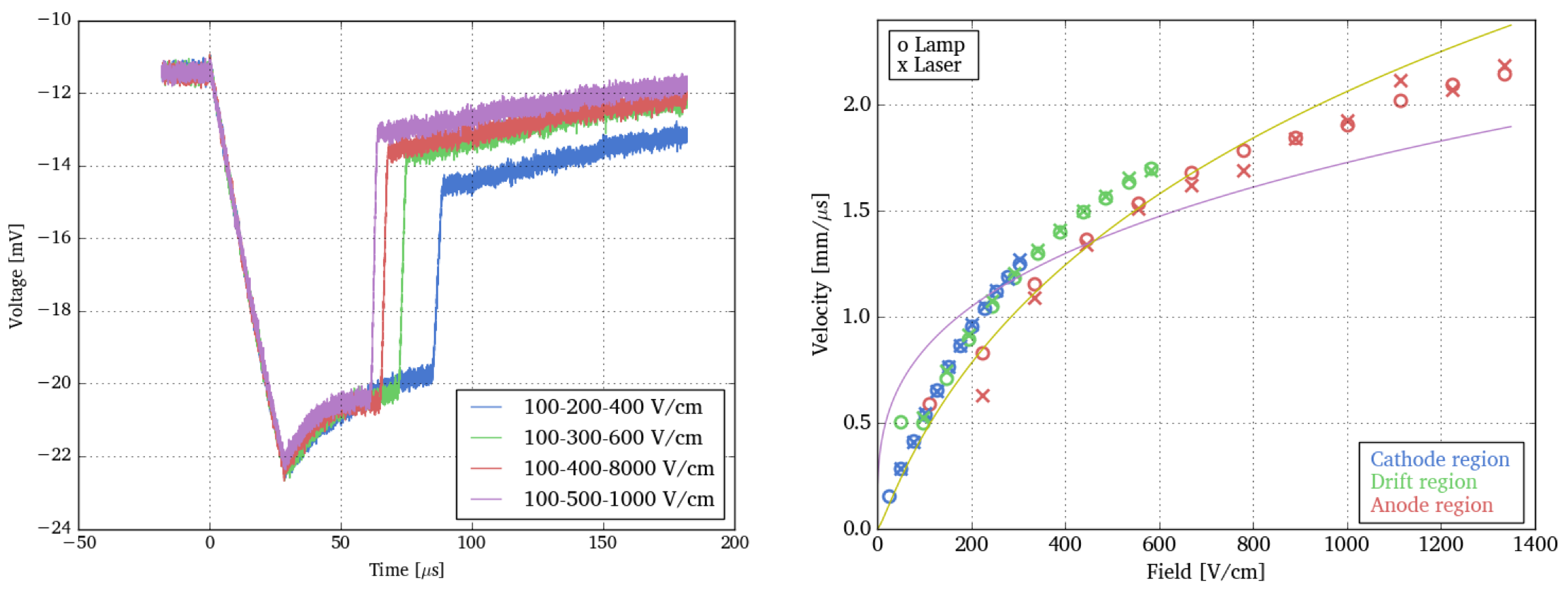}
\par\end{centering}
\caption[Electron velocity in liquid argon]{\textbf{Left:} Holding the field in the cathode region constant but
changing electric fields in the drift and anode regions demonstrates
the dependence of electron velocity on the electric field. \textbf{Right:}~Electron
drift velocity in LAr. Drift velocities can be obtained from all three
regions of the purity monitor indicated by different colored markers.
The purple line is a fit from~\cite{kalinin1996LAr_v_temperature}
at 87.3 K, and the yellow line is a fit from~\cite{he2009electron}
with parametrization at 85 K. Since electron velocity in LAr varies
with temperature it is not expected that these fits will agree. \label{fig:Electron-drift-velocity}}
\end{figure}

\subsubsection{Electron velocity in liquid argon}

Since the purity monitor can be operated at a wide range of electric
fields, it can be used to measure the velocity of electrons in LAr
as shown in Figure~\ref{fig:Electron-drift-velocity}. Fits from~\cite{kalinin1996LAr_v_temperature,he2009electron}
are included for comparison. It is clear that the observed electron
velocity is non-linear as a function of the field~\cite{miller1968charge}.
It has also been demonstrated that electron velocity in LAr varies
with temperature~\cite{walkowiak2000drift} and the amount of impurities~\cite{swan1964drift}.
In fact, some of those experiments used detectors nearly identical
to the purity monitor described here for their measurements.

\subsection{Alternative design ideas}

Initially, an alternative purity monitor design was pursued. Instead
of generating electrons by the photoelectric effect on the cathode,
a radioactive source with a PMT acting as a trigger was used. This
would eliminate the need for the fairly expensive and impractical
optical fiber needed to deliver light to the photocathode. This section
describes the basic idea of this design, which was inspired by~\cite{BARRELET2002204,ADAMS2005613,Badertscher:2004bm}.
Unfortunately, due to the move to LBNL and with a tight deadline in
sight in order to deliver a working purity monitor for the LZ HV feedthrough
tests, the radioactive source design was abandoned due to the large
amounts of R\&D\nomenclature{R\&D}{Research \& Development} needed
to get a working prototype.

\begin{figure}
\begin{centering}
\includegraphics[scale=0.65]{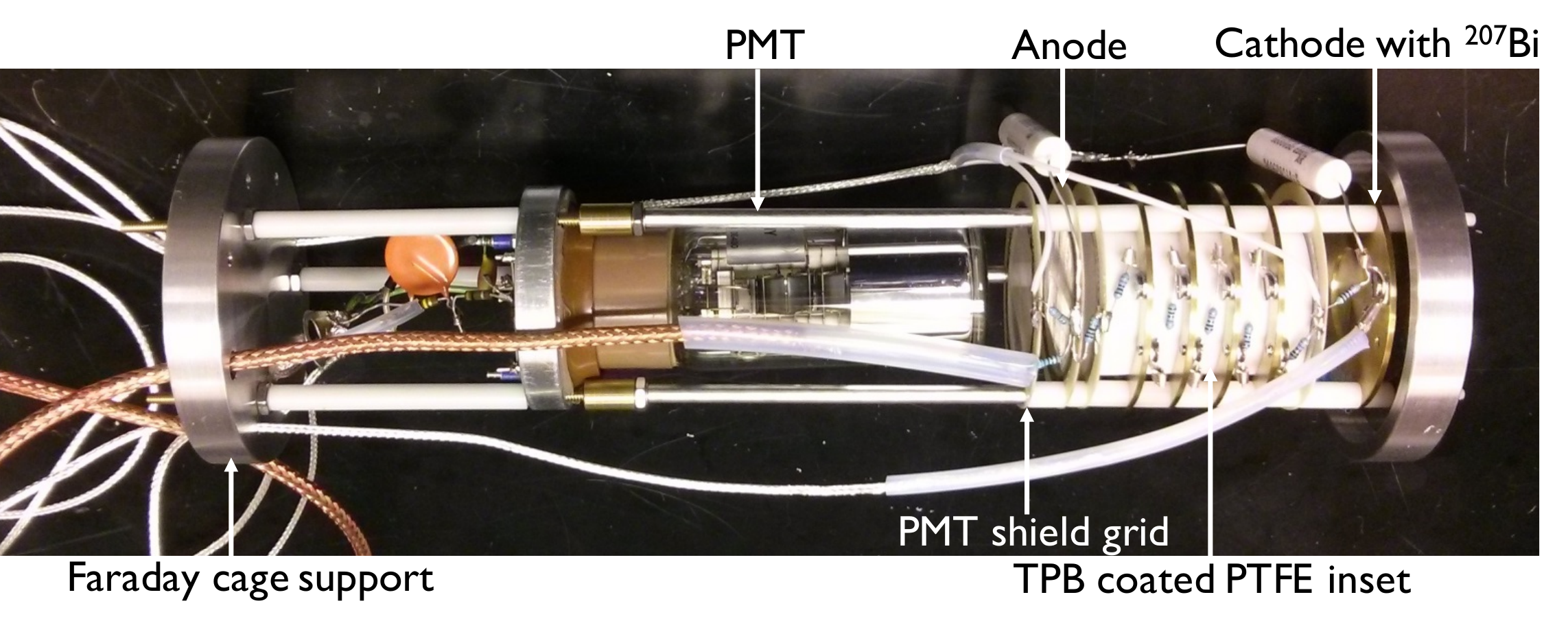}
\par\end{centering}
\caption[Photograph of the original purity monitor with a PMT]{Photograph of the original purity monitor built at Yale University
with $^{207}$Bi deposited on the cathode and a PMT serving as a trigger.
\label{fig:original_PM}}

\end{figure}

A photograph of the original purity monitor is shown in Figure~\ref{fig:original_PM}.
A radioactive source located on the cathode creates ionization and
scintillation signals as it decays. The ionization electrons drift
through the purity monitor to the anode and constitute the signal
seen on the oscilloscope. The scintillation signal is detected by
the PMT and used as a trigger. As discussed in Section~\ref{subsec:Feedthrough-tests-LBNL},
the $\unit[128]{nm}$ scintillation light emitted in LAr is not visible
to the bialkali PMT used (with a range of $\unit[\sim300-600]{nm}$).
Therefore, $\unit[0.11]{mg/cm^{3}}$ of TPB was deposited on a sheet
of PTFE rolled inside the field cage re-emitting the light within
the PMT range.

Two different sources were used for tests. $^{207}$Bi that decays
to $^{207}$Pb with a half-life of 32.9 years. It emits 1 MeV internal
conversion electrons 20\% of times. However, light is also created
by radioactive backgrounds, which makes triggering selectively on
the stochastic radioactive decay complicated. Subsequently, $^{210}$Pb
source with a higher energy decay was used instead. It disintegrates
to $^{206}$Pb via a 5.3 MeV $\alpha$ decay with a half-life of 138
days. However, neither option created a detectable PMT signal, likely
because the radioactive source was located too far from the PMT. Therefore,
radioactive background events near the PMT could create an overwhelming
amount of triggers, flooding the signal from the $^{210}$Pb decay
and making it hard to detect.

A possible design improvement would be to relocate the PMT closer
to the cathode to increase signal brightness and to improve trigger
efficiency. This would require changing the cathode design to a grid
to make it transparent to light. For example, an aqueous solution
of $^{210}$Pb could be deposited on wires as has been done in~\cite{Sorensen:2017kpl}
and the TPB could be deposited directly on the PMT face. Alternatively,
a transparent conductor, such as indium tin oxide (ITO) could be used.
In this case, the TPB-coated PMT would be located directly behind
the transparent cathode, looking into the purity monitor.

Ultimately, the advantage of the design that includes a radioactive
source is the lack of a photocathode and the optical fiber which is
delicate and can be impractical. Furthermore, if many feedthroughs
are used between the light source and the cathode, this can lead to
large losses of light. However, this problem can be avoided since
a sufficient amount of light was delivered to the purity monitors
developed in this work despite using three optical fibers connected
by two feedthroughs.

\section{Purity monitor for XeBrA\label{sec:Purity-monitor-for-XeBrA}}

A second purity monitor used in the XeBeA detector was built by Glenn
Richardson, my undergraduate advisee at UC Berkeley. The design closely
followed the purity monitor for LZ described earlier but features
a few changes described here.

\begin{table}
\begin{centering}
\begin{tabular}{lrl}
\hline 
\textbf{Purity monitor geometry} & \textbf{Value} & \textbf{Unit}\tabularnewline
Cathode \& anode diameter & 52.3 & mm\tabularnewline
Cathode - cathode shield grid & 10{\small{}.0} & mm\tabularnewline
Cathode shield grid - anode shield grid & 70{\small{}.0} & mm\tabularnewline
Anode shield grid - anode & 5{\small{}.0} & mm\tabularnewline
Cathode - anode & 85{\small{}.0} & mm\tabularnewline
Number of field shaping rings & 6 & \tabularnewline
\hline 
\textbf{Shield grids} &  & \tabularnewline
Grid wire width & 0.046 & mm\tabularnewline
Grid wire height & 0.038 & mm\tabularnewline
Grid wire spacing & 1.000 & mm\tabularnewline
\hline 
\end{tabular}
\par\end{centering}
\caption[Parameters of the XeBrA purity monitor]{Parameters of the XeBrA purity monitor. The ground shield grid that
serves to isolate the purity monitor from the main detector volume
electrostatically has a pitch of 0.2 mm. \label{tab:XeBrA_PM_details}}
\end{table}

\begin{figure}
\begin{centering}
\includegraphics[scale=0.29]{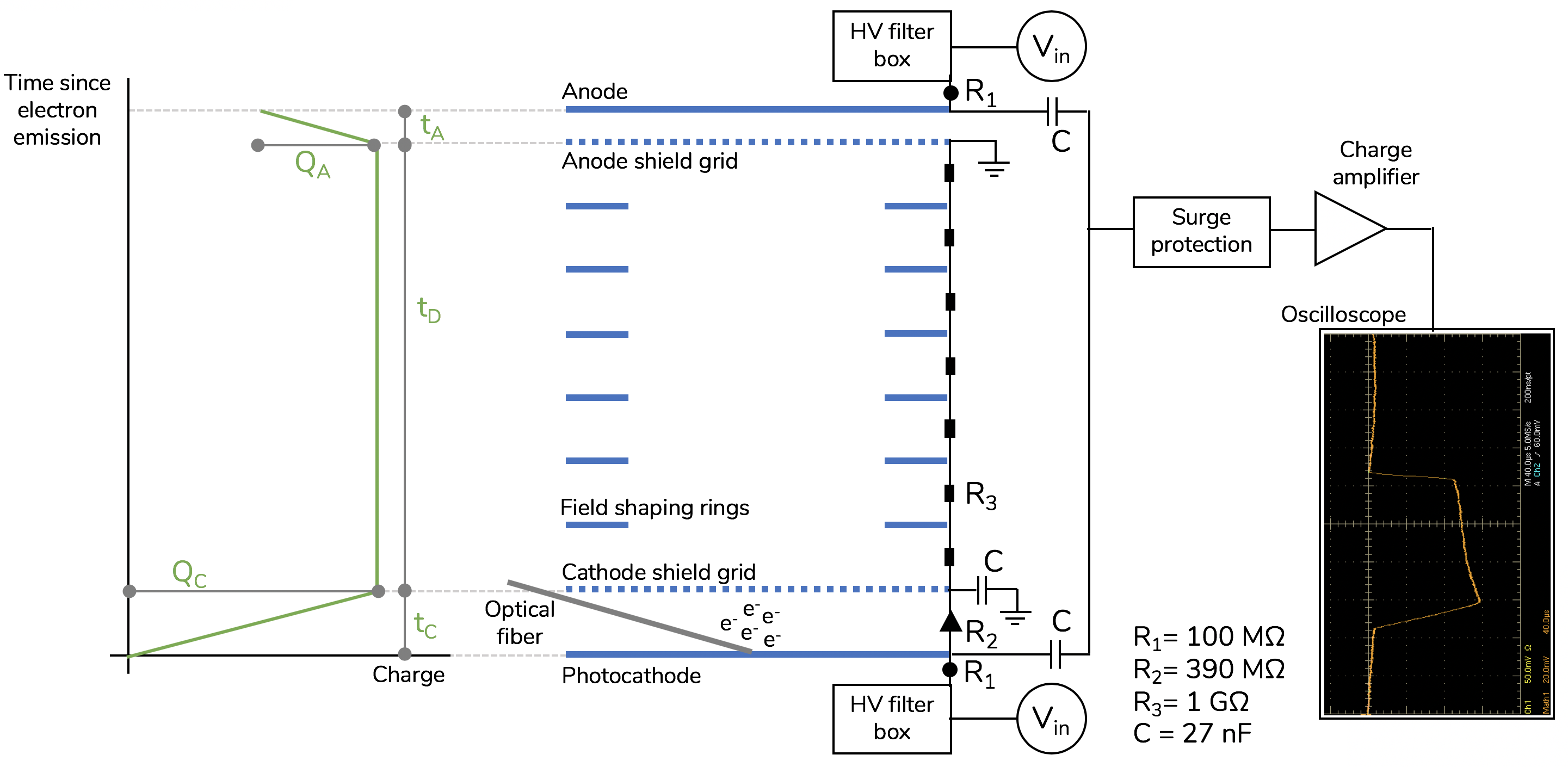}
\par\end{centering}
\caption[Schematics of the purity monitor deployed in XeBrA]{Schematics of the purity monitor deployed in XeBrA with a trace obtained
in LXe.\label{fig:Schematics-XeBrA-PM}}

\end{figure}

First, instead of having an interchangeable photocathode disk, the
entire copper cathode was coated in gold. The field shaping rings
and the anode were made out of brass. To reduce outgassing, the three
supporting rods were manufactured from PEEK and spacers were made
from alumina ceramic. All shield grids were custom made by photochemical
etching of T301 stainless steel. The cathode and anode grids were
designed to ensure that doubling electric fields would result in 100\%
transparency for drifting electrons. The ground shield disk located
at the end of the purity monitor serves as a Faraday cage to insulate
the purity monitor from the rest of the experimental region. Table~\ref{tab:XeBrA_PM_details}
includes a summary of the purity monitor parameters. The electronics
were designed similarly to the LZ purity monitor but with the cathode,
instead of the anode, biased to high voltage as shown in Figure~\ref{fig:Schematics-XeBrA-PM}.
NPO type capacitors were used since their capacitance is not temperature
dependent~\cite{teyssandier2010commercial_capacitors}.

\begin{figure}
\begin{centering}
\includegraphics[scale=0.55]{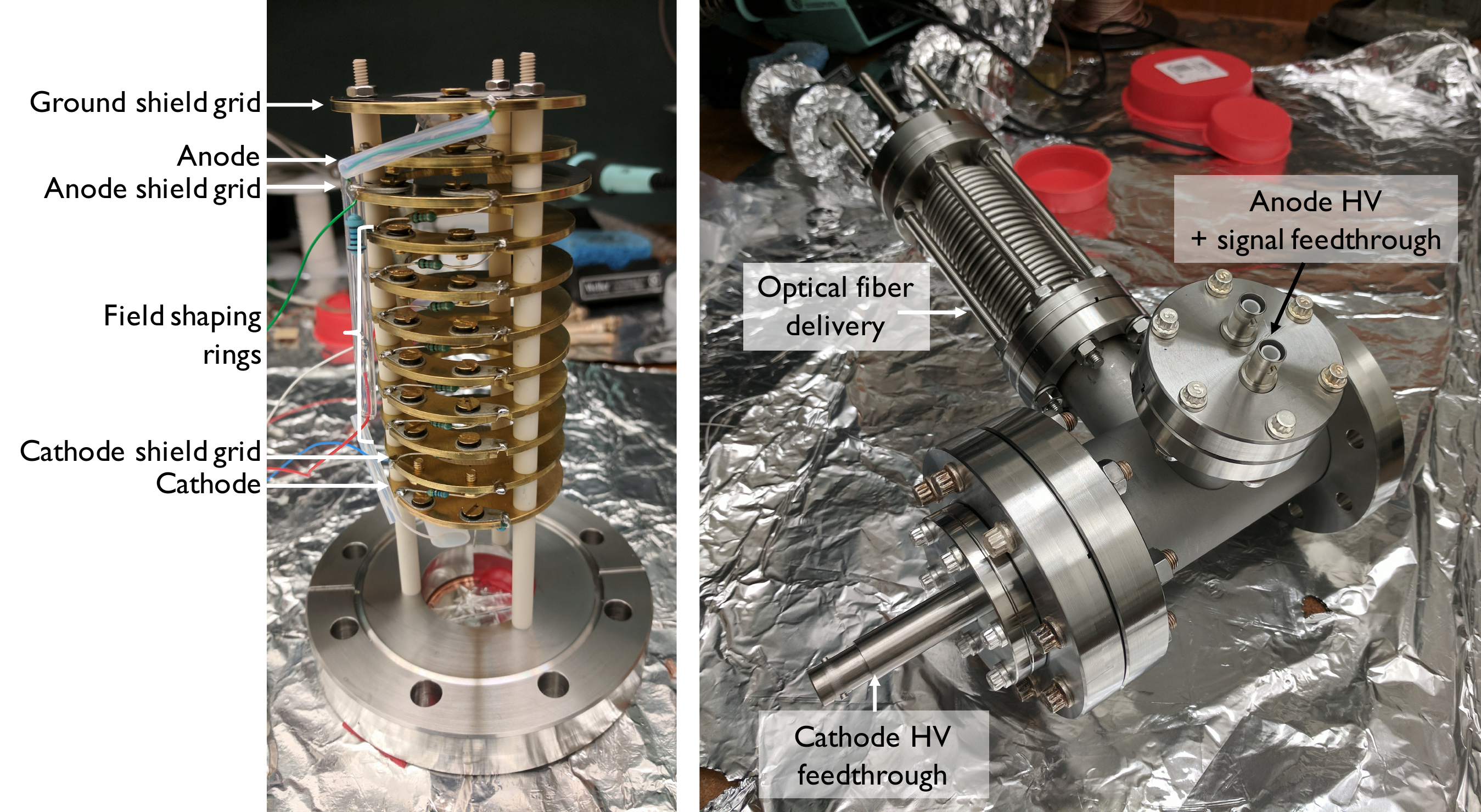}
\par\end{centering}
\caption[Photograph of purity monitor used in XeBrA]{\textbf{Left:} Photograph of purity monitor used in XeBrA. \textbf{Right:}
The purity monitor is entirely contained inside  CF flanges that also
act as a Faraday cage. The optical fiber is delivered from the side.
Since the cathode is at high voltage, it is conservatively designed
to have a separate feedthrough.\label{fig:XeBrA-PM-picture}}
\end{figure}

A photograph of the purity monitor is shown in Figure~\ref{fig:XeBrA-PM-picture}.
The purity monitor was tested in LAr using the setup described in
Section~\ref{subsec:Purity-monitor-testing} and has been successfully
operated in XeBrA in both LAr and LXe.

A third purity monitor that very closely follows the design of the
XeBrA purity monitor is currently being built at LBNL to be used in
the IBEX optical experiment. 

\section{Conclusion}

The LZ detector is a two-phase xenon TPC currently under construction
at SURF in Lead, South Dakota. The collaboration aims to start physics
data collection in 2020. The LZ detector will probe a significant
amount of the remaining parameter space above the neutrino floor for
the direct detection of SI and SD WIMP-nucleus elastic scattering.
Tests of various subsystems are ongoing at all of the collaborating
institutions, and UC Berkeley is responsible for the cathode HV delivery.
To validate results from the cathode HV feedthrough tests a purity
monitor was successfully developed to provide a reliable and straightforward
way to measure LAr purity \textit{in situ} in real time. Due to the
success of the LAr purity monitor a second purity monitor was built
to monitor LAr and LXe purity in XeBrA, followed by a third monitor
currently in development for the IBEX experiment.

\subsection{Contributions to LZ}

As a member of the LZ collaboration, I contributed to the cathode
HV feedthrough development at Yale and LBNL by designing, building,
and testing the purity monitor as described in this chapter. The purity
monitor served an essential role in the high voltage feedthrough tests.
I also assisted with the feedthrough tests themselves and optimized
electric field models of an earlier version of the LZ feedthrough.
Because of my experience modeling electric fields of the LUX detector,
I advised a graduate student at UC Berkeley developing an axially-symmetric
LZ field model used in data analysis. 

I also spent time on site at SURF to assist with the installation
of the radon reduction system used in the surface cleanroom where
the LZ detector is being assembled. The radon reduction system was
manufactured by the Ateko company based in Hradec Králové, Czech Republic~\cite{Ateko}.
I provided English-Czech translation and helped with safety training
of the workers. The system has been shown to reduce the level of $^{222}$Rn
in the air from $\unit[\sim50]{Bq/m^{3}}$ by a factor of 1,000~\cite{Benato:2017kdf}.
Similar systems have been installed in Yangyang Laboratory in South
Korea, Canfranc-Estacion in Spain, and in LNGS at Gran Sasso, Italy. 

Furthermore, I founded the LZ Equity \& Inclusion committee that has
since become an integral part of the LZ collaboration. More details
about the foundation of the committee and its activities are described
in Appendix~\ref{chap:LZ-E=000026I}.

\chapter{\textsc{Xenon Breakdown Apparatus (XeBrA)}\label{chap:XeBrA}}

The goal of the Xenon Breakdown Apparatus (XeBrA)\nomenclature{XeBrA}{Xenon Breakdown Apparatus}
at the Lawrence Berkeley National Laboratory (LBNL) is to characterize
the high voltage (HV) behavior of liquid xenon (LXe) and liquid argon
(LAr). As will be discussed in this chapter, the probability of a
dielectric breakdown\footnote{A dielectric breakdown happens when there is a current through the
LAr or LXe volume, which are electrical insulators. This usually results
in a quick visible and audible spark (or a series of sparks). Most
power supplies have a safety mechanism that interrupts the voltage
when a current exceeds a given threshold (usually $\mathcal{O}\left(\unit[10]{\mu s}\right)$
in detector applications).} increases for large electrode surfaces and depends on other factors,
such as the noble purity. This is a concern, since as noble liquid
detectors grow larger in size, achieving stable HV on the detector's
cathode becomes increasingly challenging\footnote{The details of R\&D for HV delivery in two-phase TPCs was discussed
in more detail in Section~\ref{subsec:LZ-High-voltage-delivery}.}. A breakdown in a two-phase TPC is a concern because it can damage
detector hardware. Even before a full breakdown occurs, spurious light
and charge emission can compromise detector performance at lower voltages.
Therefore the results from XeBrA will not only improve our understanding
of dielectric breakdown, but they will serve to inform the design
of the next-generation noble liquid time projection chambers (TPCs). 

The original measurements of liquid argon breakdown were done over
micrometer gap distances~\cite{swan1960influence,swan_LAr}, skewing
the community\textquoteright s outlook on the plausible HV performance
of the detectors. As larger detectors were developed, they were not
able to reach the voltages predicted by these studies. Later measurements
in LAr at more realistic centimeter scales revealed behavior pointing
to electric field breakdown dependence on many variables such as electrode
area, material, surface finish, LAr purity, and more. Due to the similarities
between LAr and LXe, this behavior is likely also to be seen in LXe.
However, there is a relative lack of such measurements in LXe. 

HV behavior in liquid nobles is only beginning to be explored, but
there have been studies of dielectric breakdown behavior in other
media. Section~\ref{sec:Breakdown-studies-primer} provides a brief
overview of selected studies and identifies trends that may be of
interest to study with XeBrA. Dependence of breakdown on electrode
area has been widely agreed upon in the community; however, a theoretical
explanation of the process has still not been identified. So far,
the best answers have arisen from extreme value (weak link) theory
discussed in Section~\ref{sec:Extreme-value-theory}. It should be
noted that breakdown dependence on electrode size is not a unique
effect: there are many products whose failure rates are often assumed
to be proportional to their size. Some examples include capacitor
dielectric (proportional to area), cable insulation (proportional
to length), electrical insulation (proportional to thickness), and
conductors in microelectronics (proportional to length and number
of bends)~\cite{nelson2009accelerated}.

The initial studies in XeBrA were focused on exploring the area effect.
The largest measured area in the literature achieved in LAr is only
$\unit[3]{cm^{2}}$, limited by the geometry of the electrodes. To
overcome this limitation, XeBrA utilizes Rogowski electrodes optimized
for field uniformity over large areas (Section~\ref{subsec:Rogowski-electrodes});
thereby providing insight into areas that are more relevant to the
large experiments. Additionally, XeBrA can be filled with either LAr
or LXe and can test electrodes with areas up to $\unit[58]{cm^{2}}$
by varying their separation remotely. This makes XeBrA the first apparatus
to enable a direct comparison of HV behavior between LAr and LXe,
providing new insight into HV behavior in noble liquids useful in
both dark matter and neutrino detection. The design of the apparatus
is detailed in Section~\ref{sec:Detector-design}, and HV considerations
for optimizing the main liquid volume to ensure breakdown occurred
between electrodes are discussed in Section~\ref{subsec:High-voltage-design}. 

There is a purity monitor (Section~\ref{subsec:Purity-monitor})
and a PMT (Section~\ref{subsec:PMT}) directly adjacent to the main
active volume. The purity monitor measures the electron lifetime,
which provides information about the liquid purity. The PMT is used
for the study of the onset of electroluminescence. Additionally, two
current monitor devices are attached to the anode. Along with the
PMT, they serve to look for signs of early breakdown development prior
to the formation of the spark. If a predictive pattern could be identified,
then breakdowns in future detectors could be prevented by stopping
the HV ramp in a timely manner. 

A dedicated gas system was designed and built to be shared by XeBrA
and IBEX\footnote{An experiment studying the reflectivity of PTFE in liquid xenon.}.
Its development is described in Section~\ref{sec:Gas-system}, and
a procedure developed for its operation is included in Appendix~\ref{chap:XeBrA-procedures}.
There are two more Appendices related to XeBrA included in this work.
Appendix~\ref{chap:Intro-to-fittings} serves as a brief introduction
to the various types of fittings used in the construction of both
the apparatus and the gas system mentioned throughout the text. The
fittings appendix can also serve as a general reference for anyone
building a system for handling noble gases. A list of parts used both
in the gas system and in XeBrA is included for completeness, and can
be found in Appendix~\ref{chap:Parts-list}.

Section~\ref{sec:Detector-operations} discusses XeBrA operations,
and Section~\ref{sec:Data-collection} provides an overview of the
four data acquisitions made during 2018: two made in LAr and two in
LXe. The analysis of these data along with preliminary results is
presented in Section~\ref{sec:Results}. A publication of results
from the 2018 scientific operation is in preparation. Looking ahead,
Section~\ref{sec:Discussion} includes a discussion of future studies
for which XeBrA would be suitable.

\section{Breakdown studies primer \label{sec:Breakdown-studies-primer}}

There are no published comprehensive studies of dielectric breakdown
behavior for large area electrodes in LXe. These measurements are
desirable not only to improve our understanding of the physical processes
involved in the breakdown but also to inform the future of TPC engineering.
These are the primary motives for XeBrA. However, the dielectric breakdown
has been of interest for many decades in various other materials due
to their application as insulators in electric power generation and
transmission. This section provides a brief overview of the literature
for selected materials and highlights relevant effects that might
be of interest in LXe. Studies of dielectric breakdown as a function
of both alternating current (AC)\nomenclature{AC}{Alternating Current}
and direct current (DC)\nomenclature{DC}{Direct Current} are available
in the literature, and the breakdown behavior appears to vary between
AC and DC. Only DC HV is currently of interest for LXe detectors,
which dictates the focus of this primer. 

All studies agree that dielectric breakdown depends on the electrode
area with high electric fields. This dependence can be justified experimentally,
although a simple thought experiment can provide good intuition about
the effect~\cite{weber1956nitrogen}. A sketch of the thought experiment
is as follows. Assume that breakdown for a given area $A_{0}$ is
measured and the breakdown field $E_{0}$ is found as the mean of
$N$ measurements. Consider an experimental apparatus consisting of
$n$ copies of the $A_{0}$ experiment connected in parallel, resulting
in a larger area $A=n\times A_{0}$. A single data point can be generated
for the apparatus with area $A$ by taking the minimum of a random
selection of $n$ data points from the original experiment. Repeat
this random sampling $m=N/n$ times to generate a full dataset, ensuring
subsequent choices are disjoint. The mean of this new dataset  will
represent the dielectric breakdown for an area $A$. This mean (since
it is a mean of minima) will be less than $E_{0}$, the mean of the
$A_{0}$ experiment. This justifies that larger areas result in dielectric
breakdowns at lower electric field values. This effect is also known
as the stability postulate in probability theory, which is leveraged
in the weak link theory.

\subsection*{Transformer oil}

Transformer oil is commonly used for its excellent electrical insulation
properties. Most tests in transformer oil were made using AC. Several
studies observed decreasing breakdown performance with increasing
electrode area using Rogowski electrodes\footnote{A wide array of electrode geometries have been used in breakdown studies,
such as two spherical electrodes, a spherical or needle-like electrode
facing a plane, sets of parallel electrodes, or Rogowski electrodes.
We do not consider geometrical effects in this study.} of multiple sizes~\cite{weber1956nitrogen,weber1955areaEffect}.
Impurities in oil increased the chance of breakdown, and the breakdown
values depended on the velocity of contaminants moved by oil circulation~\cite{ikeda_movingOil}.
Dependence of breakdown on the oil volume subject to high electric
stress was observed as well. 

\subsection*{Liquid nitrogen}

Reference~\cite{KAWASHIMA1974217} observed an area effect using
multiple parallel plate electrodes. The study also notes that dielectric
breakdown occurred at lower voltages in the presence of small bubbles.
Reference~\cite{goshima1995nitrogen} observed that increasing electrode
area and stressed volume, having a mutual correlation, decreased HV
performance (sphere to plane electrode geometry). Electrode polarity
was also observed to affect breakdown values.

\subsection*{Liquid helium }

Extensive studies have been performed in liquid helium (LHe)\nomenclature{LHe}{Liquid Helium}
for applications in neutron electric dipole moment searches~\cite{nEDM,Ito:2014jca,long2006high,SPSnEDM}.
Reference~\cite{GERHOLD1994579} contains a comprehensive review
of breakdown test conditions in LHe to that date. Reference~\cite{Gerhold1989}
identified dependence of dielectric breakdown on electrode area and
spacing, liquid purity, electrical field conditions, test procedure,
liquid pressure and temperature, and electrode surface condition.
The surface roughness of the cathode was identified as the most critical
parameter~\cite{Gerhold1989}. Addition of oxygen or conducting impurities
had an adverse effect on breakdown strength, as dirty LHe was observed
to break down at lower fields. Bubble formation occurred during any
breakdown in LHe~\cite{Gerhold1989,schwenterly1974}. Additionally,
bubbles were found to cause breakdown due to their lower dielectric
strength\footnote{Gasses generally hold lower fields than their corresponding liquids.}.
A polarity effect was observed~\cite{schwenterly1974}, with the
breakdown initiated on the cathode~\cite{Gerhold1989}. Reference~\cite{GERHOLD1994579}
observed a volume power law when considering only the half of the
stressed volume adjacent to the cathode.

\subsection*{Liquid argon}

\begin{figure}
\begin{centering}
\includegraphics[scale=0.25]{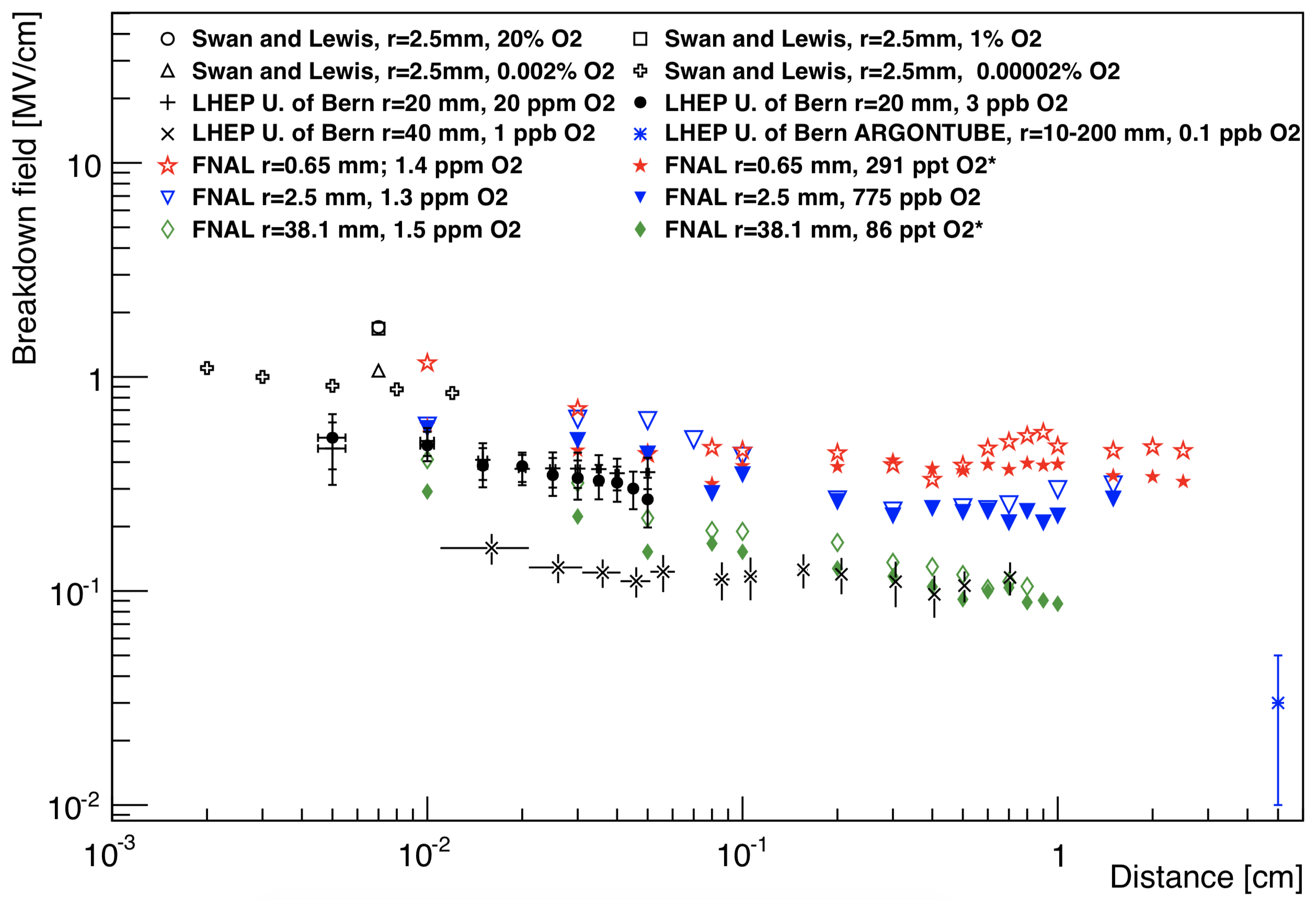}
\par\end{centering}
\caption[Previous measurements performed in LAr as a function of electrode
separation]{A comparison of various measurements performed in liquid argon as
a function of electrode separation. Dependence of breakdown on purity
can be seen. Additionally, larger electrodes were observed to perform
worse than smaller ones at a given electrode separation. Figure from~\cite{Acciarri:2014ica}.\label{fig:lar-impurities}}
\end{figure}

There are abundant studies performed in liquid argon (LAr) for applications
in neutrino interaction and oscillation experiments. One of the first
available studies in LAr used 5 mm diameter electrodes with varying
separations and tested six different electrode materials that were
observed to affect the dielectric breakdown behavior~\cite{swan_LAr}.
An effect of material oxidation was observed as well. Based on the
limited sample size of that study, stainless steel (SS)\nomenclature{SS}{Stainless Steel}\footnote{The exact SS alloy is not known. Unfortunately, most authors of the
breakdown studies mentioned throughout this chapter do not include
the type of SS alloy used for electrode manufacturing.} performed the best. Reference~\cite{swan_LAr} also observed an
increase in breakdown strength with higher oxygen impurity concentrations
(an effect opposite to what was observed in LHe). The effect of impurities
was later confirmed by~\cite{Acciarri:2014ica} as shown in Figure~\ref{fig:lar-impurities}.
Reference~\cite{Acciarri:2014ica} also confirmed that the location
of sparks varied for consecutive sparks. Using Rogowski electrodes
for their studies, Reference~\cite{Bay:2014jwa} confirmed that LAr
bubbles worsen the breakdown performance.

There were two notable efforts in recent years to investigate breakdown
behavior in LAr. One was at the Laboratory for High Energy Physics
(LHEP)\nomenclature{LHEP}{Laboratory for High Energy Physics} at
the University of Bern, Switzerland~\cite{1748-0221-9-07-P07023,Auger:2015xlo,blatter2014experimental}.
Measurements at LHEP were performed at $\sim\unit[1.2]{bar}$\footnote{Throughout this chapter, units of ``bar'' refer to the absolute
pressure. Gauge pressure is indicated by ``g'' following the units,
i.e., ``barg.''}~\cite{Auger:2015xlo}, using SS electrodes with mainly a mechanical
polish of $\unit[0.1-0.4]{\mu m}$. Their studies confirmed the area
effect. They also managed to suppress the field emission using an
emulsified water solution to deposit a layer of natural polyisoprene
(latex rubber) onto the cathode surface. However, the performance
of the coating degraded and became very similar in performance to
SS electrodes after the first breakdown that damages the rubber. They
also monitored spark development using a camera and a spectrometer
leading to a detailed description of breakdown development. 

The other effort was at the Fermi National Accelerator Laboratory
(FNAL)\nomenclature{FNAL}{Fermi National Accelerator Laboratory}.
Measurements there were taken at $\unit[1-1.6]{bar}$~\cite{Acciarri:2014ica}.
Their electrodes were polished to a mirror finish, then electropolished,
but the exact value of the electrode finish is not known.

\subsection*{Liquid xenon}

There is one published study in LXe performed at SLAC~\cite{Rebel:2014uia}
as part of the EXO-200 experiment~\cite{Albert:2017owj}. The setup
used two 1.5~cm diameter spheres separated by~1 mm for the measurement
with area $\unit[3.1]{mm^{2}}$ ($\unit[0.031]{cm^{2}}$ ) reaching
90\% of the maximum electric field. Breakdowns were observed at $\unit[250-300]{kV/cm}$.
Neither the purity nor the pressure at which the data point was obtained
is known. 

\begin{figure}
\centering{}\includegraphics[scale=0.75]{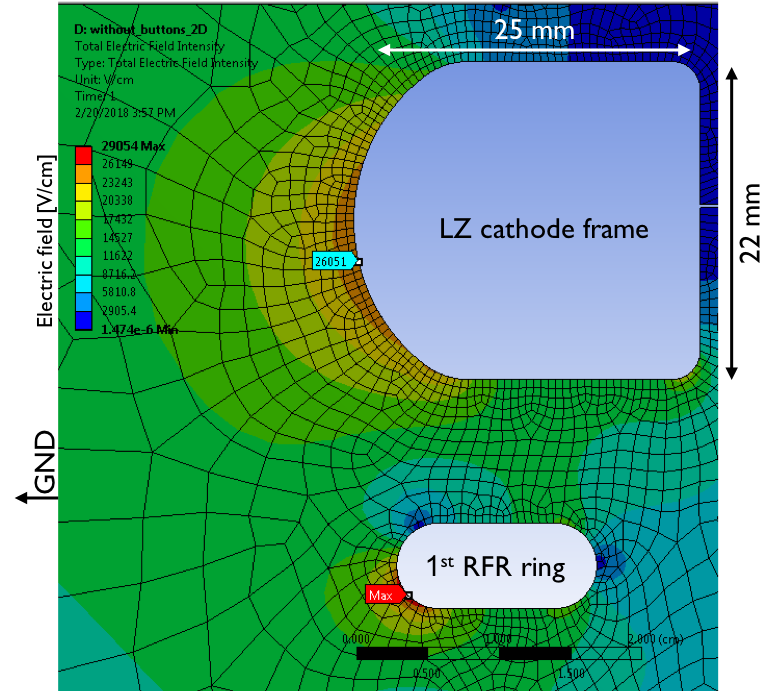}\caption[Simulation of electric fields around the cathode of the LZ detector]{Simulation of electric fields around the cathode of the LZ detector.
The inner cryostat wall is located to the left of the picture. The
outer diameter of the cathode frame is 150~cm. The turquoise arrow
indicates the point with the highest field (26~kV/cm) on the cathode
frame. The red arrow indicates the highest electric field in the field
cage (29~kV/cm), located on the reverse field region (RFR)\nomenclature{RFR}{Reverse Field Region}
ring. Figure from~\cite{LZcathodeSim}. \label{fig:lz-cathode-field}}
\end{figure}

The stressed area in LZ will be much larger than what was studied
at SLAC. According to simulations and assuming -100~kV on the cathode,
the highest electric field in the LZ field cage is on the field shaping
ring directly below the cathode, in the reverse field region, as shown
in Figure~\ref{fig:lz-cathode-field}. The electrode area with a
field equal to or larger than 90\% of the maximum field (29~kV/cm)
corresponds to $\unit[150]{cm^{2}}$. The largest area at high voltage
is the cathode frame, with a maximum field of 26~kV/cm over an area
of $\unit[500]{cm^{2}}$. The LZ electrodes are manufactured from
304~SS.

\subsection{Summary of parameters affecting breakdown behavior}

The adverse effect on dielectric breakdown caused by increasing electrode
area has been agreed on in studies in many materials. However, other
effects have been identified to influence dielectric breakdown behavior.
A list of some of these effects with references to the relevant literature
follows.
\begin{itemize}
\item \textbf{Area:} The most universally agreed upon factor affecting breakdown
behavior. See for example~\cite{weber1956nitrogen,weber1955areaEffect,KAWASHIMA1974217,goshima1995nitrogen,GERHOLD1994579,Gerhold1989,Acciarri:2014ica,Auger:2015xlo,goshime1995weibull,Gerhold_LiquidHelium,toya1981}.
\item \textbf{Volume:} Some studies also report breakdown strength decrease
with an increase of the stressed liquid volume~\cite{goshima1995nitrogen,GERHOLD1994579,goshime1995weibull,Gerhold_LiquidHelium}.
\item \textbf{Surface finish:} Smoother finish has been shown to break down
at higher voltages in LHe~\cite{GERHOLD1994579,Gerhold1989} and
in LN~\cite{goshima1995nitrogen}. 
\item \textbf{Material:} Varying materials, as well as surface oxidation,
affected dielectric breakdown in LAr~\cite{swan_LAr}. Material dependence
was also observed in LHe~\cite{GERHOLD1994579,Gerhold1989}.
\item \textbf{Pressure and temperature:} Breakdown performance improved
with increasing pressure in LN~\cite{KAWASHIMA1974217} and LHe~\cite{Gerhold1989,long2006high}.
However, Reference~\cite{KAWASHIMA1974217} notes that earlier studies
in LN saw a decrease in breakdown values with increasing temperature
(which is directly correlated with pressure).
\item \textbf{Purity:} Impurities increased the chance of a breakdown in
transformer oil~\cite{ikeda_movingOil} and LHe~\cite{Gerhold1989}.
However, in LAr, the presence of impurities improved HV performance
of the experimental setup~\cite{swan_LAr,Acciarri:2014ica,blatter2014experimental}.
\item \textbf{Polarity:} It is believed that the spark during a dielectric
breakdown is initialized on the cathode surface and affected by space
charge effects on the cathode surface and in the stressed volume~\cite{Acciarri:2014ica,blatter2014experimental}.
Previously, the effect of polarity was observed in a vacuum where
the breakdown performance was worse when strong fields were applied
to the cathode rather than the anode~\cite{toya1981}. The same behavior
was also observed in LN~\cite{goshima1995nitrogen} and LHe~\cite{Gerhold1989,schwenterly1974}.
\item \textbf{Conditioning:} Conditioning is frequently used in vacuum applications
to increase the breakdown voltage of surfaces. Applying high electric
fields to critical regions serves to burn off impurities and surface
irregularities that might act as field emission sites during breakdown
otherwise. Note that this is only the case if the current is not high
enough to damage the electrodes. However, the effect of conditioning
electrode surfaces is unclear in liquid nobles. It appears to be a
valid mechanism in a vacuum, particularly with impulse voltage~\cite{toya1981,ballat,yasuokaelectrode}
and transformer oil~\cite{ikeda_movingOil}, but its effects are
uncertain in LHe~\cite{Gerhold1989,7464486}.
\item \textbf{Circulation speed:} Breakdown voltage in transformer oil has
been shown to depend on circulation speeds. The impact on breakdown
varies at various velocities~\cite{ikeda_movingOil}.
\item \textbf{Capacitance:} References~\cite{cross1982,mazurek1987} have
argued that a decrease in dielectric strength in a vacuum can be caused
by changes in the energy stored in the system. They observed that
a capacitor placed in parallel with the electrodes reduces dielectric
strength (and also makes electrode damage more significant).
\end{itemize}
This section does not attempt to provide an exhaustive list of parameters
affecting breakdown behavior in liquids. XeBrA's construction and
design allow for all of these parameters to be explored in the future,
for both LAr and LXe. 

Aside from identifying trends caused by dielectric breakdown, the
physics of the initiation of dielectric breakdown and the succeeding
discharge evolution also remains a mystery. 

There are two leading plasma-initiating mechanisms in water. One,
the electronic mechanism, considers electrical discharge to be generated
purely by ionization followed by electron multiplication, only later
followed by a bubble. The other one, the bubble mechanism, assumes
that plasma initiation occurs in the gas phase, in the naturally present
nanobubbles. Characterizing this behavior is an active field of research,
but there is no universal agreement so far. The answer likely depends
on the precise experimental parameters. Reference~\cite{plasma_physics}
provides an excellent overview of the state of the plasma physics
of liquids to date.

\section{Review of statistical analysis of area effects\label{sec:Extreme-value-theory}}

Extreme value theory aims to find a probability of more extreme events
than previously observed. A thorough introduction to the topic is,
for example, in~\cite{gumbel1954statistical}. It is commonly applied
in engineering to predict failures of products based on a small tested
sample, in hydrology to assess likelihoods of 100-year floods, or
in traffic prediction. Also known as weak link theory\footnote{Extreme value theory arose from efforts to make cotton thread stronger
in the 1950's, when Leonard Tippett realized that the strength of
weakest fibers controlled the strength of a thread. }, in application to dielectric breakdown it is expected that continuously
increasing the area or volume subjected to stress will increase the
probability of finding a weaker link causing a breakdown. These weak
links are presumed to be particles, active emission sites on electrodes,
or surface irregularities\footnote{All these parameters are constant in time. In this unclear how time-dependent
variables, such as muons passing through the apparatus, affect this
behavior.}. For extreme value theory to be applicable, there should be a substantial
number of flaws distributed at random and independently. Additionally,
any statistical model used to describe this behavior should follow
the stability postulate. 

An early application of extreme value theory to breakdown probability
in materials (and many other real-life applications) was demonstrated
in~\cite{epstein1948,gumbel1954statistical}. Several cumulative
distribution functions satisfy the stability postulate in the extreme
value theory field, such as Fréchet, Weibull, or Gumbel functions\footnote{The normal distribution does not obey the stability postulate.};
Gumbel and Weibull distributions have been widely used in the field,
but the Weibull function is particularly popular since it is commonly
used in reliability engineering. The Weibull function is a special
case of the generalized extreme value distribution. Representing the
breakdown probability as a 2- or 3-parameter Weibull function takes
one of the forms:
\begin{align}
p= & 1-e^{\left(\frac{x}{\lambda}\right)^{k}}\label{eq:weibull}\\
p= & 1-e^{-\left(\frac{x-\theta}{\lambda}\right)^{k}}\label{eq:weibull_2params}
\end{align}
for $x\geq0$ where $x$ is a random variable (breakdown voltage or
field), $k>0$ is the shape parameter, $\lambda>0$ is the scale parameter,
and $\theta$ is the location parameter of the distribution. For $\theta=0$,
Equation~\ref{eq:weibull_2params} reduces to the 2-parameter Weibull
distribution in Equation~\ref{eq:weibull}. The shape parameter $k$
is correlated to the stressed area as
\[
A=\underset{S}{\iint}\left(\frac{E_{i}}{E_{max}}\right)^{k}dS
\]
where $E_{i}$ is the electric field at an individual unit on the
surface of an electrode and $E_{max}$ is the maximum electric field
and $dS$ is the area differential~\cite{8124618}. The authors of~\cite{8124618}
show that the 3-parameter Weibull distribution is a good method for
approximating breakdown stressed area in transformer oil. 

From the 2-parameter Weibull function, assuming the stability postulate,
a general power law for an area effect can be derived
\begin{equation}
E_{max}=E_{0}\left(\frac{A}{A_{0}}\right)^{-\nicefrac{1}{k}}\label{eq:weibull_lazy}
\end{equation}
where $A_{0}$ is a hypothetical unit area, and $E_{0}$ is the electric
field that breaks down at that area. Note that $E_{0}$, $A_{0}$,
and $k$ are all constants in this formula. This formula has been
simplified further and is commonly used to fit the area effect (and
the volume effect):
\begin{equation}
E_{max}=C\times A^{-b}\label{eq:area-effect-fit}
\end{equation}
where $b=1/k$. 

Furthermore, if the breakdown distribution at area $A_{0}$ is known,
the mean $E_{0}$ may either be computed directly from the sample
or calculated as the analytical mean of the fitted Weibull function.
Equation~\ref{eq:weibull_lazy} then becomes

\begin{equation}
E_{max}=\lambda\Gamma\left(1+\frac{1}{k}\right)\left(\frac{A}{A_{0}}\right)^{-\nicefrac{1}{k}}\label{eq:weibull-prediction}
\end{equation}
and can, in theory, be used to attempt to predict breakdown behavior
as is done in Section~\ref{subsec:Fit-to-weibull}. Here $\Gamma\left(x\right)$
is the gamma function. Moreover, the weak link theory predicts that
the location of sparks will vary for each breakdown.

Reference~\cite{wangACoil} demonstrated that the scale parameter
of the Weibull function decreases as a function of electrode area,
confirming the area effect for AC studies in oil. They also demonstrated
that the shape parameter of the Weibull distribution is independent
of the stressed area of the electrode and instead is a constant determined
by the liquid type. The weakest link theory has also been applied
to breakdown in LHe~\cite{Gerhold_LiquidHelium,1996812,suehiro_LHe},
LN~\cite{weber1956nitrogen,goshime1995weibull,goshima_LN,haywaka_LN,gerhold_LN},
alumina~\cite{alumina_breakdown}, polymer films~\cite{Ulhaq_polymer},
and more~\cite{Hill_breakdown,choulkov}. Further refinement and
applications of these statistical techniques of data generated by
XeBrA is a promising area of future study with the apparatus.

\section{Apparatus design\label{sec:Detector-design}}

\begin{figure}
\begin{centering}
\includegraphics[scale=0.4]{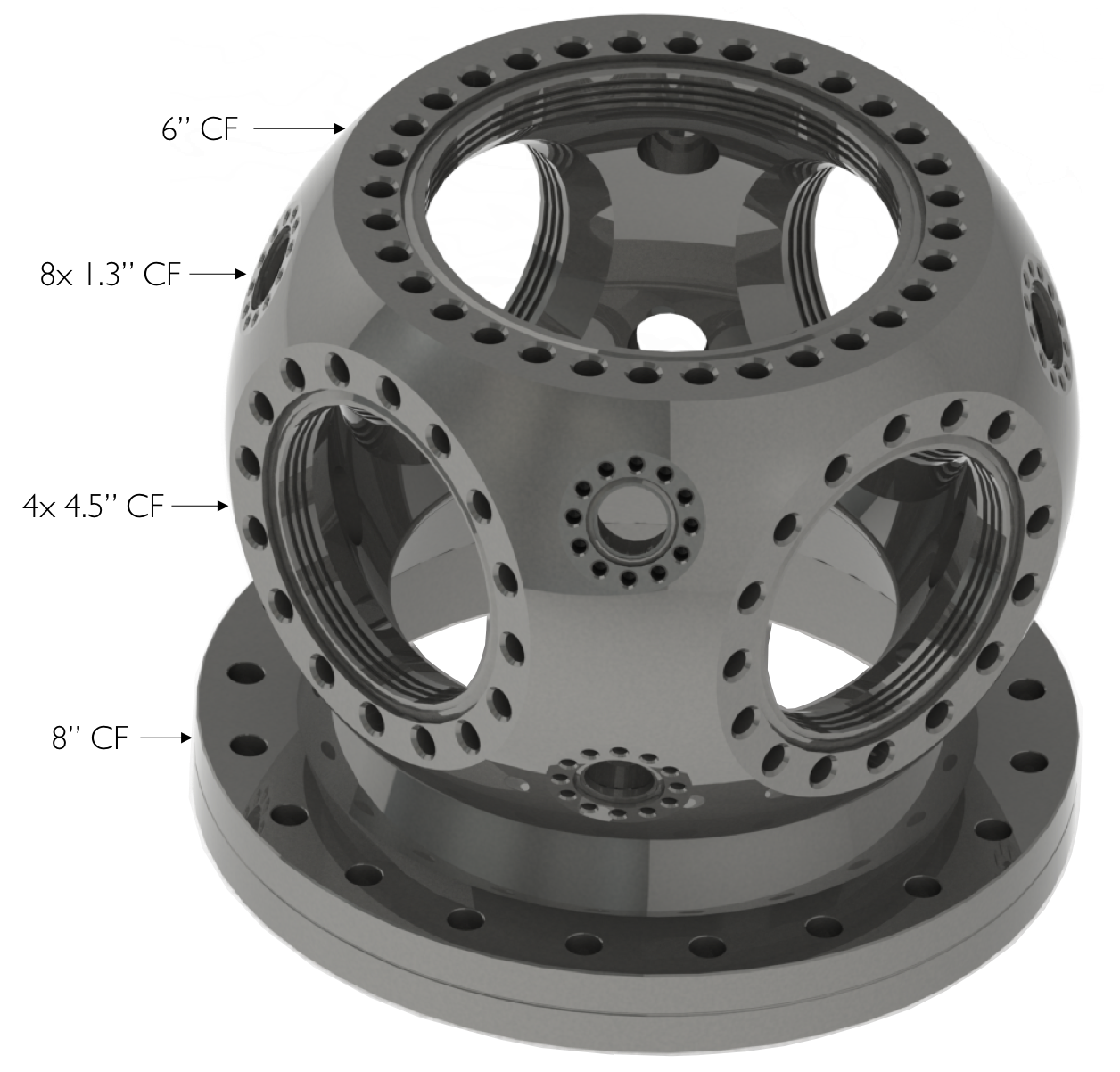}
\par\end{centering}
\caption[Modified spherical square used in XeBrA]{Modified spherical square used in XeBrA with annotated ports. Its
spherical inner diameter is 149.86~mm (5.90~inches) with an internal
volume of $\unit[2028]{cm^{3}}$.\label{fig:Modified-spherical-square}}

\end{figure}

The goal of XeBrA is to study HV behavior in LAr and LXe. The apparatus
design was optimized to use the largest possible electrode areas inside
a given volume. All parts were chosen after deliberate considerations
as outlined below and all parts submerged in LXe were selected to
minimize the amount of impurities they will release into the LXe.
Overview of the detector is shown in Figure~\ref{fig:Modified-spherical-square}.

\begin{figure}
\begin{centering}
\includegraphics[scale=0.75]{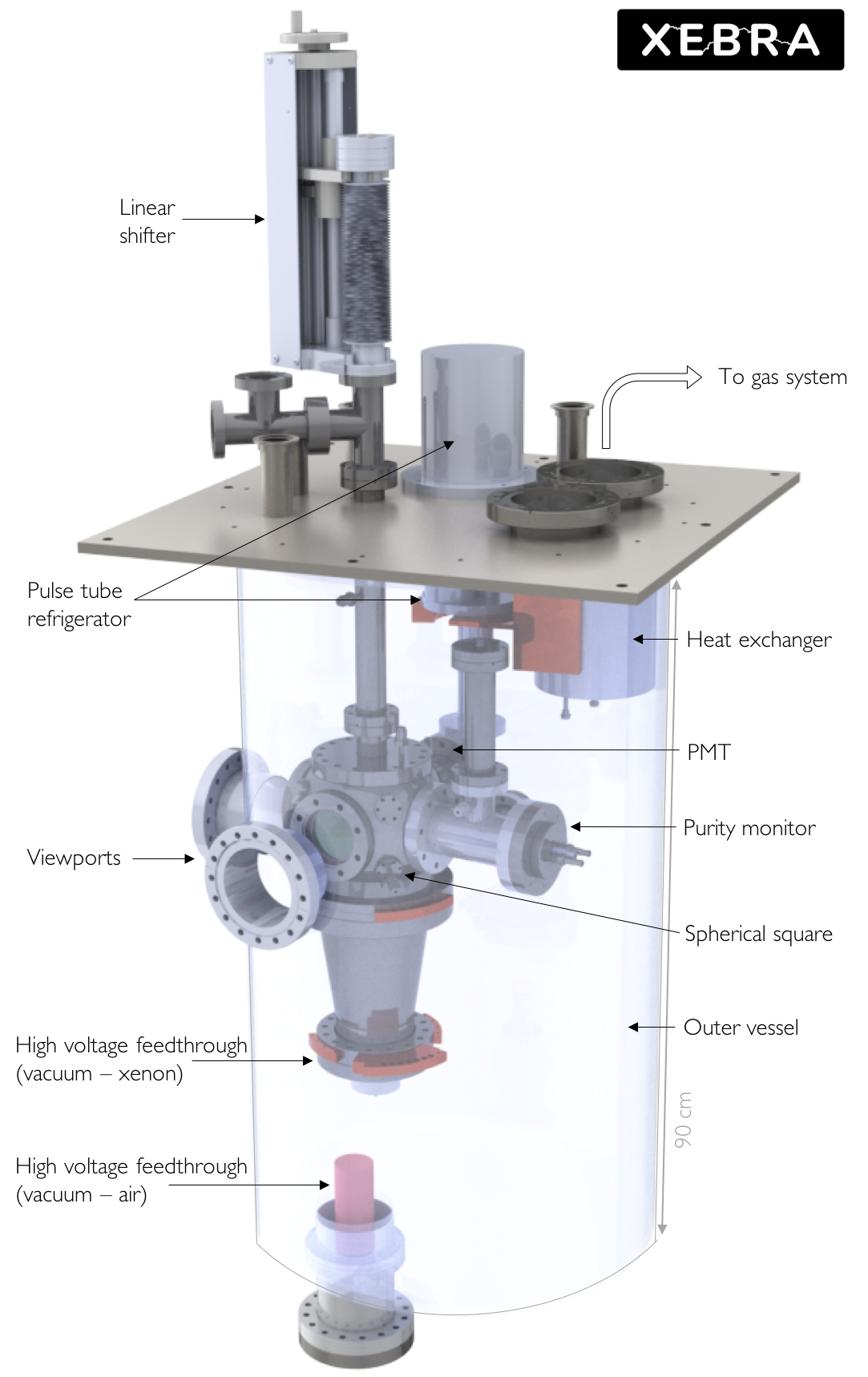}
\par\end{centering}
\caption[CAD rendering XeBrA]{Annotated CAD rendering of XeBrA.\label{fig:CAD-rendering-XeBrA}}
\end{figure}

XeBrA uses $\unit[\sim16]{kg}$ of LXe. Practical considerations limited
the amount of xenon: the apparatus is expected to enable multiple
measurements while varying parameters between data acquisitions, so
it is vital that xenon condensation, cleaning, and recovery take at
most a few days. For comparison, an apparatus with 250~kg in the
active volume, the size of LUX, requires months to condense and clean
the xenon. On the other end, a detector that is too small does not
make it possible to probe dielectric breakdown over large areas. This
optimization led to a selection of a 6-inch stainless steel spherical
square vacuum chamber from Kimball physics, shown in Figure~\ref{fig:CAD-rendering-XeBrA},
as the central part of the apparatus. The spherical square used in
XeBrA was modified and includes one 8-in CF\footnote{CF is a type of vacuum fitting. For information, consult Appendix
\ref{chap:Intro-to-fittings}.}, one 6-in CF, four 4.5-in CF, and eight 1.33-in CF ports. 

\begin{figure}
\begin{centering}
\includegraphics[scale=0.7]{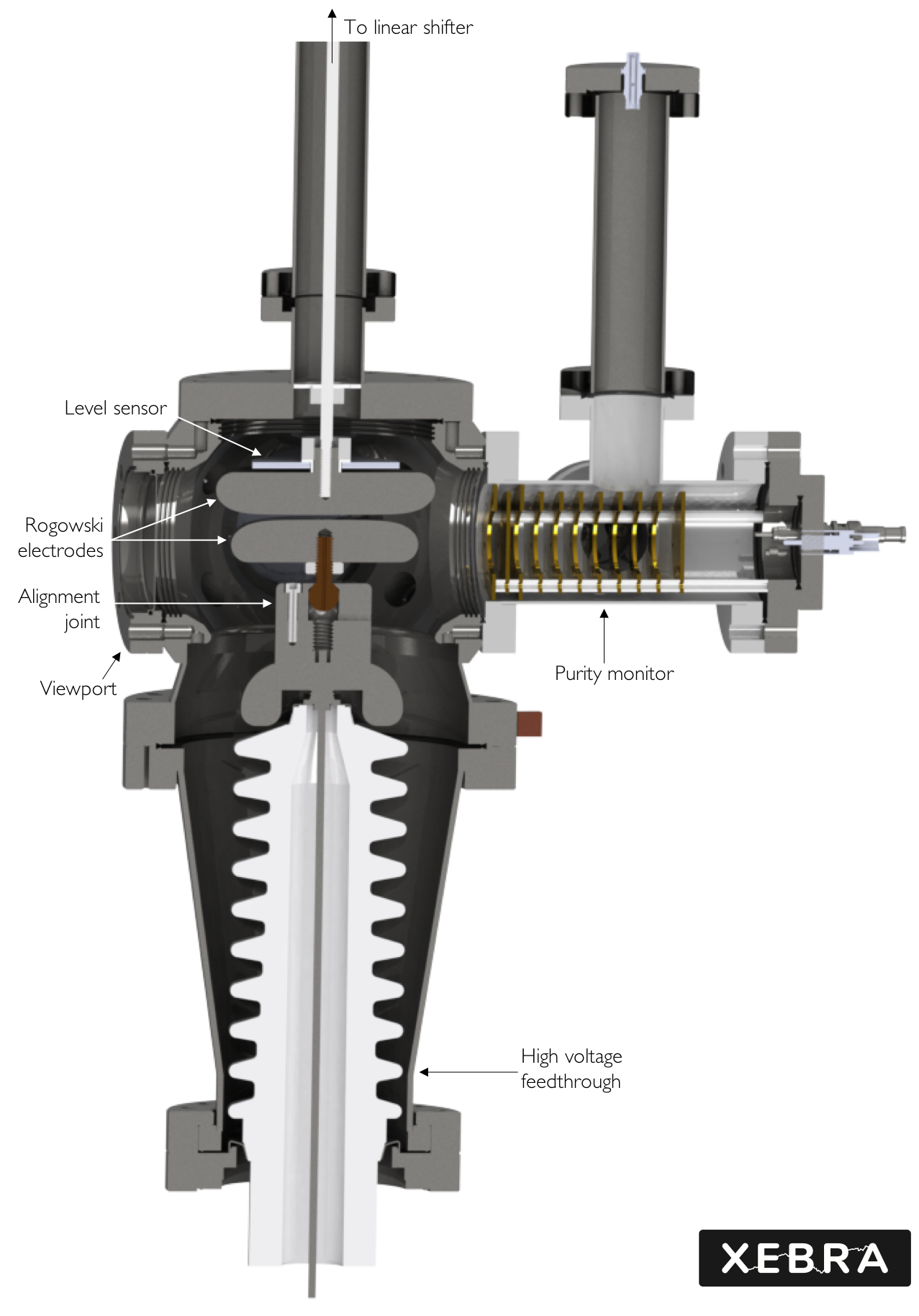}
\par\end{centering}
\caption[CAD rendering of the active volume of XeBrA]{CAD rendering of the active volume of XeBrA. The level sensor is
fastened to the top electrode. The PMT is not visible. \label{fig:CAD-XeBrA-inner}}
\end{figure}

Measurements were performed using Rogowski electrodes to maximize
the electrode area. The Rogowski electrodes, discussed in Section~\ref{subsec:Rogowski-electrodes},
were located in the center of the spherical square as shown in Figure~\ref{fig:CAD-XeBrA-inner}.
The top electrode was at 0~V and attached to a linear shifter that
changed the separation of the electrodes. The bottom electrode was
connected via an alignment connector to an HV feedthrough as discussed
in Section~\ref{subsec:Feedthroughs-+-alignement}. Two of the 4.5
in CF ports were used as viewports; a purity monitor (Section~\ref{subsec:Purity-monitor})
and a PMT (Section~\ref{subsec:PMT}) were attached to the other
two ports. On top, the spherical square attached to the top flange
of the outer vessel (OV)\nomenclature{OV}{Outer Vessel} via a CF
nipple to create vertical clearance for the heat exchanger. The OV
was evacuated for thermal insulation as described in Section~\ref{subsec:Vacuum-system}.
The apparatus was cooled through a heat exchanger attached to a pulse
tube refrigerator (PTR)\nomenclature{PTR}{Pulse Tube Refrigerator}
as specified in Section~\ref{subsec:Cooling}. 

\subsection{High voltage design\label{subsec:High-voltage-design}}

\begin{figure}
\begin{centering}
\includegraphics[scale=0.63]{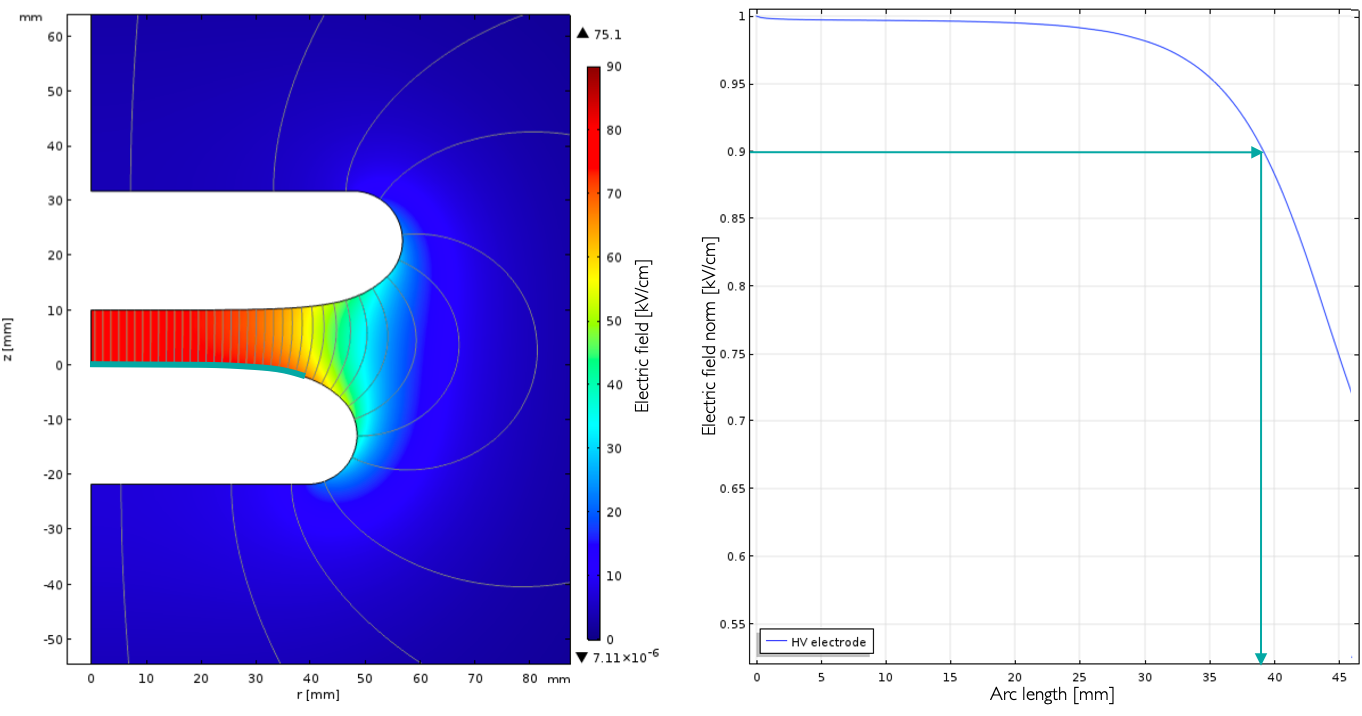}\caption[Illustration of cathode stressed area]{Illustration of cathode stressed area. \textbf{Left:}~Electric field
between two Rogowski electrodes. The teal line illustrates the arc
length along the cathode surface where the electric field magnitude
is being evaluated. \textbf{Right}:~Plot of the electric field for
the setup on the left normalized to the maximum electric field. The
teal arrows illustrate 90\% of the maximum field. \label{fig:Illustration-of-cathode-SA}}
\par\end{centering}
\end{figure}

Along with electric fields, a key quantity of interest is the stressed
cathode (or electrode) area (SEA)\nomenclature{SEA}{Stressed Electrode Area}.
In this work, the SEA is defined as the area of the cathode surface
with an electric field magnitude exceeding 90\% of the maximum electric
field, as illustrated in Figure~\ref{fig:Illustration-of-cathode-SA}.
The radius was obtained from simulations in COMSOL Multiphysics v5.0$^{\circledR}$~\cite{comsolRef}.
The stressed area was calculated assuming that the Rogowski electrodes
are perfectly flat within the region of interest. The choice of 90\%
SEA was influenced by recent studies done in LAr~\cite{Auger:2015xlo}.
However, most breakdown values reported in the literature define SEA
as either 80\% or 90\% of the maximum electric field. Reference~\cite{goshime1995weibull}
found that 80-85\% SEA was appropriate (in LN), while a more recent
work of Reference~\cite{8124618} found that Weibull distribution
corresponded to 90\% of the electrode surface (in transformer oil).

\subsubsection{Rogowski electrodes\label{subsec:Rogowski-electrodes}}

\begin{figure}
\begin{centering}
\includegraphics[scale=0.51]{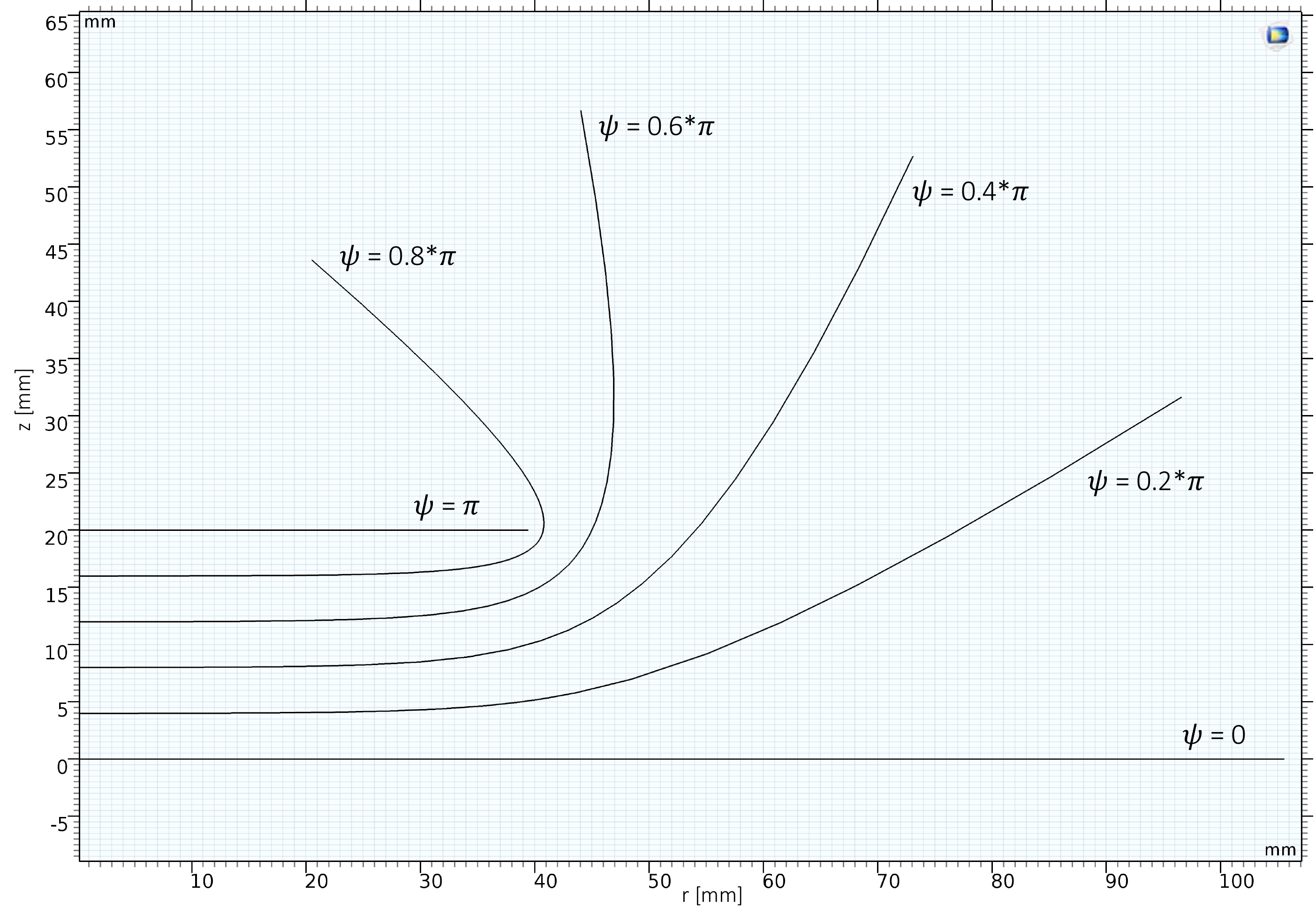}
\par\end{centering}
\caption[Equipotential lines illustrating the effect of varying $\psi$]{Equipotential lines for a finite plate 2~cm above an infinite ground
plate illustrating different values of $\psi$. \label{fig:equipotentials}}

\vspace{0.8cm}
\begin{centering}
\includegraphics[scale=0.62]{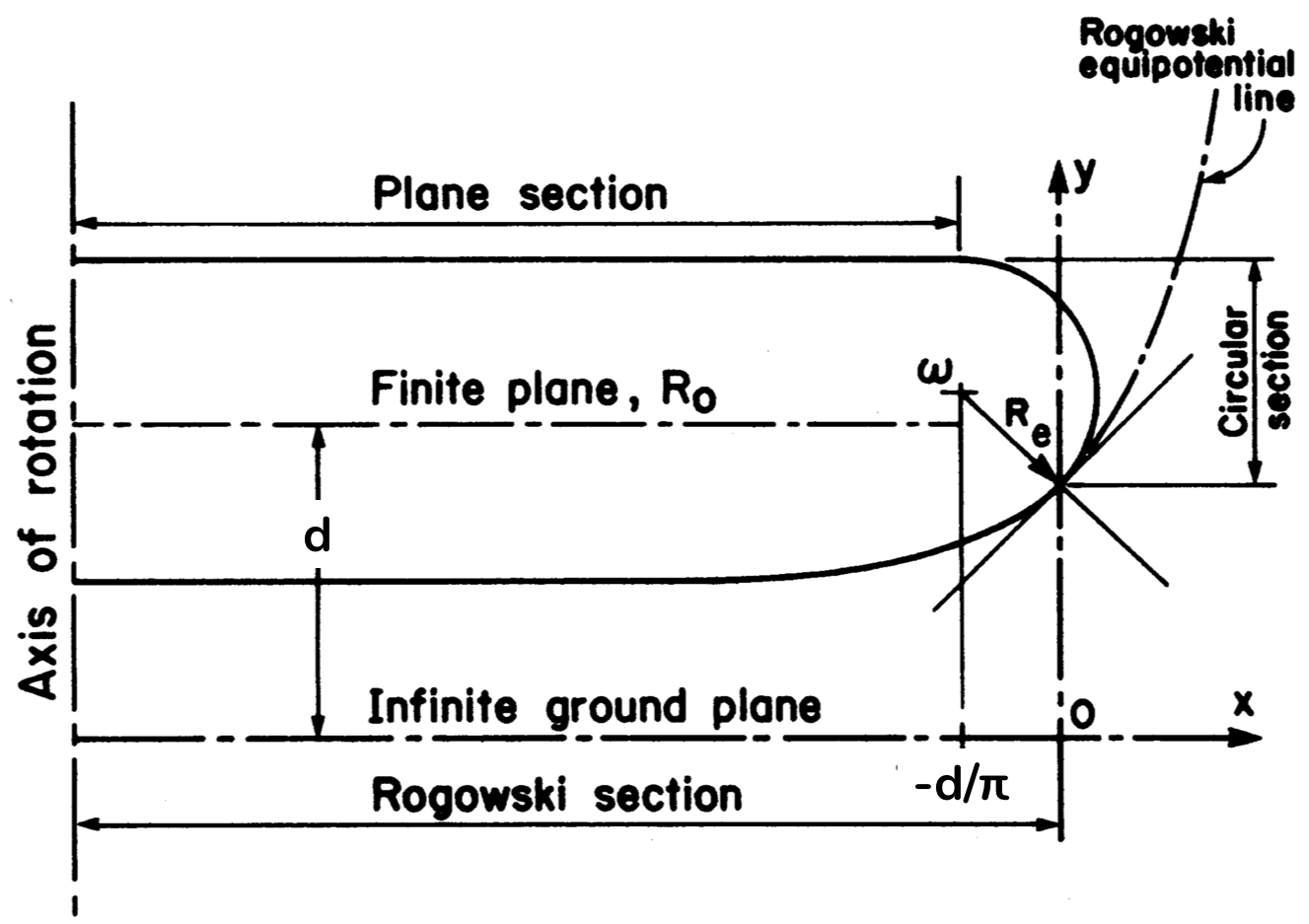}
\par\end{centering}
\caption[Rogowski electrode profile]{Rogowski electrode profile with $\psi=\pi/2$. Figure modified from~\cite{rogowski}.\label{fig:Rogowski-electrode-profile.}}
\end{figure}

Previous experiments in LAr have used a grounded flat plane as one
electrode and spheres of varying diameters as the second electrode.
These setups do not achieve very large electrode stressed areas. Instead,
XeBrA uses Rogowski electrode profiles, as calculated in~\cite{rogowski}.
Originally conceived by W. Rogowski~\cite{Rogowski1923}, the Rogowski
equipotential lines in a gap formed by a finite plane above an infinite
ground plane are given by 
\begin{equation}
\begin{cases}
x=\frac{d}{\pi}\left(\phi+e^{\phi}\cos\psi\right)\\
y=\frac{d}{\pi}\left(\psi+e^{\phi}\sin\psi\right)
\end{cases}\label{eq:xy}
\end{equation}
where $\psi\in\left[0,\pi\right]$ is the equipotential surface, $\phi$
is the line of electrostatic force and $d$ is the distance between
the plane electrode and the infinite ground plane. This expression
can be obtained via a conformal mapping in complex analysis~\cite{ERNST1984275}.
The values of $\phi\in\left[-\infty,\infty\right]$; for $\phi\rightarrow-\infty$,
the expression approaches a constant, corresponding to the center
of the electrode. For $\phi\rightarrow\infty$, the expression diverges.
The effect of varying $\psi$ is illustrated in Figure~\ref{fig:equipotentials}. 

The conventional Rogowski profile corresponds to an equipotential
surface at $\psi=\pi/2$. This choice was verified experimentally
to impede arcing at the edges of the electrodes\footnote{This value of $\psi$ is one of many possible choices that impede
arcing at the electrode edges. It is possible that values other than
$\pi$/2 can achieve better performance.}. This simplifies Equation~\ref{eq:xy} to
\[
\begin{cases}
x=\frac{d}{\pi}\phi\\
y=\frac{d}{\pi}\left(\frac{\pi}{2}+e^{\phi}\right)
\end{cases}
\]
and results in the Rogowski equipotential line shown in Figure~\ref{fig:Rogowski-electrode-profile.}.
To manufacture such an electrode, it is necessary for the electrode
to be finite. A circular section can be used to terminate the electrode.
The calculation leading to the design shown in Figure~\ref{fig:Rogowski-electrode-profile.}
used in XeBrA can be found in~\cite{rogowski}. Since the electrode
needs to be enclosed inside of a grounded vessel, it is also important
to consider electric field interactions between the electrodes and
the vessel. Field enhancement due to the proximity of a grounded  wall
begins when the radial distance separating the electrode from the
vessel is about twice the main gap spacing.

Conveniently, the shape of the electrodes depends only on $d$, the
separation between the electrodes, rather than on their potential
difference. If two Rogowski electrodes are used, the electrode thickness
should be twice the maximum gap space $d$ to avoid excessive field
enhancement at the electrode edges. This fully defined the electrode
design. Therefore, electrodes with 2~cm thickness can be used to
study electrode areas up to $\unit[58]{cm^{2}}$ with up to 1~cm
electrode separation. Since it is challenging to ensure that the two
electrodes are perfectly aligned, the ground electrode (anode) was
manufactured with a slightly larger radius ($r=\unit[56.6]{mm}$)
compared to the HV electrode (cathode) with $r=\unit[48.4]{mm}$.
The slightly larger size of the ground electrode ensures that if the
two electrodes are somewhat off-axis, all the electric field benefits
of Rogowski design would be maintained. The maximum size of the HV
electrode was determined from HV simulations in COMSOL constrained
by the size of the spherical square used. The electrodes were manufactured
from 303 stainless steel. The cathode had a surface finish with roughness
average (Ra)\nomenclature{Ra}{Roughness average} of $\unit[0.05]{\mu m}$
and anode with $\mathrm{Ra}=\unit[0.07]{\mu m}$.

\begin{figure}
\begin{centering}
\includegraphics[scale=0.5]{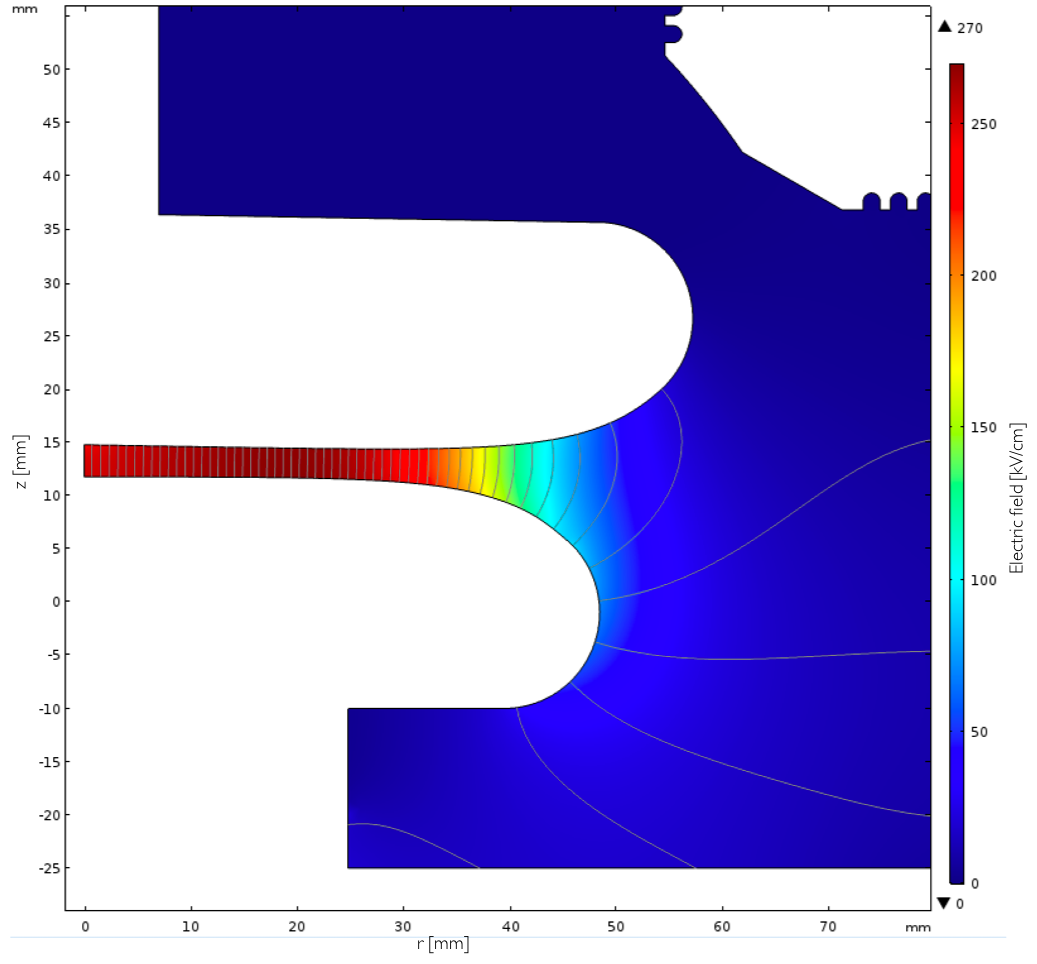}
\par\end{centering}
\caption[Effect of electrode tilt on electric field]{Effect of electrode tilt on electric field obtained from an axially
symmetric simulation performed in COMSOL. The simulation was made
in LXe for a 3~mm electrode separation with a 1$^{\circ}$ tilt.\label{fig:Effect-of-tilt} }
\end{figure}

\begin{table}
\begin{centering}
\begin{tabular}{>{\raggedleft}p{1.6cm}>{\raggedleft}p{2.3cm}>{\raggedleft}p{2.3cm}>{\raggedleft}p{2.3cm}>{\raggedleft}p{1.8cm}>{\raggedleft}p{2.4cm}}
\hline 
Separation {[}mm{]} & Nominal field {[}kV/cm{]} & Max field at 0.4$^{\circ}$ tilt & \% increase from nominal field & Max field at 1$^{\circ}$ tilt & \% increase from nominal field\tabularnewline
\hline 
\hline 
1.0 & 750 & 789 & 5 & 957 & 28\tabularnewline
1.4 & 536 & 555 & 4 & 634 & 18\tabularnewline
2.0 & 375 & 385 & 3 & 421 & 12\tabularnewline
3.0 & 250 & 254 & 2 & 270 & 8\tabularnewline
4.0 & 188 & 190 & 1 & 199 & 6\tabularnewline
5.0 & 150 & 152 & 1 & 158 & 5\tabularnewline
6.0 & 125 & 126 & 1 & 131 & 5\tabularnewline
7.0 & 107 & 108 & 1 & 111 & 4\tabularnewline
\hline 
\end{tabular}
\par\end{centering}
\caption[Effects of electrode tilt]{Effects of electrode tilt. \label{tab:Effects-of-tilt}}
\end{table}

\begin{figure}
\begin{centering}
\includegraphics[scale=0.44]{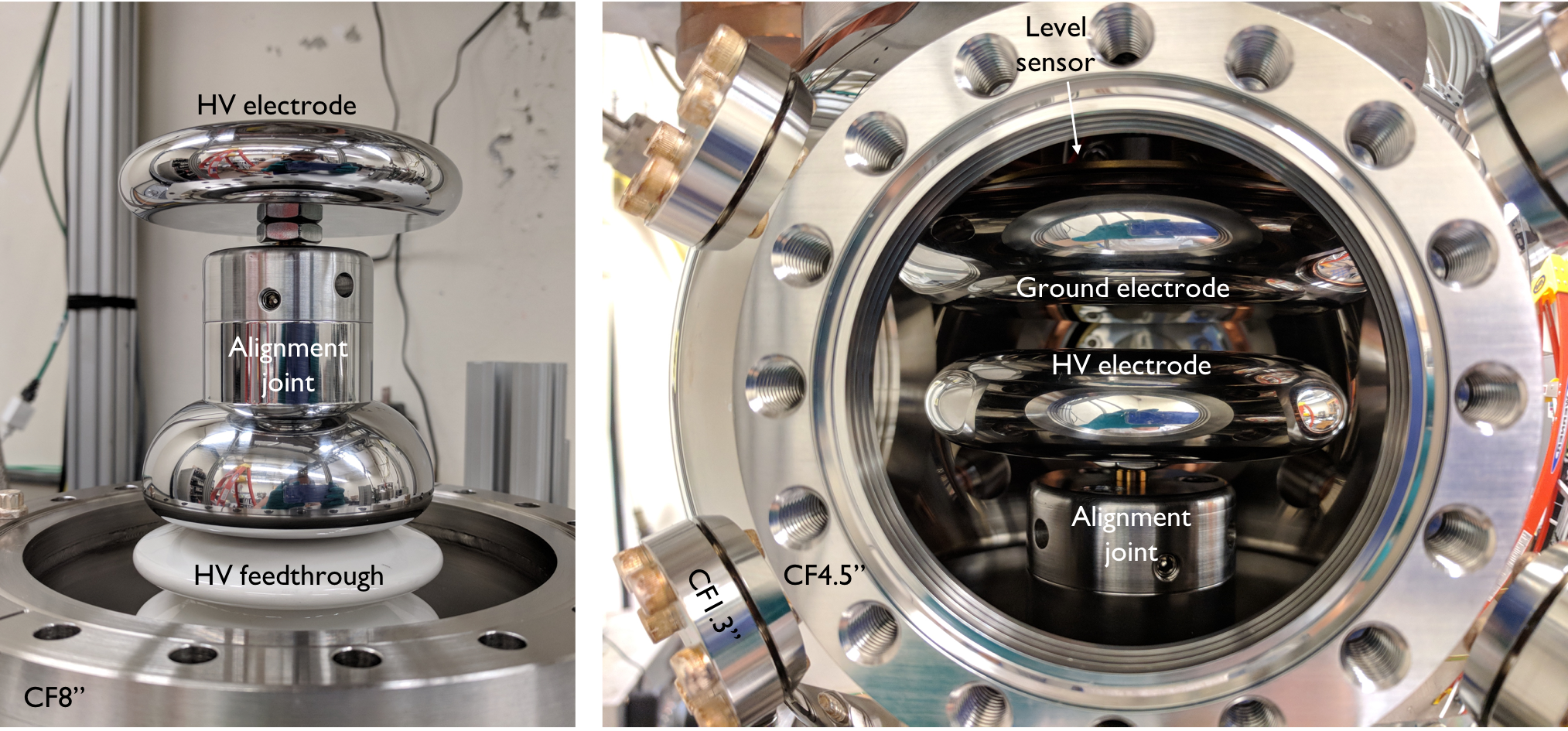}
\par\end{centering}
\caption[Image of the HV assembly]{Image of the HV assembly. \textbf{Left:} Image of the HV electrode
attached via the alignment joint mounted on top of the Ceramtec feedthrough.
\textbf{Right:} Image of the assembly inside the spherical square.
\label{fig:HV-assembly}}
\end{figure}

A pressure recording film\footnote{The film had a grayscale sensitive to 2-20 psi. See Appendix~\ref{chap:Parts-list}
for the part number.} was used to align the electrodes to ensure that they are parallel
inside the apparatus. Nevertheless, it is possible that the electrodes
will be tilted with respect to each other. A COMSOL simulation was
performed to study the effects of a 0.4$^{\circ}$ and 1$^{\circ}$
tilt as summarized in Table~\ref{tab:Effects-of-tilt}. A 1$^{\circ}$
tilt would be clearly visible  as illustrated in Figure~\ref{fig:Effect-of-tilt}.
Based on a perception study, a 0.3$^{\circ}$ tilt would be visible
by eye in the setup. Since there are two directions of possible tilt,
this results in a 0.4$^{\circ}$ systematic error.

\subsubsection{HV feedthrough connection \label{subsec:Feedthroughs-+-alignement}}

The HV electrode was attached to an alumina ceramic feedthrough made
by CeramTec (model 6722-01-CF), rated up to 100~kV of 6.5~A DC in
vacuum\footnote{This HV feedthrough has been previously used for tests of the LUX
detector feedthrough.}. The connection of the HV electrode to the feedthrough was carefully
designed to ensure that the electric fields caused by this connection
did not exceed the electric field expected between the electrodes.
A special connector was designed with its dimensions optimized in
COMSOL, and its implementation was engineered at LBNL. The resulting
alignment joint is shown in Figure~\ref{fig:HV-assembly}. The alignment
joint served two purposes. First, it served as an HV connection between
the HV electrode and the feedthrough. The radius of the bulge at the
bottom of the joint was designed to minimize the high electric fields
present at this interface. Second, a ball spindle located near the
top of the joint allowed the electrodes to be aligned to be parallel
to each other once inside the apparatus.

\begin{figure}[t]
\begin{centering}
\includegraphics[scale=0.8]{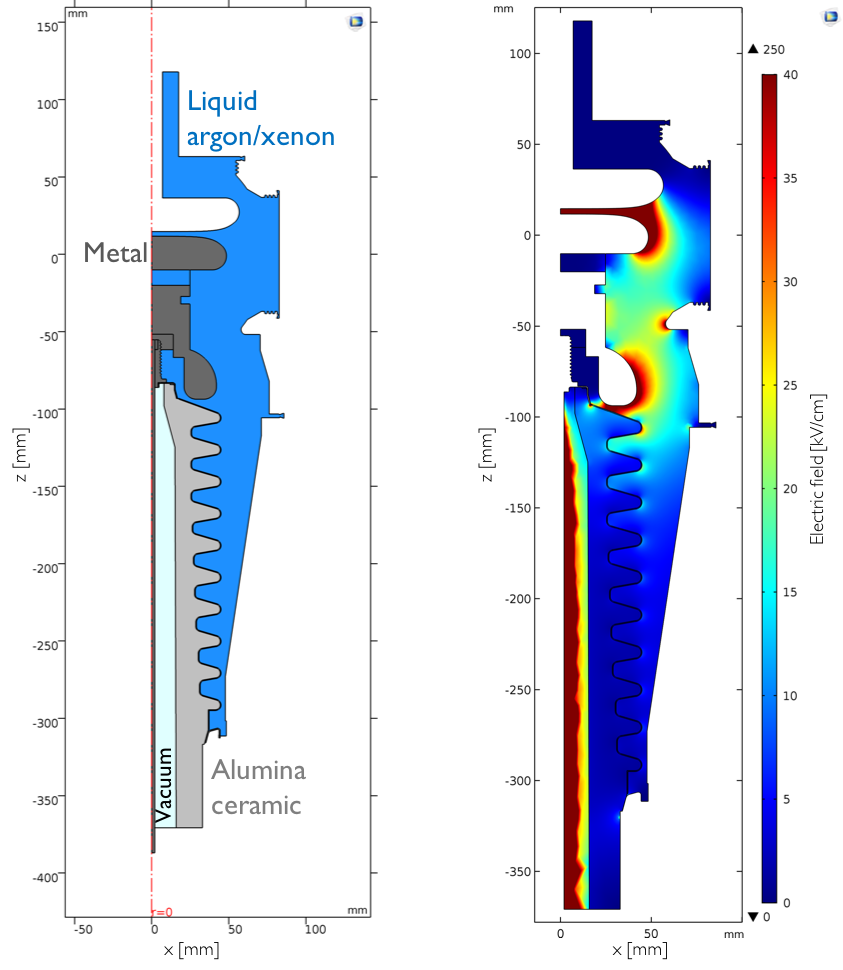}
\par\end{centering}
\caption[An axially symmetric simulation of the active volume of XeBrA ]{An axially symmetric simulation of the active volume of XeBrA performed
in COMSOL. \textbf{Left:}~Illustration of the geometry and materials
used in the simulation. \textbf{Right:}~Results of an electric field
simulation in LXe with -75~kV on the cathode (bottom electrode).
Maximum field occurs between the two electrode faces. At 5~mm electrode
separation, the maximum field between the electrode faces is 150~kV/cm,
while the second highest field in the apparatus (54~kV/cm) is at
the rounded part of the alignment joint. \label{fig:XeBrA-active-axial}}
\end{figure}

The final simulation of the electric field inside the active volume
is shown in Figure~\ref{fig:XeBrA-active-axial}. The DC dielectric
constant of liquid xenon is assumed to be 1.85, which is typical of
measured values in the scientific literature~\cite{doi:10.1063/1.1724850,doi:10.1139/p70-033,doi:10.1142/5113,SAWADA2003449}.
The dielectric constant of alumina ceramic was set to 9.4.

A power supply was purchased from Spellman (model SL100N10/CMS/LL20)
with a range up to -100~kV with $\unit[100]{\mu A}$ DC. This power
supply was previously tested for its stability and is very similar
to the power supply used in the LZ detector. To comply with the Lab's
Environment, Health \& Safety (EH\&S) radiation regulations, the power
supply was limited to -75~kV during operations with a maximum current
of $\unit[10]{\mu A}$. The power supply can be controlled either
manually or remotely through \textsc{LabView }in the Slow Control. 

\begin{figure}
\begin{centering}
\includegraphics[scale=0.4]{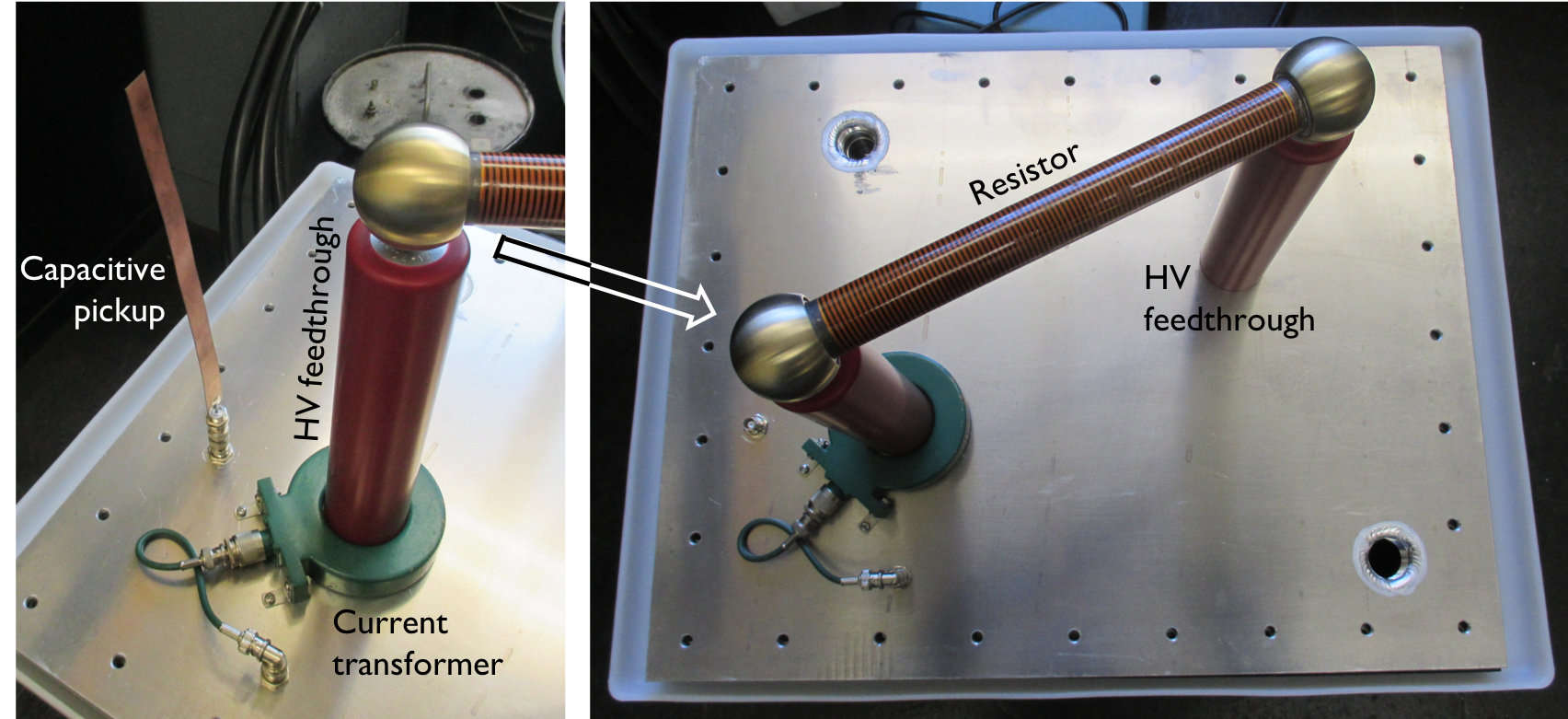}
\par\end{centering}
\caption[Oil tank electronics]{Electronics of the oil tank attached to an upside down lid. The parts
are submerged in transformer oil during operations. \textbf{Left:}~Current
transformer and capacitive pickup next to one of the HV feedthroughs
made by Isolation Products. \textbf{Right:}~A $\unit[956]{k\Omega}$
resistor attached to two HV feedthroughs.\label{fig:oil-tank}}
\end{figure}

A $\unit[956]{k\Omega}$ resistor is installed between the power supply
and the apparatus as shown in Figure~\ref{fig:oil-tank} to limit
the current and minimize the stored energy that can go into a breakdown.
The resistor isolates the load from the stored energy in the power
supply and some of the cable, which limits the amount of potential
damage caused by a spark\footnote{For the same reason, inside the Spellman power supply, there is a
series resistance of $\unit[24.2]{k\Omega}$ to limit the stored energy
from the 0.8~nF output capacitance (that limits the energy to 4~J
at 100~kV).}. The stored energy that is in the cable connecting the oil tank to
the HV feedthrough will go into the breakdown. The resistor is located
inside an oil tank instrumented with a current transformer to identify
very large events and a capacitive pickup that can be connected to
an oscilloscope. The capacitive pickup can detect a change in voltage
as a result of a partial discharge or a full breakdown. If the HV
were to break down to this pickup, a box with a spark gap to ground
limits the signal level to 150~V at the output. Two HV feedthroughs
(custom-made model D-102-10 from Isolation Products) are used to make
the connection in the oil. They are rated to 100~kV DC and made from
cast epoxy. 

A cable (Dielectric Sciences model 2121) is used to connect the power
supply, the oil tank, and the apparatus. The center of the cable is
made from gauge 12 AWG tin-plated copper surrounded by semiconductive
polyethylene. The conductive core is enclosed by low density high
molecular weight (LDHMW)\nomenclature{LDHMW}{Low Density High Molecular Weight | AcronymAttic}
polyethylene. The cable is rated to 150~kV DC with a capacitance
of 29~pF/ft. The total length of the cable between the oil tank and
the apparatus is 2.1~m (7~ft), resulting in 
\[
E=\frac{1}{2}CV^{2}=\unit[0.6]{J}
\]
of stored energy at -75~kV.

\subsection{Purity monitor\label{subsec:Purity-monitor}}

Very high purity levels are desired for noble liquids in XeBrA for
two reasons. First, the breakdown field strength is expected to depend
on the presence of impurities in the liquid. Second, the results from
XeBrA should inform direct dark matter and neutrino experiments, which
typically operate at very high purities. To that end, a purity monitor
capable of detecting very low levels of impurities was built to serve
inside the XeBrA experiment. For more details about the design, development,
and operating principles of the purity monitor, refer to Section~\ref{sec:Purity-monitor-for-XeBrA}.

The purity monitor was directly connected to the active  volume as
shown in Figure~\ref{fig:XeBrA-insulated}. It was mounted to one
of the 4.5~in CF ports of the spherical square as shown in Figure~\ref{fig:CAD-XeBrA-inner}. 

An anode signal was never seen in the purity monitor while operating
in LXe. A similar problem was observed during the construction of
the IBEX purity monitor. It was discovered that the alumina ceramic
spacers likely had small amounts of conductive impurities on their
surfaces, making them slightly conductive. The ceramic spacers separating
the anode and anode shield grid are the shortest so the number of
lost charge would be most pronounced around the anode. Even though
the purity monitor was operated successfully in LAr, it is anticipated
that this effect was also present. It is possible that either this
effect is temperature dependent or the purity achieved in LAr was
higher than that of LXe, and the anode detected some of the charges.
If this is true, then the purity of LAr reported in this work presents
a lower limit on the electron lifetime, since some of the drifting
electrons would not have been detected. Disassembly of the purity
monitor and thorough cleaning of the ceramic spacers is planned before
future operations. 

\subsection{PMT\label{subsec:PMT}}

As discussed in Section~\ref{subsec:LZ-High-voltage-delivery}, one
of the concerns of dual-phase LXe detectors is light production before
a dielectric breakdown. Therefore, a PMT was installed to monitor
the onset of electroluminescence prior to a breakdown. Similar to
the purity monitor, the PMT is mounted to one of the 4.5~in CF ports
of the spherical square adjacent to the active volume. 

Electroluminescence is different for LAr and LXe, so two different
PMTs were used for the two liquids. Each PMT was mounted inside a
4.5~in CF nipple so the entire assembly could be easily replaced
between data acquisitions with different liquids. When collecting
data with the PMT, the two viewports located on the OV are covered,
making the active  volume light tight. Then the cathode HV is slowly
ramped up to a voltage that is very unlikely to result in a spark\footnote{This value is determined from previous breakdown studies at that separation
without a PMT.} for a given electrode separation. The voltage is held constant for
a few minutes, before being slowly ramped back down to the ground.
Data are collected throughout the entire procedure to observe the
onset of electroluminescence.

\begin{figure}
\centering{}\includegraphics[scale=0.5]{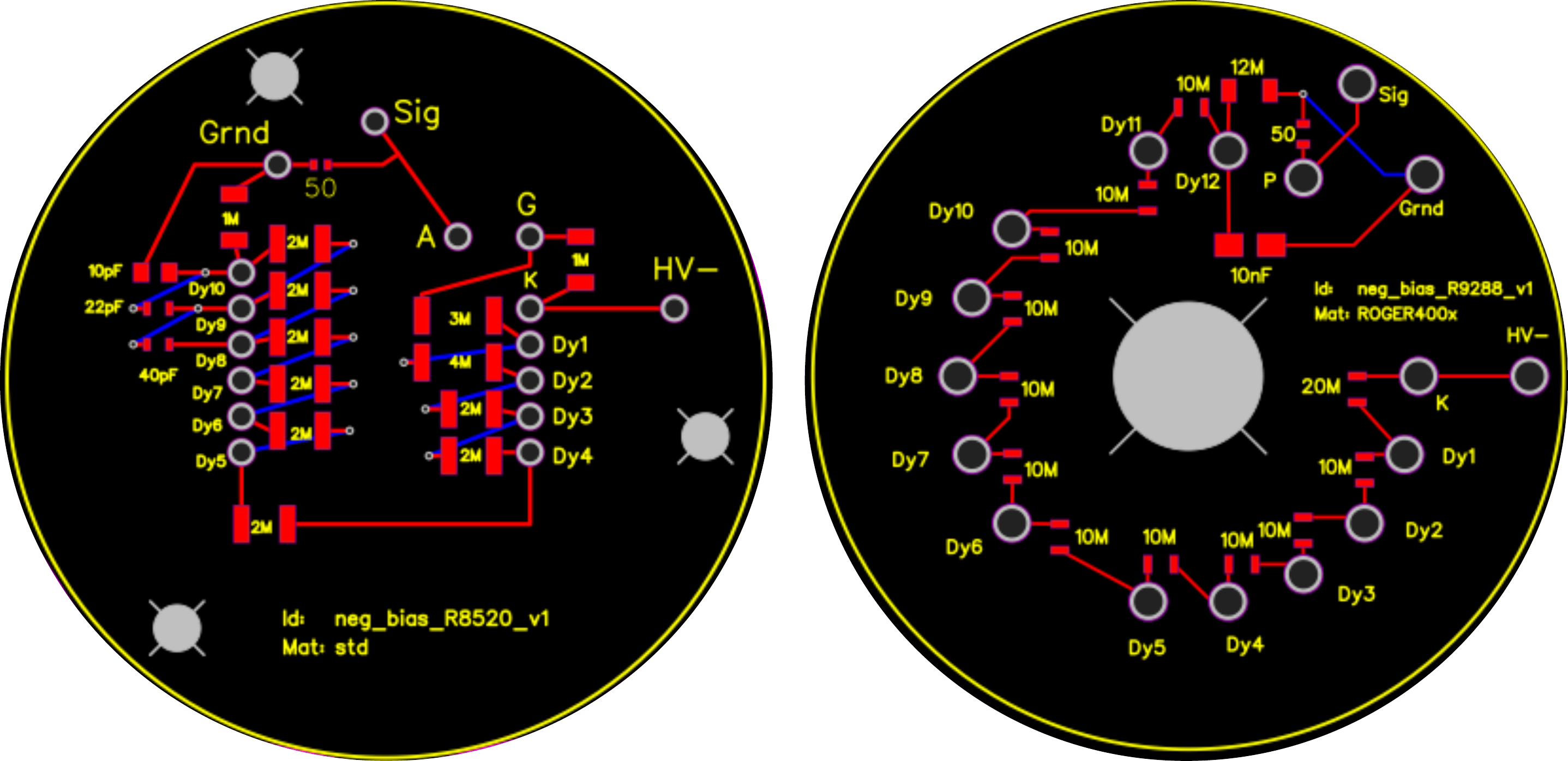}\caption[Circuit diagrams of voltage dividers used for PMT operations]{Schematics of PCBs used for PMT operations. Connections to the PMT
dynodes are indicated by ``Dy.'' \textbf{Right:}~For LAr PMT. \textbf{Left:}~For
LXe PMT. \label{fig:Circuit-diagrams-PMTs}}
\end{figure}

Note that the base for the PMT was designed to have low capacitance.
Even though the PMT is not expected to be biased during breakdowns,
if an unanticipated spark happens during PMT operations, this small
capacitance should prevent PMT damage. However, the increase in the
dark rate on the photocathode would likely last several hours at XeBrA's
operating temperatures preventing further data acquisition. The PMT
bases were manufactured by Sunstone Circuits using Roger400x boards
to minimize outgassing. Schematics of the printed circuit boards (PCBs)\nomenclature{PCB}{Printed Circuit Board}
used for each PMT shown in Figure~\ref{fig:Circuit-diagrams-PMTs},
as designed by Quentin Riffard, a postdoctoral researcher at LBNL.

\subsubsection{Liquid argon PMT}

\begin{figure}
\begin{centering}
\includegraphics[scale=0.5]{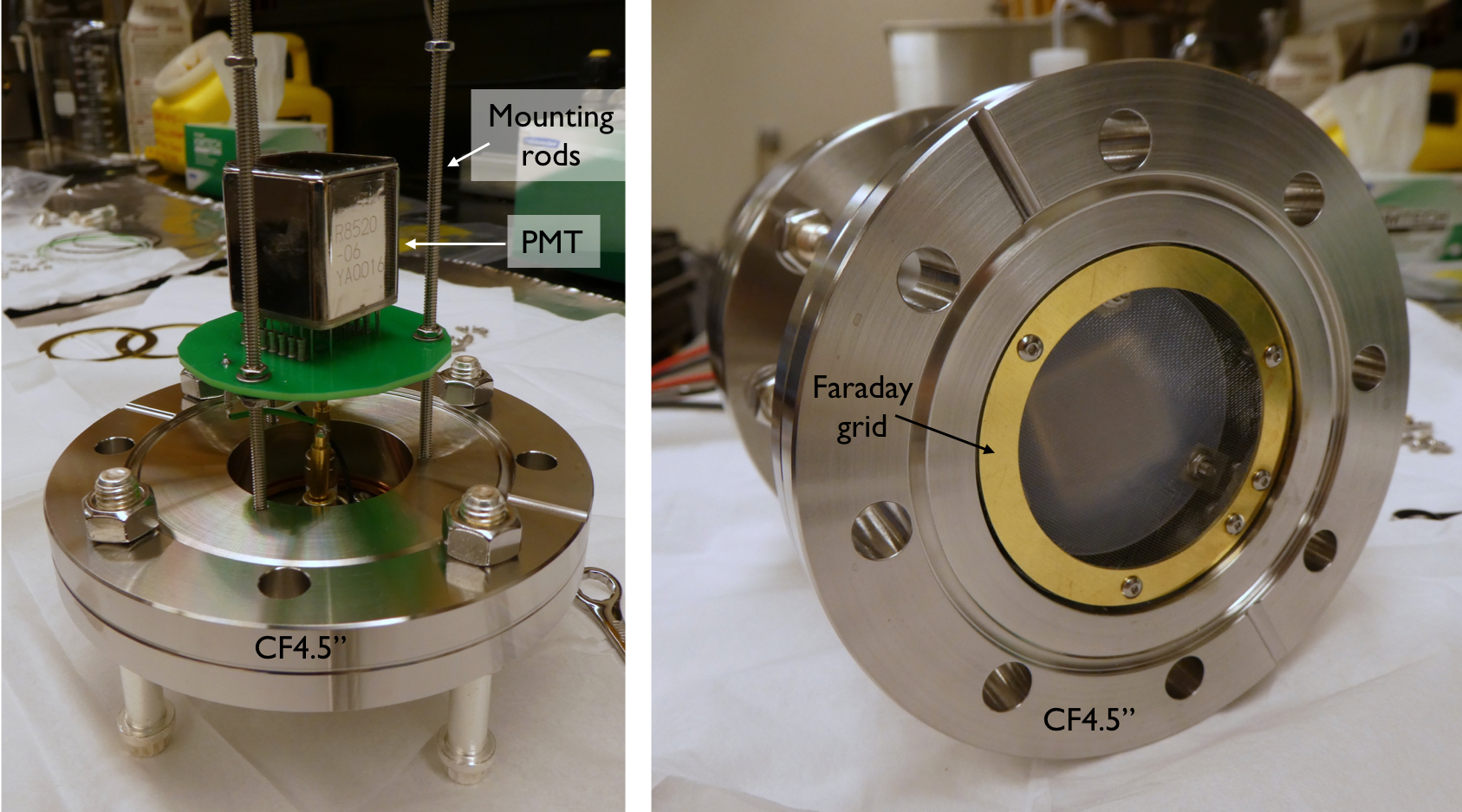}
\par\end{centering}
\caption[Pictures of the PMT assembly used in LAr]{Pictures of the PMT assembly used in LAr. \textbf{Left:}~PMT mounted
on a CF4.5'' flange. \textbf{Right:}~View of the PMT inside the
CF4.5'' nipple. The Faraday grid at the forefront electrostatically
separates the PMT from the active volume. Faintly visible just behind
the grid is a TPB-coated quartz disc. \label{fig:LAr-PMT}}
\end{figure}

For operations in LAr, a bialkali PMT from Hamamatsu (model R8520-06)
was installed as shown in Figure~\ref{fig:LAr-PMT}. It has an effective
area of $\unit[\left(20.5\times20.5\right)]{mm^{2}}$ and a spectral
response ranging from 160 to 650~nm. Tetraphenyl butadiene (TPB)-coated
quartz disc was installed in front of it facing the active volume
to wavelength-shift the 128~nm LAr scintillation light to 430 nm,
a wavelength detectable by the PMT with high quantum efficiency.

\subsubsection{Liquid xenon PMT}

PMT model R9228 from Hamamatsu with a bialkali photocathode was used
for operations in LXe. It has a spectral response ranging from 160
to 650~nm and a minimum effective area of $\unit[15.9]{cm^{2}}$.

\begin{figure}
\centering{}\includegraphics[scale=0.44]{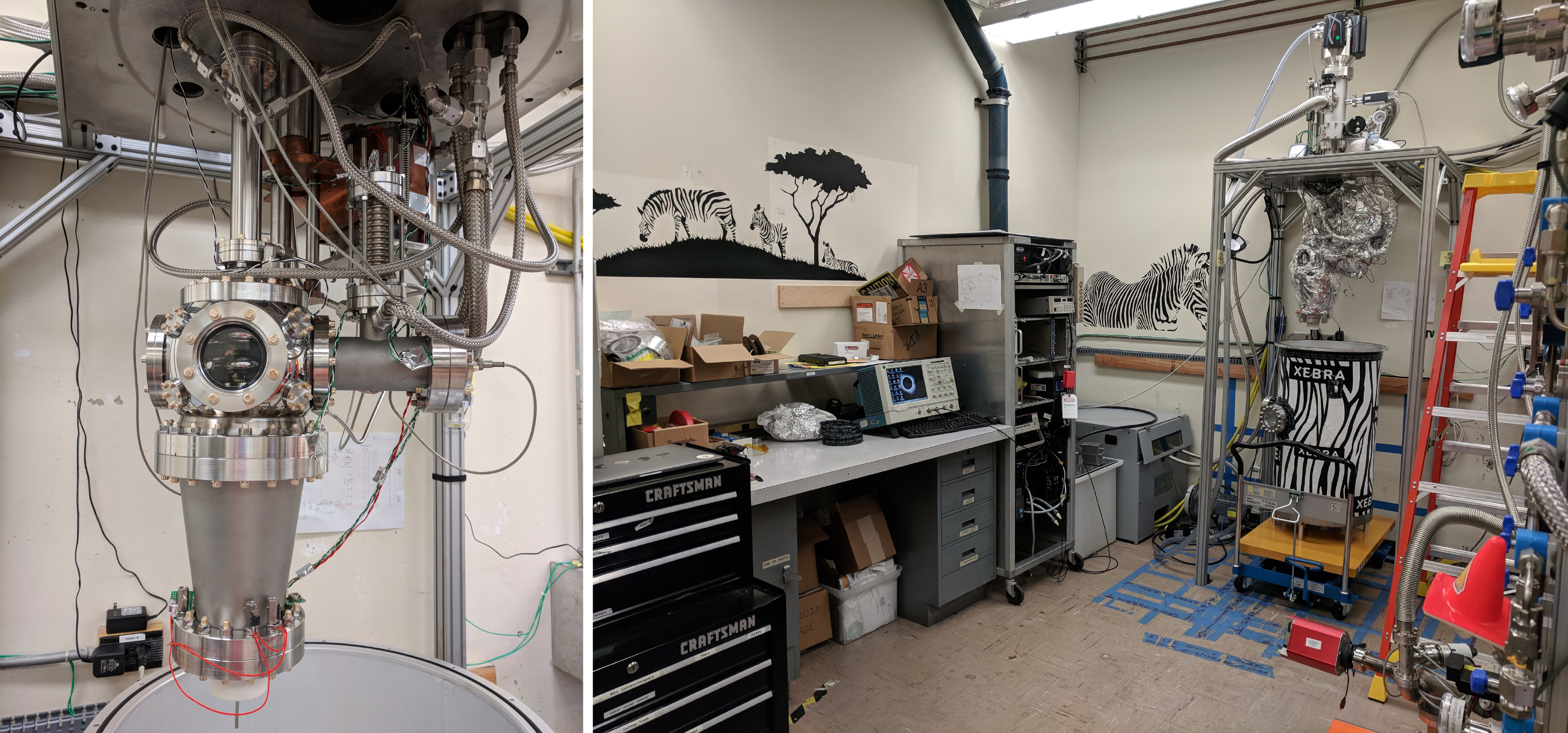}\caption[Final XeBrA assembly]{Final XeBrA assembly. \textbf{Left:}~View of the xenon volume before
the application of superinsulation with the purity monitor attached
to the right of the central volume with viewports. The PMT is attached
at the back. \textbf{Right:}~Superinsulated hardware waiting to be
enclosed in the OV. The edge of the gas system is visible on the right
of the picture. \label{fig:XeBrA-insulated}}
\end{figure}

\subsection{Outer vessel\label{subsec:Vacuum-system}}

The experimental chamber is wrapped in multiple layers of superinsulation
and located inside an outer vacuum vessel (OV) for thermal insulation
from the room as shown in Figure~\ref{fig:XeBrA-insulated}. Nevertheless,
a small amount of heat $\mathcal{O}\left(\unit[10]{W}\right)$ still
reaches the chamber\footnote{See Section~\ref{subsec:Power-radiated-xebra} for the calculation.},
which causes xenon bubbles to be present in the apparatus. A 0.14~barg
(2 psig) pressure relief valve prevents overpressure of the OV\footnote{Unfortunately, that came in handy during LXe Run~3.}.
An HV feedthrough identical to the ones used in the oil tank electronics
is installed in an ISO100 port\footnote{Consult Appendix~\ref{chap:Intro-to-fittings} for details about
this type of vacuum fitting.} and connects the power supply to the apparatus.

\begin{figure}[t]
\begin{centering}
\includegraphics[scale=0.58]{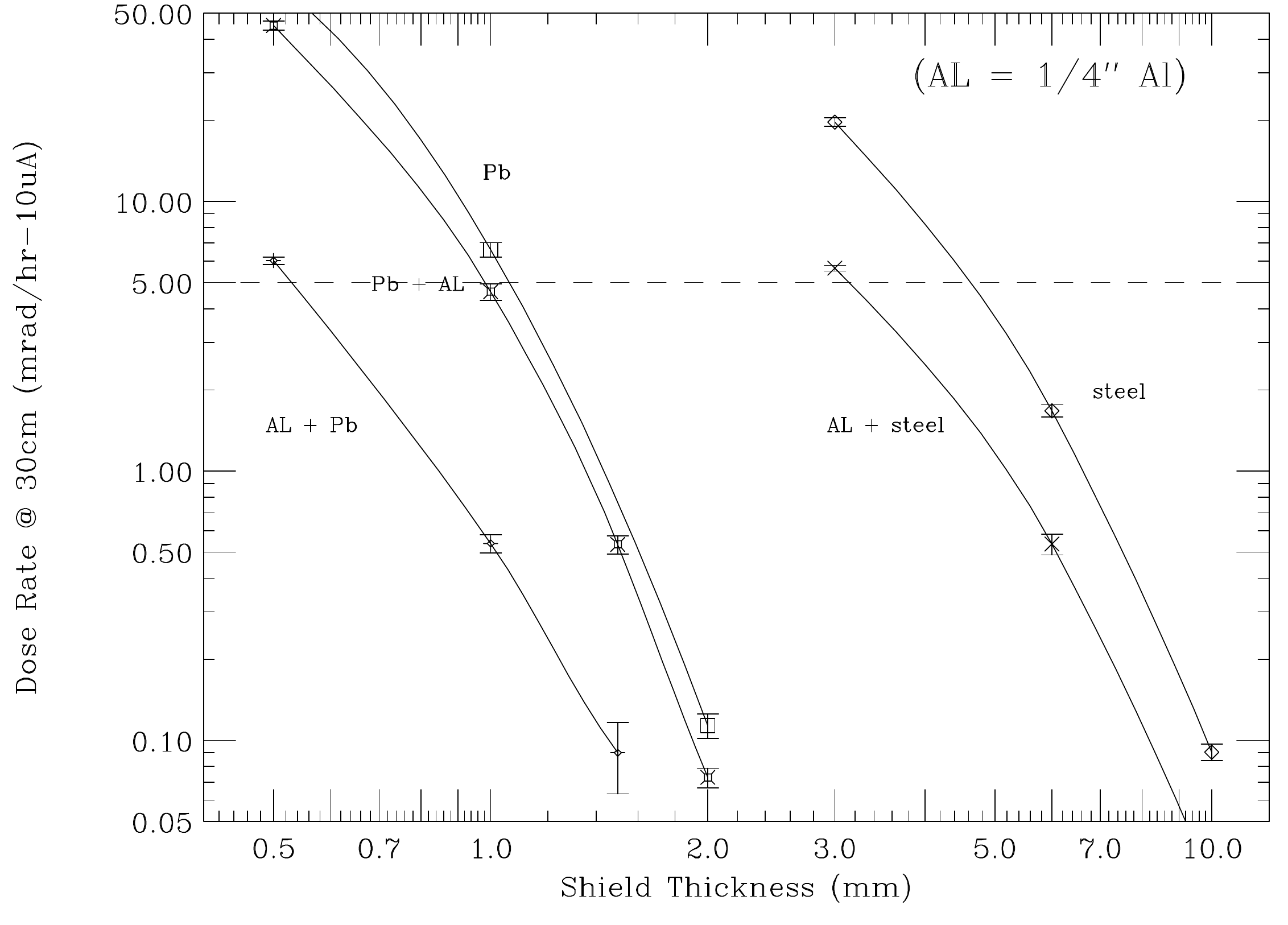}
\par\end{centering}
\caption[Expected dose rate for 75 kV on the cathode]{Expected dose rate for 75~kV on the cathode evaluated at 30~cm
outside of the OV for a focused electron beam with normal incidence
on the inside wall. Various combinations of shielding material were
considered. A combination of 1/4-in of aluminum (Al) and 1/32-in (0.8~mm)
of lead (Pb) located on the outside of the OV was chosen based on
its best attenuation property. The shielding limits the dose rate
to $\unit[\sim1]{mrem/hr}$, below the $\unit[5]{mrem/hr}$ threshold
for posting a radiation area. Note that for x- and $\gamma$-rays
$\unit[1]{rad}=\unit[1]{rem}$. \label{fig:XeBrA-bremsstrahlung} }
\end{figure}

The HV feedthrough in the vacuum space can cause an x-ray hazard if
a breakdown occurs between the feedthrough and the OV. When this fault
condition occurs, electrons generated during the discharge produce
Bremsstrahlung photons while scattering off the aluminum of the OV.
LBNL\textquoteright s engineers modeled this effect using the MCNP
software~\cite{briesmeister2000mcnptm}. The model evaluates the
amount of radiation emitted if a fault condition occurs over 100~hours
per year. The result of the simulation is shown in Figure~\ref{fig:XeBrA-bremsstrahlung}.
To ensure workers' safety, the assumptions going into the simulation
were maximally pessimistic. Since both the current and voltage are
limited by the power supply, and the power supply automatically ramps
down to 0~V when it trips, a more realistic approximation for the
fault condition might be $\mathcal{O}\left(10\,\mathrm{s}\right)$
per year, were a breakdown to occur.

The simulation in Figure~\ref{fig:XeBrA-bremsstrahlung} evaluates
various combinations of shielding material. It shows that placing
Pb on the outside of the Al OV is most effective at attenuating the
radiation. This is because for the x-ray energies $E_{\gamma}$, the
dominant reaction in the lead (Pb) is photoelectric absorption. The
cross section of the photoelectric effect $\propto\mathrm{Z^{4}}/E_{\gamma}^{3.5}$~\cite{photoabsorption}
so it is particularly strong in Pb ($Z=82$) compared to Al ($Z=13$)
and very effective at lower x-ray energies. Placing the Al before
the Pb softens the x-ray spectrum via Compton scattering. This softer
x-ray spectrum is then more easily absorbed via photoelectric absorption
in Pb. 

This led to a design and installation of a 1/32~in (0.8~mm) Pb shield\footnote{The lead was manufactured by Santa Rosa Lead Products and cut into
the final shape by the LBNL machine shop. Buy local!} on the outside of the OV. Additionally, because lead is a toxin,
the shield was wrapped in a vinyl coating, giving XeBrA its unique
zebra print shown in Figure~\ref{fig:XeBrA-insulated}.

\subsection{Cooling\label{subsec:Cooling}}

The apparatus was cooled by a pulse tube refrigerator manufactured
by Cryomech with cold head model PT805 and compressor model CP950.
A pulse tube refrigerator is commonly used in cryogenic applications
as it is a type of cryocooler that does not contain any moving parts
in the low-temperature part of the device, which results in high reliability
and low mechanical vibrations. An overview of PTR operations can be
found in~\cite{DEWAELE2000479}. The cold head has two stages: it
delivers 40~W of cooling power at 80~K at its $1^{\mathrm{st}}$
stage, and even though the $2^{\mathrm{nd}}$ stage was not used,
it can provide 1.5~W at 10~K, well below the desired XeBrA operating
temperatures. The compressor is cooled using the building's low conductivity
water\footnote{To ensure the compressor's longevity, a minimum water quality is required.
To that end, a chemical analysis of the water was performed with the
following results: total alkalinity $\leq\unit[5]{ppm}$, calcium
carbonate $\leq\unit[3]{ppm}$, pH = 7.2, and conductivity = 4.4~mS/m.
This quality is sufficient for compressor operations. Additionally,
to prevent large particles from accidentally entering the water stream
a water filter was installed at the inlet.}. 

The PTR removes heat from the xenon to maintain a constant temperature
inside the apparatus. This is done through a heat exchanger. The heat
exchanger is made from SS brazed to a copper plate that makes a connection
with the $1^{\mathrm{st}}$ stage of the PTR. The incoming room temperature
gas enters the condenser where it is cooled by the outgoing gas on
the inside surface and the PTR on the outside surface. That causes
the xenon to liquefy and fall into the apparatus. The inner part of
the heat exchanger is made from an electroformed bellows located inside
an SS cylinder to prevent mixing of the two streams. The outgoing
cold gas exits through the evaporator as illustrated in Figure~\ref{fig:inside-plumbing}. 

A set of heaters (power resistors) connected to a programmable power
supply were installed on the brazed copper part of the heat exchanger
to enable temperature control. During data taking the heater was controlled
through the Slow Control and enabled stable pressure in the apparatus.
A simple see-saw control was used: when the pressure in the apparatus
decreased below a given threshold, the heater turned on, and once
a higher threshold was crossed, the heater turned off. 

Throughout the operations, bubbles were continually formed at the
bottom of the apparatus at the base of the HV feedthrough. That is
the location of the highest heat load since it was not superinsulated
for two reasons. One was to not disturb electric fields in the region
and two, due to the proximity of boil-off heaters\footnote{Boil-off heaters are used during recovery. They increase the temperature
at the bottom of the apparatus, which increases the rate of evaporation
of the liquid, making a recovery go faster.}. Bubbles are undesirable since they are known to decrease breakdown
voltage. In an attempt to reduce the formation of bubbles, two thermal
links were designed from oxygen-free high-conductivity copper and
installed in the apparatus. They deliver $\unit[\sim1.5]{W}$ of cooling
power each to two different locations in the apparatus. One is fastened
to the bottom flange of the spherical square, and the other is attached
to the bottom flange of the HV feedthrough. Each thermal link takes
part of the cooling power away from the heat exchanger and instead
delivers it to a specific location. That place is therefore cooled
below the xenon boiling point. Unfortunately, even though the thermal
links reduced bubble formation, they did not eliminate bubbles entirely.

\section{Gas system\label{sec:Gas-system}}

The gas system delivers gas to and from storage bottles to XeBrA and
enables circulation of gas during  operations for continuous removal
of impurities. It serves other purposes such as injection of calibration
sources or gas sample analysis. Since XeBrA's gas system is designed
to handle clean noble gases, it is made from all-metal parts where
possible, with most parts made out of stainless steel. It can be controlled
either manually or via the Slow Control (see Section~\ref{subsec:Slow-control}).

The primary goals of XeBrA's gas system were the ability to perform
warm vacuum pumping, slow cooldown and gas condensation, gas circulation,
and both regular and emergency recoveries. It was also designed to
be user-friendly and easy to operate. The tubing diameter was determined
by the required gas flow rate and from experiences using other xenon
gas circulations systems (such as LUX, see Chapter~\ref{chap:The-LUX-experiment},
and PIXeY~\cite{Edwards:2017emx}). Due to a lack of data for gasses
other than air, it is very hard to calculate flow for noble gases;
what follows is an outline of a quick back-of-the-envelope flow calculation
for xenon gas. 

To find the pressure drop per unit length of tubing, Reynolds number
$Re$ is calculated. Reynolds number is a dimensionless quantity used
to predict whether flow will be laminar ($Re<2300$), intermittent,
or turbulent ($Re>4000$):
\[
Re=\frac{QD_{H}}{\nu A}.
\]
Here $Q$ is the volumetric flow rate, $D_{H}$ is the hydraulic diameter,
$\nu$ is the dynamic viscosity of the fluid, and $A$ is the pipe's
cross-sectional area. Next, the pressure drop $\Delta P$ due to friction
per unit length $L$ for a given mean fluid velocity $\left\langle v\right\rangle $
can be calculated using the Darcy-Weisbach equation:
\[
\frac{\Delta P}{L}=f_{D}\cdot\frac{\rho\left\langle v\right\rangle ^{2}}{2D_{H}}
\]
where $\rho$ is the density of the fluid and $f_{D}$ is the Darcy
friction factor that can be looked up in a Moody chart~\cite{MoodyChart}.
This provides the pressure drop across some idealized parts, such
as a piece of straight tubing. To find the pressure drop across a
bent piece of tubing, a minor loss can be calculated using
\[
\Delta P_{minor}=\xi\frac{\rho}{2}\left(\frac{Q}{A}\right)^{2}
\]
where $\xi$ is the minor loss coefficient that is usually determined
empirically. For a $90^{\circ}$ elbow $\xi=0.3$. 

Additionally, some parts purchased from manufacturers quote a flow
coefficient, which describes the relationship between the pressure
drop and flow rate. Pressure drops for different gas system components
at 20 standard liters per minute (SLPM) are shown in Table~\ref{tab:Calculated-pressure-drops}.

\begin{table}
\begin{centering}
\begin{tabular}{>{\raggedright}p{1.8cm}>{\raggedleft}p{3cm}>{\raggedleft}p{2.5cm}}
\hline 
Tubing diameter {[}in{]} & Drop along straight tube {[}mbar/m{]} & Drop due to $90^{\circ}$ bent {[}mbar/m{]}\tabularnewline
\hline 
\hline 
1/4 & 16.0 & 3.50\tabularnewline
3/8 & 1.4 & 0.40\tabularnewline
1/2 & 0.2 & 0.06\tabularnewline
\hline 
\end{tabular}
\par\end{centering}
\bigskip{}
\begin{centering}
\begin{tabular}{l>{\raggedleft}p{5cm}}
\hline 
Part & Drop due to component {[}mbar{]}\tabularnewline
\hline 
\hline 
LUX getter (from data) & 700\tabularnewline
Swagelok 1/4 in valve & 10\tabularnewline
Flow meter & 30\tabularnewline
\hline 
\end{tabular}\caption[Pressure drops across gas system components]{\textbf{Top:}~Calculated pressure drops across tubing of varying
diameter at 20~SLPM. \textbf{Bottom:}~Pressure drop across varying
gas system components. \label{tab:Calculated-pressure-drops}}
\par\end{centering}
\end{table}

Based on the above calculations and experience from other detectors
the main gas system panel consists of 3/8-in diameter tubing. Most
flexible hoses connected to the panel are 1/2-in diameter, except
for the hose connecting the apparatus outlet to the pump inlet, which
is 3/4-in. That is the region of lowest pressure in the system, and
a larger diameter tubing causes smaller pressure drops (see Table~\ref{tab:Calculated-pressure-drops}).
Additionally, purges are connected to the gas system using 1/4-in
flexible hoses.

\begin{figure}
\centering{}\includegraphics[angle=90,scale=0.9]{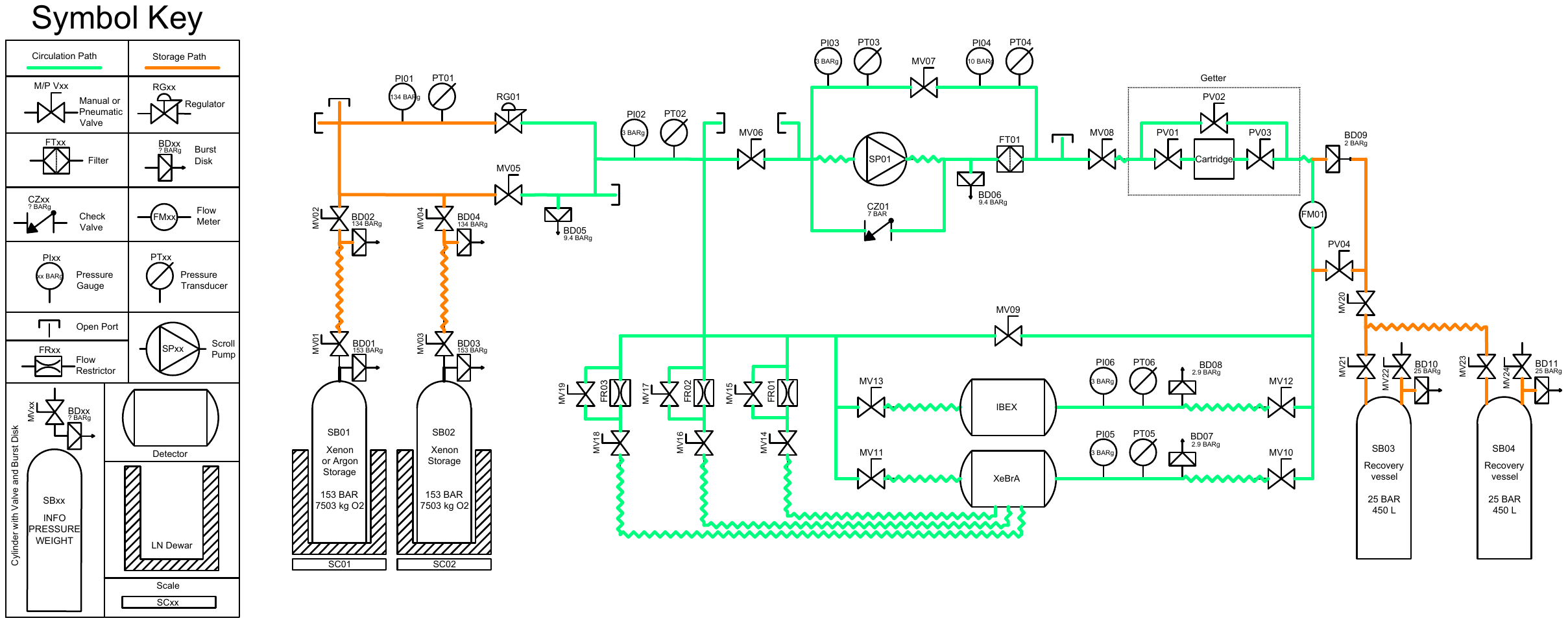}\caption[XeBrA's P\&ID]{XeBrA's P\&ID.\label{fig:XeBrA's-PID.}}
\end{figure}

\begin{figure}
\begin{centering}
\includegraphics[scale=0.65]{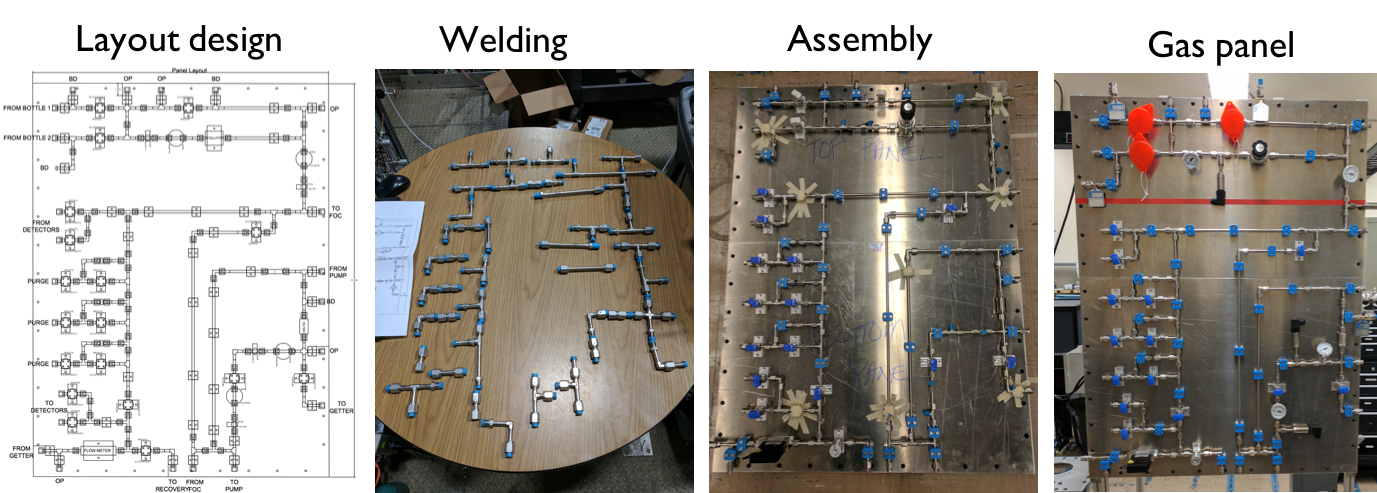}
\par\end{centering}
\caption[Gas system design]{From left to right: First, the layout was optimized in AutoCAD. Then,
various parts were welded as needed. The gas system was then assembled
attached to an aluminum panel for support and leak checked. \label{fig:gas_panel_layout}}
\end{figure}

The final XeBrA piping and instrumentation diagram (P\&ID)\nomenclature{P\&ID}{Piping and Instrumentation Diagram}
is shown in Figure~\ref{fig:XeBrA's-PID.}. Once the P\&ID was complete,
the design of the physical construction of the hardware began as outlined
in Figure~\ref{fig:gas_panel_layout}. First, a sketch of the physical
location of the parts was created in AutoCAD. The individual components
were welded using Swagelok's orbital welder as needed. Then the gas
system parts were pre-assembled on an aluminum panel to locate holes
for attachment of valves and other components. Once the holes were
drilled, the gas system was assembled and thoroughly helium leak checked.
The final gas system panel was mounted on a seismically anchored 80-20
frame that also houses all of the other gas system components, such
as the pump and the getter.

\subsection{Gas system extensibility}

Even though XeBrA was the first to benefit from this easy-to-operate
gas system, the gas system was designed to serve two experiments,
XeBrA and IBEX, as can be seen from Figure~\ref{fig:XeBrA's-PID.}.
Only one apparatus can use the gas system at a time, but both can
remain plumbed in. Therefore, once one experiment finishes taking
data, the other one can start operating soon afterward (there is a
short delay as the frozen recovered xenon must return to room temperature).
There are a few particular components of the gas system worth highlighting.

The gas system uses an oil-free scroll pump for circulation made to
order by Air Squared. A ceramic-based filter removes particles smaller
than $\unit[0.003]{\mu m}$ from the flow downstream of the pump.
Operating the pump in a vacuum can damage it, so a check valve is
installed leading from the pump outlet to the pump inlet. If a pressure
difference between the pump ends exceeds the threshold set on the
check valve, the valve will open, thus equalizing the pressure and
preventing pump damage.

Three purge lines were installed in XeBrA to prevent stagnation of
xenon gas in parts not directly connected to the main circulation.
If xenon is not circulated through the getter on a regular basis,
outgassing of  components can contaminate the gas. One purge line
each leads out of the xenon gas space from the purity monitor arm
and the PMT arm. The third purge line is connected to the xenon gas
space near the linear shifter. Only a small gas flow is desired through
the purges to ensure sufficient flow through the heat exchanger. This
was achieved by installing flow restrictors inline as shown in Figure~\ref{fig:XeBrA's-PID.}.

\begin{figure}
\begin{centering}
\includegraphics[scale=0.7]{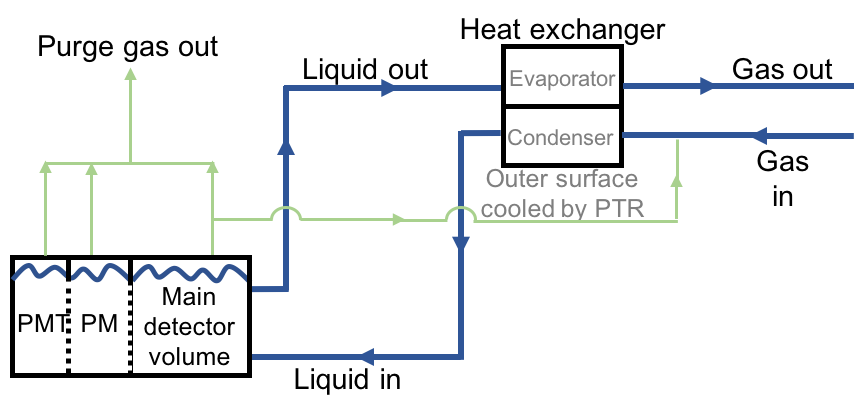}
\par\end{centering}
\caption[Schematics of the plumbing inside the vacuum space]{Schematics of the plumbing inside the vacuum space.\label{fig:inside-plumbing}}
\end{figure}

Figure~\ref{fig:inside-plumbing} shows the connections of the plumbing
lines inside the OV. The blue lines indicate the main flow. The incoming
warm gas enters through the condenser. Most of the xenon that evaporates
from the apparatus exits through the evaporator. A small fraction
of gas exits through the purge lines (green). Additionally, there
is a line that connects the purge lines to the gas inlet which ensures
that the pressure of the gas entering the condenser matches the pressure
in the gas above the main bath. This makes the liquid in the condenser
fall into the main bath.

A brass parallel plate capacitor located directly above the ground
electrode as shown in Figure~\ref{fig:HV-assembly} monitors the
liquid level height during the condensation phase. The change in capacitance
between the brass discs acts as a level sensor and determines when
both electrodes are fully submerged in liquid. The relative dielectric
constant of vacuum $\varepsilon_{r\,vacuum}=1$, while that of LXe
$\varepsilon_{r\,LXe}=1.85$ and $\varepsilon_{r\,LAr}=1.5$. The
expected change in capacitance in LXe is given by
\begin{align*}
\Delta C_{LXe} & =\varepsilon_{0}\frac{A}{d}\left(\varepsilon_{r\,LXe}-\varepsilon_{r\,vacuum}\right)\\
 & =\unit[48]{pF}
\end{align*}
where $A=\unit[48.33]{cm^{2}}$ is the area of the capacitor, $d=\unit[0.76]{mm}$
(0.03~in) is the separation of the two capacitor discs, and $\varepsilon_{0}=\unit[8.85\times10^{-12}]{F/m}$
is the permittivity of vacuum. A similar calculation reveals a change
in capacitance for argon $\Delta C_{LAr}=\unit[28]{pF}$. The wires
connecting the level sensor to a capacitance meter contribute to the
total capacitance of the system, which was measured to be $\sim160$~pF
in LXe. The change in capacitance between the brass discs was large
enough to suffice for the operation of the level sensor. 

\subsection{Xenon recovery via cryopumping}

\begin{figure}[t]
\begin{centering}
\includegraphics[scale=0.55]{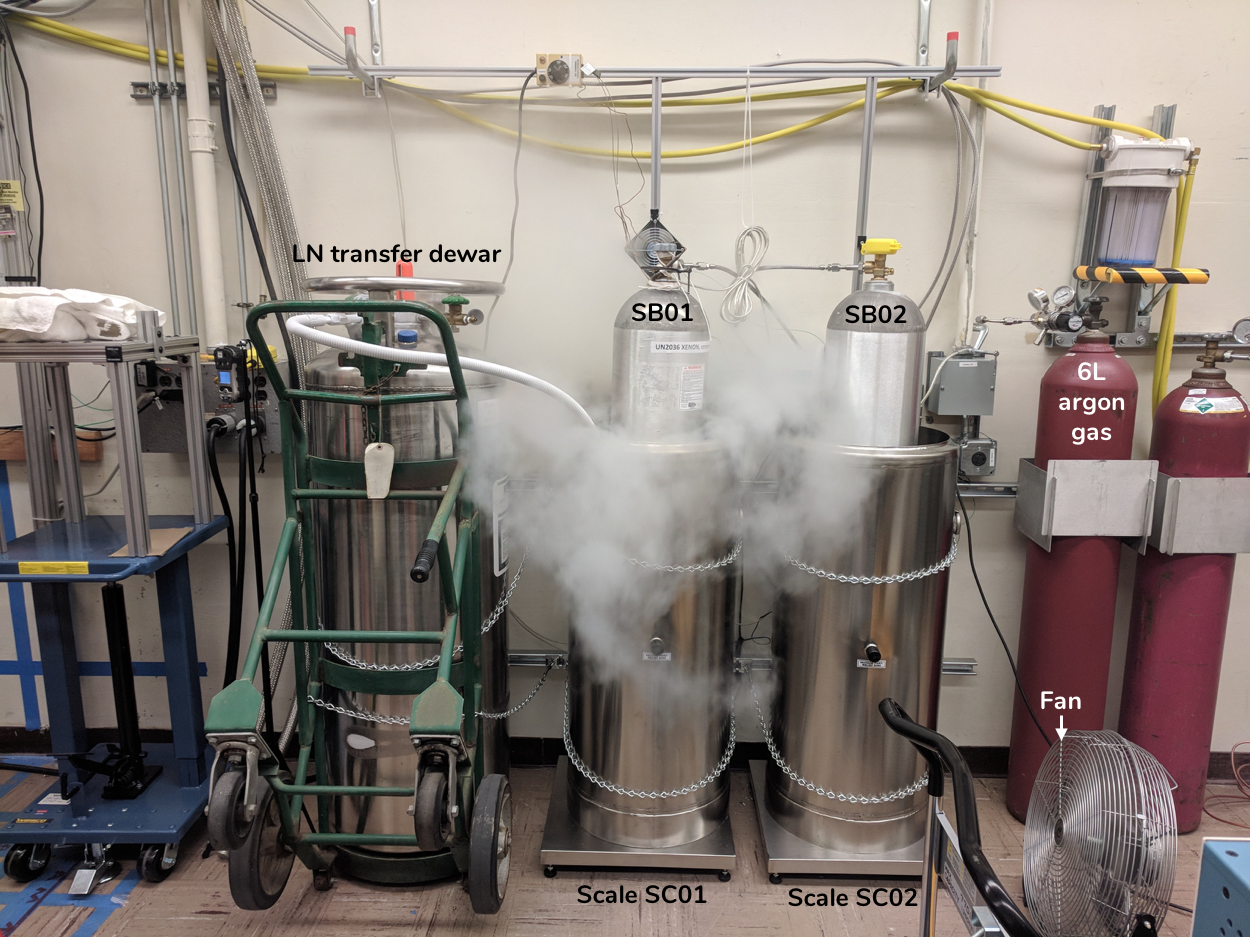}
\par\end{centering}
\caption[Xenon recovery via cryopumping]{Xenon recovery via cryopumping in progress. Xenon is being recovered
from the apparatus into SB01.\label{fig:cryopumping}}
\end{figure}

At the end of regular  operations, xenon is recovered via cryopumping.
Cryopumping uses a cold surface to produce a vacuum in the enclosed
space. In this particular case when xenon recovery is desired, a high-pressure
aluminum storage cylinder located inside a stainless steel open top
dewar is cooled with liquid nitrogen (LN)\nomenclature{LN}{Liquid Nitrogen}\footnote{An excessive amount of cryogens in enclosed spaces can lead to an
oxygen deficiency hazard (ODH)\nomenclature{ODH}{Oxygen Deficiency Hazard},
so an oxygen monitor (series 1300 from Alpha Omega) was installed
in the room.}, and xenon from the apparatus is allowed to flow into this cylinder
as shown in Figure~\ref{fig:cryopumping}. There, the xenon freezes
to the inner walls of the cooled cylinder since the boiling temperature
of LN is 77~K while the melting point of xenon is 161 K. It is important
to keep the pressure in the apparatus above the xenon triple point
throughout this operation to prevent xenon freezing inside the apparatus.
To speed the recovery, two 100~W heaters are installed at the bottom
of the apparatus to encourage xenon boiling. Additionally, heat should
be provided to the top portion of the aluminum cylinder, including
the valve, to prevent it from freezing. Because both the valve and
the seal between the valve and the cylinder contain elastomer seals,
extreme temperatures (both hot and cold) might damage those and allow
xenon to leak into the room.

Once all the xenon has left the apparatus and is recovered into the
cylinder, the valve on the top of the cylinder is closed, and the
cylinder is allowed to warm. The xenon is contained within the aluminum
storage cylinder until its next use. To recover the entire amount
of xenon contained in XeBrA takes about one full working day. Further
details and particular steps of this procedure are included in Appendix~\ref{chap:XeBrA-procedures}.

At the end of operations with liquid argon, argon is slowly released
into the room since argon is not very expensive. 

\subsection{Xenon emergency recovery}

The standard recovery via cryopumping requires LN and needs to be
planned. If the xenon cannot be recovered using the standard recovery
procedure, a safety burst disc (BD07 in Figure~\ref{fig:XeBrA's-PID.})
would rupture and vent xenon into the room once the pressure in the
apparatus rises above 3.9~bar. Such a situation might arise if a
hardware or a power failure occurs. A power failure halts apparatus
cooling, and the increased heat load results in a rise of pressure
inside the apparatus within $\sim1$~hour. The room ventilation is
also interrupted during a power failure. The lack of air circulation
in the room would prevent xenon recovery via cryopumping as it would
pose an ODH hazard. A low-pressure passive recovery system was implemented
to avoid this xenon loss.

\begin{figure}[t]
\begin{centering}
\includegraphics[scale=0.6]{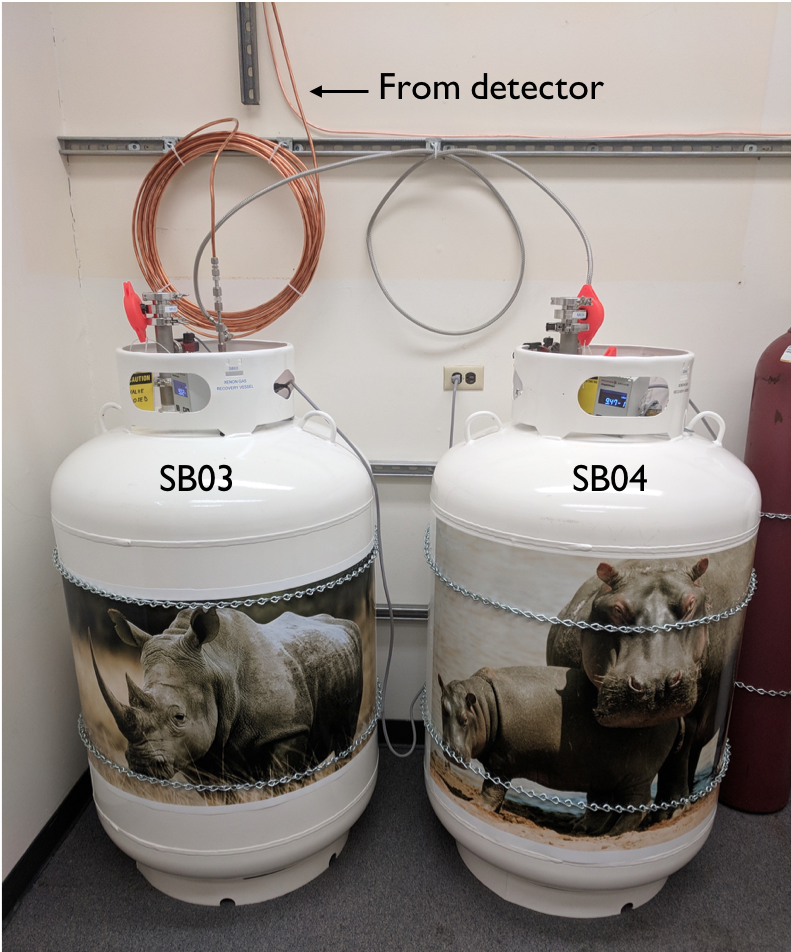}\caption[Passive xenon recovery system (Hippo and Rhino)]{Passive xenon recovery system, known as Hippo (SB04) and Rhino (SB03).
The tanks are connected through a flexible hose and to XeBrA via 100~ft
of 5/8-in clean copper tubing. Each tank has a vacuum gauge for pressure
monitoring and is chained to the wall for seismic safety (to prevent
wildlife escape).\label{fig:Passive-xenon-recovery}}
\par\end{centering}
\end{figure}

Two large (450~l) tanks are plumbed in parallel and act as a single
900~l volume, as shown in Figure~\ref{fig:Passive-xenon-recovery}.
Affectionately known as Hippo and Rhino, these two tanks can store
the entire quantity of xenon contained in XeBrA during operations
at room temperature at low pressure. The 900~l volume of the recovery
tanks is large enough to hold $\unit[\sim16]{kg}$ of xenon after
it has boiled and warmed to room temperature at a pressure of 3.3~bar.
This is below the 4~bar absolute pressure at which the BD07 burst
disk would otherwise vent the xenon into the 70A-2263 lab.

The combined volume is evacuated before operations and plumbed into
the gas handling system as shown in Figure~\ref{fig:XeBrA's-PID.}.
In normal operation, a 3~barg burst disk (BD09) separates the apparatus
from the recovery tanks. If the pressure inside the apparatus at any
time exceeds 2~barg\footnote{The tanks are kept at a vacuum.}, BD09
will open, connecting it to the 900~l evacuated volume of the low-pressure
recovery tanks. However, a burst disc rupture is irreversible, so
a pneumatic valve (PV04)\footnote{The pneumatic valve can be opened or closed by controlling the flow
of high-pressure gas (nitrogen). The nitrogen flow is regulated by
a solenoid valve, which in the actuated state opens a path of nitrogen
to the pneumatic valve. This compresses a spring and opens the pneumatic
valve.} is installed in parallel with the burst disc. The valve opens automatically
if a high-pressure alarm threshold in the Slow Control is exceeded.
For this to be effective, the alarm threshold must be below 2~barg
and is typically set to 1.8~barg. The valve also automatically closes
once the pressure drops below the alarmed threshold. 

Once the emergency is over, xenon from Hippo and Rhino can be cryopumped
back to the storage bottles SB01 and SB02.

\section{Operations\label{sec:Detector-operations}}

The core components of XeBrA are the apparatus and the gas system.
There are multiple other supporting components required for successful
operation. Since the apparatus and the sub-components take up physical
space, their intelligent placement within the lab can affect operations.
Detailed procedures were developed to ensure safe apparatus operation
without loss of valuable xenon. Several electrical components were
installed for apparatus stability monitoring, and a software system
was implemented to enable data collection and 24/7 and remote operations.

\begin{figure}[t]
\begin{centering}
\includegraphics[scale=0.61]{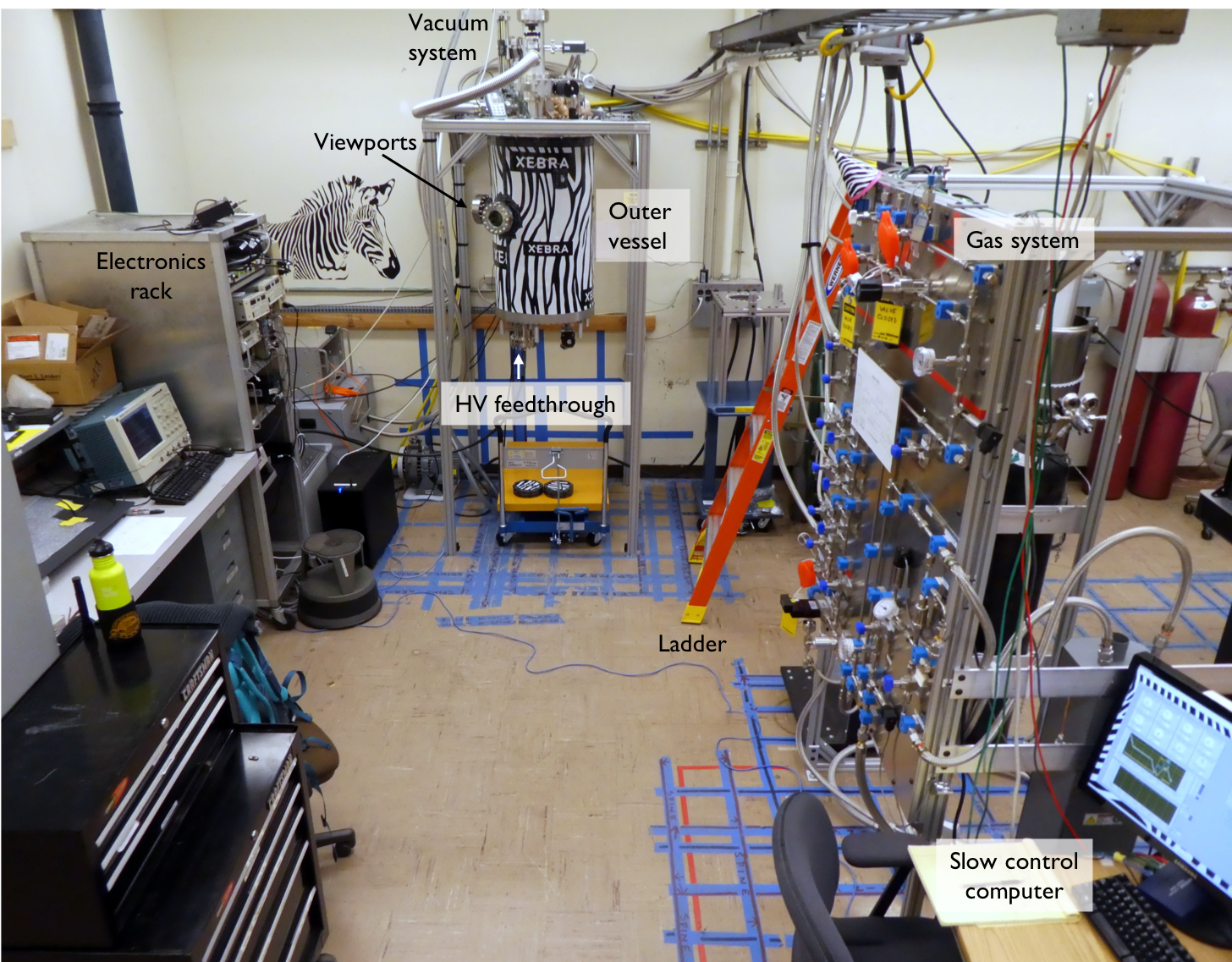}
\par\end{centering}
\caption[Overview of the lab]{Overview of the lab. \label{fig:lab-view}}
\end{figure}

\begin{figure}
\begin{centering}
\includegraphics[angle=90,scale=0.48]{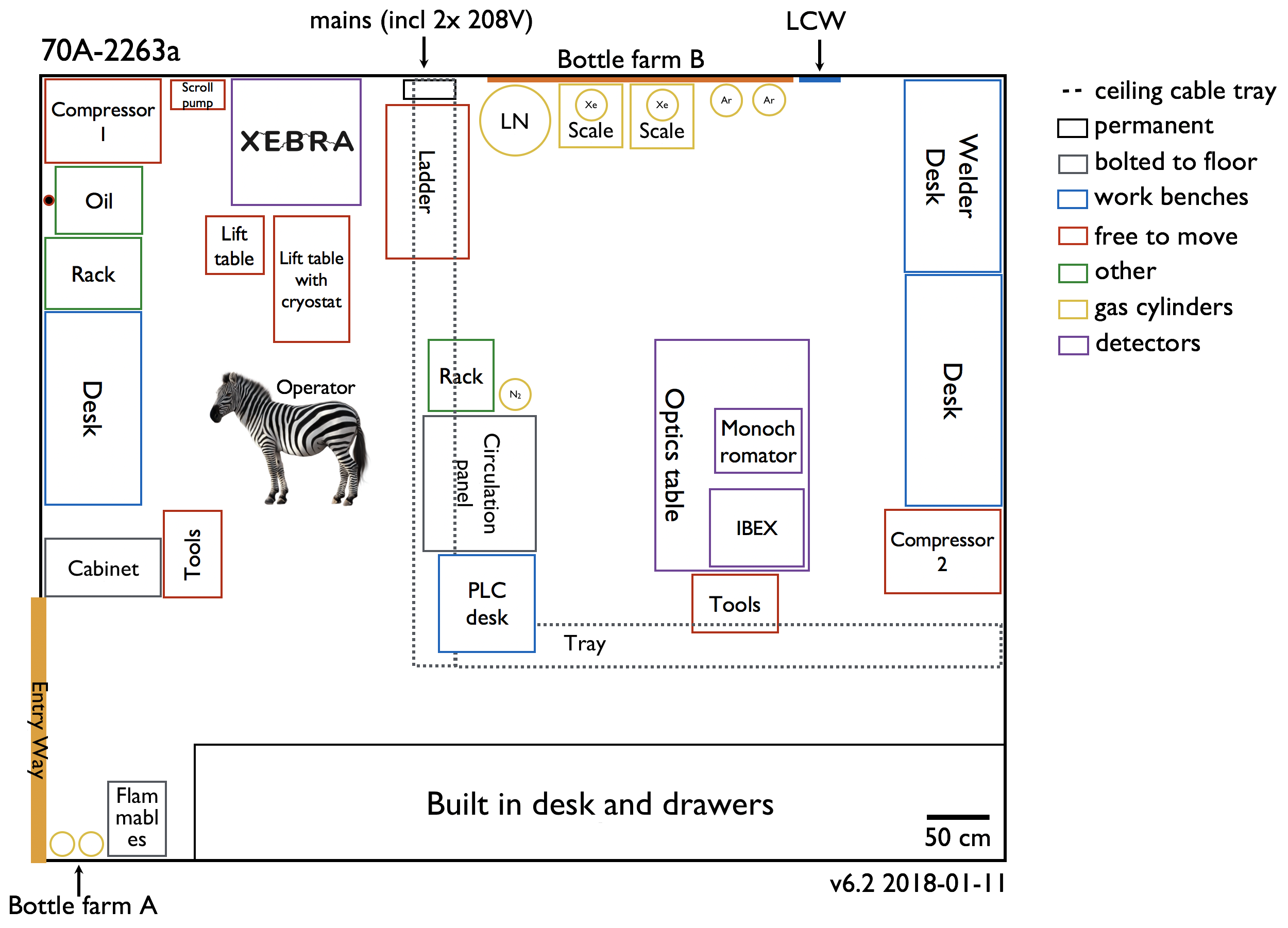}
\par\end{centering}
\caption[XeBrA lab layout]{The layout of the lab space in 70A-2263 at LBNL.\label{fig:lab_layout}}
\end{figure}

The limited floor space at LBNL in lab 70A-2263 where XeBrA is located
required careful optimization of equipment placement, ultimately resulting
in the floor plan illustrated in Figure~\ref{fig:lab_layout}. A
picture of the final layout of the lab is shown in Figure~\ref{fig:lab-view}.
This final layout achieves two real-world practical goals: both XeBrA
and IBEX fit in the space available, and an individual attempting
to operate XeBrA has all of the controls and monitors available to
herself within a few steps. And of course, there are clear paths for
emergency escape.

\begin{figure}[t]
\begin{centering}
\includegraphics[scale=0.42]{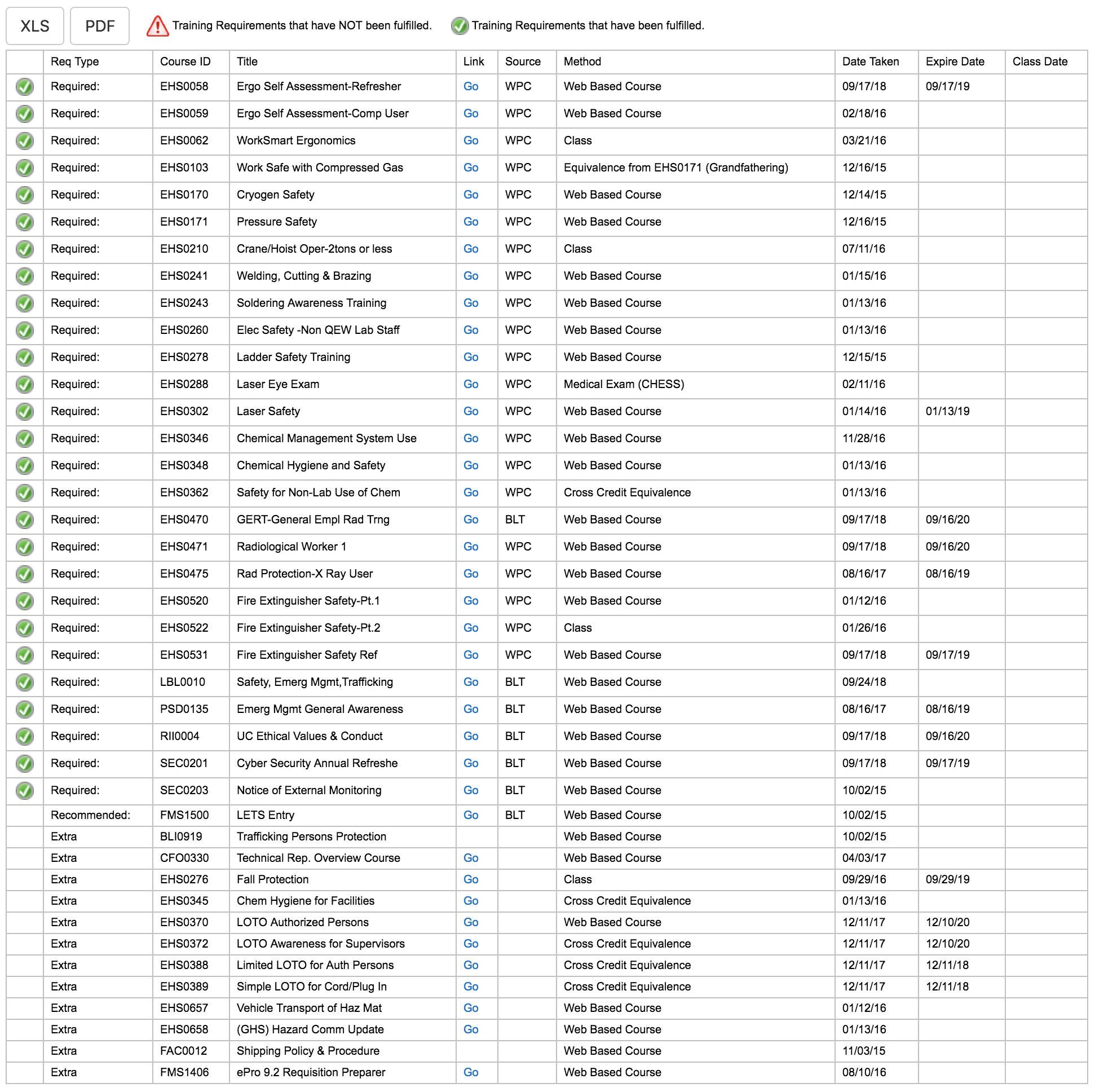}
\par\end{centering}
\caption[List of training courses passed while at LBNL]{List of training courses passed while at LBNL.\label{fig:List-of-training}}

\end{figure}

An integral part of working at a national lab is a long list of training
required to do anything from purchasing apparatus parts, standing
on ladders, to cryogen handling. During my time at the lab, this has
resulted in forty various training courses listed in Figure~\ref{fig:List-of-training}.
This does not include many, many more courses passed for on-site LUX
shifts at SURF and safety training at UCB.

\subsection{Procedures }

Procedures for operating the gas system and the apparatus were developed
for use in cases with both LAr and LXe. Appendix~\ref{chap:XeBrA-procedures}
contains the procedure for xenon condensing, circulation, and recovery
in XeBrA.

\subsection{Performance monitoring}

To monitor  parameters during operations and data taking, the following
components were installed in the apparatus. All of them were monitored
in the Slow Control described in the next section.
\begin{itemize}
\item \textbf{Resistance temperature detectors (RTDs):}\nomenclature{RTD}{Resistance Temperature Detector}
RTDs consist of a wire of pure platinum housed in a protective probe.
The material has an accurate resistance vs. temperature relationship,
which is used to indicate temperature. There are 8 RTDs installed
throughout the apparatus to monitor temperatures of the various parts. 
\item \textbf{Pressure gauges and pressure transducers:} Pressure gauges
have physical dials while pressure transducers are read out in the
Slow Control. They serve to monitor pressure throughout the apparatus
as indicated on the P\&ID (Figure~\ref{fig:XeBrA's-PID.}). An additional
pressure transducer is located on the nitrogen gas bottle that supplies
gas needed for pneumatic valve operation.
\item \textbf{Xenon scales:} Each of the two storage bottles is located
on a scale which enables direct measurement of the amount of xenon
that has entered the  apparatus.
\item \textbf{Vacuum gauges:} Several vacuum gauges monitor pressure inside
the OV and the recovery vessels.
\item \textbf{Picoammeter \& charge amplifier:} These were connected to
the ground electrode (anode) and provided monitoring of current during
data taking. The charge amplifier was read out by a DAQ system rather
than the Slow Control.
\end{itemize}

\subsubsection{Slow Control\label{subsec:Slow-control}}

\begin{figure}
\begin{centering}
\includegraphics[angle=90,scale=0.95]{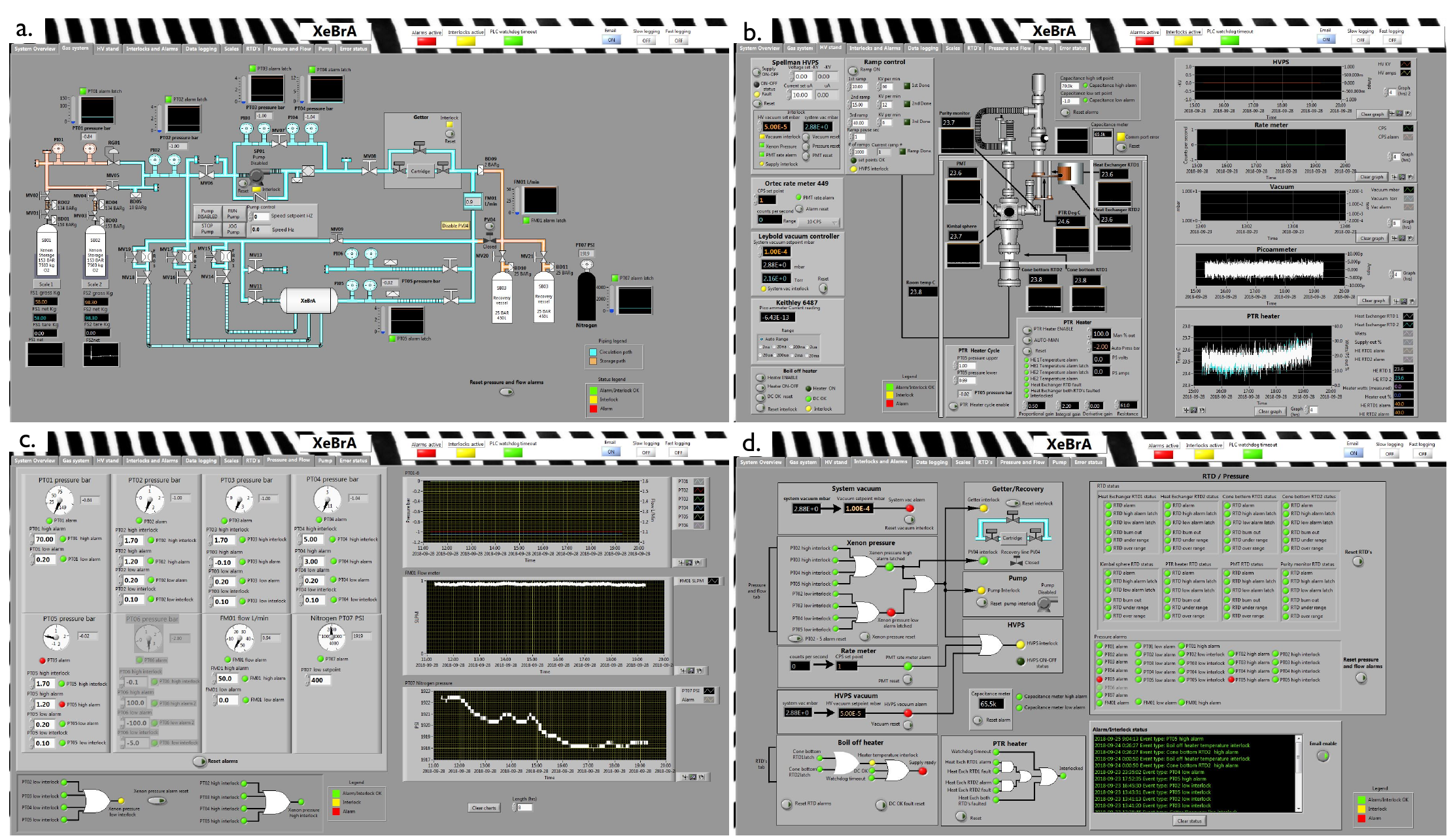}
\par\end{centering}
\caption[Slow control screenshots]{Slow control screenshots. \textbf{a:}~Overview of the gas system.
\textbf{b:}~Overview of the HV stand with controls for HV data taking.
\textbf{c:}~Pressure monitoring. \textbf{d:}~Interlock screen. \label{fig:Slow-control}}
\end{figure}

Maintaining stability in XeBrA's day-to-day operations requires manipulation
of a few critical components. The pressure, temperature, and gas circulation
speed all increase and decrease depending on the settings of the heaters,
pump, and other regulators in the system. Slow control is a well-tested
solution for monitoring of gas flow and apparatus status, automation
of  controls, and remote  operations at time scales ranging from 0.1~seconds
to weeks. Low and high alarms can be set on various parameters. If
a set of conditions is met, interlocks trigger automatic shut-off
of selected parts of the system (getter, pump) or ultimately start
xenon emergency recovery. A user can be informed about any alarms
through an email system which can be forwarded to a mobile phone to
sound noisy alarms at any time. This mitigates risks resulting from
apparatus operations that span multiple day-night cycles. Additionally,
the Slow Control also provides two data logging modes: a slow mode
that can log apparatus stability every few seconds, and a fast mode
with an ability to save data every 100~ms used for HV breakdown data
collection. Several screenshots from XeBrA's slow control are included
in Figure~\ref{fig:Slow-control}.

The Slow Control backend runs on programmable logic controllers (PLCs)\nomenclature{PLC}{Programmable Logic Controller},
flexible, easily programmable digital computers known for their high-reliability
automation. The frontend runs on \textsc{LabView}~\cite{LabVIEW},
a visual programming language often used for data acquisition, instrument
control, and automation. XeBrA's Slow Control was developed by Mark
Regis, an engineer at LBNL. 

\section{Data collection\label{sec:Data-collection}}

Data collection started in February 2018 and continued until September
2018. During that time, XeBrA was filled twice with argon, and twice
with xenon, resulting in four different data acquisitions described
in this section and summarized in Table~\ref{tab:Summary-of-data}.

\begin{table}
\begin{centering}
\begin{tabular}{llrrrr}
\hline 
 & Liquid & Electrode separation & Pressure & Purity & Purity\tabularnewline
 &  & {[}mm{]} & {[}bar{]} & {[}$\mu$s{]} & {[}ppb{]}\tabularnewline
\hline 
\hline 
Run 0 & Ar & 1-8 & N/A & 223 & 2\tabularnewline
Run 1 & Ar & 1-6 & 1.5, 2 & 400 & 1\tabularnewline
Run 2 & Xe & 1-5 & 2 & N/A & >ppm\tabularnewline
Run 3 & Xe & 1-2 & 2 & 2.2 & 200\tabularnewline
\hline 
\end{tabular}
\par\end{centering}
\caption[Summary of data acquisitions used in analysis]{Summary of data acquisitions. Only Runs 1, 2, and 3 were used in
analyses since pressure during data acquisition varied in Run 0. The
temperature of the liquid is not known due to the poor thermal coupling
of RTDs. Additionally, there was a temperature gradient present in
LXe and LAr during data acquisitions.\label{tab:Summary-of-data}}
\end{table}

\subsection{Run 0 (liquid argon)}

Data collection in XeBrA started in early 2018. Careful planning occasionally
pays off, and all the  components worked as expected during the first
cool down with LAr in February 2018 (Run~0). However, it was discovered
that LAr was forming gas bubbles near the bottom of the HV feedthrough
cone. These bubbles rose through the apparatus, flowed through the
active volume, and distorted breakdown datasets. To avoid bubbles
during data acquisition, the apparatus was kept at low pressures overnight
($\unit[\sim1.3]{bar}$). Then, during the data acquisition, the pressure
was slowly increased in the apparatus by increasing the heat load
on the heat exchanger. This enabled several hours of bubble-free data
taking, but at varying pressure. 

\begin{figure}
\begin{centering}
\includegraphics[scale=0.13]{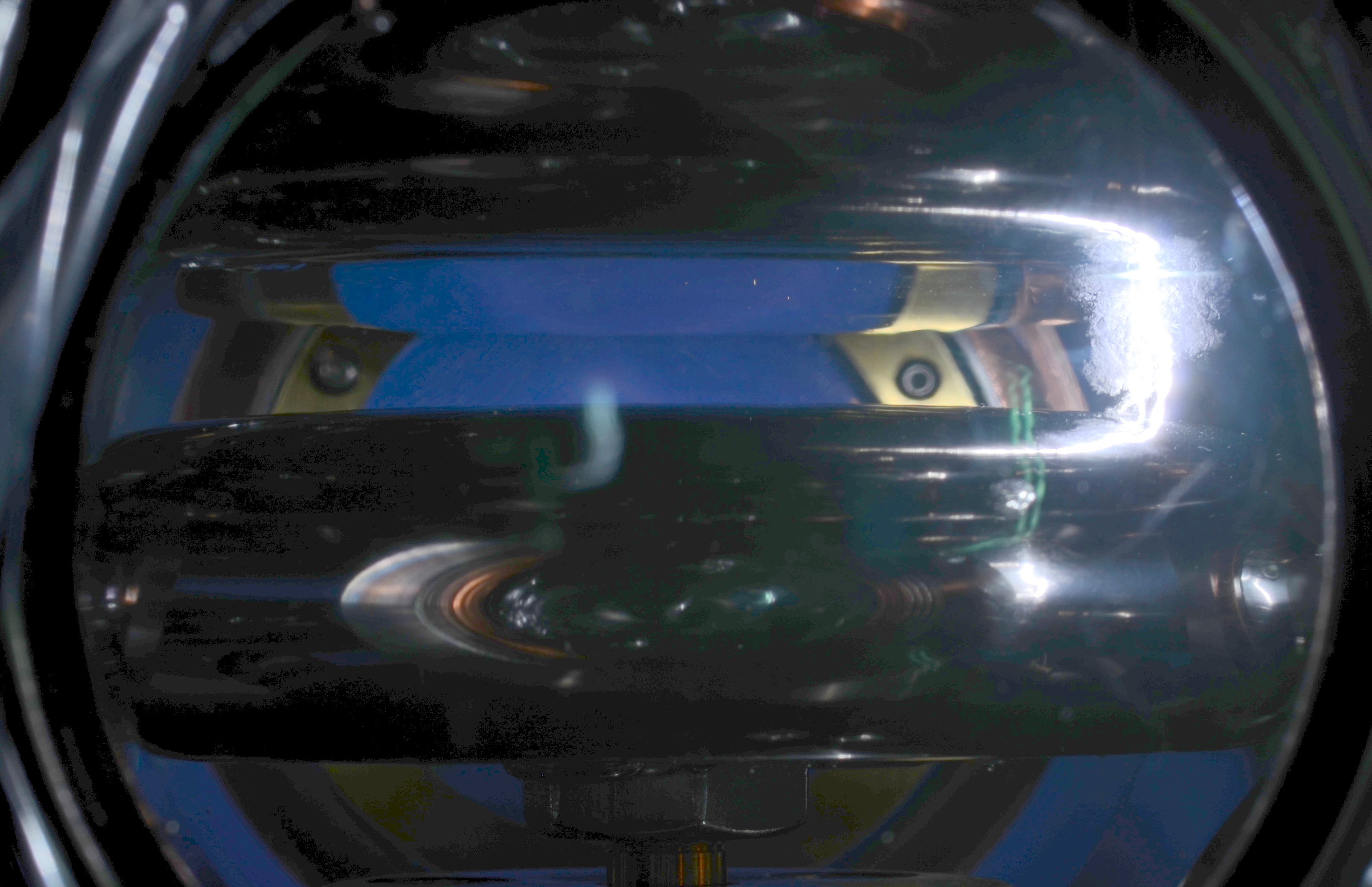}\caption[Photograph of a spark in LAr]{Photograph of a spark near the edge of the electrodes during data
acquisition in LAr in Run~0. Photograph by Ethan Bernard.\label{fig:Photograph-of-spark}}
\par\end{centering}
\vspace{0.8cm}
\begin{centering}
\includegraphics[scale=0.13]{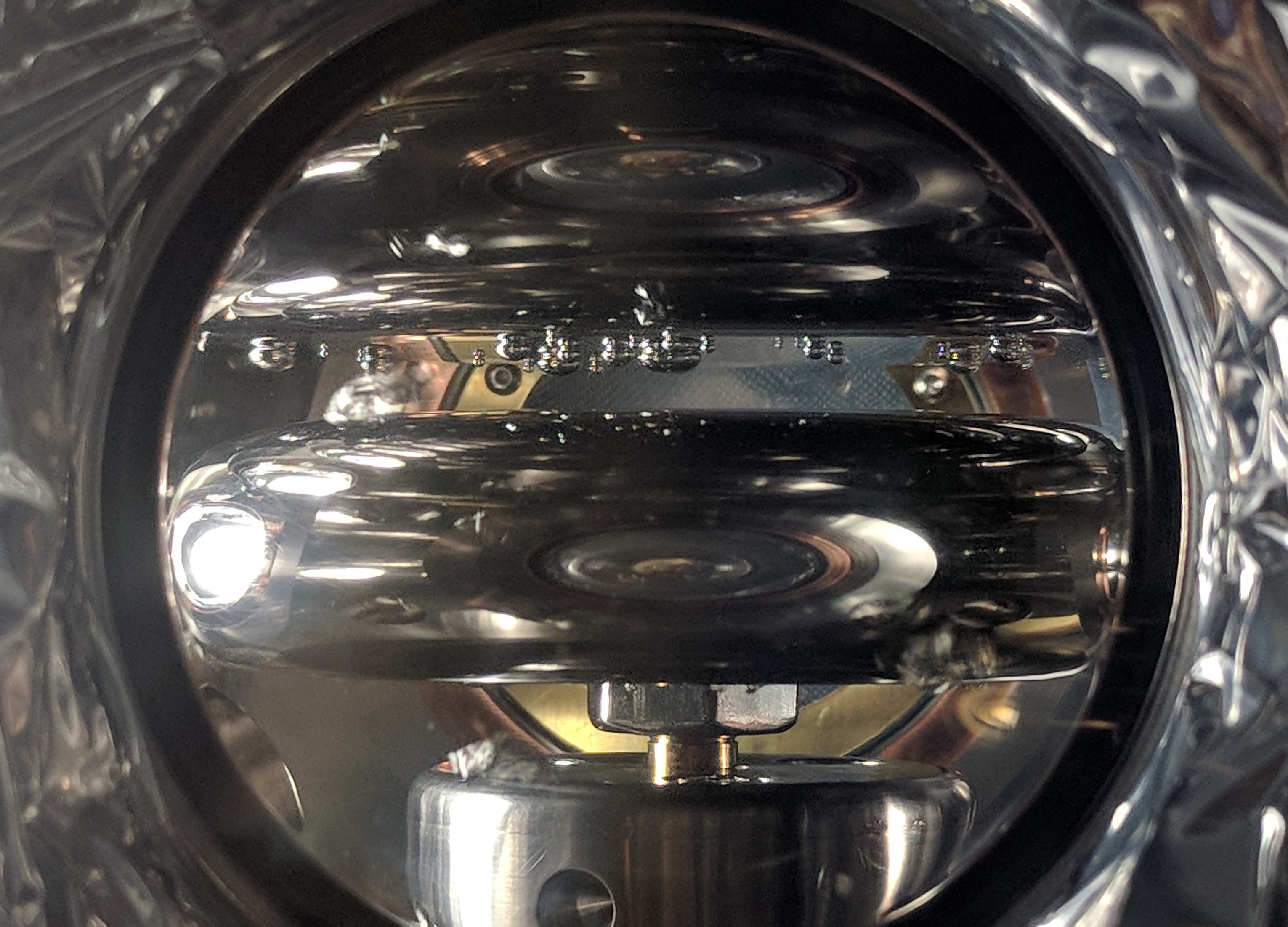}
\par\end{centering}
\caption[Photograph of bubbles in the LAr active volume]{Photograph of bubbles in the LAr active volume between data acquisitions
in Run~1. A couple forming bubbles are visible at the top of the
alignment joint and behind the electrodes. \label{fig:Photograph-of-bubbles}}
\end{figure}

Data were acquired at 1, 5, 7, and 8 mm electrode separations between
February 17-19, 2018. Figure~\ref{fig:Photograph-of-spark} shows
a photograph of a spark during data acquisition in LAr in Run 0. The
purity of LAr was $\sim\unit[223]{\mu s}$ as measured by the purity
monitor described in Section~\ref{subsec:Purity-monitor}. This corresponds
to $\sim\unit[1]{ppb}$ oxygen equivalent impurities. The conversion
from $\mu s$ to ppb was described in Section~\ref{subsec:Electron-Lifetime-to-ppb}.
Because of the varying pressure, the data were not included in any
analyses discussed in this work. 

\subsection{Run 1 (liquid argon)}

In an attempt to reduce the number of bubbles in the apparatus a thermal
link, described in Section~\ref{subsec:Cooling}, was introduced
to XeBrA for the next data acquisition in LAr (Run~1). This improved
the bubble situation, but not entirely: bubbles that were visible
in the apparatus are shown in Figure~\ref{fig:Photograph-of-bubbles}.
There is evidence in the literature that pressure affects breakdown~\cite{KAWASHIMA1974217,Gerhold1989},
so unlike in Run~0, a procedure was developed to achieve data acquisition
at a stable pressure. The apparatus was kept at lower pressures overnight.
Then, throughout the data acquisition, argon from a storage bottle
was slowly added. This caused a brief pressure increase at the beginning
after which the pressure remained constant for a couple of hours while
the additional argon was condensing and the apparatus was free from
bubbles. 

\begin{figure}[t]
\begin{centering}
\includegraphics[scale=0.47]{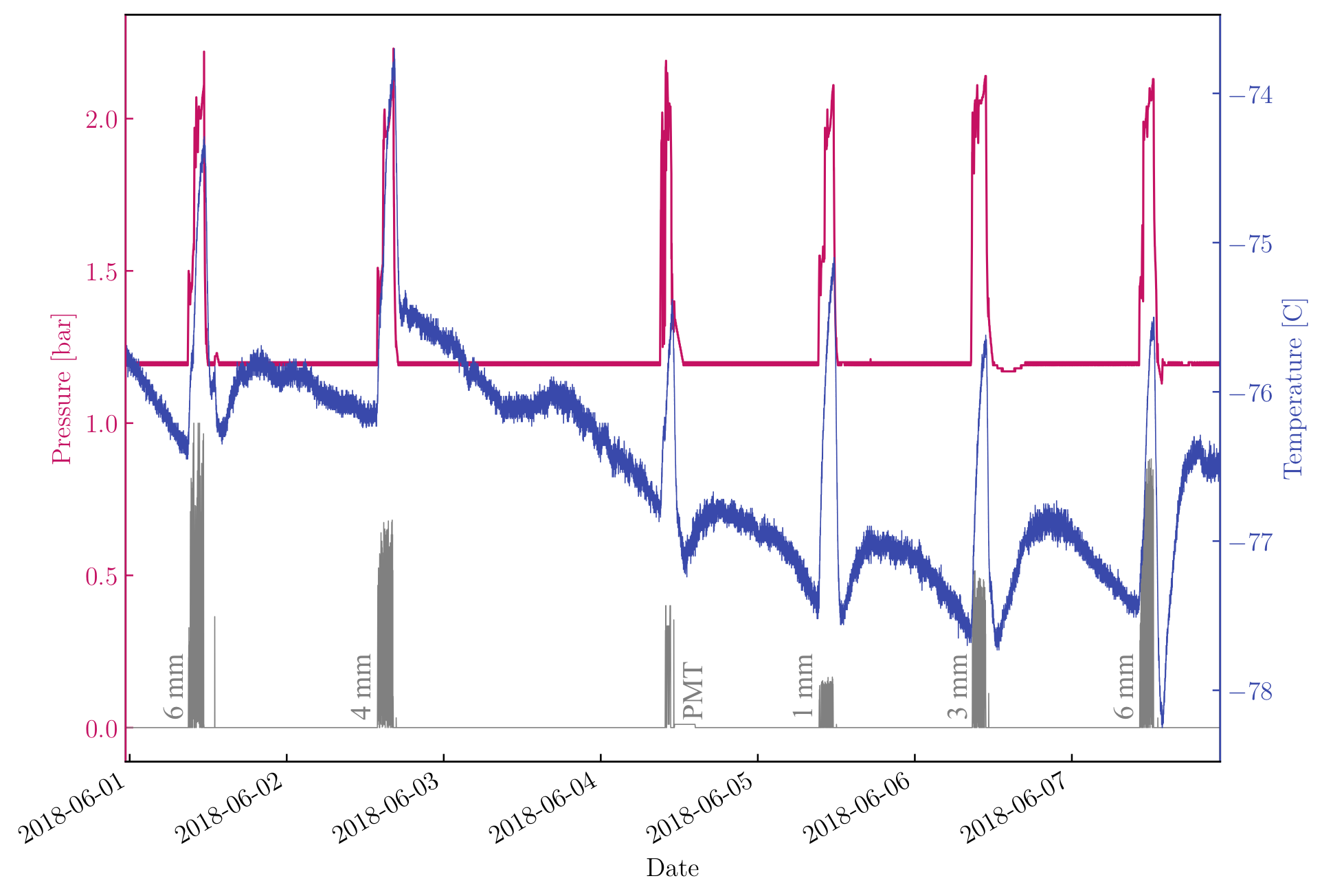}
\par\end{centering}
\caption[Pressure and temperature during operations in LAr]{Pressure and temperature during operations in LAr. Pressure, as measured
by PT05, is shown in pink. Due to the poor thermal coupling of RTDs,
the absolute temperature in the apparatus is not known, but changes
in temperature as measured at the bottom of the HV cone are shown
in blue. For illustration, the gray line shows the normalized voltage
of the power supply to indicate the period of data taking at various
electrode separations.\label{fig:lar-temp-pressure}}
\end{figure}

Therefore, data were collected at two different pressures, 1.5 and
2.0~bar. Usually, data taken at 1.5 bar were obtained first in the
morning, while data at 2 bar were collected in the afternoon since
a pressure increase temporarily suppressed bubble formation. Figure~\ref{fig:lar-temp-pressure}
illustrates the behavior of pressure and temperature throughout one
of the periods of data acquisition. LAr data used in analyses were
collected between May 26 and June 8, 2018. The purity of LAr was $\sim\unit[400]{\mu s}$,
which corresponds to $\sim\unit[1]{ppb}$ oxygen equivalent impurities.

\subsection{Run 2 (liquid xenon)}

After the successful data acquisition in LAr, the apparatus proceeded
to LXe operation (Run~2) in July 2018. Only the PMT was changed between
data acquisitions in LAr and LXe, everything else remained intact.
There were significantly fewer bubbles present in the apparatus when
operated in LXe. Unfortunately, even a small amount of bubbles is
disruptive to HV breakdown measurements, so the same pressure trick
employed in LAr in Run~1 was used to collect LXe data. The data taking
procedure is detailed in Section~\ref{sec:Data-taking} of the xenon
procedure included in Appendix~\ref{chap:XeBrA-procedures}. 

\begin{figure}
\begin{centering}
\includegraphics[scale=0.48]{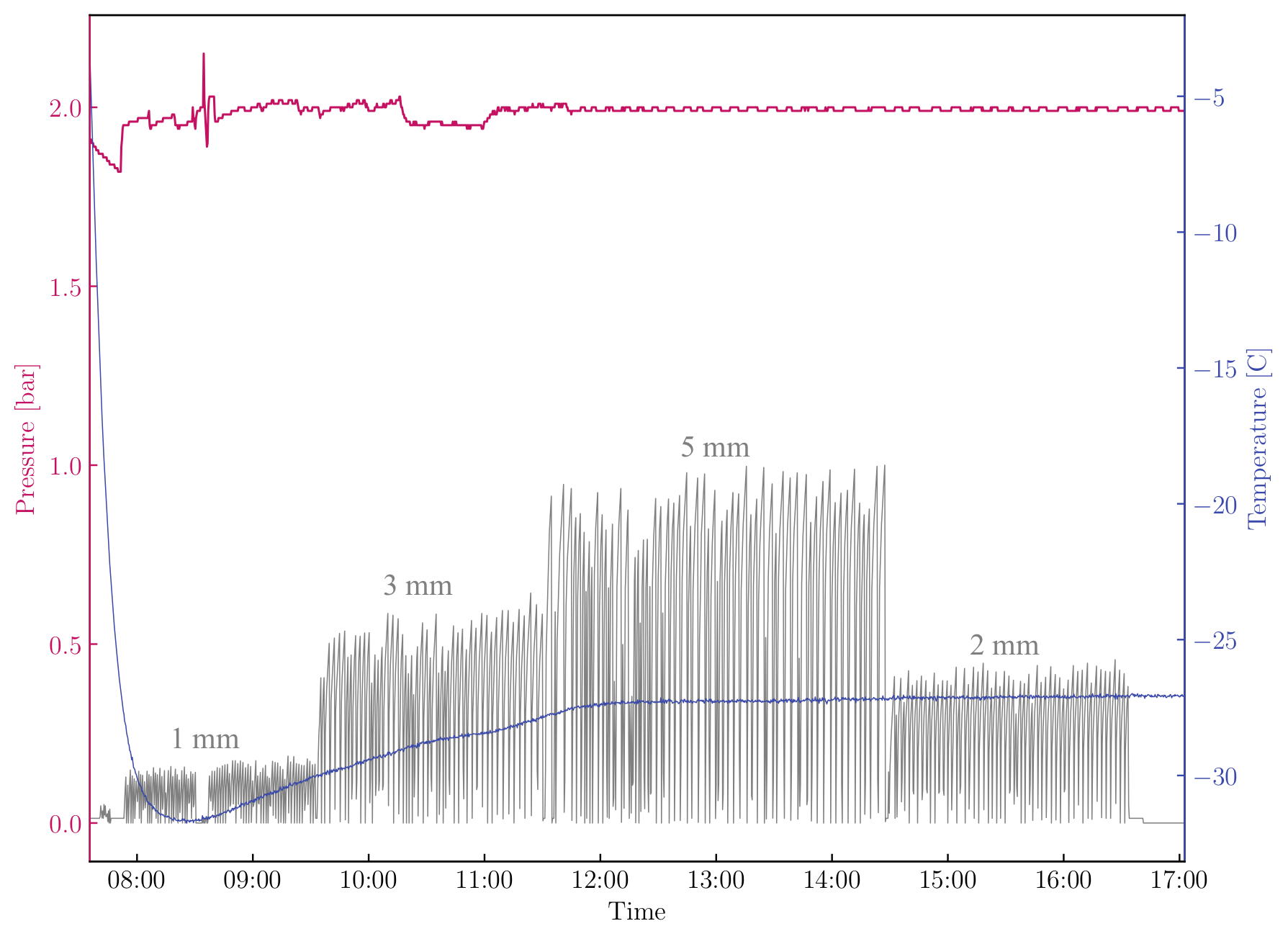}
\par\end{centering}
\begin{centering}
\includegraphics[scale=0.48]{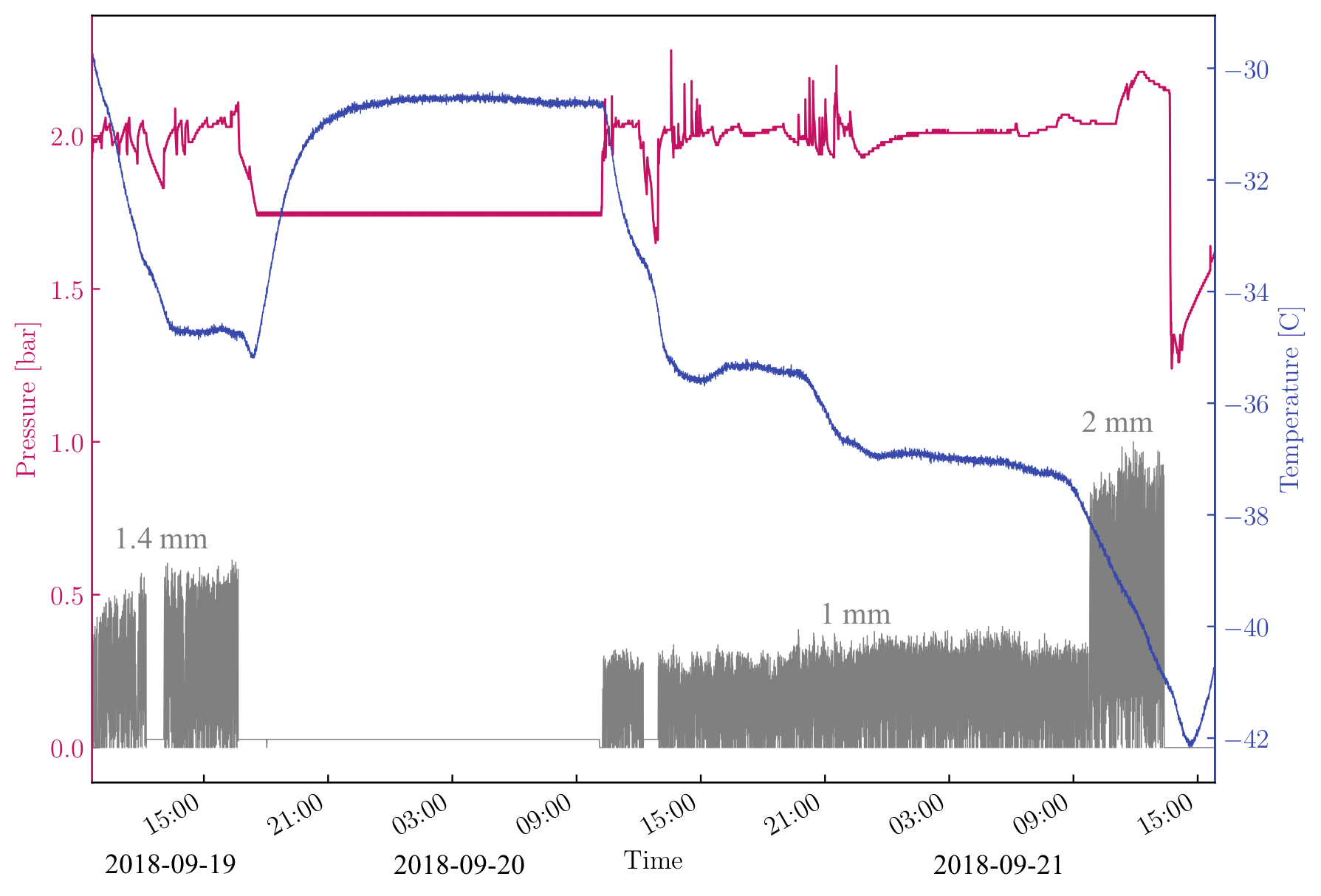}
\par\end{centering}
\caption[Pressure and temperature during operations in LXe]{Pressure and temperature in the apparatus during operations in LXe.
Pressure, as measured by PT05, is shown in pink. Due to the poor thermal
coupling of RTDs, the absolute temperature in the apparatus is not
known, but changes in temperature as measured at the bottom of the
HV cone are shown in blue. For illustration, the gray line shows the
normalized voltage of the power supply to indicate the period of data
taking at various electrode separations. \textbf{Top:}~Run~2. \textbf{Bottom:}~Run~3.
There is a noticeable drop in pressure between HV data at 1.4 and
1~mm, which shows the preparation period for the 1~mm data acquisition.
\label{fig:lxe-temp-pressure}}
\end{figure}

Data were acquired on July 12, 2018, at 2~bar. The behavior of the
detector is shown in Figure~\ref{fig:lxe-temp-pressure}. Since the
purity monitor did not work in liquid xenon, a residual gas analyzer
(RGA)\nomenclature{RGA}{Residual Gas Analyzer}, a small quadrupole
mass spectrometer, was used to estimate the amounts of impurities
in LXe during Run~2. Based on that measurement, we estimate that
the LXe purity had $>\mathcal{O}\left(\mathrm{ppm}\right)$ oxygen
equivalent impurities during Run~2. It was later discovered that
the xenon used was polluted by air during cylinder filling at Yale. 

\subsection{Run 3 (liquid xenon)}

A new getter cartridge and two chemical inline purifiers\footnote{The model of the getter and inline purifiers used in the experiment
can be found in Appendix~\ref{chap:Parts-list}.} were installed prior to the second xenon data acquisition (Run~3)
in September 2018. The apparatus was filled with clean xenon from
the LUX detector\footnote{It was traded with LZ at SLAC. All xenon used in LZ uses chromatographic
separation system based on adsorption on activated charcoal to reduce
levels of radioactive $^{85}$Kr. This also removes other impurities,
such as air, so air pollution prior to purification is not an issue~\cite{Akerib:2016hcd}.}. Additionally, the location of one of the thermal links was moved
to the bottom of the HV cone in an attempt to further minimize bubbles.
The xenon procedure included in Appendix~\ref{chap:XeBrA-procedures}
was used for data collection as in Run~2. However, this time the
xenon was added through a series of inline purifiers which have a
high resistivity. This caused the circulation speeds to be much lower
than in previous runs. That ultimately led to the formation of solid
xenon in the apparatus. A small amount of solid xenon near the bottom
of the HV feedthrough where all the bubbles were formed would be desirable
as it would suppress the bubble formation in the apparatus, but frozen
xenon also likely formed in the heat exchanger, which hindered its
performance. 

\begin{figure}
\begin{centering}
\includegraphics[scale=0.47]{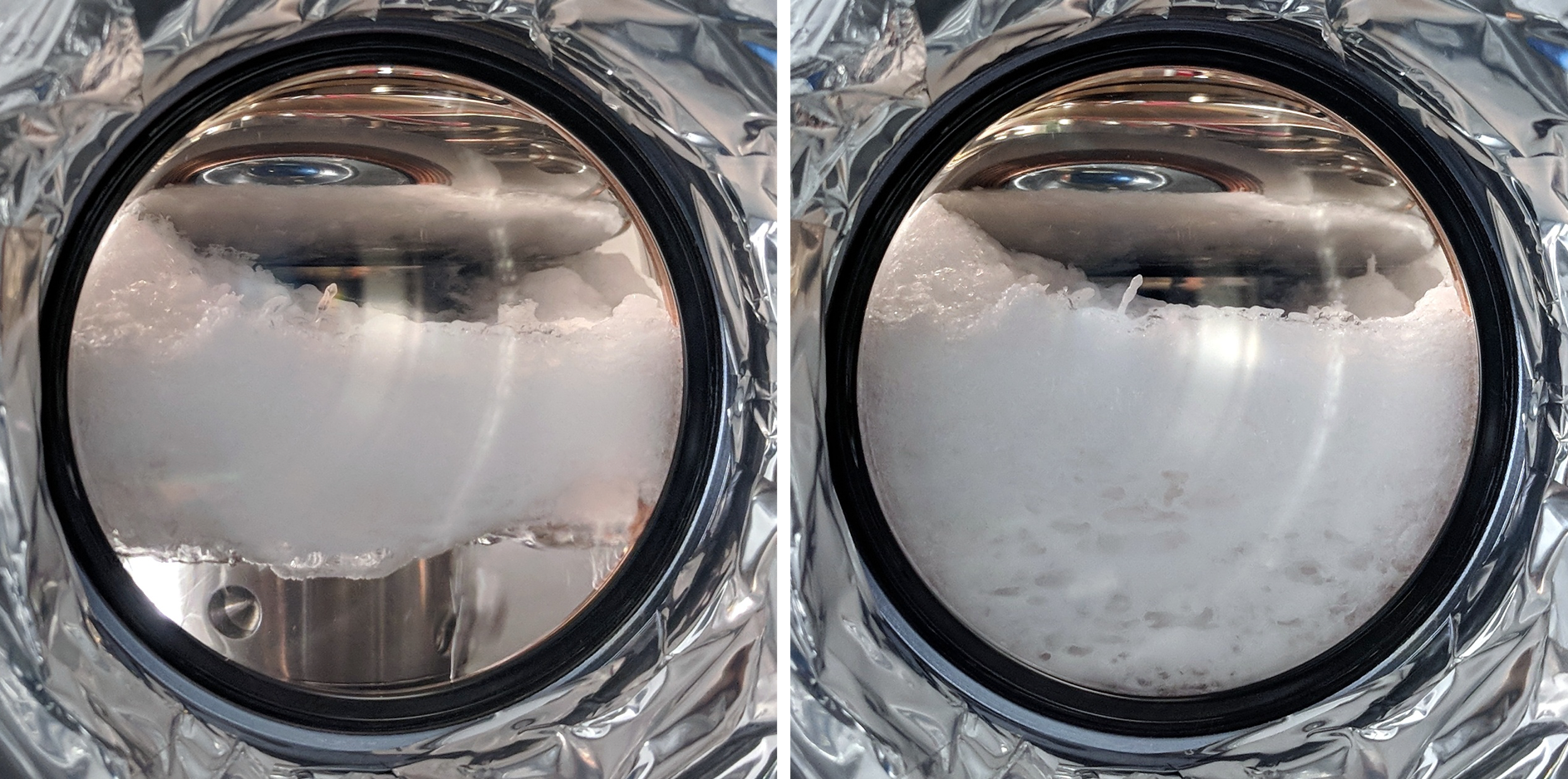}
\par\end{centering}
\caption[Pictures of frozen xenon]{\textbf{Left:} Layer of frozen xenon being formed on top of liquid
xenon. \textbf{Right:} Xenon slushee.\label{fig:xice}}
\end{figure}

During an attempt to slowly warm up these parts of the apparatus,
a weak weld of the heat exchanger burst open, which stopped XeBrA
operations on September 23, 2018. Figure~\ref{fig:xice} shows frozen
xenon that formed in the apparatus during the emergency xenon recovery.
Consultation with Joe Saba, an LBNL engineer, suggested that the operational
parameters of the heat exchanger were incorrectly specified before
manufacturing, which eventually led to a rupture of a weld due to
material fatigue. Plans to fix XeBrA to prepare it for future data
acquisitions are underway.

Some data were collected before the detector stopped working, on September
19-21, 2018. All data were obtained at 2~bar absolute pressure. In
an attempt to obtain a larger sample size to study the spread of the
data, many breakdown voltages were recorded for one fixed electrode
separation. Figure~\ref{fig:lxe-temp-pressure} illustrates the behavior
of pressure and temperature throughout data acquisitions.

To estimate purity of LXe during Run~3, XeBrA was temporarily transformed
into a small TPC during the initial LXe condensation. The two electrodes
were separated by 20~mm, the maximum possible distance allowed by
hardware design. The fill was stopped once the liquid level was 5~mm
below the top electrode. The electrodes could then be biased to create
drift and electroluminescence regions between them. Due to EH\&S regulations,
we were unable to use radioactive sources with short notice, so instead
naturally occurring radioactivity from old thoriated welding rods
was used\footnote{The welding rods are made of 98\% tungsten and 2\% thorium. The thorium
is present in its natural abundance during the rod's fabrication,
but the strongest signal comes from the $\beta$ decay of $^{208}$Tl,
which is near the end of the $^{232}$Th series. $^{208}$Tl has a
half-life of 3.1 minutes and emits a $\gamma$ with 2614 keV, which
is the highest energy strong gamma line from naturally occurring sources.
The welding rods maintain a relatively high concentration of $^{208}$Tl
because $^{228}$Ra cannot escape the rod as it decays $\left(T_{\nicefrac{1}{2}}=\unit[5.7]{years}\right)$
since it is trapped in the solid.}. Several such rods were attached to the outside of the OV, and the
PMT was used to record the data overnight. A fit to the plot of $\log_{10}(\mathrm{S2/10)}$
vs. drift time revealed an electron lifetime of $\mathrm{\unit[2.2]{\mu s}}$.
Using Equation~\ref{eq:lxe-oxygen-equivalen}, this corresponds to
$\sim200$~ppb oxygen equivalent impurities.

\subsection{Review of apparatus parameters}

This section serves as an overview of parameters that pertain to data
collection. 

The voltage delivered by the Spellman power supply was monitored by
the Slow Control every 100~ms, along with pressure measurements  (PT05)
and readings from the picoammeter. Additionally, the signal from a
charge amplifier connected to the ground electrode was recorded by
the DAQ computer. HV breakdown was defined as the voltage where the
current drawn by the power supply exceeded the $\unit[10]{\mu s}$
current limit imposed on the power supply. After a current trip, the
power supply automatically ramps down to 0~V. During data acquisition,
HV was slowly ramped up using an automated system in Slow Control.
This allowed three different ramping speeds as shown in Figure~\ref{fig:Slow-control}.b.
Table~\ref{tab:Summary-of-detector} includes the value of the ramp
parameters used during data acquisitions along with other  parameters
relevant to the analysis. 

\begin{table}
\begin{centering}
\begin{tabular}{lrl}
\hline 
Parameter & Value & Unit\tabularnewline
\hline 
\hline 
Electrode material & 303 SS & \tabularnewline
Cathode surface finish (Ra) & $0.05$ & $\mu$m\tabularnewline
Anode surface finish (Ra) & $0.07$ & $\mu$m\tabularnewline
Power supply limit & -75 & kV\tabularnewline
Trip current & 10 & $\mu$s\tabularnewline
Cable capacitance & 203 & pF\tabularnewline
Ramp speed 1 & 30-60 & kV/min\tabularnewline
Ramp speed 2 & 12 & kV/min\tabularnewline
Ramp speed 3 & 6 & kV/min\tabularnewline
Wait period & 10-20 & s\tabularnewline
\hline 
\end{tabular}
\par\end{centering}
\caption[Summary of apparatus parameters]{Summary of apparatus parameters that pertain to data collection.
Cable capacitance refers to the length of the cable between the oil
box and the apparatus.\label{tab:Summary-of-detector}}
\end{table}

The electrode separation was found by zeroing the caliper on the linear
shifter as the electrodes touched faces. At each separation, dozens
to hundreds of breakdown events were measured. This analysis uses
the 90\% stressed electrode area as estimated using COMSOL Multiphysics
simulations of the system. The values of 90\% stressed area and volume
at different electrode separations are summarized in Table~\ref{tab:Electrode-areas}.

To achieve a comparable  state for measurements at each electrode
separation, usually, data at a given separation were collected during
one day after the apparatus was held stable overnight. This resulted
in more data points taken at smaller separations since it takes less
time to ramp up to lower voltages. The location of the spark on the
electrode appeared to vary. Since the detector was not assembled in
a clean room, a small amount of debris or speckles of dust were visible
on the electrode surfaces, usually near the beginning of each run.
The debris moved away from the electrode surfaces due to liquid circulation,
effects of breakdown, and bubbles.

\begin{table}
\begin{centering}
\begin{tabular}{rrrrrr}
\hline 
Separation & Radius & Area & Volume & $\mathrm{C_{LXe}}$ & $\mathrm{C_{LAr}}$\tabularnewline
{[}mm{]} & {[}mm{]} & $\left[\mathrm{cm^{2}}\right]$  & $\left[\mathrm{cm^{3}}\right]$ & {[}pF{]} & {[}pF{]}\tabularnewline
\hline 
\hline 
1.0 & 18.9 & 11.22 & 1.12 & 18.4 & 14.9\tabularnewline
1.4 & 21.0 & 13.85 & 1.94 & 16.2 & 13.1\tabularnewline
2.0 & 23.3 & 17.06 & 3.41 & 14.0 & 11.3\tabularnewline
3.0 & 26.0 & 21.24 & 6.37 & 11.6 & 9.4\tabularnewline
4.0 & 28.2 & 24.98 & 9.99 & 10.2 & 9.3\tabularnewline
5.0 & 30.2 & 28.65 & 14.33 & 9.4 & 7.6\tabularnewline
6.0 & 32.2 & 32.57 & 19.54 & 8.9 & 7.2\tabularnewline
7.0 & 34.2 & 36.75 & 27.73 & 8.6 & 7.0\tabularnewline
\hline 
\end{tabular}
\par\end{centering}
\caption[Values of 90\% stressed electrode area and volume used in data analysis]{Values of 90\% stressed electrode area and volume as simulated by
COMSOL Multiphysics. Also tabulated is the capacitance caused by LAr
and LXe between the electrodes at various separations. \label{tab:Electrode-areas}}
\end{table}

\section{Sparks}

Before we examine the data in aggregate, it is interesting to investigate
the behavior of individual sparks visually. Several videos of sparks
at different electrode separations were recorded in LAr and LXe with
a point-and-shoot camera and a cell phone. Despite the slow frame
rate of these devices, some videos reveal information about spark
development. 

\subsection{Liquid argon}

\begin{figure}
\begin{centering}
\includegraphics[angle=90,scale=0.8]{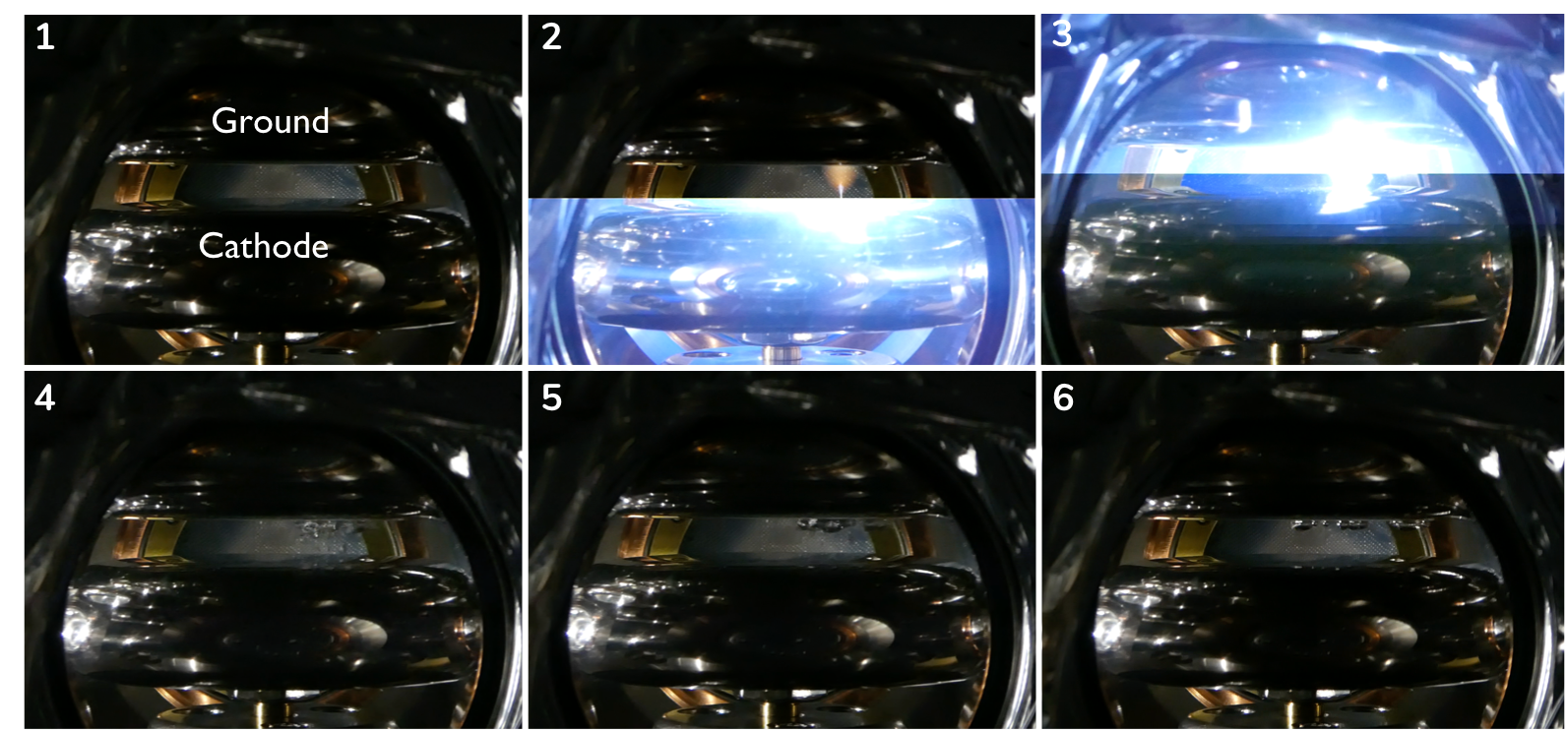}
\par\end{centering}
\caption[Spark development in LAr at 7 mm electrode separation]{Spark development in LAr at 7~mm electrode separation. Frame~2
captures the initial light cone, the subsequent filament, and the
final emission of light. The individual images were not taken at regular
intervals. The sequence was captured by Panasonic Lumix DMC-ZS60 camera.\label{fig:Spark-development-LAr}}
\end{figure}

Breakdown behavior of sparks in LAr has been studied in detail previously
and was published in~\cite{Auger:2015xlo}. The spark registered
in XeBrA shown in Figure~\ref{fig:Spark-development-LAr}, appears
to have very similar behavior. The work of~\cite{Auger:2015xlo}
can be used to explain its evolution.

\begin{figure}
\centering{}\includegraphics[angle=90,scale=0.75]{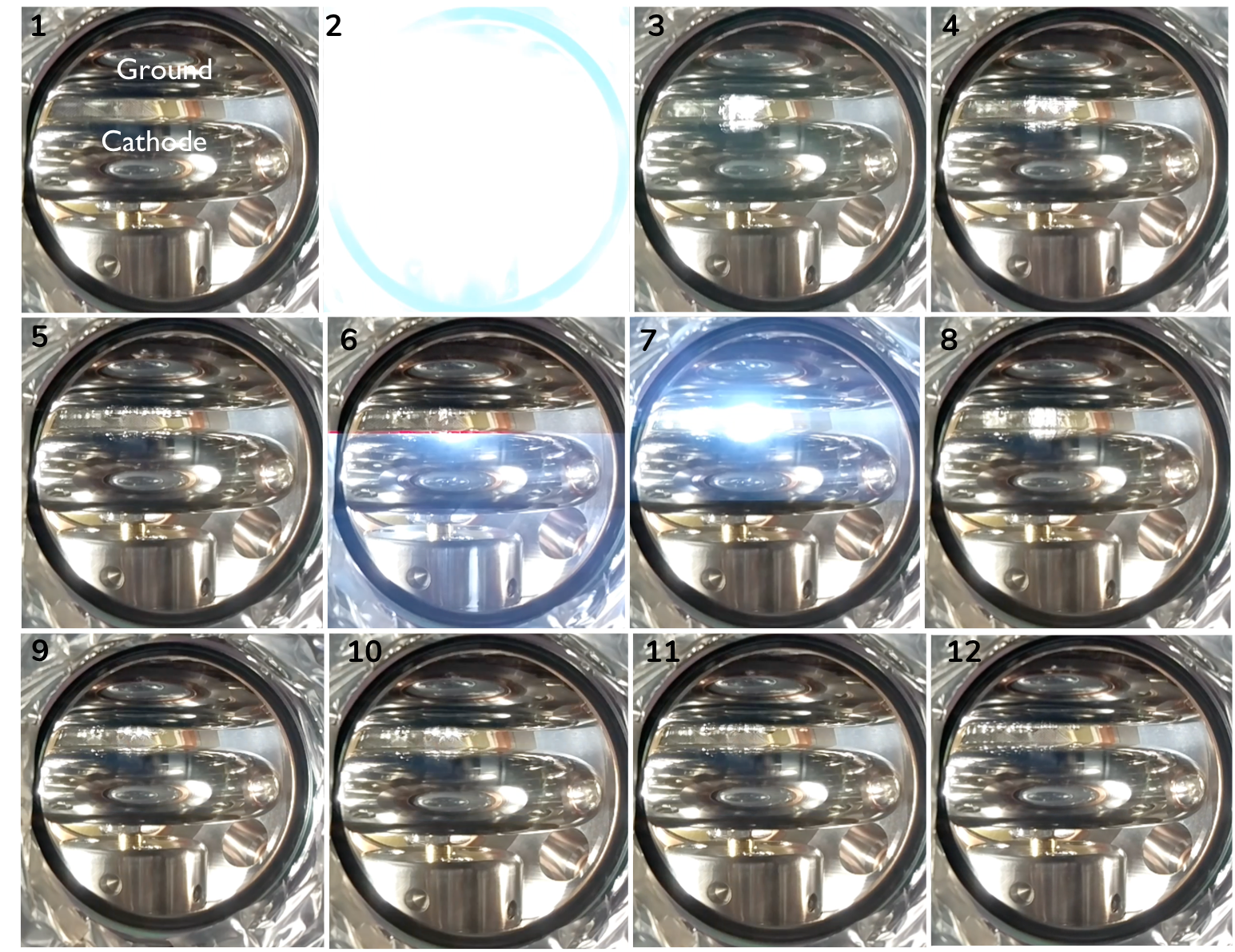}\caption[Spark development in LXe at 5~mm electrode separation]{Step-by-step development of breakdown in LXe at 5~mm electrode separation
with HV electrode (cathode) at $\unit[\sim-53]{kV}$. Two sparks arise
from the breakdown. The first, larger spark (frames~1-5) creates
bubbles that in turn likely cause the second spark (frames~6-12).
The duration of the entire event is $\unit[\sim0.9]{s}$. The second
spark happens 0.6~s after the first spark. After the spark is over,
the bubbles slowly leave the area. Screenshots were taken from a video
recorded by a Pixel XL phone at 60~fps at irregular intervals. The
red line in frame 6 is an artifact from the phone recording rather
than a result of the spark. \label{fig:spark-development-LXe}}
\end{figure}

Frame~2 shows the field emission of electrons that ionize and excite
argon atoms on their way to the anode. This creates the visible light
cone, whose brightness increases from the cathode toward the anode.
Positive ions are produced during this process. Once the positive
ions reach the cathode surface, they raise the local temperature,
and the liquid near the initial discharge point forms a bubble. This
leads to a formation of plasma with an increased conductivity resulting
in a decrease of the electric field near the initial field emission
point. This suppresses further growth of the initial field emission
current, while the electrons of the gas plasma force the bubble to
elongate and grow. The streamer then collapses to its well-known filament
shape. As the streamer propagates toward the anode, it is subject
to thermodynamic fluctuations at its forefront at the liquid-gas boundary,
which results in the random path of the filament. 

When the streamer reaches the anode a short light emission is registered
(frames~2-3). The exact mechanism for this light emission is unknown,
but it might be triggered once the filament loses its thermodynamic
stability and expands into a gas bubble, where an arc discharge forms.
This is accompanied by an acoustic shock and an extensive production
of gas bubbles on the anode near the location of the discharge (frames~4-6)\footnote{Reference~\cite{Auger:2015xlo} notes that this third phase of the
spark is not always observed in their experiment. However, based on
my experience, an audible spark was heard for the majority (if not
all) of the sparks occurring between the two electrodes in XeBrA.}. 

For further details about the development of the spark, along with
a spectral analysis of the light emitted throughout the process and
a well-documented time evolution of each step, consult Reference~\cite{Auger:2015xlo}.

\subsection{Liquid xenon}

Figure~\ref{fig:spark-development-LXe} shows the evolution of a
spark in LXe. The spark is created on the cathode and propagates up
to the grounded anode. A second spark follows shortly, likely caused
by bubbles of the first spark. 

\section{Results\label{sec:Results}}

We now proceed to the analysis of the data. There are multiple sources
of data that might contain interesting information. Here, we focus
only on HV data from the power supply recorded by the Slow Control.
A more detailed analysis that includes data from the picoammeter,
the charge amplifier, and the PMT will follow in the future. 

First, the expected errors of the measurements are discussed in Section~\ref{subsec:Statistical-and-systematic}.
Results from LAr are presented in Section~\ref{subsec:Liquid-argon}
and results from LXe are presented in Section~\ref{subsec:Liquid-xenon}.
Due to XeBrA's unique construction, data from LAr and LXe can be compared
directly. This was done in Section~\ref{subsec:Comparison-of-liquid-data}.
Finally, a fit to Weibull function as predicted by the weak link theory
is attempted in Section~\ref{subsec:Fit-to-weibull}. 

\subsection{Statistical and systematic errors\label{subsec:Statistical-and-systematic}}

The statistical error was low at each separation so systematic errors
dominate the results presented in this work. The statistical error
was calculated as the standard deviation of the mean breakdown value
divided by the square root of the number of breakdowns at each electrode
separation. Several sources of systematic error were considered. A
0.1~mm uncertainty was estimated for the electrode separation based
on the mechanical error of zeroing the electrodes. Additionally, the
effect of the tilt of the electrodes was considered. Based on a perception
study, a 0.3$^{\circ}$ tilt of the electrodes should be visible to
a user. Since there are two uncorrelated axes along which the electrodes
might be tilted (and two perpendicular viewports), this results in
the 0.4$^{\circ}$ systematic error. The effect of this tilt on the
electric field was derived in Table~\ref{tab:Effects-of-tilt} and
is summarized in Table~\ref{fig:errors}. A follow-up study could
use two high-resolution photographs through XeBrA viewports to place
more precise bounds on the alignment error. A 100~V uncertainty on
the breakdown voltage was estimated based on the minimum power supply
voltage increment. 

Errors from pressure fluctuations and conditioning effects were also
taken into account. To estimate the systematic error caused by the
variation of pressure, the data were divided into low- and high-pressure
subsets for each electrode separation. The difference in the average
breakdown value for the lower and higher pressure datasets was then
taken as a systematic error on the breakdown voltage. Similarly, to
account for conditioning effects, the data at each electrode separation
was split into two groups based on the time of data collection. The
difference in breakdown voltage for the earlier and later halves of
data was treated as a systematic error. All the systematic errors
were treated as uncorrected uncertainties in the analysis. For the
illustration of the size of each systematic error and the statistical
error in LAr, the contributions are shown in Table~\ref{fig:errors}. 

The values of both statistical and systematic errors along with the
combined error for LAr and LXe measurements are documented in Tables~\ref{tab:LAr_error-0.5barg},~\ref{tab:LAr_error},
and~\ref{tab:LXe-error}. 

\begin{table}
\begin{centering}
\begin{tabular}{r|>{\raggedleft}p{1.5cm}rr>{\raggedleft}p{1.3cm}>{\raggedleft}p{1.5cm}>{\raggedleft}p{1.7cm}>{\raggedleft}p{1.7cm}}
\hline 
$d$ & $\Delta$ separation & $\Delta$ tilt & $\Delta$ voltage & $\Delta$ pres\-sure & $\Delta$ conditioning & Statistical error & Combined error\tabularnewline
{[}mm{]} & \% & \% & \% & \% & \% & \% & \%\tabularnewline
\hline 
1 & 10 & 5 & 0.9 & 3.8 & 2.1 & 1.2 & 12.1\tabularnewline
2 & 5 & 3 & 0.5 & 4.4 & 2.6 & 2.0 & 8.0\tabularnewline
3 & 3 & 2 & 0.3 & 4.5 & 3.1 & 1.5 & 6.9\tabularnewline
4 & 3 & 1 & 0.2 & 2.1 & 4.8 & 1.9 & 6.2\tabularnewline
5 & 2 & 1 & 0.2 & 2.1 & 4.3 & 2.9 & 6.0\tabularnewline
6 & 2 & 1 & 0.2 & 16.7 & 14.8 & 2.2 & 22.5\tabularnewline
\hline 
\end{tabular}
\par\end{centering}
\caption[Summary of systematic and statistical errors in LAr measurements]{Summary of systematic and statistical errors in liquid argon measurements.
All errors are quoted as a percent error on the breakdown field. \label{fig:errors}}
\end{table}

\subsection{Liquid argon data analysis\label{subsec:Liquid-argon}}

\begin{figure}
\begin{centering}
\includegraphics[scale=0.5]{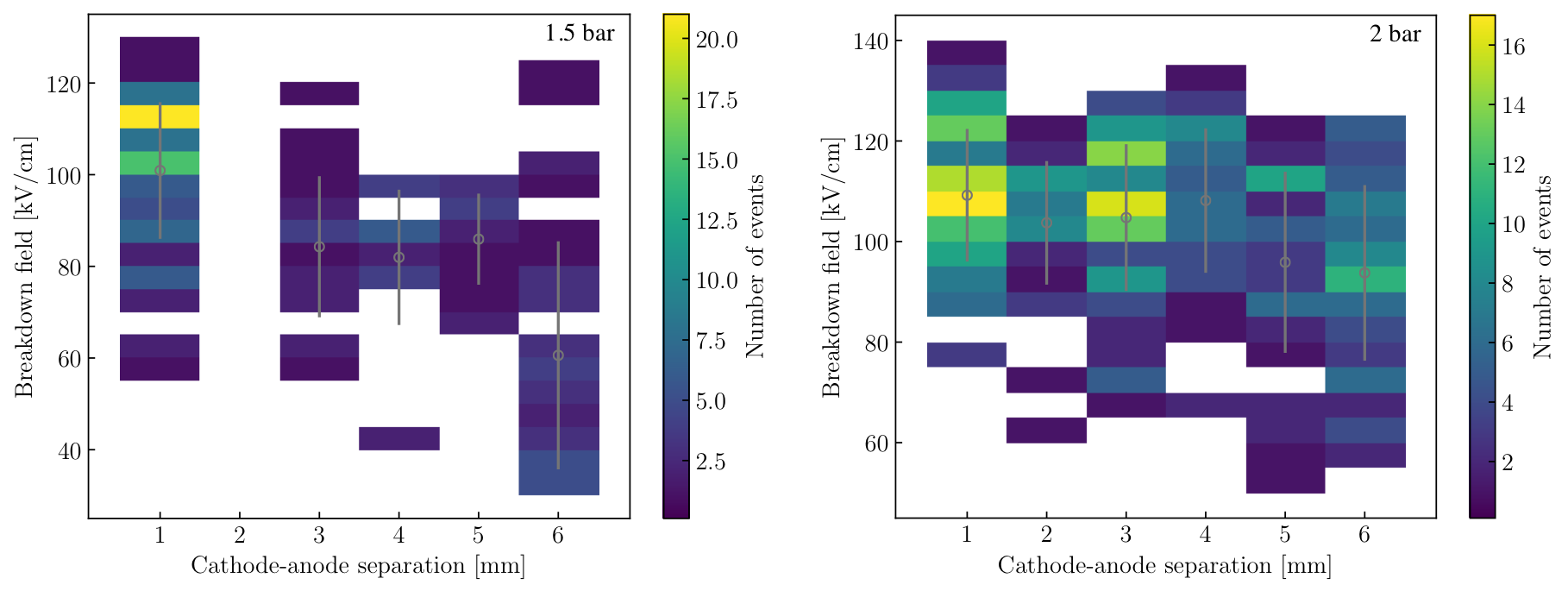}
\par\end{centering}
\caption[Histogram of breakdowns in LAr as a function of electrode separation]{Histogram of breakdowns in LAr as a function of electrode separation.
Data were taken at 1.5~bar (left) and 2~bar (right). The vertical
bars indicate the standard deviation at each separation. \label{fig:lar-distance}}
\end{figure}

Figure~\ref{fig:lar-distance} shows a histogram of breakdown events
recorded in XeBrA for varying electrode separation at 1.5 and 2~bar.
Tables~\ref{tab:LAr_error-0.5barg} and~\ref{tab:LAr_error} summarize
the information about data acquired in LAr. 

\begin{table}[p]
\begin{centering}
\begin{tabular}{r|>{\raggedleft}p{1cm}rr>{\raggedleft}p{1.6cm}>{\raggedleft}p{1.7cm}>{\raggedleft}p{1.6cm}rr}
\hline 
$d$ & \#  & V  & E & Statistical error & Systematic error & Combined error & P  & $\sigma_{\mathrm{P}}$\tabularnewline
{[}mm{]} &  & {[}kV{]} & {[}kV/cm{]} & {[}kV{]} & {[}kV{]} & {[}kV{]} & {[}bar{]} & {[}bar{]}\tabularnewline
\hline 
\hline 
1.0 & 85 & 10.1 & 100.9 & 1.6 & 11.4 & 11.5 & 1.5 & 0.03\tabularnewline
3.0 & 18 & 25.3 & 84.3 & 3.6 & 34.7 & 34.9 & 1.5 & 0.04\tabularnewline
4.0 & 18 & 32.8 & 82.0 & 3.5 & 13.1 & 13.5 & 1.5 & 0.08\tabularnewline
5.0 & 14 & 43.0 & 85.9 & 2.7 & 12.7 & 13.0 & 1.5 & 0.07\tabularnewline
6.0 & 38 & 36.4 & 60.6 & 4.0 & 13.5 & 14.1 & 1.5 & 0.06\tabularnewline
\hline 
\end{tabular}
\par\end{centering}
\caption[Summary of breakdown data collected in LAr at 1.5 bar ]{Summary of breakdown data collected in XeBrA in LAr at 1.5~bar.
Each data point is the mean value of all measurements from one run
taken at the same gap distance. The electrode separation is referred
to as $d$, \# is the number of data points recorded at each separation,
and $\sigma_{\mathrm{P}}$ is the standard deviation of the pressure.
\label{tab:LAr_error-0.5barg} }

\vspace{0.8cm}
\begin{centering}
\begin{tabular}{r|>{\raggedleft}p{1cm}rr>{\raggedleft}p{1.6cm}>{\raggedleft}p{1.7cm}>{\raggedleft}p{1.6cm}rr}
\hline 
$d$ & \#  & V  & E & Statistical error & Systematic error & Combined error & P  & $\sigma_{\mathrm{P}}$\tabularnewline
{[}mm{]} &  & {[}kV{]} & {[}kV/cm{]} & {[}kV{]} & {[}kV{]} & {[}kV{]} & {[}bar{]} & {[}bar{]}\tabularnewline
\hline 
\hline 
1.0 & 104 & 10.9 & 109.2 & 1.3 & 13.1 & 13.2 & 2.0 & 0.05\tabularnewline
2.0 & 35 & 20.7 & 103.7 & 2.1 & 8.1 & 8.3 & 2.0 & 0.06\tabularnewline
3.0 & 91 & 31.4 & 104.8 & 1.5 & 7.0 & 7.2 & 2.0 & 0.07\tabularnewline
4.0 & 47 & 43.3 & 108.2 & 2.1 & 6.4 & 6.7 & 2.0 & 0.06\tabularnewline
5.0 & 41 & 47.9 & 95.9 & 2.8 & 5.0 & 5.8 & 2.0 & 0.06\tabularnewline
6.0 & 73 & 56.3 & 93.8 & 2.0 & 21.0 & 21.1 & 2.0 & 0.07\tabularnewline
\hline 
\end{tabular}
\par\end{centering}
\caption[Summary of breakdown data collected in LAr at 2 bar]{Summary of breakdown data collected in XeBrA in LAr at 2~bar. Each
data point is the mean value of all measurements from one run taken
at the same gap distance. The electrode separation is referred to
as $d$, \# is the number of data points recorded at each separation,
and $\sigma_{\mathrm{P}}$ is the standard deviation of the pressure.
\label{tab:LAr_error} }
\begin{centering}
\vspace{0.8cm}
\begin{tabular}{>{\raggedleft}p{1.1cm}|rrr>{\raggedleft}p{1.6cm}>{\raggedleft}p{1.7cm}>{\raggedleft}p{1.6cm}rr}
\hline 
$d$ {[}mm{]} & \#  & V  & E & Statistical error & Systematic error & Combined error & P  & $\sigma_{\mathrm{P}}$\tabularnewline
Run 2 &  & {[}kV{]} & {[}kV/cm{]} & {[}kV{]} & {[}kV{]} & {[}kV{]} & {[}bar{]} & {[}bar{]}\tabularnewline
\hline 
\hline 
1.0 & 75 & 10.0 & 99.5 & 1.1 & 20.1 & 20.1 & 2.0 & 0.02\tabularnewline
2.0 & 55 & 25.5 & 127.5 & 1.2 & 8.3 & 8.3 & 2.0 & 0.00\tabularnewline
3.0 & 51 & 32.6 & 108.7 & 1.7 & 13.5 & 13.7 & 2.0 & 0.02\tabularnewline
5.0 & 68 & 52.5 & 104.9 & 1.7 & 14.0 & 14.1 & 2.0 & 0.00\tabularnewline
\hline 
Run 3 &  &  &  &  &  &  &  & \tabularnewline
1.0 & 1215 & 9.6 & 95.9 & 0.4 & 18.7 & 18.7 & 2.0 & 0.03\tabularnewline
1.4 & 234 & 15.2 & 108.3 & 1.1 & 25.5 & 25.5 & 2.0 & 0.03\tabularnewline
2.0 & 108 & 26.4 & 131.8 & 1.3 & 16.3 & 16.4 & 2.0 & 0.03\tabularnewline
\hline 
\end{tabular}
\par\end{centering}
\caption[Summary of breakdown data collected in LXe at 2 bar]{Summary of breakdown data collected in XeBrA in LXe at 2~bar. Each
data point is the mean value of all measurements from one run taken
at the same gap distance. The electrode separation is referred to
as $d$, \# is the number of data points recorded at each separation,
and $\sigma_{\mathrm{P}}$ is the standard deviation of the pressure.\label{tab:LXe-error}}
\end{table}

\begin{figure}[p]
\begin{centering}
\includegraphics[scale=0.72]{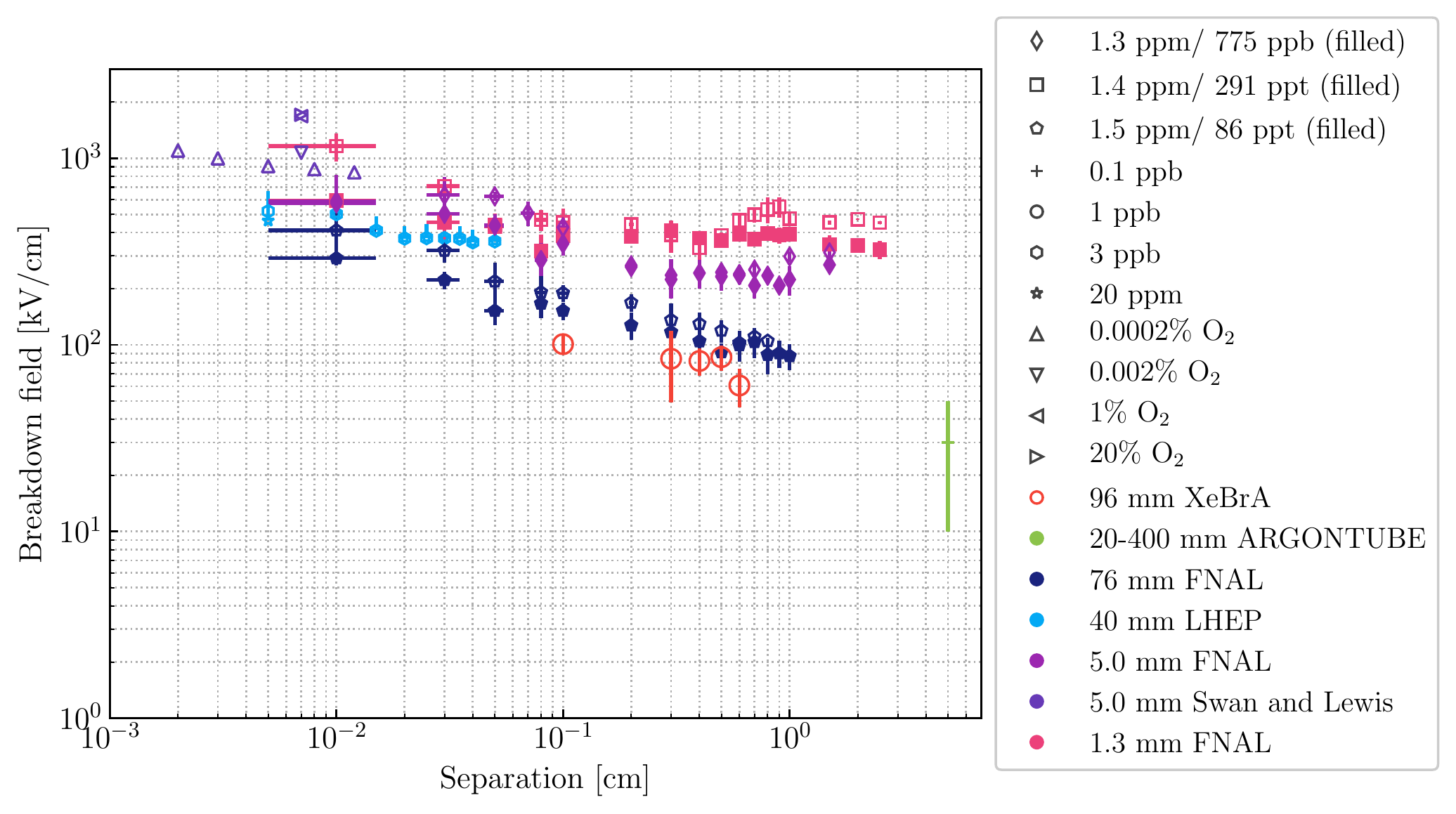}
\par\end{centering}
\caption[A plot of breakdown field vs. electrode separation in LAr]{A plot of breakdown field vs. electrode separation in LAr at $\unit[\sim1.5]{bar}$.
Larger electrodes perform worse, indicating that the breakdown likely
depends on another factor, such as the stressed electrode area. The
different colors indicate the total cathode diameter used in each
experiment, while the various shapes indicate LAr purity. Data from
FNAL were published in~\cite{Acciarri:2014ica}. Measurements from
LHEP were published in~\cite{Auger:2015xlo}. Results from Swan \&
Lewis can be found in~\cite{swan_LAr}. ARGONTUBE result was taken
from~\cite{Argontube}. The pressure at which data points from Swan
\& Lewis and ARGONTUBE were obtained is unknown. \label{fig:lar-field-distance}}

\vspace{0.7cm}
\begin{centering}
\includegraphics[scale=0.8]{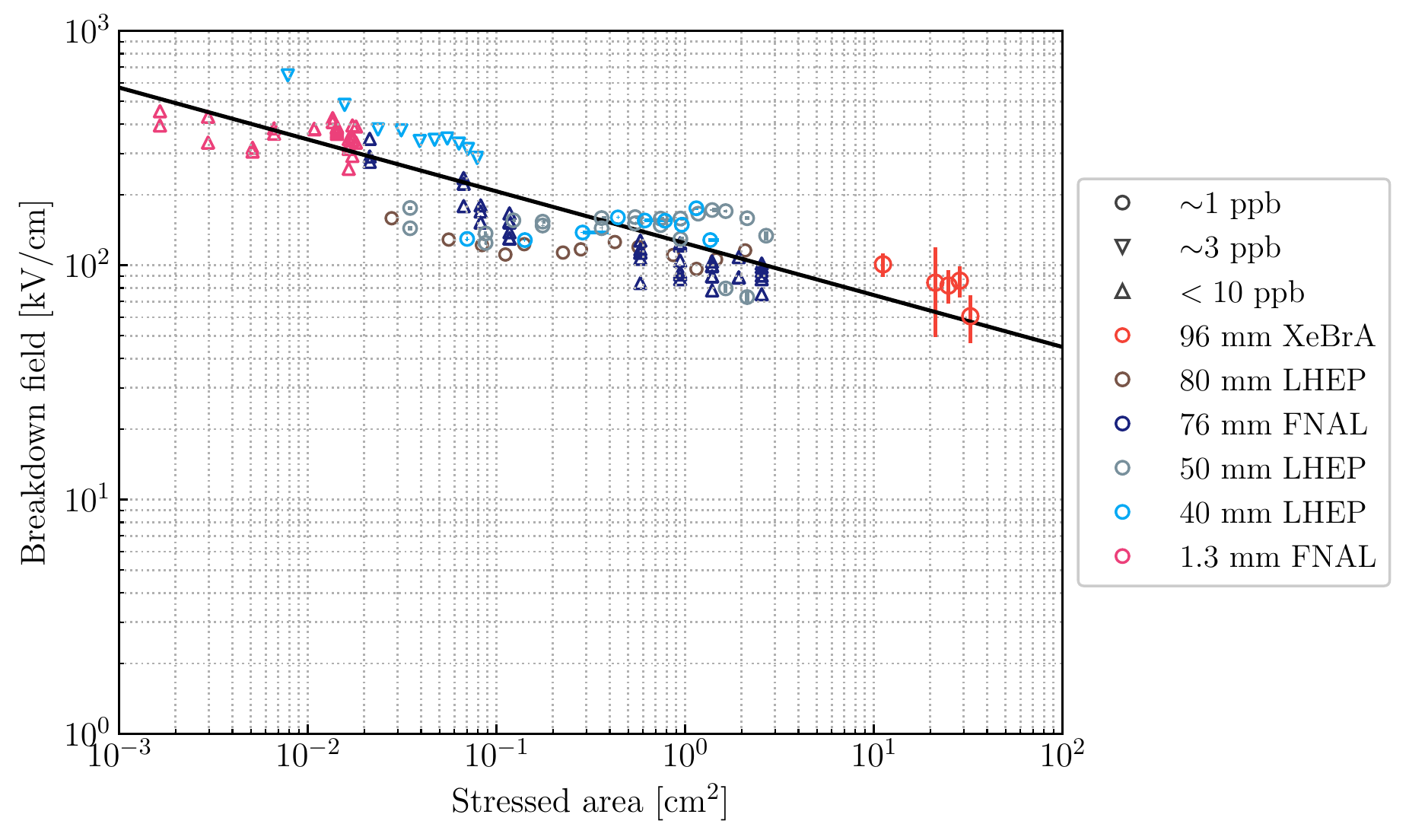}
\par\end{centering}
\caption[A plot of breakdown field vs. 90\% stressed area of the cathode in
LAr]{A plot of breakdown field vs. 90\% stressed area of the cathode in
LAr at $\unit[\sim1.5]{bar}$. The different colors indicate the total
cathode diameter used in each experiment, while the various shapes
indicate LAr purity. The fit line corresponds to $E_{\mathrm{max}}=C\cdot A^{-b}$/cm$^{2}$
where $C=124.26\pm0.09$~kV/cm and $b=0.2214\pm0.0002$ with $\chi^{2}=5\times10^{5}$
and 129 degrees of freedom (DOF)\nomenclature{DOF}{Degrees Of Freedom}.
Data from FNAL were published in~\cite{Acciarri:2014ica}. Measurements
from LHEP were published in~\cite{Auger:2015xlo}. \label{fig:lar-field-area}}
\end{figure}

\begin{figure}[p]
\begin{centering}
\includegraphics[scale=0.85]{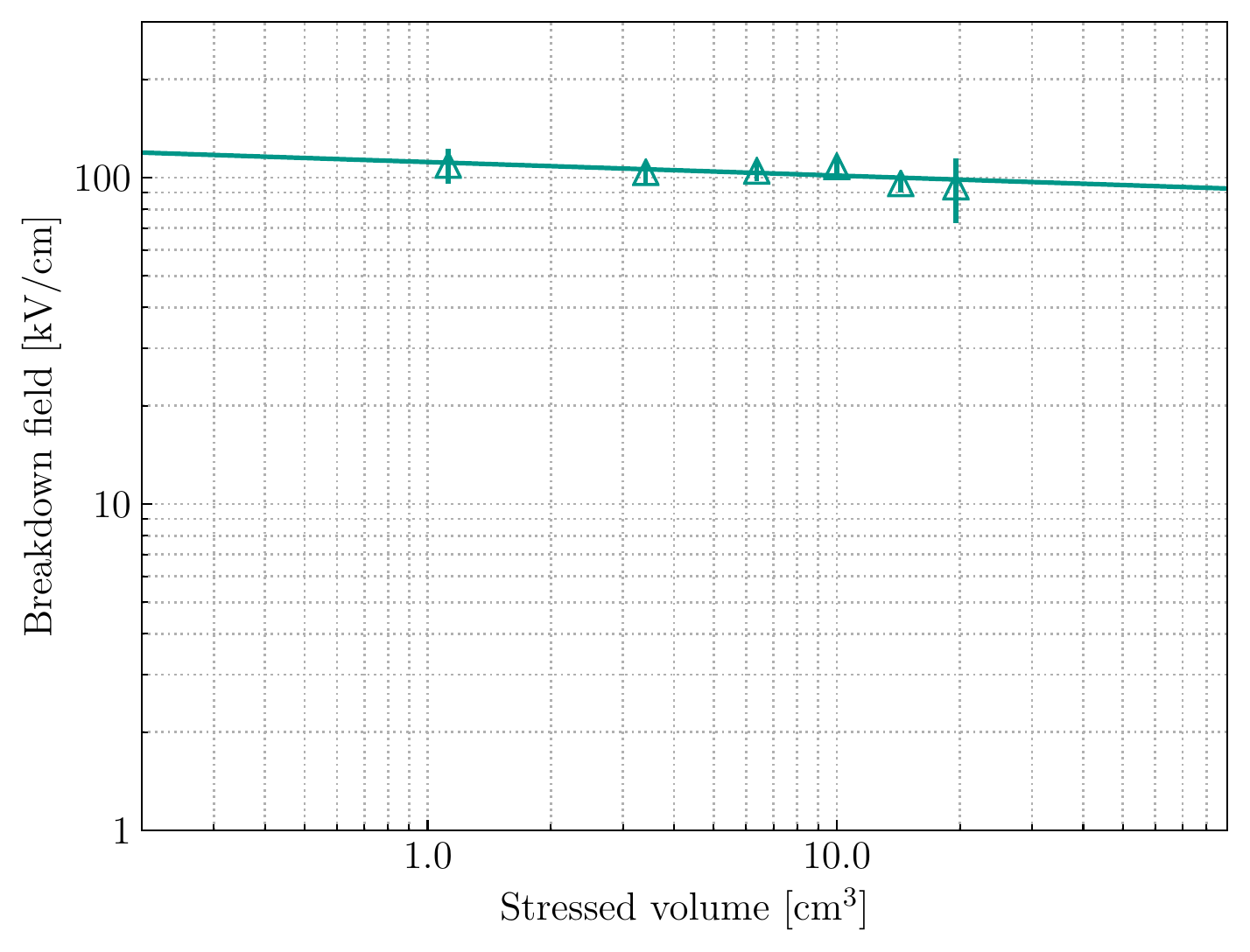}
\par\end{centering}
\caption[A plot of breakdown field vs. 90\% stressed volume in LAr]{A plot of breakdown field vs. 90\% stressed volume in liquid argon
at 2~bar. The fit line corresponds to $E_{\mathrm{max}}=C\cdot V^{-b}$/cm$^{3}$
where $C=112\pm1$~kV/cm and $b=0.041\pm0.043$ with $\chi^{2}=1.7$
and DOF = 3. \label{fig:lar-field-volume}}
\begin{centering}
\vspace{0.7cm}
\includegraphics[scale=0.85]{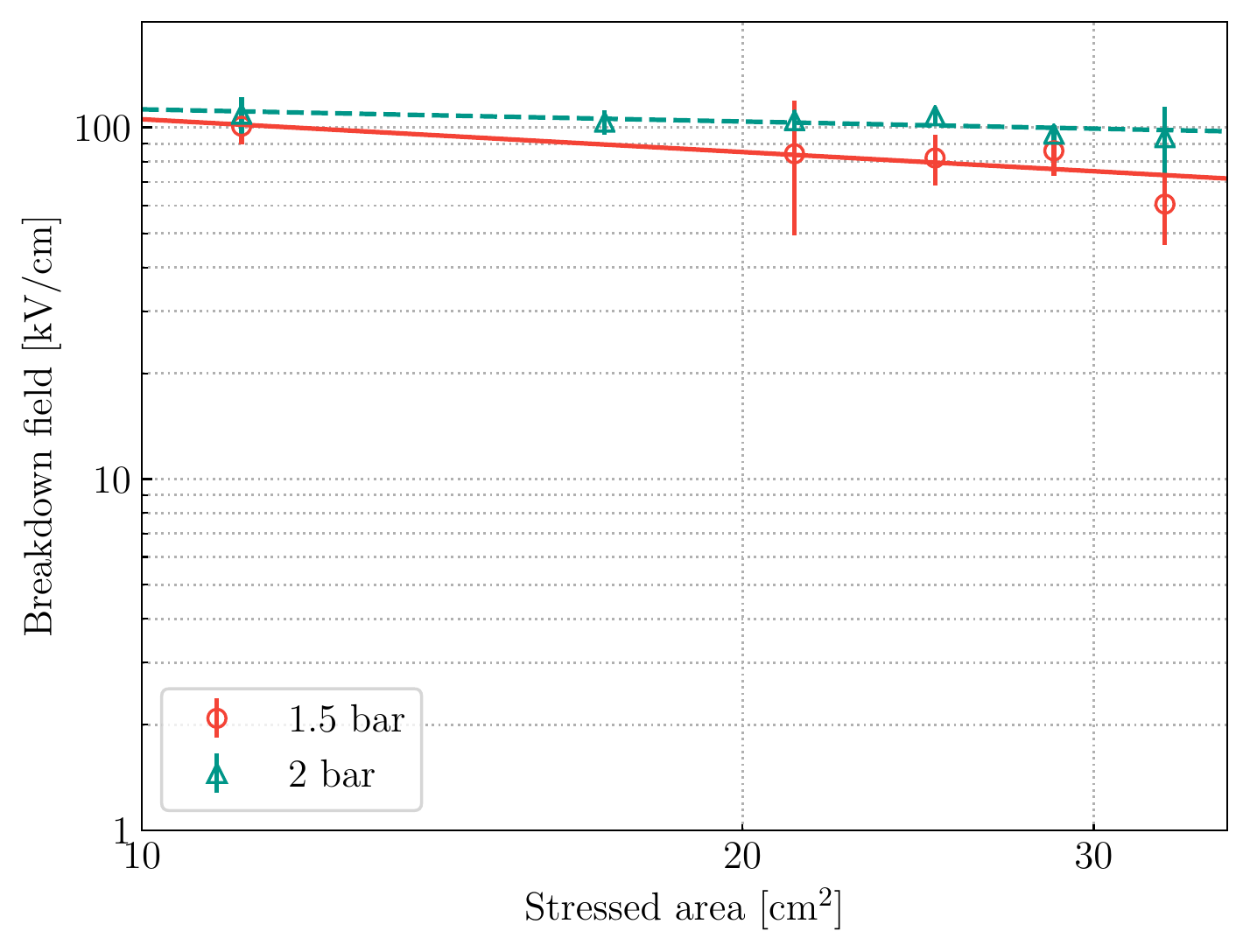}
\par\end{centering}
\caption[Comparison of LAr data taken at 1.5 bar and 2 bar]{Comparison of breakdown performance in LAr for data taken at 1.5
bar and 2 bar as a function of stressed electrode area. Fit at each
pressure was performed to $E_{\mathrm{max}}=C\cdot A^{-b}$/cm$^{2}$.
At 1.5~bar $C=216\pm103$~kV/cm and $b=0.31\pm0.16$ (solid red
line) with $\chi^{2}=1.4$ and DOF = 2, while at 2~bar $C=147\pm54$~kV/cm
and $b=0.11\pm0.12$ (dashed teal line) with $\chi^{2}=1.7$ and DOF
= 3. \label{fig:Lar-field-area-pressure}}
\end{figure}

The characteristic behaviors of dielectric breakdown as a function
of separation, stressed electrode area, and stressed volume are shown
in Figures~\ref{fig:lar-field-distance},~\ref{fig:lar-field-area},
and~\ref{fig:lar-field-volume}, respectively. Figure~\ref{fig:lar-field-distance}
combines experimental data available in the literature with data obtained
in XeBrA at 1.5~bar. 

Each data point in Figure~\ref{fig:lar-field-distance} is the mean
value of all measurements from one run taken at the same gap distance
as summarized in Table~\ref{tab:LAr_error}. Larger electrode diameters
are likely to break down at lower fields. Figure~\ref{fig:lar-field-area}
shows the breakdown field as a function of 90\% stressed cathode area
using data with a purity better than 10~ppb as available in the literature.
A fit to Equation~\ref{eq:area-effect-fit} $E_{max}=C\times A^{-b}$/cm$^{2}$
was performed where $C=124.26\pm0.09$~kV/cm and $b=0.2214\pm0.0002$.
Figure~\ref{fig:lar-field-volume} shows a fit to the volume effect
using data collected in XeBrA at 2~bar alone since data from other
experiments are not available in the literature. A fit to $E_{max}=C\cdot V^{-b}$/cm$^{3}$
was performed where $C=112\pm1$~kV/cm and $b=0.041\pm0.043$. 

Since data in LAr in XeBrA were collected at two different pressures,
Figure~\ref{fig:Lar-field-area-pressure} compares the breakdown
performance at 1.5~bar and 2~bar. Data at each pressure were fit
to Equation~\ref{eq:area-effect-fit}. At 1.5~bar the slope was
$b=0.31\pm0.16$ while at 2~bar $b=0.11\pm0.12$, so the measured
slope of the breakdown dependence on the area at 1.5~bar is steeper
than for 2~bar. Note that temperature inside the apparatus was changing
throughout the measurements as illustrated in Figure~\ref{fig:lar-temp-pressure}.

\subsection{Liquid xenon data analysis\label{subsec:Liquid-xenon}}

\begin{figure}
\begin{centering}
\includegraphics[scale=0.51]{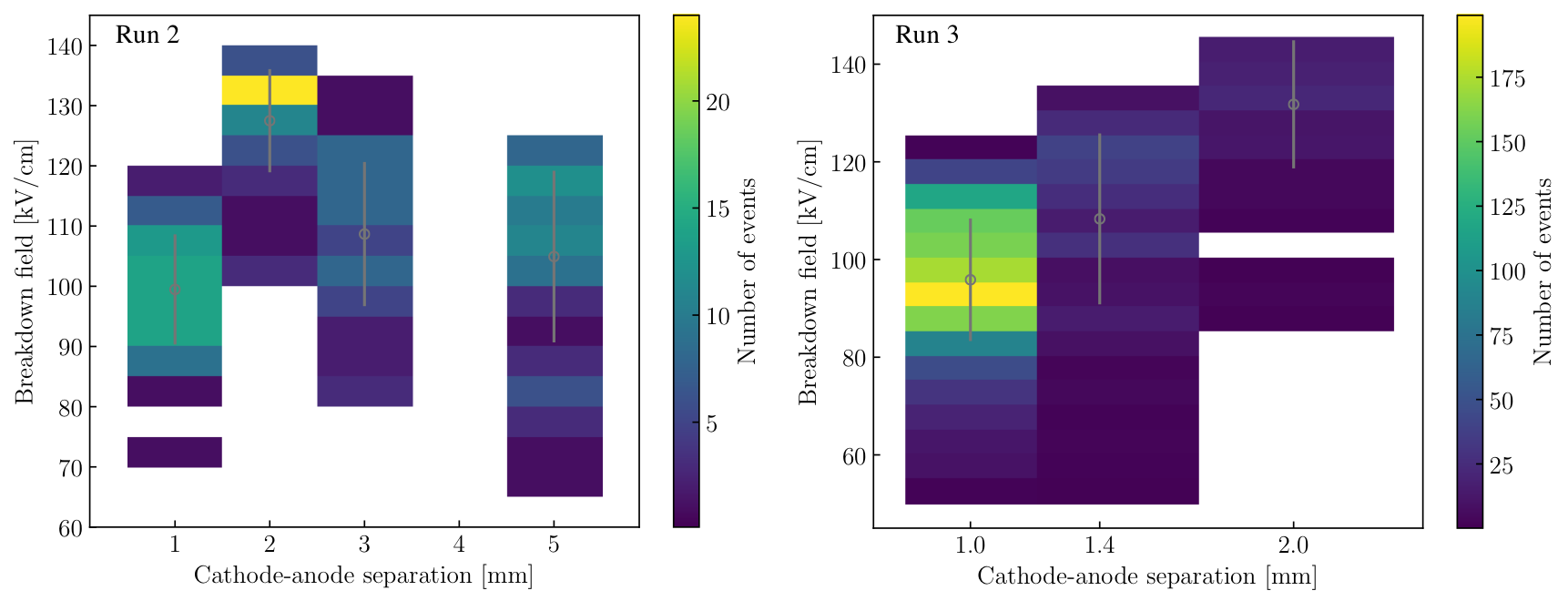}
\par\end{centering}
\caption[Histogram of breakdowns in LXe as a function of electrode separation]{Histogram of breakdowns in LXe as a function of electrode separation.
Data were taken at 1.5~bar (left) and 2~bar (right). The vertical
bars indicated the standard deviation at each separation. \label{fig:lxe-distance}}
\end{figure}

Figure~\ref{fig:lxe-distance} shows a histogram of breakdown events
for each electrode separation during Runs~2 and 3. Table~\ref{tab:LXe-error}
summarizes information about data acquired in LXe. 

\begin{figure}[p]
\begin{centering}
\includegraphics[scale=0.82]{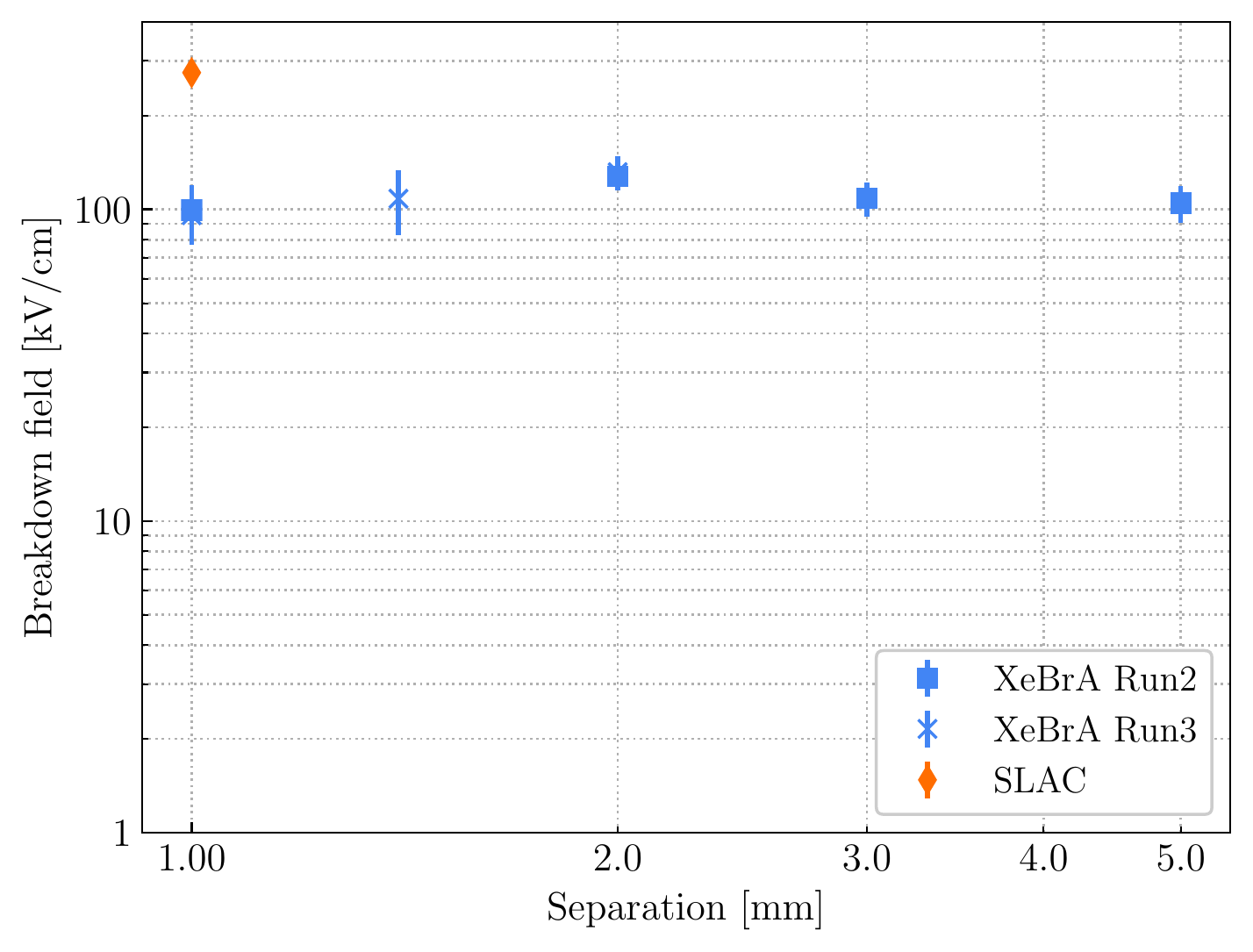}
\par\end{centering}
\caption[A plot of breakdown field vs. electrode separation in LXe]{A plot of breakdown field vs. electrode separation in LXe. XeBrA's
large electrodes perform worse than the 1.5~cm diameter spheres employed
in the SLAC measurement, indicating that the breakdown likely depends
on another factor, such as the stressed electrode area. The SLAC result
is taken from~\cite{Rebel:2014uia}. \label{fig:lxe-field-separation}}
\begin{centering}
\vspace{0.7cm}
\includegraphics[scale=0.82]{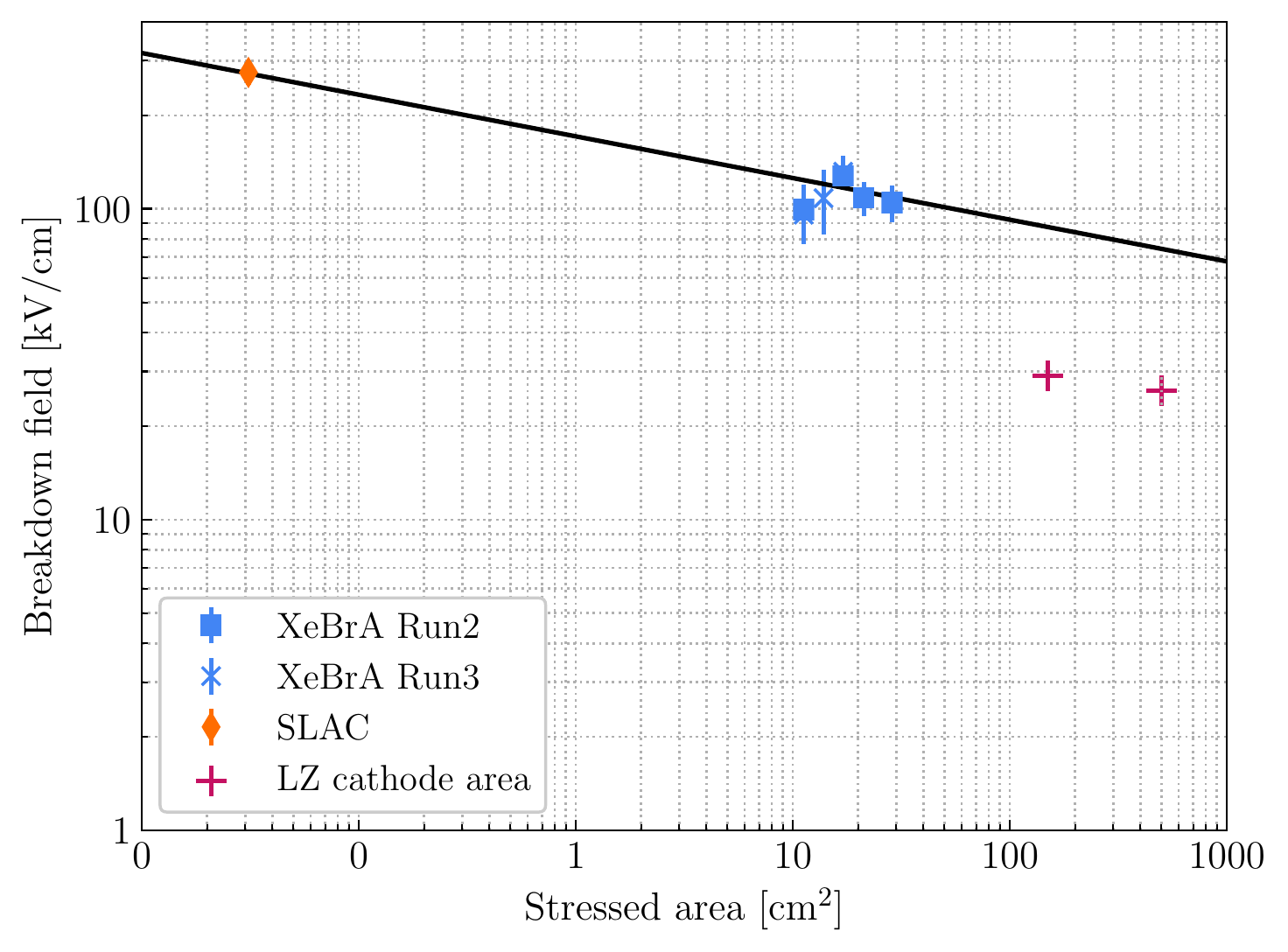}
\par\end{centering}
\caption[A plot of breakdown field vs. 90\% stressed area of the cathode in
LXe]{A plot of breakdown field vs. 90\% stressed area of the cathode in
LXe. The fit line corresponds to $E_{\mathrm{max}}=C\cdot A^{-b}$/cm$^{2}$
where $C=171\pm8$~kV/cm and $b=0.13\pm0.02$ with $\chi^{2}=6.5$
and DOF = 5. The SLAC result is taken from~\cite{Rebel:2014uia}.
Also shown (not a measurement) are the expected area and the peak
electric field in LZ for the largest surface cathodic surface (the
cathode ring) and the electrode with the highest field, for a cathode
voltage of -100~kV.\label{fig:lxe-field-area}}
\end{figure}

\begin{figure}
\begin{centering}
\includegraphics[scale=0.85]{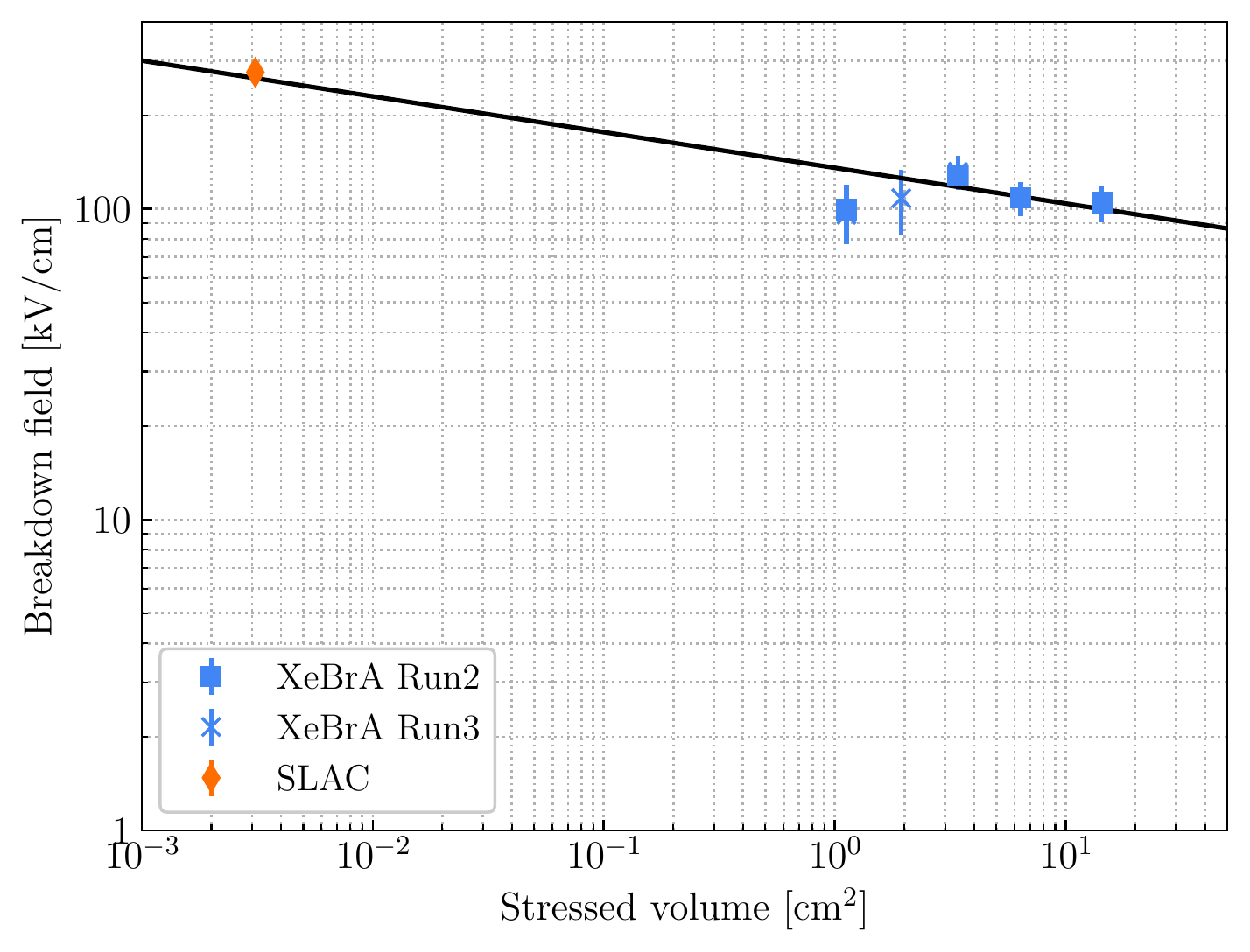}
\par\end{centering}
\caption[A plot of breakdown field vs. 90\% stressed volume in LXe]{A plot of breakdown field vs. 90\% stressed volume in liquid xenon.
The fit line corresponds to $E_{\mathrm{max}}=C\cdot V^{-b}$/cm$^{3}$
where $C=136\pm6$~kV/cm and $b=0.11\pm0.01$ with $\chi^{2}=10.0$
and DOF = 5. The SLAC result is taken from~\cite{Rebel:2014uia}.
\label{fig:lxe-field-volume}}
\end{figure}

The data collected in Run~2 and Run~3 did not exhibit dependence
on purity, so both acquisitions are included in the following analysis.
The characteristic behaviors of dielectric breakdown as a function
of separation, stressed electrode area, and stressed volume are shown
in Figures~\ref{fig:lxe-field-separation},~\ref{fig:lxe-field-area},
and~\ref{fig:lxe-field-volume}, respectively. Figure~\ref{fig:lxe-field-separation}
shows that smaller electrodes perform better at equal separation in
LXe. Figure~\ref{fig:lxe-field-area} shows the breakdown field as
a function of 90\% stressed cathode area. A fit to Equation~\ref{eq:area-effect-fit}
$E_{max}=C\times A^{-b}$/cm$^{2}$ was performed where $C=171\pm8$~kV/cm
and $b=0.13\pm0.02$. Figure~\ref{fig:lxe-field-volume} shows a
fit of the volume effect to $E_{\mathrm{max}}=C\cdot V^{-b}$/cm$^{3}$
where $C=136\pm6$~kV/cm and $b=0.11\pm0.01$. 

\subsection{Conditioning}

As mentioned in Section~\ref{sec:Breakdown-studies-primer}, surface
conditioning might be present during breakdown studies. Conditioning
is frequently used in vacuum applications to increase the breakdown
voltage of surfaces. Reference~\cite{ballat} saw that the breakdown
voltage between the open contacts of vacuum circuit breakers immediately
after manufacturing is low and shows high scatter, an effect that
goes away after conditioning. Note that conditioning process will
only dominate as long as the spark energy $\left(E=\frac{1}{2}CV^{2}\right)$
is low enough not to lead to pronounced conditioning damage right
at the first, low breakdown value~\cite{7464486}.

\begin{figure}[p]
\begin{centering}
\includegraphics[scale=0.52]{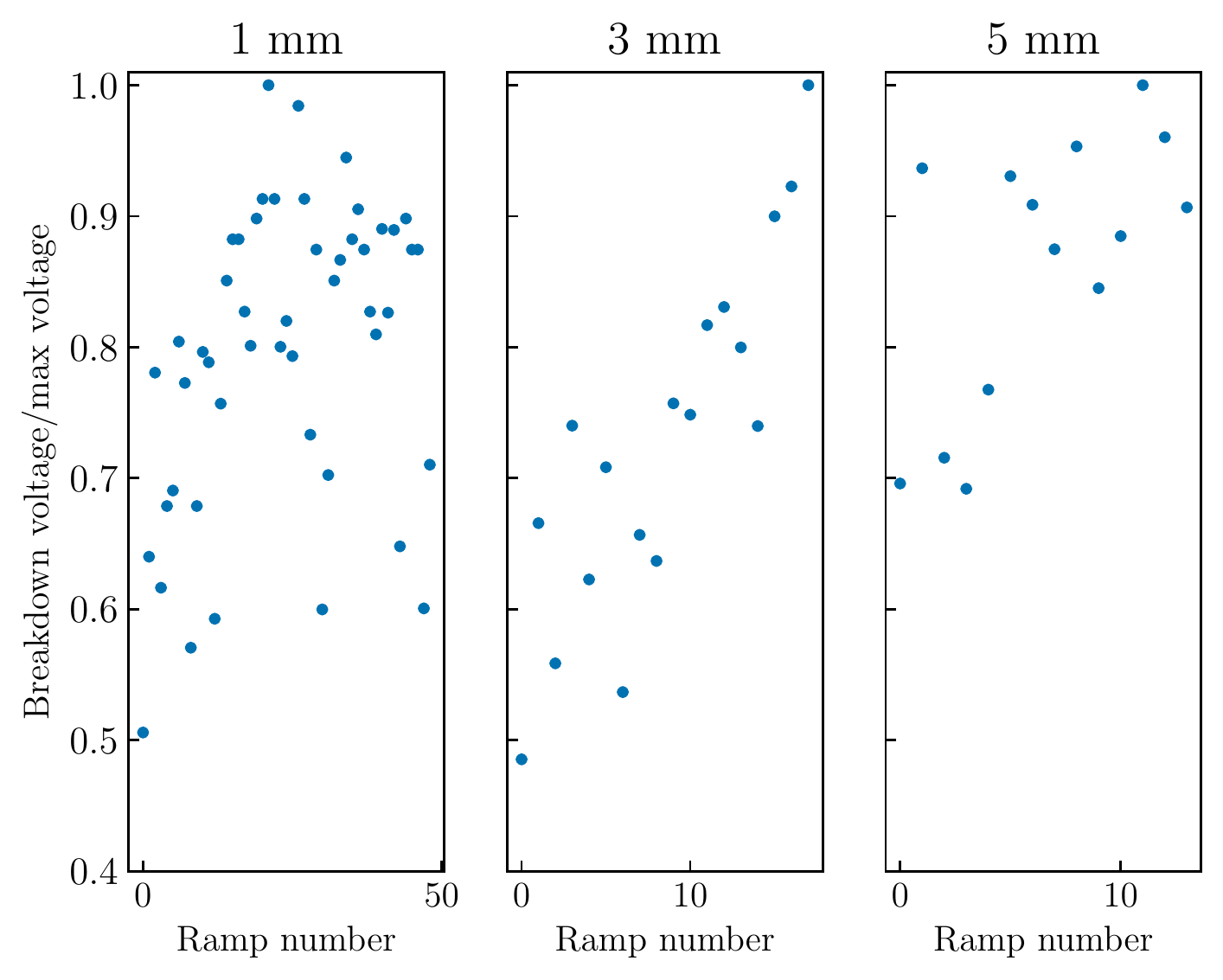}\includegraphics[scale=0.52]{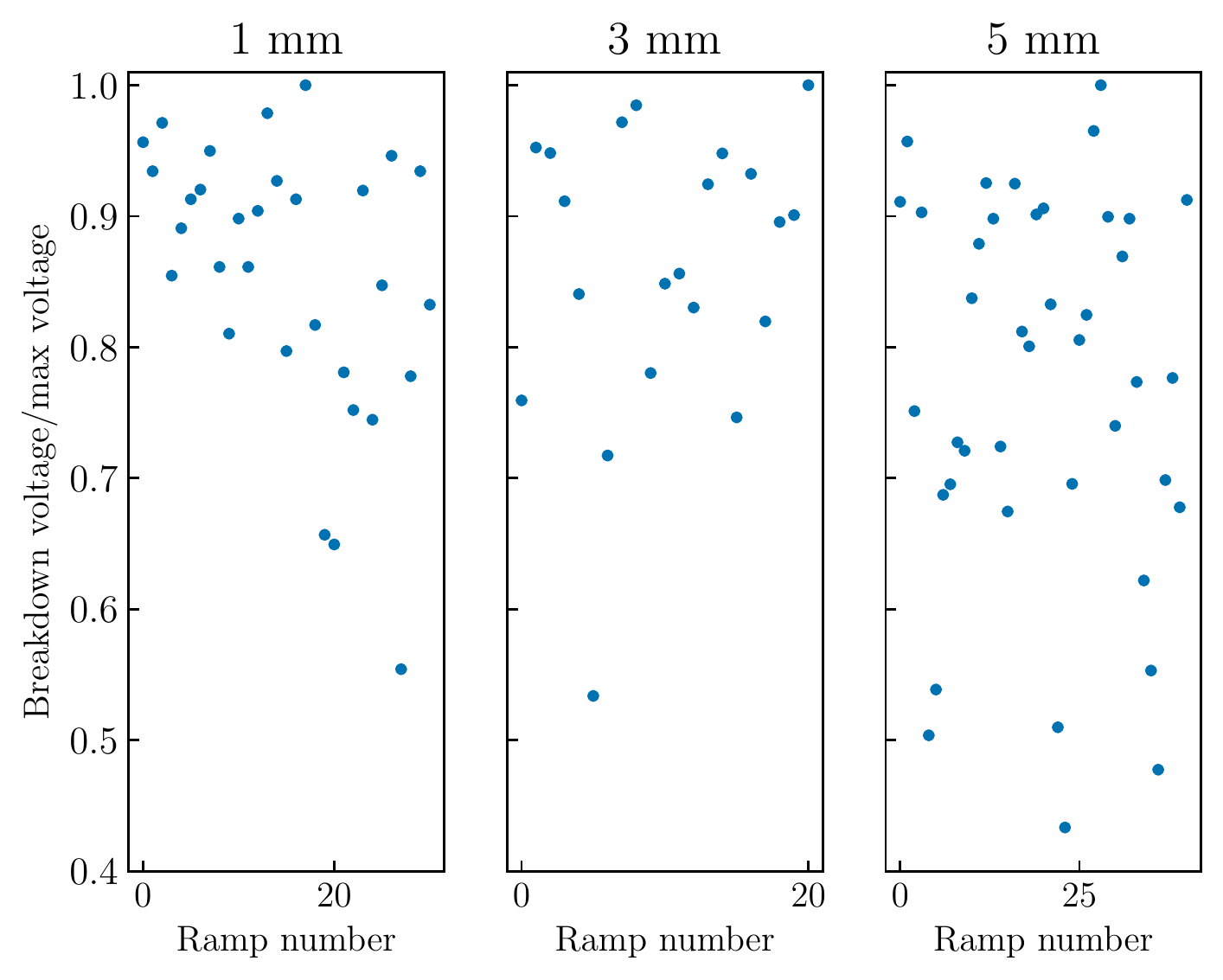}
\par\end{centering}
\caption[Normalized breakdown value as a function of ramp number in LAr]{Normalized breakdown value as a function of ramp number in LAr for
three different electrode separations. Data were normalized to the
maximum breakdown voltage at each separation. \textbf{Left:}~Data
collected in the morning of May 26, 29, and 31, 2018, at 1.5~bar.
\textbf{Right:}~Data collection starting $\sim0.5$ hour after the
end of the data on the left. Data were collected at 2~bar.\label{fig:Conditioning-lar}}

\vspace{0.8cm}
\begin{centering}
\includegraphics{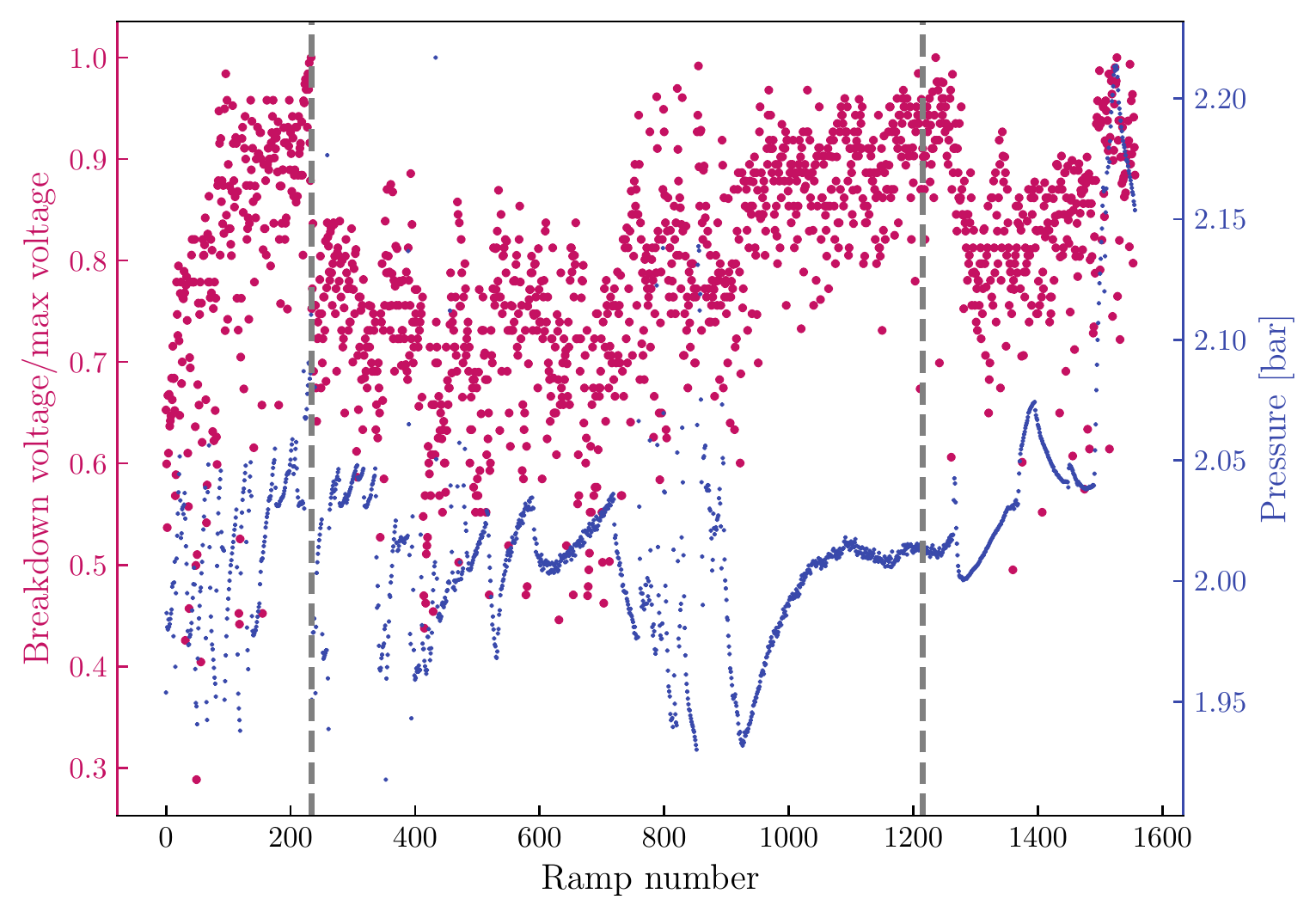}
\par\end{centering}
\caption[Normalized breakdown value as a function of of ramp number in LXe
in Run~3]{Normalized breakdown value as a function of ramp number in LXe in
Run~3. Data were collected at three different electrode separations
indicated by the gray dashed lines; 1.4~mm (left), 1~mm (center),
and 2~mm (right). Data at each separation were recorded without interruption.
There was a half-day break between data acquisitions at 1.4~mm and
1~mm. Data collection at 2~mm immediately followed the 1~mm dataset
as shown in Figure~\ref{fig:lxe-temp-pressure}. \label{fig:Conditioning-lxe}}
\end{figure}

To observe whether any conditioning effect was present in studies
with XeBrA, the successive breakdowns at three different electrode
separations were plotted in Figure~\ref{fig:Conditioning-lar} for
the morning (1.5~bar) and afternoon (2~bar) data acquisitions that
were separated by $\sim0.5$~hour. Due to small sample size, it is
hard to draw any conclusions from the data.

More data were available in LXe during Run~3 as shown in Figure~\ref{fig:Conditioning-lxe}
for three different electrode separations. At each separation, the
breakdown voltage was normalized to the maximum breakdown value at
each separation. There was a half-day break between data acquisitions
at 1.4~mm and 1~mm. Data collection at 2~mm immediately followed
the 1~mm dataset as illustrated in Figure~\ref{fig:lxe-temp-pressure}.
The breakdown value appears to improve throughout data acquisition
at 1.4~mm and 1~mm. 

\subsection{Comparison of liquid argon and liquid xenon data\label{subsec:Comparison-of-liquid-data}}

\begin{figure}[p]
\begin{centering}
\includegraphics[scale=0.85]{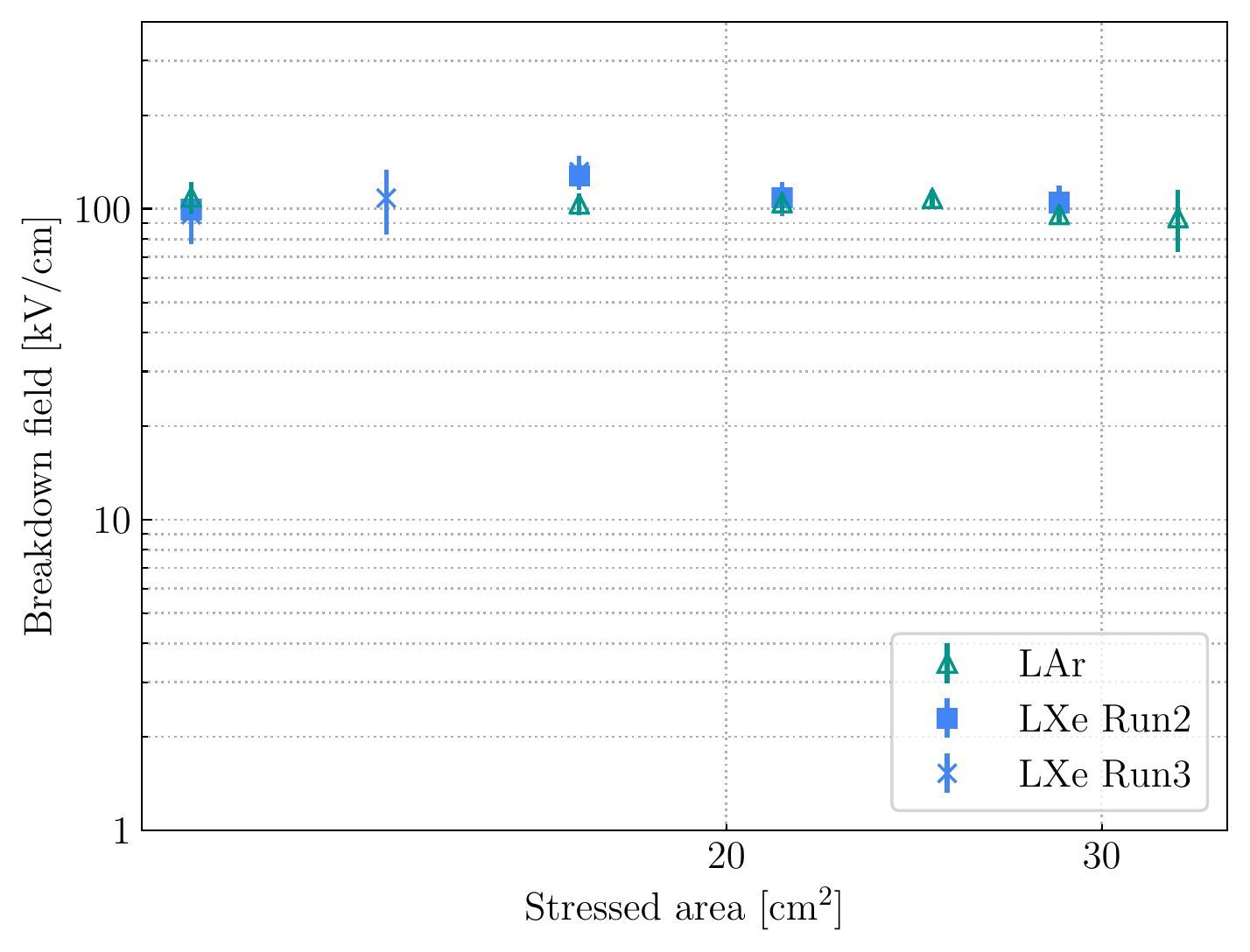}
\par\end{centering}
\caption[Comparison of LAr and LXe data acquired in XeBrA]{Comparison of LAr and LXe data acquired in XeBrA at 2~bar. \label{fig:Comparison-lar-lxe}}
\begin{centering}
\vspace{0.8cm}
\includegraphics[scale=0.78]{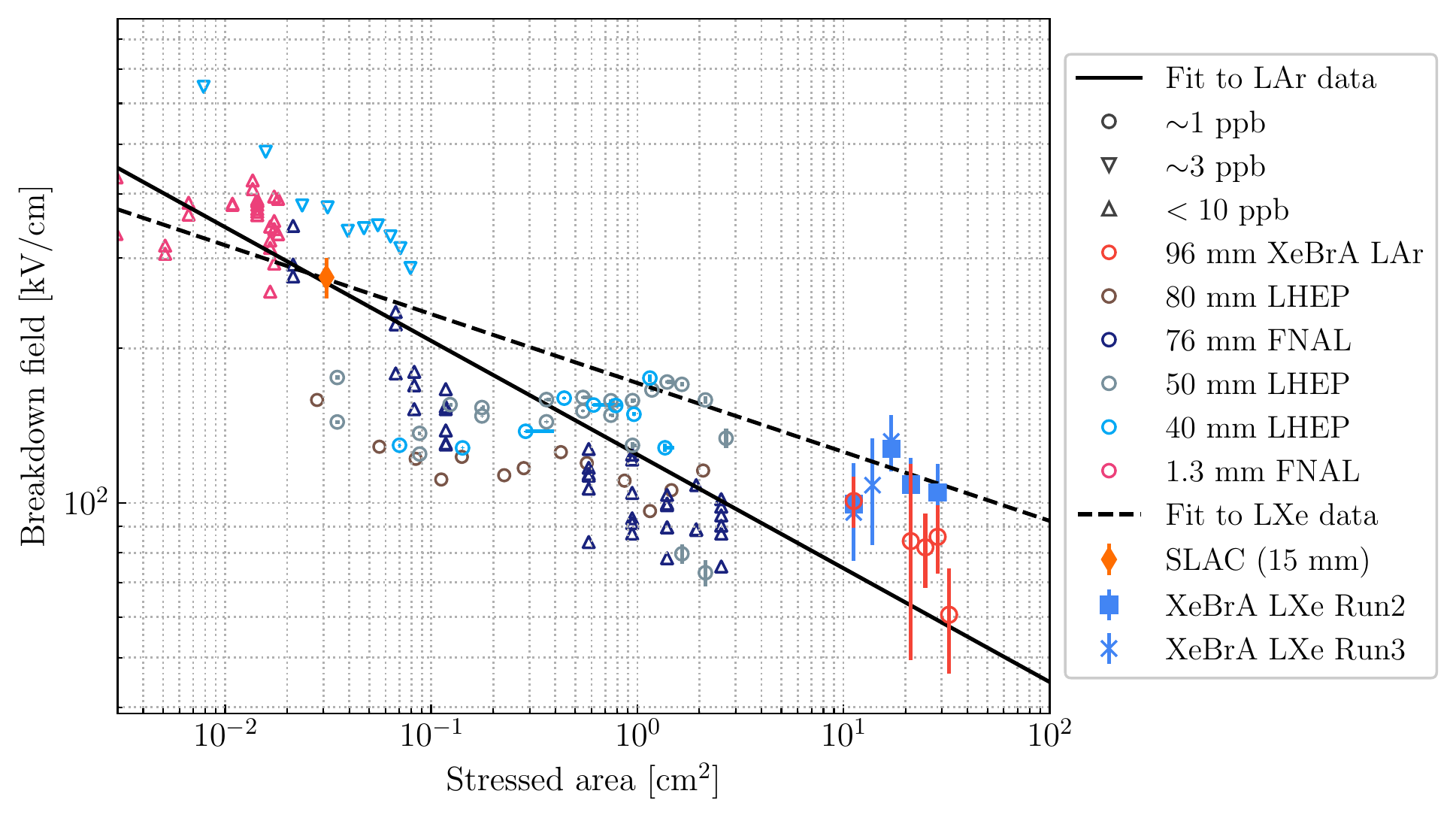}
\par\end{centering}
\caption[Comparison of breakdown field vs. stressed electrode area in LAr and
LXe]{Comparison of breakdown field vs. 90\% stressed area of the cathode
in liquid argon (empty markers) and liquid xenon (solid markers).
The different colors indicate the total cathode diameter used in each
experiment in LAr, while the various shapes indicate LAr purity. Both
LAr and LXe data are fitted to $E_{\mathrm{max}}=C\cdot A^{-b}$/cm$^{2}$.
As stated previously, for LAr $C=124.26\pm0.09$~kV/cm and $b=0.2214\pm0.0002$,
while for LXe $C=171\pm8$~kV/cm and $b=0.13\pm0.02$. Note that
the plot combines data at 1.5~bar in LAr with 2~bar data from LXe.
\label{fig:Comparison-lar-lxe-allData}}
\end{figure}

The unique construction of XeBrA enables direct comparison of data
acquired in LXe and LAr with the same geometry and electrodes. Figure~\ref{fig:Comparison-lar-lxe}
shows all data acquired in XeBrA in LXe and LAr as a function of stressed
electrode area. There does not appear to be a significant difference
between breakdown behavior in LXe and LAr. The literature suggests
that electrode surface is one of the major factors influencing breakdown
behavior, which might be supported by this result.

Figure~\ref{fig:Comparison-lar-lxe-allData} combines data acquired
in XeBrA with other data available in the literature in an attempt
to identify a trend in the breakdown behavior. Fits to Equation~\ref{eq:area-effect-fit}
as obtained in Sections~\ref{subsec:Liquid-argon} and~\ref{subsec:Liquid-xenon}
are also included. Comparison of data from the literature seems to
suggest a steeper slope of the breakdown field with respect to SEA
in LAr than for LXe. However, due to a lack of data in LXe at a wide
range of areas, this comparison should be taken with caution.

\subsection{Fitting breakdown data to a Weibull function\label{subsec:Fit-to-weibull}}

\begin{figure}[t]
\begin{centering}
\includegraphics[scale=0.9]{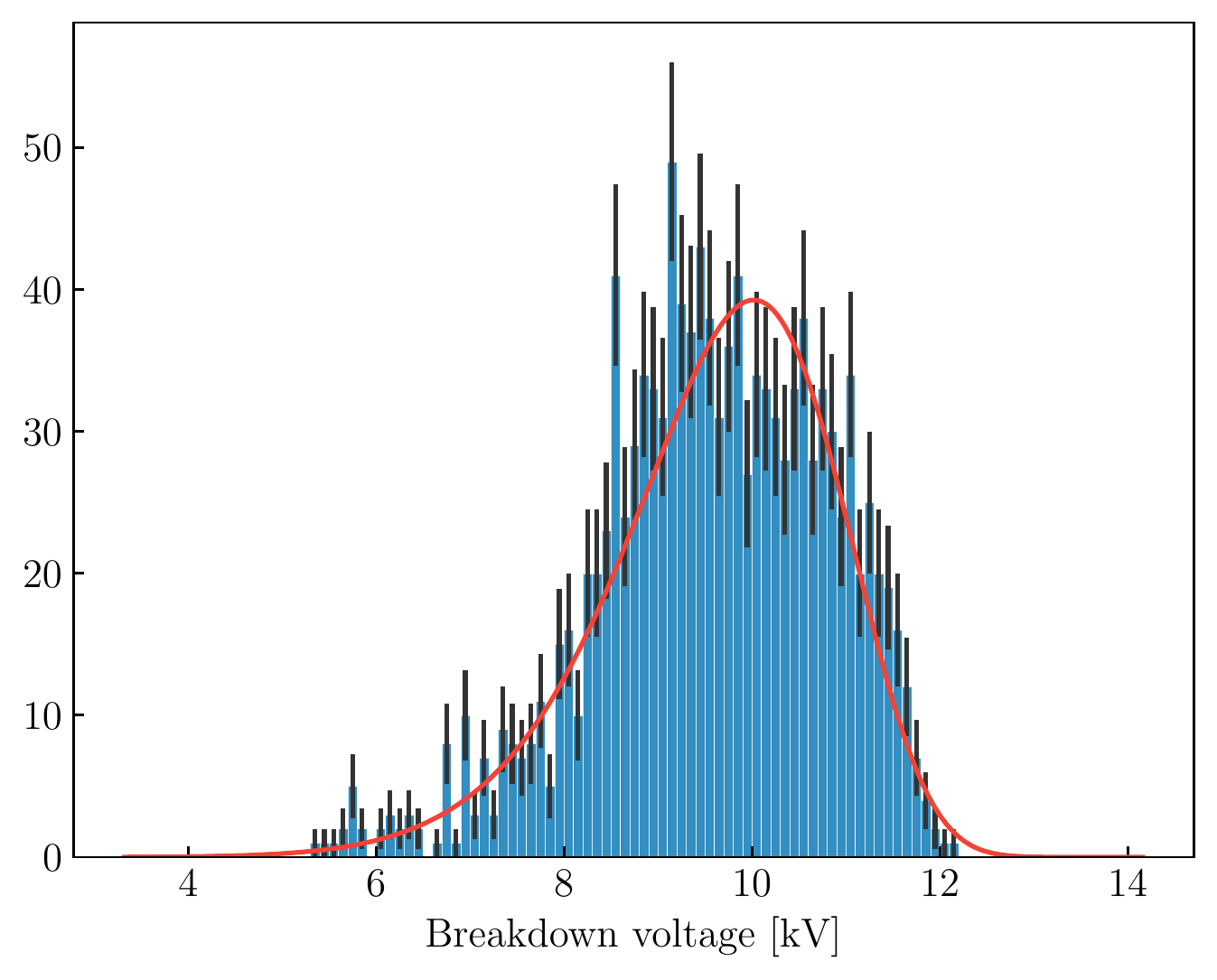}
\par\end{centering}
\caption[Histogram of breakdown values for 1 mm electrode separation in LXe]{Histogram of breakdown values for 1 mm electrode separation in LXe.
The solid red line shows a fit to the Weibull function with $k=9.6\pm0.2$
and $\lambda=10.13\pm0.03$. The $\chi^{2}$ value of the fit is 99.8,
with 63 degrees of freedom. \label{fig:weibull-fit-lxe}}
\end{figure}

Assuming the power law expressed in Equation~\ref{eq:weibull-prediction}
holds in this application, fitting a Weibull distribution to breakdown
distributions at various areas should yield an agreeable description
of breakdown behavior $E_{max}$. Since most data at each separation
was acquired in Run~3 in LXe, these data are used for this test of
the weak link theory. The 2-parameter Weibull function 
\begin{align*}
f(x) & =\frac{k}{\lambda}\left(\frac{x}{\lambda}\right)^{k-1}e^{-\left(\frac{x}{\lambda}\right)^{k}}
\end{align*}
was used to fit data at $x>0$, where $k>0$ is the shape parameter,
and $\lambda>0$ is the scale parameter of the distribution. The resulting
fit for 1~mm electrode separation is shown in Figure~\ref{fig:weibull-fit-lxe}.
The values of the fitted parameters were $k=9.6\pm0.2$ and $\lambda=10.13\pm0.03$. 

The value of $k$ can be compared to the fit obtained from the area
effect in Figure~\ref{fig:lxe-field-area} since $k=1/b$ as stated
in Equation~\ref{eq:area-effect-fit}. This results in $k=7.7\pm1.2$.
Ideally, a fit to a breakdown distribution at any electrode separation
would yield a similar result. Therefore, this fit was also tried for
the breakdown distributions of other electrode configurations. The
distributions appeared Weibull-like. However, due to the small sample
size at all other electrode configurations in both LAr and LXe, the
error on the fit parameters was too large for a meaningful result\footnote{There were 1215 data points collected in Run~3 at 1 mm electrode
separation, and only 234 at 1.4~mm and 108 at 2~mm. Data acquisitions
in both LAr and LXe Run~2 contained less than 100~samples at each
electrode separation.}. A future study with larger sample sizes and therefore more opportunity
for statistical rigor could attempt to confirm the weak link theory.
A fit to a Gumbel function was also attempted but did not yield good
agreement with data.

\section{Discussion\label{sec:Discussion}}

We now turn our attention to possible apparatus upgrades and future
studies with XeBrA. The first data from XeBrA shed some light on the
characteristics of HV breakdown in LAr and LXe. Earlier measurements
performed in other liquids suggest that breakdown behavior is a complex
process that depends on many parameters. Various parameters, such
as electrode area, material, and surface finish, stressed liquid volume
and its purity, system capacitance, and pressure might be contributing
to dielectric breakdown. The contributions from those different parameters
can be hard to distinguish from one another particularly since it
can be hard to design electrodes or operate XeBrA in a way that does
not strongly correlate with these parameters. 

XeBrA's measurement has further validated the existence of an area
scaling effect for breakdown behavior. HV breakdown in XeBrA in LAr
occurred at slightly higher fields than what has been predicted by
previous LAr studies. This can be due to many uncontrolled or untested
parameters mentioned previously. The measured slope of the breakdown
dependence on the SEA in LAr at 1.5~bar was steeper than for 2~bar.
This observation is consistent with the effects observed in LN and
LHe~\cite{Gerhold1989,KAWASHIMA1974217,long2006high}. Additionally,
breakdown performance as a function of the SEA in LAr was observed
to have a steeper slope than in LXe. Further studies at a wider range
of areas should be conducted to confirm this observation.

The capacitance between electrodes is higher for smaller gap configurations
(cf. Table~\ref{tab:Electrode-areas}), so the stored energy available
for the breakdown is higher. References~\cite{cross1982,mazurek1987}
argue that an increase of the locally stored energy increases the
chance of a weak initiating event to grow to a breakdown point. They
correlate the increase of the local energy to the increased SEA. However,
in XeBrA the capacitance between the electrodes decreases as a function
of the SEA as shown in Table~\ref{tab:Electrode-areas} and yet the
area effect is observed. Additionally, the energy stored between electrodes
is only a fraction of the energy stored in the HV cable, thus is unlikely
to affect measurements at various electrode separations strongly.
However, it is possible that the capacitance for an equivalent area
is higher in the systems used for breakdown measurements at FNAL and
LHEP compared to XeBrA, which could contribute to their lower observed
breakdown value.

\begin{figure}[t]
\begin{centering}
\includegraphics[scale=0.6]{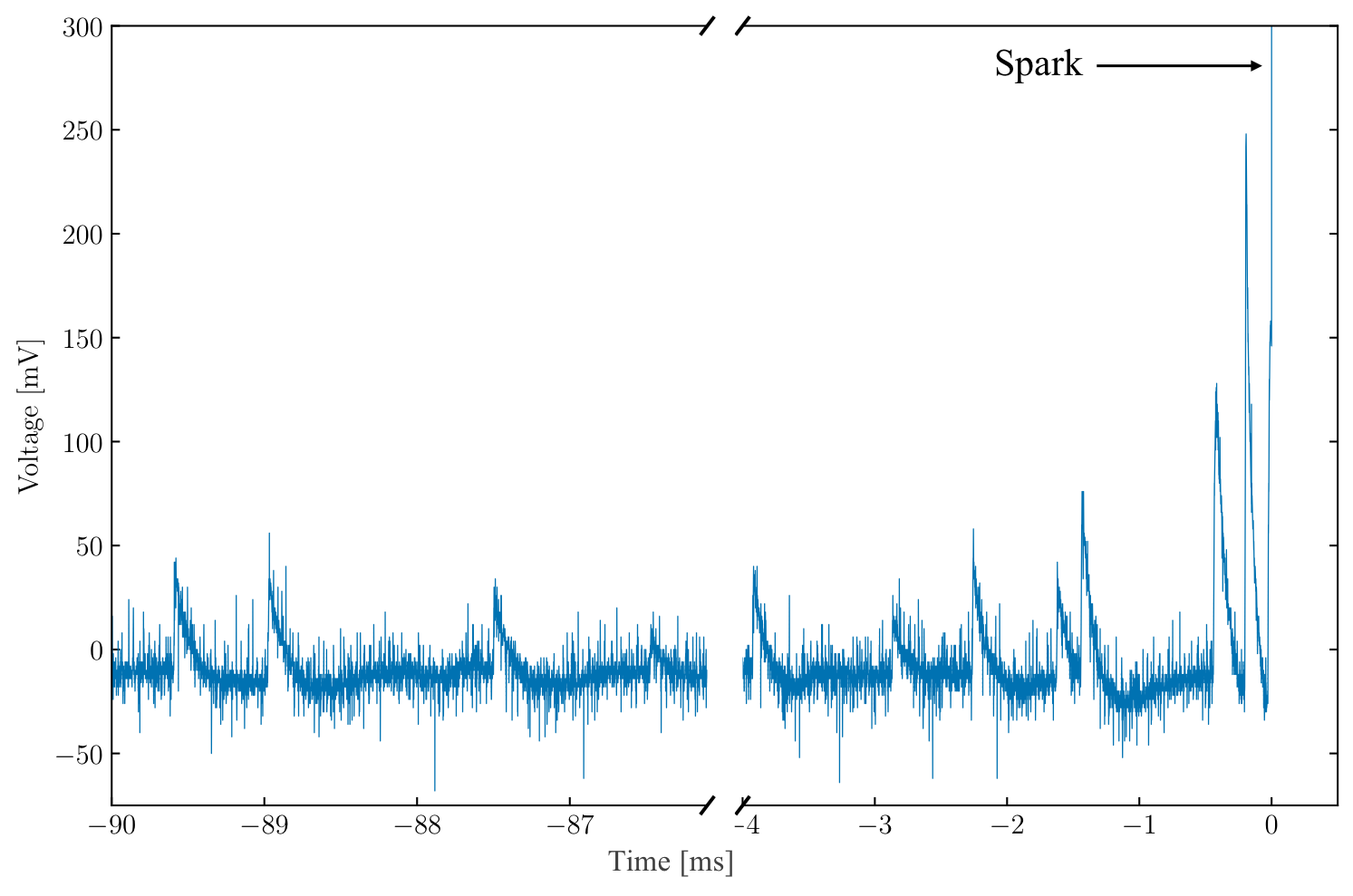}
\par\end{centering}
\caption[Signal recorded by the charge amplifier before a spark in LXe]{Signal recorded by the charge amplifier before one of the sparks
in LXe at 1.4 mm electrode separation. Behavior between -86 and -4~ms
is similar to the trace shown. \label{fig:charge-amp}}

\end{figure}

There were also data collected with a PMT in LAr for 4 mm electrode
separation. These data will be analyzed to study the onset of electroluminescence
prior to the spark. Furthermore, all data collected in LXe during
Run~3 contain information recorded by both a picoammeter and a charge
amplifier. Signals as early as 90~ms before the breakdown were seen
in the charge amplifier, an example of which is shown in Figure~\ref{fig:charge-amp}.
These data need to be analyzed to identify any patterns that might
contain helpful information.

The remainder of this section discusses future  upgrades in Section~\ref{subsec:Future-detector-upgrades},
and possible effects that could be studied with XeBrA in Section~\ref{subsec:Future-possible-studies}.

\subsection{Future  XeBrA upgrades\label{subsec:Future-detector-upgrades}}

Operations of XeBrA in LXe revealed some necessary upgrades to the
apparatus that will ensure seamless operation in the future. These
can be split into two broad categories. First, some upgrades must
occur prior to subsequent operations. Second, there are a few upgrades
that will improve  operations, but their absence does not prevent
future data acquisition. This should not be considered a complete
list since it is anticipated that the upgrades needed for the apparatus
will evolve through time as we gain a better understanding of the
breakdown behavior.

\subsubsection*{Near-term}

Most imminently, the broken heat exchanger needs to be re-welded and
pressure tested. Additionally, the ceramic spacers of the purity monitor
need to be cleaned. A few other minor fixes include installing a new
power supply to increase the heat delivered to the heat exchanger\footnote{Currently, there is 40~W of heating power there, but that should
be increased to $\sim100$~W.} and improving the thermal coupling between the RTDs and the apparatus.
Attaching RTDs to breadboards was a mistake since the breadboards
made it very hard to achieve proper thermal contact of the RTD with
the apparatus.

\subsubsection*{Long-term}

The latest data acquisition in LXe revealed that a small amount of
xenon could be frozen near the HV feedthrough, which suppresses unwanted
bubbles during data taking. However, this approach cannot be used
for data collection in LAr due to a lower freezing temperature. Instead,
a small heater could be installed directly in the LAr space. Turning
the heater on would cause a steep increase in pressure, making the
apparatus temporarily bubble-free for data taking. This approach is
similar to adding gas from the bottle as was done in this work but
does not require the addition of gas into the apparatus. 

Breakdowns near the power supply's maximum of -75~kV did not always
occur between the two electrodes. Instead, the breakdown sometimes
took place between the air-vacuum epoxy HV feedthrough and the ISO100
spool at ground. In order to measure the breakdown between electrodes
near this -75~kV threshold, the ISO100 port should be enlarged. This
will reduce the value of electric fields near the feedthrough, thus
diminishing this issue. 

A two camera system is being developed to monitor the location of
the spark to improve understanding of the spark development. One camera
will be installed at each of the two viewports. This will enable a
2D position reconstruction of the spark to determine whether the breakdown
occurs within the 90\% SEA and if the location varies between each
spark. This should improve the understanding of the stochastic nature
of the breakdown and indicate whether the weak link theory is appropriate.
It will also clarify whether using 90\% SEA is appropriate or if a
larger area should be used, similar to study~\cite{8124618}. Furthermore,
a high-speed camera with a frame rate of 1,000~fps or better could
be installed at one of the viewports to document the development of
the breakdown in xenon. The estimate for the camera frame rate is
based on a similar study performed in LAr~\cite{Auger:2015xlo}. 

\subsection{Future possible studies with XeBrA\label{subsec:Future-possible-studies}}

As discussed in Section~\ref{sec:Breakdown-studies-primer}, there
are many parameters suspected to affect dielectric breakdown in liquids.
XeBrA can be used to attempt to explore many of those in the future. 

\subsubsection*{Electrode material and surface finish}

Since the electrodes in XeBrA can be easily changed, electrodes from
various materials could be manufactured, and their performance tested.
Apart from different materials, effects caused by various surface
finishes can be studied to better understand breakdown performance.
A high-pressure rinse could be tested as a method to achieve a very
smooth surface finish. It has been observed to improve electrode performance
in other applications~\cite{highPressureWash,Bernard:1991du}. Power
washing involves using high pressure $\left(\geq\unit[1200]{psi}\right)$
ultrapure water with resistivity $\geq\unit[18]{M\Omega/cm}$. The
procedure needs to be performed inside a clean room to maintain the
electrode cleanliness. 

Electrodes can also be coated with various materials. Reference~\cite{1748-0221-9-07-P07023}
examined performance of coating SS electrodes in latex. This improved
the performance of the electrodes but only until the first breakdown.
In an upcoming test, XeBrA will test electrodes with a passivated
surface to inform the design of the LZ detector's cathode surface.
Passivation causes some of the chromium in the SS to oxidize and form
a layer of $\mathrm{Cr_{2}O_{3}}$ on the material's surface. This
has been found to improve the breakdown resistance of electrodes by
suppressing the field emitted electrons~\cite{Koontz:1990greening}.
However, in pulsed HV applications, it has been observed that once
a hole develops in the coating, its performance will decrease. We
want to confirm whether the performance of the passivated surface
worsens after the first breakdown prior to installing such surfaces
in LZ.

Additionally, to minimize surface roughness, a layer of graphite or
other conductive material could be deposited to ensure a smooth electrode
structure, which might improve electrode performance. Graphite dispersed
in distilled water, known as aquadag\texttrademark, could be used.
After application, once the surface has dried, aquadag leaves a conducting
layer of pure graphite.

\subsubsection*{Effects of different kinds of impurities}

There is evidence that impurities affect breakdown in LAr. However,
no such measurement has been done in LXe. Additionally, the effect
of impurities on electron lifetime in LXe has not been well described.
Since XeBrA contains a purity monitor, a known amount of various impurities
could be introduced to the system to study their effect on dielectric
breakdown and electron lifetime. This is predicated on the assumption
that XeBrA can reach purity levels that are measurable with the existing
purity monitor design, which is not yet the case but should be achievable.

\subsubsection*{Other}

More data should be collected to determine whether the value of breakdown
is indeed correlated with pressure, as indicated by XeBrA's measurement
in LAr. Since at the current configuration XeBrA cannot accommodate
larger electrodes, smaller electrodes can be installed to evaluate
breakdown at a broader range of SEA to identify breakdown trends in
LXe. Furthermore, larger sample sizes of breakdown data at each electrode
separation could be used to evaluate the validity of the weak link
theory. Additionally, the effect of muons on breakdown could be examined.
Muons passing through the  active volume will ionize the liquid, potentially
creating a region with enhanced breakdown probability. 

\section{Conclusion}

XeBrA is the first experiment to provide systematic studies of dielectric
breakdown in LAr and LXe at electrode areas greater than $\unit[3]{cm^{2}}$.
Results from XeBrA obtained in LAr and LXe with the same electrode
also enable direct comparison of breakdown behavior in those two noble
liquids. These early results indicate that a steeper dependence on
the SEA may be expected for breakdowns in LAr than in LXe. However,
the validity of the measurement needs to be confirmed with a more
extensive dataset in LXe. Results from XeBrA also indicate a stronger
area dependence of breakdown field at lower pressures in LAr, an effect
that is worth future examination.

This study also attempted to fit a single breakdown distribution to
examine the validity of the weak link theory prediction. A single
data point agreed with the prediction. However, a bigger sample is
needed to evaluate the validity of the Weibull hypothesis.

Further studies will also incorporate information acquired by a picoammeter
and a charge amplifier connected to the ground electrode during data
acquisition. Additionally, PMT data will be examined to study the
onset of electroluminescence prior to a spark. 

Given the complex nature of dielectric breakdown, it is difficult
to predict whether the results from XeBrA can be extrapolated for
application in LZ. Nevertheless, the current results from XeBrA bear
good news for the LZ detector since the predicted breakdown strength
in LXe is well above the LZ design parameters. Investigation of a
diverse set of parameters over a wider range of areas will need to
be performed to better understand the contribution of various parameters
to the dielectric breakdown probability. The results presented in
this chapter started this investigation. Future studies with XeBrA
will further improve our understanding of breakdown behavior in LAr
and LXe.

\begin{figure}[p]
\begin{centering}
\includegraphics[angle=90,scale=0.14]{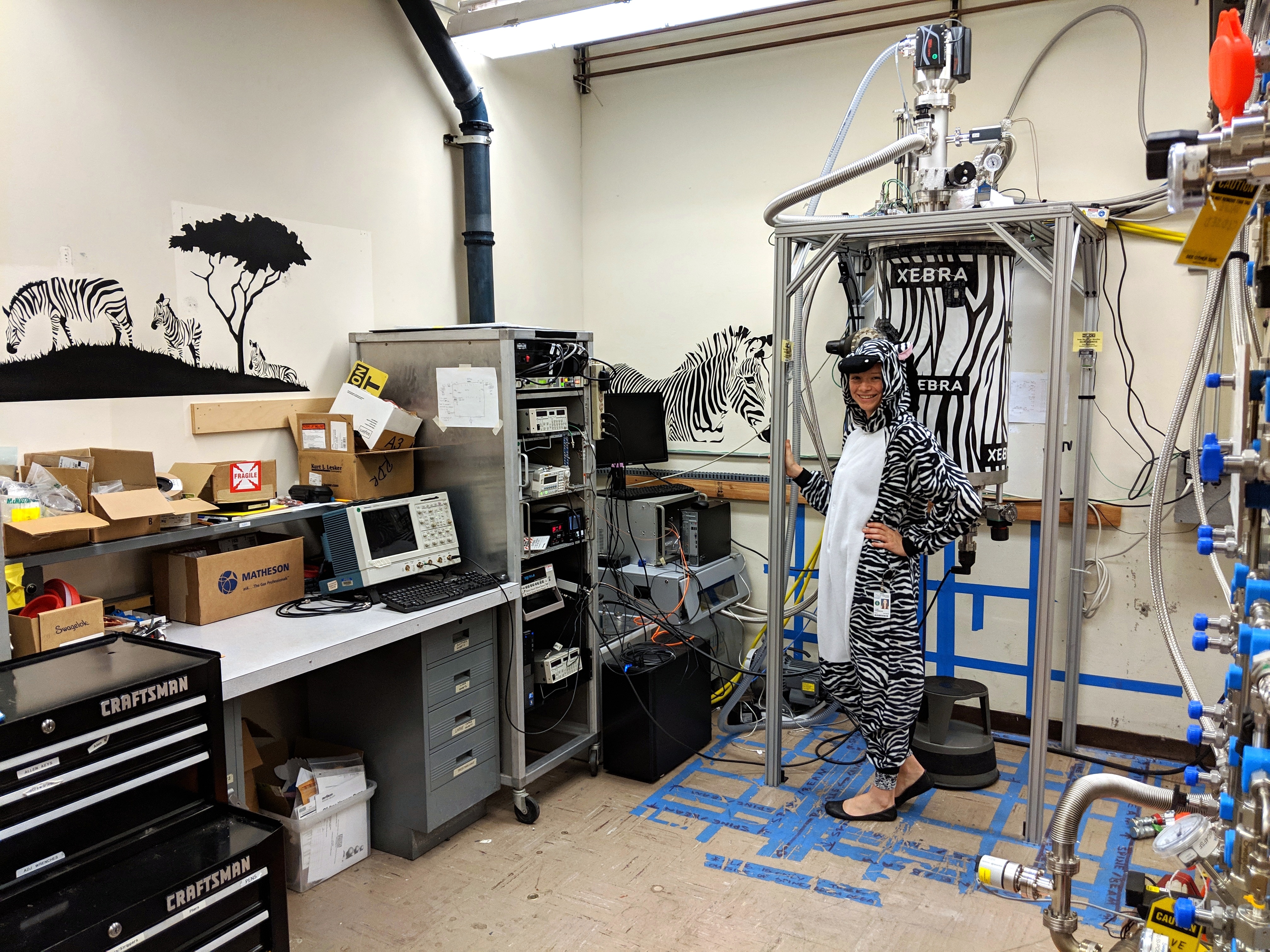}
\par\end{centering}
\caption[Z/XeBrAs]{Z/XeBrAs.}
\end{figure}

\appendix

\chapter{\textsc{LZ Equity \& Inclusion}\label{chap:LZ-E=000026I}}

\begin{figure}
\begin{centering}
\includegraphics[scale=0.55]{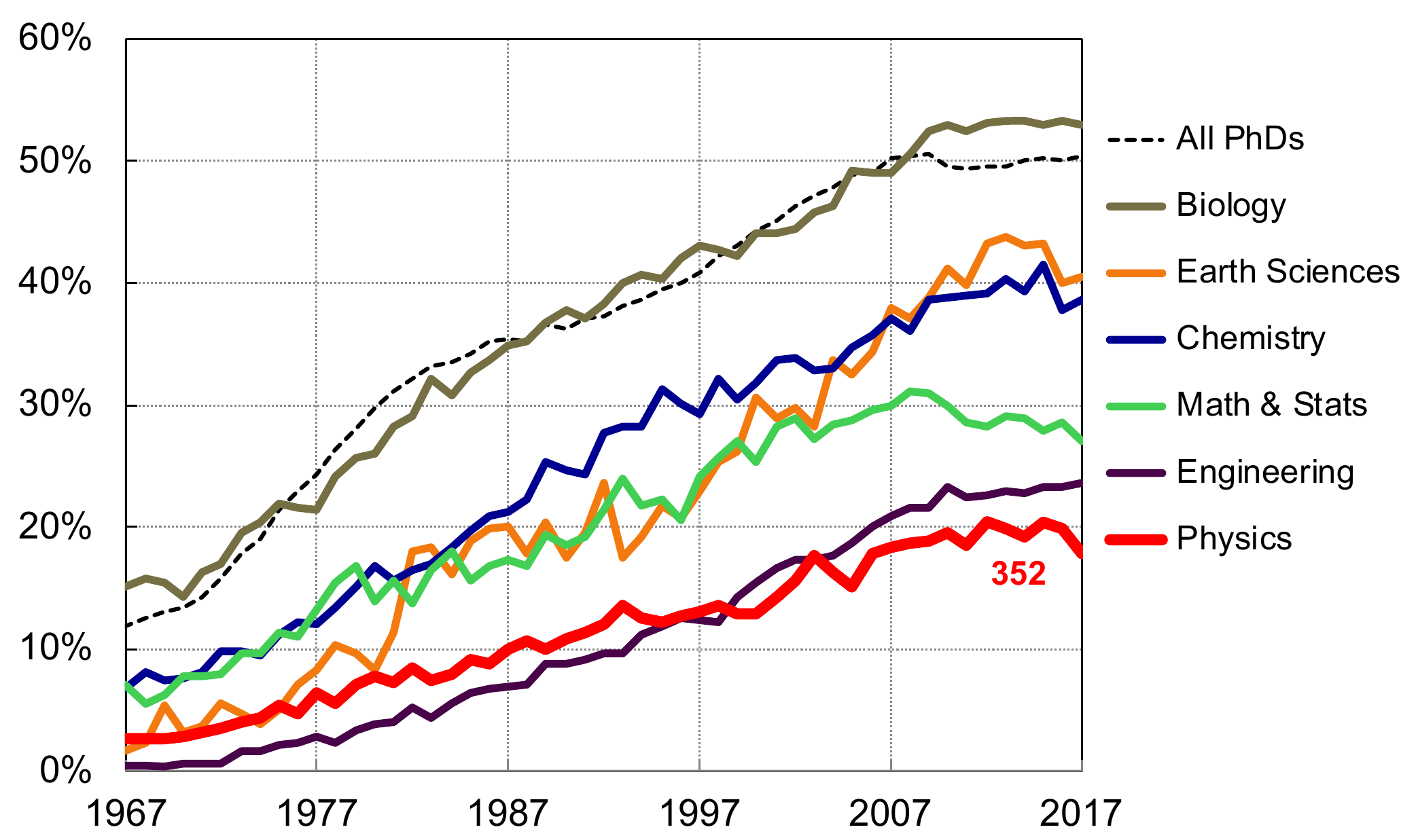}
\par\end{centering}
\caption[Fraction of U.S. doctoral degrees awarded to women in STEM fields]{Fraction of U.S. doctoral degrees awarded to women in Science, Technology,
Engineering, and Mathematics (STEM)\nomenclature{STEM}{Science, Technology, Engineering, and Mathematics}
fields. There were 352 doctoral degrees in total awarded to women
in 2017. Overall, more doctoral degrees were awarded to women than
to men. Women received the majority of doctoral degrees in popular
fields, such as health sciences (70.3\%), education (68.8\%), and
social and behavioral sciences (61.1\%)~\cite{phd_survey}. Figure
from IPEDS and APS~\cite{aps_stats}. \label{fig:physics_phds}}
\end{figure}

Only about $20\%$ of doctoral degrees in physics are awarded to women
as illustrated in Figure~\ref{fig:physics_phds}, and one factor
may be that women are very likely to experience sexism and microaggressions
in their career~\cite{PhysRevPhysEducRes.12.020119}. In fact, research
shows that $20\%$ or more of undergraduate and graduate female science
students have experienced sexual harassment from faculty or staff~\cite{NAP24994}.
The LZ Equity and Inclusion (E\&I)\nomenclature{E\&I}{Equity and Inclusion}
committee was formed to help address these problems more broadly.
It attempts to address these problems within the scope of the LZ collaboration
and to create a model for future collaborations. This is not only
a moral imperative but also a strategic investment for the collaboration
both in attracting new talent and in preventing attrition due to unsavory
working conditions for women and minorities. Solving issues faced
not only by women, but by minorities in general, creates a healthier
and more productive~\cite{SAXENA201476} working environment for
all collaboration members that enables the LZ collaboration to achieve
its highest potential. 

The LZ E\&I working group was formed in spring 2017 after it was realized
that some of the issues mentioned above have surfaced within the LUX
collaboration in the past and we wanted to provide a safeguard to
the even larger, more diverse LZ collaboration. Harry Nelson, an LZ
professor, and I attended a workshop about the Impacts of Sexual Harassment
in Academia organized by the US National Academy of Sciences, Engineering,
and Medicine in June~2017 at UC Irvine. This committee very recently
released an extensive report~\cite{NAP24994}, featuring fifteen
concrete recommendations on improving the current climate - many of
which were adopted by LZ prior to the release of this report on the
recommendation of the E\&I committee. 

One of the recommendations asserts that academic institutions should
consider sexual harassment to be equally important as research misconduct
in terms of its effect on the integrity of research. Academic misconduct
is a broad umbrella topic with a lot of complexity and nuance for
which institutions have spent years implementing systems and policies
in an attempt to protect academic integrity. Similarly, sexual harassment
likely cannot be solved with one simple, black-and-white rule change,
but instead requires a change of climate and an entire paradigm shift.
It is important to avoid thinking about this problem with a too narrow
mindset and to set expectations corresponding to a long and nuanced
journey toward excellence.

Efforts to reduce sexual harassment and discrimination in the workplace
have been ongoing for decades; laws have been put in place addressing
these issues, albeit not very successfully. In particular, Title IX,
that prohibits discrimination on the basis of sex in any federally
funded program, and Title VII that prohibits employers from discriminating
based on sex, race, color, national origin, and religion, spurred
the creation of policies and sexual harassment training in order to
comply with the law and avoid liability. While laws, regulation, and
policy frequently reflect a real need for change, the specific implementations
can cause their efficacy to vary wildly or even induce harm. Indeed,
research suggests that voluntary sexual harassment or diversity training
has proven to be effective, while mandatory training can have no or
even adverse effects~\cite{dobbin2016diversity}. Furthermore since
academia relies so heavily on its hierarchical structure with junior
members often needing recommendations from their superiors to advance
their careers, reporting any harassment can be impractical due to
worries of retaliation. Reporting has even been shown to harm the
victims instead of providing the much-needed support~\cite{smith2014institutional}
and frequently the institutional response shields the star faculty
and protects the institution from liability rather than looking after
victims~\cite{mackinnon2015their}. 

Therefore the LZ E\&I committee set out to implement a more thoughtful
and comprehensive approach to address both sexual harassment and a
broader set of workplace-related issues. These issues include anything
that might prevent people from joining or staying in the collaboration
and range anywhere from implicit biases and stereotype threats, hostile
workplace, and sexual misconduct, to microaggression, inappropriate
jokes, being spoken over during meetings, paper authorship issues,
and beyond. Our attempts to address all of these issues started in
the spring of 2017 in a small group that included me, Dan Akerib,
and Harry Nelson, later also joined by Rachel Mannino. We met weekly,
consulted many researchers in psychology, sociology, and other physicists
with experience in similar projects, and ultimately presented a comprehensive
set of policies to the LZ collaboration in March 2018. Section~\ref{sec:LZ-policies}
chronicles that year of hard work and describes the actions and policies
that resulted therefrom. In addition, Section~\ref{sec:double-blind}
summarizes the findings of a side project that delves into the benefits
of double-blind review for academic publications as it presents yet
another piece of the puzzle that has been shown to benefit minorities.

\section{LZ policies\label{sec:LZ-policies}}

The goal of the policies presented to the collaboration was to develop
positive, aspirational procedures and for the collaboration as a whole
to make a positive statement that a fair working environment is an
important factor toward delivering great science. Since membership
in the LZ collaboration is a privilege and should be approached as
such, the policies strive to set expectations for a professional environment
protecting equal research grounds for everyone. Given the breadth
of in-person interactions within the LZ collaboration at a wide variety
of locations throughout the world, the policies address the specific
needs of the collaborations acknowledging different cultural backgrounds
of collaborators and the variety of interactions encountered during
both remote and in-person interactions. Unlike universities and research
labs, LZ is not a legal entity (and therefore cannot be sued), so
the policies developed are not overly legal and restrictive, but rather
serve to inform collaborators of resources available at their home
institutions and provide additional support. 

The documents presented here were co-authored with the Equity \& Inclusion
committee and adopted as official LZ policies by the LZ Institutional
Board (IB)\nomenclature{IB}{Institutional Board} in March 2018. They
were presented to the LZ collaboration at the inaugural Equity \&
Inclusion plenary session at a collaboration-wide meeting in Coimbra,
Portugal in April 2018. These efforts generated very positive feedback,
and following the Coimbra collaboration meeting, the Equity \& Inclusion
committee grew to 15 members and proceeded to nominate its first two
Ombudspersons.

There were three documents adopted by the LZ collaboration, along
with a modification to the LZ governance document that made the E\&I
committee an official part of the collaboration. As the name suggests,
the LZ Equity \& Inclusion Scope of Activities presented in Section~\ref{subsec:LZ-Equity-scope}
defines the charge of the committee. The LZ Code of Conduct is an
aspirational 1-page long document that sets expectations for collaborators
regarding their behavior and scientific misconduct as presented in
Section~\ref{subsec:LZ-Code-of-conduct}. It also links to the LZ
Institutional resource list, maintained by the E\&I committee, that
features an up-to-date list of resources available at each of the
collaborating institutions. 

LZ also appointed two ombudspersons. The role of the LZ Ombudspersons
is to help and refer people who come forward with any issues that
might arise within the collaboration as outlined in Section~\ref{subsec:LZ-Ombudspersons-Guide}.
Information shared with the LZ Ombudspersons is assumed confidential
within the collaboration, hoping to encourage members who might worry
of retaliation otherwise. The ombudspersons are also expected to identify
systemic problems and organizational issues within LZ and assist in
finding solutions as needed. At any time there are two ombudspersons
from different institutions serving on a 2-year staggered term. This
avoids potential interference with any mandatory reporting rules that
might be in place at either of the ombudspersons' institutions. 

Last, but not least, every in-person LZ collaboration meeting now
features a plenary and a parallel session on diversity-related topics
to encourage discourse within collaborators on this topic, as discussed
in Section~\ref{subsec:LZ-speaker}. 

The E\&I efforts are expected to change throughout time as we learn
more and better understand the best practices for promoting a genuinely
inclusive working environment. The E\&I committee also considered
deploying a collaboration-wide survey, such as the ARC3 Campus Climate
survey~\cite{arc3}. This effort was postponed due to a worry of
ability to provide anonymity to women who might have experienced harassment
within the collaboration because of potentially low survey response
rate (and due to the low number of women in the collaboration in general),
lack of experience analyzing this type of data, and funding reasons.
However, the survey should be considered in the future in order to
acquire data to help track the impact of the committee's efforts.
Other possible future projects include a journal club on the topic
of diversity, so that committee members remain up-to-date on the most
recent advances in the field, establishing a mentoring program for
graduate students, postdoctoral fellows and junior faculty within
the collaboration, and more. Since there are a few other physics collaborations
with ombudspersons, the committee is also considering connecting all
those ombudspersons to facilitate sharing insights and leverage each
other's experiences in solving problems that come up. Depending on
the success of the E\&I efforts within LZ, the committee might also
consider helping other physics collaborations set up their E\&I ventures.

\subsection{LZ Equity \& Inclusion Scope of Activities\label{subsec:LZ-Equity-scope}}
\begin{enumerate}
\item \textbf{Introduction}: This document describes the LZ Equity \& Inclusion
Committee (EIC), which was established by the LZ Institutional Board
(IB) as described in the LZ Governance Document. The committee aspires
to ensure that the climate in the LZ collaboration is positive, respectful,
and supportive of all its member. It will work towards this goal by
providing resources, establishing and managing the role of LZ ombudspersons,
developing a Code of Conduct, organizing talks and other educational
activities at collaboration meetings, and occasionally providing assessment
and advice on the climate to the collaboration leadership. 
\item \textbf{Code of Conduct}: The EIC will submit to the IB for approval
a Code of Conduct to set out expectations for behavior among collaborators
across the variety of circumstances that they may encounter. By its
very nature, the Code is an aspirational document and aims to encourage
positive behavior towards all members of the collaboration. The Code
will be described in a separate document and is subject to approval
and modification by the IB. All active members of the collaboration
must agree to follow the Code. 
\item \textbf{Ombudspersons}: The EIC will propose 2 candidates for the
role of ombudspersons appointed by the LZ Spokesperson to provide
collaboration members with a confidential resource to turn to if they
encounter situations with other members of the collaboration that
are counter to the Code of Conduct, or other troubling situations
in which they require support, including but not limited to sexual
harassment, hostility, coercion, or other negative behaviors. The
role and structure of the ombudspersons are described more fully in
a separate document to be approved by the IB. At large gatherings,
such as collaboration meetings or analysis workshops, the EIC and
ombudspersons will appoint a meeting ally in cases where the ombudspersons
are not in attendance in order to provide in-person support. 
\item \textbf{Resources}: The EIC will promote a respectful environment
for all collaboration members during collaboration interactions and
maintain a resource page on the LZ Twiki. This will include general
equity \& inclusion resources as well as resources specific to each
member institution. For example, each US educational institution employs
a Title IX coordinator to enforce sexual harassment prevention, and
other policies pertaining to hostility, coercion and other prohibited
practices, as well as providing support to victims. Many institutions
also maintain an ombudsperson or similar to provide a range of support,
including confidential support. 
\item \textbf{Collaboration Meetings}: The EIC will be allocated plenary
time at each collaboration meeting to program at its discretion. Such
programming may include outside speakers on matters pertinent to the
committee\textquoteright s work, occasional presentation and discussion
of the Code, resources, ombudspersons, or other matters aimed at informing
the collaboration within the EIC\textquoteright s scope. 
\item \textbf{Assessment}: The EIC may occasionally conduct surveys of the
collaboration to assess the needs of the collaboration within the
committee\textquoteright s scope. These will be conducted in consultation
with the collaboration leadership and reported to the IB and the collaboration.
The committee also welcomes broad input from the collaboration on
any matters within the committee\textquoteright s scope on ways to
maintain and improve a positive and supportive culture. 
\item \textbf{Structure}: The EIC committee will consist of members from
all ranks and titles of the collaboration, but will include at least
one graduate student and one postdoc member. The Spokesperson is an
ex-officio member. The committee will be structured as a committee
of the whole. The committee shall meet before every collaboration
meeting and more frequently as needed.
\end{enumerate}

\subsection{LZ Code of Conduct\label{subsec:LZ-Code-of-conduct}}

The LZ Code of Conduct~\cite{CoC} aspires to strengthen trust and
mutual respect among collaborators to foster an inclusive collaboration
environment supportive of sound scientific research. Every LZ member
should read the Code, and follow both its spirit and letter, always
bearing in mind that each of us has a personal responsibility to incorporate,
and to encourage others to incorporate, the principles of the Code
into our work.

The LZ collaboration is committed to a supportive work environment,
where everyone has the opportunity to reach their fullest potential
dedicated to the advancement, application, and transmission of knowledge
through academic excellence. This policy applies to all interactions
between LZ collaborators although the formal policies of individual
institutions take precedence. Some examples of interactions include
collaborators\textquoteright{} in-person, virtual, or phone interactions,
or any social or professional gathering both at formal institutions
and during social events associated with the collaboration. LZ collaborators
should act in concordance with a professional culture that fosters
respectful interactions and a safe environment free of harassment,
intimidation, bullying, bias, discrimination, or mistreatment of any
kind. LZ collaborators must refrain from any actions or statements
that denigrate other collaborators on the basis of personal characteristics
or beliefs. This includes, among other actions, intimidation, disruptive,
or harassing behavior of any kind, sexual or crude jokes and comments,
offensive images, or any unwelcome physical contact. Please be aware
that behaviors and language acceptable to one person may not be to
another.

The conduct of LZ collaborators must be based on dedication to the
highest scientific and technical standards. Collaborators are expected
to perform research in a well documented and ethically sound manner.
Falsification of data or results, plagiarism, violations of the LZ
Publication Policy, taking credit for others\textquoteright{} work,
or any other scientific misconduct will not be tolerated.

Unacceptable conduct may warrant further action. For example, code
violations could result in consequences, up to sanctions by the IB
according to the LZ Governance Policy.

If you have a question or ever think that one of your colleagues or
the collaboration as a whole may be falling short of our commitment,
don\textquoteright t be silent. Your collaborators and leadership
need to hear from you. You may report any misconduct to your supervisor,
collaboration\textquoteright s spokesperson, ombudspersons, meeting
allies, or the Title IX or diversity officer of either your institution
or the institution where the incident took place (see the LZ institutions
resource list for details). LZ does not tolerate retaliation against
any collaborator who reports or participates in an investigation of
a possible violation of our Code, policies, or the law.

The intention of this Code of Conduct is to ensure that the collaboration
is a welcoming place for everyone so if you see something that you
think isn\textquoteright t right \textendash{} speak up!

\subsection{LZ Ombudspersons Guide\label{subsec:LZ-Ombudspersons-Guide}}

This document summarizes reach and responsibilities of the LZ Ombudspersons~\cite{ombuds}.
Further information about the Ombudspersons reach and responsibility
can be obtained from the International Ombuds Association.

The LZ Ombudspersons provide confidential, informal, independent,
and impartial assistance for any LZ collaborators on matters pertaining
to collaboration values, including dispute resolution services. The
LZ Ombudspersons are available to LZ collaborators who are experiencing
conflicts or disputes as part of their LZ activities rather than with
problems internal to an academic institution, which can usually be
mediated by the appropriate offices in that institution (including
the institutional ombudsperson). The LZ Ombudspersons will work together
with the members who consult them to identify options for managing
and resolving disputes and conflicts. This includes providing advice
and support, finding out information about, or referring individuals
to, appropriate resources, and facilitating mediation.

An LZ Ombudsperson may report systemic issues or patterns of concern
in the collaboration to the LZ Equity \& Inclusion committee (EIC)
and the Spokesperson but may do so only without disclosing individual
names or other aspects that would identify parties. Confidentiality
is a privilege of the LZ Ombudspersons. The only circumstances that
there may be an exception to confidentiality is if the Ombudsperson
is concerned by an imminent risk of serious harm, or if required by
law in the applicable jurisdiction to anonymously report credible
evidence of fraud, waste, or abuse concerning the use of government
funds. Please note there is no recognized privilege, similar to attorney-client
privilege, for ombudspersons in U.S. courts.

To avoid mandatory reporting requirements of an ombudsperson to their
home institution (where their confidential role may not be recognized),
LZ shall have two Ombudspersons from two different LZ institutions;
LZ scientists can contact whichever Ombudsperson they feel is appropriate.
The LZ Ombudspersons should not have any other collaboration leadership
role that may compromise their impartiality or confidentiality.

The Ombudspersons are required to become familiar with the Title IX
rules of their institution (ideally by meeting with their institution\textquoteright s
Title IX or Diversity officer), with other code of conduct policies
at their institution, and with the organizational structure of the
LZ collaboration. This will allow them to provide current information
about services, programs, policies, and procedures. By having two
LZ Ombudspersons, it is expected collaboration members will have access
to an ombudsperson not at their home institution, providing them with
a confidential resource outside of their research group. Additionally,
for matters arising within an institution, LZ members are encouraged
to seek out the ombudsperson or other resources at their home institution.

If neither one of the Ombudspersons is able to attend an in-person
collaboration meeting, the Ombudspersons and the EIC will appoint
two \textquotedblleft meeting allies\textquotedblright{} for the event.
These persons will temporarily act as Ombudspersons for the duration
of the event.

To prevent recurring events and maintain institutional memory given
the 2-year term of the Ombudspersons, they are expected to keep a
log. Each ombudsperson may share the identity of the complainant with
other ombudspersons (current and future) only with explicit permission
of the complainant. The EIC will explore the use of other ombuds tools,
such as Callisto, but will employ such only after appropriate consultation
with the IB and with full knowledge by the collaboration.

\subsection{LZ collaboration meeting plenary presentations\label{subsec:LZ-speaker}}

Research shows that discussion of diversity can have a very positive
effect on retaining female (and potentially minorities in general)
in physics~\cite{PhysRevSTPER.9.020115}. Therefore an essential
part of the adopted policies are the in-person presentations at the
bi-annual collaboration-wide meetings, which are intended to enhance
collaborators' knowledge on a diversity topic and to encourage a discussion
on the topic. The first plenary session in Coimbra in April 2018 was
devoted to the introduction of the LZ E\&I efforts. The second plenary
session at Brandeis University in July 2018 hosted Prof Frank Dobbin,
a sociology researcher from Harvard University, whose talk was very
well received and set high expectations for future meetings.

The parallel sessions enable the Equity \& Inclusion committee to
gather feedback from the collaboration members on topics the committee
is working on at the moment and provide an excellent opportunity to
discuss thought-provoking diversity manuscripts in a smaller group
than the plenary session. The well-attended, inaugural parallel session
at Brandeis University featured a presentation highlighting recommendations
relevant to the collaboration from the report on Sexual Harassment
in Academia~\cite{NAP24994} and a discussion about the shared collaboration
housing in Lead, South Dakota. 

\section{Case for the double-blind peer review\label{sec:double-blind}}

The text below was also published in~\cite{tvrznikova2018case}.

Peer review is a process designed to produce a fair assessment of
research quality prior to the publication of scholarly work in a journal.
The process consists of sending a candidate work such as a journal
article to a number of similarly qualified professionals for feedback
and informs publishers of academic journals whether a work should
be accepted or rejected. Peer review is used as a gold standard in
assessing manuscripts and is essential for academics since tenure,
grant applications, and many other career-related decisions are based
upon the quality of their work and publications making the stakes
are high. Most journals today use a single-blind review method, in
which the reviewer's identity is concealed to the author of the article,
but not vice versa. However, some journals use alternative review
strategies~\cite{blank1991effects,baggs2008blinding}, such as double-blind
review, where both the author's and the reviewer's identities are
concealed, or open review~\cite{Hopewell_2014,nature_openReview},
where all information is public. 

It is well known that implicit biases influence people's decision
making, but ideally, decisions regarding the importance of scientific
contributions would be free of these biases. Nevertheless, demographics,
nepotism, and seniority have been shown to affect reviewer behavior
suggesting the most common, single-blind review method (or the less
common open review method) might be biased~\cite{nature_nepoticm,Link_1998}.
Blinding both the author's and reviewer's identity can reduce exposure
to extraneous information thus bypassing bias and optimizing decision
making. Although this article focuses on academic journals, blinding
has the potential to affect decision making in many areas: it has
been demonstrated that blinding applicants in the first stages of
grant applications reduces biases related to researcher\textquoteright s
characteristics~\cite{solans2017blinding} and blind musical auditions
have significantly increased the number of women in symphony orchestras~\cite{goldin2000orchestrating}.
Moss \textit{et al.} also showed that blinding the gender of job applicants
can reduce unconscious and unintended bias~\cite{moss2012science}. 

Many characteristics can be subject to biases that might skew the
reviewer\textquoteright s decisions as discussed below. For example,
there is substantial evidence for biases in favor of famous authors
and world-renown institutions, but there are only limited data on
the impact of review on women and minorities. Budden \textit{et al.}
found that switching to a double-blind review in an ecology journal
led to a small increase of female first author paper~\cite{budden2008double}.
However, these findings were later rebutted~\cite{whittaker2008journal,webb2008does,engqvist2008double}
and this increase was attributed to a proportional increase of women
in the workforce overall. Journal of Neurophysiology also found no
gender bias in their publications, but carefully notes that at the
time of the analysis 5 out of 9 members of the editorial panel were
women and only 191 of 713 submissions (26\%) were from women-first
authors~\cite{lane2009there}. Blank found that although women perform
slightly better under a double-blind system, the data were statistically
not significant~\cite{blank1991effects}. It should be noted that
women published only 150 out of 1,223 analyzed papers (12\%). Other
journals also found author identity did not influence acceptance rates
or quality ratings in review~\cite{chung2015double,borsuk2009name}. 

Conversely, after the Modern Language Association (MLA) switched to
double-blind review a substantial increase in acceptances for female-authored
publications was observed, eventually comparable to that for men~\cite{largent2016blind}.
Tomkins \textit{et al.} did not find any gender bias in their study
of the review process in computer science using their data alone,
but combining their results with previous studies showed a statistically
significant gender effect bias~\cite{tomkins2017single}. Studying
this topic, Helmer \textit{et al.} demonstrated that women are currently
underrepresented in the peer review process and that editors of both
genders preferred same-gendered authors~\cite{helmer2017gender}.
 This behavior highlights the need to employ review methods that combat
subtler forms of gender bias in scholarly publishing.

Despite the mixed data regarding the effect of double-blind review
on gender bias, it has been demonstrated that double-blind review
can mitigate biases arising from researcher's popularity or location.
Tomkins \textit{et al.} found that single-blind reviewers were significantly
more likely to recommend for acceptance papers from famous authors
and top institutions~\cite{tomkins2017single}. Moreover, if reviewers
knew the author's identity, they disfavored authors that were not
sufficiently embedded in their research community~\cite{seeber2017does}.
Link discovered that in a medical journal the location of the authors
mattered as US reviewers ranked US papers much more favorably compared
to the non-US ones~\cite{Link_1998}. Additionally, Blank found that
under a double-blind review acceptance rates were lower and reviewers
were more critical~\cite{blank1991effects}. This suggests that a
double-blind system might lead to more critical feedback from reviewers
and a mix of accepted papers from more diverse authorship. 

It should be noted that before a manuscript reaches the review stage,
journal editors who have access to authors' information reject many,
sometimes most, of the papers submitted to a journal before a reviewer
can see them. While motives will differ between editors and reviewers,
there is no reason to assume that editors are less susceptible to
bias than reviewers~\cite{garvalov2015stands}. This issue could
be overcome with the use of double-blind review since editors would
also lose access to authorship information.

A common concern for double-blind review is the identification of
author or institution from self-citations, nature of work, or personal
connections~\cite{Yankauer_1991,katz2002incidence}. Hill and Provost
found that even using the best method to identify authors based on
discriminative self-citations, authors were identified correctly only
40-45\% of the time suggesting that even in the worst-case scenario
blinding is successful most of the time~\cite{Hill_2003}. Other
arguments against double-blind review such as administrative inconvenience,
posting articles to other websites prior to their publications, possible
conflicts of interest, or tradition can be overcome~\cite{goues2017effectiveness}.
In physics, the issue of identification becomes significantly more
relevant due to the high number of large international collaborations
of various scales ranging from dozens to thousands of members. Often,
publications from these collaborations will be trivial to identify.
However, it is also unlikely they will suffer any disadvantages while
the double-blind review could still benefit researchers who might
currently be affected by the biases in the review process.

A scientist's career should not be influenced by stereotypes, but
rather depend on an individual's proven track record to perform. Reviewers
are people too, and since biases can affect anyone the community should
strive toward a peer review system that makes its best effort to overcome
the susceptibility to bias. Current research suggests that double-blind
review reduces bias that might arise while assessing the quality of
the research presented in a manuscript, without imposing any major
downsides. Double-blind review offers a solution to many biases stemming
from author's gender, seniority, and institution and should be a foundation
of any journal that strives to evaluate author's work strictly based
on the quality of the research presented in the manuscript.

\chapter{\textsc{A very brief introduction to fittings}\label{chap:Intro-to-fittings}}

There are many different types of fittings one encounters during a
day in a lab. The world of fittings can be very confusing and overwhelming
at first due to similar names and very slight differences in use.
This appendix attempts to offer a brief, high-level overview of standard
fittings one might encounter in a lab that is concerned with noble
(and not so noble) gasses and liquids. 

An overview of the fittings and their applications is provided in
Table~\ref{tab:Overview-of-fittings}. For clean noble gasses, all-metal
connections such as Swagelok VCR (Section~\ref{sec:Swagelok-VCR})
or CF (Section~\ref{sec:ConFlat-(CF)}) should be used whenever possible.
Other connections that can be encountered when working with clean
noble gasses at higher pressures are Swagelok tube fitting (Section~\ref{sec:Swagelok-tube-fittings})
and CGA 580 cylinder connections (Section~\ref{sec:CGA580-(bottle-connection)}).
For vacuum application CF (Section~\ref{sec:ConFlat-(CF)}), KF (Section~\ref{sec:KF/QF/ISO}),
or Swagelok VCR (Section~\ref{sec:Swagelok-VCR}) are among the most
common connections available. The NPT standard (Section~\ref{sec:NPT})
is most often used for water connections. 

Unfortunately, it is very hard to build a system that uses a single
fitting standard, but parts that convert between nearly all types
of fittings discussed in this appendix can usually be purchased. If
they cannot be purchased, they can usually be custom welded to create
a leak-tight connection. All parts should be cleaned with lint-free
wipes and isopropyl alcohol or ethanol before use. If the parts are
very dirty, a sonicator can be used as well.

The information presented in this section is based on lab experience
and documentation available on the websites of Swagelok and Kurt J.
Lesker companies.

\begin{table}
\begin{centering}
\begin{tabular}{lccccccc}
\hline 
Fitting type & VCR & CF & KF & Tube & NPT & Flare & CGA580\tabularnewline
Section & \ref{sec:Swagelok-VCR} & \ref{sec:ConFlat-(CF)} & \ref{sec:KF/QF/ISO} & \ref{sec:Swagelok-tube-fittings} & \ref{sec:NPT} & \ref{sec:Flare-fitting} & \ref{sec:CGA580-(bottle-connection)}\tabularnewline
\hline 
\hline 
All-metal seal & Y & Y & N & Y & N & Y & Y\tabularnewline
Durability of seal & Y & Y & Y & N & N & N & Y\tabularnewline
Leak tight & Y & Y & Y & Y & Y & Y & Y\tabularnewline
Vacuum & Y & Y & Y & Y & Y & Y & Y\tabularnewline
Positive pressure & Y & Y & N & Y & Y & Y & Y\tabularnewline
Recommended for cryogenic & Y & Y & N & Y & N & N & N\tabularnewline
Recommended for baking & Y & Y & Y & Y & N & N & N\tabularnewline
\hline 
\end{tabular}
\par\end{centering}
\caption[Overview of fittings and their application]{Overview of fittings and their application. Note that this presents
the most common uses of each fitting. For example, KF can be used
with a metal O-ring thus create an all-metal, bakeable seal. Consult
the individual sections for details. \label{tab:Overview-of-fittings}}
\end{table}

\section{Swagelok VCR$^{\circledR}$\label{sec:Swagelok-VCR}}

Swagelok Vacuum Coupling Radiation (VCR)\nomenclature{VCR}{Vacuum Coupling Radiation}
fittings offer a high purity metal-to-metal leak-tight seal from vacuum
to positive pressures. VCR fittings are particularly useful for clean
gas system design due to their compact size and a wide variety of
parts available. Similarly to other metal-to-metal seals, VCR can
be baked to remove internal contamination. A variety of VCR face seal
glands and bodies is available with controlled surface finishes and
specially cleaned to meet ultrahigh purity system requirements. The
most common sizes of fittings are 1/4~in, 3/8~in\footnote{Fittings for 3/8~in tubing are also referred to as high flow VCR
(HVCR)\nomenclature{HVCR}{High flow VCR}. The nuts are compatible
with 1/4~in parts, except that the internal diameters of the nuts
are larger. }, and 1/2~in, referring to the outer diameter (OD)\nomenclature{OD}{Outer Diameter}
of the tubing. Fittings are also available for 1/8~in, 5/8~in, and
1~in tubing but with a smaller selection of compatible parts and
adaptors, so where possible should be avoided. 

\begin{figure}
\begin{centering}
\includegraphics[scale=0.55]{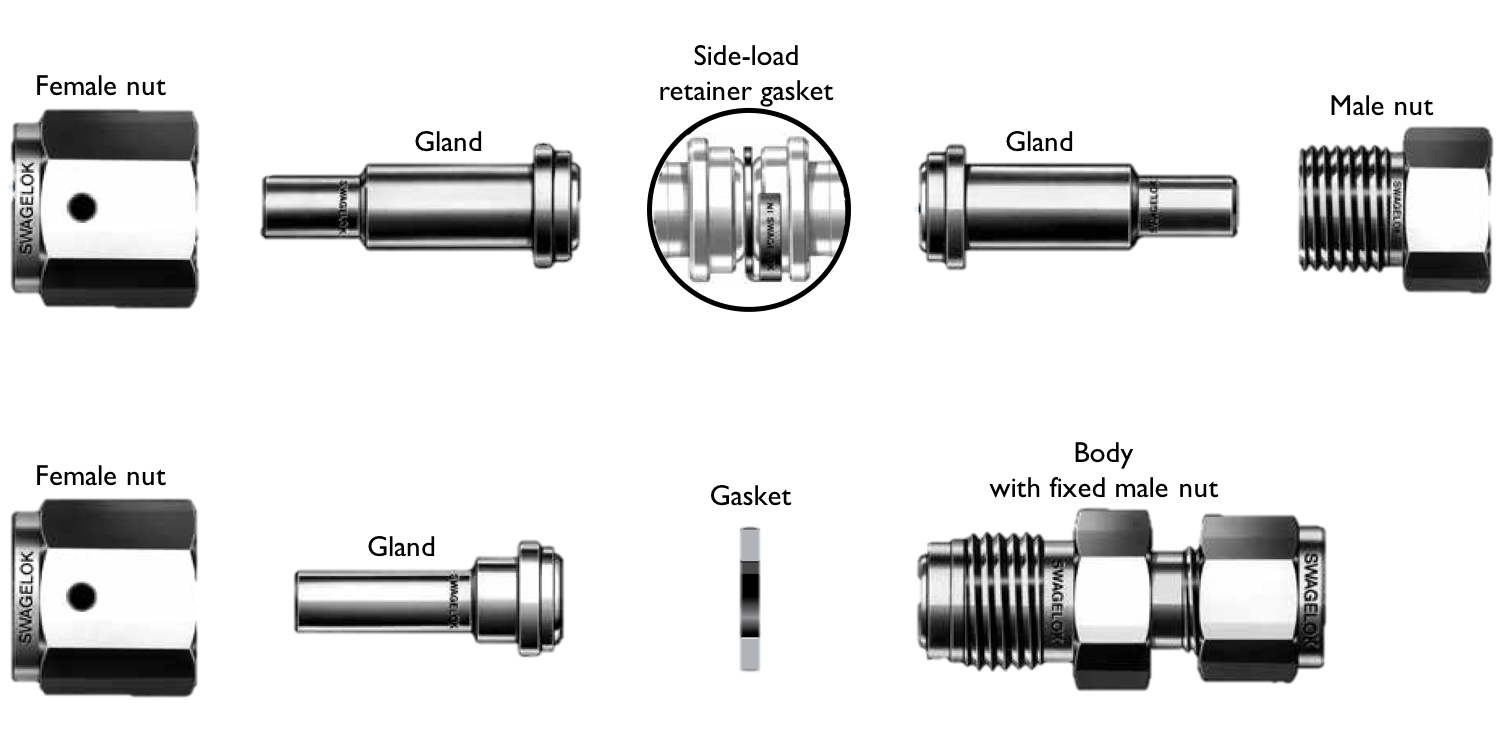}
\par\end{centering}
\caption[Illustration of components needed for VCR assembly]{Illustration of components needed for VCR assembly. \textbf{Top:}
Rotatable male and female nuts with a gasket retainer are used to
create a seal. \textbf{Bottom:} Fixed hex male body and a rotatable
female nut are used to create a seal. Figure modified from~\cite{Swagelok}.\label{fig:VCR_bodies}}
\end{figure}

Two glands compressing a gasket make the seal of a VCR assembly. The
two glands are pushed together by either a male nut or a male body
hex and a female nut or a female body hex as shown in Figure~\ref{fig:VCR_bodies}.
As the two nuts are tightened, the glands crush the gasket, which
forms a leak-tight seal.

There are several types of gaskets available made from stainless steel
(SS), silver-plated SS, copper, and nickel. Additionally, gasket retainers
are available to make insertion of gaskets between the glands in awkward
places easier. The use of silver-plated SS gasket is recommended for
uses in noble gas systems, while the use of pure SS gaskets should
be avoided. The unplated SS gasket is likely to form a permanent seal
with the glands by cold welding\footnote{Cold welding causes two clean, flat surfaces of similar metal to strongly
adhere if brought into contact under vacuum.} making the connection inseparable. For the same reason, the female
threads are silver plated to prevent galling, ensuring smooth, consistent
assembly. Therefore, chemical processes such as electropolishing and
passivation should be avoided to prevent the removal of the plating. 

\begin{figure}
\begin{centering}
\includegraphics[scale=0.58]{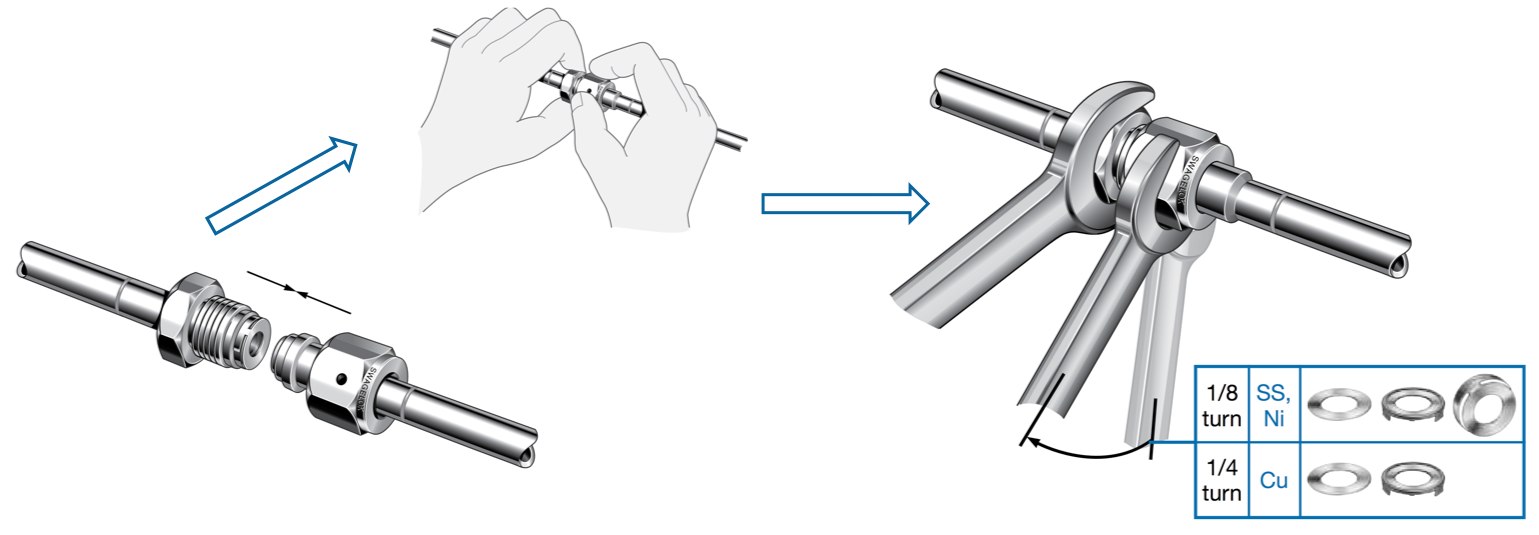}
\par\end{centering}
\caption[Illustration of steps for Swagelok VCR assembly]{Illustration of steps for Swagelok VCR assembly. Figure modified
from~\cite{Swagelok}.\label{fig:VCR-assembly}}
\end{figure}

\begin{figure}
\begin{centering}
\includegraphics[scale=0.4]{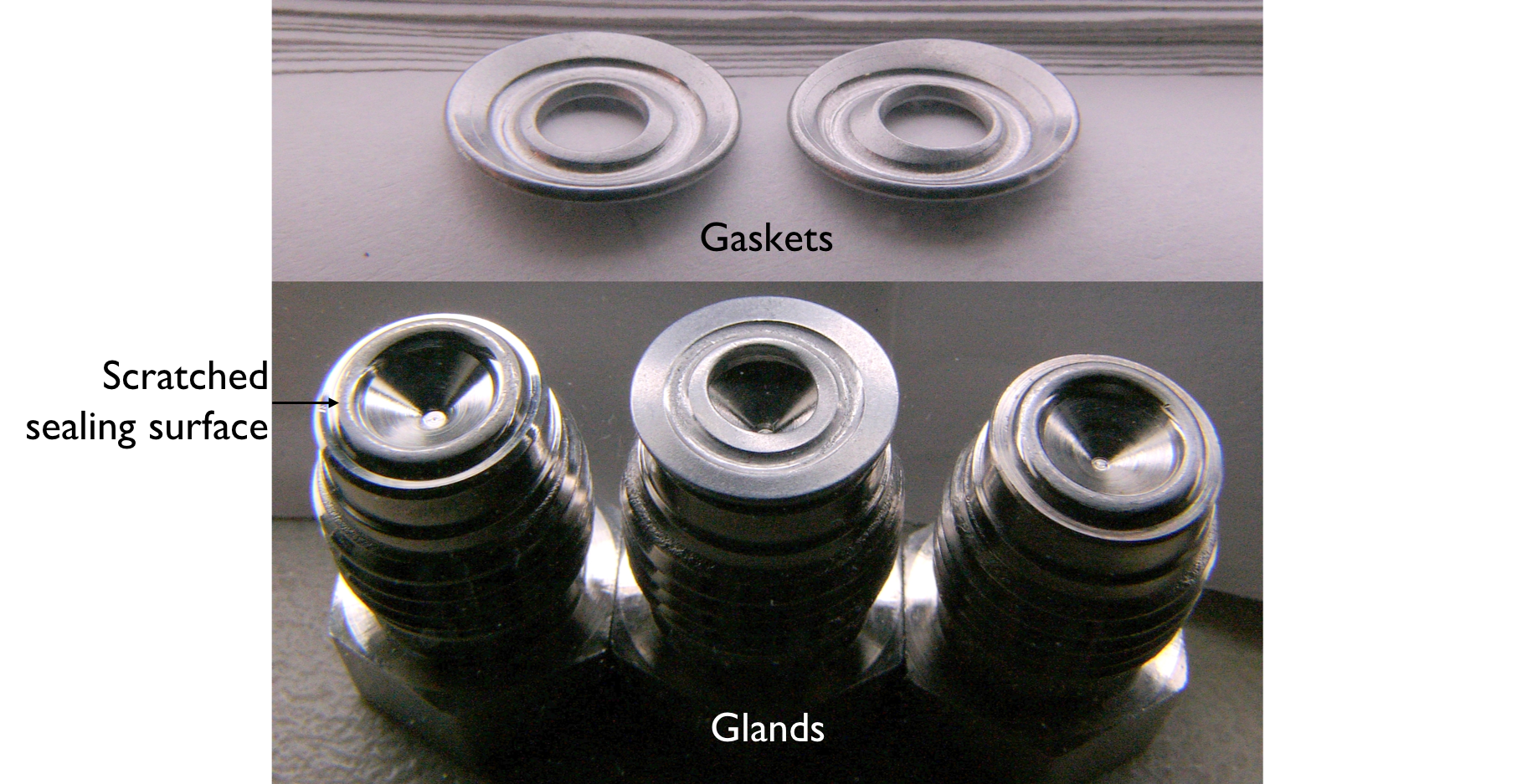}
\par\end{centering}
\caption[Failed VCR connections]{Failed VCR connections. Overtightening the connection causes damage
to the sealing surface and may cause a permanent connection between
the gasket and the gland. Photograph from Ethan Bernard.\label{fig:Failed-VCR-connections}}
\end{figure}

To make a VCR connection, the mating surfaces should first be inspected
to ensure they are free from scratches. Lint-free gloves should be
worn during VCR assembly since the slightest fingerprint or other
residues can significantly influence the quality of the seal. A gasket
is then installed between the two glands as shown in Figure~\ref{fig:VCR-assembly}.
To make the connection, at least one of the nuts needs to be free
to spin, and non-rotatable nuts must not move. The two parts should
then be threaded together until finger-tight. The final seal is made
by using two wrenches and tightening them by 1/8~turn (45 degrees)
when using a silver-plated SS gasket. Beware of overtightening the
connection as it will cause permanent damage to the sealing surface
as illustrated in Figure~\ref{fig:Failed-VCR-connections}. VCR connections
can be remade; however, a new gasket has to be used for each new connection.
The male and female parts should be well aligned before tightening
to avoid excessive strain, which can prevent the seals from being
leak-tight.

Each female nut has two test ports for easy leak testing. The VCR
fitting design can achieve a maximum leak rate of $\unit[4\times10^{-9}]{cm^{3}/s}$
with silver-plated and copper gaskets, and $\unit[4\times10^{-11}]{cm^{3}/s}$
with unplated gaskets~\cite{Swagelok_catalog}.

Note that VCR is a registered trademark of Swagelok company. Other
VCR-like products can be purchased, but the fittings may not necessarily
be compatible across different brands. Generically, the glands are
called ``face seal.''

Additionally, an orbital welding system can be purchased from Swagelok
to produce precise gas tungsten arc welds and used with VCR face seal
fittings. It provides automated autogenous welding that can be used
in the lab to provide welds with good quality and consistency in a
timely manner\footnote{Based on my experience from the XeBrA gas system construction.}.
Most importantly, one only needs a minimal amount of training to produce
high-quality welds.

\section{ConFlat (CF)\label{sec:ConFlat-(CF)}}

The ConFlat (CF) flange is a genderless design where both flanges
are identical. The flanges can be manufactured from either SS or surface
hardened aluminum. When using SS flanges with a copper gasket, the
connection creates an all-metal seal, which makes CF appropriate for
clean applications. The seal is created when a soft metal (usually
copper) gasket is inserted between the annular knife-edge grooves
of the two flanges, and a set of bolts is used to compress the flanges
together. The conformed metal of the gasket yields a leak-tight seal.
The CF seal is appropriate for vacuum applications across a wide range
of temperatures. Although CF is not officially recommended for operation
at positive pressures, they have been shown to remain leak-tight at
pressures $\mathcal{O}\left(100\right)$ bar depending on the CF diameter. 

\begin{figure}
\begin{centering}
\includegraphics[scale=0.65]{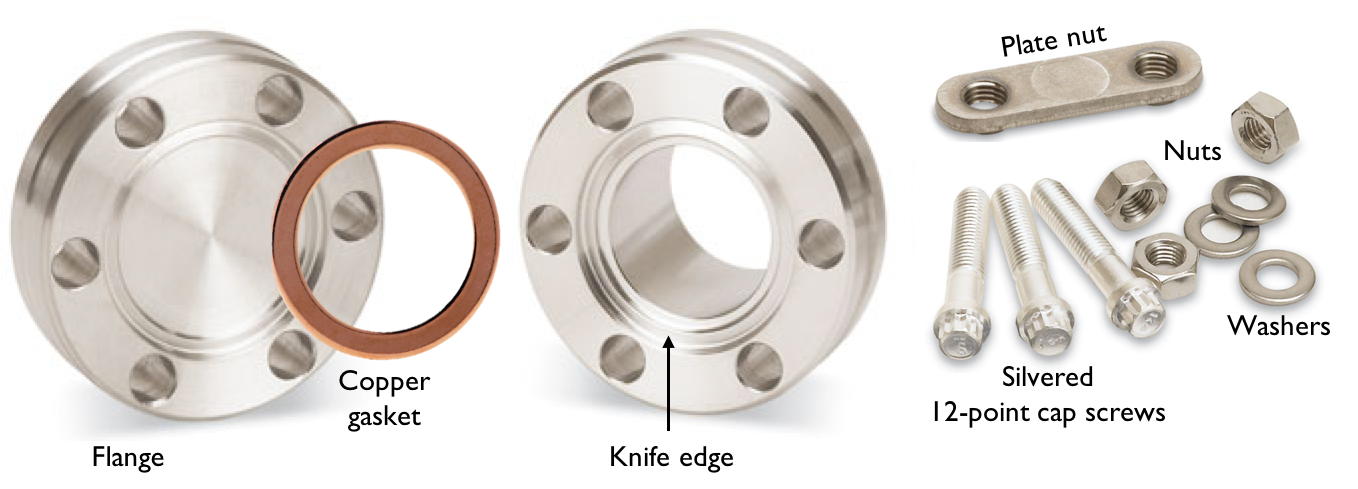}
\par\end{centering}
\caption[Illustration of components needed for CF assembly]{Illustration of components needed for CF assembly. Figure modified
from~\cite{KJLesker}. \label{fig:CF-assembly}}
\end{figure}

Flange sizes in North America\footnote{A different standard is used in Europe and Asia. }
are listed by their outside diameter. The flanges are available in
many sizes; the more common ones are 1.3~in (also called CF mini\footnote{I prefer CF adorable.}$^{,}$\footnote{The ZEPLIN experiment \cite{Lebedenko:2008gb} built a gas system
constructed entirely from CF mini.}), 2.75~in, 4.5~in, 6~in, or 8~in. CF flanges can be either fixed
or rotatable. A rotatable flange has an outer bolt ring that is free
to rotate around an inner weld ring, which provides flexibility during
assembly. Both fixed and rotatable flanges can have either clearance
holes or tapped holes. Flanges with clearance holes use nuts or plate
nuts to secure bolts going through the holes. Flanges with tapped
holes can be connected without the use of nuts, but when using a flange
with tapped holes, the other flange must have clearance holes. Figure~\ref{fig:CF-assembly}
shows all the various parts used for a CF assembly. Clearance holes
are usually preferred in designs because a damaged thread in a nut
does not require replacement of the flange.

\begin{figure}
\begin{centering}
\includegraphics[scale=0.75]{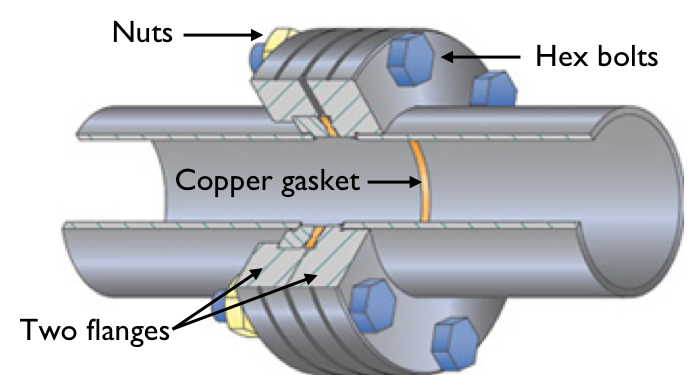}
\par\end{centering}
\caption[Schematics of a CF seal]{Schematics of a CF seal. Figure from~\cite{KJLesker}. \label{fig:CF-seal}}
\end{figure}

To make a CF connection, the mating surfaces should be carefully inspected
to ensure they are free from scratches. Lint-free gloves should be
worn during CF assembly since the slightest fingerprint or other residues
can significantly influence the quality of the seal. A new copper
gasket should be installed in the recess of the flange and then mated
to the other flange. Bolts are then inserted into all of the holes.
The flange faces should remain parallel while the bolts are finger
tightened. The bolts should be tightened one at a time by $1/4-1/2$~turn
using a crisscross pattern to the appropriate torque value, or until
the flange faces are touching as shown in Figure~\ref{fig:CF-seal}.
Recommended torque for proper assembly can be found on the manufacturer's
website. Correctly assembled flanges can yield vacuum pressures down
to $\unit[1.3\times10^{-13}]{cm^{3}/s}$~\cite{KJLesker}.

CF connections can be remade; however, a new gasket must be used for
each new connection. If frequent reassemblies are expected, silver-plated
12-point cap screws are recommended. It is strongly recommended that
all bolt, nuts, or plate nut combinations are lubricated to avoid
thread galling, either by one component being silver-plated or by
applying a suitable thread lubricant. For an assembly in a limited
space, plate nuts can be used instead of individual nuts. These are
metal plates with two threaded holes spaced apart so that they match
the corresponding bolt holes. The plate nut act as its backing wrench
so only one wrench is needed to tighten the bolts, which is desired
if the other side of the flange cannot be accessed easily. Note that
aluminum CF flanges must use either aluminum gaskets or elastomer
gaskets. The use of elastomer gaskets is not appropriate for clean
applications.

\section{KF/ISO\label{sec:KF/QF/ISO}}

\begin{figure}
\centering{}\includegraphics[scale=0.65]{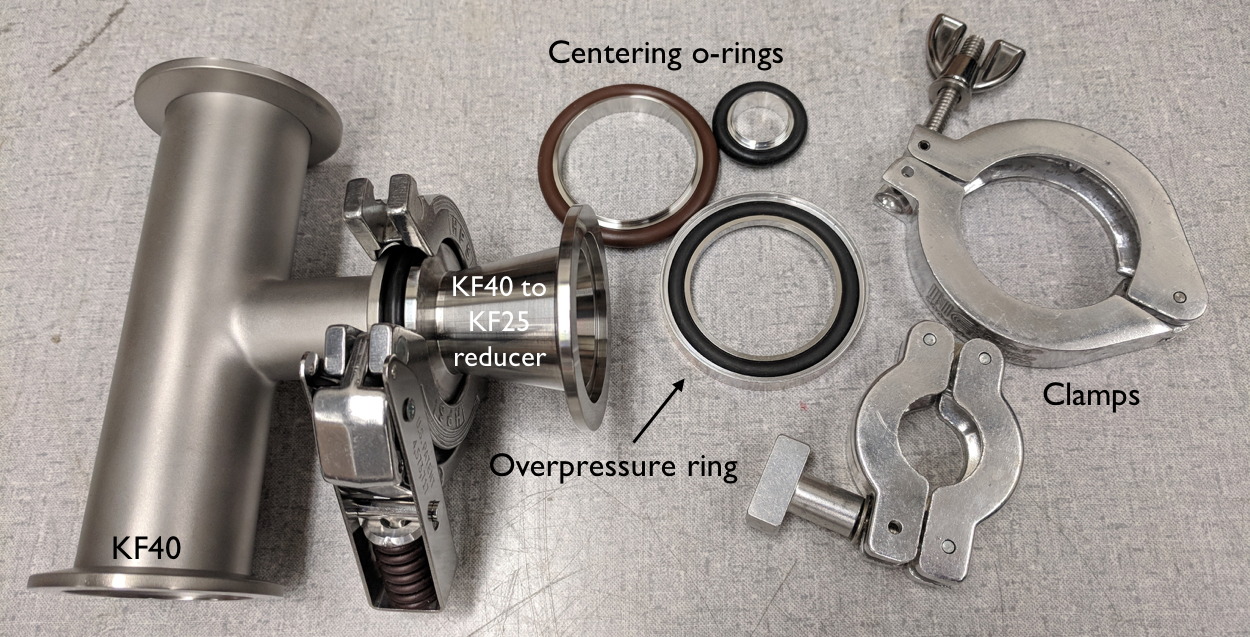}\caption[Illustration of parts needed for a KF connection]{Illustration of parts needed for a KF connection.\label{fig:KF}}
\end{figure}

The Klein Flansche (KF), also referred to as QF, connection permits
rapid fitting and replacement of components in vacuum systems. The
name KF was adopted by ISO, DIN, and other standards organizations.
The connection is made with two symmetrical genderless flanges, a
centering ring with an O-ring gasket, and a clamping ring as shown
in Figure \ref{fig:KF}. KF flanges are usually made from SS, aluminum,
or brass. The properties of the O-ring limit the operating pressure
and temperature ranges of the connection. The standards sizes are
KF10, KF16, KF25, KF40, and KF50 where the number refers to the largest
internal diameter (ID)\nomenclature{ID}{Internal Diameter} (in mm)
of a tube that can be welded to it. 

To make the connection, a metal centering ring with an elastomeric
O-ring around its OD is inserted between the two flanges. The centering
ring fits into flanges' counter-bore. A vacuum-tight connection is
made without the use of tools by merely finger-tightening the circumferential
clamping ring\footnote{The ring can have many shapes from wing-nut, thumbscrew, bolt, to
over-center lever.}. Overtightening the clamp can deform the KF flange and can lead to
galling of the screw-thread on the clamp if that style of clamp is
used.

Metal overpressure rings can be installed around the O-rings if the
system needs to operate at positive pressures. Depending on the KF
diameter, this can increase the operating range up to 7 bar. Additionally,
metal seals can be used instead of the standard elastomer O-rings.
These metal seals require high contact pressures, but achieve a helium
leak-tight seal and can be baked repeatedly up to 450$^{\circ}$C
depending on the O-ring material.

The design principles of ISO and KF seals are identical, but the ISO
standard is used for tubing with a larger diameter. Because of this
larger diameter, the ISO O-ring is more likely to roll off, so a spring-loaded
retainer holds it externally. A variety of bolt and clamp options
are used to mate ISO flanges.

\section{Swagelok tube fittings\label{sec:Swagelok-tube-fittings}}

Swagelok tube fittings are a type of compression fittings available
for all standard tubing diameters. The most appropriate use is in
applications where the fitting will not be disturbed and will not
be subjected to flexing or bending after assembly. Swagelok fittings
can be used for both metal and plastic tubing. When used with metal
tubing, Swagelok tube fitting forms an all-metal seal. Special non-metallic
ferrules and tubing inserts have to be used with plastic tubing.

A significant advantage of Swagelok tube fittings is that it allows
connection of two tubes without any welding or other work. However,
once the joint is made, it cannot be disassembled and reassembled
satisfactorily. The ferrules and part of the tube must be cut off,
and a new ferrule should be used on a new piece of tubing. This assures
a leak proof connection.

\begin{figure}[p]
\begin{centering}
\includegraphics[scale=0.4]{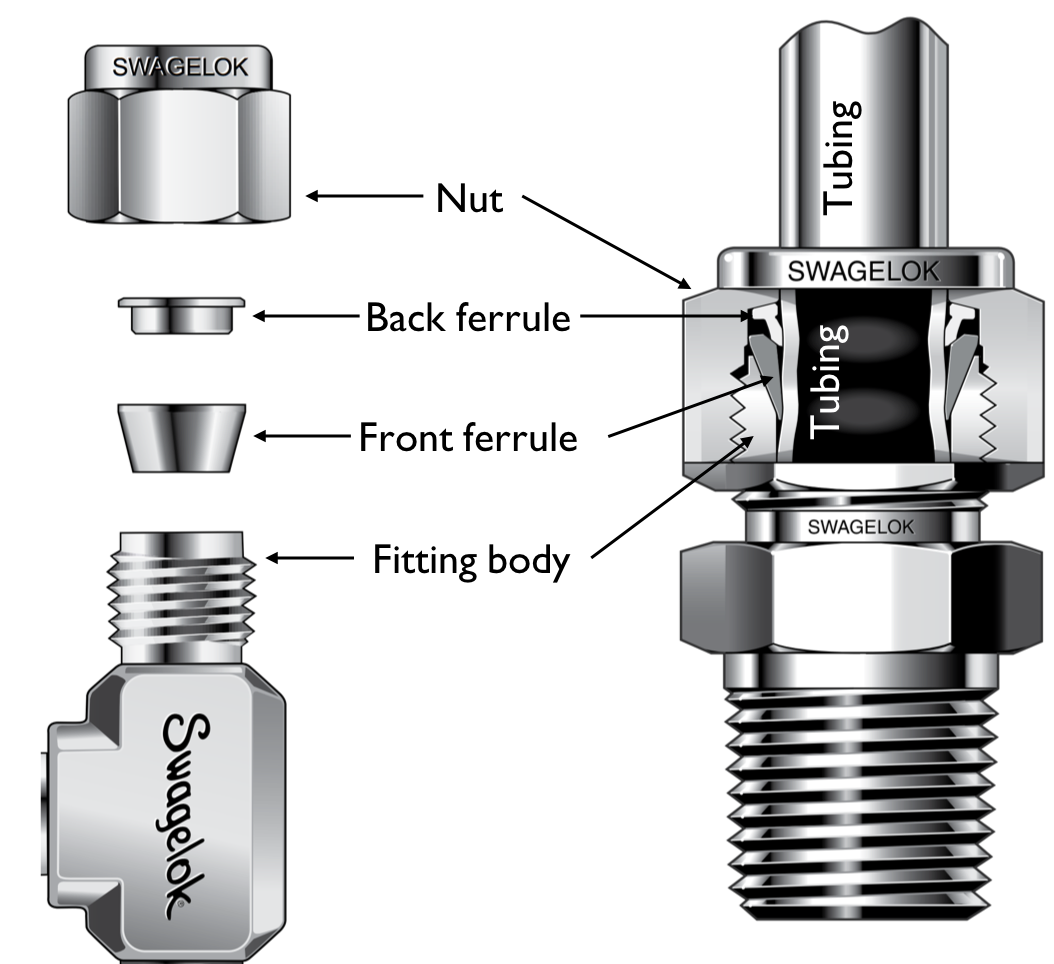}
\par\end{centering}
\caption[Two-ferrule Swagelok tube fitting]{Two-ferrule Swagelok tube fitting. The front ferrule creates a seal
against the fitting body and to the outside diameter of the tubing.
The back ferrule pushes the front ferrule forward and provides a grip
on the tubing. Figure modified from~\cite{Swagelok}.\label{fig:Two-ferrule-Swagelok}}

\vspace{4cm}
\begin{centering}
\includegraphics[scale=0.45]{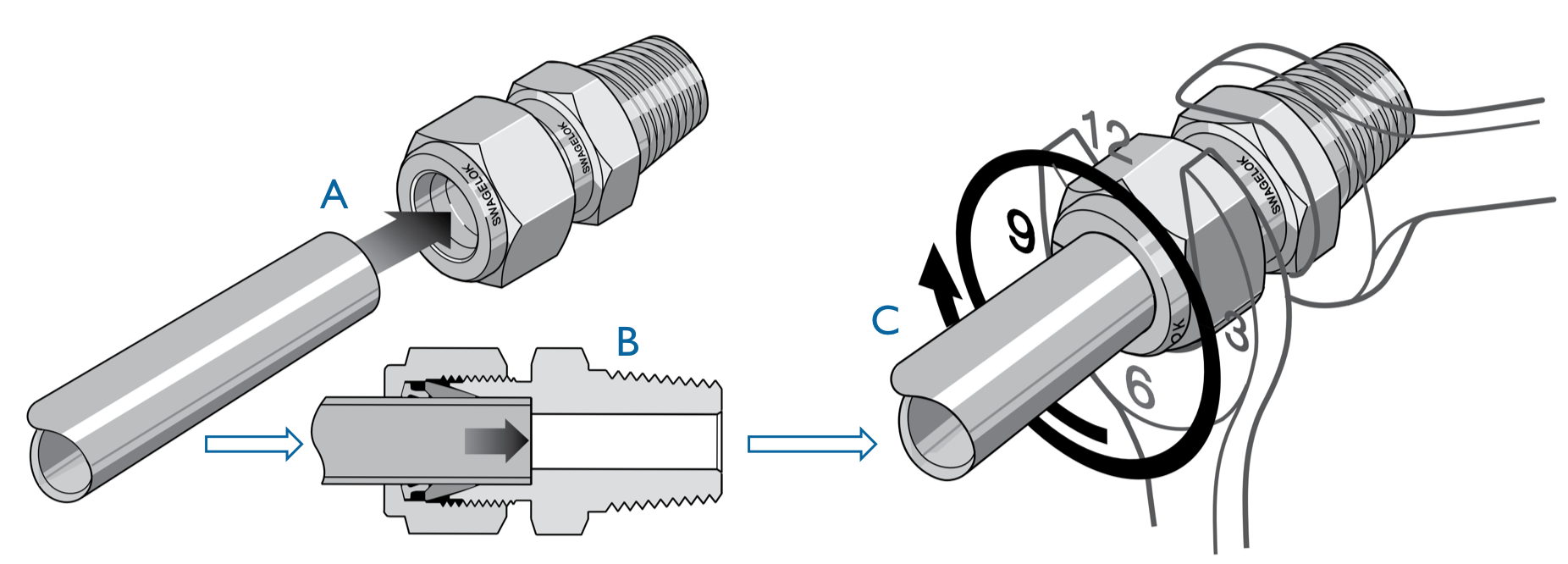}
\par\end{centering}
\caption[Illustration of steps to assemble Swagelok tube fitting]{Illustration of steps to assemble Swagelok tube fitting. First, slip
a female nut a piece of tubing (A), then press the tube against the
ferrules inside the male part (B). Rotate the nut finger-tight and
then tighten the connection using a pair of wrenches (C). Figure modified
from~\cite{Swagelok}.\label{fig:compression-assembly}}
\end{figure}

To assemble fittings for tubes below 1~in OD, first, slip a female
nut over a piece of tubing and press the tube against the ferrules
inside the male part. Then rotate the nut finger-tight. For a reliable
connection, tighten the nut until the tube will not turn by hand or
slide out of the fitting, then mark the nut at the 6~o'clock position.
Hold the male fitting body and tighten the female nut 1.25 additional
turns to the 9~o'clock position as shown in Figure~\ref{fig:compression-assembly}.
For tube fittings smaller than 3/16~in, tighten the connection by
a pair of wrenches only by an additional 3/4~turn (to the 3 o'clock
position). During the assembly, the front ferrule creates a primary
seal against the fitting body and to the outside diameter of the tubing,
which results in permanent deformation of the tubing as shown in Figure~\ref{fig:Two-ferrule-Swagelok}.
The back ferrule pushes the front ferrule forward and provides a grip
on the tubing. 

Swagelok does not recommend mixing components from various manufacturers,
but the design is generically known as a compression fitting.

\section{NPT\label{sec:NPT}}

\begin{figure}[b]
\begin{centering}
\includegraphics[scale=0.4]{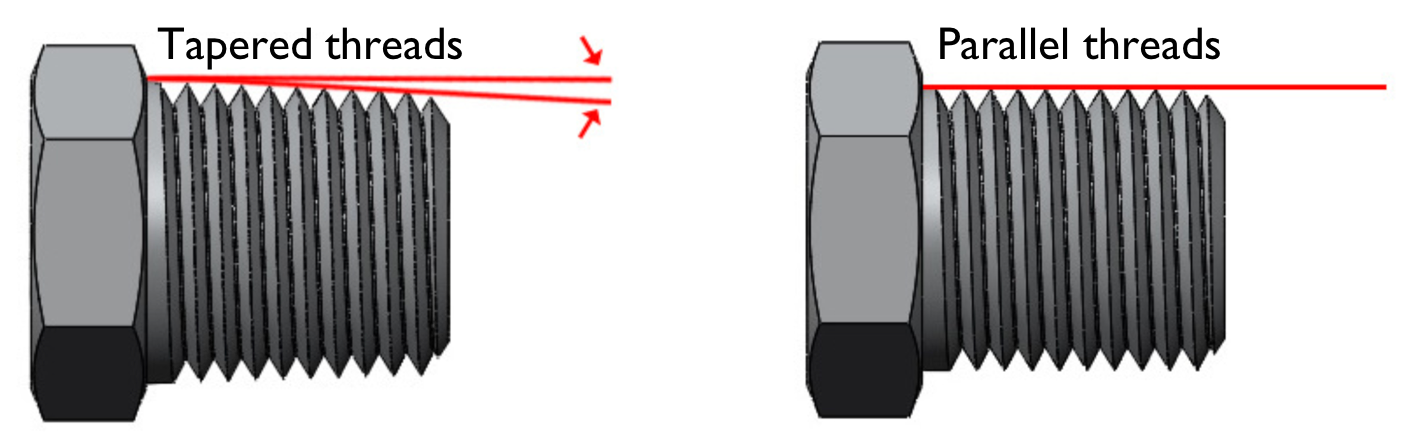}
\par\end{centering}
\caption[Tapered vs parallel threads]{Illustration of tapered (left) and parallel (right) threads. NPT
uses tapered threads. Figure from~\cite{NPT}.\label{fig:NPT}}
\end{figure}

The National Pipe Tapered (NPT)\nomenclature{NPT}{National Piper Tapered thread}
thread is a standard used in the USA to join pipes and fittings. It
is frequently used for pipes transporting liquids, gases, or steam.
Unlike most other plumbing fittings that have straight or parallel
threads, NPT fittings have a tapered thread profile illustrated in
Figure~\ref{fig:NPT}. The male thread has a larger OD at the base
of the fitting that tapers down toward the opening, while the female
part has the largest OD at the opening. 

Unlike all other fittings described in this appendix, NPT fittings
seal at the threads as the male and female parts compress against
each other. However, the threads may not align perfectly, which can
leave a spiral leakage path. To mitigate that, NPT fittings are made
leak free by applying a thread seal tape. Since the threads make the
connection, the fittings should screw in only partway before jamming.
If an NPT fitting bottoms out, the threads are either mis-tapped or
worn out, and a new part should be used in assembly.

To make an NPT connection, clean the threads and apply a thread sealant,
usually Teflon tape or thread, with the direction of the thread. Screw
the parts together until finger-tight and then use a pair of wrenches
to tighten the connection. Note that the nominal size of NPT fittings
has minimal relation to the actual thread size; an NPT fitting measuring
3/8~in at the middle of the taper is actually an 1/8~in NPT. Therefore,
a table should be referenced to identify the proper NPT fitting sizes. 

Sometimes NPT threads are referred to as MPT (Male Pipe Thread), or
MNPT for male threads, and FPT (Female Pipe Thread), or FNPT for female
threads. There is also a variant called the National Pipe Thread Tapered
Fine (NPTF), compatible with standard NPT fittings, designed to provide
a better leak-tight seal without the use of Teflon tape (although
Teflon tape can still be used). Note that British Standard Pipe (BSP)
threads are used commonly outside the USA and are not compatible with
NPT threads. 

\section{Flare fitting\label{sec:Flare-fitting}}

Flare fittings are another type of compression fitting usually used
with soft metal tubing. Flare fittings work well with natural gas
and propane fuel lines, but can occasionally be encountered in the
lab in high-pressure applications. The connection is straightforward
to assemble and disassemble, but flare fittings require modification
of the existing tube.

\begin{figure}[b]
\begin{centering}
\includegraphics[scale=0.8]{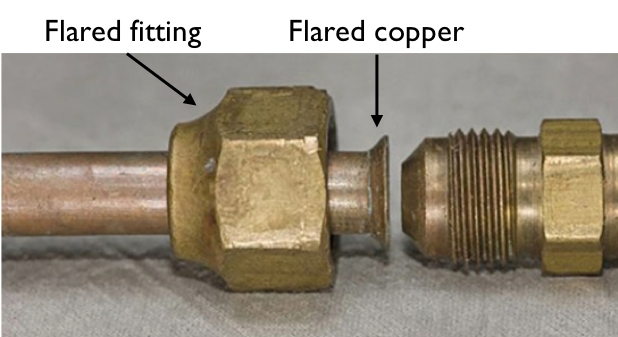}
\par\end{centering}
\caption[Flare fitting]{Flare fitting. Figure from~\cite{flare}. \label{fig:Flare-fitting}}
\end{figure}

The fitting requires the tube to be formed in a flared taper at the
end as shown in Figure~\ref{fig:Flare-fitting}. The flared taper
can be made using a special hand tool by cold working an even-cut
end of a tube. The most common flare fitting standards in use today
are the 45\textdegree{} SAE style, and the 37\textdegree{} AN style. 

During the assembly, a female flared fitting nut is slipped over the
tube and finger tightened to the male nut. Tightening the connection
using a pair of wrenches produces a high-pressure, leak-tight seal.
As usual, overtightening the nut should be avoided as it will cause
the connection to lose its shape. 

\section{CGA 580\label{sec:CGA580-(bottle-connection)}}

Compressed Gas Association (CGA)\nomenclature{CGA}{Compressed Gas Association}
580 is one of the many compressed gas cylinder outlet connections
used in USA and Canada. Different standards are used in other parts
of the world. Within the USA, slightly different standards are used
for various gasses in various cylinders and are differentiated by
a numeral; for example, CGA 540 is used for oxygen, while CGA 300
and 510 are used for acetylene. CGA 580 is used in particular for
inert gasses such as helium, neon, argon, krypton, xenon, and nitrogen.
Note that CGA connections can have either right- or left-hand threads
depending on the contents of the cylinder. 

\begin{figure}[h]
\begin{centering}
\includegraphics[scale=0.65]{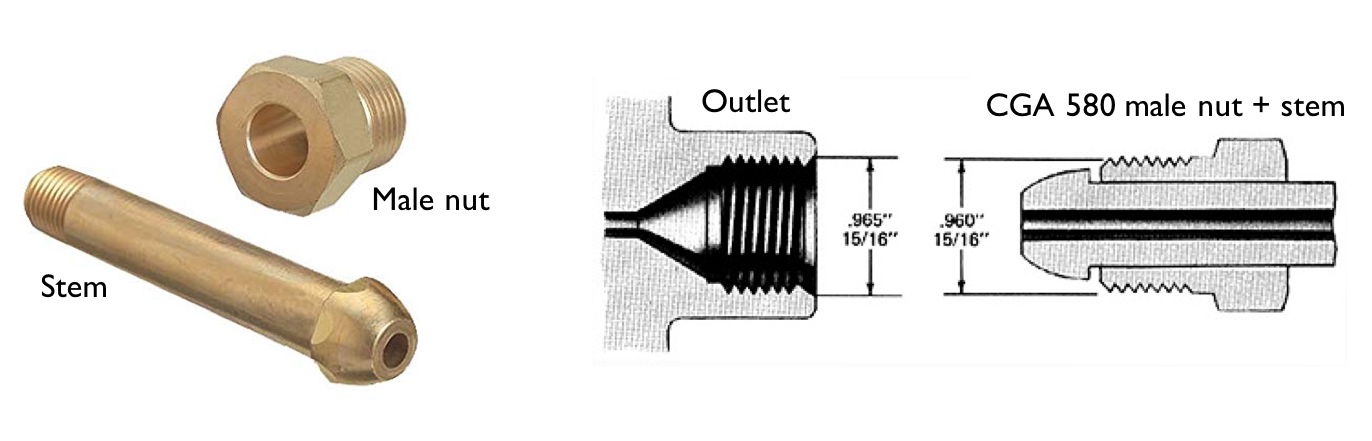}
\par\end{centering}
\caption[CGA 580 connection]{Illustration of a CGA 580 connection. The male nut is free to rotate
around the stem. When the stem is pressed into the outlet, it creates
a seal. Figure modified from~\cite{CGA}. \label{fig:CGA580}}

\end{figure}

A connection is made when a male nut presses the stem into the cylinder
outlet as illustrated in Figure~\ref{fig:CGA580}. The connection
should be tightened using a pair of wrenches.

\chapter{\textsc{XeBrA xenon condensing, circulation, and recovery procedure}\label{chap:XeBrA-procedures}}

This document outlines the procedure for condensation, circulation,
and recovery of xenon in XeBrA. The detector is located at LBNL in
room 70A-2263. The procedure assumes two bottles of xenon gas are
attached to the gas system, and two recovery vessels in room 70A-2255
are connected to the gas system. 

To begin a xenon fill, the gas system and detector are evacuated and
prepared for cool down. A small amount of gas is introduced into the
detector, and a constant pressure of gas is then maintained inside
the detector while it is being cooled. Eventually, xenon starts condensing
and fills the active volume. This process takes $\sim18-48$ hours.
Circulation should be started once the detector is full of liquid
xenon to start removing impurities. Once a stable detector state is
achieved, the detector is ready for data taking.

Since a stable detector state involves a constant stream of bubbles,
it is necessary to slowly keep adding xenon during data taking to
suppress bubbling as outlined below. 

\begin{figure}[p]
\centering{}\includegraphics[angle=90,scale=0.9]{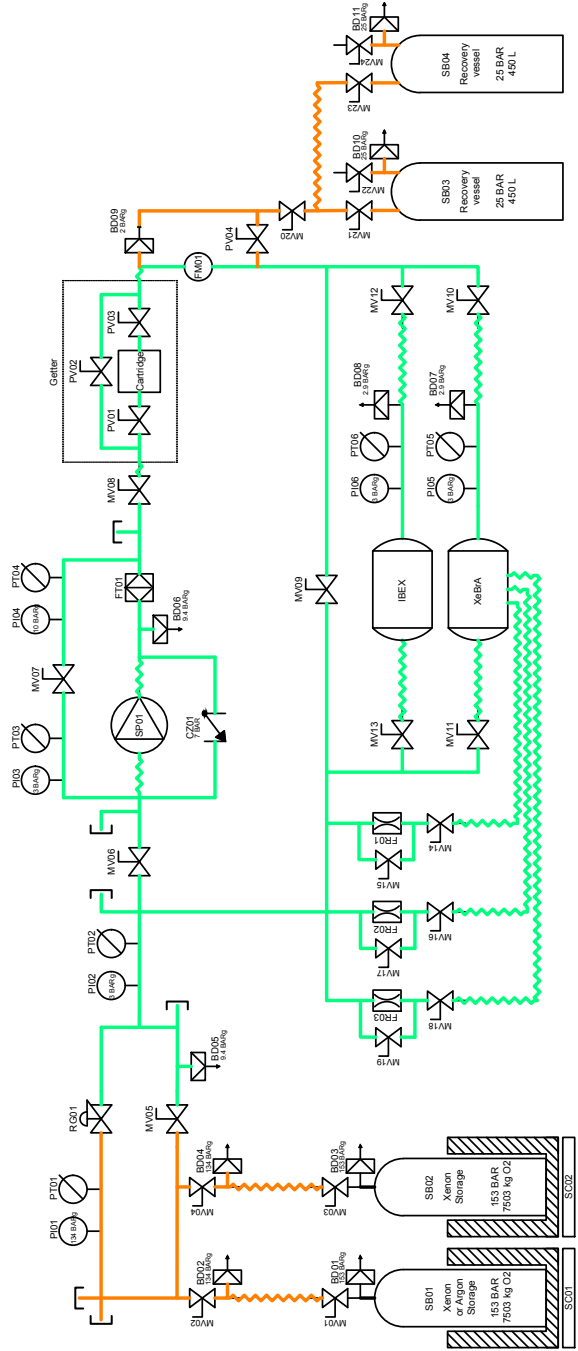}\caption[Gas system P\&ID]{XeBrA's gas system P\&ID.\label{fig:XeBrA's-PID-appendix}}
\end{figure}

The P\&ID can be found on the gas system panel and in Figure~\ref{fig:XeBrA's-PID-appendix}. 

The following steps are performed during each cool down: 
\begin{enumerate}
\item Leak checking
\item System evacuation
\item Introduction of xenon gas to the detector 
\item Detector cool down and xenon condensation 
\item Xenon circulation 
\item Data taking 
\item Xenon recovery 
\end{enumerate}
This document borrows from several LUX Critical Procedures: the LUX
Detector Cooling Procedure, the LUX Condensing Procedure, and the
LUX Circulation Critical Procedure.

\section{Leak checking }

This section outlines the procedure for leak checking the detector\textquoteright s
inner volume, the gas system, and the recovery vessels. Before using
any helium for leak checking, turn on a big fan to prevent helium
pile up in the room.

\textcolor{red}{Do not operate pump SP01 under vacuum. The pressure
ratio on input/output should never be larger than 5 (measured in absolute,
not gauge, pressure). }

\subsection{Gas system leak check}
\begin{enumerate}
\item Ensure manual valves (MV)\nomenclature{MV}{Manual Valve} MV01 and
MV03 are closed to disconnect storage bottles. 
\item Open regulator (RG)\nomenclature{RG}{Regulator} RG01. 
\item Close MV10, MV11, MV12 and purge lines MV14, MV16 and MV18 to isolate
XeBrA. 
\item Close MV13 (connection to the IBEX experiment). 
\item Attach leak checker to port behind MV12. 
\item Open MV02, MV04, MV05, MV06, MV07, MV08, MV09, MV15, MV17, and MV19
to open the main circulation path. 
\item Open MV12. 
\item Start pumping. 
\item Ensure the getter heater is off. As the pump is spinning up, switch
the state of pneumatic valves (PV)\nomenclature{PV}{Pneumatic Valve}
PV01, PV02, and PV03 a couple of times to release trapped air by using
the \textquotedblleft valve control\textquotedblright{} button on
the front of the getter housing. 
\item Perform leak checking. 
\item While leak checking the getter, ensure to switch the states once to
check all connections. 
\item Close MV12. 
\item Turn off and disconnect the leak checker. 
\end{enumerate}

\subsection{Circulation pump purge}

\begin{figure}
\begin{centering}
\includegraphics[scale=0.4]{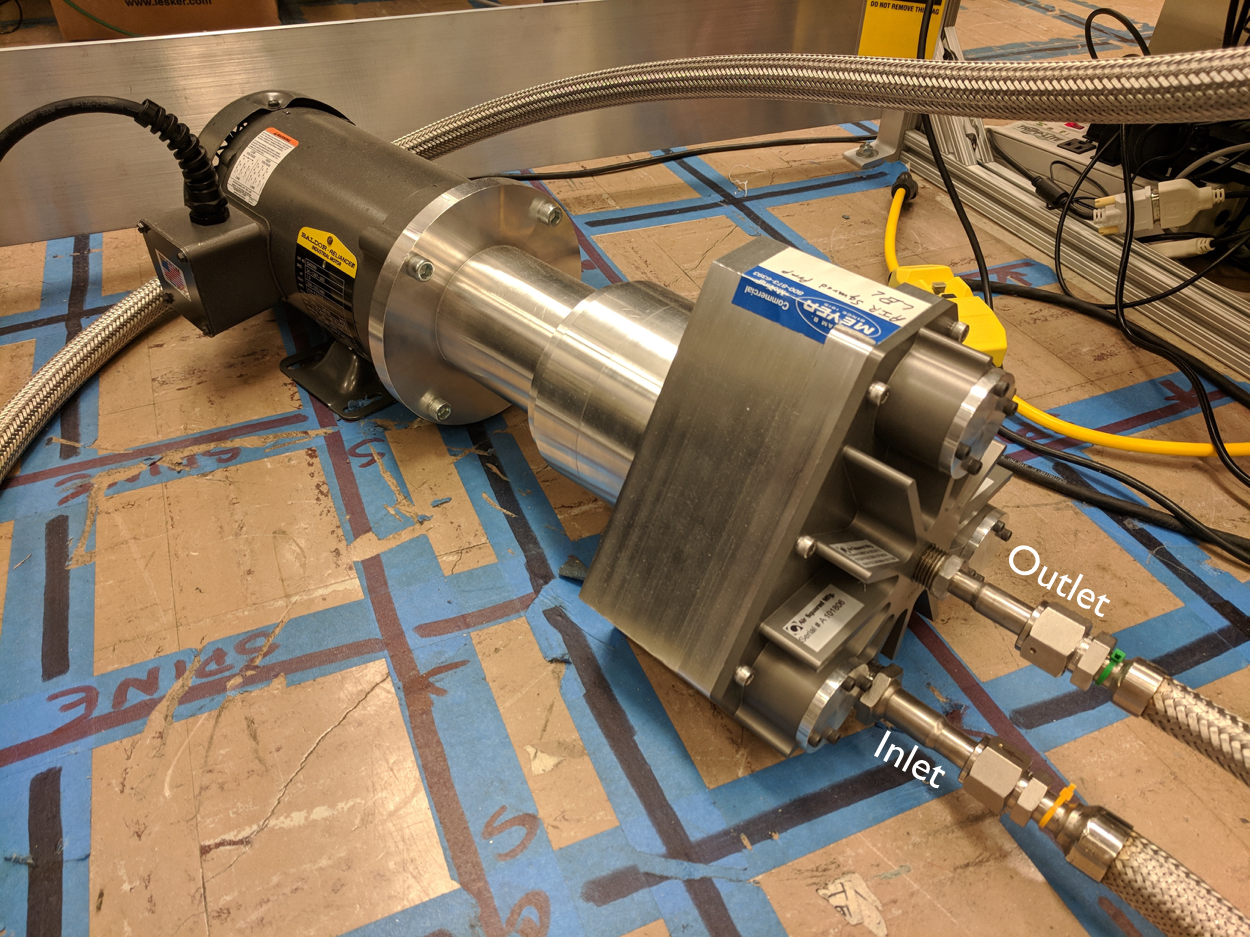}
\par\end{centering}
\caption[Circulation pump]{Circulation pump.\label{fig:Circulation-pump}}

\end{figure}

The circulation pump is shown in Figure~\ref{fig:Circulation-pump}.
If the pump has not been exposed to air since previous detector fill,
this step can be skipped. Purging the pump with clean argon gas ensures
that any potential trapped volumes will be cleaned. 
\begin{enumerate}
\item Attach argon gas cylinder with at least 99.999\% purity as storage
bottle (SB)\nomenclature{SB}{Storage Bottle} SB01. 
\item To ensure the getter is in the bypass mode (PV02 open), switch the
getter bypass mode off using the \textquotedblleft valve control\textquotedblright{}
button on the front of the getter housing. 
\item Close MV07, MV10, MV11, MV12 and MV14, MV16 and MV18. 
\item Open a circulation path by opening MV06, MV08, MV09. 
\item Fill the gas system with argon up to 1 barg (anything above 0 barg
works). Follow the procedure for opening and closing xenon fill path
from the gas bottle (Section~\ref{subsec:Opening-xenon-fill-path}). 
\item Run circulation pump SP01 at 5 SLPM or less for a minute. 
\item Turn off circulation scroll pump (SP)\nomenclature{SP}{Scroll Pump}
SP01. 
\item Open MV12 and carefully vent argon down to $\sim0.1$ barg. 
\item Close MV12. 
\item Attach vacuum pump to MV12. 
\item Open MV12. 
\item Using vacuum pump attached to MV12, pump out the argon. MV01, MV02,
MV05, MV06, MV07, MV08, MV15, MV17, and MV19 should be open. MV04,
MV10, MV11, MV12, MV13, MV14, MV16, and MV18 should be closed. 
\item Close MV12. 
\end{enumerate}

\subsection{XeBrA main volume leak check }
\begin{figure}
\begin{centering}
\includegraphics[scale=0.6]{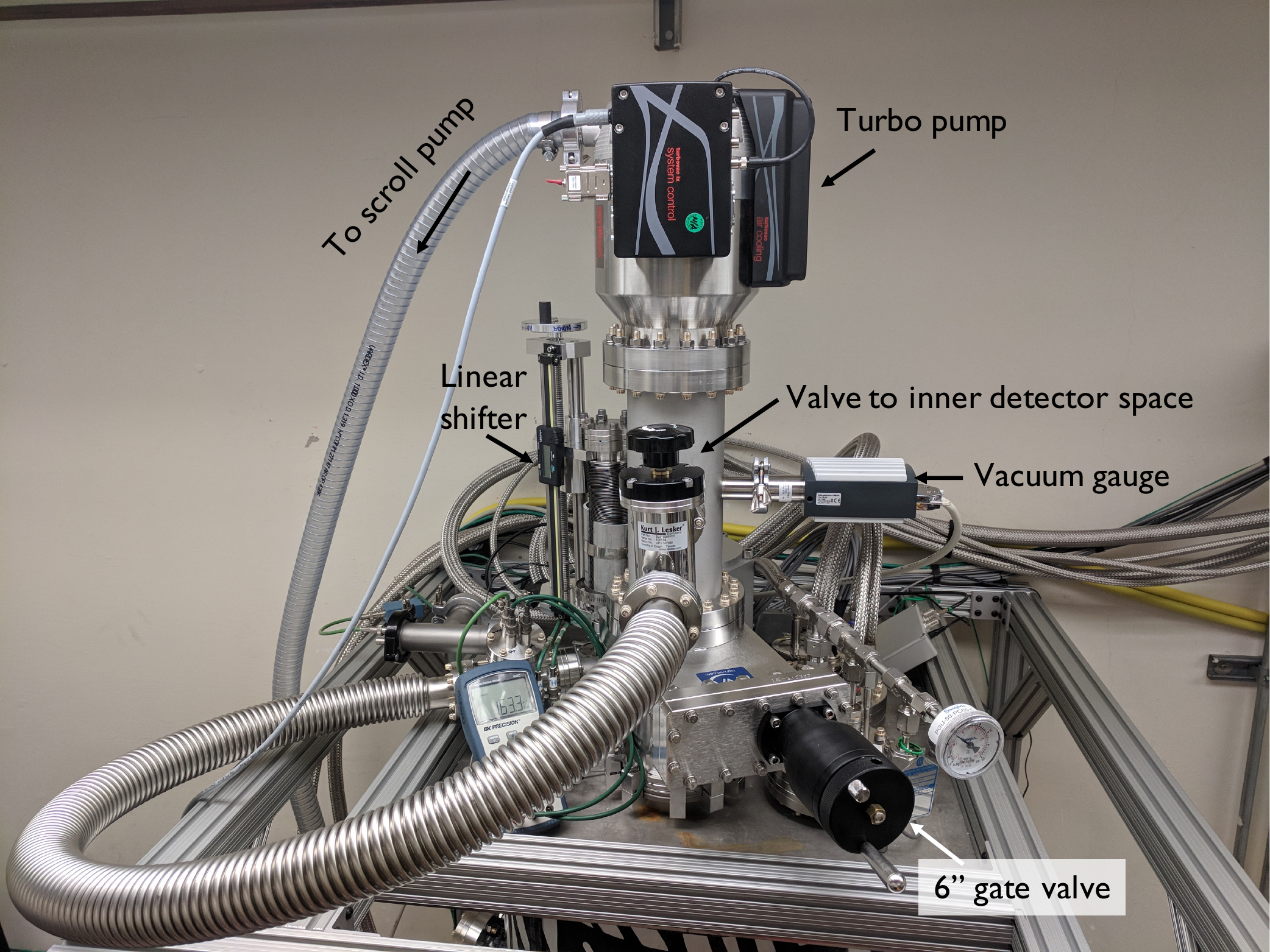}
\par\end{centering}
\caption[Annotated photograph of the top of the XeBrA test stand]{Annotated photograph of the top of the XeBrA test stand.\label{fig:xebra_top}}
\end{figure}

\begin{figure}
\begin{centering}
\includegraphics[scale=0.7]{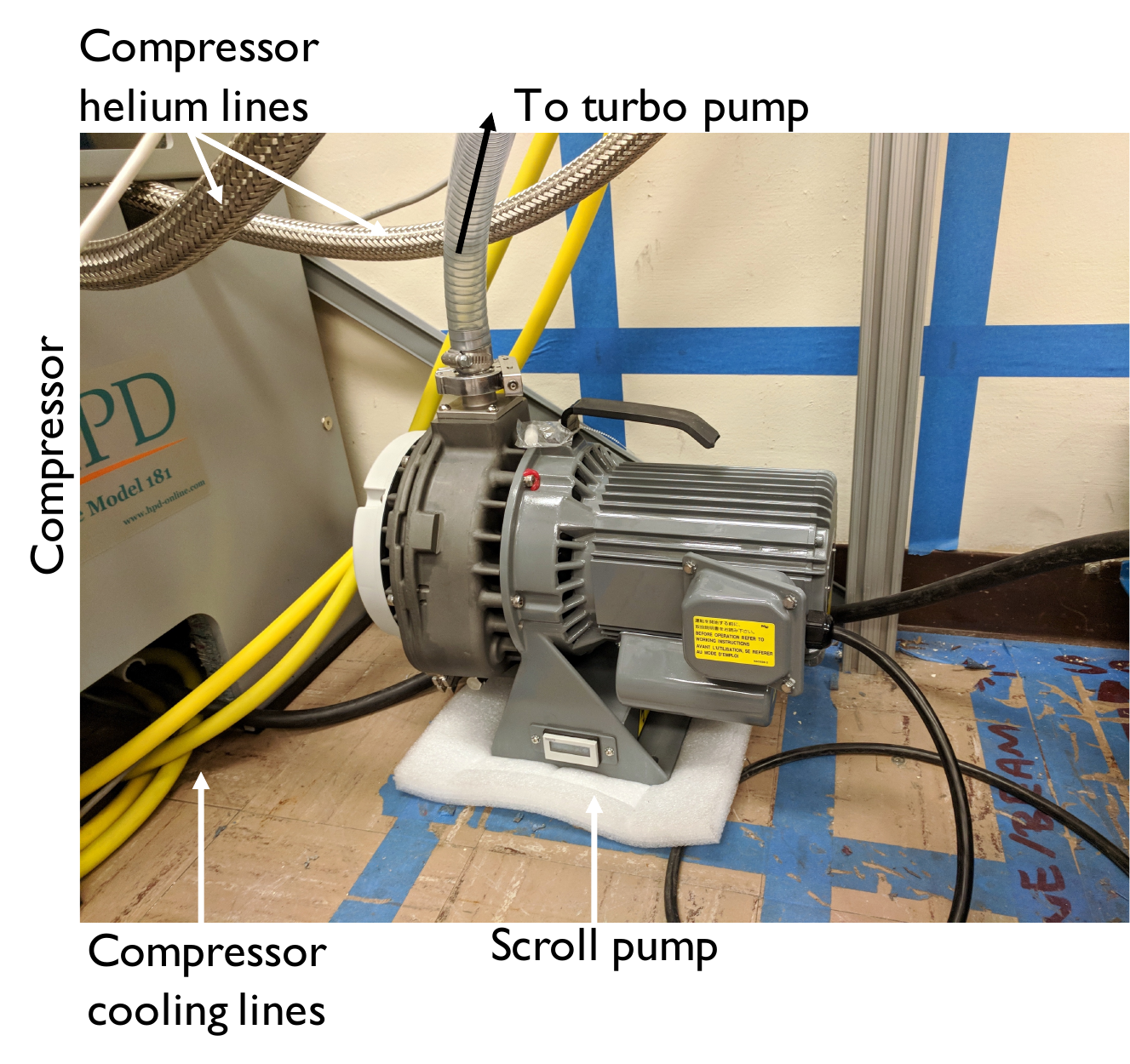}
\par\end{centering}
\caption[XeBrA's scroll pump and compressor]{XeBrA's scroll pump and compressor. \label{fig:XeBrA's-scroll-pump} }

\end{figure}

\begin{enumerate}
\item Ensure MV10, MV11, and MV14, MV16, and MV18 are closed to isolate
XeBrA. 
\item Close the 6\textquotedblright{} gate valve (see Figure~\ref{fig:xebra_top}
for location). 
\item Make sure the turbo pump is off. 
\item Attach the leak checker in place of the scroll pump backing XeBrA\textquoteright s
turbo pump (see Figure~\ref{fig:XeBrA's-scroll-pump}). 
\item Leak check. 
\item Disconnect leak checker. 
\end{enumerate}

\subsection{Xenon recovery leak check}
\begin{enumerate}
\item Ensure PV04 is closed. 
\item Open MV20, MV21, and MV23. 
\item Leak check each of the recovery vessels separately. 
\begin{enumerate}
\item To leak check SB03 (Rhino) connect a pump and a leak checker to MV22.
Close MV23, open MV22 and perform the leak check. Close MV22 and disconnect
the leak checker. 
\item To leak check SB04 (Hippo) attach a pump and a leak checker to MV24.
Close MV23, open MV24 and perform the leak check. Close MV24 and disconnect
the leak checker. 
\end{enumerate}
\end{enumerate}

\section{Evacuating XeBrA\label{sec:Evacuating-XeBrA}}

This section outlines the procedure for evacuating the entire volume
that will be filled with xenon. It is to be performed before every
detector cool down. 
\begin{enumerate}
\item Attach two xenon bottles as SB01 and SB02. 
\item Ensure MV01 and MV03 are closed to isolate storage bottles. 
\item Open MV02, MV04, MV05, MV06, MV07, MV08, MV09, MV10, MV11, MV14, MV15,
MV16, MV17, MV18, and MV19. This opens the entire detector volume. 
\item Open RG01 all the way. 
\item Close and tag MV13 (connection to IBEX). 
\item Close and tag MV12 (leak checking port). 
\item Close the 6\textquotedblright{} gate valve (see Figure~\ref{fig:xebra_top}
for location). 
\item Open valve to inner space (see Figure~\ref{fig:xebra_top} for location). 
\item Turn on XeBrA\textquoteright s scroll pump and once the vacuum has
reached 0.1~mbar turn on the turbo pump. 
\item Pump out the entire detector and gas system volume for at least a
day. 
\item To pump out the getter, switch the getter bypass mode off using the
\textquotedblleft valve control\textquotedblright{} button on the
front of the getter housing. Make sure the getter heater is off. After
10~minutes, enable the bypass mode again using the same button. 
\item Pump until the vacuum pressure is lower than $1\times10^{-6}$~mbar
as read by the vacuum gauge near the turbo pump. Since the gauge is
so close to the pump, the pressure in the rest of the volume will
be higher, but sufficiently low. 
\item Close the turbo pump\textquoteright s access to inner space. 
\item Turn off the turbo pump and let it spin down. 
\item Once the turbo has spun down, start pumping on the outer vacuum vessel
(OV). 
\item Open the 6\textquotedblright{} gate valve and ensure that the valve
to inner space (see Figure~\ref{fig:xebra_top} for locations) is
closed. 
\item Turn on the scroll pump, and once the vacuum has reach 0.1~mbar turn
on the turbo pump. 
\item Leave the scroll and the turbo pumping on the OV throughout xenon
operations. 
\item Separately, evacuate the two recovery vessels SB03 and SB04. Make
sure that PV04 is closed. Attach a pump to MV22 and evacuate SB03
to least $1\times10^{-2}$~mbar. Then attach a pump to MV24 and evacuate
SB04 to the same pressure. If the vessels have been exposed to air
between data runs, purge the vessels with argon gas prior to evacuating
them to prevent contamination of potentially recovered xenon. 
\end{enumerate}

\section{Introduction of xenon gas to the detector}

This section introduces an initial amount of xenon gas to facilitate
detector cooling. It assumes the detector has been leak checked and
evacuated.

\textcolor{red}{Exceeding 2.0 barg (29 psig) in the detector will
result in the rupture of a burst disc (BD)\nomenclature{BD}{Burst Disc}
BD07. If there is a severe problem during the condensing procedure
follow the xenon recovery procedure in Section}~\textcolor{red}{\ref{sec:Xenon-recovery}. }

\subsection{Emergency precautions}

This section describes precautions which must be in place whenever
there is liquid xenon in the detector making overpressure possible. 
\begin{enumerate}
\item Turn on to the Slow Control and verify it is functional. 
\item Turn on email forwarding of alarms. 
\item Set appropriate alarms as outlined in Section~\ref{subsec:Alarm-settings}. 
\begin{enumerate}
\item Ensure that high high alarms are at least 20\% below the nearest burst
disc value. 
\end{enumerate}
\item Appoint trained personnel with reliable communication devices to be
on call within 1~hour of travel time while the detector contains
any liquid xenon. 
\item The Oxygen Deficiency Monitor (ODH) must be operational. 
\begin{enumerate}
\item There is a remote sensor in 70A-2263 and a local sensor inside the
main control unit 70A-2255. 
\end{enumerate}
\item Ensure PV04 is closed and MV20, MV21, and MV23 are opened for a clear
path to the recovery vessels. 
\begin{enumerate}
\item If needed, PV04 can be opened in Slow Control. It is password protected
to avoid accidentally opening the valve.
\begin{enumerate}
\item Password: XeBrA
\end{enumerate}
\end{enumerate}
\end{enumerate}

\subsubsection{SciNote logs }

Make sure to log important milestones in SciNote, the electronic lab
notebook~\cite{SciNote}. Items to log include the amount of xenon
left in SB01 \& SB02, steady-state settings of the circulation pump
(frequency setting and the resulting flow rate), and any other actions
that differ from the regular procedures. 

\subsection{Gas system preparation\label{subsec:Gas-system-preparation}}

In this section, we ensure the gas system is in a known state, with
appropriate valves locked to ensure that high pressure gas does not
enter and break the hardware in the rest of the gas system. 
\begin{enumerate}
\item Ensure MV01, MV02, MV03, and MV04 are closed to isolate the xenon
storage bottles and prevent the introduction of xenon into the system
before it is in the correct state. 
\item To prevent high pressure from being accidentally introduced to the
low pressure portion of the gas system, valve MV05 must be closed
and tagged. 
\item MV12 should be covered and tagged closed. 
\item Close MV13 (connection to IBEX). 
\item Ensure PV04 is closed and BD09 is intact (these open to the recovery
line leading to 70A-2255). 
\item Ensure MV20, MV21 and MV23 are opened to connect recovery vessels
properly. 
\item Ensure regulator RG01 is fully closed (i.e., in the zero-pressure
position). 
\item Ensure the Slow Control is properly activated and email forwarding
of alarms is enabled. 
\item Start slow data logging using the Slow Control (a 20 s save interval
should be sufficient). 
\item Ensure the outer vacuum volume is under vacuum maintained by the main
turbo and scroll pumps as established at the end of the evacuating
XeBrA procedure (Section~\ref{sec:Evacuating-XeBrA}). 
\item Ensure that recovery vessels SB03 and SB04 (Hippo and Rhino) have
been evacuated to at least $1\times10^{-2}$ mbar. 
\end{enumerate}

\subsection{Initial xenon gas into detector }

This procedure assumes the xenon volume has been evacuated. The fill
is done using SB01. This is to be done while the detector is still
warm.

\begin{figure}
\begin{centering}
\includegraphics[scale=0.5]{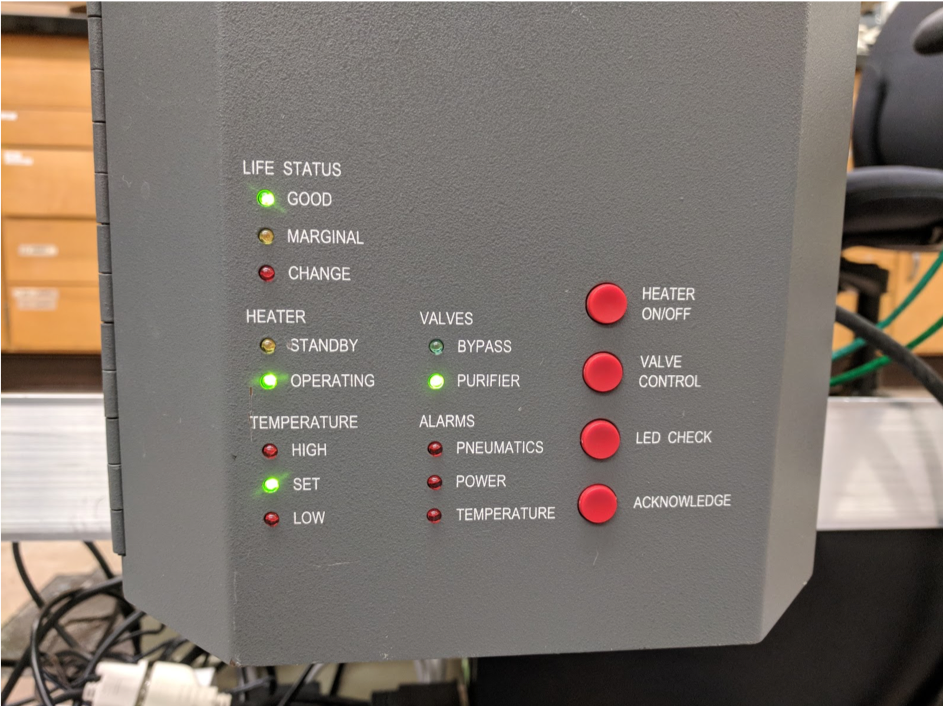}
\par\end{centering}
\caption[Getter settings during operation]{Getter settings during operation. ``Valves $\rightarrow$ Purifier''
light should be green and ``Heater $\rightarrow$ Operating'' light
should be blinking while the heater is on. \label{fig:Getter-settings}}

\end{figure}

\textcolor{red}{Exceeding 2.0 barg (29 psig) in the detector will
result in the rupture of a burst disc BD07.}
\begin{enumerate}
\item Log the initial amount of gas in the bottle to SciNote by noting the
weight of SC01. 
\item Follow the gas system preparation procedure (Section~\ref{subsec:Gas-system-preparation}). 
\item Close MV09, MV11, MV13, and MV14 through MV19. 
\item Close and tag MV04 and MV12. 
\item Open MV06, MV07, MV08, MV10. 
\item Turn on the getter using the indicator LEDs and buttons on the getter
front panel. The getter should be warm for 1 hour before gas introduction. 
\item Make sure getter is not bypassed. 
\begin{enumerate}
\item The correct setting for a getter in operation is illustrated in Figure~\ref{fig:Getter-settings}. 
\end{enumerate}
\item Introduce xenon gas into the system by following the opening xenon
fill path procedure (Section~\ref{subsec:Opening-xenon-fill-path}). 
\item Wait until the detector is full. This should only take a few minutes.
Pressure indicator (PI)\nomenclature{PI}{Pressure Gauge} PI05 and
pressure transducer (PT)\nomenclature{PT}{Pressure Transducer} PT05
should stabilize at the desired fill pressure of 0.5 barg. 
\item Once the desired pressure is reached, follow the procedure for closing
the xenon fill path (Section~\ref{subsec:Closing-xenon-fill-path})
unless you are proceeding directly to the detector cooling (Section~\ref{subsec:Detector-cooling}).
\end{enumerate}

\subsubsection{Opening the xenon filling path from the gas bottle\label{subsec:Opening-xenon-fill-path}}
\begin{enumerate}
\item Ensure RG01 is fully closed (i.e., in the zero-pressure position). 
\item Close and tag MV05. 
\item Log the xenon mass in the bottle in SciNote. 
\item Open MV01 and MV02. 
\item This opens the path from the xenon gas storage bottle to the gas system
panel. 
\item Slowly open RG01 to let the desired amount of gas into the low pressure
part of the gas system. Watch PI02/PT02 and PI05/PT05 for the desired
pressure. 
\end{enumerate}

\subsubsection{Closing the xenon filling path from the gas bottle \label{subsec:Closing-xenon-fill-path}}
\begin{enumerate}
\item Log the xenon mass in the bottle in SciNote. 
\item Close the bottle with MV01. 
\item Slowly close RG01. 
\item Close MV02. 
\begin{enumerate}
\item There might be pressure build up by PT01. This pressure will slowly
leak into the detector through RG01 afterward and is expected. The
pressure decrease over time indicated by PT01 does not (necessarily)
mean there is a leak in that section of the gas system. 
\end{enumerate}
\end{enumerate}

\section{Detector cooling \label{subsec:Detector-cooling}}

Keep a close eye on temperatures all over the detector throughout
this procedure. This section assumes the system is filled with 0.5
barg (7.3 psig) of xenon gas. Be sure to log important milestones
in SciNote.

\begin{figure}[t]
\begin{centering}
\includegraphics[scale=0.55]{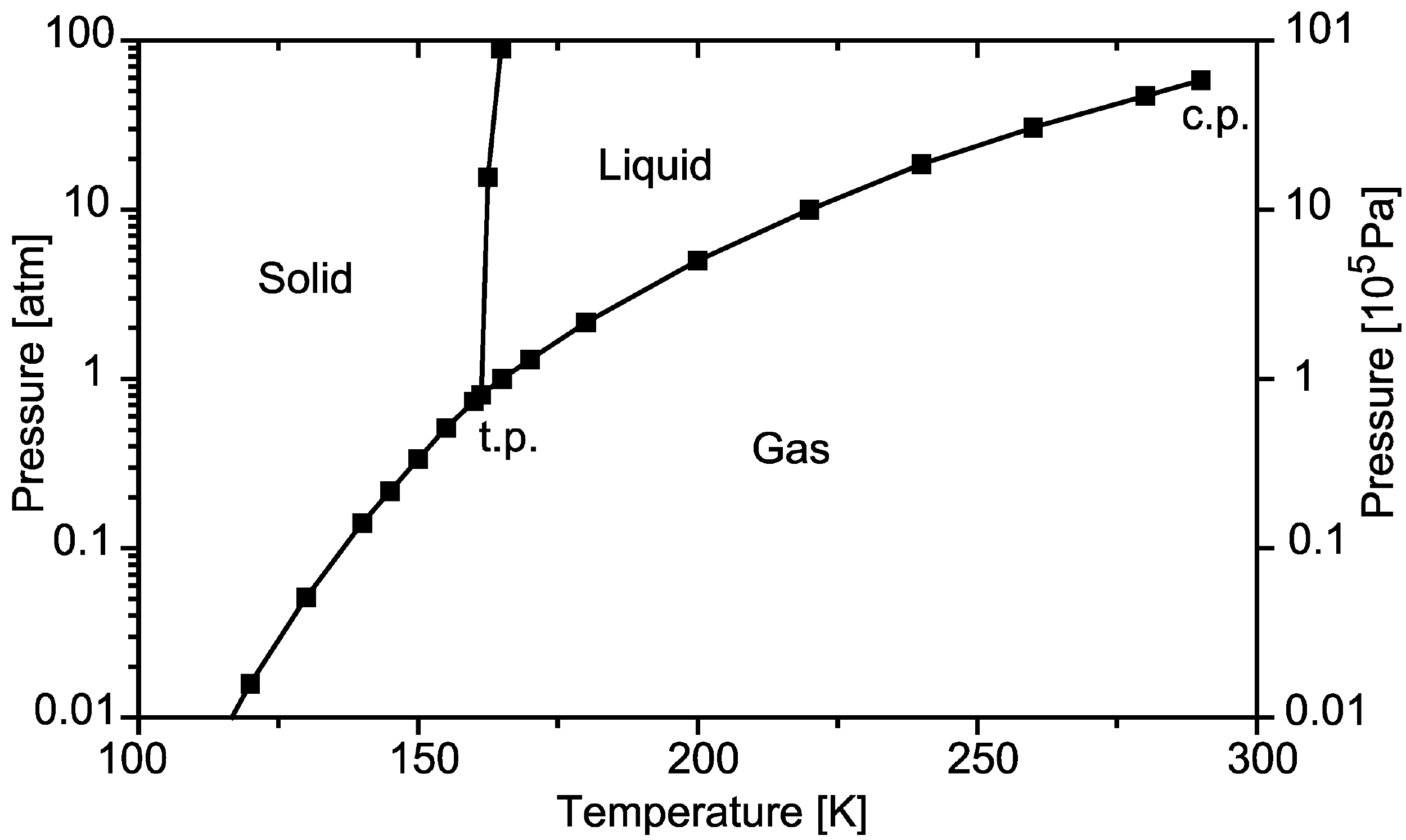}
\par\end{centering}
\caption[Phase diagram of xenon]{Phase diagram of xenon. Triple point (t.p.): 161.3 K, 81.6 kPa. Critical
point (c.p.): 289.7 K, 5.88 MPa. Unit conversions: 1 atm = 101,325
Pa; 0 C = 273.15 K. Figure from~\cite{CHERUBINI20031}.\label{fig:Phase-diagram-xenon}}

\end{figure}

\textcolor{red}{Monitor the cooling rate and pressure and ensure that
the transfer gas in the detector does not freeze by maintaining the
pressure above the triple point of xenon (0.8 bar - see Figure}~\textcolor{red}{\ref{fig:Phase-diagram-xenon}).
Exceeding 2.0 barg in the detector will result in the rupture of a
burst disc BD07. }

\subsection{Starting the pulse tube refrigerator }

It is not advised to fill a detector that does not have a functioning
heater. Check both heaters attached at the bottom of the detector
by turning them on briefly in the Slow Control and seeing the temperature
on RTD3 and RTD4 rise.
\begin{enumerate}
\item Ensure that acceptable vacuum in the OV has been achieved ($1\times10^{-4}$~mbar
for cooling, and $1\times10^{-5}$ or better for HV operations). 
\begin{enumerate}
\item This can be read off from the \textquotedblleft System vacuum\textquotedblright{}
in the Slow Control. 
\end{enumerate}
\item Make sure helium lines are between the Pulse Tube Refrigerator (PTR)
and the compressor. Make sure that cooling lines are plugged to the
compressor and the water supply and return. 
\item Ensure helium pressure in compressor is at $200\pm5$~psig by checking
the gauge on the front of the compressor. If the pressure is lower/higher
top up/relieve helium. Procedure for this is included in the compressor
manual. 
\item Turn the water supply and return fully on. 
\item Plug in the compressor into the mains. 
\item Turn on the compressor.
\end{enumerate}

\subsection{Xenon condensing }
\begin{figure}
\begin{centering}
\includegraphics[scale=0.5]{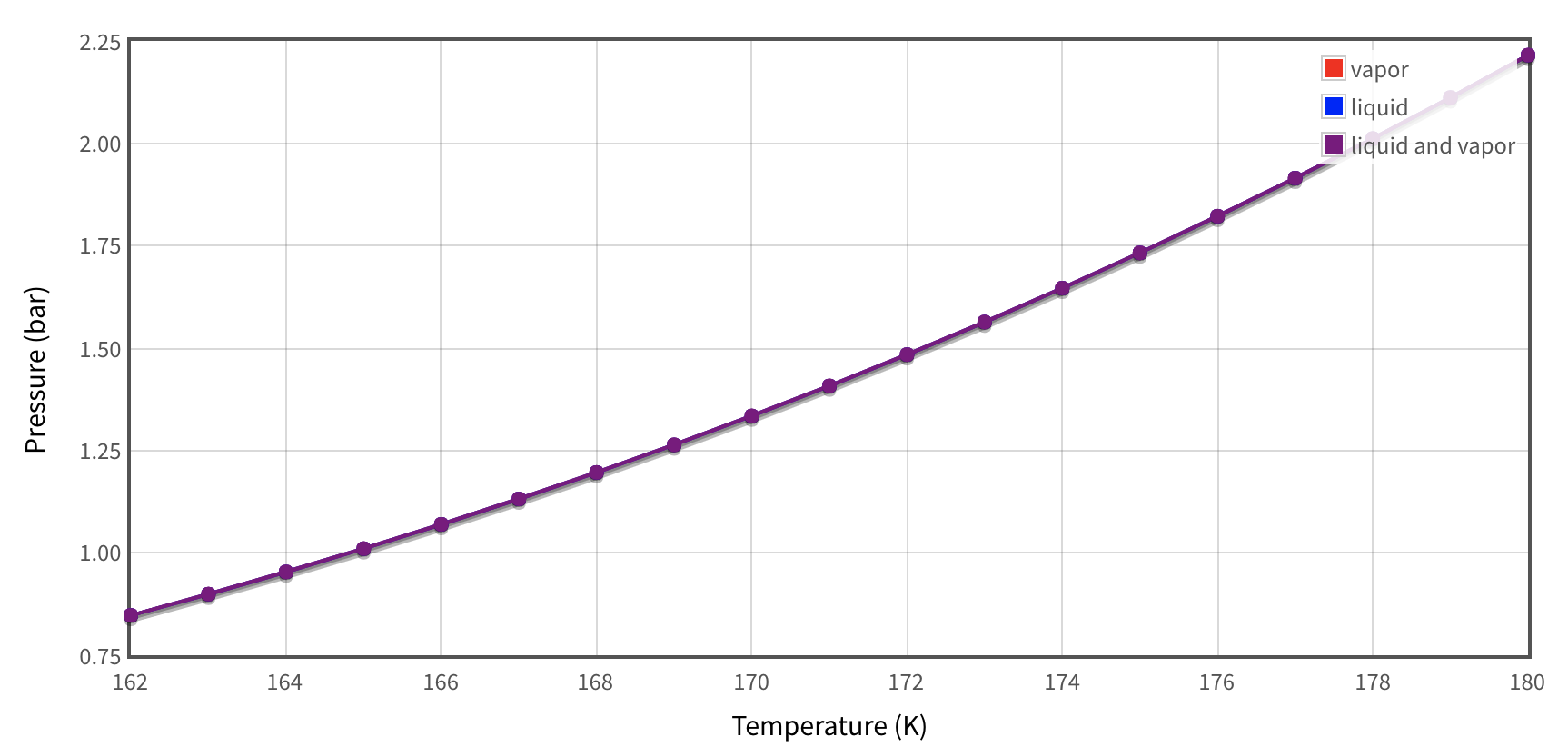}
\par\end{centering}
\caption[Xenon vapor curve.]{Xenon vapor curve. Figure from~\cite{NIST}.\label{fig:Xenon-vapor-curve}}

\end{figure}

\begin{enumerate}
\item Follow the procedure for opening the xenon fill path from the bottle
(Section~\ref{subsec:Opening-xenon-fill-path}). 
\item Set RG01 to 0.5~barg. 
\begin{enumerate}
\item To check the settings of the regulator, ensure that MV11, MV14, MV16,
and MV18 are closed. 
\item Temporarily close MV06. Carefully monitor PT03, PT04, and PT05 to
not over-pressurized the detector. 
\item Open RG01 to the desired pressure by observing PT02. 
\item Reopen MV06. 
\end{enumerate}
\item Monitor the temperature of the heat exchanger. For a given pressure
in the detector, use the vapor curve to calculate the condensation
temperature. The rate of cooling of the detector will increase once
an initial amount of xenon has condensed. 
\item The thermometers on the PTR and heat exchanger will be the first to
register cooling. These are: 
\begin{enumerate}
\item RTD1 \& RTD2 on heat exchanger 
\item RTD6 on the PTR.
\item The rest of the RTDs will follow over time. 
\end{enumerate}
\item Expect a total cooling time of the metal bulk of several hours (see
Section~\ref{sec:Calculations} for details). 
\item The condensation phase begins once enough of the detector reaches
the vapor curve. 
\begin{enumerate}
\item Find the condensation temperature from the vapor curve of xenon in
Figure~\ref{fig:Xenon-vapor-curve} (-101~C at 0.5~barg).
\end{enumerate}
\item It is important to keep the pressure well above 0.8~bar (see Figure~\ref{fig:Phase-diagram-xenon})
at the beginning of the condensation phase to prevent freezing. The
open regulator RG01 should prevent this. 
\item Every few hours, determine the amount of gas condensed in the detector
and log it in SciNote. Based on this calculation one can extrapolate
expected fill time. 
\begin{enumerate}
\item Use the scale SC01 to determine the amount of xenon that has already
condensed in the detector. 
\item Once the detector is nearly full, liquid level should be visible through
the detector window. 
\end{enumerate}
\item Assuming the condensation proceeds as expected one should see the
level sensor changing value. When the capacitance value has stopped
increasing, condense an additional 0.5~l of xenon, then proceed to
start circulation. This guarantees that the liquid exit is submerged. 
\item Close xenon filling path from the gas bottle (Section~\ref{subsec:Closing-xenon-fill-path}).
\end{enumerate}

\section{Circulation start}

At this point, the detector is filled with liquid, and the evaporator
should be at a similar temperature to the condenser. The purge lines
should be opened when using the PMT or the purity monitor.

\textcolor{red}{Monitor the detector pressure (PT05) and ensure it
does not drop below 0.8}~\textcolor{red}{bar by adjusting the heater
in the Slow Control. Exceeding 2.0 barg (29 psig) in the detector
will result in the rupture of a burst disc BD07. }
\begin{enumerate}
\item Close MV01, MV02, RG01, MV09, and MV14 through MV19. 
\item Open purge valves MV14, MV16, and MV18. 
\item Open MV06, MV08, MV10, and MV11. 
\item Close the pump bypass valve MV07. 
\item Start the pump by slowly increasing the pump speed by 1~Hz until
FM01 reads 2~SLPM. 
\item Note the pressure difference between PT03 and PT04. The pressure difference
at the pump outlet (PT04) should not be more than 5~times as large
as the pressure at the pump inlet (PT03). 
\begin{enumerate}
\item Note: Running the circulation pump fast will add a heat burden. 
\end{enumerate}
\item Slowly increase the pump speed and observe the pressures and temperatures
throughout the circulation path. If the circulation is going too fast,
a stream of bubbles will be visible in the detector at the location
of xenon inlet. This means that xenon is not being liquefied inside
the heat exchanger. If the pump is going too slow, the pressure might
continue to decrease and cause the detector to start freezing xenon
because not enough heat is being introduced into the system. If you
can circulate faster than $\sim5$~SLPM while maintaining stable
pressure, you have established an efficient heat exchange. 
\begin{enumerate}
\item The settings are likely to differ slightly for each cool down, based
on the amount of xenon condensed to the detector. In 2018 stable circulation
was achieved with the pump at $\unit[8-10]{Hz}$, circulation speeds
of $\unit[\sim20]{SLPM}$, and the PTR heater at 100\% while circulating
through the getter. 
\end{enumerate}
\item Set up the PTR heater on its seesaw setting and circulate happily
ever after. 
\begin{enumerate}
\item If the detector pressure keeps dropping, the bottom heaters can be
turned on to help raise detector pressure. 
\item It usually takes 12-24 hours for the detector to settle into a stable
state. Many adjustments may be necessary in the interim. 
\end{enumerate}
\item Once the detector appears stable, adjust the circulation to the desired
flow rate. 
\begin{enumerate}
\item Note: Faster circulation speeds achieve less efficient thermal exchange
in the heat exchanger.
\item Note: The recommended circulation speed by the getter to achieve efficient
removal of impurities is 15 SLPM. However, the getter's internal temperature
has been increased since xenon has a lower thermal conductivity so
optimal purifying speeds are likely around (or slightly below) this
value.
\end{enumerate}
\item Set alarms as outlined in Section~\ref{subsec:Alarm-settings}. 
\item To collect HV breakdown data follow Section~\ref{sec:Data-taking}. 
\end{enumerate}

\subsection{Slow control settings }

\subsubsection{Settings for stable circulations }
\begin{itemize}
\item Pump between 6-10 Hz (the flow rate on FM01 should be 20-25~SLPM). 
\item PTR Heater cycle is on with pressure boundaries on PT05 set with 0.01~bar
pressure difference. 
\end{itemize}

\subsubsection{\textsc{LabView} interruption}

If \textsc{LabView} is stopped during detector operations (which can
happen if Windows forces a restart to install updates), the pump and
getter will also shut off, but otherwise stopping \textsc{LabView}
alone should not change detector operations. Turning off the PLC is
strongly discouraged during detector operations. 

After \textsc{LabView} code is restarted ensure that: 
\begin{enumerate}
\item The getter is not bypassed. 
\begin{enumerate}
\item The correct setting for a getter in operation is illustrated in Figure
\ref{fig:Getter-settings}.
\end{enumerate}
\item The pump and PTR heater cycle are on correct settings and operational. 
\item Turn slow data logging back on and ensure email alarm forwarding is
on. 
\item Check back in a few hours to make sure the detector is in a stable
state. 
\end{enumerate}

\subsubsection{Alarm settings during circulation\label{subsec:Alarm-settings}}

This list of alarms assumes that the detector is operating at 1.0~barg.
The alarms should be modified during condensation or recovery as needed.
Low and high alarms only trigger an email notification. Low low and
high high pressure alarms trigger email notification and interlocks.
In general, for setting high alarms, the user should find the nearest
burst disc to inform the threshold selection. High high alarms should
be at least $20\%$ below the nearest burst disc rating to avoid a
rupture. 

\begin{table}
\begin{centering}
\begin{tabular}{l>{\raggedleft}p{1.2cm}>{\raggedleft}p{1.2cm}>{\raggedleft}p{1.2cm}>{\raggedleft}p{1.2cm}>{\raggedleft}p{1.2cm}>{\raggedleft}p{1.2cm}}
\hline 
 & PT01 {[}barg{]} & PT02 {[}barg{]} & PT03 {[}barg{]} & PT04 {[}barg{]} & PT05 {[}barg{]} & PT07 {[}psi{]}\tabularnewline
\hline 
\hline 
Low low & N/A & 0.1 & 0.1 & 0.1 & 0.1 & N/A\tabularnewline
Low & 0.1 & 0.7 & 0.3 & 0.7 & 0.7 & 200\tabularnewline
High & 10.0 & 1.3 & 1.3 & 3.0 & 1.3 & N/A\tabularnewline
High high & N/A & 1.8 & 1.8 & 5.0 & 1.8 & N/A\tabularnewline
\hline 
\end{tabular}
\par\end{centering}
\bigskip{}
\begin{centering}
\begin{tabular}{l>{\raggedleft}p{1.5cm}>{\raggedleft}p{2.8cm}>{\raggedleft}p{2.8cm}}
\hline 
 & FM01 {[}SLPM{]} & System vacuum {[}mbar{]} & HV vacuum {[}mbar{]}\tabularnewline
\hline 
\hline 
Low & 0 & N/A & N/A\tabularnewline
High & 30 & $1\times10^{-4}$ & $1\times10^{-5}$\tabularnewline
\hline 
\end{tabular}
\par\end{centering}
\caption[Alarm settings for XeBrA detector operations while circulating]{Alarm settings for detector operations while circulating. PT03 is
at the pump inlet, so it reaches lower pressures than the rest of
the system. Similarly, PT04 is at the pressure outlet and can reach
higher pressures than the rest of the system. PT07 is installed on
the compressed nitrogen cylinder used for pneumatic valve actuation.
The low alarm notifies the user to purchase and install a new cylinder.
FM01 is rated to 50 SLPM. Since the thermal coupling of the RTDs to
the system is not optimal, the temperature set points need to be adjusted
after each fill. }

\end{table}

\section{Data taking\label{sec:Data-taking}}

Once the detector is stable (usually about one day after completing
the fill), one can start taking HV data. This requires suppressing
bubbles, which can be achieved by adding xenon into the detector.
Assume that target detector pressure for data taking is 1.0 barg.

\textcolor{red}{Monitor the detector temperatures to avoid excessive
xenon ice formation in the detector. }
\begin{enumerate}
\item Prior to data taking, keep the detector $\sim0.3$ barg below target
pressure for $\sim12$~hours. 
\item Slow down the circulation slightly. 
\item With the detector circulating, ensure RG01 is closed. 
\item Open MV01 and MV02 to open path of xenon to the detector. 
\item Slowly open RG01. The pressure will rise fast at first until the target
pressure of 1.0~barg is reached. 
\begin{enumerate}
\item Monitor temperature in the detector throughout data taking. If the
temperature of the PTR and the heat exchanger start dropping dramatically,
ice has started forming in the heat exchanger. Stop data taking and
carefully begin warming the heat exchanger using the attached heater. 
\item If a small amount of ice starts forming at the bottom of the cone,
but not in the heat exchanger, this is okay. This might enable stable
data taking without the need to add xenon. 
\end{enumerate}
\item This creates a stable, bubble-free detector for HV data taking. 
\end{enumerate}
To start taking HV data: 
\begin{enumerate}
\item Zero electrodes and note the distance on the caliper. 
\item Set electrodes to the desired separation. 
\item Ensure the PMT is off. 
\item Turn off purity monitor\textquoteright s charge amp and ensure surge
protector is grounded. 
\item Disconnect the capacitor by unplugging the banana plugs. 
\item Ensure the top electrode is grounded. 
\begin{enumerate}
\item Either plug a shorted cable to the signal SHV channel,
\item or plug in the charge sensing box. 
\end{enumerate}
\item Start fast data logging in Slow Control.
\end{enumerate}
Ensure the detector is bubble free throughout the data taking and
that the spark is occurring between the two electrodes. 

\section{Xenon recovery\label{sec:Xenon-recovery}}

This procedure empties all xenon from the entire system. First, the
recovery vessels are cooled down to below xenon freezing point, with
heaters on the storage bottle's valve to prevent valve damage. Then,
the gas system is prepared for recovery before pulling all the xenon
from XeBrA into one of the aluminum storage cylinders.

Ensure that the ODH monitor in the room is working and wear a portable
ODH monitor. Expect to take a full day to recover the entire detector
volume ($\sim10$ hours). Never walk away from the detector until
you are certain all the xenon is in the storage bottle and that bottle
is closed.

\textcolor{red}{Note the start and end of detector warm up in SciNote.
Always have nitrogen in the transfer dewar throughout this procedure.
A single 160 l transfer dewar should be sufficient for one recovery. }
\begin{enumerate}
\item Slowly turn off the pump SP01 by stepping down the frequency 1~Hz
at the time.. 
\item Turn off the getter and set to bypass using the buttons on the front
of the getter housing. 
\item Monitor oxygen levels during the liquid nitrogen (LN) fill and turn
on the big fan to encourage air circulation. 
\item Figure out which bottle you want to use for recovery (SB01 or SB02)
and make sure it contains either vacuum or xenon. This procedure assumes
recovery into SB01. Fill the dewar that contains your bottle of choice
with LN about half way and wait for the temperature to equalize. This
will take 0.5-1 hour depending on the quantity of xenon left in the
aluminum bottle. 
\begin{enumerate}
\item MV01 contains elastomer seals so it is important to find the Goldilocks
equilibrium. To prevent valve damage a heater or a heating tape should
be installed on the top of SB01 and around MV01. It is also important
to not overheat the valve.
\item Once you are sure that SB01 is cold enough (nitrogen is no longer
violently boiling) you can start recovery. 
\end{enumerate}
\item Open MV06, MV07, MV08, MV09, MV10, MV11, and MV14 through MV 19. 
\item Close RG01, MV03, and MV04. 
\item Open MV01 and MV02. This will open up the path to SB01. 
\item Check PT01 to verify that SB01 is at negligible pressure before opening
MV05. 
\item Open MV05. This opens the path from detector to SB01 and xenon should
start freezing in the bottle. 
\item Set up high alarms for the bottom cone heaters in Slow Control. 
\item Turn on the heater at the bottom of the cone. Monitor the thermometers
and ensure the detector does not get too hot once the xenon has evaporated
- the alarms should trigger an interlock turning off the heaters once
the threshold is exceeded. 
\item Ensure the pressure does not go below 0.1~barg during recovery to
prevent xenon contamination. 
\item Once you are sure the pressure is falling steadily, turn off PTR and
its water cooling. 
\item After a while, you can turn off the turbo and later also the scroll
pump to increase the heat load on the detector. 
\begin{enumerate}
\item If you want to increase the heat load, introduce a small amount of
nitrogen gas into the OV. Avoid exposing this volume to air since
that would cause water condensation on surfaces. 
\end{enumerate}
\item Keep topping up the dewar with liquid nitrogen until you are confident
that all the xenon has been recovered into SB01. 
\item Close MV01. Do not close MV01 if it is too hot (too hot to touch)
or too cold (frosted) to prevent damaging the valve.
\item Once all liquid nitrogen has evaporated, log the final weight of SB01
in SciNote. 
\end{enumerate}

\subsection{Emergency recovery}

If the detector pressure exceeds the pressure set by an alarm on PT05,
emergency xenon recovery will trigger. In this case, PV04 will open
automatically, thus allowing xenon to enter vessels SB03 (Rhino) and
SB04 (Hippo). PV04 will stay open while the pressure is above the
defined threshold. Once the pressure falls below the alarmed threshold,
PV04 will close automatically.

If there is xenon in SB03 and SB04, it then needs to be recovered
via cryopumping at the end of a run. To do that, once step 16 in Section~\ref{sec:Xenon-recovery}
is achieved (all the xenon from the main detector volume has been
recovered): 
\begin{enumerate}
\item Close off the detector volume by closing MV10, MV11, MV14, MV16, and
MV16. 
\item Open PV04. 
\begin{enumerate}
\item PV04 can be opened in Slow Control. Password: XeBrA 
\end{enumerate}
\item Wait until all the xenon has been recovered. Then proceed to Step
17 in Section~\ref{sec:Xenon-recovery}.
\end{enumerate}

\section{Calculations\label{sec:Calculations}}

The calculations below serve only to find back-of-the-envelope numbers
so should be treated appropriately. 

\subsection{Heat needed to cool down detector}

Assume all stainless steel and the room-temp specific heat is constant
to the liquid xenon temp. This calculation does not assume the presence
of xenon gas in the detector, which is expected to improve the thermal
conductivity, so the time calculated here is an upper limit. 
\begin{itemize}
\item XeBrA\textquoteright s mass is $\sim50$~kg
\item Thermal conductivity decreases with temperature; the following calculation
uses specific heat of stainless steel at room temp of $c_{V}=\unit[490]{J/kg\cdot K}$
\item Temperatures: $T_{room}=\unit[293]{K}$ , $T_{LXe}=\unit[165]{K}$ 
\end{itemize}
Heat $Q$ needed to cool down the detector to 165 K assuming it is
made entirely of stainless steel 
\[
Q=c_{V}\cdot m\cdot dT=\unit[3.1]{MJ}.
\]
The cooling power $P$ of PTR is 40 W at 80 K. Therefore, the time
needed to cool down the detector
\[
T=\frac{Q}{P}=\unit[22]{hours}.
\]

\subsection{Estimated time to condense xenon}

This calculation considers the xenon gas introduced to the detector
and estimates the heat needed first to cool it to the vapor curve
and then condense it.
\begin{itemize}
\item Volume $V_{LXe}=\unit[5.6]{l}$
\item Density at 3 bar $\rho_{LXe}=\unit[2.94]{kg/l}$, resulting in mass
$m_{LXe}=\unit[16.5]{kg}$
\item Specific heat $c_{V_{LXe}}=\unit[97]{J/kg\cdot K}$
\item Latent heat $L_{LXe}=\unit[12.64]{kJ/mol}$
\item Molar mass of LXe is 131.3 g/mol
\end{itemize}
The heat needed to cool gaseous xenon down to 165~K is
\[
Q=c_{V}\cdot m\cdot dT=\unit[205]{kJ}.
\]
The latent heat of vaporization for LXe is
\[
Q=m\cdot L=\unit[1588]{kJ}.
\]
Therefore, the total heat needed is 1793~kJ, which results in a time
needed to cool and condense the xenon of 12.3~hours.

\subsection{Power radiated from XeBrA\label{subsec:Power-radiated-xebra}}

While the detector is filled with liquid xenon, it will absorb and
emit light (heat) from the environment. The power absorbed by the
detector is 
\begin{align*}
P & =A\sigma\epsilon\left(T^{4}-T_{0}^{4}\right)\\
 & =\unit[86.2]{W}
\end{align*}
where the Stefan-Boltzmann constant, $\sigma=\unit[5.67\times10^{-8}]{W/\left(m^{2}\cdot K^{4}\right)}$,
describes the power radiated from a black body in terms of its temperature,
$A=\unit[1.6]{m^{2}}$ is the area of the active detector volume,
and the emissivity $\epsilon$ of polished 304 stainless steel\footnote{A gray body does not absorb all incident radiation (and emits less
total energy than a black body) and is characterized by an emissivity
$\epsilon<1$.} at 87~K is 0.13. The temperature $T=\unit[293]{K}$ corresponds
to the radiation absorbed while $T_{0}=\unit[165]{K}$ corresponds
to the radiation emitted. 

To improve the cooling power available to the detector, the power
radiated by the detector can be decreased if the active volume is
wrapped in several layers of superinsulation. This helps to reduce
the heat absorbed by the detector because the outgoing radiation is
much weaker compared to the incoming radiation. Reference~\cite{baudouy2015heat}
says that under a good vacuum, a typical value of the heat transfer
for a non-compressed 20-layer multi-layer insulation system is $\sim\unit[1-3]{W/m^{2}}$
from $80-300$ K and $<\unit[100]{mW/m^{2}}$ between $4-80$ K. XeBrA
was wrapped in $\sim20$ layers of superinsulation so we can expect
a reduction of the radiated heat by a few watts.

\chapter{\textsc{XeBrA parts list}\label{chap:Parts-list}}

This appendix lists selected parts used in the gas system and XeBrA
as described in Chapter~\ref{chap:XeBrA}.

\begin{table}[h]
\begin{centering}
\begin{tabular}{lll}
\hline 
Part & Manufacturer & Model\tabularnewline
\hline 
\hline 
Spherical square vacuum chamber & Kimball physics & MCF600-SphSq-F2E4A8\tabularnewline
Linear shifter & Kurt J. Lesker & LSM38-200-H-DL\tabularnewline
HV power supply & Spellman & SL100N10/CMS/LL20\tabularnewline
HV cable  & Dielectric Sciences & 2121\tabularnewline
Vacuum-xenon feedthrough & CeramTec & 6722-01-CF\tabularnewline
Air-vacuum feedthrough & Isolation Products & D-102-10\tabularnewline
Pulse tube refrigerator & Cryomech & PT805\tabularnewline
Compressor & Cryomech & CP950\tabularnewline
Current monitor & Pearson electronics & 110A\tabularnewline
Picoammeter & Keithley & 6487\tabularnewline
Capacitance meter & BK Precision & 890C\tabularnewline
Vacuum gauges & Leybold & Ionivac ITR 90\tabularnewline
Pressure relief valve & Accu-Glass & 113160\tabularnewline
Power supply for purity monitor & Stanford Research Systems & PS350\tabularnewline
Oscilloscope & Tektronix & TDS5034B\tabularnewline
Xenon flash lamp & Hamamatsu & L9455-11\tabularnewline
Argon PMT & Hamamatsu & R8520-06\tabularnewline
Xenon PMT & Hamamatsu & R9228\tabularnewline
Pressure recording film (2-20 psi) & McMaster Carr & 3145K22\tabularnewline
\hline 
\end{tabular}
\par\end{centering}
\caption[List of parts used in XeBrA]{List of selected parts used in XeBrA. }
\end{table}

\begin{table}
\begin{centering}
\begin{tabular}{>{\raggedright}p{2cm}>{\raggedright}p{4cm}l>{\raggedright}p{4cm}}
\hline 
P\&ID label  & Part & Manufacturer & Model\tabularnewline
\hline 
\hline 
SP01 & Scroll circulation pump  & Air Squared & P15H22N4.25\tabularnewline
FT01 & Filter & Swagelok & SS-SCF3-VR4-P-225\tabularnewline
CZ01 & Check valve & Swagelok & SS-4CA-VCR-50\tabularnewline
RG01 & Regulator & APTech & AP1202S-2PW-FV4-FV4\tabularnewline
FR01-03 & Flow restrictor & Mott Corp & 5140-1/4-SS-50\tabularnewline
FM01 & Flow meter & Aalborg & GFM37S-VCL6-A0\tabularnewline
PV04 & Pneumatic valve & Swagelok & 6LVV-DPFR4-P-C\tabularnewline
MV02, MV04-05 & Manual high pressure valve & Swagelok & 6LVV-DPHFR4-P\tabularnewline
MV06-08, MV10-20 & Manual valve & Swagelok & 6LVV-DPFR4-P\tabularnewline
PI02-03, PI05-06 & Pressure gauge & Swagelok & PGU-50-PC60-C-4FSF\tabularnewline
PT02-03, PT05-06 & Pressure transducer & GE & PTX5072-TA-A1-CA-H0-RA\tabularnewline
BD07-08 & Burst disc & Continental Disc & CD30669A0126\tabularnewline
SC01, SC02 & Scale & Adam & CPWplus 200M\tabularnewline
SB01, SB02 & Aluminum cylinder & Evergreen & I-265-580\tabularnewline
SB03, SB04 & Xenon recovery vessel (i.e., a propane tank) & Manchester Tank & 1499TC.11HL\tabularnewline
 & Open top nitrogen dewar & Cryofab & CF1636\tabularnewline
 & Vacuum gauges & Pfeiffer Vacuum & TPG361\tabularnewline
 & Getter & SAES Pure Gas & PF4-C15-R-1\tabularnewline
 & High current GFCI\nomenclature{GFCI}{Ground Fault Current Interrupter}
for getter & North Shore Safety & PGFI-1311N-138\tabularnewline
 & PUR-Gas\texttrademark{} In-Line Purifiers & Matheson Tri-Gas & SEQPURILOMT1\tabularnewline
 & Permanent oxygen monitor & Alpha Omega & Series 1300\tabularnewline
\hline 
\end{tabular}
\par\end{centering}
\caption[List of parts used in the gas system]{List of parts used in the gas system. The P\&ID is shown in Figure
\ref{fig:XeBrA's-PID.}. If various models of the part are used for
different pressure conditions, only one of the models is listed in
the table. GFCI stands for Ground Fault Current Interrupter.}

\end{table}

\backmatter
\singlespacing

\bibliographystyle{JHEP}
\addcontentsline{toc}{chapter}{\bibname}\bibliography{LUX/references_thesis_LUX,equity_inclusion/EI_references_thesis,sub_GeV/references_subGeV_thesis,LZ_purity_monitor/references_thesis_purityMonitor,xebra/references_thesis_XeBrA,run4_fields/references_thesis_eFields,intro/references_thesis_intro}

\end{document}